\documentclass{book}

\usepackage[swapnames,norules,nouppercase]{frontespizio}

\usepackage{amsmath}
\usepackage{amssymb}

\usepackage[normalem]{ulem}

\usepackage{bm}
\usepackage{graphicx}

\usepackage{tikz}
\usetikzlibrary{positioning}

\usepackage{multirow}

\usepackage{array}

\usepackage[nottoc,notlot,notlof]{tocbibind}

\usepackage{hyperref}
\usepackage{multibib}
\newcites{RF}{Publications}

\usepackage[T1]{fontenc}

\allowdisplaybreaks

\begin{document}

\begin{frontespizio}
\Preambolo{\renewcommand{\fronttitlefont}{\fontsize{24}{24}\bfseries}}

\Margini{4cm}{3cm}{3cm}{3cm}				
\Logo[3cm]{Uniroma1.pdf}
\Istituzione{Sapienza University of Rome}
\Divisione{Physics Department}		
\Scuola{PhD in Physics}
\Titoletto{PhD Thesis}
\Titolo{Nonrelativistic limit of QFT in curved spacetime}
\Punteggiatura{}					
\NCandidato{Candidate}					
\Preambolo{\renewcommand{\frontsmallfont}[1]{\small}}
\Candidato[]{Riccardo Falcone}
\NRelatore{Thesis Advisor}{Relatori}			
\Relatore{Prof. Claudio Conti}
\Piede{Academic Year 2020-2023 (XXXVI cycle)}					
\end{frontespizio}

\frontmatter

\chapter*{Abstract}

The formalism of nonrelativistic quantum physics was originally considered in the context of inertial frames. Here, we report on a more general framework that includes noninertial frames and arbitrarily strong gravitational fields. We derive from first principles the nonrelativistic limit of quantum fields in curved spacetime.

Unique features and subtleties of the fully covariant theory in curved spacetime affect nonrelativistic quantum systems in accelerated frames when the acceleration is sufficiently high. This includes the frame dependent notion of particles, energies, vacuum states and nonrelativistic conditions. By using the algebraic approach to quantum field theory, we detail these effects in the noninertial nonrelativistic regime.

The theoretical framework developed here gives predictions about the phenomenology of nonrelativistic quantum systems that are put in noninertial motion. As an application, we derive the realistic model of accelerated nonrelativistic atoms by using Dirac field theory in Rindler spacetime. We report on the experimental constraints for the indirect observation of the Unruh effect by means of atomic detectors via hyperfine structure.

Then, we address the problem of localization of quantum states and observables in inertial and accelerated frames. We show how the Born probabilistic notion emerges in both cases as a consequence of the respective nonrelativistic limit. As a result, we find that nonrelativistic states can be described in terms of wave functions that quantify the probability to find particles in each space position. Also, we report on nonlocal effects that originate from the frame dependent nature of vacuum states and nonrelativistic conditions in the quantum field theory in curved spacetime.

\tableofcontents

\mainmatter

\chapter*{Notation and conventions}
\addcontentsline{toc}{part}{Notation and conventions}
\markboth{NOTATION AND CONVENTIONS}{NOTATION AND CONVENTIONS}

Throughout the thesis we follow the following notation and conventions.

The signature of metrics is $(-,+,+,+)$. We adopt Einstein notation over repeated indices. Greek indices $\mu, \nu, \rho, \sigma$ are for $4$ dimensional spacetime coordinates $(0,1,2,3)$, while Latin indices $i, j, k$ for $3$ dimensional space coordinates $(1,2,3)$. The index $\alpha$ is used for spinorial degrees of freedom. The other letters $l, m, n, \dots$ are used as bare indices and do not follow the Einstein rule of repeated indices. Only in the case of 1+1 dimensional spacetimes, the letters $i, j, k$ are used as summation indices as well.

$\delta(t-t')$ and $\delta^3(\vec{x}-\vec{x}')$ are Dirac deltas, while $\delta^{ij} = \delta_{ij}$ is the Kronecker delta. We use generalized deltas $\delta_{\theta \theta'}$ as well when the variables $\theta$ and $\theta'$ include continuous and discrete degrees of freedom. Also, $\theta(t)$ is the Heaviside step function.

\section*{Fundamental constants}

\begin{tabular}{l l}

$c$ & speed of light\\

$\hbar$ & reduced Planck constant\\

$k_\text{B}$ & Boltzmann constant

\end{tabular}

\section*{Acronyms}

\begin{tabular}{l l}

QFT & Quantum Field Theory\\

QFTCS & Quantum Field Theory in Curved Spacetime\\

RQM & Relativistic Quantum Mechanics\\

NRQM & NonRelativistic Quantum Mechanics\\

NRQFTCS & NonRelativistic limit of QFT in Curved Spacetime\\

AQFT & Algebraic Quantum Field Theory\\

GR & General Relativistic

\end{tabular}

\section*{Abbreviations}

\begin{tabular}{l l}

Eq. & Equation\\

Fig. & Figure\\

Chap. & Chapter\\

Sec. & Section or subSection

\end{tabular}

\chapter*{Introduction}
\addcontentsline{toc}{part}{Introduction}
\section*{State of the art and objectives}
\addcontentsline{toc}{section}{State of the art and objectives}
\markboth{INTRODUCTION}{STATE OF THE ART AND OBJECTIVES}

NonRelativistic Quantum Mechanics (NRQM) is typically defined in terms of the following postulates.
\begin{enumerate}
\item Single particle states are identified by wave functions $\psi(\vec{x})$ as elements of the Hilbert space $\mathbb{C}^n \otimes L^2(\mathbb{R}^3)$. \label{NRQM_first}
\item Due to the Born interpretation of quantum mechanics, $|\psi(\vec{x})|^2$ gives the probability to find the particle in the position $\vec{x}$, which means that states are localized in the support of their wave functions. \label{NRQM_middle}
\item The theory is supplied with a notion of time evolution by means of a Hamiltonian operator $\hat{H}$. The time evolved wave function $\psi(t,\vec{x})$ is solution of the Schrödinger equation $i \hbar \partial_0 \psi(t,\vec{x}) = H \psi(t,\vec{x})$, where $H$ is the representative of $\hat{H}$ in $\mathbb{C}^n \otimes L^2(\mathbb{R}^3)$.\label{NRQM_last}
\end{enumerate}

NRQM is regarded as the limit of a more fundamental theory, namely Quantum Field Theory (QFT). At variance with NRQM, QFT is consistent with special relativity; however, the postulates \ref{NRQM_first}-\ref{NRQM_last} do not apply there. In particular, single particles are defined as positive frequency modes of the field equation and not as elements of $\mathbb{C}^n \otimes L^2(\mathbb{R}^3)$ \cite{Wald:1995yp}. The corresponding modal wave functions are not associated to the probability to find particles in defined positions. Also, the notion of time evolution is given by the hyperbolic field equation (e.g., Klein-Gordon or Dirac equation) which can be second order in the time coordinate. The NRQM postulates \ref{NRQM_first}-\ref{NRQM_last} only emerge in the nonrelativistic regime of QFT as an approximation of the theory \cite{srednicki_2007, zee2010quantum, schwartz_2013}.

QFT describes relativistic quantum phenomena in the Minkowski frame of a flat spacetime. More general coordinate frames or curved spacetimes are, instead, considered in the so-called Quantum Field Theory in Curved Spacetime (QFTCS) \cite{Wald:1995yp}. QFTCS is a well-established theory, which led to conceptual results and phenomenological predictions, including the frame-dependent content of particles and the consequent frame-dependent notion of vacuum state, which is at the origin of the Hawking \cite{Hawking:1975vcx} and the Unruh effect \cite{PhysRevD.7.2850, Davies:1974th, PhysRevD.14.870}.

Here, we focus on QFT in Rindler spacetime describing the frame of an accelerated observer. One of the aims of the thesis is to show the existence of a regime in which postulates \ref{NRQM_first}-\ref{NRQM_last} hold for the accelerated observer. We define NonRelativistic limit of Quantum Field Theory in Curved Spacetime (NRQFTCS) as the restriction of QFTCS to states with energy very close to their mass. We demonstrate that, when such a nonrelativistic condition is satisfied, the postulates \ref{NRQM_first}-\ref{NRQM_last} hold.

An additional goal is to compare NRQFTCS with NRQM, defined as the nonrelativistic regime in Minkowski spacetime. Some unique features of QFTCS---such as the frame dependent notion of particles, energies and vacuum states---are expected to appear in the context of NRQFTCS and cannot be predicted by the sole postulates \ref{NRQM_first}-\ref{NRQM_last} of the nonrelativistic quantum theories. What are the consequences of such frame dependent effects in the noninertial quantum experiments? Do they produce measurable differences between the NRQM and the NRQFTCS regimes?

NRQFTCS represents the natural framework for the description of, e.g., nonrelativistic systems in gravitational field or particle detectors in noninertial motion. Remarkably, the regime is experimentally accessible with current technology. In particular, the weak gravitational effects produced by the Earth's gravity on nonrelativistic quantum systems was detected by experiments on ultracold neutrons \cite{article1, PhysRevD.67.102002, article2, article3, article4, PhysRevLett.112.071101, Kamiya:2014qia}. The ever-improving measurement accuracy and the theoretical interest in describing strong gravity effects in nonrelativistic quantum systems motivate the formulation of NRQFTCS from first principles.

As an example of nonrelativistic quantum system in noninertial motion, we consider an accelerated atomic detector \cite{PhysRevLett.91.243004, PhysRevA.74.023807, PhysRevLett.128.163603}. For a comprehensive description of the noninertial atom, the idealized nonrelativistic Unruh-DeWitt detector \cite{PhysRevD.14.870, hawking1980general} needs to be superseded by a more realistic model. An explicit description of the quantum system based on bound states of atomic fields---including nuclear, electronic and electromagnetic fields---in QFT and QFTCS is currently lacking; conversely, the nonrelativistic semiclassical atomic theory is well-established and gives measurable predictions in the NRQM regime. When the atom is accelerated, NRQFTCS becomes the only accessible framework for the description of the quantum system.

The nonrelativistic limit of quantum fields in Rindler spacetime unveils conceptual and experimental subtleties, such as the frame dependence of the nonrelativistic limit and particle number. Due to the different notion of time and energies---as generators of time translations---between the Minkowski and Rindler frame, the inertial and the accelerated observer generally do not agree about the nonrelativistic nature of states. Also, the frame dependent content of particles predicted by the algebraic approach to QFTCS \cite{Wald:1995yp, haag1992local} imply that the two observers do not agree about the number of particles representing the same state either. Consequently, if the atom is prepared in the inertial frame as a nonrelativistic system with a fixed number of electrons and nuclear particles, it appears as made of an indefinite number of particles with different energies in the accelerated frame. In this thesis, we show how to suppress such a frame-dependent effect and we derive the observational limits for the detection of the Unruh effect via first-quantized atomic detectors.

In the last part of the thesis, we address the problem of localization in QFT, NRQM, QFTCS and NRQFTCS. The notion of localization in QFT is characterized by the lack of independence between the preparation of states over the Minkowski vacuum $|0_\text{M} \rangle$ and the measurement of observables in disjoint regions of space \cite{Reeh:1961ujh, 10.1063/1.1703731, 10.1063/1.1703925, haag1992local}. This nonlocality originates from Bell-like quantum correlations of $|0_\text{M} \rangle$ between the local Hilbert spaces \cite{haag1992local, Redhead1995-REDMAA-2, PhysRevA.58.135} and does not violate causality \cite{Redhead1995-REDMAA-2, CLIFTON20011, VALENTE2014147, RevModPhys.90.045003}; however, it is incompatible with the notion of Born localization considered by the postulate \ref{NRQM_middle} of NRQM. Provably, in the nonrelativistic limit of QFT such a nonlocal effect is suppressed and the Born localization scheme emerges. This leads to the independence between the preparation of nonrelativistic states over $|0_\text{M} \rangle$ by an inertial experimenter (Alice) and the measurement of observables by an inertial experimenter (Bob) in disjoint regions. 

Here, we show that the same nonlocal effect seen for QFT in Minkowski spacetime is also present in QFTCS and is suppressed in the NRQFTCS regime. Hence, we find that the preparation of nonrelativistic states over the Rindler vacuum $|0_\text{L}, 0_\text{R} \rangle$ carried out by an accelerated experimenter (Rachel) do not influence measurements of nonrelativistic observables by another accelerated observer (Rob) operating in a different region.

The frame dependent notion of vacuum states and nonrelativistic conditions in QFTCS is crucial for the description of ``hybrid'' experiments. This includes Rachel-Rob scenarios with the Minkowski vacuum $|0_\text{M} \rangle$ as the background state and Alice-Rob scenarios over $|0_\text{M} \rangle$. In all of these cases, the aforementioned nonlocal effect is not suppressed. Specifically, we find that nonrelativistic measurements carried out by Rob are influenced by local preparation of nonrelativistic Rindler particles by Rachel over the Minkowski vacuum $|0_\text{M} \rangle$ or by local preparation of nonrelativistic Minkowski states by Alice. Such a result leads to measurable outcomes in the NRQFTCS regime that cannot be predicted by the sole postulates \ref{NRQM_first}-\ref{NRQM_last} of the nonrelativistic quantum theories and makes NRQFTCS conceptually and physically different from NRQM.

\section*{Structure of the thesis}
\addcontentsline{toc}{section}{Structure of the thesis}
\markboth{INTRODUCTION}{STRUCTURE OF THE THESIS}

The thesis is divided into three parts.

\begin{itemize}
\item Part \ref{Relativistic_and_nonrelativistic_quantum_fields} gives an introduction to the nonrelativistic limit of quantum fields and particles in flat and curved spacetimes.

\begin{itemize}
\item In Chap.~\ref{QFT_in_flat_and_curved_spacetimes}, we consider scalar and Dirac fields in Minkowski and Rindler spacetimes. By following the existent literature \cite{Wald:1995yp}, we show the representation of particle states as positive frequency modes of the corresponding field equation.

\item In Chap.~\ref{Non_relativistic_limit_of_QFT_and_QFTCS}, we discuss the behavior of such a modal representation in the nonrelativistic limit of each frame. By considering particles with energy very close to their mass, we show that postulates \ref{NRQM_first} and \ref{NRQM_last} are satisfied. Specifically, we find that the modal wave functions live in a $\mathbb{C}^n \otimes L^2(\mathbb{R}^3)$ Hilbert space---up to a metric dependent measure---and their dynamics is determined by a Schrödinger-like equation.
\end{itemize}

\item While Part \ref{Relativistic_and_nonrelativistic_quantum_fields} is devoted to the representation and the time evolution of nonrelativistic states, in Part \ref{Inequivalent_particle_representations_and_Unruh_effect} we focus on the notion of nonrelativistic observables by considering the algebra of creators and annihilators of nonrelativistic particles. We discuss the frame dependent content of particles in QFTCS and its consequences in the nonrelativistic regime for inertial and accelerated observers. 

\begin{itemize}
\item In Chap.~\ref{Frame_dependent_content_of_particles}, we follow the algebraic approach to QFTCS \cite{haag1992local} to show how particle representations of one frame can be related to the other. For both scalar and Dirac fields, we compute the Bogoliubov transformations relating Fock operators of one frame to the other and we derive the consequent representation of the Minkowski vacuum $| 0_\text{M} \rangle $ as an element of the Rindler-Fock space.

\item In Chap.~\ref{Minkowski_particles_in_accelerated_frame}, we provide a general procedure to compute the statistical operator representing any Minkowski-Fock state as seen by the accelerated observer. Also, we derive the explicit formulation of Minkowski particles in terms of Rindler-Wigner characteristic functions.

\item In Chap.~\ref{Framedependent_nonrelativistic_limit}, we discuss the frame dependent notion of nonrelativistic limit between the inertial and the accelerated observers. We show that the two experimenters do not always agree about the nonrelativistic nature of particles. We derive the condition in which such a frame dependent effect is suppressed by constraining the relative acceleration of the observers.

\item In Chap.~\ref{Accelerated_non_relativistic_detectors}, we study accelerated atomic detectors as a practical application for the NRQFTCS framework. By following the results of Chap.~\ref{Framedependent_nonrelativistic_limit}, we identify the regime in which a nonrelativistic atom prepared in the laboratory frame can be described as a first-quantized system in the comoving accelerated frame. We show the observational window for the detection of the Unruh effect by means of atomic hyperfine splitting.
\end{itemize}

\item Part \ref{Localization} is reserved to the problem of localization in the inertial and the accelerated frame.
\begin{itemize}
\item In Chap.~\ref{Localization_in_Quantum_Field_Theory}, we review the existing literature about the localization in QFT and its nonrelativistic limit. We follow the algebraic approach to QFT to define a fundamental notion of localization in the fully relativistic theory. Only in the nonrelativistic regime, the Born localization scheme---identified by postulate \ref{NRQM_middle}---emerges.

\item In Chap.~\ref{Localization_in_accelerated_frame}, we address the problem of localization in the context of accelerated frames. The algebraic approach to QFT provides a fundamental notion of localization in this case as well. By considering Rindler particle operators with energy close to their mass, we recover the Born localization scheme and the consequent postulate \ref{NRQM_middle} of NRQFTCS. Also, we study the general scenario in which an experimenter A prepares states over a background $| \Omega \rangle$ by means of nonrelativistic local operators and an experimenter B carries out nonrelativistic local measurements in a different region. If both experimenters are accelerated (i.e., experimenters A and B are Rachel and Rob, respectively) and if $| \Omega \rangle$ is the Rindler vacuum $|0_\text{L}, 0_\text{R} \rangle$, then the preparation of the state by Rachel does not influence measurements by Rob. Conversely, due to the frame dependent notion of vacuum states and nonrelativistic conditions, the independence between preparation of states and measurements is not guaranteed when $| \Omega \rangle$ is different from $|0_\text{L}, 0_\text{R} \rangle$ or when one of the two experimenters is inertial.

\item In Chap.~\ref{Single_particle_beyond_Rindler_horizon}, we detail the results of Chap.~\ref{Localization_in_accelerated_frame} by considering the example of a single particle that is prepared by an inertial observer and is detected by an accelerated observer. The detection occurs even if the particle is localized beyond the Rindler horizon.

\end{itemize}
\end{itemize}

\part{Fields and particles}\label{Relativistic_and_nonrelativistic_quantum_fields}

\chapter{QFT in flat and curved spacetimes}\label{QFT_in_flat_and_curved_spacetimes}

In this chapter, we review the description of fields in terms of particles as positive frequency modes of the respective field equation \cite{Wald:1995yp}. We focus on scalar and Dirac fields as the simplest examples of bosonic and fermionic fields.

In Sec.~\ref{QFT_in_Minkowski_spacetime}, we consider the Minkowski spacetime representing an inertial frame. In Sec.~\ref{QFT_in_curved_spacetime}, we generalize the results to any static spacetime. In Sec.~\ref{QFT_in_curved_spacetime_Rindler}, we study the Rindler frame as the static spacetime associated to an accelerated observer.

\section{Minkowski spacetime}\label{QFT_in_Minkowski_spacetime}

The Minkowski spacetime is defined by the coordinates $(t,\vec{x}) = (t, x, y, z)$ and by the flat metric $\eta_{\mu\nu} = \text{diag}(-c^2,1,1,1)$. We note by $\eta^{\mu\nu}$ the inverse of $\eta_{\mu\nu}$.

In this section, we study Klein-Gordon fields and Dirac fields in Minkowski spacetime. We provide the representation of Minkowski single particles as positive frequency modes of the Klein-Gordon and Dirac equation, respectively. The generalization to any Minkowski-Fock state is given by second-quantized wave functions.

\subsection{Scalar field}\label{QFT_in_Minkowski_spacetime_scalar}

We consider a free scalar field $\hat{\phi}$ that is solution of the Klein-Gordon equation
\begin{equation} \label{Klein_Gordon}
\left[ \eta^{\mu\nu} \partial_\mu \partial_\nu - \left( \frac{mc}{\hbar} \right)^2 \right] \hat{\phi} = 0,
\end{equation}
with $m$ as the mass, and satisfying the canonical commutation relation
\begin{equation}\label{phi_commutation}
\left[ \hat{\phi}(t,\vec{x}), \hat{\phi}^\dagger(t',\vec{x}') \right] = i \hbar \Delta_\text{KG}(t-t', \vec{x} - \vec{x}'),
\end{equation}
with
\begin{equation}\label{PauliJordan_function}
\Delta_\text{KG}(t,\vec{x}) = \frac{- i}{(2 \pi)^3} \int_{\mathbb{R}^3} \frac{d^3 k}{2 \omega(\vec{k})} \left[ e^{-i\omega(\vec{k})t + i\vec{k} \cdot \vec{x}} - e^{i\omega(\vec{k})t - i\vec{k} \cdot \vec{x}} \right]
\end{equation}
as the Pauli–Jordan function (i.e., retarded minus advanced propagator) and
\begin{equation} \label{omega_k}
\omega(\vec{k}) = \sqrt{\left(\frac{mc^2}{\hbar}\right)^2 + c^2 |\vec{k}|^2}
\end{equation}
as the dispersion relation. For real scalar fields, we have that $\hat{\phi}^\dagger = \hat{\phi}$; whereas, for complex scalar fields one has to consider the following additional commutation rule
\begin{equation}\label{phi_commutation_complex_2}
\left[ \hat{\phi}(t,\vec{x}), \hat{\phi}(t',\vec{x}') \right] = 0.
\end{equation}

Equations (\ref{Klein_Gordon}), (\ref{phi_commutation}) and (\ref{phi_commutation_complex_2}) are usually derived from the Lagrangian formulation of classical scalar fields and the canonical quantization. The Klein-Gordon Lagrangian density is
\begin{equation}
\mathcal{L}_\text{KG}[\hat{\phi}] = \frac{c}{2} \left[ \eta^{\mu\nu} \partial_\mu \hat{\phi} \partial_\nu \hat{\phi}^\dagger + \left( \frac{mc}{\hbar} \right)^2 \hat{\phi} \hat{\phi}^\dagger \right],
\end{equation}
which has the dimension of a density action, i.e., ET$/$L$^4$, where E is the energy, T the time and L the length. The conjugate momentum field is defined as
\begin{equation}\label{Pi_conjugate}
\hat{\pi} = c \frac{\delta \mathcal{L}_\text{KG}}{\delta (\partial_0 \hat{\phi})} = - \partial_0 \hat{\phi}^\dagger
\end{equation}
and satisfies the equal time canonical commutation relation
\begin{subequations}\label{phi_commutation_Lagrangian}
\begin{align}
& \left[ \hat{\phi}(t,\vec{x}), \hat{\pi}(t,\vec{x}') \right] = i \hbar \delta^3(\vec{x} - \vec{x}'), \label{phi_commutation_Lagrangian_a}\\
&  \left[ \hat{\phi}(t,\vec{x}), \hat{\phi}(t,\vec{x}') \right]  =  \left[ \hat{\phi}(t,\vec{x}), \hat{\phi}^\dagger(t,\vec{x}') \right] = 0, \\
&  \left[ \hat{\pi}(t,\vec{x}), \hat{\pi}(t,\vec{x}') \right] = \left[ \hat{\pi}(t,\vec{x}), \hat{\pi}^\dagger(t,\vec{x}') \right] = 0.
\end{align}
\end{subequations}
Real scalar fields are such that $\hat{\phi}^\dagger = \hat{\phi}$ and $\hat{\pi}^\dagger = \hat{\pi}$. Complex scalar fields also satisfy the commutation relation
\begin{equation}
\left[ \hat{\phi}(t,\vec{x}), \hat{\pi}^\dagger(t,\vec{x}') \right]  = 0.
\end{equation}

The equivalence between Eqs.~(\ref{phi_commutation}) and (\ref{phi_commutation_Lagrangian}) can be proved by plugging Eq.~(\ref{Pi_conjugate}) in Eq.~(\ref{phi_commutation_Lagrangian_a}) to obtain the differential equation
\begin{align}
& - \left. \frac{\partial}{\partial t'} \left[ \hat{\phi}(t,\vec{x}), \hat{\phi}^\dagger(t',\vec{x}') \right] \right|_{t'=t} = i \hbar \delta^3(\vec{x} - \vec{x}'), & \left[ \hat{\phi}(t,\vec{x}), \hat{\phi}^\dagger(t,\vec{x}') \right] = 0,
\end{align}
whose solution is precisely Eq.~(\ref{phi_commutation}).

\subsubsection{Free modes}\label{Free_modes_scalar}

Equations (\ref{Klein_Gordon}) and (\ref{phi_commutation}) lead to the familiar expression for the scalar field
\begin{equation} \label{free_field}
\hat{\phi}(t,\vec{x}) = \int_{\mathbb{R}^3} d^3 k \left[ f(\vec{k},t,\vec{x}) \hat{a}(\vec{k}) + f^*(\vec{k},t,\vec{x}) \hat{b}^\dagger(\vec{k}) \right],
\end{equation}
where $\hat{a}(\vec{k})$ is the annihilation operator for the free particle mode
\begin{equation}\label{free_modes}
f(\vec{k},t,\vec{x}) =  \sqrt{\frac{\hbar}{(2\pi)^3 2 \omega(\vec{k})}} e^{-i\omega(\vec{k})t + i\vec{k} \cdot \vec{x}}
\end{equation}
and $\hat{b}^\dagger(\vec{k})$ is the creation operator for an antiparticle with momentum $\vec{k}$. $\hat{a}(\vec{k})$ and $\hat{b}(\vec{k})$ generate the usual Minkowski-Fock space through the canonical commutation relation
\begin{subequations}\label{Minkowski_canonical_commutation}
\begin{align}
& [\hat{a}(\vec{k}),\hat{a}(\vec{k}')] = [\hat{b}(\vec{k}),\hat{b}(\vec{k}')] = [\hat{a}(\vec{k}),\hat{b}(\vec{k}')] = 0,\\
& [\hat{a}(\vec{k}),\hat{a}^\dagger(\vec{k}')] = [\hat{b}(\vec{k}),\hat{b}^\dagger(\vec{k}')] = \delta^3(\vec{k}-\vec{k}').
\end{align}
\end{subequations}
For real scalar fields, $\hat{a}(\vec{k}) = \hat{b}(\vec{k})$ and, hence, no distinction between particles and antiparticles occurs. For complex scalar fields, one has to consider the additional commutation relation
\begin{equation}
[\hat{a}(\vec{k}),\hat{b}^\dagger(\vec{k}')] = 0.
\end{equation}

The $f(\vec{k})$ modes are defined to be solutions of the Klein-Gordon equation (\ref{Klein_Gordon}) and orthonormal with respect to the Klein-Gordon scalar product
\begin{equation}\label{KG_scalar_product}
( \phi, \phi' )_\text{KG} = \frac{i}{\hbar} \int_{\mathbb{R}^3} d^3x \left[ \phi^*(t,\vec{x}) \partial_0  \phi'(t,\vec{x}) -  \phi'(t,\vec{x}) \partial_0 \phi^*(t,\vec{x}) \right].
\end{equation}
Equation (\ref{KG_scalar_product}) is time-independent for solutions of Eq.~(\ref{Klein_Gordon}). This can be proven by using Eq.~(\ref{Klein_Gordon}) and the integration by parts as
\begin{align}
\frac{d}{dt} ( \phi, \phi' )_\text{KG} = & \frac{i}{\hbar} \int_{\mathbb{R}^3} d^3x \left( \phi^* \partial^2_0  \phi'  -  \phi' \partial^2_0 \phi^* \right)\nonumber \\
 = & \frac{i}{\hbar c^2} \int_{\mathbb{R}^3} d^3x \left( \phi^* \delta^{ij} \partial_i \partial_j \phi' -  \phi' \delta^{ij} \partial_i \partial_j \phi^* \right)\nonumber \\
 = & \frac{i}{\hbar c^2} \delta^{ij} \int_{\mathbb{R}^3} d^3x \left[ -(\partial_i \phi^*) (\partial_j \phi') +  (\partial_i \phi') (\partial_j \phi^*) \right]\nonumber \\
 = & 0.
\end{align}
The orthonormality of the $f(\vec{k})$ modes with respect to $( \phi, \phi' )_\text{KG}$ reads as
\begin{subequations}\label{KG_scalar_product_orthonormality_f}
\begin{align}
& ( f(\vec{k}), f(\vec{k}') )_\text{KG} = \delta^3(\vec{k}-\vec{k}'), \\
& ( f^*(\vec{k}), f^*(\vec{k}') )_\text{KG} = -\delta^3(\vec{k}-\vec{k}'),\\
& ( f(\vec{k}), f^*(\vec{k}') )_\text{KG} = 0,
\end{align}
\end{subequations}
which can be proven from Eqs.~(\ref{free_modes}) and (\ref{KG_scalar_product}).

In the interaction-free theory, the Hilbert space of single particles is represented by the vector space generated by the $f(\vec{k})$ modes and supplemented by the Klein-Gordon scalar product (\ref{KG_scalar_product}). The $f^*(\vec{k})$ modes have to be excluded, in order for the Klein-Gordon scalar product to satisfy positive-definiteness [Eq.~(\ref{KG_scalar_product_orthonormality_f})]. The space of single antiparticle states is analogously defined from the field $\hat{\phi}^\dagger$. In this chapter, we only focus on particles.

The mode $f(\vec{k})$ is associated to the single particle state $ | \vec{k} \rangle = \hat{a}^\dagger(\vec{k}) | 0_\text{M} \rangle$, with $| 0_\text{M} \rangle$ as the Minkowski vacuum defined by $\hat{a}(\vec{k}) | 0_\text{M} \rangle = \hat{b}(\vec{k}) | 0_\text{M} \rangle = 0$. The function $f(\vec{k},t,\vec{x})$, with varying $t$ and $\vec{x}$, provides a representation for the state $| \vec{k} (t) \rangle$ evolved with respect to the free theory as
\begin{equation}\label{free_evolution_k_states}
i \hbar \partial_0 | \vec{k} (t) \rangle = \omega(\vec{k}) | \vec{k} (t) \rangle,
\end{equation}
with initial time condition $| \vec{k} (0) \rangle = | \vec{k} \rangle$.

Any particle state $| \phi \rangle$ can be expanded as
\begin{equation}\label{free_state_decomposition}
| \phi \rangle  = \sum_{n=1}^\infty \frac{1}{\sqrt{n!}} \int_{\mathbb{R}^{3n}} d^{3n} \textbf{k}_n \tilde{\phi}_n (\textbf{k}_n) \prod_{l=1}^n \hat{a}^\dagger(\vec{k}_l) | 0_\text{M} \rangle + \tilde{\phi}_0 | 0_\text{M} \rangle,
\end{equation}
where $\textbf{k}_n = (\vec{k}_1, \dots, \vec{k}_n)$ is a $3n$ vector collecting $n$ momenta and $\tilde{\phi}_n (\textbf{k}_n)$ is the $n$-particles wave function of $| \phi \rangle$ in the momentum representation, which is defined to be symmetric with respect to the momenta variables. The representative of the time-evolved state $| \phi (t) \rangle$ in the Schrödinger picture is
\begin{equation} \label{free_wave_function}
\phi_n (t, \textbf{x}_n) = \left( \frac{2 m c^2}{\hbar^2} \right)^{n/2} \int_{\mathbb{R}^{3n}} d^{3n} \textbf{k}_n \tilde{\phi}_n (\textbf{k}_n) \prod_{l=1}^n f(\vec{k}_l, t, \vec{x}_l),
\end{equation}
where $\textbf{x}_n = (\vec{x}_1, \dots, \vec{x}_n)$. When $n=0$, we assume $\phi_0 = \tilde{\phi}_0$.

Hereafter, we will refer to $\phi_n (t, \textbf{x}_n)$ of Eq.~(\ref{free_wave_function}) as the wave function of $| \phi (t) \rangle$ in position space. However, we remark that $\phi_n (t, \textbf{x}_n)$ is not associated to the probability to find the $n$ particles in $\textbf{x}_n = (\vec{x}_1, \dots, \vec{x}_n)$; the interpretation that the state $| \phi (t) \rangle$ is localized in the support of $\phi_n (t, \textbf{x}_n)$ is generally incorrect. A genuine notion of localization is only provided by the algebraic formulation of QFT, which will be detailed in Secs.~\ref{Algebraic_approach_in_QFT_and_QFTCS} and \ref{AQFT_localization_scheme}. The modal localization scheme based on the wave function $\phi_n (t, \textbf{x}_n)$ will be rigorously defined in Sec.~\ref{Modal_localization_scheme} and it will be shown to be incompatible with the fundamental localization in QFT [Sec.~\ref{Comparison_between_modal_and_AQFT_schemes}].

A genuine notion of localization for $\phi_n (t, \textbf{x}_n)$ is only recovered in the nonrelativistic limit. In such a regime, the wave function $\phi_n (t, \textbf{x}_n)$ is approximately an element of $L^2(\mathbb{R}^3)$ and follows the interpretation of Born localization, i.e., $\phi_n (t, \textbf{x}_n)$ is the probability amplitude to find the $n$ particles in $\textbf{x}_n = (\vec{x}_1, \dots, \vec{x}_n)$. These results will be shown, respectively, in Secs.~\ref{Minkowski_spacetime_Scalar_field} and \ref{Convergence_of_the_modal_scheme_to_the_Born_scheme}. Hereafter, we will assume that the particle states are localized in the support of $\phi_n (t, \textbf{x}_n)$, but only in the context of NonRelativistic Quantum Mechanics (NRQM).

While the wave functions (\ref{free_wave_function}) give a representation for particle states, the Klein-Gordon scalar product (\ref{KG_scalar_product}) represents the Hilbert product between two single particle states in terms of their wave functions as
\begin{equation} \label{scalar_product_representation}
\langle \phi | \phi' \rangle = \frac{\hbar^2}{2mc^2} ( \phi_1, \phi'_1 )_\text{KG}.
\end{equation}
This product is time independent if evaluated on wave functions of the form of Eq.~(\ref{free_state_decomposition}) and, hence, leads to constant probabilities, as expected by the quantum theory. Equation (\ref{scalar_product_representation}) can be proven by using Eqs.~(\ref{KG_scalar_product_orthonormality_f}), (\ref{free_state_decomposition}) and (\ref{free_wave_function}).

The time evolution of the state $| \vec{k} \rangle$ is described by Eq.~(\ref{free_evolution_k_states}), which states that $| \vec{k} \rangle$ is an eigenstate of the Klein-Gordon Hamiltonian $\hat{h}_\text{KG}$ with $\omega(\vec{k})$ as the eigenvalue. If we try to represents such a Hamiltonian in the representation space of the modes $f(\vec{k})$, we must rely on some kind of square root of
\begin{equation}\label{H_KG}
H_\text{KG} = - (\hbar c)^2 \delta^{ij}\partial_i \partial_j + (mc^2)^2,
\end{equation}
since each mode $f(\vec{k})$ is a solution of 
\begin{equation}
H_\text{KG} f(\vec{k}) = [\hbar \omega(\vec{k})]^2 f(\vec{k}).
\end{equation}

What we mean by square root of $H_\text{KG}$ is the fact that $H_\text{KG}$ and the representative of $\hat{h}_\text{KG}$ share the same eigenvectors, but with different eigenvalues: if $\hbar \omega$ is the eigenvalue of $| \vec{k} \rangle$ with respect to $\hat{h}_\text{KG}$, then $(\hbar \omega)^2$ is the eigenvalue of $f(\vec{k})$ with respect to $H_\text{KG}$. We define $h_\text{KG}$ as the representative of $\hat{h}_\text{KG}$ and we write the following improper expression
\begin{equation}\label{h_KG_H_KG}
h_\text{KG} = \sqrt{H_\text{KG}}.
\end{equation}

\subsubsection{General modes}\label{General_modes_scalar}

We now generalize the results obtained for the free modes $f(\vec{k})$ by considering the expansion of $\hat{\phi}$ in terms of generic modes $g(\theta)$ and $h(\theta)$ with, respectively, positive and negative frequencies. Specifically, we consider
\begin{equation}\label{free_field_positive_negative_frequencies}
\hat{\phi}(t,\vec{x}) = \sum_\theta \left[ g(\theta, t,\vec{x}) \hat{a}(\theta)  + h(\theta, t,\vec{x}) \hat{b}^\dagger(\theta) \right],
\end{equation}
where $\theta$ is a collection of quantum numbers which can be discrete, continuum or both and $ \sum_\theta$ is a generalized sum, eventually including integrals for continuum variables. In the particular case of free modes, the quantum numbers $\theta$ are momentum components (i.e., $\theta = \vec{\theta} = \vec{k}$) and $ \sum_\theta = \int_{\mathbb{R}^3}d^3 k$. The operators $\hat{a}(\theta)$ and $\hat{b}^\dagger(\theta)$ are, respectively, the annihilator of the particle mode $g(\theta)$ and the creator of the antiparticle mode $h^*(\theta)$. The function $g(\theta, t,\vec{x})$  with varying $t$ and $\vec{x}$ is, hence, the representative of the single particle state $|\theta \rangle  = \hat{a}^\dagger(\theta) | 0_\text{M} \rangle$ identified by the quantum numbers $\theta$. 

The fact that $g(\theta)$ and $h(\theta)$ have positive and negative frequencies can be expressed by
\begin{subequations}\label{positive_negative_frequencies}
\begin{align}
& g(\theta,t,\vec{x}) = \tilde{g}(\theta,\vec{x}) e^{-i \omega(\theta) t}, \label{positive_frequencies}\\
& h(\theta,t,\vec{x}) = \tilde{h}(\theta,\vec{x}) e^{i \omega(\theta) t},
\end{align}
\end{subequations}
with $\pm \omega(\theta)$ as the real frequencies. The orthonormality with respect to the Klein-Gordon scalar product (\ref{KG_scalar_product}), instead, reads as
\begin{subequations}\label{KG_scalar_product_orthonormality_g}
\begin{align}
& ( g(\theta), g(\theta') )_\text{KG} = \delta_{\theta\theta'}, \\
 & ( h(\theta), h(\theta') )_\text{KG} = -\delta_{\theta\theta'}, \\
  & ( g(\theta), h(\theta') )_\text{KG} = 0,
\end{align}
\end{subequations}
where, in this case, the deltas are generalized, as they act as Kronecker deltas for discrete indices and as Dirac deltas for continuum variables.

The decomposition of the field into real frequencies [Eq.~(\ref{positive_negative_frequencies})] is guaranteed by the fact that $g(\theta)$ and $h(\theta)$ are solutions of the Klein-Gordon equation (\ref{Klein_Gordon}). By imposing the ansatz (\ref{positive_frequencies}), Eq.~(\ref{Klein_Gordon}) becomes a Schrödinger equation for $g(\theta)$ with eigenvalues proportional to $\omega^2$, i.e.,
\begin{align} \label{Klein_Gordon_Schrodinger}
H_\text{KG} g(\theta) = [\hbar \omega(\theta)]^2 g(\theta).
\end{align}
$H_\text{KG}$ is positive with respect to the Klein-Gordon scalar product (\ref{KG_scalar_product}) for any positive frequency solution of the Klein-Gordon equation. Indeed, by defining
\begin{align}
& h_0 = mc^2 , & h_i =\hbar c \partial_i,
\end{align}
one can prove that
\begin{equation}
( \phi, H_\text{KG} \phi' )_\text{KG} = \delta^{i j} ( h_i \phi, h_j \phi' )_\text{KG} + ( h_0 \phi, h_0 \phi' )_\text{KG}.
\end{equation}
In this way one can see that if $\phi$ is a combination of $g(\theta)$ modes, then
\begin{equation}
( \phi, H_\text{KG} \phi )_\text{KG} > 0
\end{equation}
and, hence, $H_\text{KG}$ has positive eigenvalues in the space of $g(\theta)$ modes. This is compatible with the fact that the $\omega$ appearing in Eq.~(\ref{Klein_Gordon_Schrodinger}) is real. The same proof can be obtained for the $h^*(\theta)$ modes by considering the field $\hat{\phi}^\dagger$.

As in Eqs.~(\ref{free_state_decomposition}) and (\ref{free_wave_function}), we may define the wave function of any state $|\phi \rangle$ by decomposing $|\phi \rangle$ in terms of $| \theta \rangle$ states as
\begin{equation} \label{general_Fock_expansion}
| \phi \rangle  = \sum_{n=1}^\infty \frac{1}{\sqrt{n!}} \sum_{\bm{\theta}_n} \tilde{\phi}_n (\bm{\theta}_n) \prod_{l=1}^n \hat{a}^\dagger(\theta_l) | 0_\text{M} \rangle + \tilde{\phi}_0 | 0_\text{M} \rangle,
\end{equation}
where we have defined the vector $\bm{\theta}_n = (\theta_1, \dots, \theta_n)$ and the function $\tilde{\phi}_n (\bm{\theta}_n)$ is symmetric with respect to $\theta_1, \dots, \theta_n$. In the Schrödinger picture, he state $| \phi \rangle$ is represented by
\begin{equation}\label{wavefunction_g}
\phi_n (t, \textbf{x}_n) =  \left( \frac{2 m c^2}{\hbar^2} \right)^{n/2}  \sum_{\bm{\theta}_n} \tilde{\phi}_n (\bm{\theta}_n)  \prod_{l=1}^n g(\theta_l,t,\vec{x}_l).
\end{equation}

\subsection{Dirac field}\label{QFT_in_Minkowski_spacetime_Dirac}

In this subsection, we consider a Dirac field $\hat{\psi}$ in Minkowski spacetime and we derive the modal representation of the corresponding particle states. We use the Dirac representation for $\hat{\psi}$ and its modes; hence, we identify them as a $4$ dimensional vectors and any operator acting on the left as a $4 \times 4$ matrix.

Free Dirac fields in Minkowski spacetime are solutions of the Dirac equation
\begin{equation} \label{Dirac}
\left( i c \gamma^\mu \partial_\mu  - \frac{m c^2}{\hbar} \right) \hat{\psi} = 0, 
\end{equation}
where
\begin{align}\label{gamma_matrix}
& \gamma^0 = \frac{1}{c} \begin{pmatrix}
\mathbb{I} &0 \\
0 &-\mathbb{I}
\end{pmatrix}, & \gamma^i = \begin{pmatrix}
0 &\sigma^i \\
-\sigma^i &0
\end{pmatrix}
\end{align}
are gamma matrices, with $\mathbb{I}$ as $2 \times 2$ identity matrix and
\begin{align}
& \sigma^1 = \begin{pmatrix}
0 & 1 \\
1 & 0
\end{pmatrix}, & & \sigma^2 = \begin{pmatrix}
0 & -i \\
i & 0
\end{pmatrix}, & & \sigma^3 = \begin{pmatrix}
1 & 0 \\
0 & -1
\end{pmatrix}
\end{align}
as Pauli matrices. The following identities for gamma matrices will be used throughout the thesis
\begin{subequations} \label{gamma_matrices_identities}
\begin{equation}\label{gamma_matrices_anticommutating_rule}
\{ \gamma^\mu, \gamma^\nu \} = -2 \eta^{\mu\nu},
\end{equation}
\begin{align}\label{gamma_matrices_hermitianity}
& (\gamma^0)^\dagger = \gamma^0, & (\gamma^i)^\dagger = -\gamma^i.
\end{align}
\end{subequations}

\subsubsection{Free modes}

The usual decomposition of $\hat{\psi}$ in terms of modes with defined momenta and spin is
\begin{equation} \label{free_Dirac_field}
\hat{\psi}(t,\vec{x}) = \sum_{s=1}^2 \int_{\mathbb{R}^3} d^3 k \left[ u_s(\vec{k},t,\vec{x}) \hat{c}_s(\vec{k}) + v_s(\vec{k},t,\vec{x}) \hat{d}_s^\dagger(\vec{k}) \right],
\end{equation}
where $\hat{c}_s(\vec{k})$ and $\hat{d}_s^\dagger(\vec{k})$ are, respectively, the annihilation operator for the particle and the creation operator for the antiparticle with momentum $k$ and spin number $s$ satisfying the canonical anticommutation relations
\begin{subequations}\label{free_Dirac_field_anticommutating_rules}
\begin{align}
& \{\hat{c}_s(\vec{k}),\hat{c}_{s'}(\vec{k}')\} = \{\hat{d}_s(\vec{k}),\hat{d}_{s'}(\vec{k}')\} = \{\hat{c}_s(\vec{k}),\hat{d}_{s'}(\vec{k}')\} = 0, \\
& \{\hat{c}_s(\vec{k}),\hat{c}_{s'}^\dagger(\vec{k}')\} = \{\hat{d}_s(\vec{k}),\hat{d}_{s'}^\dagger(\vec{k}')\} = \delta_{ss'} \delta^3(\vec{k}-\vec{k}').
\end{align}
\end{subequations}

The modes $u_s(\vec{k})$ and $v_s(\vec{k})$ are orthonormal positive and negative frequency solutions of the Dirac equation (\ref{Dirac}) having the form of
\begin{subequations}\label{free_Dirac_field_modes}
\begin{align}
& u_s(\vec{k},t,\vec{x}) = (2\pi)^{-3/2} e^{ -i\omega(\vec{k})t + i\vec{k} \cdot \vec{x} }  \tilde{u}_s(\vec{k}), \label{u} \\
& v_s(\vec{k},t,\vec{x}) = (2\pi)^{-3/2} e^{ i\omega(\vec{k})t - i\vec{k} \cdot \vec{x} }  \tilde{v}_s(\vec{k}). \label{v}
\end{align}
\end{subequations}
The orthonormality condition is
\begin{subequations}\label{D_scalar_product_orthonormality_u_v}
\begin{align}
& ( u_s(\vec{k}), u_{s'}(\vec{k}') )_{\mathbb{C}^4 \otimes L^2(\mathbb{R}^3)} = \delta_{ss'} \delta^3(\vec{k}-\vec{k}'),\\
 & ( v_s(\vec{k}), v_{s'}(\vec{k}') )_{\mathbb{C}^4 \otimes L^2(\mathbb{R}^3)} = \delta_{ss'} \delta^3(\vec{k}-\vec{k}'),\\
  & ( u_s(\vec{k}), v_{s'}(\vec{k}') )_{\mathbb{C}^4 \otimes L^2(\mathbb{R}^3)} = 0.
\end{align}
\end{subequations}
where
\begin{equation}\label{Dirac_scalar_product}
( \psi, \psi' )_{\mathbb{C}^4 \otimes L^2(\mathbb{R}^3)} = \int_{\mathbb{R}^3} d^3x \psi^\dagger(t,\vec{x}) \psi'(t,\vec{x})
\end{equation}
is the Minkowski-Dirac product. Owing to Eq.~(\ref{gamma_matrices_identities}), one can prove that Eq.~(\ref{Dirac_scalar_product}) is time independent for any couple of solutions of Eq.~(\ref{Dirac}) as follows
\begin{align}
 \frac{d}{dt} ( \psi, \psi' )_{\mathbb{C}^4 \otimes L^2(\mathbb{R}^3)} = & \frac{d}{dt} \int_{\mathbb{R}^3} d^3x \psi^\dagger \psi' \nonumber \\
 = & c^2 \frac{d}{dt} \int_{\mathbb{R}^3} d^3x \psi^\dagger\gamma^0 \gamma^0 \psi'  \nonumber \\
 = & c^2 \int_{\mathbb{R}^3} d^3x [ (\partial_0 \psi)^\dagger \gamma^0 \gamma^0 \psi'  + \psi^\dagger \gamma^0 \gamma^0 \partial_0 \psi' ]\nonumber \\
 = & c^2 \int_{\mathbb{R}^3} d^3x [(\gamma^0 \partial_0 \psi)^\dagger \gamma^0 \psi' + \psi^\dagger \gamma^0 \gamma^0 \partial_0 \psi' ]\nonumber \\
 = & c^2 \int_{\mathbb{R}^3} d^3x \left\lbrace \left[ \left( -\gamma^i \partial_i - i \frac{mc}{\hbar} \right) \psi\right]^\dagger \gamma^0 \psi' \right. \nonumber \\
 & \left. + \psi^\dagger \gamma^0 \left( -\gamma^i \partial_i - i \frac{mc}{\hbar} \right) \psi' \right\rbrace\nonumber \\
 = & c^2 \int_{\mathbb{R}^3} d^3x [ (\partial_i  \psi^\dagger) \gamma^i  \gamma^0 \psi'  - \psi^\dagger \gamma^0 \gamma^i \partial_i \psi'] \nonumber \\
 = & c^2 \int_{\mathbb{R}^3} d^3x [(\partial_i  \psi^\dagger) \gamma^i  \gamma^0 \psi' + \psi^\dagger \gamma^i \gamma^0 \partial_i \psi']\nonumber \\
 = & 0.
\end{align}

The functions $\tilde{u}_s(\vec{k})$ and $\tilde{v}_s(\vec{k})$ are solutions of the following equations 
\begin{subequations} \label{Dirac_uv_tilde}
\begin{align}
& \left[ \omega(\vec{k}) \gamma^0 - k_i \gamma^i - \frac{mc}{\hbar} \right] \tilde{u}_s(\vec{k}) = 0,\label{Dirac_u_tilde}\\
& \left[ \omega(\vec{k}) \gamma^0 - k_i \gamma^i + \frac{mc}{\hbar} \right] \tilde{v}_s(\vec{k}) = 0,\label{Dirac_v_tilde}\\
& \tilde{u}^\dagger_s(\vec{k}) \tilde{u}_{s'}(\vec{k}) = \delta_{ss'},\label{orthonormality_u_tilde}\\
& \tilde{v}^\dagger_s(\vec{k}) \tilde{v}_{s'}(\vec{k}) = \delta_{ss'},\label{orthonormality_v_tilde}\\
& \tilde{u}^\dagger_s(\vec{k}) \tilde{v}_{s'}(-\vec{k}) = 0.\label{orthonormality_uv_tilde}
\end{align}
\end{subequations}
One can use Eq.~(\ref{free_Dirac_field_modes}) and the fact that $u_s(\vec{k})$ and $v_s(\vec{k})$ are solutions of Eq.~(\ref{Dirac}) to obtain Eqs.~(\ref{Dirac_u_tilde}) and (\ref{Dirac_v_tilde}). The orthonormality conditions (\ref{orthonormality_u_tilde}), (\ref{orthonormality_v_tilde}) and (\ref{orthonormality_uv_tilde}), instead, can be checked by plugging Eq.~(\ref{free_Dirac_field_modes}) in Eq.~(\ref{D_scalar_product_orthonormality_u_v}) and using Eq.~(\ref{Dirac_scalar_product}).

The index $s$ is associated to the two independent spin degrees of freedom. One can consider any couple of solutions of Eq.~(\ref{Dirac_uv_tilde}) and associate each solution to either $s=1$ or $s=2$. This freedom is due to the arbitrary definition of the spin basis for positive and negative frequency modes.

A possible basis is given by particles with defined spin along one direction. For instance, states with spin up and down with respect to $z$ are such that in the particle comoving frame (i.e., the one obtained by performing a Lorentz boost with opposite momentum $-\vec{k}$) the representatives have only one spinorial component different from zero. In this basis, the functions $\tilde{u}_s(\vec{k})$ and $\tilde{v}_s(\vec{k})$ are
\begin{subequations}\label{free_Dirac_field_modes_spin_up_spin_down}
\begin{align}
\tilde{u}_s(\vec{k}) = &  \frac{c \gamma^0 \omega(\vec{k}) - c \gamma^i k_i + mc^2/\hbar}{\sqrt{2 \omega(\vec{k}) [\omega(\vec{k})+mc^2/\hbar]}} \mathfrak{u}_s, \\
\tilde{v}_s(\vec{k}) = &  \frac{-c \gamma^0 \omega(\vec{k}) + c \gamma^i k_i + mc^2/\hbar}{\sqrt{ 2 \omega(\vec{k}) [\omega(\vec{k})+mc^2/\hbar]}}  \mathfrak{v}_s,
\end{align}
\end{subequations}
with
\begin{align}\label{free_Dirac_field_basis}
\mathfrak{u}_1 = \begin{pmatrix}
1  \\
0 \\
0\\
0
\end{pmatrix},
& & \mathfrak{u}_2 =   \begin{pmatrix}
0  \\
1 \\
0\\
0
\end{pmatrix},
& & \mathfrak{v}_1 =  \begin{pmatrix}
0  \\
0 \\
1\\
0
\end{pmatrix},
& & \mathfrak{v}_2 =   \begin{pmatrix}
0  \\
0 \\
0\\
1
\end{pmatrix}.
\end{align}

It can be noticed that the Dirac equation (\ref{Dirac}) is already put in a Schrödinger equation form. Indeed, by acting on Eq.~(\ref{Dirac}) with a $\hbar c \gamma^0$ matrix and using Eq.~(\ref{gamma_matrices_anticommutating_rule}), one obtains
\begin{equation} \label{Dirac_schrodinger}
i \hbar \partial_0  \hat{\psi}  = h_\text{M}   \hat{\psi},
\end{equation}
with Hamiltonian
\begin{equation}\label{Dirac_Hamiltonian}
h_\text{M}  = - i \hbar c^2 \gamma^0 \gamma^i \partial_i + m c^3 \gamma^0.
\end{equation}
It can also be noticed that $h_\text{M}$ is hermitian with respect to the $\mathbb{C}^4 \otimes L^2(\mathbb{R}^3)$ scalar product, in the sense that
\begin{equation}
( h_\text{M} \psi,  \psi' )_{\mathbb{C}^4 \otimes L^2(\mathbb{R}^3)}  = ( \psi, h_\text{M}  \psi' )_{\mathbb{C}^4 \otimes L^2(\mathbb{R}^3)}.
\end{equation}
This can be proven by using Eq.~(\ref{gamma_matrices_identities}) to obtain
\begin{align}
( h_\text{M} \psi,  \psi' )_{\mathbb{C}^4 \otimes L^2(\mathbb{R}^3)}  = & \int_{\mathbb{R}^3} d^3x \left[ \left( - i \hbar c^2 \gamma^0 \gamma^i \partial_i + m c^3 \gamma^0 \right) \psi \right]^\dagger \psi'\nonumber \\
  = & \int_{\mathbb{R}^3} d^3x [ - i \hbar c^2 ( \partial_i \psi^\dagger) \gamma^i \gamma^0 + m c^3 \psi^\dagger \gamma^0] \psi'\nonumber \\
  = & \int_{\mathbb{R}^3} d^3x [  i \hbar c^2 ( \partial_i \psi^\dagger) \gamma^0 \gamma^i + m c^3 \psi^\dagger\gamma^0 ] \psi'\nonumber \\
  = & \int_{\mathbb{R}^3} d^3x \psi^\dagger   \left( - i \hbar c^2 \gamma^0 \gamma^i \partial_i + m c^3 \gamma^0 \right)\psi'\nonumber \\
    = & ( \psi, h_\text{M}  \psi' )_{\mathbb{C}^4 \otimes L^2(\mathbb{R}^3)}.
\end{align}

The quantum states defined as $|s,\vec{k}\rangle = \hat{c}^\dagger_s (\vec{k}) | 0_\text{M} \rangle$ are orthonormal and generate the Hilbert space of single particles. This means that any Fock state $| \psi \rangle$ can be decomposed as
\begin{equation}\label{free_Dirac_state_decomposition}
| \psi \rangle = \sum_{n=1}^\infty \frac{1}{\sqrt{n!}} \sum_{\textbf{s}_n} \int_{\mathbb{R}^{3n}} d^{3n} \textbf{k}_n \tilde{\psi}_n (\textbf{s}_n,\textbf{k}_n) \prod_{l=1}^n \hat{c}_{s_l}^\dagger(\vec{k}_l) | 0_\text{M} \rangle + \tilde{\psi}_0 | 0_\text{M} \rangle,
\end{equation}
with $(\textbf{s}_n,\textbf{k}_n) = ((s_1,\vec{k}_1), \dots, (s_n,\vec{k}_n))$ as a collection of spin and momenta and with $\tilde{\psi}_n (\textbf{s}_n,\textbf{k}_n)$ as a function that is antisymmetric with respect to the spin-momenta variables. Equation (\ref{free_Dirac_state_decomposition}) is the equivalent of Eq.~(\ref{free_state_decomposition}) for Dirac particles and provides the definition of $\tilde{\psi}_n (\textbf{s}_n,\textbf{k}_n)$ as the wave function for $| \psi \rangle$ in the spin-momentum representation space.

Equivalently to Eq.~(\ref{free_wave_function}), the representative of the time-evolved state $| \psi (t) \rangle$ in the Schrödinger picture is
\begin{equation} \label{free_Dirac_wave_function}
\psi_n^{\bm{\alpha}_n} (t, \textbf{x}_n) = \sum_{\textbf{s}_n} \int_{\mathbb{R}^{3n}} d^{3n} \textbf{k}_n  \tilde{\psi}_n (\textbf{s}_n,\textbf{k}_n) \prod_{l=1}^n u^{\alpha_l}_{s_l}(\vec{k}_l, t, \vec{x}_l).
\end{equation}
where $u^\alpha_s(\vec{k}, t, \vec{x})$ is the $\alpha$-th spinorial component of $u_s(\vec{k}, t, \vec{x})$. In Eq.~(\ref{free_Dirac_wave_function}), the $4$ dimensional spinorial degrees of freedom are, thus, indicated by means of the $\alpha$ indices collected as $\bm{\alpha}_n = \alpha_1 \dots \alpha_n$. Single particle wave functions, instead, can be written in the spinorial notation without indices as
\begin{equation} \label{free_Dirac_wave_function_single}
\psi_1 (t, \vec{x}) = \sum_{s=1}^2 \int_{\mathbb{R}^3} d^3 k  \tilde{\psi}_1 (s, \vec{k}) u_{s}(\vec{k}, t, \vec{x}).
\end{equation}

The spinor $\psi_1$ is solution of Eq.~(\ref{Dirac_schrodinger}) and, hence, its time evolution is provided by $h_\text{M}$. Also, the Hilbert product between any couple of single particle states $| \psi \rangle $, $| \psi' \rangle $ can be written in terms of the $\mathbb{C}^4 \otimes L^2(\mathbb{R}^3)$ product of their wave functions as
\begin{equation}\label{scalar_product_Dirac_single_particles}
\langle \psi | \psi' \rangle = (\psi_1, \psi'_1)_{\mathbb{C}^4 \otimes L^2(\mathbb{R}^3)}.
\end{equation}
This means that the single particle content of the Dirac field can be fully described by spin-momentum wave functions (\ref{free_Dirac_wave_function_single}), the $\mathbb{C}^4 \otimes L^2(\mathbb{R}^3)$ product and the Hamiltonian $h_\text{M}$. We may think that $\mathbb{C}^4 \otimes L^2(\mathbb{R}^3)$ is the representation space of the Dirac single particle states. However, the orthonormal functions $u_s(\vec{k})$ do not provide a complete basis for $\mathbb{C}^4 \otimes L^2(\mathbb{R}^3)$, as it is possible to see from Eq.~(\ref{D_scalar_product_orthonormality_u_v}). The actual representation space is a subspace of $\mathbb{C}^4 \otimes L^2(\mathbb{R}^3)$, namely the positive frequency subspace of $\mathbb{C}^4 \otimes L^2(\mathbb{R}^3)$.

A generalization to Eqs.~(\ref{free_Dirac_wave_function_single}) and (\ref{scalar_product_Dirac_single_particles}) for more than one particle can be given by the wave functions (\ref{free_Dirac_wave_function}) and by the Hilbert product
\begin{equation}
\langle \psi | \psi' \rangle = \sum_{n=0}^\infty (\psi_n, \psi'_n)_{\mathbb{C}^{4n} \otimes L^2(\mathbb{R}^{3n})},
\end{equation}
where
\begin{subequations}
\begin{align}
& (\psi_n, \psi'_n)_{\mathbb{C}^{4n} \otimes L^2(\mathbb{R}^{3n})} = \sum_{\bm{\alpha}_n} \int_{\mathbb{R}^{3n}} d^{3n} \textbf{x}_n [\psi^*_n (t, \textbf{x}_n)]^{\bm{\alpha}_n} [\psi'_n (t, \textbf{x}_n)]^{\bm{\alpha}_n}, \\
& (\psi_0, \psi'_0)_{\mathbb{C}^0 \otimes L^2(\mathbb{R}^0)} = \psi_0^* \psi'_0.
\end{align}
\end{subequations}

\subsubsection{General modes}

We now consider the decomposition of $\hat{\psi}$ with respect to general modes
\begin{equation}\label{free_field_positive_negative_frequencies_Dirac}
\hat{\psi}(t,\vec{x}) = \sum_\theta \left[ u(\theta,t,\vec{x}) \hat{c}(\theta)  + v(\theta,t,\vec{x}) \hat{d}^\dagger(\theta) \right].
\end{equation}
Equation (\ref{free_field_positive_negative_frequencies_Dirac}) is similar to Eq.~(\ref{free_field_positive_negative_frequencies}); here, however, the spin degrees of freedom introduce an additional energy degeneracy and the modes $u(\theta)$ and $v(\theta)$ have spinorial components. The time-dependency of $u(\theta)$ and $v(\theta)$ is identical to Eq.~(\ref{positive_negative_frequencies}) and reads as
\begin{subequations}\label{positive_negative_frequencies_Dirac}
\begin{align}
& u(\theta,t,\vec{x}) = \tilde{u}(\theta,\vec{x}) e^{-i \omega(\theta) t},\label{positive_frequencies_Dirac} \\
 & v(\theta,t,\vec{x}) = \tilde{v}(\theta,\vec{x}) e^{i \omega(\theta) t}.
\end{align}
\end{subequations}
Equation (\ref{positive_negative_frequencies_Dirac}) is guaranteed by the already-proven hermicity of $h_\text{M}$. $u(\theta)$ and $v(\theta)$ are also defined to be orthonormal with respect to the $\mathbb{C}^4 \otimes L^2(\mathbb{R}^3)$ product
\begin{subequations}\label{D_scalar_product_orthonormality_u_v_generic}
\begin{align}
& ( u(\theta), u(\theta') )_{\mathbb{C}^4 \otimes L^2(\mathbb{R}^3)}  =  \delta_{\theta\theta'} ,\\
& ( v(\theta), v(\theta') )_{\mathbb{C}^4 \otimes L^2(\mathbb{R}^3)}  =  \delta_{\theta\theta'}, \\
& ( u(\theta), v(\theta') )_{\mathbb{C}^4 \otimes L^2(\mathbb{R}^3)} = 0.
\end{align}
\end{subequations}

Any Fock state $| \psi \rangle$ can be expanded with respect to the single particle basis $| \theta \rangle =  \hat{c}^\dagger(\theta) | 0_\text{M} \rangle$ as in Eq.~(\ref{general_Fock_expansion}), i.e.,
\begin{equation} \label{general_Fock_Dirac_expansion}
| \psi \rangle  = \sum_{n=1}^\infty \frac{1}{\sqrt{n!}} \sum_{\bm{\theta}_n} \tilde{\psi}_n (\bm{\theta}_n) \prod_{l=1}^n \hat{c}^\dagger(\theta_l) | 0_\text{M} \rangle +  \tilde{\psi}_0 | 0_\text{M} \rangle.
\end{equation}
Similarly to Eq.~(\ref{free_Dirac_wave_function}), the representative of state $| \psi (t) \rangle$ in the Schrödinger picture reads  as
\begin{equation}\label{Dirac_wavefunction_general}
\psi^{\bm{\alpha}_n}_n (t, \textbf{x}_n) =  \sum_{\bm{\theta}_n} \tilde{\psi}_n (\bm{\theta}_n) \prod_{l=1}^n u^{\alpha_l}(\theta_l, t, \vec{x}_l).
\end{equation}

It is straightforward to prove Eq.~(\ref{scalar_product_Dirac_single_particles}) for single particles with the new definition of $\psi_1 (t, \vec{x})$ given by Eq.~(\ref{Dirac_wavefunction_general}). As a result, we obtain the description of single particles by means of the $\mathbb{C}^4 \otimes L^2(\mathbb{R}^3)$ Hilbert space. The general basis is identified by modes with quantum numbers $\theta$. Owing to Eq.~(\ref{D_scalar_product_orthonormality_u_v_generic}) we notice again that the representation space is actually a subspace of $\mathbb{C}^4 \otimes L^2(\mathbb{R}^3)$.

\section{Curved spacetime}\label{QFT_in_curved_spacetime}

At variance with Sec.~\ref{QFT_in_Minkowski_spacetime}, here we consider a more general class of spacetimes (namely, hyperbolic static spacetimes) and we extend the discussion about the modal representation of particle states to this entire class.

Globally hyperbolic spacetimes are the class of manifolds on which the dynamics of fields can be written in terms of a Cauchy problem. Static spacetimes are defined by a Lorentzian metric $g_{\mu\nu}$ satisfying the following property
\begin{align}\label{metric_constraint_static}
& \partial_0 g^{\mu \nu} = 0, & g^{0i} = g^{i0} = 0.
\end{align}
Hyperbolic static spacetimes appear to have a well defined notion of particles as positive frequency solutions of the field equation \cite{Wald:1995yp}.

We note by $(T,\vec{X})$ the coordinate system describing the manifold and by $g_{\mu \nu}$ its metric.  Hereafter, we will use the term ``curved spacetime'' to refer to both genuinely curved spacetimes and flat spacetimes in non inertial frames (e.g., Rindler spacetime for accelerated frames). In other words, we say that the spacetime is curved if $g_{\mu\nu} \neq \eta_{\mu\nu}$.

\subsection{Scalar field}\label{QFT_in_curved_spacetime_scalar}

Here, we consider a free scalar field $\hat{\Phi}$ that is solution of the Klein-Gordon equation in curved spacetime
\begin{equation} \label{Klein_Gordon_curved}
\left[ \frac{c^2}{\sqrt{-g}} \partial_\mu \left( \sqrt{-g} g^{\mu \nu}\partial_\nu   \right) - \left( \frac{mc^2}{\hbar}\right)^2 \right] \hat{\Phi} = 0,
\end{equation}
with $g$ as the determinant of $g_{\mu\nu}$, and satisfying the following commutation relation
\begin{equation}\label{Phi_commutation}
\left[ \hat{\Phi}(T,\vec{X}), \hat{\Phi}^\dagger(T',\vec{X}') \right] = i \hbar \Delta_\text{CKG}(T, \vec{X}, T', \vec{X}'),
\end{equation}
where $\Delta_\text{CKG}(T, \vec{X})$ is the solution of
\begin{subequations}
\begin{align}
& c \sqrt{- g(T,\vec{X})} g^{0 \mu}(T,\vec{X}) \left. \frac{\partial}{\partial X^{\prime \mu}} \Delta_\text{CKG}(T, \vec{X}, T', \vec{X}') \right|_{T=T'} = \delta^3(\vec{X} - \vec{X}'),\\
& \Delta_\text{CKG}(T, \vec{X}, T, \vec{X}') = 0.
\end{align}
\end{subequations}

We also consider the curved Klein-Gordon scalar product
\begin{align}\label{KG_curved_scalar_product}
( \Phi, \Phi' )_\text{CKG} =  & -\frac{i c}{\hbar} \int_{\mathbb{R}^3} d^3X  \sqrt{-g(T,\vec{X})} g^{0 \mu}(T,\vec{X}) \nonumber \\
& \times \left[ \Phi^*(T,\vec{X}) \partial_\mu \Phi'(T,\vec{X})  -  \Phi'(T,\vec{X}) \partial_\mu \Phi^*(T,\vec{X}) \right],
\end{align}
which is time independent for solutions of the curved Klein-Gordon equation (\ref{Klein_Gordon_curved}); this can be proven by computing
\begin{align}
\frac{d}{dT}( \Phi, \Phi' )_\text{CKG}  = & -\frac{i c}{\hbar} \int_{\mathbb{R}^3} d^3X \left[ (\partial_0 \Phi^*) \sqrt{-g} g^{0 \mu}  \partial_\mu \Phi' + \Phi^* \partial_0 (\sqrt{-g} g^{0 \mu}  \partial_\mu \Phi') \right. \nonumber \\
& \left.  - (\partial_0 \Phi') \sqrt{-g} g^{0 \mu}  \partial_\mu \Phi^* -  \Phi' \partial_0 ( \sqrt{-g} g^{0 \mu}  \partial_\mu \Phi^*) \right]\nonumber \\
 = & -\frac{i c}{\hbar} \int_{\mathbb{R}^3} d^3X \left\lbrace (\partial_0 \Phi^*) \sqrt{-g} g^{0 \mu}  \partial_\mu \Phi'  + \Phi^* \left[ -\partial_i (\sqrt{-g} g^{i \mu}  \partial_\mu) \right. \right. \nonumber \\
& \left. + \sqrt{-g} \left( \frac{mc}{\hbar} \right)^2 \right] \Phi'  - (\partial_0 \Phi') \sqrt{-g} g^{0 \mu}  \partial_\mu \Phi^* \nonumber \\
& \left. -  \Phi'  \left[ -\partial_i (\sqrt{-g} g^{i \mu}  \partial_\mu) + \sqrt{-g} \left( \frac{mc}{\hbar} \right)^2 \right]  \Phi^* \right\rbrace\nonumber \\
 = & -\frac{i c}{\hbar} \int_{\mathbb{R}^3} d^3X \left[ (\partial_0 \Phi^*) \sqrt{-g} g^{0 \mu}  \partial_\mu \Phi' + (\partial_i \Phi^* ) \sqrt{-g} g^{i \mu}  \partial_\mu \Phi' \right. \nonumber \\
& \left. - (\partial_0 \Phi') \sqrt{-g} g^{0 \mu}  \partial_\mu \Phi^*  -   (\partial_i \Phi') \sqrt{-g} g^{i \mu}  \partial_\mu   \Phi^* \right]\nonumber \\
 = & -\frac{i c}{\hbar} \int_{\mathbb{R}^3} d^3X \left[ (\partial_\nu \Phi^*) \sqrt{-g} g^{\nu \mu}  \partial_\mu \Phi'  -   (\partial_\nu \Phi') \sqrt{-g} g^{\nu \mu}  \partial_\mu   \Phi^* \right]\nonumber \\
 = & 0.
\end{align}
Consequently, $( \Phi, \Phi' )_\text{CKG}$ can be used as the Hilbert product for positive frequency modes.

By expanding $\hat{\Phi}$ in terms of modes with real frequencies with respect to the time $T$, we obtain
\begin{equation}\label{Klein_Gordon_curved_Phi}
\hat{\Phi}(T,\vec{X}) = \sum_\theta\left[ G(\theta,T,\vec{X}) \hat{A}(\theta)  + H(\theta,T,\vec{X}) \hat{B}^\dagger(\theta) \right],
\end{equation}
where $\hat{A}(\theta)$ ($\hat{B}(\theta)$) is the annihilation operator associated to the particle (antiparticle) mode $G(\theta)$ ($H^*(\theta)$). The modes $G(\theta)$ and $H(\theta)$ are defined to be orthonormal with respect to the curved Klein-Gordon scalar product (\ref{KG_curved_scalar_product}), in the sense that
\begin{subequations}\label{KG_scalar_curved_product_orthonormality_G}
\begin{align}
& ( G(\theta), G(\theta'))_\text{CKG} = \delta_{\theta\theta'} , \\
 & ( H(\theta), H(\theta'))_\text{CKG}  = - \delta_{\theta\theta'} ,\\
 & ( G(\theta), H(\theta'))_\text{CKG} = 0.
\end{align}
\end{subequations}
As in Eq.~(\ref{positive_negative_frequencies}), the definition of positive and negative frequency modes is expressed by
\begin{subequations}\label{scalar_curved_modes_tilde}
\begin{align} 
& G(\theta,T,\vec{X}) = \tilde{G}(\theta,\vec{X}) e^{-i \Omega(\theta) T}, \\
 & H(\theta,T,\vec{X}) = \tilde{H}(\theta,\vec{X}) e^{i \Omega(\theta) T}.
\end{align}
\end{subequations}

At the beginning of this section, we remarked that hyperbolic static spacetimes are provided with the notion of particles as positive frequency solutions of the field equation. In general, the ansatz (\ref{scalar_curved_modes_tilde}) can be incompatible with Eq.~(\ref{Klein_Gordon_curved}) and, hence, the expansion of $\hat{\Phi}$ in positive and negative frequency modes is not always possible. Sufficient condition for the validity of Eq.~(\ref{scalar_curved_modes_tilde}) is given by Eq.~(\ref{metric_constraint_static}). Indeed, when Eq.~(\ref{metric_constraint_static}) holds, Eq.~(\ref{Klein_Gordon_curved}) becomes a Schrödinger equation for $G(\theta) $ with eigenvalues proportional to $\Omega^2$
\begin{equation}\label{scalar_curved_Schrodinger_equation_tilde}
 (\hbar \Omega)^2 G(\theta) = H_\text{CKG} G(\theta), 
\end{equation}
with Hamiltonian
\begin{align}\label{Schrodinger_equation_tilde_Hamiltonian}
H_\text{CKG} = & g_{0 0} \left[  \frac{\hbar^2}{ \sqrt{-g}} \partial_i \left( \sqrt{-g} g^{i j} \partial_j \right)  - (mc)^2 \right],
\end{align}
that is positive with respect to the curved Klein-Gordon scalar product (\ref{KG_curved_scalar_product}) for positive frequency modes. The positivity of $H_\text{CKG}$ guarantees the existence of real $\Omega$ for Eq.~(\ref{scalar_curved_Schrodinger_equation_tilde}).

It is possible to prove that $H_\text{CKG}$ is positive owing to the following identity
\begin{equation}\label{scalar_curved_Klein_Gordon_hamiltonian_positivity}
( \Phi, H_\text{CKG} \Phi' )_\text{CKG} = \delta^{i j} ( H_i \Phi, H_j \Phi' )_\text{CKG}  + ( H_0 \Phi, H_0 \Phi' )_\text{CKG},
\end{equation}
with
\begin{align}
& H_0 =  m c^2 e^0{}_0 , & H_i = \hbar c e^0{}_0 e_i{}^j \partial_j,
\end{align}
where $e_\alpha{}^\mu$ is the vierbein field defined as
\begin{equation}
e_\alpha{}^\mu e_\beta{}^\nu g_{\mu \nu} =  \eta_{\alpha \beta}
\end{equation}
and with $e^\alpha{}_\mu$ as the inverse of $e_\alpha{}^\mu$. Equation (\ref{scalar_curved_Klein_Gordon_hamiltonian_positivity}), in turn, can be proven by using the static spacetime condition (\ref{metric_constraint_static}), which in terms of the vierbein field reads as
\begin{align}\label{vierbein_constraint_static}
& \partial_0 e_\alpha{}^\mu = 0, & e_i{}^0 = e_0{}^i = 0.
\end{align}
The product (\ref{KG_curved_scalar_product}) in static spacetimes [Eq.~(\ref{metric_constraint_static})] becomes
\begin{align}\label{KG_curved_scalar_product_2}
( \Phi, \Phi' )_\text{CKG} = & -\frac{i c}{\hbar} \int_{\mathbb{R}^3} d^3X  \sqrt{-g(\vec{X})} g^{0 0}(\vec{X}) \nonumber \\
& \times \left[ \Phi^*(T,\vec{X}) \partial_0 \Phi'(T,\vec{X})  -  \Phi'(T,\vec{X}) \partial_0 \Phi^*(T,\vec{X}) \right].
\end{align}
By integrating by parts, one obtains
\begin{align}\label{KG_curved_scalar_product_2_proof}
&  ( \Phi, H_\text{CKG} \Phi' )_\text{CKG}\nonumber \\ = & -i\hbar c \int_{\mathbb{R}^3} d^3X  \{ \Phi^* [\partial_0 \partial_i \left( \sqrt{-g} g^{i j} \partial_j \right) \Phi']  -  [\partial_i \left( \sqrt{-g} g^{i j} \partial_j \right) \Phi'] \partial_0 \Phi^*  \} \nonumber \\
& + i \frac{m^2 c^3}{\hbar} \int_{\mathbb{R}^3} d^3X \sqrt{-g}  \left( \Phi^* \partial_0 \Phi' -  \Phi' \partial_0 \Phi^*  \right) \nonumber \\
  = & -i\hbar c \int_{\mathbb{R}^3} d^3X  \sqrt{-g} g^{i j} [ - (\partial_i \Phi^*) \partial_0 \partial_j \Phi' +  (\partial_j \Phi') \partial_i \partial_0 \Phi^*  ] \nonumber \\
&+ i \frac{m^2 c^3}{\hbar} \int_{\mathbb{R}^3} d^3X  \sqrt{-g}  \left( \Phi^* \partial_0 \Phi' -  \Phi' \partial_0 \Phi^*  \right) \nonumber \\
  = &  i \hbar c^3 \int_{\mathbb{R}^3} d^3X   \sqrt{-g} g^{0 0}  e^0{}_0 e^0{}_0 \eta^{i j} e_i{}^{i'} e_j{}^{j'} [ - (\partial_{i'} \Phi^*)  \partial_0 \partial_{j'} \Phi' + ( \partial_{j'} \Phi') \partial_{i'} \partial_0 \Phi^*  ]  \nonumber \\
& - i \frac{m^2c^5}{\hbar} \int_{\mathbb{R}^3} d^3X   \sqrt{-g} g^{0 0} e^0{}_0 e^0{}_0 \left( \Phi^* \partial_0 \Phi' - \Phi' \partial_0 \Phi^*  \right)\nonumber \\
=  & \delta^{i j} ( H_i \Phi, H_j \Phi' )_\text{CKG} + ( H_0 \Phi, H_0 \Phi' )_\text{CKG},
\end{align}
which proves Eq.~(\ref{scalar_curved_Klein_Gordon_hamiltonian_positivity}).

Seemingly, one can prove that $H_\text{CKG}$ is positive with respect to the following scalar product
\begin{equation}\label{KG_curved_scalar_product_2_nonrelativistic}
( \Phi, \Phi' )_{L^2_\text{S}(\mathbb{R}^3)} = - c \int_{\mathbb{R}^3} d^3X  \sqrt{-g(\vec{X})} g^{0 0}(\vec{X}) \Phi^*(\vec{X}) \Phi'(\vec{X}),
\end{equation}
which can be seen as the $L^2(\mathbb{R}^3)$ inner product with a metric dependent measure. The positivity of $H_\text{CKG}$ with respect to $( \Phi, \Phi' )_{L^2_\text{S}(\mathbb{R}^3)}$ can be obtained from
\begin{equation}
( \Phi, H_\text{CKG} \Phi' )_{L^2_\text{S}(\mathbb{R}^3)} = \delta^{i j} ( H_i \Phi, H_j \Phi' )_{L^2_\text{S}(\mathbb{R}^3)}  + ( H_0 \Phi, H_0 \Phi' )_{L^2_\text{S}(\mathbb{R}^3)}.
\end{equation}

As in Sec.~\ref{QFT_in_Minkowski_spacetime_scalar}, we interpret the Klein-Gordon equation for positive frequency solutions as a Schrödinger equation
\begin{equation}
i \hbar \partial_0 G(\theta) = h_\text{CKG} G(\theta),
\end{equation}
with Hamiltonian $h_\text{CKG}$ that is the square root of $H_\text{CKG}$. The equivalent of Eq.~(\ref{h_KG_H_KG}) in curved spacetime is
\begin{equation}\label{h_cKG_H_cKG}
h_\text{CKG} = \sqrt{H_\text{CKG}}.
\end{equation}

In summary, the single particle description of the field is defined by the Hamiltonian $h_\text{CKG}$ and the scalar product $( \Phi, \Phi' )_\text{CKG}$. Instead, general Fock states $| \Phi \rangle$ are represented in the Schrödinger picture by
\begin{equation}\label{wavefunction_g_curved}
\Phi_n (T, \textbf{X}_n) =  \left( \frac{2 m c^2}{\hbar^2} \right)^{n/2} \sum_{\bm{\theta}_n} \tilde{\Phi}_n (\bm{\theta}_n)    \prod_{l=1}^n G(\theta_l,T,\vec{X}_l),
\end{equation}
where $ \tilde{\Phi}_n (\bm{\theta}_n)$ is defined from the decomposition 
\begin{equation} \label{general_Fock_expansion_curved}
| \Phi \rangle  = \sum_{n=1}^\infty \frac{1}{\sqrt{n!}} \sum_{\bm{\theta}_n} \tilde{\Phi}_n (\bm{\theta}_n) \prod_{l=1}^n \hat{A}^\dagger(\theta_l) | 0_\text{C} \rangle + \tilde{\Phi}_0 | 0_\text{C} \rangle,
\end{equation}
and is symmetric with respect to $\theta_1, \dots, \theta_n$. The state $ | 0_\text{C} \rangle $ is the vacuum in the curved spacetime defined by $ \hat{A}(\theta_l)| 0_\text{C} \rangle = \hat{B}(\theta_l)| 0_\text{C} \rangle = 0$.

\subsection{Dirac field}\label{QFT_in_curved_spacetime_Dirac}

Here, we detail the quantum Dirac field theory on a globally hyperbolic static spacetime.

The free field $\hat{\Psi}$ is solution of the curved spacetime Dirac equation (see for instance Ref.~\cite{collas_klein_2019})
\begin{equation} \label{Dirac_curved}
\left( i c e_\alpha{}^\mu \gamma^\alpha D_\mu  - \frac{m c^2}{\hbar} \right) \hat{\Psi} = 0,
\end{equation}
with
\begin{align}
& D_\mu = \partial_\mu + \Gamma_\mu, & \Gamma_\mu (T,\vec{X}) = - \frac{1}{2} \sigma^{\alpha \beta} \omega_{\alpha \beta \mu},
\end{align}
the spin connection
\begin{equation} \label{spin_connection}
\omega_{\alpha \beta \mu} = \eta_{\alpha \gamma} e^\gamma{}_\nu (\partial_\mu e_\beta{}^\nu + \Gamma^\nu{}_{\mu\rho} e_\beta{}^\rho),
\end{equation}
the Christoffel symbols
\begin{equation}
\Gamma^\rho{}_{\mu\nu} = \frac{1}{2} g^{\rho\sigma} (\partial_\nu g_{\sigma\mu} + \partial_\mu g_{\nu\sigma} -\partial_\sigma g_{\mu\nu})
\end{equation}
and the generators of the Clifford algebra
\begin{equation}
\sigma^{\mu\nu} = \frac{1}{4} [\gamma^\mu, \gamma^\nu].
\end{equation}

We consider the product
\begin{equation}\label{Dirac_curved_scalar_product}
( \Psi, \Psi' )_{\mathbb{C}^4 \otimes L^2_\text{D}(\mathbb{R}^3)} = c \int_{\mathbb{R}^3} d^3X \sqrt{-g(T,\vec{X})} e_\alpha{}^0 (T,\vec{X})  \Psi^\dagger(T,\vec{X}) \gamma^0 \gamma^\alpha \Psi'(T,\vec{X}).
\end{equation}
It has been proven \cite{PhysRevD.22.1922} that when the metric is static, $( \Psi, \Psi' )_{\mathbb{C}^4 \otimes L^2_\text{D}(\mathbb{R}^3)}$ is time independent for solutions of Eq.~(\ref{Dirac_curved}). 

As a consequence of condition (\ref{metric_constraint_static}), Eq.~(\ref{vierbein_constraint_static}) holds, together with
\begin{align}
& \partial_0 \Gamma^\rho{}_{\mu\nu} = 0, & \Gamma^0{}_{i j} = \Gamma^i{}_{0 j} = \Gamma^i{}_{j 0} = 0.
\end{align}
Correspondingly, we have that
\begin{align}
& \partial_0 \omega_{\alpha\beta\mu} = 0, & \omega_{i j 0} = \omega_{i 0 j} = \omega_{0 i j} = 0,
\end{align}
which leads to
\begin{align} \label{omega_condition_Dirac}
& \partial_0 \Gamma_\mu = 0, & & \Gamma_0 = -\frac{1}{4} \omega_{0 i 0} \sigma^{0 i}, & & \Gamma_i = -\frac{1}{8} \omega_{j k i} \sigma^{j k}.
\end{align}
By taking in count Eq.~(\ref{gamma_matrices_hermitianity}), we also find that $\Gamma_0$ is hermitian while $\Gamma_i$ antihermitian, i.e.,
\begin{align} \label{omega_condition_Dirac_hermitianity}
&  \Gamma^\dagger_0 = \Gamma_0, & \Gamma^\dagger_i = - \Gamma_i.
\end{align}
Moreover, Eq.~(\ref{Dirac_curved_scalar_product}) becomes
\begin{equation}\label{Dirac_curved_scalar_product_2}
( \Psi, \Psi' )_{\mathbb{C}^4 \otimes L^2_\text{D}(\mathbb{R}^3)} = \frac{1}{c} \int_{\mathbb{R}^3} d^3X \sqrt{-g(\vec{X})} e_0{}^0 (\vec{X})  \Psi^\dagger(T,\vec{X}) \Psi'(T,\vec{X}),
\end{equation}
as a consequence of Eqs.~(\ref{gamma_matrices_anticommutating_rule}) and (\ref{vierbein_constraint_static}). Equation (\ref{Dirac_curved_scalar_product_2}) implies that $( \Psi, \Psi' )_{\mathbb{C}^4 \otimes L^2_\text{D}(\mathbb{R}^3)}$ can be seen as the inner product of $\mathbb{C}^4 \otimes L^2(\mathbb{R}^3)$ but with a metric dependent measure.

In a static spacetime, the Hamiltonian $h_\text{CD}$ associated to the curved Dirac equation (\ref{Dirac_curved}) is hermitian with respect to the scalar product $( \Psi, \Psi' )_{\mathbb{C}^4 \otimes L^2_\text{D}(\mathbb{R}^3)}$. Such a Hamiltonian can be obtained from
\begin{equation}\label{Dirac_curved_static}
\left[ i c e_0{}^0 \gamma^0 (\partial_0 + \Gamma_0) + i c e_i{}^j \gamma^i (\partial_j + \Gamma_j)  - \frac{m c^2}{\hbar} \right] \hat{\Psi} = 0,
\end{equation}
which leads to
\begin{equation}
h_\text{CD} =  - i \hbar c^2 e^0{}_0 e_i{}^j \gamma^0 \gamma^i (\partial_j + \Gamma_j)+ mc^3 e^0{}_0 \gamma^0 - i \hbar \Gamma_0.
\end{equation}
Indeed, by acting with $ \hbar c e^0{}_0 \gamma^0$ on the left of Eq.~(\ref{Dirac_curved_static}) and using Eq.~(\ref{gamma_matrices_anticommutating_rule}), one obtains
\begin{equation}\label{Dirac_curved_Schrodinger}
i \hbar \partial_0 \hat{\Psi} = h_\text{CD} \hat{\Psi}. 
\end{equation}

The proof for the hermicity of $h_\text{CD}$ with respect to $( \Psi, \Psi' )_{\mathbb{C}^4 \otimes L^2_\text{D}(\mathbb{R}^3)}$ arises from the fact that $( \Psi, \Psi' )_{\mathbb{C}^4 \otimes L^2_\text{D}(\mathbb{R}^3)}$ is time independent for solutions of the curved Dirac equation (\ref{Dirac_curved}) and, hence, for solutions of Eq.~(\ref{Dirac_curved_Schrodinger}), which means that
\begin{align}
0 = & i \hbar \frac{d}{dt} ( \Psi, \Psi' )_{\mathbb{C}^4 \otimes L^2_\text{D}(\mathbb{R}^3)} \nonumber \\
 = &  -( i \hbar \partial_0 \Psi, \Psi' )_{\mathbb{C}^4 \otimes L^2_\text{D}(\mathbb{R}^3)} + (\Psi,  i \hbar \partial_0 \Psi' )_{\mathbb{C}^4 \otimes L^2_\text{D}(\mathbb{R}^3)} \nonumber \\
 = & -( h_\text{CD} \Psi, \Psi' )_{\mathbb{C}^4 \otimes L^2_\text{D}(\mathbb{R}^3)} + (\Psi,  h_\text{CD} \Psi' )_{\mathbb{C}^4 \otimes L^2_\text{D}(\mathbb{R}^3)}.
\end{align}

The hermicity of $h_\text{CD}$ guarantees the separation of the field into positive and negative frequency modes as
\begin{equation}\label{Psi}
 \hat{\Psi}(T,\vec{X}) =  \sum_\theta \left[ U(\theta,T,\vec{X}) \hat{C}(\theta) + V(\theta,T,\vec{X}) \hat{D}^\dagger(\theta) \right],
\end{equation}
with
\begin{subequations}\label{Dirac_curved_modes_tilde_positive_negative}
\begin{align} 
& U(\theta,T,\vec{X}) = e^{-i \Omega(\theta) T} \tilde{U}(\theta,\vec{X}), \\
 & V(\theta,T,\vec{X}) = e^{i \Omega(\theta) T} \tilde{V}(\theta,\vec{X}).
\end{align}
\end{subequations}
The positive frequency modes $U(\theta)$, the product $( \Psi, \Psi' )_{\mathbb{C}^4 \otimes L^2_\text{D}(\mathbb{R}^3)}$ and the Hamiltonian $h_\text{CD}$ define the Dirac single particle space in curved spacetime.

A general Fock state $| \Psi \rangle$ is represented in the Schrödinger picture by
\begin{equation}\label{Dirac_curved_wavefunction_general}
\Psi^{\bm{\alpha}_n}_n (T, \textbf{X}_n)   =  \sum_{\bm{\theta}_n} \tilde{\Psi}_n (\bm{\theta}_n) \prod_{l=1}^n U^{\alpha_l}(\theta_l, T, \vec{X}_l).
\end{equation}
where $ \tilde{\Psi}_n (\bm{\theta}_n)$ comes from the decomposition
\begin{equation} \label{general_Fock_Dirac_curved_expansion}
| \Psi \rangle  = \sum_{n=1}^\infty \frac{1}{\sqrt{n!}} \sum_{\bm{\theta}_n} \tilde{\Psi}_n (\bm{\theta}_n) \prod_{l=1}^n \hat{C}^\dagger(\theta_l) | 0_\text{C} \rangle +  \tilde{\Psi}_0 | 0_\text{C} \rangle.
\end{equation}
The inner product between two states $| \Psi \rangle$, $| \Psi' \rangle$ can be obtained from a generalization of $( \Psi, \Psi' )_{\mathbb{C}^4 \otimes L^2_\text{D}(\mathbb{R}^3)}$ for states with arbitrary number of particles
\begin{equation}
\langle \Psi | \Psi' \rangle = \sum_{n=0}^\infty (\Psi_n, \Psi'_n)_{\mathbb{C}^{4n} \otimes L^2_D(\mathbb{R}^{3n})},
\end{equation}
with
\begin{subequations}
\begin{align}
 (\Psi_n, \Psi'_n)_{\mathbb{C}^{4n} \otimes L^2_D(\mathbb{R}^{3n})} = &\frac{1}{c^n} \sum_{\bm{\alpha}_n} \int_{\mathbb{R}^{3n}} d^{3n} \textbf{X}_n  \prod_{l=1}^n\left[ \sqrt{-g(\vec{X}_l)} e_0{}^0 (\vec{X}_l) \right]  \nonumber \\
& \times [\Psi^*_n (T, \textbf{X}_n)]^{\bm{\alpha}_n} [\Psi'_n (T, \textbf{X}_n)]^{\bm{\alpha}_n},\\
(\Psi_0, \Psi'_0)_{\mathbb{C}^0 \otimes L^2_D(\mathbb{R}^0)} = & \Psi_0^* \Psi'_0.
\end{align}
\end{subequations}

\section{Rindler spacetime}\label{QFT_in_curved_spacetime_Rindler}

In Sec.~\ref{QFT_in_curved_spacetime}, we gave a description of fields and particles in hyperbolic static spacetimes. In this section, we consider the Rindler spacetime as a particular example of hyperbolic static spacetime.

By definition, the Rindler coordinates $(T,\vec{X}) = (T, X, Y, Z)$ describe an accelerated frame with acceleration $\alpha$ in a flat manifold. By assuming that the acceleration of the Rindler observer is along the direction of $z$, the coordinate transformation between the inertial $(t,\vec{x})$ and the accelerated frame $(T,\vec{X})$ is 
\begin{align}\label{Rindler_coordinates_transformation_R}
& t = \frac{e^{a Z}}{c a} \sinh(c a T), && x = X, && y = Y, && z = \frac{e^{a Z}}{a} \cosh(c a T),
\end{align}
with $a = \alpha/c^2$. The corresponding metric in the accelerated frame is
\begin{equation} \label{Rindler_metric}
g_{\mu\nu}(T,\vec{X}) = \text{diag} \left( -c^2 e^{2aZ}, 1, 1, e^{2aZ} \right).
\end{equation}

We assume that $a$ is positive, so that the coordinates $(t,\vec{x})$ in Eq.~(\ref{Rindler_coordinates_transformation_R}) cover the right wedge in the Minkowski frame identified by $z > c |t|$. Then, we say that the right Rindler spacetime is defined by the coordinate transformation (\ref{Rindler_coordinates_transformation_R}) and the metric (\ref{Rindler_metric}) with $a > 0$.

The spacelike hypersurfaces with constant $T$ are Cauchy surfaces for the right Rindler frame; however, they are not Cauchy surfaces for the entire Minkowski spacetime. For this reason, one usually considers the left Rindler frame in addition to the right Rindler frame. The coordinates and metric of the left Rindler frame are defined by replacing $a$ with $-a$ in Eqs.~(\ref{Rindler_coordinates_transformation_R}) and (\ref{Rindler_metric}). This new coordinate system covers the left wedge $z<-c|t|$. The union of the left and the right Rindler frames gives the entire Rindler frame. We find that the hypersurface $t = 0$ is a Cauchy surface for both the Minkowski and the Rindler frame.

In summary, the coordinate transformation between the Minkowski and the Rindler frame is
\begin{align}\label{Rindler_coordinates_transformation}
& t = t_\nu(T,\vec{X}), & \vec{x} = \vec{x}_\nu(T,\vec{X}),
\end{align}
where $\nu \in \{ \text{L}, \text{R} \}$ is the variable associated to the left (L) and the right (R) wedges and $\vec{x}_\nu(T,\vec{X})$ is such that
\begin{align}\label{Rindler_coordinates_transformation_2}
& z = z_\nu(T,\vec{X}), & \vec{x}_\perp = \vec{X}_\perp,
\end{align}
where $\vec{x}_\perp = (x,y)$ and $\vec{X}_\perp = (X,Y)$ are the transverse coordinates in each frame. The functions $t_\nu(T,\vec{X})$ and $z_\nu(T,\vec{X})$ appearing in Eqs.~(\ref{Rindler_coordinates_transformation}) and (\ref{Rindler_coordinates_transformation_2}) are
\begin{subequations}\label{Rindler_coordinate_transformation}
\begin{align}
& t_\nu(T,\vec{X}) = \frac{e^{s_\nu a Z}}{c a} \sinh(c a T), \\
 & z_\nu(T,\vec{X}) = s_\nu \frac{e^{s_\nu a Z}}{a} \cosh(c a T), \label{Rindler_coordinate_transformation_x}
\end{align}
\end{subequations}
where $s_\nu$ is such that $s_\text{L} = -1$ and $s_\text{R} = 1$. The spacetime metric in the right wedge is given by Eq.~(\ref{Rindler_metric}); conversely, the left wedge metric is obtained by replacing $a $ with $ -a$ in Eq.~(\ref{Rindler_metric}).

Due to the decomposition of the Rindler spacetime into the left and the right wedge, the scalar field in the Rindler spacetime is described by the couple $\hat{\Phi}_\nu = (\hat{\Phi}_\text{L}, \hat{\Phi}_\text{R})$, while the Dirac field by $\hat{\Psi}_\nu = (\hat{\Psi}_\text{L}, \hat{\Psi}_\text{R})$. Fields in opposite wedges are defined independently, in the sense that scalar (Dirac) fields in one wedge always commute (anticommute) with fields in the other wedge.

In this section, we study the fields $\hat{\Phi}_\nu$ and $\hat{\Psi}_\nu$ and we show their decomposition in terms of real frequency modes. This will provide the corresponding modal representation of Rindler-Fock states.

\subsection{Scalar field}\label{QFT_in_curved_spacetime_Rindler_scalar}

In the case of scalar fields in the right Rindler spacetime, Eq.~(\ref{Klein_Gordon_curved}) becomes
\begin{equation}\label{Rindler_Klein_Gordon}
\left\lbrace - \partial_0^2 + c^2 \partial_3^2 +   c^2 e^{2 a Z} \left[ \partial_1^2 + \partial_2^2 - \left( \frac{mc}{\hbar} \right)^2 \right] \right\rbrace \hat{\Phi}_\text{R} = 0.
\end{equation}
An explicit decomposition of $\hat{\Phi}_\text{R} $ is known \cite{RevModPhys.80.787} and can be obtained by considering the frequency $\Omega$ and the transverse momentum components $\vec{K}_\perp = (K_1,K_2)$ as the quantum numbers $\theta = \vec{\theta} = (\Omega, \vec{K}_\perp)$. The decomposition is
\begin{align}
\hat{\Phi}_\text{R}(T,\vec{X}) = & \int_0^{\infty} d\Omega \int_{\mathbb{R}^2} d^2 K_\perp   \left[ F(\Omega,\vec{K}_\perp,T,\vec{X}) \hat{A}_\text{R}(\Omega,\vec{K}_\perp) \right. \nonumber \\
& \left. + F^*(\Omega,\vec{K}_\perp,T,\vec{X})\hat{B}_\text{R}^\dagger(\Omega,\vec{K}_\perp) \right],
\end{align}
with
\begin{subequations} \label{F_Rindler_all}
\begin{align}
& F(\Omega,\vec{K}_\perp,T,\vec{X}) = \tilde{F}(\Omega,\vec{K}_\perp,Z) e^{ i \vec{K}_\perp \cdot \vec{X}_\perp - i \Omega T }, \label{F_Rindler}  \\
& \tilde{F}(\Omega,\vec{K}_\perp,Z) = \frac{1}{2 \pi^2 c} \sqrt{ \frac{\hbar}{a} \left| \sinh \left( \frac{\beta \Omega}{2} \right) \right| }  K_{i \Omega /ca} \left( \kappa (\vec{K}_\perp) \frac{e^{aZ}}{a} \right), \label{F_tilde_Rindler}
\end{align}
\end{subequations}
and $\beta = 2 \pi/c a $. The functions appearing in Eq.~(\ref{F_tilde_Rindler}) are
\begin{equation}\label{kappa_k_perp}
\kappa (\vec{K}_\perp) = \sqrt{ \left( \frac{m c}{\hbar} \right)^2 + |\vec{K}_\perp|^2 }
\end{equation}
and $K_\zeta (\xi)$ as the modified Bessel function of the second kind [Appendix \ref{Bessel_functions}].

It can be noticed that, in the right Rindler spacetime, $g_{00} = -c \sqrt{-g}$ and, hence,
\begin{align}\label{Rindler_Klein_Gordon_product}
& ( \Phi, \Phi' )_\text{CKG} = ( \Phi, \Phi' )_\text{KG}, & ( \Phi, \Phi' )_{L^2_\text{S}(\mathbb{R}^3)} = ( \Phi, \Phi' )_{L^2(\mathbb{R}^3)}.
\end{align}
The modes $F(\Omega,\vec{K}_\perp)$ defined in Eq.~(\ref{F_Rindler}) are orthonormal with respect to the scalar product $( \Phi, \Phi' )_\text{CKG}$, in the sense that
\begin{subequations}\label{KG_scalar_curved_product_orthonormality_F}
\begin{align}
& ( F(\vec{\theta}), F(\vec{\theta}'))_\text{CKG} = \delta^3 (\vec{\theta}-\vec{\theta}') , \\
 & ( F^*(\vec{\theta}), F^*(\vec{\theta}'))_\text{CKG}  = - \delta^3 (\vec{\theta}-\vec{\theta}') ,\\
 & ( F(\vec{\theta}), F^*(\vec{\theta}'))_\text{CKG} = 0.
\end{align}
\end{subequations}
As a consequence of Eq.~(\ref{Rindler_Klein_Gordon_product}), the modes $F(\Omega,\vec{K}_\perp)$ are also orthonormal with respect to $( \Phi, \Phi' )_\text{KG}$.

In the left wedge, $\hat{\Phi}_\text{L} $ can be decomposed as $\hat{\Phi}_\text{R}$ with $Z $ replaced by $ -Z$. Indeed, the Klein-Gordon equation in Rindler spacetime (\ref{Rindler_Klein_Gordon}) is invariant under the transformation $a \mapsto -a$, $Z \mapsto -Z$ and the orthonormality condition (\ref{KG_scalar_curved_product_orthonormality_F})
also holds for the modes $F(\Omega,\vec{K}_\perp,T, \vec{X}_\perp, -Z)$ in the left wedge. Therefore, by considering both wedges, the field $\hat{\Phi}_\nu (T,\vec{X})$ is
\begin{align} \label{Rindler_scalar_decomposition}
\hat{\Phi}_\nu (T,\vec{X}) = & \int_0^{\infty} d\Omega \int_{\mathbb{R}^2} d^2 K_\perp \left[ F_\nu(\Omega,\vec{K}_\perp,T, \vec{X})\hat{A}_\nu(\Omega,\vec{K}_\perp)  \right. \nonumber \\
& \left. + F_\nu^*(\Omega,\vec{K}_\perp,T, \vec{X})\hat{B}_\nu^\dagger(\Omega,\vec{K}_\perp) \right],
\end{align}
with
\begin{equation}\label{F_F_nu}
F_\nu(\Omega,\vec{K}_\perp,T, \vec{X}) = F(\Omega,\vec{K}_\perp,T, \vec{X}_\perp,s_\nu Z).
\end{equation}

Particle wave functions are defined as
\begin{equation}\label{wavefunction_F}
\Phi_n (T, \bm{\nu}_n, \textbf{X}_n) =  \left( \frac{2 m c^2}{\hbar^2} \right)^{n/2}  \sum_{\bm{\theta}_n} \tilde{\Phi}_n (\bm{\nu}_n,\bm{\theta}_n)    \prod_{l=1}^n F(\vec{\theta}_l,T,X_l, Y_l, s_{\nu_l} Z_l),
\end{equation}
where $(\bm{\nu}_n,\bm{\theta}_n) = ((\nu_1,\vec{\theta}_1) \dots, (\nu_n , \vec{\theta}_n))$ is a collection of quantum numbers including wedges $\nu \in \{ \text{L}, \text{R} \}$ and energy-momentum degrees of freedom $\vec{\theta} = (\Omega, \vec{K}_\perp)$. The function $ \tilde{\Phi}_n (\bm{\nu}_n,\bm{\theta}_n)$ is defined from the decomposition
\begin{equation} \label{general_Fock_expansion_Rindler}
| \Phi \rangle  = \sum_{n=1}^\infty \frac{1}{\sqrt{n!}} \sum_{\bm{\nu}_n,\bm{\theta}_n} \tilde{\Phi}_n (\bm{\nu}_n,\bm{\theta}_n) \prod_{l=1}^n \hat{A}_{\nu_l}^\dagger(\vec{\theta}_l) | 0_\text{L}, 0_\text{R} \rangle +  \tilde{\Phi}_0 | 0_\text{L}, 0_\text{R} \rangle,
\end{equation}
with $| 0_\text{L}, 0_\text{R} \rangle$ as the Rindler vacuum.

Particle states that are only made by right Rindler particles are such that $\Phi_n (T, \bm{\nu}_n, \textbf{X}_n) = 0$ if $\nu_l = \text{L}$ for some $l \in \{ 1, \dots, n \}$. For simplicity, they can be described by the wave function
\begin{equation}\label{wavefunction_F_right}
\Phi_n (T, \textbf{X}_n) =  \left( \frac{2 m c^2}{\hbar^2} \right)^{n/2}  \sum_{\bm{\theta}_n} \tilde{\Phi}_n (\bm{\theta}_n)    \prod_{l=1}^n F(\vec{\theta}_l,T, \vec{X}_l),
\end{equation}
with $ \tilde{\Phi}_n (\bm{\theta}_n)$ defined by
\begin{equation} \label{general_Fock_expansion_right_Rindler}
| \Phi \rangle  = \sum_{n=1}^\infty \frac{1}{\sqrt{n!}} \sum_{\bm{\theta}_n} \tilde{\Phi}_n (\bm{\theta}_n) \prod_{l=1}^n \hat{A}_{\text{R}}^\dagger(\vec{\theta}_l) | 0_\text{L}, 0_\text{R} \rangle +  \tilde{\Phi}_0 | 0_\text{L}, 0_\text{R} \rangle.
\end{equation}

\subsection{Dirac field}\label{Rindler_Dirac_modes}

In this subsection, we study the Dirac field in Rindler coordinates $\hat{\Psi}_\nu (T,\vec{X})$ defined by the Rindler-Dirac equation
\begin{equation} \label{Dirac_Rindler}
\left[  e^{-s_\nu aZ} \left( i c \gamma^0 \partial_0 + s_\nu i  \frac{c a}{2} \gamma^3 +  i c \gamma^3 \partial_3 \right) + i c \gamma^1 \partial_1 + i c \gamma^2 \partial_2 - \frac{m c^2}{\hbar}  \right] \hat{\Psi}_\nu = 0.
\end{equation}
We derive the orthonormal positive and negative frequency modes $U_{\nu s}(\Omega,\vec{K}_\perp)$ and $V_{\nu s}(\Omega,\vec{K}_\perp)$ that are solutions of Eq.~(\ref{Dirac_Rindler}), have the form of
\begin{subequations}\label{UV_UV_tilde}
\begin{align}
& U_{\nu s}(\Omega,\vec{K}_\perp,T,\vec{X}) = e^{i \vec{K}_\perp \cdot \vec{X}_\perp - i \Omega T} \tilde{U}_{\nu s}(\Omega,\vec{K}_\perp,Z),\label{U_U_tilde} \\
& V_{\nu s}(\Omega,\vec{K}_\perp,T,\vec{X}) = e^{-i \vec{K}_\perp \cdot \vec{X}_\perp + i \Omega T} \tilde{V}_{\nu s}(\Omega,\vec{K}_\perp,Z)\label{V_V_tilde}
\end{align}
\end{subequations}
and generate Dirac fields in Rindler spacetime as
\begin{align} \label{free_Dirac_field_Rindler}
\hat{\Psi}_\nu(T,\vec{X}) = & \sum_{s=1}^2 \int_0^\infty d\Omega \int_{\mathbb{R}^2} d^2 K_\perp  \left[ U_{\nu s}(\Omega,\vec{K}_\perp,T,\vec{X}) \hat{C}_{\nu s}(\Omega,\vec{K}_\perp) \right. \nonumber \\
& \left. + V_{\nu s}(\Omega,\vec{K}_\perp,T,\vec{X}) \hat{D}_{\nu s}^\dagger(\Omega,\vec{K}_\perp) \right].
\end{align}
The orthonormality condition for such modes is
\begin{subequations}\label{D_scalar_product_orthonormality_U_V}
\begin{align}
& ( U_{\nu s}(\Omega,\vec{K}_\perp), U_{\nu s'}(\Omega',\vec{K}'_\perp) )_\nu = \delta_{ss'} \delta(\Omega-\Omega') \delta^2(\vec{K}_\perp-\vec{K}'_\perp),\label{D_scalar_product_orthonormality_U}\\
& ( V_{\nu s}(\Omega,\vec{K}_\perp), V_{\nu s'}(\Omega',\vec{K}'_\perp) )_\nu = \delta_{ss'} \delta(\Omega-\Omega') \delta^2(\vec{K}_\perp-\vec{K}'_\perp),\label{D_scalar_product_orthonormality_V}\\
&  (U_{\nu s}(\Omega,\vec{K}_\perp), V_{\nu s'}(\Omega',\vec{K}'_\perp) )_\nu = 0,
\end{align}
\end{subequations}
with
\begin{equation}\label{Dirac_Rindler_scalar_product}
( \Psi, \Psi' )_\nu = \int_{\mathbb{R}^3} d^3X e^{s_\nu a Z} \Psi^\dagger(T,\vec{X}) \Psi'(T,\vec{X})
\end{equation}
as Rindler-Dirac product.

The operators $\hat{C}^\dagger_{\nu s}(\Omega,\vec{K}_\perp)$ and $\hat{D}^\dagger_{\nu s}(\Omega,\vec{K}_\perp)$ appearing in Eq.~(\ref{free_Dirac_field_Rindler}) create particles and antiparticles of the $\nu$ wedge with spin number $s$, frequency $\Omega$ and transverse momentum $\vec{K}_\perp$ and satisfy the anticommutation rules
\begin{subequations}\label{anticommutating_rules}
\begin{align}
& \{ \hat{C}_{\nu s}(\Omega,\vec{K}_\perp), \hat{C}^\dagger_{\nu' s'}(\Omega',\vec{K}'_\perp) \}  = \delta_{\nu \nu'} \delta_{ss'} \delta(\Omega-\Omega') \delta^2(\vec{K}_\perp-\vec{K}'_\perp), \\
&  \{ \hat{D}_{\nu s}(\Omega,\vec{K}_\perp), \hat{D}^\dagger_{\nu' s'}(\Omega',\vec{K}'_\perp) \}  = \delta_{\nu \nu'} \delta_{ss'} \delta(\Omega-\Omega') \delta^2(\vec{K}_\perp-\vec{K}'_\perp),  \\
&  \{ \hat{C}_{\nu s}(\Omega,\vec{K}_\perp), \hat{C}_{\nu' s'}(\Omega',\vec{K}'_\perp) \}  = 0, \label{anticommutating_rules_c} \\
&  \{ \hat{D}_{\nu s}(\Omega,\vec{K}_\perp), \hat{D}_{\nu' s'}(\Omega',\vec{K}'_\perp) \}  = 0, \label{anticommutating_rules_d}\\
&   \{ \hat{C}_{\nu s}(\Omega,\vec{K}_\perp), \hat{D}_{\nu' s'}(\Omega',\vec{K}'_\perp) \}  = 0, \label{anticommutating_rules_e}\\
&   \{ \hat{C}_{\nu s}(\Omega,\vec{K}_\perp), \hat{D}^\dagger_{\nu' s'}(\Omega',\vec{K}'_\perp) \}  = 0 \label{anticommutating_rules_f}.
\end{align}
\end{subequations}

By extending the definition of the variable $\Omega$ also for negative values, one may define the function
\begin{equation} \label{W_UV}
W_{\nu s}(\Omega,\vec{K}_\perp,T,\vec{X}) = \begin{cases}
U_{\nu s}(\Omega,\vec{K}_\perp,T,\vec{X}) & \text{if } \Omega>0 \\
V_{\nu s}(-\Omega,-\vec{K}_\perp,T,\vec{X}) & \text{if } \Omega<0
\end{cases},
\end{equation}
that includes both positive and negative frequency solutions of the Rindler-Dirac equation (\ref{Dirac_Rindler}). Equation (\ref{free_Dirac_field_Rindler}) can now be written in the following equivalent ways
\begin{subequations}\label{free_Dirac_field_Rindler_2}
\begin{align} 
\hat{\Psi}_\nu(T,\vec{X}) = & \sum_{s=1}^2 \int_\mathbb{R} d\Omega \int_{\mathbb{R}^2} d^2 K_\perp W_{\nu s}(\Omega,\vec{K}_\perp,T,\vec{X})   \nonumber \\
& \times \left[ \theta(\Omega) \hat{C}_{\nu s}(\Omega,\vec{K}_\perp)+  \theta(-\Omega)\hat{D}_{\nu s}^\dagger(-\Omega,-\vec{K}_\perp) \right], \\
\hat{\Psi}_\nu(T,\vec{X}) = & \sum_{s=1}^2 \int_\mathbb{R} d\Omega \int_{\mathbb{R}^2} d^2 K_\perp W_{\nu s}(-\Omega,-\vec{K}_\perp,T,\vec{X})   \nonumber \\
& \times \left[ \theta(-\Omega) \hat{C}_{\nu s}(-\Omega,-\vec{K}_\perp)+  \theta(\Omega)\hat{D}_{\nu s}^\dagger(\Omega,\vec{K}_\perp) \right],
\end{align}
\end{subequations}
with $\theta$ as the Heaviside step function. The orthonormality condition (\ref{D_scalar_product_orthonormality_U_V}) with respect to the modes $W_{\nu s}(\Omega,\vec{K}_\perp)$ is
\begin{equation}\label{D_scalar_product_orthonormality_W}
( W_{\nu s}(\Omega,\vec{K}_\perp), W_{\nu s'}(\Omega',\vec{K}'_\perp) )_\nu = \delta_{ss'} \delta(\Omega-\Omega') \delta^2(\vec{K}_\perp-\vec{K}'_\perp).
\end{equation}

Notice that Eq.~(\ref{W_UV}) is compatible with Eq.~(\ref{UV_UV_tilde}). Indeed, one may define the function $\tilde{W}_{\nu s}(\Omega,\vec{K}_\perp,Z)$ such that
\begin{equation} \label{UVW_UVW_tilde}
W_{\nu s}(\Omega,\vec{K}_\perp,T,\vec{X}) = e^{i \vec{K}_\perp \cdot \vec{X}_\perp - i \Omega T} \tilde{W}_{\nu s}(\Omega,\vec{K}_\perp,Z)
\end{equation}
and that
\begin{equation} \label{W_tilde_UV_tilde}
\tilde{W}_{\nu s}(\Omega,\vec{K}_\perp,Z) = \begin{cases}
\tilde{U}_{\nu s}(\Omega,\vec{K}_\perp,Z) & \text{if } \Omega>0 \\
\tilde{V}_{\nu s}(-\Omega,-\vec{K}_\perp,Z) & \text{if } \Omega<0
\end{cases}.
\end{equation}

The modes $W_{\nu s}(\Omega,\vec{K}_\perp)$ are solutions of the Rindler-Dirac equation (\ref{Dirac_Rindler}); hence, $\tilde{W}_{\nu s}(\Omega,\vec{K}_\perp,Z)$ satisfies the differential equation
\begin{align} \label{Dirac_Rindler_2}
& \left[ e^{-s_\nu  aZ} \left( \Omega \gamma^0 + s_\nu i \frac{a}{2} \gamma^3 +  i \gamma^3 \partial_3  \right) \right. \nonumber \\
& \left. -  \left( K_1 \gamma^1 + K_2 \gamma^2 + \frac{m c}{\hbar} \right) \right] \tilde{W}_{\nu s}(\Omega,\vec{K}_\perp,Z) = 0.
\end{align}
By multiplying Eq.~(\ref{Dirac_Rindler_2}) with $c \gamma^0$ on the left and using Eq.~(\ref{gamma_matrices_anticommutating_rule}), one obtains
\begin{align} \label{Dirac_Rindler_3}
& \left[  e^{-s_\nu aZ} \left( \frac{\Omega}{c} + s_\nu i \frac{c a}{2} \gamma^0 \gamma^3 +  i c \gamma^0 \gamma^3 \partial_3 \right)  \right. \nonumber \\
& \left. - s_\nu i \kappa (\vec{K}_\perp) \mathfrak{G}_\nu (\vec{K}_\perp) \right] \tilde{W}_{\nu s}(\Omega,\vec{K}_\perp,Z) = 0,
\end{align}
with
\begin{equation}\label{Gamma}
\mathfrak{G}_\nu (\vec{K}_\perp) = - \frac{s_\nu i c} {\kappa (\vec{K}_\perp)} \gamma^0 \left( K_1 \gamma^1 + K_2 \gamma^2 + \frac{m c}{\hbar} \right),
\end{equation}
and $\kappa (\vec{K}_\perp)$ given by Eq.~(\ref{kappa_k_perp}).

The spinor $\tilde{W}_{\nu s}(\Omega,\vec{K}_\perp,Z)$ can be decomposed into eigenvectors of $c \gamma^0 \gamma^3$ with eigenvalues $\pm 1$ by using the projections
\begin{equation}\label{P_pm}
P_\pm = \frac{1}{2} (1 \pm c \gamma^0 \gamma^3).
\end{equation}
The projected modes $\tilde{W}^\pm_{\nu s}(\Omega,\vec{K}_\perp,Z)$ are such that
\begin{subequations}
\begin{align}
& \tilde{W}_{\nu s}(\Omega,\vec{K}_\perp,Z) = \tilde{W}^+_{\nu s}(\Omega,\vec{K}_\perp,Z) + \tilde{W}^-_{\nu s}(\Omega,\vec{K}_\perp,Z), \label{U_U_p_U_m}\\
 & \tilde{W}^\pm_{\nu s}(\Omega,\vec{K}_\perp,Z) = P_\pm \tilde{W}_{\nu s}(\Omega,\vec{K}_\perp,Z), \label{U_pm}\\ 
& c\gamma^0 \gamma^3 \tilde{W}^\pm_{\nu s}(\Omega,\vec{K}_\perp,Z) = \pm \tilde{W}^\pm_{\nu s}(\Omega,\vec{K}_\perp,Z) \label{U_pm_eigenvectors}.
\end{align}
\end{subequations}
By using Eqs.~(\ref{gamma_matrices_anticommutating_rule}) and (\ref{Gamma}) one can prove that
\begin{equation}\label{gamma_03_gamma_012}
\gamma^0 \gamma^3 \mathfrak{G}_\nu (\vec{K}_\perp) = - \mathfrak{G}_\nu (\vec{K}_\perp) \gamma^0 \gamma^3.
\end{equation}
Hence, the projections $P_\pm $ and the matrix $\mathfrak{G}_\nu (\vec{K}_\perp)$ are related by
\begin{equation}\label{P_mp_gamma_012}
P_\pm \mathfrak{G}_\nu (\vec{K}_\perp) = \mathfrak{G}_\nu (\vec{K}_\perp)  P_\mp,
\end{equation}
which can be proved by using Eqs.~(\ref{P_pm}) and (\ref{gamma_03_gamma_012}). By projecting Eq.~(\ref{Dirac_Rindler_3}) with respect to $P_\pm$ and using Eqs.~(\ref{U_pm}), (\ref{U_pm_eigenvectors}) and (\ref{P_mp_gamma_012}), one obtains the following coupled equations for $\tilde{W}^+_{\nu s}(\Omega,\vec{K}_\perp,Z)$ and $\tilde{W}^-_{\nu s}(\Omega,\vec{K}_\perp,Z)$
\begin{align} \label{Dirac_Rindler_4}
& e^{-s_\nu aZ} \left[ \frac{\Omega}{c} \pm i \left( s_\nu \frac{a}{2}  +  \partial_3 \right)  \right] \tilde{W}^\pm_{\nu s}(\Omega,\vec{K}_\perp,Z) \nonumber \\
= & s_\nu i \kappa (\vec{K}_\perp) \mathfrak{G}_\nu (\vec{K}_\perp) \tilde{W}^\mp_{\nu s}(\Omega,\vec{K}_\perp,Z).
\end{align}

Equation (\ref{Dirac_Rindler_4}) can be decoupled by applying $e^{-s_\nu aZ}[\Omega/c \mp i (s_\nu a/2 + \partial_3)]$ on the left, leading to
\begin{align} \label{Dirac_Rindler_5}
& e^{-s_\nu aZ} \left[ \frac{\Omega}{c} \mp i \left( s_\nu \frac{a}{2}  +  \partial_3 \right)  \right] \left\lbrace e^{-s_\nu aZ} \left[ \frac{\Omega}{c}  \pm i \left( s_\nu \frac{a}{2}  +  \partial_3 \right)  \right] \right\rbrace \tilde{W}^\pm_{\nu s}(\Omega,\vec{K}_\perp,Z) \nonumber \\
= & - \kappa^2 (\vec{K}_\perp)  \mathfrak{G}^2_\nu (\vec{K}_\perp)   \tilde{W}^\pm_{\nu s}(\Omega,\vec{K}_\perp,Z).
\end{align}
The derivative operator on left side of Eq.~(\ref{Dirac_Rindler_5}) can be computed as
\begin{align} \label{Dirac_Rindler_5_left}
& \left[ \frac{\Omega}{c} \mp i \left( s_\nu \frac{a}{2}  +  \partial_3 \right)  \right] \left\lbrace e^{-s_\nu aZ} \left[ \frac{\Omega}{c} \pm i \left( s_\nu \frac{a}{2}  +  \partial_3 \right)  \right] \right\rbrace \nonumber \\
= & e^{-s_\nu aZ}  \left\lbrace \pm s_\nu i a \left[ \frac{\Omega}{c} \pm i \left( s_\nu \frac{a}{2}  +  \partial_3 \right)  \right] + \left( \frac{\Omega}{c} \right)^2 + \left( s_\nu \frac{a}{2}  +  \partial_3 \right)^2  \right\rbrace \nonumber \\
= & e^{-s_\nu aZ}  \left[ \pm s_\nu i a  \frac{\Omega}{c} - \frac{a^2}{4}   + \left( \frac{\Omega}{c} \right)^2  +  \partial^2_3    \right] \nonumber \\
= & e^{-s_\nu aZ}  \left[ \left( \frac{\Omega}{c} \pm s_\nu i \frac{a}{2}  \right)^2  +  \partial^2_3    \right].
\end{align}
The right side of Eq.~(\ref{Dirac_Rindler_5}), instead, can be computed by using Eqs.~(\ref{gamma_matrices_anticommutating_rule}), (\ref{Gamma}) and (\ref{kappa_k_perp}),
\begin{align}\label{Dirac_Rindler_5_right}
\mathfrak{G}^2_\nu (\vec{K}_\perp)= & - \frac{c^2} {\kappa^2 (\vec{K}_\perp)} \gamma^0 \left( K_1 \gamma^1 + K_2 \gamma^2 + \frac{m c}{\hbar} \right) \gamma^0 \left( K_1 \gamma^1 + K_2 \gamma^2 + \frac{m c}{\hbar} \right)  \nonumber \\
= &  - \frac{c^2} {\kappa^2 (\vec{K}_\perp)} \gamma^0 \gamma^0 \left( -K_1 \gamma^1 - K_2 \gamma^2 + \frac{m c}{\hbar} \right)\left( K_1 \gamma^1 + K_2 \gamma^2 + \frac{m c}{\hbar} \right)  \nonumber \\
= & - \frac{1} {\kappa^2 (\vec{K}_\perp)} \left( -K_1 \gamma^1 - K_2 \gamma^2 + \frac{m c}{\hbar} \right)  \left( K_1 \gamma^1 + K_2 \gamma^2 + \frac{m c}{\hbar} \right)  \nonumber \\
= & - \frac{1} {\kappa^2 (\vec{K}_\perp)} \left[  -K_1^2 \gamma^1 \gamma^1 - K_2^2 \gamma^2 \gamma^2 + \left( \frac{m c}{\hbar} \right)^2  - K_1 K_2 \{\gamma^1, \gamma^2 \} \right] \nonumber \\
= & - \frac{1} {\kappa^2 (\vec{K}_\perp)} \left[  K_1^2 + K_2^2  + \left( \frac{m c}{\hbar} \right)^2 \right] \nonumber \\
= &  -1.
\end{align}
By using Eqs.~(\ref{Dirac_Rindler_5_left}) and (\ref{Dirac_Rindler_5_right}) in Eq.~(\ref{Dirac_Rindler_5}), one obtains
\begin{equation} \label{Dirac_Rindler_6}
 e^{-s_\nu 2aZ} \left[ \left( \frac{\Omega}{c} \pm s_\nu i \frac{a}{2} \right)^2 + \partial_3^2  \right] \tilde{W}^\pm_{\nu s}(\Omega,\vec{K}_\perp,Z) = \kappa^2 (\vec{K}_\perp)  \tilde{W}^\pm_{\nu s}(\Omega,\vec{K}_\perp,Z).
\end{equation}

The solutions of Eq.~(\ref{Dirac_Rindler_6}) that converge to $0$ for $s_\nu Z \rightarrow +\infty$ have the form of
\begin{equation}\label{U_pm_s_W_pm_s}
\tilde{W}^\pm_{\nu s}(\Omega,\vec{K}_\perp,Z) = \mathfrak{K} (\pm s_\nu \Omega,\vec{K}_\perp,s_\nu Z) \mathfrak{W}^\pm_{\nu s}(\Omega,\vec{K}_\perp),
\end{equation}
where
\begin{equation}\label{K_pm}
\mathfrak{K}(\Omega,\vec{K}_\perp,Z) = K_{i \Omega /ca - 1/2} \left( \kappa (\vec{K}_\perp) \frac{e^{aZ}}{a} \right)
\end{equation}
and $K_\zeta (\xi)$ is the modified Bessel function of the second kind. An integral representation for $K_\zeta (\xi)$ can be found in Appendix \ref{Bessel_functions_Integral_representation}. Notice that Eq.~(\ref{Dirac_Rindler_6}) is a necessary but not sufficient condition for the modes $\tilde{W}^\pm_{\nu s}(\Omega,\vec{K}_\perp,Z)$. Indeed, Eq.~(\ref{Dirac_Rindler_6}) is a decoupled second order differential equation originated from the first order differential equation (\ref{Dirac_Rindler_4}). Hence, we now look for the spinor functions $\mathfrak{W}^\pm_{\nu s}(\Omega,\vec{K}_\perp)$ of Eq.~(\ref{U_pm_s_W_pm_s}) such that $\tilde{W}^\pm_{\nu s}(\Omega,\vec{K}_\perp,Z)$ satisfies Eq.~(\ref{Dirac_Rindler_4}).

The first order derivatives of $\tilde{W}^\pm_{\nu s}(\Omega,\vec{K}_\perp,Z)$ that appear in Eq.~(\ref{Dirac_Rindler_4}) can be computed by using the following recurrence relation for Bessel functions \cite{abramowitz1965handbook}
\begin{equation}\label{Bessel_derivative_recursive}
\partial_\xi K_\zeta (\xi) - \frac{\zeta}{\xi} K_\zeta (\xi) = -K_{\zeta + 1} (\xi)
\end{equation}
and the fact that $K_\zeta (\xi)$ is even with respect to the order $\zeta$, which means that
\begin{equation}\label{Bessel_derivative_recursive_2}
\partial_\xi K_\zeta (\xi) - \frac{\zeta}{\xi} K_\zeta (\xi) = -K_{-\zeta - 1} (\xi).
\end{equation}
By considering $\xi = \kappa (\vec{K}_\perp) \exp(s_\nu aZ)/a$ and $\zeta = \pm s_\nu i \Omega /ca - 1/2 $, one may write Eq.~(\ref{Bessel_derivative_recursive_2}) in terms of the functions $\mathfrak{K}(\Omega,\vec{K}_\perp,Z)$ [Eq.~(\ref{K_pm})] as
\begin{equation}\label{K_recurrence}
\frac{e^{-s_\nu aZ}}{\kappa (\vec{K}_\perp)} \left(s_\nu \partial_3 \mp s_\nu  i \frac{\Omega}{c} + \frac{a}{2} \right) \mathfrak{K}(\pm s_\nu \Omega,\vec{K}_\perp,s_\nu Z) = - \mathfrak{K} (\mp s_\nu \Omega,\vec{K}_\perp,s_\nu Z) ,
\end{equation}
which, multiplied with $ \pm s_\nu i \kappa (\vec{K}_\perp)$, becomes
\begin{align}\label{K_recurrence_2}
&  e^{-s_\nu aZ} \left[\frac{\Omega}{c} \pm i \left( s_\nu \frac{a}{2} + \partial_3 \right) \right] \mathfrak{K}(\pm s_\nu \Omega,\vec{K}_\perp,s_\nu Z)  \nonumber \\
= & \mp s_\nu i \kappa (\vec{K}_\perp) \mathfrak{K} (\mp s_\nu \Omega,\vec{K}_\perp,s_\nu Z) .
\end{align}
By using Eqs.~(\ref{U_pm_s_W_pm_s}) and (\ref{K_recurrence_2}) in Eq.~(\ref{Dirac_Rindler_4}) one obtains the following linear equation for $\mathfrak{W}^\pm_{\nu s}(\Omega,\vec{K}_\perp)$
\begin{equation} \label{Dirac_Rindler_4_W}
 \mathfrak{W}^\pm_{\nu s}(\Omega,\vec{K}_\perp)= \mp  \mathfrak{G}_\nu (\vec{K}_\perp)  \mathfrak{W}^\mp_{\nu s}(\Omega,\vec{K}_\perp).
\end{equation}

The two equations appearing in Eq.~(\ref{Dirac_Rindler_4_W}) are equivalent. This can be proven by acting on Eq.~(\ref{Dirac_Rindler_4_W}) with $ \pm \mathfrak{G}_\nu (\vec{K}_\perp)$ and by using Eq.~(\ref{Dirac_Rindler_5_right}). Hence, one can consider a single spinor function $\mathfrak{W}_{\nu s}(\Omega,\vec{K}_\perp)$ such that
\begin{equation}\label{W_pm_W}
\mathfrak{W}^\pm_{\nu s}(\Omega,\vec{K}_\perp) = \left[ \mathfrak{G}_\nu(\vec{K}_\perp) \right]^{(1 \mp 1)/2}  \mathfrak{W}_{\nu s}(\Omega,\vec{K}_\perp).
\end{equation}

Notice that each spinor $\mathfrak{W}^\pm_{\nu s} (\Omega,\vec{K}_\perp)$ is eigenvector of $c \gamma^0 \gamma^3$ with eigenvalue $\pm 1$ [Eqs.~(\ref{U_pm_eigenvectors}) and (\ref{U_pm_s_W_pm_s})]. Hence the following identity must be considered together with Eq.~(\ref{W_pm_W})
\begin{equation}\label{W_pm_eigenvalues}
c\gamma^0 \gamma^3 \mathfrak{W}^\pm_{\nu s}(\Omega,\vec{K}_\perp) = \pm \mathfrak{W}^\pm_{\nu s}(\Omega,\vec{K}_\perp).
\end{equation}
Equations (\ref{W_pm_W}) and (\ref{W_pm_eigenvalues}) are outnumbered. Indeed, one may consider one of the two equations appearing in Eq.~(\ref{W_pm_eigenvalues}) and obtain the other by using Eq.~(\ref{W_pm_W}). For instance, by choosing $c\gamma^0 \gamma^3 \mathfrak{W}^+_{\nu s}(\Omega,\vec{K}_\perp) = \mathfrak{W}^+_{\nu s}(\Omega,\vec{K}_\perp)$, one can use Eq.~(\ref{gamma_03_gamma_012}) and (\ref{W_pm_W}) to prove that
\begin{align}
 c\gamma^0 \gamma^3 \mathfrak{W}^-_{\nu s}(\Omega,\vec{K}_\perp) = & c\gamma^0 \gamma^3 \mathfrak{G}_\nu(\vec{K}_\perp)  \mathfrak{W}_{\nu s}(\Omega,\vec{K}_\perp)\nonumber \\
= & c\gamma^0 \gamma^3 \mathfrak{G}_\nu(\vec{K}_\perp)  \mathfrak{W}^+_{\nu s}(\Omega,\vec{K}_\perp)\nonumber \\
= & - c \mathfrak{G}_\nu(\vec{K}_\perp) \gamma^0 \gamma^3  \mathfrak{W}^+_{\nu s}(\Omega,\vec{K}_\perp)\nonumber \\
= & - \mathfrak{G}_\nu(\vec{K}_\perp)  \mathfrak{W}^+_{\nu s}(\Omega,\vec{K}_\perp) \nonumber \\
= & - \mathfrak{G}_\nu(\vec{K}_\perp)  \mathfrak{W}_{\nu s}(\Omega,\vec{K}_\perp) \nonumber \\
= & - \mathfrak{W}^-_{\nu s}(\Omega,\vec{K}_\perp).
\end{align}
Both equations appearing in Eq.~(\ref{W_pm_eigenvalues}) are equivalent to the following single equation
\begin{equation}\label{W_p_eigenvalues}
c\gamma^0 \gamma^3 \mathfrak{W}_{\nu s}(\Omega,\vec{K}_\perp) = \mathfrak{W}_{\nu s}(\Omega,\vec{K}_\perp).
\end{equation}
 
The third identity defining $\mathfrak{W}_{\nu s} (\Omega,\vec{K}_\perp)$ comes from the orthonormality condition (\ref{D_scalar_product_orthonormality_W}). The product $( W_{\nu s}(\Omega,\vec{K}_\perp), W_{\nu {s'}}(\Omega',\vec{K}'_\perp) )_\nu$ can be computed by using Eqs.~(\ref{Dirac_Rindler_scalar_product}), (\ref{UVW_UVW_tilde}), (\ref{U_U_p_U_m}), (\ref{U_pm_s_W_pm_s}) and the orthogonality condition between eigenstates of $c \gamma^0 \gamma^3$ with different eigenvalues. Explicitly, the product reads as
\begin{align} \label{W_R_W_R_product}
& ( W_{\nu s}(\Omega,\vec{K}_\perp), W_{\nu {s'}}(\Omega',\vec{K}'_\perp) )_\nu  = e^{ i( \Omega - \Omega') T}  \sum_{\sigma = \pm} \int_{\mathbb{R}^3} d^3X e^{s_\nu aZ} e^{i (\vec{K}'_\perp-\vec{K}_\perp) \cdot \vec{X}_\perp}   \nonumber \\
 & \times  \mathfrak{K}^*(\sigma s_\nu \Omega,\vec{K}_\perp,s_\nu Z)  \mathfrak{K}(\sigma s_\nu \Omega',\vec{K}_\perp',s_\nu Z) \left[ \mathfrak{W}^\sigma_{\nu s} (\Omega,\vec{K}_\perp) \right]^\dagger \mathfrak{W}^\sigma_{\nu s'} (\Omega',\vec{K}_\perp').
\end{align}

By using Eqs.~(\ref{gamma_matrices_identities}) and (\ref{Gamma}), one can prove that $\mathfrak{G}_\nu(\vec{K}_\perp)$ is antihermitian
\begin{align}\label{G_antihermitian}
 \mathfrak{G}^\dagger_\nu(\vec{K}_\perp) = & \frac{s_\nu i c} {\kappa (\vec{K}_\perp)} \left[ \gamma^0 \left( K_1 \gamma^1 + K_2 \gamma^2 + \frac{m c}{\hbar} \right) \right]^\dagger \nonumber \\
= & \frac{s_\nu i c} {\kappa (\vec{K}_\perp)} \left(-K_1 \gamma^1 - K_2 \gamma^2 + \frac{m c}{\hbar} \right) \gamma^0 \nonumber \\
= & \frac{s_\nu i c} {\kappa (\vec{K}_\perp)} \gamma^0 \left(K_1 \gamma^1 + K_2 \gamma^2 + \frac{m c}{\hbar} \right) \nonumber \\
= & - \mathfrak{G}_\nu(\vec{K}_\perp).
\end{align}
Equations (\ref{Dirac_Rindler_5_right}) and (\ref{G_antihermitian}) imply that $\mathfrak{G}_\nu(\vec{K}_\perp)$ is also unitary
\begin{equation}\label{G_unitarity}
\mathfrak{G}^\dagger_\nu(\vec{K}_\perp) \mathfrak{G}_\nu(\vec{K}_\perp) = 1.
\end{equation}
By using Eqs.~(\ref{W_pm_W}) and (\ref{G_unitarity}) one can prove that for any $\sigma=\pm$,
\begin{equation}
 \left[ \mathfrak{W}^\sigma_{\nu s} (\Omega,\vec{K}_\perp) \right]^\dagger \mathfrak{W}^\sigma_{\nu s'} (\Omega',\vec{K}_\perp')= \mathfrak{W}^\dagger_{\nu s} (\Omega,\vec{K}_\perp) \mathfrak{W}_{\nu s'} (\Omega',\vec{K}_\perp'),
\end{equation}
which means that Eq.~(\ref{W_R_W_R_product}) reads as
\begin{align} \label{W_R_W_R_product_2}
& ( W_{\nu s}(\Omega,\vec{K}_\perp), W_{\nu {s'}}(\Omega',\vec{K}'_\perp) )_\nu = e^{ i( \Omega - \Omega') T} \mathfrak{W}^\dagger_{\nu s} (\Omega,\vec{K}_\perp) \mathfrak{W}_{\nu s'} (\Omega',\vec{K}_\perp')  \nonumber \\
 & \times \sum_{\sigma = \pm} \int_{\mathbb{R}^3} d^3X  e^{s_\nu aZ} e^{i (\vec{K}'_\perp-\vec{K}_\perp) \cdot \vec{X}_\perp}  \mathfrak{K}^*(\sigma s_\nu \Omega,\vec{K}_\perp,s_\nu Z)  \mathfrak{K}(\sigma s_\nu \Omega',\vec{K}_\perp',s_\nu Z) .
\end{align}
Furthermore, one can use the following property for the Bessel function 
\begin{equation}\label{Bessel_conjugate}
K^*_\zeta (\xi) = K_{\zeta^*} (\xi),
\end{equation}
with $\xi \in \mathbb{R}$. A proof for Eq.~(\ref{Bessel_conjugate}) can be obtained by considering the integral representation for the Bessel function [Appendix \ref{Bessel_functions_Integral_representation}]. In terms of the functions $\mathfrak{K}(\Omega,\vec{K}_\perp,Z)$, Eq.~(\ref{Bessel_conjugate}) reads as
\begin{equation}\label{K_conjugate}
\mathfrak{K}^*(\Omega,\vec{K}_\perp,Z) = \mathfrak{K}(- \Omega,\vec{K}_\perp,Z),
\end{equation}
which can be plugged in Eq.~(\ref{W_R_W_R_product_2}) to give
\begin{align} \label{W_R_W_R_product_3}
& ( W_{\nu s}(\Omega,\vec{K}_\perp), W_{\nu {s'}}(\Omega',\vec{K}'_\perp) )_\nu = e^{ i( \Omega - \Omega') T}   \mathfrak{W}^\dagger_{\nu s} (\Omega,\vec{K}_\perp) \mathfrak{W}_{\nu s'} (\Omega',\vec{K}_\perp')  \nonumber \\
 & \times \sum_{\sigma = \pm} \int_{\mathbb{R}^3} d^3X  e^{s_\nu aZ} e^{i (\vec{K}'_\perp-\vec{K}_\perp) \cdot \vec{X}_\perp}  \mathfrak{K}(-\sigma s_\nu \Omega,\vec{K}_\perp,s_\nu Z)  \mathfrak{K}(\sigma s_\nu \Omega',\vec{K}_\perp',s_\nu Z) .
\end{align}

By computing the integral with respect to $X$ and $Y$ in Eq.~(\ref{W_R_W_R_product_3}), one obtains
\begin{align} \label{W_R_W_R_product_4}
& ( W_{\nu s}(\Omega,\vec{K}_\perp), W_{\nu {s'}}(\Omega',\vec{K}'_\perp) )_\nu = 4 \pi^2  \delta^2(\vec{K}_\perp-\vec{K}_\perp')   e^{ i( \Omega - \Omega') T} \nonumber \\
 & \times \mathfrak{W}^\dagger_{\nu s} (\Omega,\vec{K}_\perp) \mathfrak{W}_{\nu s'} (\Omega',\vec{K}_\perp')  \sum_{\sigma = \pm} \int_{\mathbb{R}} dZ e^{s_\nu aZ} \nonumber \\
 & \times  \mathfrak{K}(-\sigma s_\nu \Omega,\vec{K}_\perp,s_\nu Z)  \mathfrak{K}(\sigma s_\nu \Omega',\vec{K}_\perp,s_\nu Z).
\end{align}
The integral with respect to $Z$, instead, can be computed by using the following identity for Bessel functions
\begin{equation}\label{Bessel_orthonormal}
\int_0^\infty d\xi \left[ K_{-i \zeta-1/2}(\xi) K_{i \zeta'-1/2}(\xi) + K_{i \zeta-1/2}(\xi) K_{-i \zeta'-1/2}(\xi) \right] = \frac{\pi^2 \delta(\zeta-\zeta')}{\cosh(\pi \zeta)}.
\end{equation}
A proof for Eq.~(\ref{Bessel_orthonormal}) can be found in Appendix \ref{Proof_of_Bessel_orthonormal}. By replacing $\xi$, $\zeta$ and $\zeta'$ with $\kappa (\vec{K}_\perp) e^{s_\nu aZ} /a$,  $ s_\nu \Omega/ca$ and $ s_\nu \Omega'/ca$, respectively, in Eq.~(\ref{Bessel_orthonormal}), one obtains the identity
\begin{align}\label{K_p_orthogonal}
& \sum_{\sigma = \pm} \int_{\mathbb{R}} dZ e^{s_\nu aZ}  \mathfrak{K}(- \sigma s_\nu \Omega,\vec{K}_\perp,s_\nu Z) \mathfrak{K}(\sigma s_\nu \Omega',\vec{K}_\perp,s_\nu Z) \nonumber \\
= & \frac{\pi^2 c a \delta(\Omega-\Omega')}{\kappa (\vec{K}_\perp)}  \left[ \cosh \left( \frac{\beta}{2} \Omega \right) \right]^{-1},
\end{align}
with $\beta = 2 \pi/c a $. Equation (\ref{K_p_orthogonal}) can be plugged in Eq.~(\ref{W_R_W_R_product_4}) to give
\begin{align} \label{W_R_W_R_product_5}
& ( W_{\nu s}(\Omega,\vec{K}_\perp), W_{\nu {s'}}(\Omega',\vec{K}'_\perp) )_\nu = \delta(\Omega-\Omega')  \delta^2(\vec{K}_\perp-\vec{K}_\perp') \nonumber \\
 & \times \frac{4 \pi^4  c a  }{\kappa (\vec{K}_\perp) } \mathfrak{W}^\dagger_{\nu s} (\Omega,\vec{K}_\perp) \mathfrak{W}_{\nu s'} (\Omega',\vec{K}_\perp')   \left[ \cosh \left( \frac{\beta}{2} \Omega \right) \right]^{-1},
\end{align}
which is equivalent to Eq.~(\ref{D_scalar_product_orthonormality_W}) only when the following condition is met
\begin{equation}\label{W_orthogonormal}
  \mathfrak{W}^\dagger_{\nu s} (\Omega,\vec{K}_\perp) \mathfrak{W}_{\nu s'} (\Omega',\vec{K}_\perp')  = \delta_{ss'} \frac{\kappa (\vec{K}_\perp)}{4 \pi^4 c a} \cosh \left( \frac{\beta}{2} \Omega \right).
\end{equation}

Equation (\ref{W_orthogonormal}) suggests the definition of the spinor function $\tilde{\mathfrak{W}}_{\nu s}(\Omega, \vec{K}_\perp)$ such that
\begin{equation}\label{W_W_tilde}
\mathfrak{W}_{\nu s}(\Omega,\vec{K}_\perp) = \frac{1}{2 \pi^2} \sqrt{\frac{\kappa (\vec{K}_\perp)}{c a} \cosh \left( \frac{\beta}{2} \Omega \right)} \tilde{\mathfrak{W}}_{\nu s}(\Omega, \vec{K}_\perp).
\end{equation}
The equations defining $\tilde{\mathfrak{W}}_{\nu s}(\Omega, \vec{K}_\perp)$ are given by Eqs.~(\ref{W_p_eigenvalues}) and (\ref{W_orthogonormal}) and explicitly read as
\begin{subequations} \label{W_tilde_constraints}
\begin{align}
& c\gamma^0 \gamma^3 \tilde{\mathfrak{W}}_{\nu s}(\Omega, \vec{K}_\perp) =  \tilde{\mathfrak{W}}_{\nu s}(\Omega, \vec{K}_\perp),\label{W_tilde_constraints_b} \\
&  \tilde{\mathfrak{W}}^\dagger_{\nu s} (\Omega,\vec{K}_\perp)  \tilde{\mathfrak{W}}_{\nu s'} (\Omega,\vec{K}_\perp) = \delta_{ss'} \label{W_tilde_constraints_c}.
\end{align}
\end{subequations}
For fixed $\nu$, $\Omega$ and $\vec{K}_\perp$ and for varying $s=\{ 1, 2 \}$, the spinors $\tilde{\mathfrak{W}}_{\nu s}(\Omega, \vec{K}_\perp)$ are an orthonormal basis for the eigenspace of $c \gamma^0 \gamma^3$ with eigenvalue $1$. Hence, the only freedom left by Eq.~(\ref{W_tilde_constraints}) is about the arbitrary choice for the spin basis $\tilde{\mathfrak{W}}_{\nu s}(\Omega, \vec{K}_\perp)$. 

Any change of basis $\tilde{\mathfrak{W}}_{\nu s}(\Omega, \vec{K}_\perp) \mapsto \bar{\tilde{\mathfrak{W}}}_{\nu s'}(\Omega, \vec{K}_\perp)$ is defined by an unitary matrix $\bar{M}_{\nu ss'} (\Omega, \vec{K}_\perp)$ (with matrix indices $s$ and $s'$), as
\begin{subequations}
\begin{align}
& \bar{\tilde{\mathfrak{W}}}_{\nu s}(\Omega, \vec{K}_\perp) = \sum_{s'=1}^2 \bar{M}_{\nu ss'} (\Omega, \vec{K}_\perp) \tilde{\mathfrak{W}}_{\nu s'}(\Omega, \vec{K}_\perp),\\
& \bar{M}_{\nu ss'} (\Omega, \vec{K}_\perp) = \tilde{\mathfrak{W}}^\dagger_{\nu s'}(\Omega, \vec{K}_\perp) \bar{\tilde{\mathfrak{W}}}_{\nu s}(\Omega, \vec{K}_\perp).
\end{align}
\end{subequations}

Notice that for any basis $\tilde{\mathfrak{W}}_{\nu s}(\Omega, \vec{K}_\perp)$ satisfying Eq.~(\ref{W_tilde_constraints}), also the spinor functions $\tilde{\mathfrak{W}}_{\bar{\nu} s}(-\Omega, \vec{K}_\perp)$ (with $\bar{\nu}$ as the opposite of $\nu$, i.e., $\bar{\nu}=\text{L} $ if $\nu= \text{R}$ and $\bar{\nu}= \text{R}$ if $\nu= \text{L}$) satisfy Eq.~(\ref{W_tilde_constraints}). By acknowledging this symmetry, we prove the existence of the change of basis $M_{\nu ss'}(\Omega, \vec{K}_\perp)$ such that
\begin{subequations}
\begin{align}
& \tilde{\mathfrak{W}}_{\bar{\nu} s}(-\Omega, \vec{K}_\perp) = \sum_{s'=1}^2 M_{\nu ss'}(\Omega, \vec{K}_\perp) \tilde{\mathfrak{W}}_{\nu s'}(\Omega, \vec{K}_\perp),\label{W_L_W_R}\\
& M_{\nu ss'} (\Omega, \vec{K}_\perp) =  \tilde{\mathfrak{W}}^\dagger_{\nu s'}(\Omega, \vec{K}_\perp)  \tilde{\mathfrak{W}}_{\bar{\nu} s}(-\Omega, \vec{K}_\perp).\label{M}
\end{align}
\end{subequations}
The unitarity of $M_{\nu ss'} (\Omega, \vec{K}_\perp)$ reads as
\begin{subequations}\label{M_unitary}
\begin{align}
& \sum_{s''=1}^2 M^*_{\nu s''s} (\Omega, \vec{K}_\perp)  M_{\nu s''s'} (\Omega, \vec{K}_\perp) = \delta_{ss'}, \\
& \sum_{s''=1}^2 M^*_{\nu ss''} (\Omega, \vec{K}_\perp)  M_{\nu s's''} (\Omega, \vec{K}_\perp) = \delta_{ss'}.
\end{align}
\end{subequations}

Hereafter we do not specify any particular solution of Eq.~(\ref{W_tilde_constraints}). Instead, we consider a general basis $\tilde{\mathfrak{W}}_{\nu s}(\Omega, \vec{K}_\perp)$ for the eigenspace of $c\gamma^0 \gamma^3$ with eigenvalue $1$. In Sec.~\ref{Minkowski_vacuum_in_Rindler_spacetime}, we will show that for different choices of $\tilde{\mathfrak{W}}_{\nu s}(\Omega, \vec{K}_\perp)$, different Rindler-Fock representations of the Minkowski vacuum exist. Then, by tracing the left wedge, the dependency of $\tilde{\mathfrak{W}}_{\nu s}(\Omega, \vec{K}_\perp)$ will disappear [Sec.~\ref{Unruh_effect_for_Dirac_fields}]. Only in Sec.~\ref{Basis_choice}, we will discuss different choices for the spin basis $\tilde{\mathfrak{W}}_{\nu s}(\Omega, \vec{K}_\perp)$.

In conclusion, the positive and negative frequency modes for Dirac fields in Rindler spacetimes are
\begin{subequations}\label{UV_UV_tilde_conclusion}
\begin{align}
& U_{\nu s}(\Omega,\vec{K}_\perp,T,\vec{X}) = e^{i \vec{K}_\perp \cdot \vec{X}_\perp - i \Omega T} \tilde{W}_{\nu s}(\Omega,\vec{K}_\perp,Z),\label{U_U_tilde_conclusion} \\
& V_{\nu s}(\Omega,\vec{K}_\perp,T,\vec{X}) = e^{-i \vec{K}_\perp \cdot \vec{X}_\perp + i \Omega T} \tilde{W}_{\nu s}(-\Omega,-\vec{K}_\perp,Z)\label{V_V_tilde_conclusion},
\end{align}
\end{subequations}
with
\begin{equation}\label{W_tilde_W_tilde}
\tilde{W}_{\nu s}(\Omega,\vec{K}_\perp,Z_\nu(z)) = \sum_{\sigma=\pm}  \mathfrak{K} (\sigma s_\nu \Omega,\vec{K}_\perp,s_\nu Z)   \left[ \mathfrak{G}_\nu(\vec{K}_\perp) \right]^{(1 - \sigma)/2} \mathfrak{W}_{\nu s}(\Omega,\vec{K}_\perp).
\end{equation}
To obtain Eq.~(\ref{W_tilde_W_tilde}), use Eqs.~(\ref{U_U_p_U_m}), (\ref{U_pm_s_W_pm_s}) and (\ref{W_pm_W}). The explicit expression of $\mathfrak{K} (\Omega,\vec{K}_\perp,Z) $, $  \mathfrak{G}_\nu(\vec{K}_\perp) $ and $ \mathfrak{W}_{\nu s}(\Omega,\vec{K}_\perp)$ can be obtained, respectively, from Eqs.~(\ref{K_pm}), (\ref{Gamma}) and (\ref{W_W_tilde}), with $\tilde{\mathfrak{W}}_{\nu s}(\Omega, \vec{K}_\perp)$ as solutions of Eq.~(\ref{W_tilde_constraints}).

\chapter{Non relativistic limit of QFT and QFTCS}\label{Non_relativistic_limit_of_QFT_and_QFTCS}

\textit{This chapter is based on and contains material from Ref.~\citeRF{PhysRevD.107.045012}.}

\section{Introduction}\label{Non_relativistic_limit_of_QFT_and_QFTCS_Introduction}

The study of gravitational effects in quantum mechanics is driven by the search for a bridge between general relativity and the quantum theory. In the last twenty years, a remarkable series of experiments reported evidence of gravitational effects on the discrete spectrum of neutron bouncing~\cite{article1, PhysRevD.67.102002, article2, article3, article4, PhysRevLett.112.071101, Kamiya:2014qia}. These experiments confirmed the prediction of neutron wave functions having the form of Airy functions in the presence of an homogeneous gravity field.

\begin{figure}
\center
\includegraphics[]{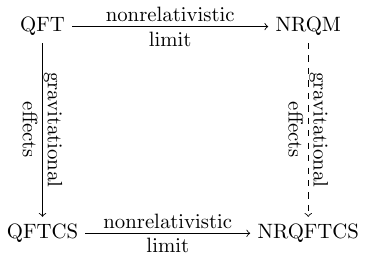}
\caption{The links between Quantum Field Theory (QFT), NonRelativistic Quantum Mechanics (NRQM), Quantum Field Theory in a Curved Space Time (QFTCS), and its nonrelativistic limit (NRQFTCS). The path NRQM $\rightarrow$ NRQFTCS is not rigorous as it ignores the relativistic nature of the fields.}\label{theory_limits}
\end{figure}

The reported observations can be explained by the NonRelativistic Quantum Mechanics (NRQM) with an external gravitational Newtonian potential. This theoretical approach is the first step to analyze phenomena in the regime of Non-Relativistic Quantum Field Theory in Curved Spacetime (NRQFTCS), ignoring the back-reaction of quantum particles on the gravitational field and any eventual quantum nature of gravity. In Fig.~\ref{theory_limits}, we represent the approach by two vertexes (NRQM and NRQFTCS).

Despite being the most direct attack on the problem, the former approach can be inconsistent or too simplified. Indeed, the NRQM description of quantum particles approximates the fully-relativistic Quantum Field Theory (QFT) into a non-covariant theory. Therefore, in NRQM, we ignore the relativistic nature of fields. As a result, we may miss some interactions between matter and gravity arising from covariance, e.g., spin-gravity couplings for Dirac fields. A nonrelativistic theory cannot furnish General Relativistic (GR) corrections. On the other hand, the experimental precision may eventually increase to the point that these GR corrections become detectable.

By looking at Fig.~\ref{theory_limits}, we identify these steps with the path QFT $\rightarrow$ NRQM $\rightarrow$ NRQFTCS. The fully-relativistic QFT is approximated by NRQM in the nonrelativistic limit, and, then, by considering gravitational effects, one studies the NRQFTCS regime. The nonrelativistic limit (QFT $\rightarrow$ NRQM) cancels out information \textit{before} the gravitational effects are introduced (NRQM $\rightarrow$ NRQFTCS).

Another way to address the problem exists. Instead of introducing the gravitational effects \textit{after} the nonrelativistic limit, we may consider them \textit{before} such a limit. In this way, we are able to take track of the GR corrections on the gravity-matter interaction avoiding the inconsistencies. The procedure relies on the Quantum Field Theory in Curved Spacetime (QFTCS), which is the description of fully-relativistic quantum fields affected by a gravitational field. QFTCS also ignores the back-reaction of the field on the metric (i.e., the gravity is not quantum), but it is the simplest attempt to a quantum theory that takes into account a non-flat metric. We identify the new approach in Fig.~\ref{theory_limits} through the path QFT $\rightarrow$ QFTCS $\rightarrow$ NRQFTCS, and corresponds to the nonrelativistic limit of a fully-relativistic quantum field theory in a curved spacetime.

The most known predictions of QFTCS are the Hawking \cite{Hawking:1975vcx}, and Unruh effect \cite{PhysRevD.7.2850, Davies:1974th, PhysRevD.14.870}, which have never been directly observed due to their inaccessible energy scales. Conversely, the neutron-bouncing experiments \cite{article1, PhysRevD.67.102002, article2, article3, article4, PhysRevLett.112.071101, Kamiya:2014qia} prove that the NRQFTCS regime is nowadays experimentally accessible. This circumstance motivates the study of the nonrelativistic limit of QFTCS. For instance, in a recent work \cite{2021IJMPD..3050098R}, the problem of quantum bouncing particles in a gravitational field is discussed in the context of QFTCS. By solving the Dirac equation in Rindler spacetime with bouncing boundary conditions, the authors found GR corrections to the energy spectrum of the neutrons in a gravitational field. Others considered related scenarios: authors in \cite{PhysRevD.22.1922, PhysRevD.25.3180} found the perturbations of the energy levels of an atom placed in curved spacetime; in \cite{PhysRevA.88.022121}, a generalized Schwarzschild metric is used to investigate GR corrections with gravitational spin-orbit coupling. These results were derived from solving the Dirac equation in Rindler spacetime in a nonrelativistic limit and, hence, by following the path QFTCS $\rightarrow$ NRQFTCS.

Here, we report on a general procedure to perform the nonrelativistic limit for bosonic and fermionic fields in a static spacetime. We consider complex scalar and Dirac fields and provide the nonrelativistic description of quantum particles in terms of wave functions, scalar product, and Hamiltonian. As detailed in the next section, the nonrelativistic limit of QFT in Minkowski spacetime is well-established. Despite early investigation \cite{PhysRevD.22.1922, PhysRevD.25.3180, PhysRevD.84.085018, PhysRevA.88.022121}, the case of curved spacetime has not been extensively considered.

It is known that, in the Minkowski spacetime, the time evolution of free nonrelativistic single particles can be approximately described by the free Hamiltonian, which has the same form for both scalar and Dirac fields. Indeed, the Klein-Gordon and the Dirac equation asymptotically lead to the same nonrelativistic Schrödinger equation. For a Dirac field, the spinorial components---obeying the same Schrödinger equation---are decoupled and can be treated as spectral degeneracy. Therefore, without a spin-dependent interaction or enough experimental precision, a Minkowski observer cannot distinguish the time evolution of a nonrelativistic scalar particle from a Dirac particle. This also happens if one introduces a first-order correction due to a weak gravitational field. In the case of a Rindler spacetime with a nearly flat metric and for nonrelativistic particles, the first correction introduced in the Schrödinger equation corresponds to the Newtonian gravitational potential, with no difference between scalar and Dirac field.

By considering GR corrections, the difference between scalar and Dirac fields appears. In this chapter, we show that metrics not approximated by the flat spacetime lead to a non-vanishing difference between the Schrödinger equations arising from the Klein-Gordon and Dirac equation in curved spacetime. A spin-metric coupling occurs, and the observer can distinguish between a scalar and a Dirac particle. We also show that for approximately flat metrics, such coupling can be observed at different orders. For sufficiently large curvature, the precision required to distinguish between scalar and Dirac fields is lower than the one needed in flat spacetime.

The chapter is organized as follows. In Sec.~\ref{Minkowski_spacetime} we give a review for the nonrelativistic limit of scalar and Dirac fields in the Minkowski spacetime. We also show how nonrelativistic particles are approximately solutions to the same Schrödinger equation. Sec.~\ref{NonMinkowksi_spacetime}, is devoted to the curved case. We derive the nonrelativistic limit of fields in a static spacetime and we show how the approximated Schrödinger equations differ in the two cases. Finally, we detail these results for the case of Rindler metric in Sec.~\ref{Rindler_frame}. Conclusions are drawn in Sec.~\ref{Conclusions}.

\section{Minkowski spacetime}\label{Minkowski_spacetime}

In the standard formulation of NRQM, states are represented by time-evolved wave functions satisfying the Schrödinger equation. If we ignore internal degrees of freedom, representatives of states are elements of the $L^2(\mathbb{R}^3)$ Hilbert space. This is known as the Schrödinger position representation. We wonder if the same representation can be obtained in the nonrelativistic regime of QFT.

The nonrelativistic limit of QFT in Minkowski spacetimes is well understood and discussed in the literature. In different textbooks one can find the usual procedure to recover NRQM as the limiting case of QFT \cite{srednicki_2007, zee2010quantum, schwartz_2013}. In these works, one can see how the fully-relativistic Klein-Gordon equation can be approximated by the familiar Schrödinger equation with vanishing potential. Consequently, one may assume that solutions of the Klein-Gordon equation can be identified as relativistic wave functions (i.e., representatives of relativistic states), in analogy to the standard approach to NRQM. However, this is not possible, since the Klein-Gordon equation is second order in the time derivative and, hence, does not provide conservation of probability \cite{srednicki_2007}. Also, one cannot use the Klein-Gordon scalar product as the Hilbert product between relativistic states, because it is positive definite only for positive frequency solutions [Eq.~(\ref{KG_scalar_product_orthonormality_f})]. The issue is solved if one only considers positive frequency modes as representatives of particle states.

In Sec~\ref{QFT_in_Minkowski_spacetime}, we derived the representation space of relativistic particles in terms of positive frequency modes. There, the Hilbert product between states is represented by the Klein-Gordon scalar product between the respective modes. In the present section, we show how the standard representation of states in NRQM is recovered by the nonrelativistic limit. In particular, we demonstrate that positive frequency solutions of the Klein-Gordon equation are approximated by solutions of the Schrödinger equation and that the Klein-Gordon scalar product is approximated by the standard $L^2(\mathbb{R}^3)$ scalar product between wave functions.

In addition to spin-less scalar fields $\hat{\phi}$, we consider Dirac fields $\hat{\psi}$ \cite{srednicki_2007} as well. Also, we extend the results to interacting relativistic fields and obtain Schrödinger equations with non-vanishing potential.

Additionally, we revise the nonrelativistic limit of flat QFT by considering generic solutions of the Klein-Gordon equation (i.e., modes with not defined momentum). In this way, we address a problem that arises when one switches from the flat to the curved case. In Ref.~\cite{Padmanabhan2017ObtainingTN}, it has been argued that the difficulty around the definition of wave functions in QFT comes from the problematic definition of position particle states. Conversely, states with defined momentum are well-defined in both NRQM and QFT and have been used to connect the two theories. This fact does not occur in curved spacetimes, where particles are defined as solutions of the curved Klein-Gordon equation and, hence, do not have defined momentum. Therefore, one may be interested in recovering the nonrelativistic limit of QFT by avoiding modes with defined momentum. In this section, we perform such limit with generic modes in flat spacetimes. The curved scenario will be then discussed in the next section.

\subsection{Scalar field}\label{Minkowski_spacetime_Scalar_field}

Here, we study the scalar field $\hat{\phi}$. We start by considering the case without interaction and we use the decomposition in positive and negative frequency modes with fixed momenta which was detailed in Sec.~\ref{QFT_in_Minkowski_spacetime_scalar}. The positive frequency modes $f(\vec{k})$ form a basis for the Hilbert space of single particles, while the Klein-Gordon product between modes represents the inner product between the corresponding states. By considering the nonrelativistic limit, we show that the Klein-Gordon scalar product can be approximated by the usual $L^2(\mathbb{R}^3)$ inner product of NRQM and the evolution of states is approximately described by the free Schrödinger equation. In this way, we recover the nonrelativistic description of free particles in terms of wave functions, scalar product and free Hamiltonian.

Then, we derive the analogue description of nonrelativistic particles starting from the general decomposition of the field in terms of positive and negative frequency modes $g(\theta)$ and $h(\theta)$, which are not necessarily associated to particles with fixed momenta. Also, we estimate the errors for the nonrelativistic approximations.

Finally, we describe the interacting case by a non-vanishing external potential. We represent a generic particle state as a time-dependent combination of free-evolving modes. We show that in the nonrelativistic limit, these wave functions are approximately solutions of a Schrödinger equation with a potential. Also, we show that the product of two single particle states can be approximated by the $L^2(\mathbb{R}^3)$ product of their wave functions.

\subsubsection{Free modes}

In Sec.~\ref{QFT_in_Minkowski_spacetime_scalar}, we derived the representation of particles in terms of the wave functions $\tilde{\phi}_n (\textbf{k}_n)$ and $\phi_n (t, \textbf{x}_n)$ by means of Eqs.~(\ref{free_state_decomposition}) and (\ref{free_wave_function}). One can use this representation to define the nonrelativistic condition as a constraint for the support of $\tilde{\phi}_n (\textbf{k}_n)$.

The nonrelativistic regime is characterized by energies that are very close to the mass energy and, hence, by momenta $\vec{k}$ satisfying
\begin{equation}\label{non_relativistic_limit}
\left| \frac{\hbar \omega(\vec{k})}{mc^2} - 1 \right| \lesssim \epsilon.
\end{equation}
Here, $\epsilon \ll 1$ is the nonrelativistic parameter representing the maximum ratio between the nonrelativistic energy $E=\hbar \omega - mc^2$ and the mass energy $mc^2$. We say that $| \phi \rangle $ is nonrelativistic if the corresponding wave function $\tilde{\phi}_n (\textbf{k}_n)$ is non-vanishing only for momenta $\vec{k}$ satisfying Eq.~(\ref{non_relativistic_limit}). Explicitly, this means that
\begin{equation}\label{non_relativistic_limit_wavefunction}
\tilde{\phi}_n (\textbf{k}_n) \approx 0 \text{ if there is an } l \in \{ 1, \dots , n \} \text{ such that } \left| \frac{\hbar \omega(\vec{k}_l)}{mc^2} - 1 \right| \gg \epsilon.
\end{equation}

When Eq.~(\ref{non_relativistic_limit}) holds, the frequency dispersion relation (\ref{omega_k}) can be approximated by
\begin{equation} \label{omega_k_approximation}
\omega(\vec{k}) \approx \frac{m c^2}{\hbar} + \frac{\hbar |\vec{k}|^2}{2m},
\end{equation}
and, hence, Eq.~(\ref{free_modes}) becomes
\begin{equation}\label{Minkowski_modes_nonrelativistic}
f(\vec{k},t,\vec{x}) \approx \frac{\hbar}{\sqrt{(2\pi)^3 2 m c^2}} \exp \left( -i\frac{m c^2 t}{\hbar} - i \frac{\hbar |\vec{k}|^2 t}{2m} + i\vec{k} \cdot \vec{x} \right).
\end{equation}
This means that $f(\vec{k})$ is approximately solution of
\begin{equation} \label{Schrodinger_free_mass}
i \hbar \partial_0 f(\vec{k}) \approx H_\text{M} f(\vec{k}),
\end{equation}
with Hamiltonian
\begin{equation}
H_\text{M} = m c^2  - \frac{\hbar^2}{2 m} \delta^{ij}\partial_i \partial_j.
\end{equation}

Equivalently, Eq.~(\ref{Schrodinger_free_mass}) can be obtained by using Eq.~(\ref{non_relativistic_limit}) to approximate the second-order time derivative of $f(\vec{k},t,\vec{x})$ as\footnote{Hereafter we use the notation such that $|\mathcal{O}(\epsilon)| \lesssim \epsilon$.}
\begin{align} \label{Klein_Gordon_approximation_second_derivative}
- \partial_0^2 f(\vec{k},t,\vec{x}) = & \omega^2(\vec{k})  f(\vec{k},t,\vec{x}) \nonumber \\
 = & \left( \frac{mc^2}{\hbar} \right)^2 \left\lbrace 1 + \left[ \frac{\hbar \omega(\vec{k})}{mc^2} - 1 \right] \right\rbrace^2 f(\vec{k},t,\vec{x})  \nonumber \\
 = & \left( \frac{mc^2}{\hbar} \right)^2\left\lbrace 1 + 2 \left[ \frac{\hbar \omega(\vec{k})}{mc^2}  - 1 \right] + \mathcal{O}(\epsilon^2) \right\rbrace f(\vec{k},t,\vec{x}) \nonumber \\
= &  \frac{mc^2}{\hbar}\left[ 2 i \partial_0 - \frac{mc^2}{\hbar} + \frac{mc^2}{\hbar} \mathcal{O}(\epsilon^2) \right] f(\vec{k},t,\vec{x}).
\end{align}
The mode $f(\vec{k})$ is solution of the Klein-Gordon equation (\ref{Klein_Gordon}). By plugging Eq.~(\ref{Klein_Gordon_approximation_second_derivative}) into Eq.~(\ref{Klein_Gordon}) we obtain Eq.~(\ref{Schrodinger_free_mass}).

As a result, we find that any state satisfying Eq.~(\ref{non_relativistic_limit_wavefunction}) is characterized by a time-dependent wave function (\ref{free_wave_function}) that is solution of the Schrödinger equation
\begin{equation}\label{Schrodinger_free_mass_n_particles}
i \hbar \partial_0  \phi_n   \approx \sum_{l=1}^n \left( m c^2  - \frac{\hbar^2}{2 m} \nabla^2_{\vec{x}_l} \right) \phi_n,
\end{equation}
where
\begin{equation}
\nabla^2_{\vec{x}} = \delta^{ij}\frac{\partial}{\partial x^i}\frac{\partial}{\partial x^j} .
\end{equation}
In this way, we obtain the familiar Schrödinger description of time evolution in NRQM.

It can be noticed that $H_\text{M}$ is hermitian with respect to both the Klein-Gordon scalar product (\ref{KG_scalar_product}) and the $L^2(\mathbb{R}^3)$ inner product, which is defined as
\begin{equation}\label{L2_inner_product}
( \phi, \phi' )_{L^2(\mathbb{R}^3)} = \int_{\mathbb{R}^3} d^3x \phi^*(t,\vec{x}) \phi'(t,\vec{x}).
\end{equation}
Indeed, by integrating by parts, one can prove that
\begin{subequations}
\begin{align}
& ( H_\text{M} \phi,  \phi' )_\text{KG}  = ( \phi, H_\text{M}  \phi' )_\text{KG}, \\
 & ( H_\text{M} \phi,  \phi' )_{L^2(\mathbb{R}^3)}  = ( \phi, H_\text{M}  \phi' )_{L^2(\mathbb{R}^3)}.
\end{align}
\end{subequations}

By using Eq.~(\ref{non_relativistic_limit}), it is straightforward to prove that for nonrelativistic modes, the Klein-Gordon scalar product (\ref{KG_scalar_product}) can be approximated by the $L^2(\mathbb{R}^3)$ inner product, with the only exception given by a $2 m c^2 / \hbar^2$ factor, i.e.,
\begin{equation}\label{KG_scalar_product_f_nonrelativistic}
( f(\vec{k}), f(\vec{k}') )_\text{KG} \approx \frac{2m c^2}{\hbar^2} ( f(\vec{k}), f(\vec{k}') )_{L^2(\mathbb{R}^3)}. 
\end{equation}
By using Eqs.~(\ref{free_wave_function}), (\ref{scalar_product_representation}), (\ref{non_relativistic_limit_wavefunction}) and (\ref{KG_scalar_product_f_nonrelativistic}), we can derive the same approximation for nonrelativistic single particle states
\begin{equation}\label{KG_scalar_product_single_particle_nonrelativistic}
\langle \phi | \phi' \rangle \approx ( \phi_1 , \phi'_1 )_{L^2(\mathbb{R}^3)}. 
\end{equation}
This approximation can be also extended to the case of general numbers of particles
\begin{equation}\label{KG_scalar_product_nonrelativistic}
\langle \phi | \phi' \rangle \approx \sum_{n=0}^\infty( \phi_n , \phi'_n )_{L^2(\mathbb{R}^{3n})},
\end{equation}
where
\begin{subequations}
\begin{align}
& ( \phi_n , \phi'_n )_{L^2(\mathbb{R}^{3n})} = \int_{\mathbb{R}^{3n}} d^{3n} \textbf{x}_n \phi^*_n (t, \textbf{x}_n)  \phi'_n (t, \textbf{x}_n), \\
& ( \phi_0 , \phi'_0 )_{L^2(\mathbb{R}^0)} = \phi_0^* \phi'_0.
\end{align}
\end{subequations}

\begin{table}
\begin{center}
\begin{tabular}{|l||c|c|}
\hline
 & QFT & NRQM \\
\hline
\hline
Inner product $\langle \phi | \phi' \rangle $ & $(\hbar^2/2mc^2) ( \phi_1, \phi'_1 )_\text{KG}$ & $ ( \phi_1 , \phi'_1 )_{L^2(\mathbb{R}^3)}$\\
\hline
Hamiltonian & $h_\text{KG}$ & $H_\text{M}$ \\
\hline
\end{tabular}
\end{center}
\caption{Inner product (first line) and Hamiltonian (second line) for free scalar single particles. The left column is for the fully relativistic theory (QFT), while the right one is for the nonrelativistic limit (NRQM).} \label{table_scalar_Minkowski}
\end{table}

While in the fully relativistic theory, single particles are described by the inner product (\ref{scalar_product_representation}) and the Hamiltonian $h_\text{KG}$ [Eqs.~(\ref{H_KG}) and (\ref{h_KG_H_KG})], nonrelativistic single particles can be approximately described by the $L^2(\mathbb{R}^3)$ inner product and the Hamiltonian $H_\text{M}$. This difference is shown schematically by Table \ref{table_scalar_Minkowski}. The Schrödinger equation (\ref{Schrodinger_free_mass_n_particles}) and the inner product (\ref{KG_scalar_product_nonrelativistic}) are the familiar ingredients for the description of free Fock states in the position representation according to NRQM.

\subsubsection{General modes}

It can be noticed that, in order to obtain Eqs.~(\ref{Schrodinger_free_mass}) and (\ref{KG_scalar_product_f_nonrelativistic}), we used the explicit form of the free modes $f(\vec{k})$ and considered the nonrelativistic limit of such functions. Conversely, it is possible to obtain the same result without looking at the explicit expression of $f(\vec{k},t,\vec{x})$ and rely on the general definition of positive frequency modes $g(\theta)$.

Here, we prove that, in the nonrelativistic limit, $\phi_n$ is approximately solution of the free Schrödinger equation (\ref{Schrodinger_free_mass_n_particles}) by showing that $g(\theta)$ is approximately solution of the free single particle Schrödinger equation
\begin{equation} \label{Schrodinger_free_mass_positive_negative_frequencies}
i \hbar \partial_0 g(\theta) \approx H_\text{M} g(\theta).
\end{equation}
Owing to Eqs.~(\ref{wavefunction_g}) and (\ref{Schrodinger_free_mass_positive_negative_frequencies}), one can check that Eq.~(\ref{Schrodinger_free_mass_n_particles}) holds also for wave functions defined by Eq.~(\ref{wavefunction_g}). 

The proof for Eq.~(\ref{Schrodinger_free_mass_positive_negative_frequencies}) follows from the fact that $g(\theta)$ is solution of Eq.~(\ref{Klein_Gordon}) and, in the nonrelativistic limit
\begin{equation}\label{non_relativistic_limit_general}
\left| \frac{\hbar \omega(\theta)}{mc^2} - 1 \right| \lesssim \epsilon,
\end{equation}
the second-order time derivative of Eq.~(\ref{Klein_Gordon}) acting on $g(\theta)$ is approximately replaced by a first-order time derivative. Indeed, by using Eq.~(\ref{positive_frequencies}), we obtain
\begin{align} \label{Klein_Gordon_approximation_positive_negative_frequencies}
- \partial_0^2 g(\theta) = & \omega^2(\theta)  g(\theta) \nonumber \\
 = & \left( \frac{mc^2}{\hbar} \right)^2 \left\lbrace 1 + \left[ \frac{\hbar \omega(\theta)}{mc^2} - 1 \right] \right\rbrace^2 g(\theta)  \nonumber \\
 = & \left( \frac{mc^2}{\hbar} \right)^2\left\lbrace 1 + 2 \left[ \frac{\hbar \omega(\theta)}{mc^2}  - 1 \right] + \mathcal{O}(\epsilon^2) \right\rbrace g(\theta) \nonumber \\
= &  \frac{mc^2}{\hbar}\left[ 2 i \partial_0 - \frac{mc^2}{\hbar} + \frac{mc^2}{\hbar} \mathcal{O}(\epsilon^2) \right] g(\theta).
\end{align}
Finally, by using Eq.~(\ref{Klein_Gordon_approximation_positive_negative_frequencies}) in the Klein-Gordon equation (\ref{Klein_Gordon}), we obtain
\begin{equation}\label{Schrodinger_free_mass_positive_negative_frequencies_error}
i \hbar \partial_0 g(\theta) = \left[ H_\text{M} + m c^2 \mathcal{O}(\epsilon^2) \right] g(\theta),
\end{equation}
which leads to the Schrödinger equation (\ref{Schrodinger_free_mass_positive_negative_frequencies}).

From Eq.~(\ref{Schrodinger_free_mass_positive_negative_frequencies_error}) one can also derive the error associated to the approximation (\ref{Schrodinger_free_mass_positive_negative_frequencies}). The difference between the nonrelativistic Hamiltonian $H_\text{M}$ and the exact fully-relativistic Hamiltonian $h_\text{KG}$ acting on nonrelativistic states is of the order of
\begin{equation} \label{H_M_h_KG_error}
H_\text{M} - h_\text{KG} = mc^2 \mathcal{O}(\epsilon^2).
\end{equation}

The equivalent of Eq.~(\ref{KG_scalar_product_f_nonrelativistic}) for the $g(\theta)$ modes is
\begin{equation}\label{KG_scalar_product_g_nonrelativistic}
( g(\theta), g(\theta') )_\text{KG} \approx  \frac{2 m c^2}{\hbar^2} ( g(\theta), g(\theta') )_{L^2(\mathbb{R}^3)}, 
\end{equation}
which can be obtained by using Eq.~(\ref{positive_frequencies}) and the approximation (\ref{non_relativistic_limit_general}). The error associated to Eq.~(\ref{KG_scalar_product_g_nonrelativistic}) comes directly from having replaced the time derivative of the modes with $m c^2/\hbar$ times such modes.  The relative error is, hence, of the order of $\epsilon$, in the sense that
\begin{align} \label{scalar_product_error_order}
( g(\theta), g(\theta') )_\text{KG}  = \frac{2 m c^2}{\hbar^2} ( g(\theta), g(\theta') )_{L^2(\mathbb{R}^3)} [ 1 + \mathcal{O}(\epsilon) ].
\end{align}

Equations (\ref{Schrodinger_free_mass_positive_negative_frequencies}) and (\ref{KG_scalar_product_g_nonrelativistic}) result again in the familiar description of free single particle states in the position representation, as before. In this case, however, $g(\theta)$ represents a generic basis $|\theta \rangle $ for the single particles space. The description of nonrelativistic Fock states is given again by Eqs.~(\ref{Schrodinger_free_mass_n_particles}) and (\ref{KG_scalar_product_nonrelativistic}), with the definition of wave functions in a generic basis provided by Eq.~(\ref{wavefunction_g}).

\subsubsection{Interaction}

Here, we consider interacting scalar fields. We adopt the interaction picture in QFT, which means that the field $\hat{\phi}(t,\vec{x})$ is free (i.e. solution of the Klein-Gordon equation (\ref{Klein_Gordon})), while any quantum state $| \phi(t) \rangle $ is time-evolved with respect to an interacting potential $\hat{V}(t)$ as
\begin{equation} \label{Schrodinger_interaction}
i \hbar \partial_0 | \phi(t) \rangle = \hat{V}(t) | \phi(t) \rangle.
\end{equation}
In the interaction picture, the field $\hat{\phi}$ can still be expanded in terms of $g(\theta)$ and $h(\theta)$ modes as in Eq.~(\ref{free_field_positive_negative_frequencies}) and the Hilbert state can still be defined as the Fock space generated by the orthonormal free single particle states $|\theta \rangle $.

We show that, in the nonrelativistic limit, states, scalar products and Hamiltonian can be represented identically to the free case, with the only exception given by an extra term in the Hamiltonian. To see this, we use the modes $g(\theta,t,\vec{x})$ as representatives of $| \theta (t) \rangle$ evolved with respect to the free theory.

A generic particle state $| \phi (t) \rangle$ is expanded as
\begin{equation}
| \phi (t) \rangle  = \sum_{n=1}^\infty \frac{1}{\sqrt{n!}} \sum_{\bm{\theta}_n} \tilde{\phi}_n (\bm{\theta}_n, t) \prod_{l=1}^n \hat{a}^\dagger(\theta_l) | 0_\text{M} \rangle + \tilde{\phi}_0 (t) | 0_\text{M} \rangle.
\end{equation}
Here, the $n$-particles wave function $\tilde{\phi}_n$ is time dependent as a consequence of the time evolution of $| \phi (t) \rangle$ in the interaction picture given by Eq.~(\ref{Schrodinger_interaction}). This leads to a differential equation for $\tilde{\phi}_n$, that reads as
\begin{equation} \label{Schrodinger_interaction_wavefunction}
i \hbar \partial_0 \tilde{\phi}_n (\bm{\theta}_n, t)  = \sum_{m=0}^\infty   \sum_{\bm{\theta}'_m}  \langle \bm{\theta}_n | \hat{V}(t) | \bm{\theta}'_m \rangle  \tilde{\phi}_m (\bm{\theta}'_m, t).
\end{equation}
The representative of the state $| \phi (t) \rangle$ in the Schrödinger picture is
\begin{equation}
\phi_n (t, \textbf{x}_n) = \left( \frac{2 m c^2}{\hbar^2} \right)^{n/2} \sum_{\bm{\theta}_n} \tilde{\phi}_n (\bm{\theta}_n, t) \prod_{l=1}^n g(\theta_l,t,\vec{x}_l).
\end{equation}
Here, differently from Eq.~(\ref{wavefunction_g}), $\tilde{\phi}_n $ is time dependent accordingly to Eq.~(\ref{Schrodinger_interaction_wavefunction}).

For interacting particles we still define nonrelativistic states as the ones such that $\tilde{\phi}_n (\bm{\theta}_n, t) $ is non-vanishing only for nonrelativistic frequencies $\omega(\theta)$. However, we also require potential energies that are very small with respect to the mass term. We therefore consider the following condition
\begin{equation}\label{non_relativistic_potential}
|\langle \bm{\theta}_n | \hat{V}(t) | \bm{\theta}'_m \rangle|  \lesssim \epsilon m c^2,
\end{equation}
so that Eq.~(\ref{Schrodinger_interaction_wavefunction}) is of order $\epsilon m c^2 \tilde{\phi}_n $.

Owing to Eq.~(\ref{Schrodinger_interaction_wavefunction}), it is straightforward to prove that Eq.~(\ref{Schrodinger_free_mass_n_particles}) still holds, but with an additional potential term, i.e.,
\begin{align}\label{Schrodinger_interactive_mass_n_particles}
& i \hbar \partial_0  \phi_n (t, \textbf{x}_n)  \approx \sum_{l=1}^n \left( m c^2  - \frac{\hbar^2}{2 m} \nabla^2_{\vec{x}_l} \right) \phi (t, \textbf{x}_n)  \nonumber \\
& + \sum_{\bm{\theta}_n} \left( \frac{2 m c^2}{\hbar^2} \right)^{n/2}  \sum_{m=0}^\infty   \sum_{\bm{\theta}'_m} \langle \bm{\theta}_n | \hat{V}(t) | \bm{\theta}'_m \rangle   \tilde{\phi}_m (\bm{\theta}'_m, t) \prod_{l=1}^n g(\theta_l,t,\vec{x}_l) .
\end{align}
Equation (\ref{Schrodinger_interactive_mass_n_particles}) can be identified as the NRQM Schrödinger equation for particles with potential. It can be noticed that the error associated to Eq.~(\ref{Schrodinger_interactive_mass_n_particles}) is still of the order $\epsilon^2 mc^2$ [Eq.~(\ref{H_M_h_KG_error})], since the interacting part of Eq.~(\ref{Schrodinger_interactive_mass_n_particles}) has been exactly derived and the error associated to the time evolution only comes from the free part.

It is also possible to prove that Eq.~(\ref{KG_scalar_product_single_particle_nonrelativistic}) holds for nonrelativistic interacting single particles. In this case, Eq.~(\ref{non_relativistic_potential}) plays an important role. Indeed, it suppresses the terms coming from the time derivative of $\tilde{\phi}_n $ [Eq.~(\ref{Schrodinger_interaction_wavefunction})] that appear as extra terms in Eq.~(\ref{KG_scalar_product_single_particle_nonrelativistic}). Moreover, the fact that $\hbar \partial_0 \tilde{\phi}_n$ is of order $\epsilon mc^2 \tilde{\phi}_n $ means that the relative error associated to the approximation (\ref{KG_scalar_product_single_particle_nonrelativistic}) is still of order $\epsilon$, as for the free case [Eq.~(\ref{scalar_product_error_order})].

The need for Eq.~(\ref{non_relativistic_potential}) implies that in order to have the same description of nonrelativistic particles for free and interacting systems, we have to assume that the energy potential is small if compared to the mass term. The fact that the energy of the particles is close to their mass energy [Eq.~(\ref{non_relativistic_limit_general})] and that the potential energy is very small with respect to the mass [Eq.~(\ref{non_relativistic_potential})] means that also the kinetic energy of the particles is small. In this way we recover the definition of nonrelativistic particles in terms of their velocity.

\subsection{Dirac field}\label{Minkowski_spacetime_Dirac_field}

In the previous subsection, we derived the familiar position representation of states, scalar product and Hamiltonian in the nonrelativistic limit of scalar fields. A very similar result holds for Dirac fields $\hat{\psi}$.

The fully relativistic theory of Dirac fields has been provided in Sec.~\ref{QFT_in_Minkowski_spacetime_Dirac}. Here, we show that nonrelativistic Dirac particles can be described by wave functions, scalar product and Hamiltonian as prescribed by NRQM. Specifically, the representation space of single particles is $\mathbb{C}^2 \otimes L^2(\mathbb{R}^3) $ and the time evolution is given by a Schrödinger equation similar to Eq.~(\ref{Schrodinger_interactive_mass_n_particles}). The difference with the scalar theory relies on the two spin degrees of freedom and the possibility to have interaction-spin coupling in the energy potential.

This subsection is organized as Sec.~\ref{Minkowski_spacetime_Scalar_field}. We start from the free theory and derive the NRQM description of nonrelativistic particles with fixed momenta. We also show that the time evolution of these states can be approximately described by the Schrödinger equation (\ref{Schrodinger_free_mass_n_particles}). Then, we use the general decomposition of the field in positive and negative frequencies to derive the same $\mathbb{C}^2 \otimes L^2(\mathbb{R}^3)$ representation space, but with a general basis. Finally, we detail the interacting case and show that Eq.~(\ref{Schrodinger_interactive_mass_n_particles}) still holds, but with a potential operator that can generally break the spin degeneracy.

\subsubsection{Free modes}

We start by considering the free Dirac modes defined by Eqs.~(\ref{free_Dirac_field_modes}) and (\ref{free_Dirac_field_modes_spin_up_spin_down}). By taking the limit (\ref{omega_k_approximation}), we obtain
\begin{equation}\label{Dirac_Minkowski_modes_nonrelativistic}
u_s(\vec{k},t,\vec{x}) \approx  \frac{1}{\sqrt{(2\pi)^3 }} \exp \left( -i\frac{m c^2 t}{\hbar} - i \frac{\hbar k^2 t}{2m} + i\vec{k} \cdot \vec{x} \right) \mathfrak{u}_s.
\end{equation}
From Eq.~(\ref{Dirac_Minkowski_modes_nonrelativistic}) it is immediate to see that the $u_s(\vec{k})$ modes cover the subspace of $\mathbb{C}^4 \otimes L^2(\mathbb{R}^3)$ with vanishing third and fourth spinorial components. More specifically, one can prove that
\begin{equation}\label{Dirac_representation_reduction_nonrelativistic_limit}
 \mathfrak{v}_{s'}^\dagger u_s(\vec{k}) = \mathfrak{u}_{s''}^\dagger u_s(\vec{k}) \mathcal{O}(\epsilon^{1/2}).
\end{equation}
This leads to a new representation for nonrelativistic particle states, where the wave functions (\ref{free_Dirac_wave_function}) and the Hilbert product (\ref{Dirac_scalar_product}) can be considered with spinorial $\alpha$ indices running through only the first two components. The representation space for nonrelativistic particles can, then, be identified with $\mathbb{C}^2 \otimes L^2(\mathbb{R}^3)$.

Moreover, the time evolution of $u_s(\vec{k})$ becomes
\begin{align} \label{Schrodinger_Dirac_free_mass}
i \hbar \partial_0 u_s(\vec{k}) \approx H_\text{M}  u_s(\vec{k}),
\end{align}
which means that the spinorial components of $u_s(\vec{k})$ are approximately decoupled and are solutions of Eq.~(\ref{Schrodinger_free_mass}). It is also possible to notice that $H_\text{M}$ is hermitian with respect to the scalar product $( \psi, \psi' )_{\mathbb{C}^4 \otimes L^2(\mathbb{R}^3)} $, in the sense that
\begin{equation}
( H_\text{M} \psi,  \psi' )_{\mathbb{C}^4 \otimes L^2(\mathbb{R}^3)}  = ( \psi, H_\text{M}  \psi' )_{\mathbb{C}^4 \otimes L^2(\mathbb{R}^3)}.
\end{equation}
This can be seen from the fact that
\begin{equation}\label{H_h_M}
H_\text{M} = \frac{h_\text{M}^2}{2 m c^2} + \frac{mc^2}{2},
\end{equation}
and, hence,
\begin{align}
 ( H_\text{M} \psi,  \psi' )_{\mathbb{C}^4 \otimes L^2(\mathbb{R}^3)}  = & \frac{1}{2 m c^2} ( h_\text{M} h_\text{M} \psi,  \psi' )_{\mathbb{C}^4 \otimes L^2(\mathbb{R}^3)} + \frac{mc^2}{2} ( \psi,  \psi' )_{\mathbb{C}^4 \otimes L^2(\mathbb{R}^3)} \nonumber \\
= & \frac{1}{2 m c^2} (  h_\text{M} \psi, h_\text{M} \psi' )_{\mathbb{C}^4 \otimes L^2(\mathbb{R}^3)} + \frac{mc^2}{2} ( \psi,  \psi' )_{\mathbb{C}^4 \otimes L^2(\mathbb{R}^3)} \nonumber \\
= & \frac{1}{2 m c^2} ( \psi,  h_\text{M} h_\text{M} \psi' )_{\mathbb{C}^4 \otimes L^2(\mathbb{R}^3)} + \frac{mc^2}{2} ( \psi,  \psi' )_{\mathbb{C}^4 \otimes L^2(\mathbb{R}^3)} \nonumber \\
= &( \psi, H_\text{M}  \psi' )_{\mathbb{C}^4 \otimes L^2(\mathbb{R}^3)}.
\end{align}
Equation (\ref{H_h_M}), on the other hand, can be derived from Eq.~(\ref{gamma_matrices_anticommutating_rule}) as
\begin{align}
\frac{h_\text{M}^2}{2 m c^2} + \frac{mc^2}{2} = &  - \frac{(\hbar c)^2}{2 m} \gamma^0 \gamma^i \gamma^0 \gamma^j \partial_i  \partial_j + \frac{m c^4}{2} \gamma^0 \gamma^0 \nonumber \\
&- i \frac{\hbar c^3}{2} (\gamma^0 \gamma^i \gamma^0 + \gamma^i \gamma^0 \gamma^0) \partial_i + \frac{mc^2}{2}\nonumber \\
 = &  \frac{(\hbar c)^2}{2 m} \gamma^0 \gamma^0 \gamma^i \gamma^j \partial_i  \partial_j  + mc^2 \nonumber \\
 = &  -\frac{\hbar^2}{2 m} \eta^{i j} \partial_i  \partial_j  + mc^2\nonumber \\
 = & H_\text{M}.
\end{align}

\begin{table}
\begin{center}
\begin{tabular}{|l||c|c|}
\hline
 & QFT & NRQM \\
\hline
\hline
Inner product $\langle \psi | \psi' \rangle$ & $(\psi_1, \psi'_1)_{\mathbb{C}^4 \otimes L^2(\mathbb{R}^3)}$ & $(\psi_1, \psi'_1)_{\mathbb{C}^2 \otimes L^2(\mathbb{R}^3)}$\\
\hline
Hamiltonian & $h_\text{M}$ & $H_\text{M}$ \\
\hline
\end{tabular}
\end{center}
\caption{Inner product (first line) and Hamiltonian (second line) for free Dirac single particles. The left column is for the fully relativistic theory (QFT), while the right one is for the nonrelativistic limit (NRQM).} \label{table_Dirac_Minkowski}
\end{table}

The result is that, in the nonrelativistic limit, single particles are described as elements of $\mathbb{C}^2 \otimes L^2(\mathbb{R}^3)$, where $u_s^\alpha(\vec{k},t,\vec{x})$ is the wave function of a particle with momentum $\vec{k}$ and spin number $s$ and with spinorial index $\alpha$ running through the first two values. The states are also approximately evolved with respect to the Hamiltonian $H_\text{M}$. These facts are listed in Table \ref{table_Dirac_Minkowski} on the right column and can be compared with the relativistic case, which is shown on the left column.

We have been able to derive the familiar description of Dirac particles in NRQM. General Fock states can be obtained from the singe particle representation space $\mathbb{C}^2 \otimes L^2(\mathbb{R}^3)$ and from the Schrödinger equation
\begin{equation}\label{Schrodinger_Dirca_free_mass_n_particles}
i \hbar \partial_0  \psi_n^{\bm{\alpha}_n} \approx \sum_{l=1}^n \left( m c^2  - \frac{\hbar^2}{2 m} \nabla^2_{\vec{x}_l} \right) \psi_n^{\bm{\alpha}_n}.
\end{equation}

\subsubsection{General modes}

We now want to provide the same description of nonrelativistic states but starting from the general real frequency Dirac modes (\ref{positive_negative_frequencies_Dirac}).

We identify the representation space of nonrelativistic particles as the one in which the third and fourth spinorial components are always vanishing. Indeed, it is possible to prove the equivalent of Eq.~(\ref{Dirac_representation_reduction_nonrelativistic_limit}) for the $u(\theta)$ modes, that reads as
\begin{equation}\label{Dirac_representation_reduction_nonrelativistic_limit_general}
 \mathfrak{v}_s^\dagger u(\theta) = \mathfrak{u}_{s'}^\dagger u(\theta)  \mathcal{O}(\epsilon^{1/2}).
\end{equation}
The proof of Eq.~(\ref{Dirac_representation_reduction_nonrelativistic_limit_general}) follows from the fact that $u(\theta)$ is solution of the Dirac equation
\begin{equation} \label{Dirac_representation_reduction_nonrelativistic_limit_general_Dirac}
 \left[ c \gamma^0 \omega(\theta) + i c \gamma^i \partial_i  - \frac{m c^2}{\hbar} \right] u(\theta) = 0
\end{equation}
and is combination of modes with nonrelativistic momenta, which means that
\begin{equation}\label{non_relativistic_momenta_epsilon}
\hbar \partial_i u(\theta) =  m c u(\theta) \mathcal{O}(\epsilon^{1/2}).
\end{equation}
By acting with $\mathfrak{u}_s^\dagger$ on the left of Eq.~(\ref{Dirac_representation_reduction_nonrelativistic_limit_general_Dirac}), one obtains
\begin{equation} \label{Dirac_representation_reduction_nonrelativistic_limit_general_Dirac_2}
\frac{mc^2}{\hbar} \epsilon \mathfrak{u}_s^\dagger  u(\theta) + i c \sum_{s'=1}^2 \mathfrak{u}_s^\dagger \gamma^i \mathfrak{v}_{s'}  \mathfrak{v}_{s'}^\dagger \partial_i   u(\theta) = 0,
\end{equation}
which together with Eq.~(\ref{non_relativistic_momenta_epsilon}) leads to Eq.~(\ref{Dirac_representation_reduction_nonrelativistic_limit_general}).

It is known that the components of any solution of the Dirac equation are also solution of the Klein-Gordon equation (\ref{Klein_Gordon}) with the same mass. This fact can be proven by multiplying Eq.~(\ref{Dirac}) with $ i c \gamma^\mu \partial_\mu + m c^2/\hbar$ on the left and using the anticommutation relation (\ref{gamma_matrices_anticommutating_rule}). Consequently, the modes $u(\theta)$ are solutions of the Klein-Gordon equation 
\begin{equation}\label{Klein_Gordon_Dirac}
\left[ c^2 \eta^{\mu\nu} \partial_\mu \partial_\nu - \left( \frac{mc^2}{\hbar} \right)^2 \right] u(\theta) = 0.
\end{equation}
We can, at this point, use the same arguments of Sec.~\ref{Minkowski_spacetime_Scalar_field} that led to Eq.~(\ref{Schrodinger_free_mass_positive_negative_frequencies}) in order to prove that
\begin{equation}\label{Schrodinger_Dirac_free_mass_positive_negative_frequencies}
i \hbar \partial_0 u(\theta) \approx H_\text{M}  u(\theta),
\end{equation}
which is the equivalent of Eq.~(\ref{Schrodinger_Dirac_free_mass}) for the mode $u(\theta)$. The error associated to the approximation (\ref{Schrodinger_Dirac_free_mass_positive_negative_frequencies}) can be identified with the equivalent of Eq.~(\ref{H_M_h_KG_error}) for Dirac fields
\begin{equation} \label{H_M_h_M_error}
H_\text{M} - h_\text{M} =  mc^2 \mathcal{O}(\epsilon^2).
\end{equation}

From Eqs.~(\ref{H_M_h_KG_error}) and (\ref{H_M_h_M_error}) one can derive the error made by considering scalar and Dirac states identical in their time evolution [Eqs.~(\ref{Schrodinger_free_mass_positive_negative_frequencies}) and (\ref{Schrodinger_Dirac_free_mass_positive_negative_frequencies})], which is of the order of
\begin{equation} \label{h_KG_h_M_error}
h_\text{KG} - h_\text{M} =  mc^2 \mathcal{O}(\epsilon^2).
\end{equation}
Equation (\ref{h_KG_h_M_error}) implies that corrective terms of Eqs.~(\ref{Schrodinger_free_mass_positive_negative_frequencies}) and (\ref{Schrodinger_Dirac_free_mass_positive_negative_frequencies}) that spoil the difference between scalar and Dirac fields in the Minkowski spacetime can be found at the order of $\epsilon^2 $.

An additional error associated to the nonrelativistic limit comes from considering the third and the fourth spinorial component of $u(\theta)$ as vanishing quantities. Such an approximation allowed us to replace the exact $\mathbb{C}^4 \otimes L^2(\mathbb{R}^3)$ scalar product with the $\mathbb{C}^2 \otimes L^2(\mathbb{R}^3)$ scalar product. The relative error can be obtained from Eq.~(\ref{Dirac_representation_reduction_nonrelativistic_limit_general}) and is of the order of $\epsilon$, as in the scalar case [Eq.~(\ref{scalar_product_error_order})].

\subsubsection{Interaction}

Finally, we consider interacting Dirac fields and by following the same steps of Sec.~\ref{Minkowski_spacetime_Scalar_field} we conclude that interacting particle states can still be described in the representation space of free particles. The difference between the interacting and the free theory is only given by the presence of a potential energy in the approximated Schrödinger equation. Such a term may introduce spin interactions that cannot appear in the scalar theory.

The representative of any state $| \psi(t) \rangle$ in the Schrödinger picture is
\begin{equation}\label{Dirac_wavefunction_general_interaction}
\psi^{\bm{\alpha}_n}_n (t, \textbf{x}_n) =  \sum_{\bm{\theta}_n} \tilde{\psi}_n (\bm{\theta}_n, t)  \prod_{l=1}^n u^{\alpha_l}(\theta_l, t, \vec{x}_l),
\end{equation}
where, differently from Eq.~(\ref{Dirac_wavefunction_general}), $\tilde{\psi}_n (\bm{\theta}_n, t)$ is time dependent. The time evolution of Eq.~(\ref{Dirac_wavefunction_general_interaction}) in the nonrelativistic limit is given by
\begin{align}\label{Schrodinger_Dirca_free_mass_n_particles_interaction}
& i \hbar \partial_0  \psi_n^{\bm{\alpha}_n} (t, \textbf{x}_n) \approx \sum_{l=1}^n \left( m c^2  - \frac{\hbar^2}{2 m} \nabla^2_{\vec{x}_l} \right) \psi_n^{\bm{\alpha}_n} (t, \textbf{x}_n)  \nonumber \\
&+ \sum_{\bm{\theta}_n}  \sum_{m=0}^\infty   \sum_{\bm{\theta}'_m} \langle \bm{\theta}_n | \hat{V}(t) | \bm{\theta}'_m \rangle  \tilde{\psi}_m (\bm{\theta}'_m, t)  \prod_{l=1}^n u^{\alpha_l}(\theta_l, t, \vec{x}_l),
\end{align}
where, in this case, $ \langle \bm{\theta}_n | \hat{V}(t) | \bm{\theta}'_m \rangle$ are the matrix elements of a potential $\hat{V}(t)$ that comes from a Dirac interacting Lagrangian. The quantum numbers $\theta$ contain spinorial degrees of freedom as well, and, hence, $\langle \bm{\theta}_n | \hat{V}(t) | \bm{\theta}'_m \rangle$ may break the spin degeneracy present in the free theory. As explained in Sec.~\ref{Minkowski_spacetime_Scalar_field}, we obtain Eq.~(\ref{Schrodinger_Dirca_free_mass_n_particles_interaction}) by considering the nonrelativistic condition for potential energies (\ref{non_relativistic_potential}).

\section{Curved spacetime}\label{NonMinkowksi_spacetime}

At variance with Sec.~\ref{Minkowski_spacetime}, here we consider the coordinates $(T,\vec{X})$ and the metric $g_{\mu\nu}$ representing a hyperbolic static spacetime. The aim of this section is to derive a description for nonrelativistic states of scalar ($\hat{\Phi}$) and Dirac ($\hat{\Psi}$) fields in curved spacetime.

We use the description of fully-relativistic particle states presented in Sec.~\ref{QFT_in_curved_spacetime} and we derive the corresponding nonrelativistic limit. We show how the representation of nonrelativistic states changes from the Minkowski to the curved case. We also derive the Schrödinger equation for particles affected by the metric and the consequent precision needed to distinguish between scalar and Dirac fields.

\subsection{Scalar field}\label{NonMinkowksi_spacetime_Scalar_field}

The field considered in the present subsection is scalar. As in Sec.~\ref{Minkowski_spacetime_Scalar_field}, we start by considering the relativistic theory of particles for the free scalar field $\hat{\Phi}$. We derive the nonrelativistic limit and show that the product between states can be approximated by the $L^2(\mathbb{R}^3)$ inner product with a metric-dependent measure. Also, we show that the quantum states are solutions of a metric-dependent Schrödinger equation. Finally, we extend the theory to the interacting case by introducing a potential energy in the Schrödinger equation.

\begin{table}
\begin{center}
\begin{tabular}{|l||c|c|}
\hline
 & QFTCS & NRQFTCS \\
\hline
\hline
Inner product $\langle \Phi | \Phi' \rangle $ & $(\hbar^2/2m c^2) ( \Phi_1, \Phi'_1 )_\text{CKG}$ & $ ( \Phi_1 , \Phi'_1 )_{L^2_\text{S}(\mathbb{R}^3)}$\\
\hline
Hamiltonian & $h_\text{CKG}$ & $H_\text{CS}$ \\
\hline
\end{tabular}
\end{center}
\caption{Inner product (first line) and Hamiltonian (second line) for free scalar single particles in curved spacetime. The left column is for the fully relativistic theory (QFTCS), while the right one is for the nonrelativistic limit (NRQFTCS).} \label{table_scalar_nonMinkowski}
\end{table}

In Sec.~\ref{QFT_in_curved_spacetime_scalar}, we showed that each positive frequency mode $G(\theta)$ is associated to a single particle state and the Klein-Gordon product in curved spacetime $( \Phi_1, \Phi'_1 )_\text{CKG}$ is used as the Hilbert product for the single particle space. Furthermore, $h_\text{CKG}$ is the Hamiltonian describing the dynamics. This result is summarized in the left column of Table \ref{table_scalar_nonMinkowski}.

In the curved spacetime, we characterize the nonrelativistic regime by means of the condition
\begin{equation}\label{non_relativistic_limit_curved}
\left| \frac{\hbar \Omega(\theta)}{mc^2} - 1 \right| \lesssim \epsilon,
\end{equation}
with $\epsilon \ll 1$ as the nonrelativistic parameter. Here, we demonstrate that $G(\theta)$ is approximately solution to the Schrödinger equation
\begin{equation} \label{Schrodinger_curved}
i \hbar \partial_0 G(\theta) \approx H_\text{CS} G(\theta),
\end{equation}
with Hamiltonian
\begin{equation}\label{H_nM}
H_\text{CS} = \frac{m c^2}{2} \left( 1 - \frac{g_{00}}{c^2} \right) + \frac{\hbar^2 g_{00}}{2 m c^2 \sqrt{-g}} \partial_i \left( \sqrt{-g} g^{i j}\partial_j \right)
\end{equation}
and that the curved Klein-Gordon scalar product $( \Phi, \Phi' )_\text{CKG}$ is approximated by
\begin{equation}\label{cKG_scalar_curved_product_G_nonrelativistic}
( G(\theta), G(\theta') )_\text{CKG}  \approx \frac{2 m c^2}{\hbar^2} ( G(\theta), G(\theta') )_{L^2_\text{S}(\mathbb{R}^3)}.
\end{equation}
In this way, we show that the nonrelativistic single particle description of the field is defined by the Hamiltonian $H_\text{CS}$ and the scalar product $( \Phi, \Phi' )_{L^2_\text{S}(\mathbb{R}^3)}$. The result can be seen as the equivalent of Eqs.~(\ref{Schrodinger_free_mass_positive_negative_frequencies}) and (\ref{KG_scalar_product_g_nonrelativistic}) in curved spacetime and are summarized by the right column of Table \ref{table_scalar_nonMinkowski}.

Notice that $H_\text{CS}$ is hermitian with respect to $( \Phi_1, \Phi'_1 )_\text{CKG}$ and $( \Phi, \Phi' )_{L^2_\text{S}(\mathbb{R}^3)}$, since it is equivalent to
\begin{equation}\label{H_nM_H_cKG}
H_\text{CS} = \frac{H_\text{CKG}}{2mc^2} + \frac{m c^2}{2}
\end{equation}
and $H_\text{CKG}$ is hermitian with respect to both products.

The nonrelativistic description of states with general numbers of particles is given by the wave function $\Phi_n$ of Eq.~(\ref{wavefunction_g_curved}), the Fock extension of the $( \Phi, \Phi' )_{L^2_\text{S}(\mathbb{R}^3)}$ scalar product
\begin{equation}\label{KG_scalar_product_nonrelativistic_curved}
\langle \Phi | \Phi' \rangle \approx \sum_{n=0}^\infty ( \Phi_n , \Phi'_n )_{L^2_S(\mathbb{R}^{3n})},
\end{equation}
and the Schrödinger equation
\begin{equation}\label{Schrodinger_free_mass_n_particles_curved}
i \hbar \partial_0  \Phi_n (T, \textbf{X}_n) \approx \sum_{l=1}^n \left\lbrace \frac{m c^2}{2} \left[ 1 - \frac{g_{00}(\vec{X}_l)}{c^2} \right]  + \frac{\hbar^2 g_{00}(\vec{X}_l)}{2 m c^2} \nabla^2_{\vec{X}_l} \right\rbrace \Phi_n (T, \textbf{X}_n),
\end{equation}
where, in this case,
\begin{subequations}
\begin{align}
 & ( \Phi_n, \Phi'_n )_{L^2_S(\mathbb{R}^{3n})} = (- c)^n \int_{\mathbb{R}^{3n}} d^{3n}\textbf{X}_n \prod_{l=1}^n \left[ \sqrt{-g(\vec{X}_l)} g^{0 0}(\vec{X}_l) \right] \Phi^*_n(\textbf{X}_n) \Phi'_n(\textbf{X}_n),\\
& ( \Phi_0 , \Phi'_0 )_{L^2_S(\mathbb{R}^0)} = \Phi_0^* \Phi'_0
\end{align}
\end{subequations}
and
\begin{equation}
\nabla^2_{\vec{X}} = \frac{1}{\sqrt{-g(\vec{X})}} \frac{\partial}{\partial X^i} \left[ \sqrt{-g(\vec{X})} g^{i j}(\vec{X})\frac{\partial}{\partial X^j} \right].
\end{equation}

A way to approximate Eq.~(\ref{Klein_Gordon_curved}) as a Schrödinger equation is to replace the second-order time derivative with a first-order time derivative. In the nonrelativistic limit, we find that
\begin{equation}\label{Klein_Gordon_curved_approximation_positive_negative_frequencies}
-  \partial_0^2 G(\theta)= \frac{mc^2}{\hbar}   \left[ 2 i \partial_0 - \frac{mc^2}{\hbar} + \frac{mc^2}{\hbar}\mathcal{O} (\epsilon^2) \right] G(\theta) ,
\end{equation}
which is the equivalent of Eq.~(\ref{Klein_Gordon_approximation_positive_negative_frequencies}) in curved spacetime. By using Eq.~(\ref{Klein_Gordon_curved_approximation_positive_negative_frequencies}) and the fact that $G(\theta)$ is a solution of Eq.~(\ref{Klein_Gordon_curved}), we obtain
\begin{equation} \label{Schrodinger_curved_error}
i \hbar \partial_0 G(\theta) = [ H_\text{CS} + mc^2 \mathcal{O} (\epsilon^2)] G(\theta),
\end{equation}
which leads to the Schrödinger equation (\ref{Schrodinger_curved}). The error associated to such an approximation is
\begin{equation} \label{H_S_h_CKG_error}
H_\text{CS} - h_\text{CKG} = mc^2 \mathcal{O}(\epsilon^2).
\end{equation}

Equation (\ref{cKG_scalar_curved_product_G_nonrelativistic}) can be proven from Eqs.~(\ref{scalar_curved_modes_tilde}), (\ref{KG_curved_scalar_product_2}) and by replacing the frequencies with $mc^2/\hbar$. The relative error associated to such approximation is of the order of $\epsilon$ as in Eq.~(\ref{scalar_product_error_order}), i.e.,
\begin{align} \label{scalar_product_error_order_curved}
( G(\theta), G(\theta') )_\text{CKG} =  \frac{2 m c^2}{\hbar^2} ( G(\theta), G(\theta') )_{L^2_\text{S}(\mathbb{R}^3)} [1 + \mathcal{O}(\epsilon)].
\end{align}

Finally, the interacting theory can be described similarly to Sec.~\ref{Minkowski_spacetime_Scalar_field}. The only modification from the free theory is given by wave functions $\tilde{\Phi}_n (\bm{\theta}_n, T)$ that are now time dependent and, hence, generate an extra term in the Schrödinger equation
\begin{align}\label{Schrodinger_n_particles_curved}
& i \hbar \partial_0  \Phi_n (T, \textbf{X}_n)   \approx \sum_{l=1}^n \left\lbrace \frac{m c^2}{2} \left[ 1 - \frac{g_{00}(\vec{X}_l)}{c^2} \right]  + \frac{\hbar^2 g_{00}(\vec{X}_l)}{2 m c^2} \nabla^2_{\vec{X}_l} \right\rbrace \Phi_n (T, \textbf{X}_n) \nonumber \\
& + \left( \frac{2 m c^2}{\hbar^2} \right)^{n/2}  \sum_{\bm{\theta}_n}  \sum_{m=0}^\infty   \sum_{\bm{\theta}'_m} \langle \bm{\theta}_n | \hat{V}(T) | \bm{\theta}'_m \rangle \tilde{\Phi}_m (\bm{\theta}'_m, T)  \prod_{l=1}^n G(\theta_l,T,\vec{X}_l) .
\end{align}
In order to obtain such a result, we consider the condition
\begin{equation}\label{non_relativistic_potential_curved}
\langle \bm{\theta}_n | \hat{V}(T) | \bm{\theta}'_m \rangle  = mc^2 \mathcal{O}(\epsilon).
\end{equation}
Then by following the same arguments of Sec.~\ref{Minkowski_spacetime_Scalar_field} we obtain Eq.~(\ref{Schrodinger_n_particles_curved}). 

\subsection{Dirac field}\label{NonMinkowksi_spacetime_Dirac_field}

Here, we consider Dirac fields $\hat{\Psi}$ and we derive the representation of nonrelativistic single particles. We also show that nonrelativistic Fock states are approximately solutions of a Schrödinger equation that is different from the one obtained in Sec.~\ref{NonMinkowksi_spacetime_Scalar_field} for scalar fields. Such a difference is noticeable at any order, unless the metric is almost flat and the limit $g_{\mu\nu} \rightarrow \eta_{\mu\nu}$ is controlled by the nonrelativistic parameter $\epsilon$. In that case, the difference between the scalar and Dirac Hamiltonians is not vanishing only at some orders. We discuss the situation in which these orders differ from the one seen for the Minkowski case [Eq.~(\ref{h_KG_h_M_error})].

In Sec.~\ref{QFT_in_curved_spacetime_Dirac} we showed that the single particle space is generated by the $U(\theta)$ modes and is supplemented by the $( \Psi, \Psi' )_{\mathbb{C}^4 \otimes L^2_\text{D}(\mathbb{R}^3)}$ product. It can be noticed that even in the nonrelativistic limit (\ref{non_relativistic_limit_curved}), such a representation is not equivalent to $\mathbb{C}^2 \otimes L^2(\mathbb{R}^3)$, at variance with the flat case. This occurs for two reasons: (i) $( \Psi, \Psi' )_{\mathbb{C}^4 \otimes L^2_\text{D}(\mathbb{R}^3)}$ is metric dependent; (ii) the curved Dirac equation (\ref{Dirac_curved}) in the nonrelativistic limit (\ref{non_relativistic_limit_curved}) does not lead to vanishing spinorial components for $U(\theta)$ modes. The familiar NRQM prescription of position representation through the $\mathbb{C}^2 \otimes L^2(\mathbb{R}^3)$ space cannot be restored in the curved case.

Single particles are also described by the Hamiltonian $h_\text{CD}$. Here, we find an approximation for $h_\text{CD}$ in the nonrelativistic limit by following the same steps of Sec.~\ref{Minkowski_spacetime_Dirac_field}. We use the fact that $U(\theta)$ is a solution of the curved Dirac equations (\ref{Dirac_curved}) which leads to the following Klein-Gordon-like equation \cite{Pollock:2010zz}
\begin{equation} \label{Klein_Gordon_curved_Dirac}
\left[ \frac{c^2}{\sqrt{-g}} D_\mu \left( \sqrt{-g} g^{\mu \nu} D_\nu   \right) - \left( \frac{mc^2}{\hbar}\right)^2 - \frac{c^2}{4} R \right] U(\theta) = 0,
\end{equation}
with $R$ as the Ricci scalar. A detailed proof for Eq.~(\ref{Klein_Gordon_curved_Dirac}) is reported in Appendix A of Ref.~\citeRF{PhysRevD.107.045012}. In static spacetimes, Eq.~(\ref{Klein_Gordon_curved_Dirac}) becomes
\begin{align}\label{Klein_Gordon_curved_Dirac_static}
& \left\lbrace c^2  g^{0 0} \left( \partial_0 + \Gamma_0 \right)^2  +  \frac{c^2}{\sqrt{-g}} \left(\partial_i + \Gamma_i \right) \left[ \sqrt{-g} g^{i j} \left( \partial_j + \Gamma_j \right)  \right] \right. \nonumber \\
& \left. - \left( \frac{mc^2}{\hbar}\right)^2 - \frac{c^2}{4} R \right\rbrace U(\theta) = 0.
\end{align}
By using the fact that $U(\theta)$ is a solution of the curved Dirac equation (\ref{Dirac_curved_Schrodinger}), we obtain
\begin{align}\label{Klein_Gordon_curved_Dirac_static_2}
& -\hbar^2 \partial_0^2 U(\theta) = \left\lbrace  \frac{\hbar^2 g_{0 0} }{\sqrt{-g}} \left(\partial_i + \Gamma_i \right) \left[ \sqrt{-g} g^{i j} \left( \partial_j + \Gamma_j \right)  \right] \right. \nonumber \\
& \left. - \frac{g_{0 0}}{c^2} \left[(mc^2)^2 + \frac{(\hbar c)^2}{4} R \right] - i 2 \hbar  \Gamma_0 h_\text{CD} + \hbar^2 \Gamma^2_0 \right\rbrace U(\theta).
\end{align}
The left hand side of Eq.~(\ref{Klein_Gordon_curved_Dirac_static_2}) can be computed by using again the fact that $U(\theta)$ is a solution of the curved Dirac equation (\ref{Dirac_curved_Schrodinger}), which means that
\begin{equation} \label{Klein_Gordon_curved_Dirac_static_3}
-\hbar^2 \partial_0^2 U(\theta) =  h_\text{CD}^2 U(\theta).
\end{equation}
By plugging Eq.~(\ref{Klein_Gordon_curved_Dirac_static_3}) in Eq.~(\ref{Klein_Gordon_curved_Dirac_static_2}), we obtain
\begin{align}\label{h_NM_2}
h_\text{CD}^2 =  &  \frac{\hbar^2 g_{0 0} }{\sqrt{-g}} \left(\partial_i + \Gamma_i \right) \left[ \sqrt{-g} g^{i j} \left( \partial_j + \Gamma_j \right)  \right]  - \frac{g_{0 0}}{c^2} \left[ (mc^2)^2 + \frac{(\hbar c)^2}{4} R \right] \nonumber \\
&  - i 2 \hbar  \Gamma_0 h_\text{CD} + \hbar^2 \Gamma^2_0 .
\end{align}

The second-order time derivative appearing in Eq.~(\ref{Klein_Gordon_curved_Dirac}) is the same as the one appearing in Eq.~(\ref{Klein_Gordon_curved}). Moreover, Eq.~(\ref{Klein_Gordon_curved_approximation_positive_negative_frequencies}) is valid also for Dirac modes $U(\theta)$ in the nonrelativistic limit. For these reasons, Eq.~(\ref{Klein_Gordon_curved_Dirac_static_3}) is equivalent to
\begin{equation} \label{Klein_Gordon_curved_Dirac_static_3_approximation}
mc^2 \left[ 2 i \hbar \partial_0 - mc^2 + mc^2 \mathcal{O} (\epsilon^2)  \right] U(\theta) = h_\text{CD}^2 U(\theta).
\end{equation}
If we now define the Hamiltonian
\begin{equation}\label{H_cD_h_nM}
H_\text{CD} = \frac{h_\text{CD}^2}{2 mc^2}  + \frac{mc^2}{2},
\end{equation}
Eq.~(\ref{Klein_Gordon_curved_Dirac_static_3_approximation}) becomes
\begin{equation}\label{Dirac_curved_Schrodinger_nonrelativistic_error}
i \hbar \partial_0 U(\theta) = [ H_\text{CD} + mc^2 \mathcal{O} (\epsilon^2)] U(\theta).
\end{equation}
Equation (\ref{Dirac_curved_Schrodinger_nonrelativistic_error}) leads to the Schrödinger equation
\begin{equation}\label{Dirac_curved_Schrodinger_nonrelativistic}
i \hbar \partial_0 U(\theta) \approx H_\text{CD} U(\theta),
\end{equation}
with an error given by
\begin{equation} \label{H_D_h_nM_error}
H_\text{CD} - h_\text{CD} = mc^2 \mathcal{O}(\epsilon^2).
\end{equation}
From Eq.~(\ref{H_cD_h_nM}) one can see that the Hamiltonian $H_\text{CD}$ is hermitian with respect to $( \Psi, \Psi' )_{\mathbb{C}^4 \otimes L^2_\text{D}(\mathbb{R}^3)}$ and can be used for the time evolution of nonrelativistic states.

\begin{table}
\begin{center}
\begin{tabular}{|l||c|c|}
\hline
 & QFTCS & NRQFTCS \\
\hline
\hline
Inner product $\langle \Psi | \Psi' \rangle$ & $(\Psi_1, \Psi'_1)_{\mathbb{C}^4 \otimes L^2_\text{S}(\mathbb{R}^3)}$ & $(\Psi_1, \Psi'_1)_{\mathbb{C}^4 \otimes L^2_\text{S}(\mathbb{R}^3)}$\\
\hline
Hamiltonian & $h_\text{CD}$ & $H_\text{CD}$ \\
\hline
\end{tabular}
\end{center}
\caption{Inner product (first line) and Hamiltonian (second line) for free Dirac single particles in curved spacetime. The left column is for the fully relativistic theory (QFTCS), while the right one is for the nonrelativistic limit (NRQFTCS).} \label{table_Dirac_nonMinkowski}
\end{table}

In summary, single particles are described by the inner product in curved spacetime $( \Psi, \Psi' )_{\mathbb{C}^4 \otimes L^2_\text{D}(\mathbb{R}^3)}$ and their time evolution is given by the Hamiltonian $h_\text{CD}$, which, in the nonrelativistic limit, can be replaced by $H_\text{CD}$. These results are shown schematically by Table \ref{table_Dirac_nonMinkowski}.

By comparing Eq.~(\ref{h_NM_2}) with Eq.~(\ref{Schrodinger_equation_tilde_Hamiltonian}) we can write
\begin{equation}
h_\text{CD}^2 = H_\text{CKG} + 2 m c^2 \Delta H
\end{equation}
and, hence,
\begin{equation}\label{H_D_H_S_Delta_H}
H_\text{CD} = H_\text{CS} + \Delta H,
\end{equation}
with
\begin{align}
\Delta H = & \frac{\hbar^2 g_{0 0}}{2 mc^2} \left\lbrace \frac{[ \partial_i ( \sqrt{-g} g^{i j} \Gamma_j )]}{\sqrt{-g}} +  g^{i j} \Gamma_i ( 2 \partial_j + \Gamma_j ) - \frac{R}{4} \right\rbrace \nonumber \\
& - i \frac{\hbar}{mc^2} \Gamma_0 h_\text{CD} + \frac{\hbar^2}{2 mc^2} \Gamma^2_0.
\end{align}
In the case of curved metrics (i.e., $g_{\mu \nu} \neq \eta_{\mu \nu}$), the difference between $H_\text{CS}$ and $H_\text{CD}$ is non-vanishing. At variance with the flat case, the spinorial decoupling does not occur and Dirac particles evolve differently from scalar states.

In the case of Minkowski spacetime (i.e., $g_{\mu \nu} = \eta_{\mu \nu}$), $\Delta H $ is identically vanishing and the difference between scalar and Dirac fields is detectable only at the order of $\epsilon^2$ [Eq.~(\ref{h_KG_h_M_error})]. We wonder if this is also true for a quasiflat spacetime ($g_{\mu \nu} \approx \eta_{\mu \nu}$). By considering the limit $g_{\mu \nu} \rightarrow \eta_{\mu \nu}$ regulated through the nonrelativistic parameter $\epsilon$, different scenarios occur for different orders of magnitude of $\Delta H / (mc^2)$ with respect to $\epsilon$. For instance, if $\Delta H$ is of an order lower than $\epsilon^2 m c^2$, the difference between $h_\text{CKG}$ and $h_\text{CD}$ is also of an order lower than $\epsilon^2 m c^2$. In that case, one can distinguish between scalar and Dirac fields with less precision than the one needed for the flat case [Eq.~(\ref{h_KG_h_M_error})].

In Sec.~\ref{Rindler_frame}, we will provide an example in which the difference between the two Hamiltonians can be observed. In particular, we will consider an accelerated observer whose acceleration is sufficiently high to detect nonvanishing values of $\Delta H$ with less precision than the one needed for an inertial observer.

For completeness we provide the nonrelativistic theory of states different from single particles. The Schrödinger equation for $\Psi_n (T, \textbf{X}_n)$ is equivalent to Eq.~(\ref{Schrodinger_free_mass_n_particles_curved}) with extra terms coming from a non-vanishing $\Delta H$
\begin{align}\label{Schrodinger_free_mass_n_particles_Dirac_curved}
i \hbar \partial_0  \Psi^{\bm{\alpha}_n}_n (T, \textbf{X}_n) \approx & \sum_{l=1}^n \left\lbrace \frac{m c^2}{2} \left[ 1 - \frac{g_{00}(\vec{X}_l)}{c^2} \right]  + \frac{\hbar^2 g_{00}(\vec{X}_l)}{2 m c^2} \nabla^2_{\vec{X}_l} \right\rbrace \Psi^{\bm{\alpha}_n}_n (T, \textbf{X}_n)  \nonumber \\
& + \sum_{l=1}^n \Delta H^{\alpha_l}{}_{\beta_l}(\vec{X}_l) \Psi^{\alpha_1 \dots \beta_l \dots \alpha_n}_n (T, \textbf{X}_n).
\end{align}

Finally, regarding the theory with interaction, we may use the same arguments of Sec.~\ref{Minkowski_spacetime_Dirac_field} to conclude that the only modification from the free theory is given by an extra term in the Schrödinger equation
\begin{align}\label{Schrodinger_free_mass_n_particles_Dirac_curved_interaction}
& i \hbar \partial_0  \Psi^{\bm{\alpha}_n}_n (T, \textbf{X}_n)  \approx \sum_{l=1}^n \left\lbrace \frac{m c^2}{2} \left[ 1 - \frac{g_{00}(\vec{X}_l)}{c^2} \right]  + \frac{\hbar^2 g_{00}(\vec{X}_l)}{2 m c^2} \nabla^2_{\vec{X}_l} \right\rbrace \Psi^{\bm{\alpha}_n}_n (T, \textbf{X}_n) \nonumber \\
&  + \sum_{l=1}^n \Delta H^{\alpha_l}{}_{\beta_l}(\vec{X}_l) \Psi^{\alpha_1 \dots \beta_l \dots \alpha_n}_n (T, \textbf{X}_n) \nonumber \\
 &  + \sum_{\bm{\theta}_n} \sum_{m=0}^\infty   \sum_{\bm{\theta}'_m} \langle \bm{\theta}_n | \hat{V}(T) | \bm{\theta}'_m \rangle \tilde{\Psi}_m (\bm{\theta}'_m, T) \prod_{l=1}^n U^{\alpha_l}(\theta_l, T, \vec{X}_l) .
\end{align}

\section{Rindler spacetime}\label{Rindler_frame}

As an example of hyperbolic static spacetime, here we consider the right Rindler frame, defined at the beginning of Sec.~\ref{QFT_in_curved_spacetime_Rindler}. We adopt the theory of Sec.~\ref{NonMinkowksi_spacetime} with metric $g_{\mu\nu}$ given by Eq.~(\ref{Rindler_metric}) to derive the nonrelativistic limit of Rindler particles. We discuss the cases in which the time evolution of scalar and Dirac fields differs.

\subsection{Scalar field}\label{Rindler_frame_Scalar_field}

Here, we consider the scalar field $\hat{\Phi}_\text{R}$. Firstly, we derive the nonrelativistic limit of single particles by means of the condition
\begin{equation}\label{non_relativistic_limit_curved_wavefunction_0}
\tilde{\Phi}_n (\bm{\theta}_n) \approx 0 \text{ if there is an } l \in \{ 1, \dots , n \} \text{ such that } \left| \frac{\hbar \Omega(\theta_l)}{mc^2} - 1 \right| \gg \epsilon.
\end{equation}
Owing to the particular form of the Rindler metric, we find that the scalar product between nonrelativistic single particles can be approximated by the $L^2(\mathbb{R}^3)$ inner product. In this way, we show that nonrelativistic Rindler-Fock states can be equivalently treated as if they were in a flat spacetime, but with a modified free Schrödinger equation. Such modifications depend on the magnitude of the acceleration. By constraining $\alpha$ with respect to the nonrelativistic parameter $\epsilon$, we show how the Schrödinger equation is further approximated by the familiar Schrödinger-Newton equation. 

In Sec.~\ref{QFT_in_curved_spacetime_Rindler_scalar}, we showed the set of positive frequency modes in the right Rindler frame labeled by the quantum numbers $\theta = \vec{\theta} = (\Omega, \vec{K}_\perp)$, where $\Omega$ is the frequency (i.e., $\Omega(\theta) = \Omega$) and $\vec{K}_\perp$ is the transverse momentum. The explicit form of such modes is given by Eq.~(\ref{F_Rindler_all}). Equation (\ref{non_relativistic_limit_curved_wavefunction_0}) is, then, equivalent to
\begin{equation}\label{non_relativistic_limit_curved_wavefunction}
\tilde{\Phi}_n (\bm{\theta}_n) \approx 0 \text{ if there is an } l \in \{ 1, \dots , n \} \text{ such that } \left| \frac{\hbar \vec{e}_1 \cdot \vec{\theta}_l}{mc^2} - 1 \right| \gg \epsilon,
\end{equation}
with $\vec{e}_1 = (1,0,0)$.

When $\Omega$ satisfies the nonrelativistic condition (\ref{non_relativistic_limit_curved}), $F(\Omega,\vec{K}_\perp)$ is approximately solution of
\begin{equation} \label{Schrodinger_Rindler_scalar_F}
i \hbar \partial_0 F(\Omega,\vec{K}_\perp) \approx H_\text{CS} F(\Omega,\vec{K}_\perp),
\end{equation}
with
\begin{equation}\label{H_nM_Rindler}
H_\text{CS} = - \frac{\hbar^2}{ 2 m} \left[ \partial_3^2  + e^{2aZ} \left( \partial_1^2 + \partial_2^2 \right) \right] + \frac{mc^2}{2} \left(  1  + e^{2aZ}  \right).
\end{equation}
Equations (\ref{Schrodinger_Rindler_scalar_F}) and (\ref{H_nM_Rindler}) can be checked by using Eq.~(\ref{Rindler_metric}) in Eqs.~(\ref{Schrodinger_curved}) and (\ref{H_nM}).

Moreover, in the nonrelativistic limit, the Klein-Gordon product can be approximated by Eq.~(\ref{cKG_scalar_curved_product_G_nonrelativistic}). In the case of Rindler modes, we find that
\begin{equation}\label{cKG_scalar_curved_product_F_nonrelativistic}
( F(\Omega,\vec{K}_\perp), F(\Omega',\vec{K}'_\perp) )_\text{CKG}  \approx  \frac{2 m c^2}{\hbar^2} ( F(\Omega,\vec{K}_\perp), F(\Omega',\vec{K}'_\perp) )_{L^2_\text{S}(\mathbb{R}^3)}.
\end{equation}
Owing to Eq.~(\ref{Rindler_Klein_Gordon_product}), Eq.~(\ref{cKG_scalar_curved_product_F_nonrelativistic}) can be also replaced by
\begin{equation}\label{cKG_scalar_curved_product_F_nonrelativistic_2}
( F(\Omega,\vec{K}_\perp), F(\Omega',\vec{K}'_\perp) )_\text{CKG} 
 \approx \frac{2 m c^2}{\hbar^2} ( F(\Omega,\vec{K}_\perp), F(\Omega',\vec{K}'_\perp) )_{L^2(\mathbb{R}^3)}.
\end{equation}
This means that nonrelativistic Rindler single particles can be treated identically to Minkowski particles, but with a different free Schrödinger equation (\ref{Schrodinger_Rindler_scalar_F}).

\subsubsection{Quasi-inertial regime}

A further approximation for Eq.~(\ref{Schrodinger_Rindler_scalar_F}) can be obtained by considering
\begin{subequations}\label{local_limit}
\begin{align} 
& \frac{\hbar a}{m c} \lesssim \epsilon^{3/2}, \label{quasi_inertial_limit_a}\\
& a |Z| \lesssim \epsilon ,\label{quasi_inertial_limit_X}\\
& \frac{ \hbar |\vec{K}_\perp|}{m c} \lesssim \epsilon^{1/2} \label{local_limit_momentum} 
\end{align}
\end{subequations}
in addition to the nonrelativistic condition (\ref{non_relativistic_limit_curved}), with $\epsilon $ representing the maximum ratio between the nonrelativistic energy $E=\hbar \Omega - mc^2$ and the mass energy $mc^2$. Notice that the limit expressed by Eqs.~(\ref{non_relativistic_limit_curved}) and (\ref{local_limit}) is equivalent to the regime in which the speed of light $c$ goes to infinity while the variables $Z$, $\vec{K}_\perp$, $\alpha$, $E$ are kept fixed.

Equation (\ref{quasi_inertial_limit_X}) is to be intended as the condition in which states are mostly localized close to the position $Z=0$ with respect to the Rindler length scale $a^{-1}$. Explicitly, this means that we only consider states $| \Phi \rangle$ whose wave function $\Phi_n (T, \textbf{X}_n)$ is such that
\begin{equation}\label{quasi_inertial_limit_X_wavefunction}
\Phi_n (T, \textbf{X}_n) \approx 0 \text{ if there is an } l \in (1, \dots, n) \text{ such that } a | \vec{e}_3 \cdot \vec{X}_l | \gg \epsilon.
\end{equation}
The fact that $a |Z|$ goes to zero with the order of $\epsilon$ means that
\begin{equation}
U_\text{g} = m  \alpha Z
\end{equation}
has the same magnitude of $E$ (i.e., $U_\text{g} = mc^2 \mathcal{O}(\epsilon)$) and can hence be regarded as a nonrelativistic energy. We anticipate that $U_\text{g}$ represents the potential energy for the approximated Schrödinger equation in the limits (\ref{non_relativistic_limit_curved}) and (\ref{local_limit}).

Equation (\ref{quasi_inertial_limit_a}) states that the acceleration is relatively small with respect to the mass of the particles; whereas, the limit (\ref{quasi_inertial_limit_X}) can be identified as a locality condition where curvature effects are considered small (i.e., $g_{\mu\nu} \approx \eta_{\mu\nu}$). For these reasons we refer to Eq.~(\ref{local_limit}) as the ``quasi-inertial limit''.

Equation (\ref{local_limit_momentum}) is equivalent to $\hbar^2 |\vec{K}_\perp|^2 / (2 m) = mc^2 \mathcal{O}(\epsilon)$, which is a nonrelativistic condition for the transverse kinetic energy. Similarly to Eqs.~(\ref{non_relativistic_limit_curved}) and (\ref{quasi_inertial_limit_X}), Eq.~(\ref{local_limit_momentum}) needs be interpreted as a restriction to states characterized by a wave function $\tilde{\Phi}_n (\bm{\theta}_n)$ satisfying
\begin{equation}\label{local_limit_momentum_wavefunction}
\tilde{\Phi}_n (\bm{\theta}_n) \approx 0 \text{ if there is an } l \in (1, \dots, n) \text{ such that } \frac{\hbar |\vec{e}_2 \cdot \vec{\theta}_l| + \hbar |\vec{e}_3 \cdot \vec{\theta}_l|}{m c} \gg \epsilon^{1/2}.
\end{equation}

In summary, we say that the acceleration $\alpha$ and the state $| \Phi \rangle$ are in the quasi-inertial regime if, respectively, $a$ satisfies Eq.~(\ref{quasi_inertial_limit_a}) and the wave functions $\Phi_n (T, \textbf{X}_n)$ and $ \tilde{\Phi}_n (\bm{\theta}_n)$ have support in (\ref{quasi_inertial_limit_X}) and (\ref{local_limit_momentum}). In Appendix \ref{Proof_of_compatibility_between_nonrelativistic_and_quasiinertial_conditions} we show that the nonrelativistic and the quasi-inertial condition are compatible, in the sense that there can be states that simultaneously satisfy Eqs.~(\ref{non_relativistic_limit_curved_wavefunction}), (\ref{quasi_inertial_limit_X_wavefunction}) and (\ref{local_limit_momentum_wavefunction}).

By definition, the wave function $\Phi_n (T, \textbf{X}_n)$ is equal to $ \tilde{\Phi}_n (\bm{\theta}_n)$ smeared out with the modes $F(\vec{\theta})$. Consequently, the effects of the quasi-inertial regime can be deduced from the limiting behavior of the function $F(\Omega, \vec{K}_\perp, T, \vec{X})$ with respect to Eq.~(\ref{local_limit}). When Eqs.~(\ref{non_relativistic_limit_curved}) and (\ref{local_limit}) hold, Eq.~(\ref{F_tilde_Rindler}) can be approximated by
\begin{align}\label{F_tilde_Rindler_Hankel_2}
\tilde{F}(\Omega,\vec{K}_\perp,Z) \approx & \frac{\hbar^{5/6}  }{2^{7/6} \pi  a^{1/6} m^{1/3} c^{4/3}}  \nonumber \\
& \times \text{Ai} \left( 2^{1/3} \left(\frac{m c}{\hbar a} \right)^{2/3} \left[ \frac{\hbar^2 |\vec{K}_\perp|^2}{2 m^2 c^2} + a Z - \left( \frac{\hbar \Omega}{m c^2} - 1  \right) \right] \right),
\end{align}
where $\text{Ai}(\xi)$ is the Airy function. The proof for Eq.~(\ref{F_tilde_Rindler_Hankel_2}) is provided by Appendix \ref{Proof_of_F_tilde_Rindler_Hankel_2}.

From Eq.~(\ref{F_tilde_Rindler_Hankel_2}), one can see that $F(\Omega,\vec{K}_\perp)$ is approximately solution of
\begin{equation}\label{Schrodinger_Rindler_scalar_local}
i \hbar \partial_0 F(\Omega,\vec{K}_\perp) \approx H_\text{QI} F(\Omega,\vec{K}_\perp),
\end{equation}
with
\begin{equation}\label{H_QI}
H_\text{QI} = - \frac{\hbar^2}{ 2 m} \left( \partial_1^2  + \partial_2^2 + \partial_3^2 \right) + mc^2  +  U_\text{g}.
\end{equation}
Indeed, by knowing that the Airy function is solution of the differential equation $\text{Ai}''(\xi) = \xi \text{Ai}(\xi)$, Eq.~(\ref{F_tilde_Rindler_Hankel_2}) leads to
\begin{equation}\label{Schrodinger_Rindler_local}
\partial_1^2 F(\Omega,\vec{K}_\perp,T,\vec{X}) \approx 2 \left(\frac{m c}{\hbar} \right)^2 \left[ \frac{\hbar^2 |\vec{K}_\perp|^2}{2 m^2 c^2} + a Z - \left( \frac{\hbar \Omega}{m c^2} - 1  \right) \right] F(\Omega,\vec{K}_\perp,T,\vec{X}) ,
\end{equation}
which, together with Eq.~(\ref{F_Rindler}), proves Eq.~(\ref{Schrodinger_Rindler_scalar_local}).

Equation (\ref{Schrodinger_Rindler_scalar_local}) is a Schrödinger-Newton equation with a mass term $ mc^2 $ and a potential energy $U_\text{g}$ generated by a uniform gravitational force $m \alpha$ along the $Z$ axis. This can be interpreted as the fact that an accelerated frame is locally equivalent to an observer that experiences a gravitational force. The result is hence expected by the equivalence principle of general relativity.

The error associated to Eq.~(\ref{Schrodinger_Rindler_scalar_local}) approximating Eq.~(\ref{Schrodinger_Rindler_scalar_F}) can be obtained by evaluating the difference between the Hamiltonians $H_\text{CS}$ and $H_\text{QI}$ acting on $F(\Omega,\vec{K}_\perp)$, which lead to
\begin{equation}
H_\text{CS} - H_\text{QI} = \frac{\hbar^2 |\vec{K}_\perp|^2}{ 2 m} (e^{2aZ} - 1)  + mc^2 \left( \frac{e^{2aZ} - 1}{2} - aZ \right).
\end{equation}
For nonrelativistic modes $F(\Omega,\vec{K}_\perp)$ and in the quasi-inertial limit (\ref{local_limit}), $H_\text{CS} - H_\text{QI}$ acts on $F(\Omega,\vec{K}_\perp)$ with leading order
\begin{equation}\label{H_S_H_QI}
H_\text{CS} - H_\text{QI} = mc^2 \mathcal{O}(\epsilon^2).
\end{equation}
By comparing Eq.~(\ref{H_S_H_QI}) with Eq.~(\ref{H_S_h_CKG_error}), one notices that the errors associated to Eq.~(\ref{Schrodinger_Rindler_scalar_local}) are at least of the same orders of Eq.~(\ref{Schrodinger_Rindler_scalar_F}). Therefore, no reason to prefer the Hamiltonian $H_\text{CS}$ over $ H_\text{QI}$ exists: they can be considered equivalent in the nonrelativistic quasi-inertial regime. Moreover, the difference between the Hamiltonian $H_\text{QI}$ and the exact fully-relativistic $h_\text{CKG}$ is
\begin{equation} \label{H_QI_h_CKG_error}
H_\text{QI} - h_\text{CKG} = mc^2 \mathcal{O}(\epsilon^2).
\end{equation}
Equation (\ref{H_QI_h_CKG_error}) gives an esteem of the GR corrections to the Schrödinger-Newton equation (\ref{Schrodinger_Rindler_scalar_local}) for scalar fields.

\subsubsection{High acceleration regime}

In the quasi-inertial regime (\ref{local_limit}), GR corrections are of the same order of special relativistic corrections [Eqs.~(\ref{H_S_h_CKG_error}) and (\ref{H_QI_h_CKG_error})] and, hence, unnoticeable in the nonrelativistic regime. We expect to obtain GR effects by raising the acceleration beyond the approximation (\ref{quasi_inertial_limit_a}) and by extending the support of the wave function $\Phi_n (T, \textbf{X}_n)$ outside the quasiflat region (\ref{quasi_inertial_limit_X_alternative}).

Here, we test the appearance of GR corrections by considering a parameter $\epsilon_\text{HA}$, which is larger than $\epsilon$ but less than one order of magnitude more, i.e., $\epsilon \ll \epsilon_\text{HA} \ll \epsilon^{1/2}$. We use $\epsilon_\text{HA}$ to give a slightly weaker conditions than Eqs.~(\ref{quasi_inertial_limit_a}) and (\ref{quasi_inertial_limit_X_alternative}) by considering the following regime
\begin{subequations}\label{local_limit_alternative}
\begin{align} 
& \frac{\hbar a}{m c} \lesssim \epsilon^{1/2}_\text{HA} \epsilon, \label{quasi_inertial_limit_a_alternative}\\
& a |Z| \lesssim \epsilon_\text{HA} ,\label{quasi_inertial_limit_X_alternative}\\
& \frac{ \hbar |\vec{K}_\perp|}{m c} \lesssim \epsilon^{1/2}. \label{local_limit_momentum_alternative} 
\end{align}
\end{subequations}
We refer to Eq.~(\ref{local_limit_alternative}) as the ``high acceleration condition''. Notice that Eq.~(\ref{local_limit_alternative}) cannot be obtained from the limit $c \rightarrow \infty$, with $X$, $\vec{K}_\perp$, $\alpha$, $E$ fixed.

In analogy to the quasi-inertial condition, we say that the acceleration $\alpha$ and the state $| \Phi \rangle$ are in the high acceleration regime if, respectively, $a$ satisfies Eq.~(\ref{quasi_inertial_limit_a_alternative}) and the wave functions $\Phi_n (T, \textbf{X}_n)$ and $ \tilde{\Phi}_n (\bm{\theta}_n)$ have support in (\ref{quasi_inertial_limit_X_alternative}) and (\ref{local_limit_momentum_alternative}). In Appendix \ref{Proof_of_compatibility_between_nonrelativistic_and_high_acceleration_conditions}, we demonstrate the compatibility between the nonrelativistic condition (\ref{non_relativistic_limit_curved}) and the high acceleration condition (\ref{local_limit_alternative}).

By following the proof of Appendix \ref{Proof_of_F_tilde_Rindler_Hankel_2}, one can show that in the high acceleration limit (\ref{local_limit_alternative}), the Rindler modes $F(\Omega,\vec{K}_\perp)$ can be approximated by Eq.~(\ref{F_tilde_Rindler_Hankel_2}) and, hence, are approximately solutions of Eq.~(\ref{Schrodinger_Rindler_scalar_local}). Notice that, at variance with the quasi-inertial regime, the potential energy $U_\text{g}$ may reach relativistic values at the border of the region (\ref{quasi_inertial_limit_X_alternative}), since $U_\text{g} = mc^2 \mathcal{O}(\epsilon_\text{HA})$ and $\epsilon_\text{HA} \gg \epsilon$. However, for negative values of $Z$, the potential energy $U_\text{g}$ is balanced by large values of the kinetic energy $-(\hbar/2m) \partial_3^2$ due to the rapidly oscillating behavior of the Airy function in this region; for positive values of $Z$, instead, the Airy function is exponentially vanishing.

The error associated to the Schrödinger-Newton equation (\ref{Schrodinger_Rindler_scalar_local}) is given by
\begin{equation} \label{H_S_H_QI_high_a}
H_\text{CS} - H_\text{QI} = mc^2 \mathcal{O}(\epsilon_\text{HA}^2),
\end{equation}
since the support of the wave functions is (\ref{quasi_inertial_limit_X_alternative}). In particular, errors of the order of $\epsilon_\text{HA}^2$ may be detected for negative values of $Z$ such that $ Z \sim - \epsilon_\text{HA}/a$. Inside the region $a|Z| \lesssim \epsilon$, instead, Eq.~(\ref{H_S_H_QI}) is still valid. Hence, a more precise evaluation of the errors to the Schrödinger-Newton equation (\ref{Schrodinger_Rindler_scalar_local}) is given by
\begin{equation} \label{H_S_H_QI_high_a_2}
H_\text{CS} - H_\text{QI} = \begin{cases}
mc^2 \mathcal{O}(\epsilon_\text{HA}^2) & \text{if } \epsilon \ll  a|Z| \lesssim \epsilon_\text{HA} \\
mc^2 \mathcal{O}(\epsilon^2) & \text{if } a|Z| \lesssim \epsilon
\end{cases}.
\end{equation}

GR corrections to the Schrödinger-Newton equation (\ref{Schrodinger_Rindler_scalar_local}) in the region $a|Z| \gg \epsilon$ are now dominated by Eq.~(\ref{H_S_H_QI_high_a}) and lead to
\begin{equation} \label{H_QI_h_CKG_error_high_a}
H_\text{QI} - h_\text{CKG} = mc^2 \mathcal{O}(\epsilon^2_\text{HA}).
\end{equation}
More precisely, we obtain
\begin{equation} \label{H_QI_h_CKG_error_high_a_2}
H_\text{QI} - h_\text{CKG} = \begin{cases}
mc^2 \mathcal{O}(\epsilon_\text{HA}^2) & \text{if } \epsilon \ll  a|Z| \lesssim \epsilon_\text{HA} \\
mc^2 \mathcal{O}(\epsilon^2) & \text{if } a|Z| \lesssim \epsilon
\end{cases}.
\end{equation}

\subsection{Dirac field}\label{Nonrelativistic_limit_Dirac_Rindler}

Here we discuss the case of Dirac fields $\hat{\Psi}_\text{R}$ in Rindler spacetime. Firstly, we review the nonrelativistic limit of single particles. We obtain a Schrödinger equation that is different from the one saw for the scalar case. We then show the convergence to the Schrödinger-Newton equation (\ref{Schrodinger_Rindler_scalar_local}) in the quasi-inertial limit (\ref{local_limit}). GR corrections to such an equation are of order $\epsilon^2 mc^2$, as for the scalar field. Conversely, by considering the high acceleration limit (\ref{local_limit_alternative}), we obtain GR corrections to the Schrödinger-Newton theory that can reach values $(\epsilon_\text{HA} / \epsilon)^{2/3}$ times larger than the ones obtained for the scalar field. This means that Dirac fields are better candidates for detecting GR corrections to the Schrödinger-Newton theory. Moreover, we show that the difference between scalar and Dirac Hamiltonians is $(\epsilon_\text{HA} / \epsilon)^{2/3}$ times larger than what we found for the Minkowski case [Eq.~(\ref{h_KG_h_M_error})]. In other words, the Rindler metric is able to enhance the distinguishability between scalar and Dirac fields.

In the nonrelativistic limit (\ref{non_relativistic_limit_curved}), the modes $U_{\text{R} s}(\Omega,\vec{K}_\perp)$ defined in Sec.~\ref{Rindler_Dirac_modes} are approximately solution of the Schrödinger equation (\ref{Dirac_curved_Schrodinger_nonrelativistic}), i.e.,
\begin{equation}\label{Schrodinger_Rindler_Dirac}
i \hbar \partial_0 U_{\text{R} s}(\Omega,\vec{K}_\perp) \approx ( H_\text{CS} + \Delta H ) U_{\text{R} s}(\Omega,\vec{K}_\perp),
\end{equation} 
where, in this case, $H_\text{CS}$ is given by Eq.~(\ref{H_nM_Rindler}) and $\Delta H$ by
\begin{equation}\label{Schrodinger_Rindler_Dirac_Delta_H}
\Delta H =  - i \frac{\hbar a}{2 m} \gamma^0 \gamma^3 h_\text{CD} + \frac{(\hbar a)^2}{8 m},
\end{equation} 
\begin{align}\label{Schrodinger_Rindler_Dirac_h_NM}
& h_\text{CD} =  - \hbar c \gamma^0  \left[ i \frac{ca}{2} \gamma^3 +  i c \gamma^3 \partial_3  + e^{aZ} \left(i c \gamma^1 \partial_1 + i c \gamma^2 \partial_2 - \frac{mc^2}{\hbar} \right) \right] .
\end{align} 

It can be noticed that the Schrödinger equation (\ref{Schrodinger_Rindler_Dirac}) differs from Eq.~(\ref{Schrodinger_Rindler_scalar_F}) obtained for the scalar case. The difference between the two Hamiltonians $H_\text{CD}$ and $H_\text{CS}$ is given by Eq.~(\ref{Schrodinger_Rindler_Dirac_Delta_H}), which, in the nonrelativistic limit can be approximated by
\begin{equation}\label{Delta_H_Rindler_nonrelativistic}
\Delta H \approx  - i \frac{\hbar c^2 a}{2} \gamma^0 \gamma^3 + \frac{(\hbar a)^2}{8 m}.
\end{equation}
Equation (\ref{Delta_H_Rindler_nonrelativistic}) is generally non-vanishing. As already explained in Sec.~\ref{NonMinkowksi_spacetime_Dirac_field}, this occurs because the metric is curved and $a$ can be arbitrarily large.

\subsubsection{Quasi-inertial regime}

Different scenarios are possible when $a$ varies with respect to other dimensional quantities and the nonrelativistic parameter $\epsilon$. For instance, in the case of the quasi-inertial limit defined by Eq.~(\ref{local_limit}), the scalar product (\ref{Dirac_Rindler_scalar_product}) can be approximated by the $\mathbb{C}^4 \otimes L^2(\mathbb{R}^3)$ inner product
\begin{equation}\label{Dirac_Rindler_scalar_product_approximation}
( \Psi, \Psi' )_{\mathbb{C}^4 \otimes L^2_\text{D}(\mathbb{R}^3)} \approx ( \Psi, \Psi' )_{\mathbb{C}^4 \otimes L^2(\mathbb{R}^3)}
\end{equation}
and the dynamics of the single particles is reduced to the familiar Schrödinger-Newton equation
\begin{equation}\label{Schrodinger_Newton_Rindler_Dirac}
i \hbar \partial_0 U_{\text{R} s}(\Omega,\vec{K}_\perp) \approx H_\text{QI} U_{\text{R} s}(\Omega,\vec{K}_\perp),
\end{equation} 
already defined for scalar field in Eqs.~(\ref{Schrodinger_Rindler_scalar_local}) and (\ref{H_QI}).

Equation (\ref{Dirac_Rindler_scalar_product_approximation}) is due to the fact that in the quasi-inertial limit, wave functions are localized inside the region $a|Z| \ll 1$, and, hence, $e^{aZ} \approx 1$. Equation (\ref{Schrodinger_Newton_Rindler_Dirac}), instead, can be directly proven by applying the nonrelativistic and the quasi-inertial conditions (\ref{non_relativistic_limit_curved}) and (\ref{local_limit}) on Eqs.~(\ref{U_U_tilde_conclusion}) and (\ref{W_tilde_W_tilde}). In particular, one can use the approximation 
\begin{equation}\label{K_pm_nonrelativistic_quasiinertial}
\mathfrak{K}(\Omega,\vec{K}_\perp,Z) \approx K_{i \Omega /ca} \left( \kappa (\vec{K}_\perp) \frac{e^{aZ}}{a} \right)
\end{equation}
and Eqs.~(\ref{F_tilde_Rindler}) and (\ref{F_tilde_Rindler_Hankel_2}) to prove Eq.~(\ref{Schrodinger_Newton_Rindler_Dirac}).

Alternatively, Eq.~(\ref{Schrodinger_Newton_Rindler_Dirac}) can be proven by noticing that $\Delta H$, acting on nonrelativistic states, is approximated by Eq.~(\ref{Delta_H_Rindler_nonrelativistic}) and, hence, in the quasi-inertial limit (\ref{local_limit}),
\begin{equation} \label{Delta_H_W_approximation_2}
\Delta H  = mc^2 \mathcal{O}(\epsilon^2),
\end{equation}
which is $\epsilon$ times smaller than the potential energy $U_\text{g} = mc^2 \mathcal{O}(\epsilon)$. This, together with the fact that, in the quasi-inertial limit (\ref{local_limit}), $H_\text{CS}$ can be replaced by $H_\text{QI}$ [Eq.~(\ref{H_S_H_QI})], leads to Eq.~(\ref{Schrodinger_Newton_Rindler_Dirac}).

Equation (\ref{Delta_H_W_approximation_2}) comes from the fact that in the quasi-inertial limit $\hbar c a \lesssim \epsilon^{3/2} mc^2$ and for any couple of nonrelativistic modes $U_{\text{R} s}(\Omega,\vec{K}_\perp)$, $U_{\text{R} s'}(\Omega',\vec{K}_\perp')$,
\begin{equation}\label{U_gamma1_U}
U_{\text{R} s'}^\dagger(\Omega',\vec{K}_\perp') c \gamma^0 \gamma^3 U_{\text{R} s}(\Omega,\vec{K}_\perp) = U_{\text{R} s'}^\dagger(\Omega',\vec{K}_\perp') U_{\text{R} s}(\Omega,\vec{K}_\perp) \mathcal{O}(\epsilon^{1/2});
\end{equation}
hence, $c \gamma^0 \gamma^3$ effectively acts on nonrelativistic modes as $\mathcal{O}(\epsilon^{1/2})$. Equation (\ref{U_gamma1_U}) comes from the following property
\begin{equation}\label{Dirac_representation_reduction_nonrelativistic_limit_general_Rindler}
 \mathfrak{v}_s^\dagger U_{\text{R} s'}(\Omega,\vec{K}_\perp) = \mathfrak{u}_r^\dagger U_{\text{R} r'}(\Omega,\vec{K}_\perp) \mathcal{O}(\epsilon^{1/2}),
\end{equation}
which holds for any nonrelativistic mode $U_{\text{R} s}(\Omega,\vec{K}_\perp)$. To prove Eq.~(\ref{Dirac_representation_reduction_nonrelativistic_limit_general_Rindler}), one has to consider the explicit form of $U_{\text{R} s}(\Omega,\vec{K}_\perp)$ and compare its spinorial components. We provide such a proof in Appendix \ref{Proof_of_Dirac_representation_reduction_nonrelativistic_limit_general_Rindler}.

As a consequence of Eq.~(\ref{Dirac_representation_reduction_nonrelativistic_limit_general_Rindler}), nonrelativistic modes are approximately described only by their first two spinorial components and, hence, the representation space becomes $\mathbb{C}^2 \otimes L^2(\mathbb{R}^3)$. This result is equivalent to what we found in the Minkowski spacetime [Eq.~(\ref{Dirac_representation_reduction_nonrelativistic_limit_general})].

In summary, Dirac modes are approximately solution of the Schrödinger equation for the scalar field and the product between states is represented by the $\mathbb{C}^2 \otimes L^2(\mathbb{R}^3)$ inner product. This means that nonrelativistic quasi-inertial Dirac particles can be described identically to scalar states with the exception of spin degeneracy, as it occurs in the Minkowski spacetime.

From Eqs.~(\ref{H_D_h_nM_error}), (\ref{H_S_H_QI}) and (\ref{Delta_H_W_approximation_2}), one can derive the errors associated to the Schrödinger-Newton equation (\ref{Schrodinger_Newton_Rindler_Dirac})
\begin{equation} \label{H_QI_h_nM_error}
H_\text{QI} - h_\text{CD} = mc^2 \mathcal{O}(\epsilon^2).
\end{equation}
By comparing Eq.~(\ref{H_QI_h_nM_error}) with Eq.~(\ref{H_QI_h_CKG_error}), one can deduce that GR corrections to the Schrödinger-Newton equation for Dirac fields are of the same order as the GR corrections for scalar fields. Moreover, the difference between scalar and Dirac Hamiltonians is of the same order as in the Minkowski case [Eq.~(\ref{h_KG_h_M_error})] and reads as
\begin{equation} \label{h_CKG_h_nM_error}
h_\text{CKG} - h_\text{CD} = mc^2 \mathcal{O}(\epsilon^2).
\end{equation}

\subsubsection{High acceleration regime}

A different scenario can be considered by changing the asymptotic behavior of $a$ with respect to the nonrelativistic limit. For instance, by considering the high acceleration limit (\ref{local_limit_alternative}), one obtains
\begin{equation}\label{Delta_H_relativistic}
\Delta H = mc^2 \mathcal{O}(\epsilon_\text{HA}^{2/3}\epsilon^{4/3}).
\end{equation}

Equation (\ref{Delta_H_relativistic}) can be proven similarly to Eq.~(\ref{Delta_H_W_approximation_2}) with the difference that $\hbar c a \lesssim \epsilon^{1/2}_\text{HA} \epsilon mc^2$, and that for any nonrelativistic couple of modes $U_{\text{R} s}(\Omega,\vec{K}_\perp)$, $U_{\text{R} s'}(\Omega',\vec{K}_\perp')$,
\begin{equation}\label{U_gamma1_U_high_a}
 U_{\text{R} s'}^\dagger(\Omega',\vec{K}_\perp') c \gamma^0 \gamma^3 U_{\text{R} s}(\Omega,\vec{K}_\perp) = U_{\text{R} s'}^\dagger(\Omega',\vec{K}_\perp') U_{\text{R} s}(\Omega,\vec{K}_\perp) \mathcal{O}(\epsilon_\text{HA}^{1/6}\epsilon^{1/3}).
\end{equation}
Equation (\ref{U_gamma1_U_high_a}) comes from the equivalent of Eq.~(\ref{Dirac_representation_reduction_nonrelativistic_limit_general_Rindler}) in the high acceleration limit, i.e.,
\begin{equation}\label{Dirac_representation_reduction_nonrelativistic_limit_general_Rindler_high_a}
 \mathfrak{v}_s^\dagger U_{\text{R} s'}(\Omega,\vec{K}_\perp) =  \mathfrak{u}_r^\dagger U_{\text{R} r'}(\Omega,\vec{K}_\perp) \mathcal{O}(\epsilon_\text{HA}^{1/6}\epsilon^{1/3})
\end{equation}
that is proved in Appendix \ref{Proof_of_Dirac_representation_reduction_nonrelativistic_limit_general_Rindler_high_a}.

From Eqs.~(\ref{H_D_h_nM_error}), (\ref{H_S_H_QI_high_a}) and (\ref{Delta_H_relativistic}), one can notice that the errors associated to the Schrödinger-Newton equation (\ref{Schrodinger_Newton_Rindler_Dirac}) are dominated by $H_\text{CS} - H_\text{QI}$ and are of the order of
\begin{equation} \label{H_QI_h_nM_error_large_a}
H_\text{QI} - h_\text{CD} = mc^2 \mathcal{O}(\epsilon_\text{HA}^2).
\end{equation}
This means that the GR corrections to the Schrödinger-Newton equation for Dirac fields are the same as the ones found for scalar fields. However, a more detailed analysis shows that errors of the order of $\epsilon^2_\text{HA}$ can only be achieved in the region $\epsilon \ll  a|Z| \lesssim \epsilon_\text{HA}$. Inside the region $a|Z| \lesssim \epsilon$, instead, the errors associated to the Schrödinger-Newton equation (\ref{Schrodinger_Newton_Rindler_Dirac}) are dominated by $\Delta H$ [Eqs.~(\ref{H_D_h_nM_error}), (\ref{H_S_H_QI_high_a_2}) and (\ref{Delta_H_relativistic})], which are of order of $\epsilon_\text{HA}^{2/3}\epsilon^{4/3}$. Overall, we obtain the following cases
\begin{equation} \label{H_QI_h_nM_error_large_a_2}
H_\text{QI} - h_\text{CD} = \begin{cases}
mc^2 \mathcal{O}(\epsilon_\text{HA}^2) & \text{if } \epsilon \ll  a|Z| \lesssim \epsilon_\text{HA}, \\
mc^2 \mathcal{O}(\epsilon_\text{HA}^{2/3}\epsilon^{4/3}) & \text{if } a|Z| \lesssim \epsilon
\end{cases}.
\end{equation}

By comparing Eq.~(\ref{H_QI_h_nM_error_large_a_2}) with Eq.~(\ref{H_QI_h_CKG_error_high_a_2}), one can deduce that GR corrections to the Schrödinger-Newton equation for scalar and Dirac fields are of the same order in the region $\epsilon \ll  a|Z| \lesssim \epsilon_\text{HA}$. Conversely, inside the region $ a|Z| \lesssim \epsilon$, errors for Dirac fields are larger, since $\epsilon_\text{HA}^{2/3}\epsilon^{4/3} \gg \epsilon^2$. Hence, less experimental precision is needed to spoil GR corrections. In particular, by reaching the experimental precision for energies up to the order of $\epsilon_\text{HA}^{2/3}\epsilon^{4/3}$, a term proportional to $\gamma^0 \gamma^3$ [Eq.~(\ref{Delta_H_Rindler_nonrelativistic})] appears in the Dirac case, while nothing shows up for scalar fields.

\begin{table}
\begin{center}
\begin{tabular}{|l||c|c|c|c|}
\hline
 & $\dfrac{\hbar a}{m c} $ & $a |Z| $ & $\dfrac{\hbar |\vec{K}_\perp|}{m c}  $ & $\dfrac{\Delta h}{mc^2} $ \\[1.5ex]
\hline
\hline
& & & & \\[-1em]
quasi-inertial limit & $\mathcal{O}(\epsilon^{3/2})$ & $\mathcal{O}(\epsilon)$ & $\mathcal{O}(\epsilon^{1/2})$ & $\mathcal{O}(\epsilon^2)$ \\
\hline
& & & & \\[-1em]
high acceleration limit & $\mathcal{O}(\epsilon^{1/2}_\text{HA} \epsilon)$ & $\mathcal{O}(\epsilon_\text{HA})$ & $\mathcal{O}(\epsilon^{1/2})$ & $\mathcal{O}(\epsilon_\text{HA}^{2/3}\epsilon^{4/3})$ \\
\hline
\end{tabular}
\end{center}
\caption{Asymptotic behavior with respect to the nonrelativistic parameters $\epsilon$ and $\epsilon_\text{HA}$ for different limits. The quasi-inertial and the high acceleration limit are defined by, respectively, Eqs.~(\ref{local_limit}) and (\ref{local_limit_alternative}) in terms of the position $Z$, the transverse momentum $\vec{K}_\perp$ and the acceleration $\alpha = a c^2$. The variable $\Delta h = h_\text{CKG} - h_\text{CD}$ is the difference between the scalar and Dirac Hamiltonians. The orders of $\Delta h $ for the two limits are shown in the last column. In the high acceleration limit, $\Delta h $ is larger than its equivalent for Minkowski spacetime $(h_\text{KG} - h_\text{M})/mc^2 \lesssim \epsilon^2$ [Eq.~(\ref{h_KG_h_M_error})]. This means that lower precision is needed to distinguish between the time evolution of scalar and Dirac fields.} \label{table_Rindler_limits}
\end{table}

By using Eqs.~(\ref{H_S_h_CKG_error}), (\ref{H_D_h_nM_error}), (\ref{H_D_H_S_Delta_H}) and (\ref{Delta_H_relativistic}), one can deduce that
\begin{equation} \label{h_CKG_h_nM_error_large_a}
h_\text{CKG} - h_\text{CD} = mc^2 \mathcal{O}(\epsilon_\text{HA}^{2/3}\epsilon^{4/3}),
\end{equation}
which means that the difference between scalar and Dirac Hamiltonians is visible at the order of $\epsilon_\text{HA}^{2/3}\epsilon^{4/3} $. Such an order of magnitude is lower than the one needed for the distinguishability between the two types of fields in the Minkowski spacetime [Eq.~(\ref{h_KG_h_M_error})], since $\epsilon_\text{HA}^{2/3}\epsilon^{4/3} \gg \epsilon^2$. The result is that in the Rindler frame, when the acceleration is sufficiently high, it is easier to distinguish between scalar and Dirac fields than in the Minkowski spacetime. This is a difference between the quasi-inertial (\ref{local_limit}) and the high acceleration limit (\ref{local_limit_alternative}) that is summarized in Table \ref{table_Rindler_limits}.

\section{Conclusions}\label{Conclusions}

We investigated the nonrelativistic limit of scalar and Dirac particles in curved static spacetimes. It is well known that particles in flat spacetime are approximated by the same Schrödinger equation in the nonrelativistic limit [Eqs.~(\ref{Schrodinger_free_mass_positive_negative_frequencies}) and (\ref{Schrodinger_Dirac_free_mass_positive_negative_frequencies})]. On the contrary, scalar and Dirac fields in curved spacetimes have different nonrelativistic asymptotic Hamiltonians $H_\text{CS}$ and $H_\text{CD}$. This implies that the two kinds of particles evolve differently when the gravitational field is sufficiently strong.

As an example, we considered nonrelativistic particles in a Rindler metric with acceleration $\alpha$. For an $\alpha$ sufficiently large, $\Delta H=H_\text{CD} - H_\text{CS}$ cannot be ignored and leads to noticeable differences on the time evolution of the particles. If the spacetime is almost flat [Eq.~(\ref{local_limit})], $\Delta H$ becomes negligible if compared to the gravitational potential $U_\text{g}$; in this way, one finds the usual Schrödinger-Newton equation (\ref{Schrodinger_Rindler_scalar_local}) for both scalar and Dirac fields.

We remark that the nonrelativistic limit is often regarded as the one in which $c \rightarrow \infty$. However this limit may vary in a way dependent on the acceleration. Letting $a = \alpha / c^2$, the limit $c \rightarrow \infty$ does not specify if $\alpha$ has to go to infinity with finite $a$, or $a$ has to go to zero with finite $\alpha$.

By considering an $\alpha$ sufficiently large [Eq.~(\ref{local_limit_alternative})], we find that GR corrections coming from $\Delta H $ can reach values of some orders larger than GR corrections coming from the Klein-Gordon equation (\ref{Rindler_Klein_Gordon}) [Eqs.~(\ref{H_QI_h_CKG_error_high_a_2}) and (\ref{H_QI_h_nM_error_large_a_2})]. This implies that an improved experimental precision will eventually unveil a second-order GR correction only for Dirac fields. We believe that this scaling addresses the possibility of observing spin-gravity coupling as a signal for general relativity in quantum particle phenomena.

The different dynamics of the Klein-Gordon and Dirac particles in the Minkowski and Rindler frames can be exploited for a test for the Einstein's Equivalence Principle \cite{einstein1907relativity, NORTON1985203}. The principle states that in freely falling frames, the laws of physics are the same as if there were no gravity; which means that physical phenomena can be locally described by a Minkowski spacetime. Conversely, a uniformly accelerated frame (i.e., Rindler spacetime) appears indistinguishable from an inertial reference frame affected by a uniform gravitational field.

The Einstein's Equivalence Principle has been repeatedly tested in the classical regime \cite{Will2014}; however, its application in the quantum domain is still debated \cite{Lammerzahl1996, PhysRevD.55.455, Rosi2017, Zych2018, giacomini2023einsteins}. In this chapter, we showed a novel quantum effect probing the nature of the observer's frame. The possibility to distinguish between Klein-Gordon and Dirac particles in the nonrelativistic limit and in presence of a gravitational field would validate the prediction of equivalence between uniformly accelerated frames and inertial frames affected by gravity.

\part{Minkowski and Rindler particles}\label{Inequivalent_particle_representations_and_Unruh_effect}

\chapter{Frame dependent content of particles} \label{Frame_dependent_content_of_particles}

\section{Introduction}\label{Frame_dependent_content_of_particles_Introduction}

In the previous Part, we studied the particle content of fields in flat and curved spacetimes. In particular, we considered Minkowski and Rindler spacetimes. In Sec.~\ref{QFT_in_Minkowski_spacetime}, we derived the positive frequency modes of scalar ($\hat{\phi}$) and Dirac ($\hat{\psi}$) fields in Minkowski spacetime; conversely, in Sec.~\ref{QFT_in_curved_spacetime_Rindler}, we computed the Rindler modes for Rindler scalar ($\hat{\Phi}_\nu$) and Dirac ($\hat{\Psi}_\nu$) fields. Positive frequency solutions of the field equation provide the particle content of the field in each spacetime by generating the Minkowski-Fock space $\mathcal{H}_\text{M}$ and the Rindler-Fock space $\mathcal{H}_{\text{L}, \text{R}}$. In this way, we gave the full description of particle phenomenology for inertial and accelerated observer in their respective frames.

Still, an important piece of information is missing. We know that a Minkowski single particle is described by an inertial observer as a positive frequency mode of the Klein-Gordon or Dirac equation. However, we still do know how an accelerated observer would describe such a state. We need a prescription to relate states of one spacetime representation to the other. If Alice is an inertial experimenter and Rob an accelerated observer, how would Rob describe states that are prepared by Alice?

We find an answer to the previous question in the axiomatic formulation of algebraic QFTCS \cite{wald1994quantum}. The main idea is that field operators in different frames are ontologically equivalent, but described by different coordinate systems. For instance, the scalar operators $\hat{\phi}(t,\vec{x})$ and $\hat{\Phi}_\nu (T,\vec{X})$ are the same if the coordinates $(t,\vec{x})$ and $(T,\vec{X})$ satisfy the coordinate transformation equation (\ref{Rindler_coordinates_transformation}). Explicitly, this means that
\begin{equation}\label{scalar_transformation_Rindler}
\hat{\Phi}_\nu (T,\vec{X}) = \hat{\phi} (t_\nu(T,\vec{X}), \vec{x}_\nu(T,\vec{X})).
\end{equation}

Equation (\ref{scalar_transformation_Rindler}) can be interpreted as the transformation $\hat{\phi} \mapsto \hat{\Phi}_\nu$ between scalar fields in inertial and accelerated frame, in analogy to classical physics. The inverse of Eq.~(\ref{scalar_transformation_Rindler}) is
\begin{equation}\label{scalar_transformation_Rindler_inverse_t}
\hat{\phi}(t,\vec{x}) = \sum_{\nu=\{\text{L},\text{R}\}} \theta(s_\nu z) \hat{\Phi}_\nu(T_\nu (t, \vec{x}),\vec{X}_\nu (t, \vec{x})),
\end{equation}
where the functions $T_\nu(t,\vec{x}) $ and $\vec{X}_\nu(t,\vec{x})$ map the Minkowski coordinates $(t,\vec{x})$ to the Rindler coordinates $(T,\vec{X})$ and are the inverse of Eq.~(\ref{Rindler_coordinates_transformation}).

For Dirac field, one also has to consider the spinorial degrees of freedom to let the field transform as a spinor. The explicit transformation $\hat{\Psi}_\nu \mapsto \hat{\psi}$ is \cite{Oriti}
\begin{equation}\label{scalar_transformation_Rindler_inverse_Dirac}
\hat{\psi}(t,\vec{x}) =  \sum_{\nu=\{\text{L},\text{R}\}} \theta(s_\nu z) \exp \left( \frac{1}{2} \gamma^0 \gamma^3 T_\nu(t,\vec{x}) \right)  \hat{\Psi}_\nu(T_\nu(t,\vec{x}),\vec{X}_\nu(t,\vec{x})).
\end{equation}

We now know how to relate field operators of one frame to the other. However, the question still remains: how do we relate particle states to each other? The usual prescription is to use Eq.~(\ref{scalar_transformation_Rindler_inverse_t}) or Eq.~(\ref{scalar_transformation_Rindler_inverse_Dirac}) to compute the Bogoliubov transformations associating creators/annihilators of one frame to the other. Then, by means of the Bogoliubov transformations, one is able to relate element of the Minkowski-Fock space $\mathcal{H}_\text{M}$ to elements of the Rindler-Fock space $\mathcal{H}_{\text{L}, \text{R}}$.

For instance, the procedure for scalar fields is described by the following steps:
\begin{enumerate}
\item Isolate the Minkowski annihilators $\hat{a}(\theta)$ and $\hat{b}^\dagger(\theta)$ from Eq.~(\ref{free_field}) by using the Klein-Gordon scalar product (\ref{KG_scalar_product}) and the orthonormality conditions (\ref{KG_scalar_product_orthonormality_f})\label{Bogoliubov_step_1}
\begin{align}\label{Bogoliubov_transformations_1}
& \hat{a}(\vec{k}) = ( f(\vec{k}), \hat{\phi})_\text{KG}, 
& \hat{b}^\dagger(\vec{k}) = -(f^*(\vec{k}),  \hat{\phi})_\text{KG}.
\end{align}
\item Plug Eq.~(\ref{scalar_transformation_Rindler_inverse_t}) into Eq.~(\ref{Bogoliubov_transformations_1}) and use  Eq.~(\ref{Rindler_scalar_decomposition}) to see the annihilation operators $\hat{a}(\vec{k})$ and $\hat{b}(\vec{k})$ as operators acting on the Rindler-Fock space $\mathcal{H}_{\text{L}, \text{R}}$ by means of the linear equation
\begin{subequations}\label{Rindler_Bogoliubov_transformations_1}
\begin{align}
\hat{a}(\vec{k}) = & \sum_{\nu=\{\text{L},\text{R}\}} \int_{\theta_1>0} d^3\theta \left[ \alpha_{\nu+}(\vec{k},\vec{\theta}) \hat{A}_\nu(\vec{\theta})  + \alpha_{\nu-}(\vec{k},\vec{\theta}) \hat{B}_\nu^\dagger(\vec{\theta}) \right],  \\
\hat{b}(\vec{k}) = & \sum_{\nu=\{\text{L},\text{R}\}} \int_{\theta_1>0} d^3\theta \left[ \alpha_{\nu+}(\vec{k},\vec{\theta}) \hat{B}_\nu(\vec{\theta})  + \alpha_{\nu-}(\vec{k},\vec{\theta}) \hat{A}_\nu^\dagger(\vec{\theta}) \right],
\end{align}
\end{subequations}
with $\vec{\theta} = (\Omega,\vec{K}_\perp)$.\label{Bogoliubov_step_2}
\item Plug Eq.~(\ref{Rindler_Bogoliubov_transformations_1}) into the definition of Minkowski vacuum, i.e.,
\begin{align}\label{Minkowski_vacuum}
& \hat{a}(\vec{k}) | 0_\text{M} \rangle = 0, & \hat{b}(\vec{k}) | 0_\text{M} \rangle = 0,
\end{align}
for any $\vec{k} \in \mathbb{R}^3$, to obtain the identity
\begin{equation}\label{Rindler_vacuum_to_Minkowski}
| 0_\text{M} \rangle = \hat{S}_\text{S} | 0_\text{L}, 0_\text{R} \rangle,
\end{equation}
for some unitary Rindler-Fock operator $\hat{S}_\text{S}$ and with $| 0_\text{L}, 0_\text{R} \rangle$ as the Rindler vacuum, defined by
\begin{align}\label{Rindler_vacuum_scalar}
& \hat{A}_\nu(\vec{\theta}) | 0_{\text{L}, \text{R}} \rangle = 0, & \hat{B}_\nu(\vec{\theta}) | 0_{\text{L}, \text{R}} \rangle = 0
\end{align}
for any $\nu \in \{ \text{L}, \text{R} \}$ and for any $\vec{\theta} \in (0, \infty) \otimes \mathbb{R}^2$; Eq.~(\ref{Rindler_vacuum_to_Minkowski}) gives the representation of $| 0_\text{M} \rangle$ as an element of the Rindler-Fock space.\label{Bogoliubov_step_3}\footnote{The possibility to represent the Minkowski vacuum and, in general, any Minkowski-Fock state as an element of the Rindler-Fock space $\mathcal{H}_{\text{L}, \text{R}}$ is given by a theorem that guarantees the equivalence between a state of $\mathcal{H}_\text{M}$ and a state of $\mathcal{H}_{\text{L}, \text{R}}$ up to an arbitrarily large precision with respect to any finite set of mean values. A more detailed discussion about the explicit statement of the theorem is provided in Sec.~\ref{Algebraic_approach_in_QFT_and_QFTCS}.}
\item Use the representation of the Minkowski vacuum in the Rindler-Fock space (\ref{Rindler_vacuum_to_Minkowski}) and the Bogoliubov transformations (\ref{Rindler_Bogoliubov_transformations_1}) to see any element of the Minkowski-Fock space $\mathcal{H}_\text{M}$ as an element of the Rindler-Fock space $\mathcal{H}_{\text{L}, \text{R}}$; for instance, the Minkowski single particle state
\begin{equation}\label{single_particle_Minkowski}
| \phi \rangle = \int_{\mathbb{R}^3} d^3 k \tilde{\phi}_1(\vec{k}) \hat{a}^\dagger(\vec{k}) | 0_\text{M} \rangle
\end{equation}
can be seen as an element of the Rindler-Fock space by means of
\begin{align}\label{single_particle_Minkowski_Rindler}
| \phi \rangle = & \sum_{\nu=\{\text{L},\text{R}\}} \int_{\mathbb{R}^3} d^3 k \int_{\theta_1>0} d^3\theta \tilde{\phi}_1(\vec{k}) \left[ \alpha_{\nu+}^*(\vec{k},\vec{\theta}) \hat{A}_\nu^\dagger(\vec{\theta})  + \alpha_{\nu-}^*(\vec{k},\vec{\theta}) \hat{B}_\nu(\vec{\theta}) \right]  \nonumber \\
& \times \hat{S}_\text{S} | 0_\text{L}, 0_\text{R} \rangle.
\end{align}
\label{Bogoliubov_step_4}
\end{enumerate}

Steps \ref{Bogoliubov_step_3} and \ref{Bogoliubov_step_4} raise the following questions: What does it mean to give the representation of $| 0_\text{M} \rangle$ as an element of the Rindler-Fock space? What does it mean that elements of $\mathcal{H}_\text{M}$ are seen as elements of $\mathcal{H}_{\text{L}, \text{R}}$? After all, $\mathcal{H}_\text{M}$ and $\mathcal{H}_{\text{L}, \text{R}}$ are two distinct Hilbert spaces which are not related by unitary transformation. The questions can be answered by assuming that the role of elements of $\mathcal{H}_\text{M}$ and $\mathcal{H}_{\text{L}, \text{R}}$ is to represent the same physical configurations (i.e., genuine states) in two different frames. The elements of $\mathcal{H}_\text{M}$ and $\mathcal{H}_{\text{L}, \text{R}}$ represent objective physical objects, resulting in different description for the same physical phenomena. In other words, there is nothing physical per se in the Hilbert spaces $\mathcal{H}_\text{M}$ and $\mathcal{H}_{\text{L}, \text{R}}$; the real physical content of the vectors $| \psi \rangle$ is the probabilistic notion entailed in the concept of state. We are allowed to use different representations, since two different observers should only agree about physical predictions and not how physical phenomena are represented.

The idea to relate particle states of one frame to Fock states of the other frame is put in a mathematically precise formalism by Algebraic Quantum Field Theory (AQFT). In this section, we already gave an intuition to the algebraic approach, by putting emphasis on the field operators and deriving the relation between states by means of Eqs.~(\ref{scalar_transformation_Rindler_inverse_t}) and (\ref{scalar_transformation_Rindler_inverse_Dirac}). However, a more detailed discussion will be addressed in Sec.~\ref{Algebraic_approach}, where we will give an introduction to AQFT and a mathematically precise description of unitarily inequivalent particle representations in different frames.

The only relevant information that we need here is that the fundamental objects of AQFT are elements of an abstract algebra; fields only provide a ``coordinatization'' to the algebraic system. Hence, AQFT puts fields of different frames on an equal footing. States are defined as abstract objects that act linearly on the algebra and satisfy minimal properties to entail probability notion. Operators acting on the Fock spaces $\mathcal{H}_\text{M}$ and $\mathcal{H}_{\text{L}, \text{R}}$ (e.g., creators and annihilators)  are only representatives of elements of the algebra; whereas elements of the Fock spaces $\mathcal{H}_\text{M}$ and $\mathcal{H}_{\text{L}, \text{R}}$ concretely represent abstract states. In this framework, the existence of different unitarily inequivalent representation of states appears naturally and does not lead to conceptual inconsistencies.

As a consequence of the nontrivial Bogoliubov transformation (\ref{Rindler_Bogoliubov_transformations_1}), the Minkowski and the Rindler particle description of the field are unitarily inequivalent. In other words, the notion of particles as positive frequency modes is frame dependent. A state made by $n$ Minkowski particles is not equivalent to a state made by $n$ Rindler particles. Also, the notion of vacuum is frame dependent, in the sense that the Minkowski vacuum $| 0_\text{M} \rangle$ defined by Eq.~(\ref{Minkowski_vacuum}) and the Rindler vacuum $| 0_{\text{L}, \text{R}} \rangle$ defined by (\ref{Rindler_vacuum_scalar}) are not the same state [Eq.~(\ref{Rindler_vacuum_to_Minkowski})]. This is at the origin of particle production in the Unruh effect \cite{PhysRevD.7.2850, Davies:1974th, PhysRevD.14.870}.

We remark that the procedure describe by steps \ref{Bogoliubov_step_1}-\ref{Bogoliubov_step_4} only provides a way to relate Minkowski particle states to the elements of the left and right Rindler-Fock space $\mathcal{H}_{\text{L},\text{R}}$. The Hilbert space $\mathcal{H}_{\text{L},\text{R}}$ is the tensor product of the Fock spaces $\mathcal{H}_\text{L}$ and $\mathcal{H}_\text{R}$ generated by, respectively, $\hat{A}_\text{L}$, $\hat{B}_\text{L}$ and by $\hat{A}_\text{R}$, $\hat{B}_\text{R}$, with vacuum states $| 0_\text{L} \rangle$ and $| 0_\text{R} \rangle$ defined by
\begin{align}\label{Rindler_vacuum_scalar_nu}
& \hat{A}_\nu(\vec{\theta}) | 0_\nu \rangle = 0, & \hat{B}_\nu(\vec{\theta}) | 0_\nu \rangle = 0
\end{align}
for any $\nu \in \{ \text{L}, \text{R} \}$ and for any $\vec{\theta} \in (0, \infty) \otimes \mathbb{R}^2$. In summary, we have that $\mathcal{H}_{\text{L},\text{R}} = \mathcal{H}_\text{L} \otimes \mathcal{H}_\text{R}$ and  $| 0_{\text{L},\text{R}} \rangle = | 0_\text{L} \rangle \otimes | 0_\text{R} \rangle$.

At the beginning of Sec.~\ref{QFT_in_curved_spacetime_Rindler}, we defined the left Rindler spacetime as a way to have a common Cauchy surface in both Minkowski and Rindler frame. In the formulation of quantum fields, this gave the opportunity to define a one-to-one map between fields in the Cauchy surface $t=T=0$ [Eqs.~(\ref{scalar_transformation_Rindler_inverse_t}) and (\ref{scalar_transformation_Rindler_inverse_Dirac})] and, hence, a common initial time description of quantum phenomena for both Minkowski and (left and right) Rindler frames. However, the accelerated observer (i.e., Rob) does not have access to the left wedge as a consequence of the Rindler horizon. Hence, if we want to describe physical phenomena from Bob's point of view we have to restrict to the right wedge alone.

The mathematical procedure to restrict the quantum description from the Hilbert space $\mathcal{H}_{\text{L},\text{R}} = \mathcal{H}_\text{L} \otimes \mathcal{H}_\text{R}$ to $\mathcal{H}_\text{R}$ is the partial trace with respect to $\mathcal{H}_\text{L}$, which will be indicated by $\text{Tr}_\text{L}$. Hence, by applying $\text{Tr}_\text{L}$ on representatives of Minkowski particle states in the Rindler-Fock space $\mathcal{H}_{\text{L},\text{R}}$, one is able to derive the description of Minkowski states as seen by the accelerated observer. For instance, in the case of scalar fields, the representation of the Minkowski vacuum and Minkowski single particle states in the right Rindler frame $\mathcal{H}_\text{R}$ can be obtained by applying $\text{Tr}_\text{L}$ in Eqs.~(\ref{Rindler_vacuum_to_Minkowski}) and (\ref{single_particle_Minkowski_Rindler}), respectively. This leads to
\begin{subequations}
\begin{align}
\hat{\rho}_0 = & \text{Tr}_\text{L} \left( \hat{S}_\text{S} | 0_\text{L}, 0_\text{R} \rangle \langle 0_\text{L}, 0_\text{R} | \hat{S}_\text{S}^\dagger \right),\label{Rindler_vacuum_to_Minkowski_Tr}\\
\hat{\rho}_\phi = & \sum_{\nu=\{\text{L},\text{R}\}} \int_{\mathbb{R}^3} d^3 k \int_{\theta_1>0} d^3\theta\sum_{\nu'=\{\text{L},\text{R}\}} \int_{\mathbb{R}^3} d^3 k' \int_{\theta'_1>0} d^3\theta' \tilde{\phi}_1(\vec{k})  \tilde{\phi}_1(\vec{k}') \nonumber \\
& \times \text{Tr}_\text{L} \left( \left[ \alpha_{\nu+}^*(\vec{k},\vec{\theta}) \hat{A}_\nu^\dagger(\vec{\theta}) + \alpha_{\nu-}^*(\vec{k},\vec{\theta}) \hat{B}_\nu(\vec{\theta}) \right] \hat{S}_\text{S} | 0_\text{L}, 0_\text{R} \rangle \langle 0_\text{L}, 0_\text{R} | \hat{S}_\text{S}^\dagger \right.\nonumber \\
& \times \left. \left[ \alpha_{\nu+}(\vec{k},\vec{\theta}) \hat{A}_\nu(\vec{\theta}) + \alpha_{\nu-}(\vec{k},\vec{\theta}) \hat{B}_\nu^\dagger(\vec{\theta}) \right] \right), \label{single_particle_Minkowski_Rindler_Tr}
\end{align}
\end{subequations}
where $\hat{\rho}_0 = \text{Tr}_\text{L} ( | 0_\text{M} \rangle \langle 0_\text{M} | )$ and $\hat{\rho}_\phi = \text{Tr}_\text{L} ( | \phi \rangle \langle \phi | )$ are, respectively, the Minkowski vacuum and the single particle state $| \phi \rangle$ as seen by the accelerated observer.

In Sec.~\ref{Unruh_effect_for_scalar_fields} we will show that $\hat{\rho}_0$ is a thermal state with temperature 
\begin{align}\label{tempterature_Unruh}
& T_\text{U} = \frac{\hbar \beta}{k_\text{B}}, & \beta = \frac{2 \pi}{ca},
\end{align}
i.e.,
\begin{equation}\label{thermal}
\hat{\rho}_0 \propto \exp \left( - \frac{\beta}{\hbar} \hat{H}_\text{R} \right),
\end{equation}
where
\begin{equation}\label{H_R_general}
\hat{H}_\text{R} = \int_0^\infty d \Omega \int_{\mathbb{R}^2} d^2 K_\perp \hbar \Omega \left[ \hat{A}_\text{R}^\dagger (\Omega, \vec{K}_\perp) \hat{A}_\text{R} (\Omega, \vec{K}_\perp) + \hat{B}_\text{R}^\dagger (\Omega, \vec{K}_\perp) \hat{B}_\text{R} (\Omega, \vec{K}_\perp) \right]
\end{equation}
is the Hamiltonian in $\mathcal{H}_\text{R}$. The same result will be obtained for massless scalar real field in 1+1 dimensions [Sec.~\ref{Unruh_effect_for_scalar_fields_11}] and Dirac fields in 3+1 dimensions [Sec.~\ref{Unruh_effect_for_Dirac_fields}]. This is known as the Unruh effect and gives predictions about the detection of a thermal state by an accelerated observer whenever an inertial observer sees a vacuum state.

This Chapter is organized as follows. In Sec.~\ref{Algebraic_approach} we give a brief review to the algebraic approach and we provide the mathematically precise formalism for unitarily inequivalent particle representations in different frames. In Sec.~\ref{Massless_scalar_field_in_11_spacetime} we consider massless scalar real field in 1+1 dimensions as a toy model to discuss the frame dependent content of particles in QFTCS. In Sec.~\ref{Frame_dependent_content_of_particles_scalar} we compute the Bogoliubov coefficients $\alpha_{\nu \pm}(\vec{k},\vec{\theta})$ and the Rindler-Fock representation of the Minkowski vacuum for scalar complex fields in 3+1 dimensions. The same procedure is then applied to Dirac fields in Sec.~\ref{Frame_dependent_content_of_particles_Dirac} with the appropriate modifications.

\section{Algebraic approach}\label{Algebraic_approach}

The appearance of unitarily inequivalent representations in QFTCS motivates the use of the algebraic approach. Here we give a brief introduction to the algebraic formulation of quantum mechanics [Sec.~\ref{Algebraic_approach_in_quantum_theories}] and its application to quantum fields [Sec.~\ref{Algebraic_approach_in_QFT_and_QFTCS}]. The aim is to give a mathematically precise description of the unitarily inequivalent particle representations in the inertial and the accelerated frame. We give a formal answer to the question ``What does it mean that elements of $\mathcal{H}_\text{M}$ are seen as elements of $\mathcal{H}_{\text{L}, \text{R}}$?'' that has been raised in Sec.~\ref{Frame_dependent_content_of_particles_Introduction}.

\subsection{Algebraic approach to quantum theories}\label{Algebraic_approach_in_quantum_theories}

In the standard approach to quantum theory, one typically starts with states as elements of a complex Hilbert space $\mathcal{H}$. Observables are then defined by self-adjoint operators on $\mathcal{H}$. The expected value of the observable $\hat{A}$ with respect to the state $| \psi \rangle$ is $\langle \hat{A} \rangle_\psi = \langle \psi | \hat{A} | \psi \rangle$. The quantity $\langle \hat{A} \rangle_\psi$ entails the probabilistic interpretation of quantum mechanics by Born. Indeed, the probability associated to the any outcome $a$ resulting from measuring $\hat{A}$ is given by $\langle \hat{P}_a \rangle_\psi$, where $\hat{P}_a$ is the spectral projection of the operator $\hat{A}$ associated to the eigenvalue $a$.

In the algebraic approach, the focus shifts from states to observables, and their algebraic relations. Observables are defined as self-adjoint elements of a unital $C^*$-algebra $\mathfrak{A}$, whereas the set of states $\mathcal{S}(\mathfrak{A})$ is identified by positive normalized linear functions $\omega$ on $\mathfrak{A}$. Hence, states are not defined as elements of a vector space, but as objects which act upon observables. Each state $\omega$ consists of a map associating a real number to each observable. Such a real number is interpreted as the expectation value with respect to the state in the usual approach.

By definition, $\omega(A) \in \mathbb{C}$ is linear in $A \in \mathfrak{A}$ and is such that $\omega(A^*A) \geq 0$ for any $A \in \mathfrak{A}$ and $\omega(\mathbb{I}) = 1$. For any self-adjoint element $A=A^*$ of $\mathfrak{A}$ and for any state $\omega$, the real number $\omega(A)$ is interpreted as the expected value of $A$ with respect to $\omega$. In this way, states and observables are provided with the probabilistic notion of quantum mechanics by Born, in analogy to the standard approach. However, at variance with the standard theory, the algebraic approach is based on observables that are defined independently of their action on a Hilbert space.

Notice that the definition of operators and states in the algebraic approach are an abstract generalization of their standard formulation. Indeed, elements of the Hilbert space $\mathcal{H}$ and self-adjoint bounded operators on $\mathcal{H}$ are examples of states and observables in the algebraic sense. Specifically, the bounded operators $\mathcal{B}(\mathcal{H})$ are an example of $C^*$-algebra and every $| \psi \rangle \in \mathcal{H}$ induces a state $\omega_\psi$ on $\mathfrak{A} = \mathcal{B}(\mathcal{H})$ by $\omega_\psi(\hat{A}) = \langle \psi | \hat{A} | \psi \rangle$ for any $\hat{A} \in \mathcal{B}(\mathcal{H})$. However, states $\omega \in \mathcal{S}(\mathfrak{A})$ and observables $A \in \mathfrak{A}$ do not need to be represented by elements of Hilbert spaces and bounded operators on it; thus, the algebraic approach gives a generalization of the notion of quantum states and observables.

The connection between the standard and the algebraic approach to quantum theory is given by the definition of algebraic representations $(\hat{\pi}, \mathcal{H})$, with $\hat{\pi}$ as a representation map and $\mathcal{H}$ as a representation Hilbert space. The function $\hat{\pi}$ is a continuous homomorphism that maps each element of the algebra $A \in \mathfrak{A}$ into a bounded operator $\hat{A}=\hat{\pi}(A) \in \mathcal{B}(\mathcal{H})$ acting on the Hilbert space $\mathcal{H}$. It is required that $\hat{\pi}$ preserves the algebraic structure of $\mathfrak{A}$, in the sense that $\hat{\pi}(AB) = \hat{\pi}(A)\hat{\pi}(B)$, $\hat{\pi}(a A + b B) = a \hat{\pi}(A) + b \hat{\pi}(B)$ and $\hat{\pi}(A^*) = \hat{\pi}^\dagger(A)$ for any $a,b \in \mathbb{C}$ and $A, B \in \mathfrak{A}$. The operator $\hat{\pi}(A)$ is said to represent the element of the algebra $A \in \mathfrak{A}$, while the vector $| \psi \rangle \in \mathcal{H}$ represents the state $\omega_\psi \in \mathcal{S}(\mathfrak{A})$ defined by $\omega_\psi(A) = \langle \psi | \hat{\pi}(A) | \psi \rangle$ for any $A \in \mathfrak{A}$.

Any $A \in \mathfrak{A}$ is an abstract algebraic object that can be concretely represented by operators on Hilbert spaces. Equivalently, the state $\omega_\psi \in \mathcal{S}(\mathfrak{A})$ is an abstract object defined as a linear map on $\mathfrak{A}$ and represented by $| \psi \rangle \in \mathcal{H}$ in the Hilbert representation space. The fundamental structure of the theory is fully characterized by the abstract algebra $\mathfrak{A}$; conversely, the space of bounded operators $\mathcal{B}(\mathcal{H})$ and the Hilbert space $\mathcal{H}$ only provide a concrete representation for elements of $ \mathfrak{A}$ and $ \mathcal{S}(\mathfrak{A})$.

In principle, there can be infinitely many representations for $ \mathfrak{A}$. Some of them can also be unitarily nonequivalent. Two representations $(\hat{\pi}_1, \mathcal{H}_1)$ and $(\hat{\pi}_2, \mathcal{H}_2)$ are called unitarily equivalent if there is a unitary map $\hat{U} : \mathcal{H}_1 \mapsto \mathcal{H}_2$ such that $\hat{U} \hat{\pi}_1(A) = \hat{\pi}_2(A) \hat{U}$ for any $A \in \mathfrak{A}$.

The existence of at least one representation is guaranteed by the GNS (Gel’fand, Naimark, Segal) theorem. For any state $\omega \in \mathcal{S}(\mathfrak{A})$, there is a GNS representation $(\hat{\pi}_\omega, \mathcal{H}_\omega)$ and a vector $| \Omega_\omega \rangle \in \mathcal{H}_\omega$ which are unique up to unitary equivalence and are such that $\hat{\pi}_\omega(\mathfrak{A}) | \Omega_\omega \rangle$ is dense in $\mathcal{H}_\omega$ and $| \Omega_\omega \rangle$ represents $\omega$, i.e., $\omega(A) = \langle \Omega_\omega | \hat{\pi}_\omega(A) | \Omega_\omega \rangle$ for any $A \in \mathfrak{A}$.\footnote{There is also a generalization to the GNS theorem for *-algebras that do not satisfy the $C^*$-property (see for instance Ref.~\cite{Brunetti:2015vmh} for the details). In that case, the representation map $\hat{\pi}_\omega$ is a function between the *-algebra and the space of---not necessarily bounded---operators on $\mathcal{H}_\omega$; also, the domain of $\hat{\pi}_\omega(A)$ for any $A\in \mathfrak{A}$ is not the whole Hilbert space $\mathcal{H}_\omega$, but its dense subspace $ \mathcal{D}_\omega = \hat{\pi}_\omega(\mathfrak{A}) | \Omega_\omega \rangle$.\label{footnote_generalized_GNS}}

The algebraic approach provides the following prescription to construct quantum theories. Start from an unital $C^*$-algebra encoding the algebraic relations between the observables. Then, select a state $\omega$ as a positive, normalized, linear function on the algebra. From $\omega$, construct an Hilbert space via the GNS theorem and recover the standard probabilistic interpretation of quantum theories.

\subsection{Algebraic approach to QFT and QFTCS}\label{Algebraic_approach_in_QFT_and_QFTCS}

When applied to QFT, the algebraic approach is named Algebraic Quantum Field Theory (AQFT). AQFT is an axiomatic mathematically precise formalism that emerged in the 1950s by Haag and others. The emphasis is put on the local observables, which are abstract algebraic objects concretely represented by operators on Hilbert spaces (e.g., creators and annihilators on Fock spaces).

Remarkably, the theory lacks of an intrinsic concept of particles and treats all states on an equal footing even if they come from different particle representations. Consequently, the algebraic approach is naturally applied to QFTCS and provides a solution to the conceptual issue of the inequivalence between particles in different frames. The key observation is that notwithstanding the presence of unitarily inequivalent representations of particles, the algebraic structure of the field operators is the same.

The early development of AQFT can be found in the monograph by Haag \cite{haag1992local} and his pioneering work with Kastler \cite{10.1063/1.1704187}; while, more recent developments can be found in the edited collection \cite{Brunetti:2015vmh}. Also, Ref.~\cite{10.1007/978-3-030-38941-3_1} gives an introduction to AQFT focusing on its key features and Ref.~\cite{Halvorson:2006wj} gives an extensive survey focusing on foundational aspects of the theory.

In AQFT, the basic requirement of the theory is given by the definition of local algebras $\mathfrak{A}(\mathcal{O})$. For each region of spacetime $\mathcal{O}$, there is a $C^*$-algebra $\mathfrak{A}(\mathcal{O})$, whose elements represent physical operations that can be performed in $\mathcal{O}$. In particular, self-adjoint elements of $\mathfrak{A}(\mathcal{O})$ are the observables that can be measured in $\mathcal{O}$. The global algebra $\mathfrak{A}$ is generated by elements of the local algebras $\mathfrak{A}(\mathcal{O})$.

The minimal axioms of AQFT are isotony (i.e., $\mathfrak{A}(\mathcal{O}_1) \subset \mathfrak{A}(\mathcal{O}_2)$ if $\mathcal{O}_1 \subset \mathcal{O}_2$) and microcausality (i.e., $[ A_1, A_2] = 0$ for any $A_1 \in \mathfrak{A}(\mathcal{O}_1)$, $A_2 \in \mathfrak{A}(\mathcal{O}_2)$ and for any spacelike separated regions $\mathcal{O}_1 $ and $ \mathcal{O}_2$). The former implies that any observable in $\mathcal{O}_1$ can also be measured in a larger region $\mathcal{O}_2$; whereas the latter ensures that measurements in spacelike separated regions are independent.

\subsubsection{Algebra of scalar real field}

The standard approach to QFT can be recovered by considering the algebra of local field operators. For instance, in the case of scalar real fields in Minkowski spacetime, $\mathfrak{A}(\mathcal{O})$ is generated by the Klein-Gordon real fields $\hat{\phi}(x^\mu)$ smeared out with test functions supported in the Minkowski coordinate region $\mathcal{O}_\text{M}$ representing $\mathcal{O}$, i.e., 
\begin{equation}\label{phi_f}
\hat{\phi}[f] = \sum_n \int_{\mathcal{O}_\text{M}}d^4 x_1 \cdots \int_{\mathcal{O}_\text{M}}d^4 x_n f_n(x^\mu_1, \dots, x^\mu_n) \hat{\phi}(x^\mu_1) \cdots \hat{\phi}(x^\mu_n)
\end{equation}
with $f_n (x^\mu_1, \dots, x^\mu_n)$ as test function and with $\hat{\phi}(x^\mu)$ as solution of the Klein-Gordon equation (\ref{Klein_Gordon}) satisfying the canonical commutation relation (\ref{phi_commutation}).\footnote{We remark that the operator in Eq.~(\ref{phi_f}) is not bounded. Hence, the algebra of local field operators is not a $C^*$-algebra, but only a *-algebra. To address this issue, one may consider two alternative solutions: (i) Extend the GNS theorem to include also *-algebras; (ii) Construct a $C^*$-algebra from the algebra of local field operators. Solution (i) has already been discussed in footnote \ref{footnote_generalized_GNS} at the end of Sec.~\ref{Algebraic_approach_in_quantum_theories}, where we introduced a generalization of the GNS theorem for *-algebras; the price to be paid is that some $A \in \mathfrak{A}$ are represented by unbounded operators $\hat{\pi}_\omega(A)$ and the domain of $\hat{\pi}_\omega(A)$ is restricted on the dense subspace $\mathcal{D}_\omega \subseteq \mathcal{H}_\omega$. Solution (ii), instead, consists of using the Weyl algebra $\mathfrak{W}$ associated to the field algebra $\mathfrak{A}$ and generated by the Weyl operators $\hat{W}[f] = e^{i \hat{\phi}[f]}$ (see for instance Refs.~\cite{Wald:1995yp, Brunetti:2015vmh} for the details); at variance with $\mathfrak{A}$, $\mathfrak{W}$ is a $C^*$-algebra. Hereafter, we choose (i) when the use of the algebra of fields $\mathfrak{A}$ appears more convenient than the Weyl algebra $\mathfrak{W}$ and (ii) when the $C^*$-property is necessary.\label{footnote_generalized_GNS_Weyl}} For any coordinate $x^\mu$ representing a spacetime event $\mathcal{E}$, the operator $\hat{\phi}(x^\mu)$ is an improper element of any algebra $\mathfrak{A}(\mathcal{O})$ satisfying $\mathcal{E} \in \mathcal{O}$. For all practical purposes, we say that $\hat{\phi}(x^\mu) \in \mathfrak{A}(\mathcal{E})$ if $x^\mu$ is a coordinate representing $\mathcal{E}$.\footnote{This is a mathematically imprecise statement because $\hat{\phi}(x^\mu)$ is not a proper element of the local algebra and $\mathcal{E}$ is not a region of the spacetime, but a point.}

In analogy to QFT in Minkowski spacetime, the standard approach to QFTCS can be obtained by considering the algebra of fields in curved spacetime. For instance, the elements of the algebra of scalar real fields $\hat{\Phi}(X^\mu)$ are defined as real solutions of the Klein-Gordon equation in curved spacetime (\ref{Klein_Gordon_curved}) satisfying the commutation relations (\ref{Phi_commutation}).

The coordinates $x^\mu$ and $X^\mu$ that appear as arguments of the fields $\hat{\phi}(x^\mu)$ and $\hat{\Phi}(X^\mu)$ are actual spacetime coordinates. Hence, if $x^\mu$ and $X^\mu$ are different coordinate systems representing the same manifold, both of them can be used to describe elements of the same local algebra. In particular, if $x^\mu$ and $X^\mu$ represent the same event $\mathcal{E}$ in a spacetime, then the operators $\hat{\phi}(x^\mu)$ and $\hat{\Phi}(X^\mu)$ are elements of the same local algebra $\mathfrak{A}(\mathcal{E})$, i.e., $\hat{\phi}(x^\mu) \in \mathfrak{A}(\mathcal{E})$ and $\hat{\Phi}(X^\mu) \in \mathfrak{A}(\mathcal{E})$.

It can also happen that $\hat{\phi}(x^\mu)$ and $\hat{\Phi}(X^\mu)$ are actually the same element in $\mathfrak{A}(\mathcal{E})$, i.e., $\hat{\phi}(x^\mu) = \hat{\Phi}(X^\mu)$. Indeed, the operators $\hat{\phi}(x^\mu)$ and $\hat{\Phi}(X^\mu)$ can be two different representatives of the same element in $\mathfrak{A}(\mathcal{E})$ with respect to two different coordinate representations of the spacetime. This is a possibility that does not lead to inconsistencies, since the local net $\mathcal{O} \mapsto \mathfrak{A}(\mathcal{O})$ lacks of an intrinsic concept of coordinates. The fundamental elements of $\mathfrak{A}$ are objects locally defined on events $\mathcal{E}$ which can be labeled by arbitrary coordinate systems (e.g., $\mathcal{E} \mapsto x^\mu$ or $\mathcal{E} \mapsto X^\mu$); the process of labeling give rise to fields in different coordinate systems (e.g., $\hat{\phi}(x^\mu)$ and $\hat{\Phi}(X^\mu)$). In this sense, fields naturally emerge as a coordinatization of the algebra $\mathfrak{A}$ and can have different spacetime representations depending on the arbitrary coordinate system.

An example of different coordinate representations of the same local field algebra is provided in Sec.~\ref{Frame_dependent_content_of_particles_Introduction}, where we considered fields in inertial and accelerated frames. In particular, Eq.~(\ref{scalar_transformation_Rindler}) states that the operators $\hat{\phi}(x^\mu)$ and $\hat{\Phi}_\nu(X^\mu)$ are the same algebraic objects if $x^\mu$ and $X^\mu$ represent the same event $\mathcal{E}$ in the $\nu$ wedge. The discussion we gave about the coordinatization of $\mathfrak{A}$ provides a formal justification for Eq.~(\ref{scalar_transformation_Rindler}).

We remark that the use of the term ``coordinate representation'' of the local field algebra should not be confused with the notion of algebraic representation $(\hat{\pi}, \mathcal{H})$ given in Sec.~\ref{Algebraic_approach_in_quantum_theories}. The coordinate representation of $\mathfrak{A}$ is based on the coordinate representation of the underlying spacetime and consists of a coordinatization of $\mathfrak{A}$ by means of field operators of an arbitrary frame, e.g., $\hat{\phi}(x^\mu)$ and $\hat{\Phi}_\nu(X^\mu)$. The algebraic representation $(\hat{\pi}, \mathcal{H})$, instead, uses $\hat{\pi}$ to represent elements of the algebra $\mathfrak{A}$ as operators acting on the Hilbert space $ \mathcal{H}$ and states $\omega \in \mathcal{S}(\mathfrak{A})$ as vectors in $ \mathcal{H}$. In summary, by means of a coordinate representation, one can see the elements of the algebra $\mathfrak{A}$ as objects defined locally in a coordinate patch; whereas, by means of the algebraic representation $(\hat{\pi}, \mathcal{H})$, one can see the elements of $\mathfrak{A}$ as operators acting on the Hilbert space $\mathcal{H}$.

\subsubsection{Minkowski-Fock representation}

In Sec.~\ref{QFT_in_Minkowski_spacetime_scalar}, we gave the canonical formulation of quantum scalar fields $\hat{\phi}(x^\mu)$ in terms of Minkowski particles and Minkowski-Fock space $\mathcal{H}_\text{M}$. In the algebraic formalism, the results of Sec.~\ref{QFT_in_Minkowski_spacetime_scalar} need to be interpreted as the application of an algebraic representation $(\hat{\pi}_\text{M}, \mathcal{H}_\text{M})$, where $\hat{\pi}_\text{M}$ maps elements of the abstract algebra $\mathfrak{A}$ to operators acting on the Minkowski-Fock space $\mathcal{H}_\text{M}$, e.g., creators and annihilators of Minkowski particles.

The representation $(\hat{\pi}_\text{M}, \mathcal{H}_\text{M})$ that leads to the canonical formulation of Minkowski particles is the GNS representation of the Minkowski vacuum $\Omega_\text{M} $. In AQFT, the Minkowski vacuum is defined as an element of $\mathcal{S}(\mathfrak{A})$ satisfying certain properties, whereas the vector $| 0_\text{M} \rangle$ introduced in Eq.~(\ref{Minkowski_vacuum}) is only the representative of $\Omega_\text{M} $ in $\mathcal{H}_\text{M}$. In particular, $\Omega_\text{M} $ is an example of quasifree state. Any quasifree state $\omega$ is defined by the following property: the $n$-point functions $\omega( \hat{\phi}(x^\mu_1) \dots \hat{\phi}(x^\mu_n))$ are non-vanishing only for even $n$ and are completely determined by the $2$-point function. The $2$-point function for the Minkowski vacuum is the familiar Wightman propagator
\begin{equation}\label{twopoint_function_Minkowski}
\Omega_\text{M}( \hat{\phi}(t, \vec{x}) \hat{\phi}(t', \vec{x}')) = \frac{\hbar c^2}{(2 \pi)^3} \int_{\mathbb{R}^3} \frac{d^3 k}{2 \omega(\vec{k})} e^{-i\omega(\vec{k})(t-t') + i\vec{k} \cdot (\vec{x}-\vec{x}')}.
\end{equation}

The mathematically rigorous way to canonically decompose the real scalar field $\hat{\phi}$ in Minkowski spacetime is by means of the representation map $\hat{\pi}_\text{M}$, such that
\begin{equation}\label{free_real_field_positive_negative_frequencies_algebraic}
\hat{\pi}_\text{M} ( \hat{\phi}(t,\vec{x}) ) = \int_{\mathbb{R}^3} d^3 k \left[ f(\vec{k},t,\vec{x}) \hat{a}(\vec{k}) + f^*(\vec{k},t,\vec{x}) \hat{a}^\dagger(\vec{k}) \right],
\end{equation}
where $\hat{a}(\vec{k})$ are the annihilation operators of the Fock space $\mathcal{H}_\text{M}$. Equation (\ref{free_real_field_positive_negative_frequencies_algebraic}) is a mathematically precise version of Eq.~(\ref{free_field}) for real fields in the context of AQFT and can be interpreted as follows. The field $\hat{\phi}(x^\mu)$ is only an element of the local algebra $\mathfrak{A}(\mathcal{E})$, with $\mathcal{E}$ represented by $x^\mu$, but does not act on any Hilbert space; the canonical annihilators $\hat{a}(\theta)$, instead, act on the space of Minkowski particles $\mathcal{H}_\text{M}$. The representation map $\hat{\pi}_\text{M}$ realizes the link between the algebraic and the particle content of the field, by connecting $\hat{\phi} $ with $\hat{a}(\theta) $.\footnote{The analogue reconstruction of Fock space from the Weyl algebra $\mathfrak{W}$ can be obtained from the representation $\hat{\tilde{\pi}}_\text{M}$ defined by $\hat{\tilde{\pi}}_\text{M}(\hat{W}[f]) = e^{i \hat{\pi}_\text{M}(\hat{\phi}[f])}$.\label{footnote_Weyl_particles}}

By using Eqs.~(\ref{twopoint_function_Minkowski}) and (\ref{free_real_field_positive_negative_frequencies_algebraic}), one can check that $\Omega_\text{M}( \hat{\phi}(t, \vec{x}) \hat{\phi}(t', \vec{x}')) = \langle 0_\text{M} | \hat{\pi}_\text{M}( \hat{\phi}(t, \vec{x}) \hat{\phi}(t', \vec{x}')) |  0_\text{M} \rangle$. This is in agreement with the fact that $|  0_\text{M} \rangle$ represents $\Omega_\text{M}$ in $\mathcal{H}_\text{M}$.

\subsubsection{Rindler-Fock representation}

In Sec.~\ref{QFT_in_curved_spacetime_Rindler_scalar}, we defined Rindler particles and the Rindler-Fock space $\mathcal{H}_{\text{L},\text{R}}$ by means of scalar fields in Rindler spacetimes $\hat{\Phi}_\nu(X^\mu)$. As for the Minkowski spacetime, the Hilbert space $\mathcal{H}_{\text{L},\text{R}}$ and the operators acting on it need to come from a representation $(\hat{\pi}_{\text{L},\text{R}}, \mathcal{H}_{\text{L},\text{R}})$ of the field algebra.

In analogy to the Minkowski-Fock representation, $(\hat{\pi}_{\text{L},\text{R}}, \mathcal{H}_{\text{L},\text{R}})$ is the GNS representation of the Rindler vacuum $\Omega_{\text{L},\text{R}} $. The canonical decomposition of the real scalar field in Rindler spacetime is realized by means of the representation map $\hat{\pi}_{\text{L},\text{R}}$ in analogy to Eq.~(\ref{free_real_field_positive_negative_frequencies_algebraic}),
\begin{equation}\label{real_Klein_Gordon_curved_Phi_algebraic}
\hat{\pi}_{\text{L},\text{R}}(\hat{\Phi}_\nu(T,\vec{X})) = \int_{\theta_1>0} d^3\theta  \left[ F_\nu(\theta,T, \vec{X}) \hat{A}_\nu(\theta)  + F_\nu^*(\theta,T, \vec{X}) \hat{A}_\nu^\dagger(\theta) \right].
\end{equation}
The state $\Omega_{\text{L},\text{R}} $ is defined in terms of the Rindler vacuum $2$-point function $\Omega_\text{\text{L},\text{R}}( \hat{\Phi}_\nu(T, \vec{X}) \hat{\Phi}_{\nu'}(T', \vec{X}')) = \langle 0_\text{\text{L},\text{R}} | \hat{\pi}_\text{\text{L},\text{R}}( \hat{\Phi}_\nu(T, \vec{X}) \hat{\Phi}_{\nu'}(T', \vec{X}')) |  0_\text{\text{L},\text{R}} \rangle$ and is represented by $| 0_{\text{L},\text{R}} \rangle \in \mathcal{H}_{\text{L},\text{R}}$.

\subsubsection{Unitarily inequivalent particle representations}

In Sec.~\ref{Frame_dependent_content_of_particles_Introduction}, we remarked that the Minkowski ($x^\mu$) and the Rindler ($X^\mu$) coordinates describe, respectively, the inertial and the accelerated frame of the same flat spacetime; hence, they are two different coordinate systems of the same manifold. Both frames share the same field algebra $\mathfrak{A}$ and the operators $\hat{\phi}(x^\mu)$ and $\hat{\Phi}_\nu(X^\mu)$ describe the same scalar field in each coordinate representation. In particular, $\hat{\phi}(x^\mu)$ and $\hat{\Phi}_\nu(X^\mu)$ are the same algebraic object when $x^\mu$ and $X^\mu$ represent the same event $\mathcal{E}$ in the $\nu$ wedge [Eq.~(\ref{scalar_transformation_Rindler})]. 

The existence of a single algebra $\mathfrak{A}$ describing both Minkowski and Rindler fields implies that both $(\hat{\pi}_\text{M}, \mathcal{H}_\text{M})$ and $(\hat{\pi}_{\text{L},\text{R}}, \mathcal{H}_{\text{L},\text{R}})$ can be used to represent $\mathfrak{A}$ and $\mathcal{S}(\mathfrak{A})$. In other words, $(\hat{\pi}_\text{M}, \mathcal{H}_\text{M})$ and $(\hat{\pi}_{\text{L},\text{R}}, \mathcal{H}_{\text{L},\text{R}})$ are unitarily inequivalent representations of the same algebra $\mathfrak{A}$. This shows that the inequivalent particle content of fields in different frames actually comes from an unifying notion of algebra $\mathfrak{A}$ and states $\mathcal{S}(\mathfrak{A})$.

The unifying notion of states $\mathcal{S}(\mathfrak{A})$ between different frames formally guarantees the possibility to relate particle states of one frame to the other. In particular, the vector $| \psi_\omega  \rangle \in  \mathcal{H}_\text{M}$ representing the state $\omega \in \mathcal{S}(\mathfrak{A})$ in $\mathcal{H}_\text{M}$ can be mapped into the vector $| \Psi_\omega \rangle \in  \mathcal{H}_{\text{L},\text{R}}$ representing $\omega$ in $ \mathcal{H}_{\text{L},\text{R}}$. Physically, the map $\mathcal{F}_{\text{M} \mapsto \text{L},\text{R}}: | \psi_\omega  \rangle \mapsto | \Psi_\omega  \rangle$ tells how an accelerated observer would describe Minkowski particles.

The existence of a map $\mathcal{F}_{\text{M} \mapsto \text{L},\text{R}}$ gives a formal answer to the question ``What does it mean that elements of $\mathcal{H}_\text{M}$ are seen as elements of $\mathcal{H}_{\text{L}, \text{R}}$?''~that has been raised in Sec.~\ref{Frame_dependent_content_of_particles_Introduction}. The answer lies in the idea that vectors in $\mathcal{H}_\text{M}$ can be naturally mapped in $\mathcal{H}_{\text{L}, \text{R}}$ by $\mathcal{F}_{\text{M} \mapsto \text{L},\text{R}}$ while preserving their objective physical content.

Notice that, in principle, the map $\mathcal{F}_{\text{M} \mapsto \text{L},\text{R}}$ is not always defined for any $| \psi  \rangle \in  \mathcal{H}_\text{M}$. This a consequence of the fact that for any representation $(\hat{\pi}, \mathcal{H})$, not all states are guaranteed to be represented by $(\hat{\pi}, \mathcal{H})$. Hence, the set of states that are represented by $(\hat{\pi}_\text{M}, \mathcal{H}_\text{M})$ may differ from the set of states that are represented by $(\hat{\pi}_{\text{L},\text{R}}, \mathcal{H}_{\text{L},\text{R}})$. Hereafter these sets will be called $\mathcal{S}_\text{M}$ and $\mathcal{S}_{\text{L},\text{R}}$, respectively. By definition, the domain of $\mathcal{F}_{\text{M} \mapsto \text{L},\text{R}}$ is only restricted to $\mathcal{S}_\text{M} \cap \mathcal{S}_{\text{L},\text{R}}$, which means that vectors in $\mathcal{H}_\text{M}$ representing states outside $\mathcal{S}_\text{M} \cap \mathcal{S}_{\text{L},\text{R}}$ cannot be seen as elements of $\mathcal{H}_{\text{L}, \text{R}}$. In other words, not all Minkowski particle states are guaranteed to be described in terms of Rindler particles.

Luckily, there is a theorem for Weyl algebras\footnote{Here we consider the Weyl $C^*$-algebra $\mathfrak{W}$ constructed from the field *-algebra $\mathfrak{A}$ and already discussed in footnotes \ref{footnote_generalized_GNS_Weyl} and \ref{footnote_Weyl_particles}.} that guarantees the possibility to approximate states of different representations between each other (see Sec.~4.5 of Ref.~\cite{Wald:1995yp} and references therein). Specifically, the theorem says that for any couple of representations $(\hat{\pi}_1, \mathcal{H}_1)$ and $(\hat{\pi}_2, \mathcal{H}_2)$ representing states in $\mathcal{S}_1 \subseteq \mathcal{S}(\mathfrak{A})$ and $\mathcal{S}_2 \subseteq \mathcal{S}(\mathfrak{A})$ and for any state $\omega_1 \in \mathcal{S}_1$ there is a state $\omega_2 \in \mathcal{S}_2$ that approximates $\omega_1$. The approximation works as follows: for any state $\omega_1 \in \mathcal{S}_1$ and for any finite set $A_1, \dots, A_n \in \mathfrak{A}$ and $\epsilon_1, \dots, \epsilon_n > 0$, there is an $\omega_2 \in \mathcal{S}_2$ such that $| \omega_1(A_i) - \omega_2(A_i)| < \epsilon_i$ for any $i = 1, \dots, n$. The theorem includes also statistical operators as representatives of states, hence $\omega_1$ and $\omega_2$ do not need to be represented by pure vectors in $\mathcal{H}_1$ and $\mathcal{H}_2$.

The theorem guarantees the success of the prescription introduced in Sec.~\ref{Frame_dependent_content_of_particles_Introduction} by steps \ref{Bogoliubov_step_1}-\ref{Bogoliubov_step_4} to relate statistical operators of $\mathcal{H}_\text{M}$ to statistical operators of $\mathcal{H}_{\text{L}, \text{R}}$; however, the approach only guarantees the equivalence between states in a sort of weak limit. In particular, the existence of the unitary operator $\hat{S}_\text{S}$ in Eq.~(\ref{Rindler_vacuum_to_Minkowski}) that exactly maps $| 0_\text{M} \rangle$ in $\mathcal{H}_{\text{L}, \text{R}}$ with infinite precision is not always guaranteed. In the next subsections, we will show how in all considered scenarios (i.e., scalar and Dirac fields) the unitary operator $\hat{S}_\text{S}$ exists (though for Dirac fields it will be called $\hat{S}_\text{D}$) and elements of one vector space can be exactly written in terms on elements of the other vector space.

The procedure described by steps \ref{Bogoliubov_step_1}-\ref{Bogoliubov_step_4} in Sec.~\ref{Frame_dependent_content_of_particles_Introduction} tells how to practically construct the map $\mathcal{F}_{\text{M} \mapsto \text{L},\text{R}}$. The method is based on deriving the Bogoliubov transformation (\ref{Rindler_Bogoliubov_transformations_1}) relating the Minkowski operators to the Rindler operators. We remark that the formal way to write Eq.~(\ref{Rindler_Bogoliubov_transformations_1}) in the context of AQFT is by means of the representation maps $\hat{\pi}_\text{M}$ and $\hat{\pi}_{\text{L},\text{R}}$, which encode the particle representation in each frame, as shown by Eqs.~(\ref{free_real_field_positive_negative_frequencies_algebraic}) and (\ref{real_Klein_Gordon_curved_Phi_algebraic}). The mathematically precise way to write Eq.~(\ref{Rindler_Bogoliubov_transformations_1}) is
\begin{subequations}\label{Rindler_Bogoliubov_transformations_1_AQFT}
\begin{align}
\hat{\pi}_{\text{L},\text{R}}(\hat{\pi}_\text{M}^{-1} (\hat{a}(\vec{k}))) = & \sum_{\nu=\{\text{L},\text{R}\}} \int_{\theta_1>0} d^3\theta \left[ \alpha_{\nu+}(\vec{k},\vec{\theta}) \hat{A}_\nu(\vec{\theta})  + \alpha_{\nu-}(\vec{k},\vec{\theta}) \hat{B}_\nu^\dagger(\vec{\theta}) \right],  \\
\hat{\pi}_{\text{L},\text{R}}(\hat{\pi}_\text{M}^{-1} (\hat{b}(\vec{k}))) = & \sum_{\nu=\{\text{L},\text{R}\}} \int_{\theta_1>0} d^3\theta \left[ \alpha_{\nu+}(\vec{k},\vec{\theta}) \hat{B}_\nu(\vec{\theta})  + \alpha_{\nu-}(\vec{k},\vec{\theta}) \hat{A}_\nu^\dagger(\vec{\theta}) \right].
\end{align}
\end{subequations}
Furthermore, the formal way to write Eqs.~(\ref{Rindler_vacuum_to_Minkowski}) and (\ref{single_particle_Minkowski_Rindler}) in the context of AQFT is by means of the map $\mathcal{F}_{\text{M} \mapsto \text{L},\text{R}}$ as follows
\begin{subequations}
\begin{align}
\mathcal{F}_{\text{M} \mapsto \text{L},\text{R}} (| 0_\text{M} \rangle) = & \hat{S}_\text{S} | 0_\text{L}, 0_\text{R} \rangle,\label{Rindler_vacuum_to_Minkowski_AQFT}\\
\mathcal{F}_{\text{M} \mapsto \text{L},\text{R}}(| \phi \rangle) = & \sum_{\nu=\{\text{L},\text{R}\}} \int_{\mathbb{R}^3} d^3 k \int_{\theta_1>0} d^3\theta \tilde{\phi}_1(\vec{k}) \left[ \alpha_{\nu+}^*(\vec{k},\vec{\theta}) \hat{A}_\nu^\dagger(\vec{\theta}) \right. \nonumber \\
& \left. + \alpha_{\nu-}^*(\vec{k},\vec{\theta}) \hat{B}_\nu(\vec{\theta}) \right] \hat{S}_\text{S} | 0_\text{L}, 0_\text{R} \rangle.\label{single_particle_Minkowski_Rindler_AQFT}
\end{align}
\end{subequations}

For all practical purposes, we will leave the use of representation maps $\hat{\pi}_\text{M}$ and $\hat{\pi}_{\text{L},\text{R}}$ implicit. In particular, we implicitly assume the use of representation maps whenever particle operators of one frame are treated as if they were Fock operators of the other frame. For instance, we will use Eqs.~(\ref{free_field}), (\ref{Rindler_scalar_decomposition}) and (\ref{Rindler_Bogoliubov_transformations_1}) instead of Eqs.~(\ref{free_real_field_positive_negative_frequencies_algebraic}), (\ref{real_Klein_Gordon_curved_Phi_algebraic}) and (\ref{Rindler_Bogoliubov_transformations_1_AQFT}), respectively.

Analogously, we will avoid to name the map $\mathcal{F}_{\text{M} \mapsto \text{L},\text{R}}$ when particle states of one frame are related to particles of the other. Specifically, we implicitly assume the use of $\mathcal{F}_{\text{M} \mapsto \text{L},\text{R}}$ whenever a particle state of one frame is written as an element of the other Fock space. Hence, we write $| \psi_\omega \rangle = | \Psi_\omega \rangle$ instead of $\mathcal{F}_{\text{M} \mapsto \text{L},\text{R}} (| \psi_\omega \rangle) = | \Psi_\omega \rangle$, although $| \psi_\omega \rangle$ and $| \Psi_\omega \rangle$ are formally elements of different Hilbert spaces. For instance, we will use  Eqs.~(\ref{Rindler_vacuum_to_Minkowski}) and (\ref{single_particle_Minkowski_Rindler}) instead of  Eqs.~(\ref{Rindler_vacuum_to_Minkowski_AQFT}) and (\ref{single_particle_Minkowski_Rindler_AQFT}), respectively.

\subsubsection{Fermionic and scalar complex fields}

We now want to add the following remarks for fermionic and scalar complex fields. As a consequence of microcausality, elements of local algebras $\mathfrak{A}(\mathcal{O})$ in spacelike separated regions commute. This is an apparent contradiction with the requirement that fermionic fields anticommute in causally disconnected regions; only products of even numbers of fields satisfy the microcausality condition. However, there is no contradiction between the anticommutation relation between fermionic fields and the microcausality axiom since the only observables that can be physically measured involve even numbers of fermionic fields. A spinor field by itself is not measurable, at variance with, e.g., second degree products of spinor and cospinor fields. Consequently, operators involving odd numbers of fermionic fields are not elements of the algebra $\mathfrak{A}$.

This also occurs for complex scalar theory characterized by a global $U(1)$ gauge symmetry, where the only measurable observables are invariant under the global $U(1)$ gauge action. Products of odd numbers of fields do not satisfy this condition and, hence, are not elements of the algebra $\mathfrak{A}$.

In summary, some field operators are not measurable and, hence, do not belong to $\mathfrak{A}$. However, the algebra of non measurable field operators still plays a role in AQFT. From the local net $\mathcal{O} \mapsto \mathfrak{A}(\mathcal{O})$ one can construct an algebra $\mathfrak{F}(\mathcal{O})$ which contains non-observable fields (e.g., product of odd fermionic fields and odd complex scalar fields) in addition to elements of $ \mathfrak{A}(\mathcal{O})$. The local algebra $ \mathfrak{A}(\mathcal{O})$ is made by elements of $\mathfrak{F}(\mathcal{O})$ that are invariant under the action of the gauge group $G$. 

The reconstruction of $\mathfrak{F}(\mathcal{O})$ and $G$ from the fundamental local algebra $\mathfrak{A}(\mathcal{O})$ is achieved through the DHR (Doplicher–Haag–Roberts) program and the analysis of superselection sectors. These theories are out of the scope of the present thesis and we refer the reader to, e.g., Refs.~\cite{haag1992local, araki1999mathematical} for detailed discussions. 

For all practical purposes, we consider algebraic representations of $\mathfrak{F}$, instead of $\mathfrak{A}$. For instance, we consider the Minkowski-Fock (Rindler-Fock) spaces of Dirac and charged scalar fields, which are constructed by the Minkowski (Rindler) vacuum GNS representation of $\mathfrak{F}$.

\section{Massless scalar real field in 1+1 spacetime}\label{Massless_scalar_field_in_11_spacetime}

In this section, we consider a flat spacetime in 1+1 dimensions and a massless scalar real field. Minkowski and Rindler coordinates are identified by $(t,x)$ and $(T,X)$ and are related to each other by
\begin{align}
& t = t_\nu (T, X), & x = x_\nu (T, X),
\end{align}
with
\begin{align}\label{Rindler_coordinate_transformation_11}
& t_\nu(T,X) = \frac{e^{s_\nu a X}}{c a} \sinh(c a T),  & x_\nu(T,X) = s_\nu \frac{e^{s_\nu a X}}{a} \cosh(c a T), 
\end{align}
as the coordinate transformation from the $\nu$-Rindler frame to the Minkowski frame. In this case the left and right wedges are defined, respectively, by $x < c |t|$ and $x > c |t|$. The inverse of Eq.~(\ref{Rindler_coordinate_transformation_11}) is identified by $T_\nu(t,x)$ and $X_\nu(t,x)$. The Rindler metric in the $\nu$ wedge is $\text{diag} \left( -c^2 e^{s_\nu 2aX}, e^{s_\nu 2aX} \right)$.

The Klein-Gordon equation (\ref{Klein_Gordon}) in 1+1 dimensions and with zero mass reads
\begin{equation} \label{Klein_Gordon_11}
\left(  \partial_0^2 - c^2 \partial_1^2 \right) \hat{\phi} = 0.
\end{equation}
The Rindler-Klein-Gordon equation (\ref{Rindler_Klein_Gordon}) in 1+1 dimensions and with zero mass, instead, reads
\begin{equation} \label{Rindler_Klein_Gordon_R_11}
\left(  \partial_0^2 - c^2 \partial_1^2 \right) \hat{\Phi}_\text{R} = 0.
\end{equation}
The same identity can be found in the left wedge, i.e.,
\begin{equation} \label{Rindler_Klein_Gordon_L_11}
\left(  \partial_0^2 - c^2 \partial_1^2 \right) \hat{\Phi}_\text{L} = 0.
\end{equation}
Notice that the field equations in the Rindler frame (\ref{Rindler_Klein_Gordon_R_11}) and (\ref{Rindler_Klein_Gordon_L_11}) have the same form as the Minkowski Klein-Gordon equation (\ref{Klein_Gordon_11}). This is a consequence of the fact that the 1+1 Rindler metric is conformally flat and the field is massless.

From Eqs.~(\ref{Klein_Gordon_11}), (\ref{Rindler_Klein_Gordon_R_11}) and (\ref{Rindler_Klein_Gordon_L_11}), one can derive a decomposition for the scalar field in the Minkowski and the Rindler frames in terms of Minkowski and Rindler particles with defined momentum. By defining $\hat{a}(k)$ and $\hat{A}_\nu(K)$ as, respectively, the annihilator of the Minkowski particle with momentum $k$ and the annihilator of the $\nu$-Rindler particle with momentum $K$, one can write
\begin{subequations} 
\begin{align}
& \hat{\phi}(t,x) = \int_{\mathbb{R}} dk  \left[ f(k,t,x) \hat{a}(k) + f^*(k,t,x) \hat{a}^\dagger (k) \right],\label{scalar_field_M}\\
& \hat{\Phi}_\nu(T,X) =  \int_{\mathbb{R}} dK \left[ f(K,T,X) \hat{A}_\nu(K)  + f^*(K,T,X) \hat{A}_\nu^\dagger(K) \right],\label{scalar_field_R}
\end{align}
\end{subequations}
with
\begin{equation}\label{f_11}
f(k,t,x) = \sqrt{\frac{\hbar}{4\pi c |k|}}  e^{-i c |k|t+ikx}
\end{equation}
as 1+1 dimensional free mode with zero mass.

The Klein-Gordon scalar product (\ref{KG_scalar_product}) in 1+1 dimensions now reads
\begin{equation}\label{KG_scalar_product_11}
( \phi, \phi' )_\text{KG} = \frac{i}{\hbar} \int_{\mathbb{R}} dx \left[ \phi^*(t,x) \partial_0  \phi'(t,x) -  \phi'(t,x) \partial_0 \phi^*(t,x) \right].
\end{equation}
The positive and negative frequency modes are orthonormal with respect to Klein-Gordon scalar product (\ref{KG_scalar_product_11}), in the sense that they satisfy Eq.~(\ref{KG_scalar_product_orthonormality_f}).

The two fields $\hat{\phi} $ and $\hat{\Phi}_\nu$ are related to each other by
\begin{equation}\label{scalar_transformation_Rindler_inverse_t_11}
\hat{\phi}(t,x) = \sum_{\nu=\{\text{L},\text{R}\}} \theta(s_\nu x) \hat{\Phi}_\nu(T_\nu (t, x),X_\nu (t, x)),
\end{equation}
which is the equivalent of Eq.~(\ref{scalar_transformation_Rindler_inverse_t}) in 1+1 dimensions.

Throughout this section, the massless scalar real field in 1+1 dimensions will serve as a toy model for the theory introduced in Sec.~\ref{Frame_dependent_content_of_particles_Introduction}. We follow the prescription given by steps \ref{Bogoliubov_step_1}-\ref{Bogoliubov_step_4} of Sec.~\ref{Frame_dependent_content_of_particles_Introduction} to relate Minkowski particle states to Rindler-Fock states. In particular, we compute the Bogoliubov transformation relating creators and annihilators of one frame to the other [Sec.~\ref{Bogoliubov_transformation_scalar_11}], we derive the representation of the Minkowski vacuum in the Rindler-Fock space [Sec.~\ref{Minkowski_vacuum_in_Rindler_spacetime_scalar_11}] and we detail the consequent Unruh effect [Sec.~\ref{Unruh_effect_for_scalar_fields_11}].

\subsection{Bogoliubov transformation}\label{Bogoliubov_transformation_scalar_11}

In this subsection, we compute the Bogoliubov transformations relating Minkowski operators $\hat{a}(k)$ to Rindler operators $\hat{A}_\nu(K)$ by following steps \ref{Bogoliubov_step_1} and \ref{Bogoliubov_step_2} of Sec.~\ref{Frame_dependent_content_of_particles_Introduction}.

We start by giving the explicit form of Eq.~(\ref{Bogoliubov_transformations_1}) in 1+1 dimensions by means of Eq.~(\ref{KG_scalar_product_11}), i.e.,
\begin{equation}\label{Bogoliubov_transformations_2_Rindler_scalar_11}
\hat{a}(\vec{k}) = \frac{i}{\hbar } \int_{\mathbb{R}} dx \left[  f^*(k,t,x) \partial_0 \hat{\phi}(t,x)  - \hat{\phi}(t,x) \partial_0  f^*(k,t,x)\right].
\end{equation}
The right hand side of Eq.~(\ref{Bogoliubov_transformations_2_Rindler_scalar_11}) is independent of $t$. For practical purposes we choose $t=0$; hence, we consider
\begin{equation}\label{Bogoliubov_transformations_2_Rindler_scalar_11_2}
\hat{a}(\vec{k}) = \frac{i}{\hbar } \int_{\mathbb{R}} dx \left[  f^*(k,0,x) \left. \partial_0 \hat{\phi}(t,x) \right|_{t=0} - \hat{\phi}(0,x) \left. \partial_0  f^*(k,t,x)\right|_{t=0} \right].
\end{equation}

The next steps consists of plugging Eq.~(\ref{scalar_transformation_Rindler_inverse_t_11}) into Eq.~(\ref{Bogoliubov_transformations_2_Rindler_scalar_11_2}) to obtain
\begin{align}\label{Bogoliubov_transformations_2_Rindler_scalar_11_3}
\hat{a}(\vec{k}) = & \frac{i}{\hbar } \sum_{\nu=\{\text{L},\text{R}\}} \int_{\mathbb{R}} dx \theta(s_\nu x) \left[  f^*(k,0,x) \left. \frac{\partial}{\partial t} \hat{\Phi}_\nu(T_\nu (t, x),X_\nu (t, x)) \right|_{t=0} \right. \nonumber \\
& \left. - \hat{\Phi}_\nu(0,X_\nu (x)) \left.  \partial_0  f^*(k,t,x) \right|_{t=0} \right],
\end{align}
where $X_\nu (x) = X_\nu (0, x)$. We compute  $\left. \partial \hat{\Phi}_\nu(T_\nu (t, x),X_\nu (t, x)) / \partial t \right|_{t=0}$ by means of the chain rule
\begin{equation}
\frac{\partial}{\partial T} = s_\nu ax \frac{\partial}{\partial t}  + s_\nu ac^2t  \frac{\partial}{\partial x}.
\end{equation}
and by knowing that the hypersurface $t=0$ is equivalent to the condition $T=0$ in both wedges. We obtain
\begin{equation}\label{time_derivative_scalar_transformation_Rindler_2_11}
\left. \frac{\partial}{\partial t} \hat{\Phi}_\nu(T_\nu (t, x),X_\nu (t, x)) \right|_{t=0} = \frac{s_\nu}{a x} \left. \frac{\partial}{\partial T} \hat{\Phi}_\nu(T, X_\nu (x)) \right|_{T=0},
\end{equation}
which leads to
\begin{align}\label{Bogoliubov_transformations_2_Rindler_scalar_11_4}
\hat{a}(\vec{k}) = & \frac{i}{\hbar} \sum_{\nu=\{\text{L},\text{R}\}} \int_{\mathbb{R}} dx \theta(s_\nu x) \left[  \frac{s_\nu}{a x} f^*(k,0,x) \left. \partial_0 \hat{\Phi}_\nu(T, X_\nu (x)) \right|_{T=0} \right. \nonumber \\
& \left. - \hat{\Phi}_\nu(0,X_\nu (x)) \left. \partial_0  f^*(k,t,x) \right|_{t=0} \right].
\end{align}

By plugging Eq.~(\ref{scalar_field_R}) in Eq.~(\ref{Bogoliubov_transformations_2_Rindler_scalar_11_4}) we finally obtain the Bogoliubov transformation
\begin{equation}\label{Rindler_Bogoliubov_transformations_1_11}
\hat{a}(k) = \sum_{\nu=\{\text{L},\text{R}\}} \int_{\mathbb{R}} dK \left[ \alpha_{\nu+}(k,K) \hat{A}_\nu(K)  + \alpha_{\nu-}(k,K) \hat{A}_\nu^\dagger(K) \right],
\end{equation}
with
\begin{subequations}\label{alpha_nu_pm_11}
\begin{align}
\alpha_{\nu+}(k,K) = & \frac{i}{\hbar} \int_{\mathbb{R}} dx \theta(s_\nu x) \left[ \frac{s_\nu}{a x} f^*(k,0,x) \left. \partial_0 f(K, T , X_\nu (x)) \right|_{T=0} \right. \nonumber \\
& \left. - f(K,0,X_\nu (x)) \left. \partial_0  f^*(k,t,x) \right|_{t=0} \right] \\
\alpha_{\nu-}(k,K) = & \frac{i}{\hbar} \int_{\mathbb{R}} dx \theta(s_\nu x) \left[ \frac{s_\nu}{a x} f^*(k,0,x) \left. \partial_0 f^*(K, T,X_\nu (x)) \right|_{T=0} \right. \nonumber \\
& \left. - f^*(K,0,X_\nu (x)) \left. \partial_0  f^*(k,t,x) \right|_{t=0} \right].
\end{align}
\end{subequations}
Equation (\ref{Rindler_Bogoliubov_transformations_1_11}) is the equivalent of Eq.~(\ref{Rindler_Bogoliubov_transformations_1}) for massless scalar real fields in 1+1 dimensions.

The explicit form of the Bogoliubov coefficients can be obtained by plugging Eq.~(\ref{f_11}) in Eq.~(\ref{alpha_nu_pm_11}). This leads to
\begin{equation}\label{alpha_nu_pm_11_2}
\alpha_{\nu\pm}(k,K) = \int_{\mathbb{R}} dx \frac{\theta(s_\nu x)}{4 \pi \sqrt{|kK|}} \left( \pm s_\nu \frac{|K|}{a x} + |k| \right) e^{\pm i K X_\nu (x) - i kx }.
\end{equation}
By performing the coordinate transformation $x \mapsto X = s_\nu X_\nu (x)$, Eq.~(\ref{alpha_nu_pm_11_2}) becomes
\begin{equation}\label{alpha_nu_pm_11_3}
\alpha_{\nu\pm}(k,K) = \int_{\mathbb{R}} dX \frac{ \pm |K| + |k| e^{a X} }{4 \pi \sqrt{|kK|}} \exp \left(  \pm s_\nu i K X - s_\nu i \frac{k}{a} e^{a X} \right).
\end{equation}

By defining
\begin{equation}\label{F_k_K}
F(k,K) = \int_{\mathbb{R}}  \frac{dX}{2 \pi}  \exp \left( - i K X + i \frac{k}{a} e^{a X} \right),
\end{equation}
we can write Eq.~(\ref{alpha_nu_pm_11_3}) in the following simpler form
\begin{equation}\label{alpha_nu_pm_11_4}
\alpha_{\nu\pm}(k,K) = \frac{1 }{2 \sqrt{|kK|}} \left. \left( \pm |K| - i a |k| \frac{\partial}{\partial \kappa} \right) F(\kappa, \mp s_\nu K) \right|_{\kappa = - s_\nu k}.
\end{equation}
$F(k,K)$ is a distribution that can be obtained from the following distributional limit \cite{Mukhanov2007}
\begin{equation}\label{F_regularized}
F(k,K) =  \lim_{\epsilon \rightarrow 0^+} \int_{\mathbb{R}}\frac{dX}{2\pi}  \exp\left((-i K + \epsilon a) X + \left(i \frac{k}{a} - \epsilon\right) e^{aX} \right).
\end{equation}
The derivative of $F(k,K)$ with respect to $k$ appearing in Eq.~(\ref{alpha_nu_pm_11_4}) can be obtained by using integration by parts in Eq.~(\ref{F_regularized}) and by taking the distributional limit $\epsilon \rightarrow 0^+$:
\begin{align} \label{dFdk}
& \frac{\partial}{\partial k} F(k,K) \nonumber \\
 = & \lim_{\epsilon \rightarrow 0^+} \int_{\mathbb{R}}dX \frac{i e^{aX}}{2\pi a} \exp\left((-i K + \epsilon a) X + \left(i \frac{k}{a} - \epsilon\right) e^{aX} \right) \nonumber \\
 = & \lim_{\epsilon \rightarrow 0^+} \int_{\mathbb{R}}dX \frac{i}{2\pi a^2} e^{(-i K + \epsilon a) X } \left(i \frac{k}{a} - \epsilon\right)^{-1} \frac{d}{dX} \exp \left(\left(i \frac{k}{a} - \epsilon\right) e^{aX} \right) \nonumber \\
= & \lim_{\epsilon \rightarrow 0^+} \left[ \left. \frac{i}{2 \pi a^2} \left(i \frac{k}{a} - \epsilon\right)^{-1} \exp\left((-i K + \epsilon a) X + \left(i \frac{k}{a} - \epsilon\right) e^{aX} \right) \right|_{-\infty}^{+\infty} \right. \nonumber \\
& \left. - \int_{\mathbb{R}}dX\frac{K+ i \epsilon  a}{2 \pi a^2} \left(i \frac{k}{a} - \epsilon\right)^{-1}  \exp\left((-i K + \epsilon a) X + \left(i \frac{k}{a} - \epsilon\right) e^{aX} \right) \right] \nonumber \\
= & \lim_{\epsilon \rightarrow 0^+} \left[ \frac{iK-\epsilon a}{a (k + i \epsilon a)} \int_{\mathbb{R}} \frac{dX}{2 \pi} \exp\left((-i K + \epsilon a) X + \left(i \frac{k}{a} - \epsilon\right) e^{aX} \right) \right] \nonumber \\
= & \frac{iK}{ak} F(k,K).
\end{align}
Equation (\ref{dFdk}) can be plugged in Eq.~(\ref{alpha_nu_pm_11_4}) to obtain
\begin{align}\label{alpha_nu_pm_11_5}
\alpha_{\nu\pm}(k,K) = &  \pm \frac{1}{2 \sqrt{|kK|}}  \left( |K| +  \frac{|k|K}{k} \right) F(- s_\nu k, \mp s_\nu K) \nonumber \\
 = &  \pm  \frac{ 1 + \text{sign}(kK)}{2} \sqrt{ \left| \frac{K}{k} \right|} F(- s_\nu k, \mp s_\nu K)\nonumber \\
 = &  \pm \theta(kK) \sqrt{ \left| \frac{K}{k} \right|}  F(- s_\nu k, \mp s_\nu K) .
\end{align}

The distribution $F(k,K)$ of Eq.~(\ref{F_regularized}) can be proven to be equal to \cite{Mukhanov2007}
\begin{equation} \label{F_11}
F(k,K) = \frac{1}{2 \pi a} \Gamma \left( -\frac{i K}{a} \right) \exp \left( i \frac{K}{a} \ln \left( \frac{|k|}{a} \right) + \text{sign} \left( k \right) \frac{c \beta K}{4} \right),
\end{equation}
with $\Gamma(\xi)$ as the gamma function and $\beta = 2 \pi/ca$. Two important properties that can be deduced from Eq.~(\ref{F_11}) and will be used in the rest of this subsection are
\begin{subequations}
\begin{align}
& F(-k,-K) = F^*(k,K), \label{F_11_conjugate} \\
& F(-k,K) = \exp \left(- \text{sign}(k) \frac{c \beta K}{2} \right)F(k,K). \label{alpha_beta_to_alpha_beta_proof}
\end{align}
\end{subequations}
In particular, Eq.~(\ref{F_11_conjugate}) can be used in Eq.~(\ref{alpha_nu_pm_11_5}) to prove that
\begin{equation}
\alpha_{\text{R}\pm}(k,K) = \alpha_{\text{L}\pm}^*(k,K).
\end{equation}
Hence, one can define the couple $\alpha(k,K)$ and $\beta(k,K)$ such that
\begin{subequations}
\begin{align}
& \alpha_{\text{L}+}(k,K) = \alpha(k,K), && \alpha_{\text{L}-}(k,K) = -\beta^*(k,K), \\
& \alpha_{\text{R}+}(k,K) = \alpha^*(k,K), && \alpha_{\text{R}-}(k,K) = -\beta(k,K).
\end{align}
\end{subequations}

In conclusion, we obtain the Bogoliubov transformation
\begin{align}\label{Bogolyubov_transformation}
\hat{a}(k) = & \int_{\mathbb{R}}dK \left[ \alpha(k,K)\hat{A}_\text{L}(K) - \beta^*(k,K) \hat{A}^\dagger_\text{L}(K) \right. \nonumber \\
& \left. + \alpha^*(k,K) \hat{A}_\text{R}(K) - \beta(k,K) \hat{A}^\dagger_\text{R}(K) \right],
\end{align}
with Bogoliubov coefficients
\begin{align}\label{Bogolyubov_coefficients}
& \alpha(k,K) = \theta(kK) \sqrt{\frac{K}{k}}F(k,K), & \beta(k,K) = \theta(kK) \sqrt{\frac{K}{k}}F(-k,K)
\end{align}
and with $F(k,K)$ defined by Eq.~(\ref{F_11}). As a consequence of Eq.~(\ref{alpha_beta_to_alpha_beta_proof}), we also obtain the following identity relating $\alpha(k,K)$ to $\beta(k,K)$
\begin{equation} \label{alpha_to_beta}
\beta(k,K) = \exp \left(-\frac{c \beta |K|}{2} \right) \alpha(k,K),
\end{equation}
which will be used in the next subsection to derive the Rindler-Fock representation of the Minkowski vacuum.

\subsection{Minkowski vacuum in the left and the right Rindler frame}\label{Minkowski_vacuum_in_Rindler_spacetime_scalar_11}

In this subsection, we follow step \ref{Bogoliubov_step_3} of Sec.~\ref{Frame_dependent_content_of_particles_Introduction} to derive the Rindler-Fock representation of the Minkowski vacuum.

The vector state representing the Minkowski vacuum in the Minkowski-Fock space $\mathcal{H}_\text{M}$ is defined by means of Eq.~(\ref{Minkowski_vacuum}) adapted for the 1+1 dimensional case, i.e.,
\begin{equation} \label{vacuum_state_H_M_in_tilde_H'_M_epsilon}
\hat{a}(k) |0_\text{M}\rangle = 0,
\end{equation}
for any $k \in \mathbb{R}$. By plugging Eq.~(\ref{Bogolyubov_transformation}) in Eq.~(\ref{vacuum_state_H_M_in_tilde_H'_M_epsilon}) and by using Eq.~(\ref{alpha_to_beta}) we obtain
\begin{align}\label{vacuum_state_H_M_in_tilde_H'_M_epsilon_2}
& \int_{\mathbb{R}}dK \left\lbrace \alpha(k,K) \left[ \hat{A}_\text{L}(K) - \exp \left(-\frac{c \beta |K|}{2} \right) \hat{A}^\dagger_\text{R}(K) \right] \right. \nonumber \\
& \left. + \alpha^*(k,K) \left[ - \exp \left(-\frac{c \beta |K|}{2} \right) \hat{A}^\dagger_\text{L}(K)  +  \hat{A}_\text{R}(K) \right] \right\rbrace |0_\text{M}\rangle = 0,
\end{align}
which holds for any real $k$. Necessary and sufficient condition for Eq.~(\ref{vacuum_state_H_M_in_tilde_H'_M_epsilon_2}) is
\begin{equation} \label{vacuum_state_property}
\hat{A}_\nu(K) |0_\text{M}\rangle = \exp \left(- \frac{c \beta |K|}{2} \right) \hat{A}_{\bar{\nu}}^\dagger(K) |0_\text{M} \rangle ,
\end{equation}
which holds for any $\nu = \{ \text{L}, \text{R} \}$ and for any $K \in \mathbb{R}$. The variable $\bar{\nu}$ is defined as the opposite of $\nu$ in the sense that $\bar{\nu}=\text{L}$ if $\nu=\text{R}$ and $\bar{\nu}=\text{R}$ if $\nu=\text{L}$.

The Rindler-Fock state that satisfies Eq.~(\ref{vacuum_state_property}) is given by
\begin{equation}\label{Rindler_vacuum_to_Minkowski_SS}
| 0_\text{M} \rangle \propto \exp(\hat{s}_\text{S}) | 0_\text{L}, 0_\text{R} \rangle,
\end{equation}
with
\begin{equation} \label{SS_S}
\hat{s}_\text{S} = \int_\mathbb{R} dK \exp \left( -\frac{c \beta |K|}{2} \right) \hat{A}^\dagger_\text{L}(K) \hat{A}^\dagger_\text{R}(K).
\end{equation}
The vector $| 0_\text{L}, 0_\text{R} \rangle$ represents the Rindler vacuum in the Rindler-Fock space $\mathcal{H}_{\text{L}, \text{R}}$ and is defined by the equivalent of Eq.~(\ref{Rindler_vacuum_scalar}) in 1+1 dimensions, i.e.,
 \begin{equation}
\hat{A}_\nu(K) |0_\text{L}, 0_\text{R}\rangle = 0,
\end{equation}
for any $\nu = \{ \text{L}, \text{R} \}$ and for any $K \in \mathbb{R}$. 

One can prove that Eq.~(\ref{Rindler_vacuum_to_Minkowski_SS}) is the solution of Eq.~(\ref{vacuum_state_property}) by noticing that
$\hat{s}_\text{S}$ satisfies the commutation relation
\begin{equation}\label{A_ss_commutation}
\left[ \hat{A}_\nu(K) , \hat{s}_\text{S} \right] = \exp \left( -\frac{c \beta |K|}{2} \right) \hat{A}^\dagger_{\bar{\nu}}(K).
\end{equation}
By using the exponential series $\exp(\hat{s}_\text{S}) = \sum_{n=0}^\infty \hat{s}_\text{S}^n/n!$ and Eq.~(\ref{A_ss_commutation}) one can prove by induction the following commutation relation
\begin{equation}\label{A_SS_commutation}
\left[ \hat{A}_\nu(K) , \exp(\hat{s}_\text{S}) \right] = \exp \left( -\frac{c \beta |K|}{2} \right) \hat{A}^\dagger_{\bar{\nu}}(K) \exp(\hat{s}_\text{S}).
\end{equation}
Equation (\ref{A_SS_commutation}) implies that
\begin{align}
\hat{A}_\nu(K) \exp(\hat{s}_\text{S}) | 0_\text{L}, 0_\text{R} \rangle = & \left[ \hat{A}_\nu(K) , \exp(\hat{s}_\text{S}) \right]| 0_\text{L}, 0_\text{R} \rangle + \exp(\hat{s}_\text{S}) \hat{A}_\nu(K) | 0_\text{L}, 0_\text{R} \rangle \nonumber \\
 = & \exp \left( -\frac{c \beta |K|}{2} \right) \hat{A}^\dagger_{\bar{\nu}}(K) \exp(\hat{s}_\text{S}) | 0_\text{L}, 0_\text{R} \rangle.
\end{align}

By means of Eq.~(\ref{Rindler_vacuum_to_Minkowski_SS}), we have been able to write the Rindler-Fock representation of the Minkowski vacuum as prescribed by Eq.~(\ref{Rindler_vacuum_to_Minkowski}). However, the state $\exp(\hat{s}_\text{S}) |0_\text{L}, 0_\text{R}\rangle$ appearing in the right hand side of Eq.~(\ref{Rindler_vacuum_to_Minkowski_SS}) is not normalizable. Provably, this is a consequence of the continuous nature of the variables $K$. Consequently, Equation (\ref{Rindler_vacuum_to_Minkowski_SS}) needs to be interpreted in some regularization limit.

In Appendix \ref{Two_mode_squeezed_state_Massless_scalar_field_in_11_spacetime}, we study the state $\exp(\hat{s}_\text{S}) |0_\text{L}, 0_\text{R}\rangle$ in the regularized discrete theory, we derive its normalized version and then we perform the continuum limit. Such a limit leads to the normalized state $\hat{S}_\text{S} |0_\text{L}, 0_\text{R}\rangle$ with unitary operator
\begin{equation} \label{D}
\hat{S}_\text{S} = \exp \left( 2 \int_\mathbb{R} dK \zeta(c |K|) \left[ \hat{A}^\dagger_\text{L}(K) \hat{A}^\dagger_\text{R}(K) \right]^\text{A} \right),
\end{equation}
with
\begin{equation}\label{zeta_omega}
\zeta(\Omega) = \tanh^{-1} \left( \exp \left( -\frac{\beta \Omega}{2} \right) \right)
\end{equation}
and where $\hat{O}^\text{A} = (\hat{O} - \hat{O}^\dagger)/2$ is the antihermitian part of $\hat{O}$. This proves that Eqs.~(\ref{Rindler_vacuum_to_Minkowski_SS}) and (\ref{SS_S}) are equivalent to Eqs.~(\ref{Rindler_vacuum_to_Minkowski}) and (\ref{D}). 

The explicit particle content of the Minkowski vacuum in $\mathcal{H}_{\text{L},\text{R}}$ can be obtained by defining the states 
\begin{equation}\label{LR_Rindler_basis}
| (K'_1, \dots, K'_m), (K_1, \dots, K_n) \rangle = \prod_{j=1}^m \hat{A}_\text{L}^\dagger (K'_j) \prod_{i=1}^n \hat{A}_\text{R}^\dagger (K_i)  | 0_\text{L}, 0_\text{R} \rangle,
\end{equation}
which result in
\begin{align} \label{SS_Minkowski_vacuum}
\exp(\hat{s}_\text{S}) |0_\text{L}, 0_\text{R}\rangle = & \sum_{n=1}^\infty  \frac{1}{n!} \int_\mathbb{R} dK_1 \dots \int_\mathbb{R} dK_n \exp \left( -\frac{c \beta}{2} \sum_{i=1}^n |K_i| \right) \nonumber \\
& \times | (K_1, \dots, K_n), (K_1, \dots, K_n) \rangle + | 0_\text{L}, 0_\text{R} \rangle.
\end{align}
Notice that $| (K'_1, \dots, K'_m), (K_1, \dots, K_n) \rangle$ is the Fock basis of $\mathcal{H}_{\text{L},\text{R}}$ up to the repetition of equivalent states due to the bosonic symmetry between the elements of $\{ K_i \}_{i=1}^n$ and between the elements of $\{ K'_j \}_{j=1}^m$.

From Eqs.~(\ref{Rindler_vacuum_to_Minkowski_SS}) and (\ref{SS_Minkowski_vacuum}) one can see that the Minkowski vacuum appears as a state populated by particles when represented in the Rindler-Fock space. This is at the origin of the particle production in accelerated frames and the Unruh effect, which will be detailed in the next subsection. In particular, we will use Eq.~(\ref{SS_Minkowski_vacuum}) to derive the thermal state (\ref{thermal}).

Notice that the operator $\hat{S}_\text{S}$ of Eq.~(\ref{D}) is composed by two modes squeezed operators \cite{Barnett2002}, where each couple of modes is a left and a right Rindler mode with defined momentum $K$. The partial trace of a two mode squeezed operator is precisely a thermal state.

\subsection{Minkowski vacuum in the right Rindler frame}\label{Unruh_effect_for_scalar_fields_11}

As discussed in Sec.~\ref{Frame_dependent_content_of_particles_Introduction}, the representation of the Minkowski vacuum in the accelerated frame $\hat{\rho}_0$ is given by the partial trace $\text{Tr}_\text{L}$ over the left wedge. Also, we have anticipated that $\hat{\rho}_0$ is the thermal state (\ref{thermal}) with Unruh temperature $T_\text{U} = \hbar \beta/k_\text{B}$.

For massless scalar real fields in 1+1 dimensions the right Rindler Hamiltonian is
\begin{equation}\label{H_R_11}
\hat{H}_\text{R} = \int_{\mathbb{R}} d K \hbar c |K| \hat{A}_\text{R}^\dagger (\vec{K}) \hat{A}_\text{R} (\vec{K})
\end{equation}
and can be expanded with respect to the states $| K_1, \dots, K_n \rangle$ defined by
\begin{equation}
| K_1, \dots, K_n \rangle = \prod_{i=1}^n \hat{A}_\text{R}^\dagger (K_i)  | 0_\text{R} \rangle
\end{equation}
in analogy to Eq.~(\ref{LR_Rindler_basis}). The expansion of Eq.~(\ref{H_R_11}) with respect to the states $| K_1, \dots, K_n \rangle$ is
\begin{equation}\label{H_R_11_expansion}
\hat{H}_\text{R} = \sum_{n=1}^\infty \frac{1}{n!}  \int_\mathbb{R} dK_1 \dots \int_\mathbb{R} dK_n \left( \hbar c \sum_{i=1}^n |K_i| \right) | K_1, \dots, K_n \rangle \langle K_1, \dots, K_n |.
\end{equation}
By plugging Eq.~(\ref{H_R_11_expansion}) into Eq.~(\ref{thermal}), we obtain the expansion of the thermal state
\begin{align}\label{rho_0_11_expansion}
\hat{\rho}_0 \propto & \sum_{n=1}^\infty \frac{1}{n!}  \int_\mathbb{R} dK_1 \dots \int_\mathbb{R} dK_n \exp \left( - \frac{\beta}{\hbar} \sum_{i=1}^n |K_i| \right) | K_1, \dots, K_n \rangle \langle K_1, \dots, K_n |  \nonumber \\
& +  | 0_\text{R} \rangle \langle  0_\text{R} |.
\end{align}

It is straightforward to prove that $\rho_0 \propto \text{Tr}_\text{L}(\exp(\hat{s}_\text{S}) |0_\text{L}, 0_\text{R}\rangle \lbrace 0_\text{L}, 0_\text{R} | \exp(\hat{s}_\text{S})^\dagger )$ by means of Eqs.~(\ref{SS_Minkowski_vacuum}) and (\ref{rho_0_11_expansion}) and by knowing that there are the $n!$ combination of permutations to write $| K_1, \dots, K_n \rangle$ as the same state. In this way, one formally derives the Unruh effect that predicts the presence of a thermal state in the accelerated frame whenever an inertial observer detects the vacuum.

\section{Scalar field}\label{Frame_dependent_content_of_particles_scalar}

In this section, we consider a massive scalar complex field in 3+1 dimensions, which has been described in Sec.~\ref{QFT_in_Minkowski_spacetime_scalar} for the Minkowski spacetime and Sec.~\ref{QFT_in_curved_spacetime_Rindler_scalar} for the Rindler spacetime. We follow steps \ref{Bogoliubov_step_1}-\ref{Bogoliubov_step_4} of Sec.~\ref{Frame_dependent_content_of_particles_Introduction} to derive the Bogoliubov transformation (\ref{Rindler_Bogoliubov_transformations_1}) and the representation of the Minkowski in the Rindler-Fock space (\ref{Rindler_vacuum_to_Minkowski}). In particular, we compute the Bogoliubov coefficients $\alpha_{\nu\pm}(\vec{k},\vec{\theta})$ and the unitary operator $\hat{S}_\text{S}$ relating the Minkowski vacuum $| 0_\text{M} \rangle$ to the Rindler vacuum $| 0_{\text{L},\text{R}} \rangle$. The procedure is very similar to the one seen in Sec.~\ref{Massless_scalar_field_in_11_spacetime}; few changes include the presence of three spatial dimensions and the Rindler-Klein-Gordon modes (\ref{F_Rindler_all}) replacing the free modes (\ref{f_11}). Also, we prove Eq.~(\ref{thermal}), which implies that the Minkowski vacuum is seen by accelerated observers as a thermal state; hence, we detail the Unruh effect for scalar fields.

\subsection{Bogoliubov transformation}

Here, we prove Eq.~(\ref{Rindler_Bogoliubov_transformations_1}) and we compute the Bogoliubov coefficients $\alpha_{\nu\pm}(\vec{k},\vec{\theta})$ by following steps \ref{Bogoliubov_step_1} and \ref{Bogoliubov_step_2} of Sec.~\ref{Frame_dependent_content_of_particles_Introduction}.

We start by considering Eq.~(\ref{Bogoliubov_transformations_1}), which explicitly reads as
\begin{subequations}\label{Bogoliubov_transformations_2_Rindler_scalar_0}
\begin{align}
& \hat{a}(\vec{k}) = \frac{i}{\hbar} \int_{\mathbb{R}^3} d^3x \left[  f^*(\vec{k},t,\vec{x}) \partial_0 \hat{\phi}(t,\vec{x}) - \hat{\phi}(t,\vec{x}) \partial_0  f^*(\vec{k},t,\vec{x})\right], \\
& \hat{b}^\dagger(\vec{k}) = - \frac{i}{\hbar} \int_{\mathbb{R}^3} d^3x \left[  f(\vec{k},t,\vec{x}) \partial_0 \hat{\phi}(t,\vec{x})   - \hat{\phi}(t,\vec{x}) \partial_0  f(\vec{k},t,\vec{x})\right].
\end{align}
\end{subequations}

For $t=0$, Equation (\ref{Bogoliubov_transformations_2_Rindler_scalar_0}) becomes
\begin{subequations}\label{Bogoliubov_transformations_2_Rindler_scalar}
\begin{align}
& \hat{a}(\vec{k}) = \frac{i}{\hbar} \int_{\mathbb{R}^3} d^3x \left[  f^*(\vec{k},0,\vec{x}) \left. \partial_0 \hat{\phi}(t,\vec{x}) \right|_{t=0} - \hat{\phi}(0,\vec{x}) \left. \partial_0  f^*(\vec{k},t,\vec{x}) \right|_{t=0} \right], \\
& \hat{b}^\dagger(\vec{k}) = - \frac{i}{\hbar} \int_{\mathbb{R}^3} d^3x \left[  f(\vec{k},0,\vec{x}) \left. \partial_0 \hat{\phi}(t,\vec{x}) \right|_{t=0}   - \hat{\phi}(0,\vec{x}) \left. \partial_0  f(\vec{k},t,\vec{x}) \right|_{t=0} \right],
\end{align}
\end{subequations}
whereas Eq.~(\ref{scalar_transformation_Rindler_inverse_t}) becomes
\begin{equation}\label{scalar_transformation_Rindler_inverse_2}
\hat{\phi}(0,\vec{x}) = \sum_{\nu=\{\text{L},\text{R}\}} \theta(s_\nu x) \hat{\Phi}_\nu(0,\vec{X}_\nu(\vec{x})),
\end{equation}
where $\vec{X}_\nu(\vec{x}) = \vec{X}_\nu(0, \vec{x})$. The time derivative of Eq.~(\ref{scalar_transformation_Rindler_inverse_t}) for $t=0$, instead, can be computed by considering the chain rule
\begin{equation}\label{chain_rule_Rindler}
\frac{\partial}{\partial T} = s_\nu az \frac{\partial}{\partial t}  + s_\nu ac^2t  \frac{\partial}{\partial z},
\end{equation}
which leads to
\begin{equation}\label{time_derivative_scalar_transformation_Rindler_2}
\left. \partial_0 \hat{\phi}(t,\vec{x}) \right|_{t=0} = \sum_{\nu=\{\text{L},\text{R}\}} \theta(s_\nu z) \frac{s_\nu}{a z} \left. \partial_0 \hat{\Phi}_\nu(T,\vec{X}_\nu(\vec{x})) \right|_{T = 0}.
\end{equation}

By plugging Eqs.~(\ref{scalar_transformation_Rindler_inverse_2}) and (\ref{time_derivative_scalar_transformation_Rindler_2}) in Eq.~(\ref{Bogoliubov_transformations_2_Rindler_scalar}) one obtains
\begin{subequations}\label{Bogoliubov_transformations_3_Rindler}
\begin{align}
\hat{a}(\vec{k}) = & \frac{i}{\hbar} \sum_{\nu=\{\text{L},\text{R}\}} \int_{\mathbb{R}^3} d^3x \theta(s_\nu z)  \left[ \frac{s_\nu}{a x} f^*(\vec{k},0,\vec{x}) \left. \partial_0 \hat{\Phi}_\nu(T,\vec{X}_\nu(\vec{x})) \right|_{T=0}  \right. \nonumber \\
& \left. - \hat{\Phi}_\nu(0,\vec{X}_\nu(\vec{x})) \left. \partial_0  f^*(\vec{k},t,\vec{x}) \right|_{t=0}   \right], \\
\hat{b}^\dagger(\vec{k}) = & -\frac{i}{\hbar} \sum_{\nu=\{\text{L},\text{R}\}} \int_{\mathbb{R}^3} d^3x \theta(s_\nu z)   \left[ \frac{s_\nu}{a z} f(\vec{k},0,\vec{x}) \left. \partial_0 \hat{\Phi}_\nu(T,\vec{X}_\nu(\vec{x}))  \right|_{T=0}  \right. \nonumber \\
& \left. - \hat{\Phi}_\nu(0,\vec{X}_\nu(\vec{x})) \left. \partial_0  f(\vec{k},t,\vec{x})\right|_{t=0}   \right].
\end{align}
\end{subequations}
Equation (\ref{Rindler_scalar_decomposition}) can be used in Eq.~(\ref{Bogoliubov_transformations_3_Rindler}) to obtain the Bogoliubov transformation (\ref{Rindler_Bogoliubov_transformations_1}) with Bogoliubov coefficients
\begin{subequations}\label{alpha_nu_pm}
\begin{align} 
\alpha_{\nu+}(\vec{k},\vec{\theta}) = &   \frac{i}{\hbar} \int_{\mathbb{R}^3} d^3x \theta(s_\nu x)   \left[ \frac{s_\nu}{a x} f^*(\vec{k},0,\vec{x})  \left. \partial_0 F_\nu(\vec{\theta}, T, \vec{X}_\nu(\vec{x})) \right|_{T=0} \right. \nonumber \\
& \left. - F_\nu(\vec{\theta}, 0, \vec{X}_\nu(\vec{x})) \left. \partial_0  f^*(\vec{k},t,\vec{x}) \right|_{t=0} \right],\\
\alpha_{\nu-}(\vec{k},\vec{\theta}) =  & \frac{i}{\hbar} \int_{\mathbb{R}^3} d^3x \theta(s_\nu x)  \left[ \frac{s_\nu}{a x} f^*(\vec{k},0,\vec{x}) \left. \partial_0  F_\nu^*(\vec{\theta}, T, \vec{X}_\nu(\vec{x}))   \right|_{T=0} \right. \nonumber \\
& \left. - F_\nu^*(\vec{\theta}, 0, \vec{X}_\nu(\vec{x})) \left. \partial_0  f^*(\vec{k},t,\vec{x}) \right|_{t=0} \right].
\end{align}
\end{subequations}

By using Eqs.~(\ref{free_modes}), (\ref{F_Rindler}) and (\ref{F_F_nu}) and the fact that $\tilde{F}$ is real, Eq.~(\ref{alpha_nu_pm}) can be written in a more compact form
\begin{equation}\label{alpha_nu_pm_2}
\alpha_{\nu \pm}(\vec{k},\vec{\theta}) = \int_{\mathbb{R}^3} d^3 x \frac{\theta(s_\nu z)}{\hbar} \left[ \pm  \frac{s_\nu \theta_1}{a z} + \omega(\vec{k}) \right]  f^*( \vec{k}, 0, \vec{x} )  \tilde{F}(\vec{\theta},  s_\nu Z_\nu(z)) e^{ \pm i \vec{\theta}_\perp \cdot \vec{x}_\perp},
\end{equation}
where $\vec{\theta}_\perp$ is such that $\vec{\theta} = (\theta, \vec{\theta}_\perp) = (\Omega, \vec{K}_\perp)$ and, hence, $\vec{\theta}_\perp = \vec{K}_\perp$, or, equivalently $\vec{\theta}_\perp = (\theta_2, \theta_3)$. By knowing that $\tilde{F}(\vec{\theta}, X)$ is invariant under $\vec{\theta} \mapsto - \vec{\theta}$ [Eq.~(\ref{F_tilde_Rindler})], Eq.~(\ref{alpha_nu_pm_2}) is such that
\begin{equation}\label{alpha_nu_pm_4}
\alpha_{\nu \pm}(\vec{k},\vec{\theta}) = \alpha_\nu (\vec{k},\pm \vec{\theta}),
\end{equation}
with
\begin{equation} \label{alpha}
\alpha_\nu(\vec{k},\vec{\theta}) = \int_{\mathbb{R}^3} d^3 x \frac{\theta(s_\nu z)}{\hbar}  \left[  \frac{s_\nu \theta_1}{a z} + \omega(\vec{k}) \right] f^*( \vec{k}, 0,  \vec{x} )  \tilde{F}(\vec{\theta},  s_\nu Z_\nu(z)) e^{ i \vec{\theta}_\perp \cdot \vec{x}_\perp}.
\end{equation}

Owing to Eq.~(\ref{alpha_nu_pm_4}), the Bogoliubov transformation (\ref{Rindler_Bogoliubov_transformations_1}) can be simplified by
\begin{subequations}\label{Rindler_Bogoliubov_transformations}
\begin{align}
& \hat{a}(\vec{k}) =   \sum_{\nu=\{\text{L},\text{R}\}}\int_{\theta_1>0} d^3 \theta \left[ \alpha_\nu(\vec{k},\vec{\theta})  \hat{A}_\nu(\vec{\theta}) + \alpha_\nu(\vec{k},-\vec{\theta}) \hat{B}^\dagger_\nu(\vec{\theta}) \right], \\
& \hat{b}(\vec{k}) =  \sum_{\nu=\{\text{L},\text{R}\}}\int_{\theta_1>0} d^3 \theta \left[ \alpha_\nu(\vec{k},\vec{\theta})  \hat{B}_\nu(\vec{\theta})  + \alpha_\nu(\vec{k},-\vec{\theta}) \hat{A}^\dagger_\nu(\vec{\theta})  \right],
\end{align}
\end{subequations}
where $\alpha_\nu(\vec{k},\vec{\theta})$ is defined by Eq.~(\ref{alpha}). Equation (\ref{Rindler_Bogoliubov_transformations}) can be written in a more compact form by
\begin{align}\label{Rindler_Bogoliubov_transformations_compact}
& \hat{a}(\vec{k}) = \sum_{\nu=\{\text{L},\text{R}\}}   \int_{\mathbb{R}^3} d^3 \theta \alpha_\nu(\vec{k},\vec{\theta})  \hat{\mathcal{A}}_\nu(\vec{\theta}), & \hat{b}(\vec{k}) = \sum_{\nu=\{\text{L},\text{R}\}}   \int_{\mathbb{R}^3} d^3 \theta \alpha_\nu(\vec{k},\vec{\theta})  \hat{\mathcal{B}}_\nu(\vec{\theta}),
\end{align}
where
\begin{align}\label{Rindler_Bogoliubov_transformations_compact_AB}
& \hat{\mathcal{A}}_\nu(\vec{\theta}) = \begin{cases}
 \hat{A}_\nu(\vec{\theta}) & \text{if } \theta_1 > 0 \\
 \hat{B}^\dagger_\nu(-\vec{\theta}) & \text{if } \theta_1 < 0 \\
\end{cases} , & \hat{\mathcal{B}}_\nu(\vec{\theta}) = \begin{cases}
 \hat{B}_\nu(\vec{\theta}) & \text{if } \theta_1 > 0 \\
 \hat{A}^\dagger_\nu(-\vec{\theta}) & \text{if } \theta_1 < 0 \\
\end{cases}.
\end{align}

We are now interested in giving the explicit form of the Bogoliubov coefficient $\alpha_\nu(\vec{k},\vec{\theta})$ by computing the integral in Eq.~(\ref{alpha}). By using Eq.~(\ref{free_modes}) in Eq.~(\ref{alpha}) and by computing the derivative with respect to $\vec{x}_\perp$, one obtains
\begin{equation} \label{alpha_2}
\alpha_\nu(\vec{k},\vec{\theta}) = \delta^2(\vec{k}_\perp-\vec{\theta}_\perp) \chi_\nu (\vec{k},\theta_1),
\end{equation}
with
\begin{equation} \label{chi}
\chi_\nu(\vec{k},\Omega) = \sqrt{ \frac{\pi}{\hbar \omega(\vec{k})} } \int_{\mathbb{R}} dz \theta(s_\nu z)   \left[  \frac{s_\nu \Omega}{a z} + \omega(\vec{k}) \right]  e^{-i k_3 z} \tilde{F}(\Omega, \vec{k}_\perp, s_\nu Z_\nu(z))
\end{equation}
and $\vec{k}_\perp = (k_1, k_2)$ as transverse coordinates of momentum $\vec{k}$.

Equation (\ref{chi}) can also be written as
\begin{align} \label{chi_derivative}
& \chi_\nu(\vec{k},\Omega) = \sqrt{ \frac{\pi}{\hbar \omega(\vec{k})} } \left[-s_\nu i \frac{\Omega}{a} \lim_{\epsilon \rightarrow 0^+} \int_{-s_\nu \infty}^{k_3} d \eta \int_{\mathbb{R}} dz \theta(s_\nu z)    e^{\eta (s_\nu \epsilon - i z )} \right. \nonumber \\
&  \left. \times \tilde{F}(\Omega, \vec{k}_\perp, s_\nu Z_\nu(z))+  \omega(\vec{k}) \int_{\mathbb{R}} dz \theta(s_\nu z)    e^{-i k_3 z} \tilde{F}(\Omega, \vec{k}_\perp, s_\nu Z_\nu(z)) \right]
\end{align}
Now, by using Eq.~(\ref{F_tilde_Rindler}), one can write Eq.~(\ref{chi}) in terms of Fourier transform of Bessel functions
\begin{align}\label{chi_derivative_2}
& \chi_\nu(\vec{k},\Omega) = \frac{1}{2 \pi^2 c} \sqrt{ \frac{\pi}{a \omega(\vec{k})} \left| \sinh \left( \frac{\beta \Omega}{2} \right) \right| } \left[-s_\nu i \frac{\Omega}{a} \lim_{\epsilon \rightarrow 0^+} \int_{-s_\nu \infty}^{k_3} d \eta \int_{\mathbb{R}} dz \theta(s_\nu z)    \right. \nonumber \\
& \left. \times e^{\eta ( s_\nu \epsilon - i z )}  K_{i \Omega / c a} \left( s_\nu \kappa (\vec{k}_\perp) z \right) + \omega(\vec{k}) \int_{\mathbb{R}} dz \theta(s_\nu z)    e^{-i k_3 z} K_{i \Omega / c a} \left( s_\nu \kappa (\vec{k}_\perp) z \right) \right].
\end{align}

Notice that Eqs.~(\ref{omega_k}) and (\ref{kappa_k_perp}) are such that
\begin{align}
& \frac{\omega^2(\vec{k})}{c^2}  - k_3^2 = \kappa^2 (\vec{k}_\perp), && \omega (\vec{k})> 0, && \kappa (\vec{k}_\perp) > 0.
\end{align}
This suggests the definition of the function $\vartheta(\vec{k})$ such that
\begin{subequations}\label{theta_k}
\begin{align}
& \omega (\vec{k}) = c \kappa (\vec{k}_\perp) \cosh(\vartheta(\vec{k})),\label{omega_theta} \\
& k_3 = \kappa (\vec{k}_\perp) \sinh(\vartheta(\vec{k})) \label{k_3_theta}.
\end{align}
\end{subequations}
By means of Eqs.~(\ref{theta_k}), Eq.~(\ref{chi_derivative_2}) can be written as
\begin{align}\label{chi_derivative_3}
& \chi_\nu(\vec{k},\Omega) = \frac{1}{2 \pi^2 c} \sqrt{ \frac{\pi}{c a \kappa (\vec{k}_\perp) \cosh(\vartheta(\vec{k}))} \left| \sinh \left( \frac{\beta \Omega}{2} \right) \right| } \nonumber \\
& \times \left[-s_\nu i \frac{\Omega}{a} \lim_{\epsilon \rightarrow 0^+} \int_{-s_\nu \infty}^{\kappa (\vec{k}_\perp) \sinh(\vartheta(\vec{k}))} d \eta \int_{\mathbb{R}} dz \theta(s_\nu z) e^{\eta (s_\nu \epsilon - i z )}  K_{i \Omega / c a} \left( s_\nu \kappa (\vec{k}_\perp) z \right)   \right. \nonumber \\
& \left.   + c \kappa (\vec{k}_\perp) \cosh(\vartheta(\vec{k})) \int_{\mathbb{R}} dz \theta(s_\nu z)    e^{-i \kappa (\vec{k}_\perp) \sinh(\vartheta(\vec{k})) z} K_{i \Omega / c a} \left( s_\nu \kappa (\vec{k}_\perp) z \right) \right].
\end{align}

In Appendix \ref{Proof_of_Bessel_integral_representation_final} we compute the Fourier transform of $K_\zeta (|\xi|)$ and obtain the following result
\begin{equation}\label{Bessel_integral_representation_final}
\int_{\mathbb{R}} d\xi \theta(\xi) e^{- i \xi \sinh(\tau)} K_\zeta (\xi)   =  \frac{\pi \sin \left( \zeta \left( \frac{\pi}{2} - i \tau \right) \right)}{ \sin(\pi \zeta) \cosh(\tau)}.
\end{equation}
From Eq.~(\ref{Bessel_integral_representation_final}) we can also compute the following identity
\begin{align}\label{Bessel_integral_representation_final_integral}
& \lim_{\varepsilon \rightarrow 0^+} \int_{- \infty}^\tau d \tau' \cosh(\tau') \int_{\mathbb{R}} d\xi \theta(\xi) e^{  \sinh(\tau')( \epsilon - i \xi)} K_\zeta (\xi)  \nonumber \\
 = & \frac{\pi}{ \sin(\pi \zeta)} \lim_{\varepsilon \rightarrow 0^+} \int_{-\infty}^\tau d \tau' \sin \left( \zeta \left( \frac{\pi}{2} - i \tau' \right) \right)e^{ \epsilon \sinh(\tau')} \nonumber \\
 = & \frac{i \pi}{\zeta \sin(\pi \zeta)} \cos \left( \zeta \left( \frac{\pi}{2} - i \tau \right) \right).
\end{align}
By choosing $\xi = s_\nu \kappa (\vec{k}_\perp) z $, $\zeta = i \Omega / c a$, $\tau = s_\nu \vartheta(\vec{k})$, $\sinh(\tau') = s _\nu \eta / \kappa (\vec{k}_\perp)$ and $\varepsilon = \epsilon \kappa (\vec{k}_\perp)$ and by plugging Eqs.~(\ref{Bessel_integral_representation_final}) and (\ref{Bessel_integral_representation_final_integral}) in Eq.~(\ref{chi_derivative_3}) we obtain
\begin{align}\label{chi_derivative_4}
& \chi_\nu(\vec{k},\Omega) = \frac{i}{2 } \sqrt{ \frac{1}{\pi c a  \kappa (\vec{k}_\perp) \cosh(\vartheta(\vec{k}))} \left| \sinh \left( \frac{\beta \Omega}{2} \right) \right| } \left[ \sinh \left( \frac{\beta \Omega}{2} \right) \right]^{-1} \nonumber \\
& \times \left[ - i \cos \left( \frac{i \Omega}{c a} \left( \frac{\pi}{2} - s_\nu i \vartheta(\vec{k}) \right) \right)    + \sin \left( \frac{i \Omega}{c a} \left( \frac{\pi}{2} - s_\nu i \vartheta(\vec{k}) \right) \right) \right],
\end{align}
which leads to
\begin{align}\label{chi_derivative_5}
\chi_\nu(\vec{k},\Omega) = & \text{sign}(\Omega) \frac{1}{2 } \left[ \pi c a  \kappa (\vec{k}_\perp) \cosh(\vartheta(\vec{k})) \left| \sinh \left( \frac{\beta \Omega}{2} \right) \right| \right]^{-1/2}  \nonumber \\
& \times \exp \left( \frac{ \Omega}{c a} \left( \frac{\pi}{2} - s_\nu i \vartheta(\vec{k}) \right) \right) .
\end{align}

Equations (\ref{alpha_2}) and (\ref{chi_derivative_5}) give the explicit expression for the Bogoliubov coefficients
\begin{align} \label{alpha_final}
 \alpha_\nu(\vec{k},\Omega, \vec{K}_\perp) = & \text{sign}(\Omega)  \frac{1}{2 } \left[ \pi c a  \kappa (\vec{k}_\perp) \cosh(\vartheta(\vec{k})) \left| \sinh \left( \frac{\beta \Omega}{2} \right) \right| \right]^{-1/2}  \nonumber \\
& \times \exp \left( \frac{ \Omega}{c a} \left( \frac{\pi}{2} - s_\nu i \vartheta(\vec{k}) \right) \right) \delta^2(\vec{k}_\perp-\vec{K}_\perp).
\end{align}
Notice that Eq.~(\ref{alpha_final}) satisfies
\begin{equation}\label{alpha_to_beta_31}
\alpha_\nu(\vec{k},-\Omega, \vec{K}_\perp) = -\exp \left(-\frac{\beta |\Omega|}{2} \right) \alpha_{\bar{\nu}}(\vec{k},\Omega, \vec{K}_\perp),
\end{equation}
with $\beta = 2 \pi / c a$ and $\bar{\nu}$ is the opposite of $\nu$.

\subsection{Minkowski vacuum in the left and the right Rindler frame}\label{Unruh_effect_for_scalar_fields}

Equation (\ref{alpha_to_beta_31}) is the equivalent of Eq.~(\ref{alpha_to_beta_31}) in 3+1 dimensions as it relates Bogoliubov coefficients of one frame to the coefficients of the other by means of an $\exp( - \beta |\Omega|/2 )$ factor. In Sec.~\ref{Minkowski_vacuum_in_Rindler_spacetime_scalar_11}, we used Eq.~(\ref{alpha_to_beta_31}) to derive the representation of the Minkowski vacuum in the Rindler-Fock space as prescribed by Eq.~(\ref{Rindler_vacuum_to_Minkowski}) with unitary operator $\hat{S}_\text{S}$ given by Eq.~(\ref{D}). In particular, as a consequence of Eq.~(\ref{alpha_to_beta_31}), we obtained an equation for the Minkowski vacuum [Eq.~(\ref{vacuum_state_property})] that admits a solution in the Rindler-Fock space. In this subsection, we show that a similar equation holds for massive scalar complex fields in 3+1 dimensions.

By following step \ref{Bogoliubov_step_3} of Sec.~\ref{Frame_dependent_content_of_particles_Introduction}, we consider the definition of the Minkowski vacuum [Eq.~(\ref{Minkowski_vacuum})]. By using Eqs.~(\ref{Rindler_Bogoliubov_transformations}) and (\ref{alpha_to_beta_31}) and by transforming the summation index $\nu \mapsto \bar{\nu}$ and the integration variable $\vec{K}_\perp \mapsto -\vec{K}_\perp$ when needed we obtain
\begin{subequations}\label{vacuum_state_H_M_in_tilde_H'_M_epsilon_2_31}
\begin{align}
&  \sum_{\nu=\{\text{L},\text{R}\}} \int_0^\infty d\Omega \int_{\mathbb{R}^2} d^2\vec{K}_\perp \alpha_\nu(\vec{k},\Omega, \vec{K}_\perp) \nonumber \\
& \times \left[  \hat{A}_\nu(\Omega, \vec{K}_\perp) - \exp \left(-\frac{\beta \Omega}{2} \right) \hat{B}^\dagger_{\bar{\nu}}(\Omega, -\vec{K}_\perp) \right] | 0_\text{M} \rangle = 0, \\
&  \sum_{\nu=\{\text{L},\text{R}\}} \int_0^\infty d\Omega \int_{\mathbb{R}^2} d^2\vec{K}_\perp  \alpha_\nu(\vec{k},\Omega, \vec{K}_\perp) \nonumber \\
& \times \left[  \hat{B}_\nu(\Omega, \vec{K}_\perp)  - \exp \left(-\frac{\beta \Omega}{2} \right) \hat{A}^\dagger_{\bar{\nu}}(\Omega,-\vec{K}_\perp)  \right] | 0_\text{M} \rangle = 0,
\end{align}
\end{subequations}
which holds for any real $k$. Equation (\ref{vacuum_state_H_M_in_tilde_H'_M_epsilon_2}) is equivalent to
\begin{subequations} \label{vacuum_state_property_31}
\begin{align}
& \hat{A}_\nu(\Omega, \vec{K}_\perp) |0_\text{M}\rangle = \exp \left(-\frac{\beta \Omega}{2} \right) \hat{B}^\dagger_{\bar{\nu}}(\Omega,-\vec{K}_\perp) |0_\text{M} \rangle , \\
 & \hat{B}_\nu(\Omega, \vec{K}_\perp) |0_\text{M}\rangle = \exp \left(-\frac{\beta \Omega}{2} \right) \hat{A}^\dagger_{\bar{\nu}}(\Omega,-\vec{K}_\perp) |0_\text{M} \rangle
\end{align}
\end{subequations}
for any $\nu = \{ \text{L}, \text{R} \}$ and for any $(\Omega,\vec{K}_\perp) \in (0,\infty) \otimes \mathbb{R}^2$.

The solution of Eq.~(\ref{vacuum_state_property_31}) is (\ref{Rindler_vacuum_to_Minkowski_SS}), with
\begin{equation} \label{SS_S_31}
\hat{s}_\text{S} = \sum_{\nu=\{\text{L},\text{R}\}} \int_0^\infty d\Omega \int_{\mathbb{R}^2} d^2\vec{K}_\perp \exp \left( - \frac{\beta \Omega}{2} \right)  \hat{A}^\dagger_\nu(\Omega,\vec{K}_\perp) \hat{B}^\dagger_{\bar{\nu}}(\Omega,-\vec{K}_\perp).
\end{equation}
In analogy to the 1+1 massless scalar real field, the state $\exp(\hat{s}_\text{S}) | 0_\text{L}, 0_\text{R} \rangle$ can be normalized in the regularized discrete theory [Appendix \ref{Two_mode_squeezed_state_Frame_dependent_content_of_particles_scalar}]; the continuum limit leads to Eq.~(\ref{Rindler_vacuum_to_Minkowski}), with unitary operator
\begin{equation}\label{Rindler_vacuum_to_Minkowski_unitary_operator}
\hat{S}_\text{S} = \exp\left(  2 \sum_{\nu=\{\text{L},\text{R}\}} \int_0^\infty d\Omega \int_{\mathbb{R}^2} d^2\vec{K}_\perp \zeta(\Omega)  \left[ \hat{A}^\dagger_\nu(\Omega,\vec{K}_\perp) \hat{B}^\dagger_{\bar{\nu}}(\Omega,-\vec{K}_\perp) \right]^\text{A} \right),
\end{equation}
where $\zeta(\Omega)$ is defined by Eq.~(\ref{zeta_omega}) and the superscript A indicates the antihermitian part.

As a result, we obtain the representation of the Minkowski vacuum in the right and left Rindler frame for massive scalar complex fields in 3+1 dimensions. Notice that the operator $\hat{S}_\text{S}$ of Eq.~(\ref{Rindler_vacuum_to_Minkowski_unitary_operator}) is composed by two modes squeezed operators with defined energy $\Omega$ and opposite transverse momentum $\vec{K}_\perp$. This is in analogy to the case of massless scalar real fields in 1+1 dimensions studied in Sec.~\ref{Massless_scalar_field_in_11_spacetime}. The only difference between Eq.~(\ref{D}) and (\ref{Rindler_vacuum_to_Minkowski_unitary_operator}) is given by the Rindler quantum numbers, which in the former case is just $K$, whereas in the latter they are $\Omega$ and $\vec{K}_\perp$.

\subsection{Minkowski vacuum in the right Rindler frame}\label{Unruh_effect_for_scalar_fields_31}

In Sec.~\ref{Unruh_effect_for_scalar_fields_11} we proved for 1+1 massless scalar real fields that the representative of the Minkowski vacuum in the accelerated frame is the thermal state. The prove is based on performing the partial trace $\text{Tr}_\text{L}$ of the pure vector $\exp (\hat{s}_\text{S}) | 0_\text{L}, 0_\text{R} \rangle$.

Given the similarities between Eq.~(\ref{SS_S}) and Eq.~(\ref{SS_S_31}), it is straightforward to prove the same result for 3+1 massive scalar complex fields as well. The only difference is given by the Rindler quantum numbers, as we have detailed at the end of Sec.~\ref{Unruh_effect_for_scalar_fields}.

In conclusion, the Minkowski vacuum of massive scalar complex fields in 3+1 dimensions is seen by the accelerated frame as a thermal state $\hat{\rho}_0$ with Unruh temperature $T_\text{U} = \hbar \beta/k_\text{B}$. This result proves the Unruh effect for general scalar fields.

\section{Dirac field}\label{Frame_dependent_content_of_particles_Dirac}

\textit{This section is based on and contains material from Ref.~\citeRF{PhysRevD.107.105021}.}

\subsection{Introduction}

The Unruh effect is the prediction that an accelerated observer detects Rindler particles in the Minkowski vacuum \cite{PhysRevD.7.2850, Davies:1974th, PhysRevD.14.870}. The phenomenon was originally studied in the context of scalar fields. As we have seen in Secs.~(\ref{Massless_scalar_field_in_11_spacetime}) and (\ref{Frame_dependent_content_of_particles_scalar}), scalar particles in the accelerated frame are expected to follow the bosonic thermal distribution $(e^{\beta \Omega}-1)^{-1}$, where $\hbar \Omega$ is the particles energy and $\beta = 2 \pi / c a$ is inversely proportional to the acceleration of the observer $\alpha = c^2 a$.

More recent works considered Dirac fields \cite{Oriti, PhysRevD.103.125005}. The result is a fermionic thermal distribution $(e^{\beta \Omega}+1)^{-1}$ for Rindler-Dirac particles in the Minkowski vacuum. Despite these investigations, a complete description of the Minkowski vacuum in terms of Dirac Rindler-Fock states is missing.

The Bogoliubov coefficients relating Minkowski operators to Rindler operators have been derived in Ref.~\cite{Oriti}. Here, we give the explicit algebraic representation of the Minkowski vacuum in the Rindler frame and its thermal representation in one wedge.

We rederive the Bogoliubov transformation relating the Minkowski operators to the Rindler operators and we derive the Minkowski vacuum as an element of the Rindler-Fock space. We, hence, use the Bogoliubov coefficients to give a complete description of the Minkowski vacuum in the Rindler spacetime.

We obtain different Rindler-Fock representations depending of the choice for the spin basis in each wedge. The dependence of the spin basis is due to the presence of a spin coupling between modes of opposite wedges.

We also derive the statistical operator describing the Minkowski vacuum seen by the accelerated observer. We compute the partial trace with respect to the left wedge $\text{Tr}_\text{L}(|0_\text{M} \rangle \langle 0_\text{M} |)$ by adopting a many-body approach for Dirac particles. The result is a fermionic thermal state that completely describes the Minkowski vacuum in the right Rindler spacetime.

The section is organized as follows. In Sec.~\ref{Bogoliubov_transformation}, we compute the Bogoliubov transformations relating Minkowski to Rindler operators. The Bogoliubov coefficients are then used in Sec.~\ref{Minkowski_vacuum_in_Rindler_spacetime} to give the representation of the Minkowski vacuum in the Rindler spacetime. We compute the partial trace with respect to the left wedge and obtain the fermionic thermal state in Sec.~\ref{Unruh_effect_for_Dirac_fields}. In Sec.~\ref{Basis_choice} we discuss the dependence of the results with respect to the spin basis choice. Conclusions are drawn in Sec.~\ref{Frame_dependent_content_of_particles_Dirac_Conclusions}.

\subsection{Bogoliubov transformation}\label{Bogoliubov_transformation}

In Sec.~\ref{QFT_in_Minkowski_spacetime_Dirac} and in Sec.~\ref{Rindler_Dirac_modes} we considered the Minkowski $(t,\vec{x})$ and the Rindler $(T,\vec{X})$ spacetimes and we studied the respective Dirac fields $\hat{\psi}(t,\vec{x})$ and $\hat{\Psi}_\nu (T,\vec{X})$. We defined the operators $\hat{c}_s(\vec{k})$, $\hat{d}_s(\vec{k})$, $\hat{C}_{\nu s}(\Omega,\vec{K}_\perp)$ and $\hat{D}_{\nu s}(\Omega,\vec{K}_\perp)$ as the annihilators of positive and negative frequency modes for each spacetime.

Here, we consider both the Minkowski $(t,\vec{x})$ and the Rindler $(T,\vec{X})$ spacetimes to describe the inertial and the accelerated frame of a flat spacetime. The operators $\hat{\psi}(t,\vec{x})$ and $\hat{\Psi}_\nu (T,\vec{X})$ define the same Dirac field in each coordinate system. We compute the Bogoliubov transformation relating Minkowski ($\hat{c}_s(\vec{k})$ and $\hat{d}_s(\vec{k})$) and Rindler ($\hat{C}_{\nu s}(\Omega,\vec{K}_\perp)$ and $\hat{D}_{\nu s}(\Omega,\vec{K}_\perp)$) operators. We follow steps \ref{Bogoliubov_step_1}-\ref{Bogoliubov_step_4} of Sec.~\ref{Frame_dependent_content_of_particles_Introduction} and the same method presented in Secs.~\ref{Massless_scalar_field_in_11_spacetime} and \ref{Frame_dependent_content_of_particles_scalar} for scalar fields. A
different approach can be found in \cite{Oriti}.

Equation (\ref{D_scalar_product_orthonormality_u_v}) can be used to invert Eq.~(\ref{free_Dirac_field}) as
\begin{align}\label{Bogoliubov_transformations_1_Rindler}
& \hat{c}_s(\vec{k}) = ( u_s(\vec{k}), \hat{\psi})_{\mathbb{C}^4 \otimes L^2_\text{D}(\mathbb{R}^3)}, & \hat{d}_s^\dagger(\vec{k}) = (v_s(\vec{k}),  \hat{\psi})_{\mathbb{C}^4 \otimes L^2_\text{D}(\mathbb{R}^3)}.
\end{align}
Equation (\ref{Bogoliubov_transformations_1_Rindler}) explicitly reads as
\begin{align}\label{Bogoliubov_transformations_2_Rindler}
& \hat{c}_s(\vec{k}) = \int_{\mathbb{R}^3} d^3x u^\dagger_s(\vec{k},t,\vec{x}) \hat{\psi}(t,\vec{x}), & \hat{d}_s^\dagger(\vec{k}) = \int_{\mathbb{R}^3} d^3x v^\dagger_s(\vec{k},t,\vec{x}) \hat{\psi}(t,\vec{x}) .
\end{align}

The Dirac field transforms as a spinor field under diffeomorphisms [Eq.~(\ref{scalar_transformation_Rindler_inverse_Dirac})]. When $t=0$, the transformation (\ref{scalar_transformation_Rindler_inverse_Dirac}) reads as
\begin{equation}\label{scalar_transformation_Rindler_inverse_Dirac_2}
\hat{\psi}(0,\vec{x}) = \sum_{\nu=\{\text{L},\text{R}\}} \theta(s_\nu z) \hat{\Psi}_\nu(0,\vec{X}_\nu(\vec{x})),
\end{equation}
with $\vec{X}_\nu(\vec{x}) = \vec{X}_\nu(0, \vec{x})$. By choosing $t=0$ and using Eq.~(\ref{scalar_transformation_Rindler_inverse_Dirac_2}) in Eq.~(\ref{Bogoliubov_transformations_2_Rindler}) one obtains
\begin{subequations}\label{Bogoliubov_transformations_3_Rindler_Dirac}
\begin{align}
\hat{c}_s(\vec{k}) = & \sum_{\nu=\{\text{L},\text{R}\}}  \int_{\mathbb{R}^3} d^3x \theta(s_\nu z) u^\dagger_s(\vec{k},0,\vec{x}) \hat{\Psi}_\nu(0,\vec{X}_\nu(\vec{x})),\label{Bogoliubov_transformations_3_Rindler_a} \\
\hat{d}_s^\dagger(\vec{k}) = & \sum_{\nu=\{\text{L},\text{R}\}}  \int_{\mathbb{R}^3} d^3x \theta(s_\nu z) v^\dagger_s(\vec{k},0,\vec{x}) \hat{\Psi}_\nu(0,\vec{X}_\nu(\vec{x}))\label{Bogoliubov_transformations_3_Rindler_b}.
\end{align}
\end{subequations}

By plugging Eq.~(\ref{free_Dirac_field_Rindler_2}) in Eq.~(\ref{Bogoliubov_transformations_3_Rindler_Dirac}) one is able to related the Minkowski operators $\hat{c}_s(\vec{k})$ and $\hat{d}_s(\vec{k})$ to the Rindler operators $\hat{C}_{\nu s}(\Omega,\vec{K}_\perp)$ and $\hat{D}_{\nu s}(\Omega,\vec{K}_\perp)$ through the Bogoliubov transformation
\begin{subequations}\label{Bogoliubov_transformations_3_Rindler_2}
\begin{align}
\hat{c}_s(\vec{k}) = & \sum_{\nu=\{\text{L},\text{R}\}}  \sum_{s'=1}^2 \int_\mathbb{R} d\Omega \int_{\mathbb{R}^2} d^2 K_\perp  \nonumber \\
& \times  \int_{\mathbb{R}^3} d^3x \theta(s_\nu z) u^\dagger_s(\vec{k},0,\vec{x}) W_{\nu s'}(\Omega,\vec{K}_\perp,0,\vec{X}_\nu(\vec{x}))  \nonumber \\
& \times \left[ \theta(\Omega) \hat{C}_{\nu s'}(\Omega,\vec{K}_\perp)+  \theta(-\Omega)\hat{D}_{\nu s'}^\dagger(-\Omega,-\vec{K}_\perp) \right], \\
\hat{d}_s^\dagger(\vec{k}) = & \sum_{\nu=\{\text{L},\text{R}\}}  \sum_{s'=1}^2 \int_\mathbb{R} d\Omega \int_{\mathbb{R}^2} d^2 K_\perp    \nonumber \\
& \times  \int_{\mathbb{R}^3} d^3x \theta(s_\nu z) v^\dagger_s(\vec{k},0,\vec{x}) W_{\nu s'}(-\Omega,-\vec{K}_\perp,0,\vec{X}_\nu(\vec{x}))\nonumber \\
& \times \left[ \theta(-\Omega) \hat{C}_{\nu s'}(-\Omega,-\vec{K}_\perp)+  \theta(\Omega)\hat{D}_{\nu s'}^\dagger(\Omega,\vec{K}_\perp) \right].
\end{align}
\end{subequations}
By using Eqs.~(\ref{free_Dirac_field_modes})\footnote{In Sec.~\ref{QFT_in_Minkowski_spacetime_Dirac}, we chose the basis of particles with defined spin along one direction. A basis of modes with the same property is not available in the Rindler spacetime. Indeed, the translational symmetry with respect to the direction of the acceleration is absent, and particles with defined energy do not have defined momentum component along such a direction. Hence, no Lorentz boost leads to the comoving frame of these particles. For this reason, Rindler-Dirac modes with defined frequency and spin cannot be considered. Since our aim is to relate Minkowski modes with Rindler modes, there is no reason to prefer the basis with defined spin. Hereafter, we consider the general solutions of Eq.~(\ref{Dirac_uv_tilde}) and we do not choose any particular basis for the spin degrees of freedom.} and (\ref{UVW_UVW_tilde}) and by performing the integration with respect to $x$ and $y$, the Bogoliubov transformation (\ref{Bogoliubov_transformations_3_Rindler_2}) becomes
\begin{subequations}\label{Bogoliubov_transformations_3_Rindler_3}
\begin{align}
& \hat{c}_s(\vec{k}) = \sqrt{2\pi} \sum_{\nu=\{\text{L},\text{R}\}}  \sum_{s'=1}^2 \int_\mathbb{R} d\Omega \int_{\mathbb{R}^2} d^2 K_\perp  \delta^2(\vec{k}_\perp - \vec{K}_\perp)  \int_{\mathbb{R}} dz \theta(s_\nu z) e^{ - ik_3 z } \nonumber \\
& \times \tilde{u}^\dagger_s(\vec{k}) \tilde{W}_{\nu s'}(\Omega,\vec{k}_\perp,Z_\nu(z)) \left[ \theta(\Omega) \hat{C}_{\nu s'}(\Omega,\vec{K}_\perp)+  \theta(-\Omega)\hat{D}_{\nu s'}^\dagger(-\Omega,-\vec{K}_\perp) \right] , \label{Bogoliubov_transformations_3_Rindler_3_a}\\
& \hat{d}_s^\dagger(\vec{k}) = \sqrt{2\pi} \sum_{\nu=\{\text{L},\text{R}\}}  \sum_{s'=1}^2 \int_\mathbb{R} d\Omega \int_{\mathbb{R}^2} d^2 K_\perp  \delta^2(\vec{k}_\perp - \vec{K}_\perp)   \int_{\mathbb{R}} dz \theta(s_\nu z)e^{ ik_3 z } \nonumber \\
& \times \tilde{v}^\dagger_s(\vec{k}) \tilde{W}_{\nu s'}(-\Omega,-\vec{k}_\perp,Z_\nu(z)) \left[ \theta(-\Omega) \hat{C}_{\nu s'}(-\Omega,-\vec{K}_\perp)+  \theta(\Omega)\hat{D}_{\nu s'}^\dagger(\Omega,\vec{K}_\perp) \right]. \label{Bogoliubov_transformations_3_Rindler_3_b}
\end{align}
\end{subequations}

To obtain $\tilde{u}^\dagger_s(\vec{k}) \tilde{W}_{\nu s'}(\Omega,\vec{k}_\perp,Z_\nu(z))$ and $\tilde{v}^\dagger_s(\vec{k}) \tilde{W}_{\nu s'}(-\Omega,-\vec{k}_\perp,Z_\nu(z))$ appearing in Eqs.~(\ref{Bogoliubov_transformations_3_Rindler_3_a}) and (\ref{Bogoliubov_transformations_3_Rindler_3_b}), we compute $\tilde{u}^\dagger_s(\vec{k}) \mathfrak{G}_\nu(\vec{k}_\perp)\mathfrak{W}_{\nu s'}(\Omega,\vec{k}_\perp)$ and $\tilde{v}^\dagger_s(\vec{k}) \mathfrak{G}_\nu(-\vec{k}_\perp)\mathfrak{W}_{\nu s'}(-\Omega,-\vec{k}_\perp)$. The former can be obtained by using Eqs.~(\ref{gamma_matrices_identities}), (\ref{Dirac_u_tilde}), (\ref{Gamma}), (\ref{W_p_eigenvalues}) and (\ref{G_antihermitian})
\begin{align} \label{u_tilde_Gamma_W_tilde}
& \tilde{u}^\dagger_s(\vec{k}) \mathfrak{G}_\nu(\vec{k}_\perp)\mathfrak{W}_{\nu s'}(\Omega,\vec{k}_\perp)\nonumber \\
 = & - \left[ \mathfrak{G}_\nu(\vec{k}_\perp) \tilde{u}_s(\vec{k}) \right]^\dagger \mathfrak{W}_{\nu s'}(\Omega,\vec{k}_\perp) \nonumber \\
 = &\frac{- s_\nu i c} {\kappa (\vec{k}_\perp)} \left[ \gamma^0 \left( k_1 \gamma^1 + k_2 \gamma^2 + \frac{m c}{\hbar} \right) \tilde{u}_s(\vec{k}) \right]^\dagger \mathfrak{W}_{\nu s'}(\Omega, \vec{k}_\perp) \nonumber \\
 = &\frac{-s_\nu i c} {\kappa (\vec{k}_\perp)} \left\lbrace \gamma^0 \left[ \omega(\vec{k}) \gamma^0 - k_3 \gamma^3 \right] \tilde{u}_s(\vec{k}) \right\rbrace^\dagger \mathfrak{W}_{\nu s'}(\Omega, \vec{k}_\perp)\nonumber \\
 = &\frac{-s_\nu i} {\kappa (\vec{k}_\perp)} \left\lbrace \left[ \frac{\omega(\vec{k})}{c} -  c k_3 \gamma^0 \gamma^3 \right] \tilde{u}_s(\vec{k}) \right\rbrace^\dagger \mathfrak{W}_{\nu s'}(\Omega, \vec{k}_\perp)\nonumber \\
 = &\frac{-s_\nu i} {\kappa (\vec{k}_\perp)}  \tilde{u}^\dagger_s(\vec{k}) \left[ \frac{\omega(\vec{k})}{c} - c k_3 \gamma^0 \gamma^3  \right] \mathfrak{W}_{\nu s'}(\Omega, \vec{k}_\perp) \nonumber \\
 = &\frac{-s_\nu i} {\kappa (\vec{k}_\perp)}  \left[ \frac{\omega(\vec{k})}{c} - k_3 \right] \tilde{u}^\dagger_s(\vec{k}) \mathfrak{W}_{\nu s'}(\Omega, \vec{k}_\perp).
\end{align}
Similarly, for the second scalar product one can use Eq.~(\ref{Dirac_v_tilde}) instead of Eq.~(\ref{Dirac_u_tilde})
\begin{align} \label{v_tilde_Gamma_W_tilde}
& \tilde{v}^\dagger_s(\vec{k}) \mathfrak{G}_\nu(-\vec{k}_\perp)\mathfrak{W}_{\nu s'}(-\Omega,-\vec{k}_\perp) \nonumber \\
= & - \left[ \mathfrak{G}_\nu(-\vec{k}_\perp) \tilde{v}_s(\vec{k}) \right]^\dagger \mathfrak{W}_{\nu s'}(-\Omega,-\vec{k}_\perp) \nonumber \\
 = &\frac{s_\nu i c} {\kappa (\vec{k}_\perp)} \left[ \gamma^0 \left( k_1 \gamma^1 + k_2 \gamma^2 - \frac{m c}{\hbar} \right) \tilde{v}_s(\vec{k}) \right]^\dagger  \mathfrak{W}_{\nu s'}(-\Omega, -\vec{k}_\perp) \nonumber \\
 = &\frac{s_\nu i c} {\kappa (\vec{k}_\perp)} \left\lbrace \gamma^0 \left[ \omega(\vec{k}) \gamma^0 - k_3 \gamma^3 \right] \tilde{v}_s(\vec{k}) \right\rbrace^\dagger \mathfrak{W}_{\nu s'}(-\Omega, -\vec{k}_\perp)\nonumber \\
 = &\frac{s_\nu i} {\kappa (\vec{k}_\perp)} \left\lbrace \left[ \frac{\omega(\vec{k})}{c} -  c k_3 \gamma^0 \gamma^3 \right] \tilde{v}_s(\vec{k}) \right\rbrace^\dagger \mathfrak{W}_{\nu s'}(-\Omega, -\vec{k}_\perp)\nonumber \\
 = &\frac{s_\nu i} {\kappa (\vec{k}_\perp)}  \tilde{v}^\dagger_s(\vec{k}) \left[ \frac{\omega(\vec{k})}{c} - c k_3 \gamma^0 \gamma^3  \right] \mathfrak{W}_{\nu s'}(-\Omega, -\vec{k}_\perp) \nonumber \\
 = &\frac{s_\nu i} {\kappa (\vec{k}_\perp)}  \left[ \frac{\omega(\vec{k})}{c} - k_3 \right] \tilde{v}^\dagger_s(\vec{k}) \mathfrak{W}_{\nu s'}(-\Omega, -\vec{k}_\perp).
\end{align}

By using Eq.~(\ref{theta_k}), Eqs.~(\ref{u_tilde_Gamma_W_tilde}) and (\ref{v_tilde_Gamma_W_tilde}) turn into
\begin{subequations}\label{uv_tilde_Gamma_W_tilde_2}
\begin{align}
& \tilde{u}^\dagger_s(\vec{k}) \mathfrak{G}_\nu(\vec{k}_\perp)\mathfrak{W}_{\nu s'}(\Omega,\vec{k}_\perp)  = \exp \left(- s_\nu i \frac{\pi}{2} -\vartheta(\vec{k}) \right)   \tilde{u}^\dagger_s(\vec{k})  \mathfrak{W}_{\nu s'}(\Omega, \vec{k}_\perp), \\
& \tilde{v}^\dagger_s(\vec{k}) \mathfrak{G}_\nu(-\vec{k}_\perp)\mathfrak{W}_{\nu s'}(-\Omega,-\vec{k}_\perp)  = \exp \left(s_\nu i \frac{\pi}{2} -\vartheta(\vec{k}) \right)  \tilde{v}^\dagger_s(\vec{k}) \mathfrak{W}_{\nu s'}(-\Omega, -\vec{k}_\perp),
\end{align}
\end{subequations}
which means that
\begin{subequations}\label{uv_tilde_Gamma_W_tilde_3}
\begin{align}
\tilde{u}^\dagger_s(\vec{k}) \left[ \mathfrak{G}_\nu(\vec{k}_\perp) \right]^{(1 - \sigma)/2} \mathfrak{W}_{\nu s'}(\Omega,\vec{k}_\perp) = & \exp \left( \frac{\sigma-1}{2} \left[s_\nu i \frac{\pi}{2} +\vartheta(\vec{k})\right] \right) \nonumber \\
 & \times \tilde{u}^\dagger_s(\vec{k})  \mathfrak{W}_{\nu s'}(\Omega, \vec{k}_\perp), \\
\tilde{v}^\dagger_s(\vec{k})\left[ \mathfrak{G}_\nu(-\vec{k}_\perp) \right]^{(1 - \sigma)/2} \mathfrak{W}_{\nu s'}(-\Omega,-\vec{k}_\perp)  = & \exp \left(\frac{\sigma-1}{2} \left[-s_\nu i \frac{\pi}{2} +\vartheta(\vec{k})\right]  \right)   \nonumber \\
 & \times \tilde{v}^\dagger_s(\vec{k}) \mathfrak{W}_{\nu s'}(-\Omega, -\vec{k}_\perp).
\end{align}
\end{subequations}
Equations (\ref{K_conjugate}), (\ref{W_tilde_W_tilde}) and (\ref{uv_tilde_Gamma_W_tilde_3}) and the fact that $\kappa (\vec{K}_\perp)$ and $ \mathfrak{K} (\Omega,\vec{K}_\perp,Z)$ are even with respect to $\vec{K}_\perp$ [Eqs.~(\ref{kappa_k_perp}) and (\ref{K_pm})] allow to compute the scalar products
\begin{subequations}\label{uv_tilde_Gamma_W_tilde_4}
\begin{align}
& \tilde{u}^\dagger_s(\vec{k}) \tilde{W}_{\nu s'}(\Omega,\vec{k}_\perp,Z_\nu(z))  = \sum_{\sigma=\pm}  \mathfrak{K} (\sigma s_\nu \Omega,\vec{k}_\perp,Z_\nu(z))  \nonumber \\
 &\times  \exp \left( \frac{\sigma-1}{2} \left[s_\nu i \frac{\pi}{2} +\vartheta(\vec{k})\right] \right)  \tilde{u}^\dagger_s(\vec{k})  \mathfrak{W}_{\nu s'}(\Omega, \vec{k}_\perp), \\
& \tilde{v}^\dagger_s(\vec{k}) \tilde{W}_{\nu s'}(-\Omega,-\vec{k}_\perp,Z_\nu(z)) = \sum_{\sigma=\pm} \mathfrak{K}^* (\sigma s_\nu \Omega,\vec{k}_\perp,Z_\nu(z))  \nonumber \\
 &\times \exp \left(\frac{\sigma-1}{2} \left[-s_\nu i \frac{\pi}{2} +\vartheta(\vec{k})\right]  \right)  \tilde{v}^\dagger_s(\vec{k}) \mathfrak{W}_{\nu s'}(-\Omega, -\vec{k}_\perp).
\end{align}
\end{subequations}

By plugging Eq.~(\ref{uv_tilde_Gamma_W_tilde_4}) in Eq.~(\ref{Bogoliubov_transformations_3_Rindler_3}) and using Eq.~(\ref{W_W_tilde}), one obtains
\begin{subequations}\label{Bogoliubov_transformations_3_Rindler_4}
\begin{align}
\hat{c}_s(\vec{k}) = & \sum_{\nu=\{\text{L},\text{R}\}}  \sum_{s'=1}^2 \int_\mathbb{R} d\Omega \int_{\mathbb{R}^2} d^2 K_\perp    \alpha_{\nu}(\vec{k},\Omega,\vec{K}_\perp) \tilde{u}^\dagger_s(\vec{k})  \tilde{\mathfrak{W}}_{\nu s'}(\Omega, \vec{K}_\perp) \nonumber \\
& \times \left[ \theta(\Omega) \hat{C}_{\nu s'}(\Omega,\vec{K}_\perp)+  \theta(-\Omega)\hat{D}_{\nu s'}^\dagger(-\Omega,-\vec{K}_\perp) \right] ,\label{Bogoliubov_transformations_3_Rindler_4_a}\\
\hat{d}_s^\dagger(\vec{k}) = & \sum_{\nu=\{\text{L},\text{R}\}}  \sum_{s'=1}^2 \int_\mathbb{R} d\Omega \int_{\mathbb{R}^2} d^2 K_\perp    \alpha^*_{\nu}(\vec{k},\Omega,\vec{K}_\perp) \tilde{v}^\dagger_s(\vec{k})  \tilde{\mathfrak{W}}_{\nu s'}(-\Omega, -\vec{K}_\perp) \nonumber \\
& \times \left[ \theta(-\Omega) \hat{C}_{\nu s'}(-\Omega,-\vec{K}_\perp)+  \theta(\Omega)\hat{D}_{\nu s'}^\dagger(\Omega,\vec{K}_\perp) \right],\label{Bogoliubov_transformations_3_Rindler_4_b}
\end{align}
\end{subequations}
with the Bogoliubov coefficient
\begin{align}\label{Bogoliubov_coefficient}
\alpha_{\nu}(\vec{k},\Omega,\vec{K}_\perp) = & \frac{1}{\pi} \delta^2(\vec{k}_\perp - \vec{K}_\perp) \sqrt{\frac{\kappa (\vec{k}_\perp)}{2 \pi c a} \cosh \left( \frac{\beta}{2} \Omega \right)} \exp \left( -s_\nu i \frac{\pi}{4} - \frac{\vartheta(\vec{k})}{2} \right) \nonumber \\
& \times  I_\nu (\vec{k},\Omega)
\end{align}
and with
\begin{subequations}
\begin{align}
& I_\nu (\vec{k},\Omega) = \sum_{\sigma = \pm} \tilde{I}_\nu (\vec{k},\sigma \Omega) \exp \left( \sigma s_\nu i \frac{\pi}{4} + \sigma \frac{\vartheta(\vec{k})}{2} \right), \label{I_I_tilde} \\
&  \tilde{I}_\nu (\vec{k},\Omega) =  \int_{\mathbb{R}} dz \theta(s_\nu z) e^{ - ik_3 z }  \mathfrak{K} (s_\nu \Omega,\vec{k}_\perp, s_\nu Z_\nu(z)). \label{I_tilde}
\end{align}
\end{subequations}

The integral appearing in Eq.~(\ref{I_tilde}) can be computed by considering Eq.~(\ref{Bessel_integral_representation_final}). By replacing the variables $\xi$, $\zeta$ and $\tau$ with, respectively, $s_\nu \kappa(\vec{k}_\perp) z$, $ s_\nu i \Omega/ca - 1/2$ and $s_\nu \vartheta(\vec{k})$ and by dividing the equation with $\kappa(\vec{k}_\perp)$, one obtains
\begin{align}\label{Bessel_integral_representation_6}
& \int_{\mathbb{R}} dz \theta(s_\nu z) \exp( -  i \kappa(\vec{k}_\perp) \sinh(\vartheta(\vec{k})) z ) K_{s_\nu i \Omega / c a - 1/2} \left( s_\nu \kappa (\vec{k}_\perp) z \right)  \nonumber \\
 = &  \frac{\pi \sin \left( s_\nu i \frac{\pi \Omega}{2 c a} - \frac{\pi}{4} +  \frac{\vartheta(\vec{k}) \Omega}{c a} + s_\nu i \frac{\vartheta(\vec{k})}{2}  \right) }{ \kappa(\vec{k}_\perp) \sin\left( s_\nu i \frac{\pi \Omega}{c a} - \frac{\pi}{2}  \right) \cosh(\vartheta(\vec{k})) }.
\end{align}
By using Eqs.~(\ref{K_pm}), (\ref{k_3_theta}), (\ref{I_tilde}) and (\ref{Bessel_integral_representation_6}), one can compute
\begin{align}\label{I_tilde_2}
&  \tilde{I}_\nu(\vec{k},\Omega) \nonumber \\
= & \int_{\mathbb{R}} dz \theta(s_\nu z) \exp( - i \kappa(\vec{k}_\perp) \sinh(\vartheta(\vec{k})) z )   K_{s_\nu i \Omega /ca - 1/2} \left( \kappa (\vec{k}_\perp) \frac{e^{s_\nu a Z_\nu(z)}}{a} \right)  \nonumber \\
= & \int_{\mathbb{R}} dz \theta(s_\nu z) \exp( - i \kappa(\vec{k}_\perp) \sinh(\vartheta(\vec{k})) z )  K_{s_\nu i \Omega /ca - 1/2} \left( s_\nu \kappa (\vec{k}_\perp) z \right)  \nonumber \\
 = &  \frac{\pi \sin \left( s_\nu i \frac{\beta \Omega}{4} - \frac{\pi}{4} +  \frac{\vartheta(\vec{k}) \Omega}{c a} + s_\nu i \frac{\vartheta(\vec{k})}{2}  \right)}{\kappa(\vec{k}_\perp) \sin\left( s_\nu i \frac{\beta \Omega}{2} - \frac{\pi}{2}  \right) \cosh(\vartheta(\vec{k})) } \nonumber \\
= & - \frac{s_\nu \pi \sin \left( i \frac{\beta \Omega}{4} - s_\nu \frac{\pi}{4} + s_\nu \frac{\vartheta(\vec{k}) \Omega}{c a} + i \frac{\vartheta(\vec{k})}{2} \right) }{\kappa(\vec{k}_\perp) \cosh\left( \frac{\beta \Omega}{2}  \right) \cosh(\vartheta(\vec{k})) }.
\end{align}
By plugging Eq.~(\ref{I_tilde_2}) in Eq.~(\ref{I_I_tilde}) one obtains
\begin{align}\label{I}
& I_\nu (\vec{k},\Omega) \nonumber \\
= &- s_\nu \pi \left[ \kappa(\vec{k}_\perp) \cosh\left( \frac{\beta \Omega}{2}  \right) \cosh(\vartheta(\vec{k})) \right]^{-1} \sum_{\sigma = \pm} \exp \left( \sigma s_\nu i \frac{\pi}{4} + \sigma \frac{\vartheta(\vec{k})}{2} \right) \nonumber \\
& \times \sin \left( \sigma i \frac{\beta \Omega}{4} - s_\nu \frac{\pi}{4} + \sigma s_\nu \frac{\vartheta(\vec{k}) \Omega}{c a} + i \frac{\vartheta(\vec{k})}{2} \right) \nonumber \\
= &s_\nu i \pi \left[ 2 \kappa(\vec{k}_\perp)  \cosh\left( \frac{\beta \Omega}{2}  \right) \cosh(\vartheta(\vec{k})) \right]^{-1} \nonumber \\
& \times  \sum_{\sigma = \pm} \left[ \exp \left( s_\nu i \frac{\sigma - 1}{4} \pi  + \frac{\sigma-1}{2} \vartheta(\vec{k})  - \sigma \frac{\beta \Omega}{4} + \sigma s_\nu i \frac{\vartheta(\vec{k}) \Omega}{c a}\right) \right.\nonumber \\
&  \left. - \exp \left( s_\nu i \frac{\sigma + 1}{4} \pi+ \frac{\sigma+1}{2} \vartheta(\vec{k})  + \sigma  \frac{\beta \Omega}{4}  - \sigma s_\nu i \frac{\vartheta(\vec{k}) \Omega}{c a} \right)  \right]\nonumber \\
= & s_\nu i \pi \left[ 2 \kappa(\vec{k}_\perp) \cosh\left( \frac{\beta \Omega}{2}  \right) \cosh(\vartheta(\vec{k})) \right]^{-1}  \left[ - \exp \left(  s_\nu i \frac{\pi}{2} + \vartheta(\vec{k}) +  \frac{\beta \Omega}{4} \right. \right.\nonumber \\
& \left. \left. - s_\nu i \frac{\vartheta(\vec{k}) \Omega}{c a}  \right) + \exp \left( - s_\nu i \frac{\pi}{2} - \vartheta(\vec{k}) + \frac{\beta \Omega}{4} - s_\nu i  \frac{\vartheta(\vec{k}) \Omega}{c a} \right)\right] \nonumber \\
= &  s_\nu i \pi \left[ 2 \kappa(\vec{k}_\perp)  \cosh\left( \frac{\beta \Omega}{2}  \right) \cosh(\vartheta(\vec{k})) \right]^{-1}  \exp \left(  \frac{\beta \Omega}{4} - s_\nu i  \frac{\vartheta(\vec{k}) \Omega}{c a} \right)   \nonumber \\
 &  \times    \left[ - s_\nu i e^{\vartheta(\vec{k})}  - s_\nu i e^{-\vartheta(\vec{k})} \right] \nonumber \\
= &  \pi \left[ \kappa(\vec{k}_\perp)  \cosh\left( \frac{\beta \Omega}{2}  \right) \right]^{-1}   \exp \left(  \frac{\beta \Omega}{4} - s_\nu i  \frac{\vartheta(\vec{k}) \Omega}{c a} \right).
\end{align}

Equation (\ref{I}) can be used in Eq.~(\ref{Bogoliubov_coefficient}) to obtain the final expression for the Bogoliubov coefficients
\begin{equation}\label{Bogoliubov_coefficient_2}
\alpha_{\nu}(\vec{k},\Omega,\vec{K}_\perp) = \delta^2(\vec{k}_\perp - \vec{K}_\perp) \frac{\exp \left( - s_\nu i \frac{\pi}{4} - \frac{\vartheta(\vec{k})}{2} + \frac{\beta \Omega}{4} - s_\nu i  \frac{\vartheta(\vec{k}) \Omega}{c a} \right)}{\sqrt{2 \pi c a \kappa (\vec{k}_\perp) \cosh \left( \frac{\beta}{2} \Omega \right)}}.
\end{equation}
By using the fact that $s_{\bar{\nu}}=-s_\nu$, Eq.~(\ref{Bogoliubov_coefficient_2}) leads to
\begin{align}\label{alpha_beta_Unruh}
 \alpha_{\bar{\nu}}(\vec{k},-\Omega,\vec{K}_\perp) = & s_\nu i e^{-\beta \Omega / 2}  \alpha_{\nu}(\vec{k},\Omega,\vec{K}_\perp),
\end{align} 
which can be used in Eq.~(\ref{Bogoliubov_transformations_3_Rindler_4}) to relate operators of opposite frequency and wedge. By inverting the variables $\Omega \mapsto -\Omega$ and $\nu \mapsto \bar{\nu}$ when  $\Omega < 0$, Eq.~(\ref{Bogoliubov_transformations_3_Rindler_4}) becomes
\begin{subequations}\label{Bogoliubov_transformations_3_Rindler_5}
\begin{align}
\hat{c}_s(\vec{k}) = & \sum_{\nu=\{\text{L},\text{R}\}}  \sum_{s'=1}^2 \int_0^\infty d\Omega \int_{\mathbb{R}^2} d^2 K_\perp   \nonumber \\
& \times  \tilde{u}^\dagger_s(\vec{k}) \left[ \alpha_{\nu}(\vec{k},\Omega,\vec{K}_\perp) \tilde{\mathfrak{W}}_{\nu s'}(\Omega, \vec{K}_\perp) \hat{C}_{\nu s'}(\Omega,\vec{K}_\perp) \right. \nonumber \\
&  \left. +  \alpha_{\bar{\nu}}(\vec{k},-\Omega,\vec{K}_\perp) \tilde{\mathfrak{W}}_{\bar{\nu} s'}(-\Omega, \vec{K}_\perp) \hat{D}_{\bar{\nu} s'}^\dagger(\Omega,-\vec{K}_\perp) \right] ,\\
\hat{d}_s^\dagger(\vec{k}) = & \sum_{\nu=\{\text{L},\text{R}\}}  \sum_{s'=1}^2 \int_0^\infty d\Omega \int_{\mathbb{R}^2} d^2 K_\perp   \nonumber \\
& \times   \tilde{v}^\dagger_s(\vec{k}) \left[ \alpha^*_{\bar{\nu}}(\vec{k},-\Omega,\vec{K}_\perp) \tilde{\mathfrak{W}}_{\bar{\nu} s'}(\Omega, -\vec{K}_\perp)  \hat{C}_{\bar{\nu} s'}(\Omega,-\vec{K}_\perp)\right. \nonumber \\
&  \left. +  \alpha^*_{\nu}(\vec{k},\Omega,\vec{K}_\perp) \tilde{\mathfrak{W}}_{\nu s'}(-\Omega, -\vec{K}_\perp) \hat{D}_{\nu s'}^\dagger(\Omega,\vec{K}_\perp) \right].
\end{align}
\end{subequations}
By plugging Eq.~(\ref{alpha_beta_Unruh}) in Eq.~(\ref{Bogoliubov_transformations_3_Rindler_5}),
\begin{subequations}\label{Bogoliubov_transformations_3_Rindler_6}
\begin{align}
\hat{c}_s(\vec{k}) = & \sum_{\nu=\{\text{L},\text{R}\}}  \sum_{s'=1}^2 \int_0^\infty d\Omega \int_{\mathbb{R}^2} d^2 K_\perp   \alpha_{\nu}(\vec{k},\Omega,\vec{K}_\perp) \nonumber \\
& \times  \tilde{u}^\dagger_s(\vec{k}) \left[ \tilde{\mathfrak{W}}_{\nu s'}(\Omega, \vec{K}_\perp) \hat{C}_{\nu s'}(\Omega,\vec{K}_\perp) \right. \nonumber \\
&  \left. +  s_\nu i e^{-\beta \Omega / 2}  \tilde{\mathfrak{W}}_{\bar{\nu} s'}(-\Omega, \vec{K}_\perp) \hat{D}_{\bar{\nu} s'}^\dagger(\Omega,-\vec{K}_\perp) \right] , \\
\hat{d}_s^\dagger(\vec{k}) = & \sum_{\nu=\{\text{L},\text{R}\}}  \sum_{s'=1}^2 \int_0^\infty d\Omega \int_{\mathbb{R}^2} d^2 K_\perp  \alpha^*_{\nu}(\vec{k},\Omega,\vec{K}_\perp)  \nonumber \\
& \times   \tilde{v}^\dagger_s(\vec{k}) \left[ -s_\nu i e^{-\beta \Omega / 2}  \tilde{\mathfrak{W}}_{\bar{\nu} s'}(\Omega, -\vec{K}_\perp)  \hat{C}_{\bar{\nu} s'}(\Omega,-\vec{K}_\perp)\right. \nonumber \\
&  \left. +   \tilde{\mathfrak{W}}_{\nu s'}(-\Omega, -\vec{K}_\perp) \hat{D}_{\nu s'}^\dagger(\Omega,\vec{K}_\perp) \right].
\end{align}
\end{subequations}
Finally, by using Eq.~(\ref{W_L_W_R}),
\begin{subequations}\label{Bogoliubov_transformations_3_Rindler_7}
\begin{align}
\hat{c}_s(\vec{k}) = & \sum_{\nu=\{\text{L},\text{R}\}}  \sum_{s'=1}^2 \int_0^\infty d\Omega \int_{\mathbb{R}^2} d^2 K_\perp   \alpha_{\nu}(\vec{k},\Omega,\vec{K}_\perp) \nonumber \\
& \times  \tilde{u}^\dagger_s(\vec{k}) \tilde{\mathfrak{W}}_{\nu s'}(\Omega, \vec{K}_\perp) \left[ \hat{C}_{\nu s'}(\Omega,\vec{K}_\perp) \right. \nonumber \\
&  \left. +  s_\nu i e^{-\beta \Omega / 2}  \sum_{s''=1}^2 M_{\nu s''s'}(\Omega, \vec{K}_\perp) \hat{D}_{\bar{\nu} s''}^\dagger(\Omega,-\vec{K}_\perp) \right] ,\label{Bogoliubov_transformations_3_Rindler_7_a} \\
\hat{d}_s^\dagger(\vec{k}) = & \sum_{\nu=\{\text{L},\text{R}\}}  \sum_{s'=1}^2 \int_0^\infty d\Omega \int_{\mathbb{R}^2} d^2 K_\perp  \alpha^*_{\nu}(\vec{k},\Omega,\vec{K}_\perp)  \nonumber \\
& \times   \tilde{v}^\dagger_s(\vec{k})\tilde{\mathfrak{W}}_{\nu s'}(-\Omega, -\vec{K}_\perp) \left[   \hat{D}_{\nu s'}^\dagger(\Omega,\vec{K}_\perp)  \right. \nonumber \\
&  \left. -s_\nu i e^{-\beta \Omega / 2}  \sum_{s''=1}^2 M_{\nu s''s'}(-\Omega, -\vec{K}_\perp)  \hat{C}_{\bar{\nu} s''}(\Omega,-\vec{K}_\perp)\right].\label{Bogoliubov_transformations_3_Rindler_7_b}
\end{align}
\end{subequations}

From the definition of $M_{\nu ss'}(\Omega, \vec{K}_\perp)$ [Eq.~(\ref{M})], one can compute its complex conjugate, i.e.,
\begin{equation}\label{M_conjugate}
M^*_{\nu ss'}(\Omega, \vec{K}_\perp) = M_{\bar{\nu} s's}(-\Omega, \vec{K}_\perp).
\end{equation}
By using Eq.~(\ref{M_conjugate}), one can also conjugate Eq.~(\ref{Bogoliubov_transformations_3_Rindler_7_b}) to obtain
\begin{align}\label{Bogoliubov_transformations_3_Rindler_7_b_conjugate}
\hat{d}_s(\vec{k}) = & \sum_{\nu=\{\text{L},\text{R}\}}  \sum_{s'=1}^2 \int_0^\infty d\Omega \int_{\mathbb{R}^2} d^2 K_\perp  \alpha_{\nu}(\vec{k},\Omega,\vec{K}_\perp)  \nonumber \\
& \times   \tilde{\mathfrak{W}}^\dagger_{\nu s'}(-\Omega, -\vec{K}_\perp) \tilde{v}_s(\vec{k})\left[   \hat{D}_{\nu s'}(\Omega,\vec{K}_\perp)  \right. \nonumber \\
&  \left. + s_\nu i e^{-\beta \Omega / 2}  \sum_{s''=1}^2 M_{\bar{\nu} s's''}(\Omega, -\vec{K}_\perp)  \hat{C}^\dagger_{\bar{\nu} s''}(\Omega,-\vec{K}_\perp)\right].
\end{align}

In conclusion, we computed the Bogoliubov transformations relating Minkowski and Rindler operators [Eq.~(\ref{Bogoliubov_transformations_3_Rindler_4})]. The explicit form of the Bogoliubov coefficient $\alpha_{\nu}(\vec{k},\Omega,\vec{K}_\perp)$ is reported in Eq.~(\ref{Bogoliubov_coefficient_2}). The symmetry between Bogoliubov coefficients of opposite wedge [Eq.~(\ref{alpha_beta_Unruh})] resulted in a coupling between Rindler operators of opposite wedge and frequency in the Bogoliubov transformation [Eqs.~(\ref{Bogoliubov_transformations_3_Rindler_7_a}) and (\ref{Bogoliubov_transformations_3_Rindler_7_b_conjugate})]. A similar coupling has been found for scalar fields, as it can be noticed by comparing Eq.~(\ref{alpha_beta_Unruh}) with Eq.~(\ref{alpha_to_beta}) or Eq.~(\ref{alpha_to_beta_31}). In the next subsection, we will show how this coupling is involved in the Rindler-Fock representation of the Minkowski vacuum, in analogy to the bosonic case.

\subsection{Minkowski vacuum in the left and the right Rindler frame}\label{Minkowski_vacuum_in_Rindler_spacetime}

In Sec.~\ref{Bogoliubov_transformation} we derived the Bogoliubov transformations relating Minkowski and Rindler operators. We obtained an expression in which operators of opposite wedge and frequency are coupled. Here, we will use these transformations to show how the Minkowski vacuum can be represented as an element of the Rindler-Fock space. The proof is similar to the one seen for scalar fields [Secs.~\ref{Minkowski_vacuum_in_Rindler_spacetime_scalar_11} and \ref{Unruh_effect_for_scalar_fields}] and involves the coupling between operators of opposite wedges [Eqs.~(\ref{alpha_to_beta}), (\ref{alpha_to_beta_31}) and (\ref{alpha_beta_Unruh})]. We will obtain two-modes squeezed states where each Rindler mode is paired with the mode with opposite wedge and frequency. The spin degrees of freedom are coupled through the matrix $M_{\nu ss'} (\Omega, \vec{K}_\perp)$ defined in Sec.~\ref{Rindler_Dirac_modes}. Hence, we will obtained different representations depending of the chosen basis $\tilde{\mathfrak{W}}_{\nu s}(\Omega, \vec{K}_\perp)$.

The Minkowski vacuum $|0_\text{M} \rangle $ is defined as the state that is always annihilated by the Minkowski operators $\hat{c}_s(\vec{k}) $ and $\hat{d}_s(\vec{k})$, i.e.,
\begin{align}\label{Minkowski_vacuum_Dirac}
& \hat{c}_s(\vec{k}) |0_\text{M} \rangle = 0, & \hat{d}_s(\vec{k}) |0_\text{M} \rangle = 0,
\end{align}
for any $s$ and $\vec{k}$. Conversely, the Rindler vacuum $ |0_\text{L},0_\text{R} \rangle$ is defined as
\begin{align}\label{Rindler_vacuum}
& \hat{C}_{\nu s}(\Omega,\vec{K}_\perp) |0_\text{L},0_\text{R} \rangle = 0, & \hat{D}_{\nu s}(\Omega,\vec{K}_\perp)|0_\text{L},0_\text{R} \rangle = 0,
\end{align}
for any $\nu$, $s$, $\Omega$ and $\vec{K}_\perp$. Equations (\ref{Minkowski_vacuum_Dirac}) and (\ref{Rindler_vacuum}) are the equivalent of Eqs.~(\ref{Minkowski_vacuum}) and (\ref{Rindler_vacuum_scalar}), respectively, for Dirac fields.

In order to see $|0_\text{M} \rangle$ as an element of the Rindler-Fock space, one has to plug the Bogoliubov transformations (\ref{Bogoliubov_transformations_3_Rindler_7_a}) and (\ref{Bogoliubov_transformations_3_Rindler_7_b_conjugate}) in Eq.~(\ref{Minkowski_vacuum_Dirac}) and look for a Rindler-Fock state such that
\begin{subequations}\label{Minkowski_vacuum_2}
\begin{align}
& \sum_{\nu=\{\text{L},\text{R}\}}  \sum_{s'=1}^2 \int_0^\infty d\Omega \int_{\mathbb{R}^2} d^2 K_\perp   \alpha_{\nu}(\vec{k},\Omega,\vec{K}_\perp)   \tilde{u}^\dagger_s(\vec{k}) \tilde{\mathfrak{W}}_{\nu s'}(\Omega, \vec{K}_\perp) \left[ \hat{C}_{\nu s'}(\Omega,\vec{K}_\perp)  \right. \nonumber \\
&  \left. +  s_\nu i e^{-\beta \Omega / 2} \sum_{s''=1}^2 M_{\nu s''s'}(\Omega, \vec{K}_\perp) \hat{D}_{\bar{\nu} s''}^\dagger(\Omega,-\vec{K}_\perp) \right]  |0_\text{M} \rangle = 0, \\
& \sum_{\nu=\{\text{L},\text{R}\}}  \sum_{s'=1}^2 \int_0^\infty d\Omega \int_{\mathbb{R}^2} d^2 K_\perp  \alpha_{\nu}(\vec{k},\Omega,\vec{K}_\perp)   \tilde{\mathfrak{W}}^\dagger_{\nu s'}(-\Omega, -\vec{K}_\perp) \tilde{v}_s(\vec{k})\left[   \hat{D}_{\nu s'}(\Omega,\vec{K}_\perp)  \right. \nonumber \\
&  \left.  + s_\nu i e^{-\beta \Omega / 2} \sum_{s''=1}^2 M_{\bar{\nu} s's''}(\Omega, -\vec{K}_\perp)  \hat{C}^\dagger_{\bar{\nu} s''}(\Omega,-\vec{K}_\perp)\right] |0_\text{M} \rangle = 0, 
\end{align}
\end{subequations}
for any $s$ and $\vec{k}$. As a consequence of the generality of $s$ and $\vec{k}$, Eq.~(\ref{Minkowski_vacuum_2}) is equivalent to
\begin{subequations}\label{Minkowski_vacuum_Rindler}
\begin{align}
& \left[ \hat{C}_{\nu s}(\Omega,\vec{K}_\perp) +  s_\nu i e^{-\beta \Omega / 2} \sum_{s'=1}^2 M_{\nu s's}(\Omega, \vec{K}_\perp)   \hat{D}_{\bar{\nu} s'}^\dagger(\Omega,-\vec{K}_\perp) \right]  |0_\text{M} \rangle = 0,\label{Minkowski_vacuum_Rindler_a}\\
& \left[   \hat{D}_{\nu s}(\Omega,\vec{K}_\perp) + s_\nu i e^{-\beta \Omega / 2}  \sum_{s'=1}^2 M_{\bar{\nu} ss'}(\Omega, -\vec{K}_\perp)  \hat{C}^\dagger_{\bar{\nu} s'}(\Omega,-\vec{K}_\perp)\right] |0_\text{M} \rangle = 0.\label{Minkowski_vacuum_Rindler_b}
\end{align}
\end{subequations}

The solution of Eq.~(\ref{Minkowski_vacuum_Rindler}) is
\begin{equation}\label{0_M_0_LR}
|0_\text{M} \rangle \propto \exp (\hat{s}_\text{D} ) |0_\text{L},0_\text{R} \rangle,
\end{equation}
with operator
\begin{align}\label{O}
\hat{s}_\text{D} = & - i \sum_{\nu=\{\text{L},\text{R}\}} s_\nu \sum_{s=1}^2 \sum_{s'=1}^2 \int_0^{+\infty} d\Omega \int_{\mathbb{R}^2} d^2 K_\perp  e^{-\beta \Omega / 2}   \nonumber \\ 
&  \times  M_{\nu s's} (\Omega, \vec{K}_\perp) \hat{C}^\dagger_{\nu s}(\Omega,\vec{K}_\perp)\hat{D}^\dagger_{\bar{\nu} s'}(\Omega,-\vec{K}_\perp).
\end{align}
Equation (\ref{0_M_0_LR}) can be seen as the equivalent of Eq.~(\ref{Rindler_vacuum_to_Minkowski_SS}) for Dirac fields and will be used in the next subsection to derive the thermal vacuum state. In Appendix \ref{Two_mode_squeezed_state_Frame_dependent_content_of_particles_Dirac}, we discuss how to normalize the state $\exp (\hat{s}_\text{D} ) |0_\text{L},0_\text{R} \rangle$ and obtain the identity
\begin{equation}\label{Rindler_vacuum_to_Minkowski_Dirac}
| 0_\text{M} \rangle = \hat{S}_\text{D} | 0_\text{L}, 0_\text{R} \rangle,
\end{equation}
with
\begin{align}\label{OO}
\hat{S}_\text{D} = & \exp \left( 2 \sum_{\nu=\{\text{L},\text{R}\}} s_\nu \sum_{s=1}^2 \sum_{s'=1}^2 \int_0^{+\infty} d\Omega \int_{\mathbb{R}^2} d^2 K_\perp \zeta(\Omega) \right. \nonumber \\ 
&  \times\left. \left[ - i  M_{\nu s's} (\Omega, \vec{K}_\perp) \hat{C}^\dagger_{\nu s}(\Omega,\vec{K}_\perp)\hat{D}^\dagger_{\bar{\nu} s'}(\Omega,-\vec{K}_\perp) \right]^\text{A} \right).
\end{align}
where $\zeta(\Omega)$ is defined by Eq.~(\ref{zeta_omega}) and the superscript A indicates the antihermitian part. Equation (\ref{Rindler_vacuum_to_Minkowski_Dirac}) is the equivalent of Eq.~(\ref{Rindler_vacuum_to_Minkowski}) for Dirac fields and gives the representation of the Minkowski vacuum in terms of left and right Rindler particles. 

We now provide a proof for Eq.~(\ref{0_M_0_LR}) as the solution of Eq.~(\ref{Minkowski_vacuum_Rindler}). By using the anticommutation properties of the Dirac operators (\ref{anticommutating_rules}), one obtains
\begin{subequations}\label{anticommutating_rules_2}
\begin{align}
& \hat{C}_{\nu s}(\Omega,\vec{K}_\perp) \hat{C}^\dagger_{\nu' s'}(\Omega',\vec{K}'_\perp) \hat{D}^\dagger_{\bar{\nu}' s''}(\Omega',-\vec{K}'_\perp) \nonumber \\
= & \delta_{\nu \nu'} \delta_{ss'} \delta(\Omega-\Omega') \delta^2(\vec{K}_\perp-\vec{K}'_\perp)\hat{D}^\dagger_{\bar{\nu} s''}(\Omega,-\vec{K}_\perp)\nonumber \\
& + \hat{C}^\dagger_{\nu' s'}(\Omega',\vec{K}'_\perp) \hat{D}^\dagger_{\bar{\nu}' s''}(\Omega',-\vec{K}'_\perp) \hat{C}_{\nu s}(\Omega,\vec{K}_\perp), \\
& \hat{D}_{\nu s}(\Omega,\vec{K}_\perp) \hat{C}^\dagger_{\nu' s'}(\Omega',\vec{K}'_\perp) \hat{D}^\dagger_{\bar{\nu}' s''}(\Omega',-\vec{K}'_\perp) \nonumber \\
= & - \delta_{\nu \bar{\nu}'} \delta_{ss''} \delta(\Omega-\Omega') \delta^2(\vec{K}_\perp+\vec{K}'_\perp)\hat{C}^\dagger_{\bar{\nu} s'}(\Omega,-\vec{K}_\perp)\nonumber \\
& + \hat{C}^\dagger_{\nu' s'}(\Omega',\vec{K}'_\perp) \hat{D}^\dagger_{\bar{\nu}' s''}(\Omega',-\vec{K}'_\perp)\hat{D}_{\nu s}(\Omega,\vec{K}_\perp),  \\
& \hat{C}^\dagger_{\nu s}(\Omega,\vec{K}_\perp) \hat{C}^\dagger_{\nu' s'}(\Omega',\vec{K}'_\perp) \hat{D}^\dagger_{\bar{\nu}' s''}(\Omega',-\vec{K}'_\perp) \nonumber \\
= & \hat{C}^\dagger_{\nu' s'}(\Omega',\vec{K}'_\perp) \hat{D}^\dagger_{\bar{\nu}' s''}(\Omega',-\vec{K}'_\perp) \hat{C}^\dagger_{\nu s}(\Omega,\vec{K}_\perp),  \\
 &  \hat{D}^\dagger_{\nu s}(\Omega,\vec{K}_\perp) \hat{C}^\dagger_{\nu' s'}(\Omega',\vec{K}'_\perp) \hat{D}^\dagger_{\bar{\nu}' s''}(\Omega',-\vec{K}'_\perp)  \nonumber \\
= &  \hat{C}^\dagger_{\nu' s'}(\Omega',\vec{K}'_\perp) \hat{D}^\dagger_{\bar{\nu}' s''}(\Omega',-\vec{K}'_\perp) \hat{D}^\dagger_{\nu s}(\Omega,\vec{K}_\perp).
\end{align} 
\end{subequations}
By using Eqs.~(\ref{O}) and (\ref{anticommutating_rules_2}) and the fact that $s_{\bar{\nu}} = - s_\nu$,
\begin{subequations}\label{anticommutating_rules_3}
\begin{align}
 \hat{C}_{\nu s}(\Omega,\vec{K}_\perp)\hat{s}_\text{D} = & -s_\nu i  e^{-\beta \Omega / 2} \sum_{s'=1}^2 M_{\nu s's} (\Omega, \vec{K}_\perp) \hat{D}^\dagger_{\bar{\nu} s'}(\Omega,-\vec{K}_\perp)\nonumber \\
& + \hat{s}_\text{D} \hat{C}_{\nu s}(\Omega,\vec{K}_\perp), \\
 \hat{D}_{\nu s}(\Omega,\vec{K}_\perp)\hat{s}_\text{D}  = & -s_\nu i  e^{-\beta \Omega / 2} \sum_{s'=1}^2 M_{\bar{\nu} ss'} (\Omega, -\vec{K}_\perp) \hat{C}^\dagger_{\bar{\nu} s'}(\Omega,-\vec{K}_\perp) \nonumber \\
& \times +\hat{s}_\text{D}   \hat{D}_{\nu s}(\Omega,\vec{K}_\perp) , \\
  \hat{C}^\dagger_{\nu s}(\Omega,\vec{K}_\perp)\hat{s}_\text{D} = & \hat{s}_\text{D} \hat{C}^\dagger_{\nu s}(\Omega,\vec{K}_\perp), \\
 \hat{D}^\dagger_{\nu s}(\Omega,\vec{K}_\perp)\hat{s}_\text{D}  =&  \hat{s}_\text{D}   \hat{D}^\dagger_{\nu s}(\Omega,\vec{K}_\perp) .
\end{align}
\end{subequations}
Recursively one may prove the following identity from Eq.~(\ref{anticommutating_rules_3})
\begin{subequations}\label{anticommutating_rules_4}
\begin{align}
\hat{C}_{\nu s}(\Omega,\vec{K}_\perp) \hat{s}_\text{D}^n =  & -n s_\nu i e^{-\beta \Omega / 2} \sum_{s'=1}^2 M_{\nu s's} (\Omega, \vec{K}_\perp) \hat{D}^\dagger_{\bar{\nu} s'}(\Omega,-\vec{K}_\perp) \hat{s}_\text{D}^{n-1}  \nonumber \\
&  + \hat{s}_\text{D}^n \hat{C}_{\nu s}(\Omega,\vec{K}_\perp) , \\
\hat{D}_{\nu s}(\Omega,\vec{K}_\perp) \hat{s}_\text{D}^n = & -n s_\nu i  e^{-\beta \Omega / 2}  \sum_{s'=1}^2 M_{\bar{\nu} ss'} (\Omega, -\vec{K}_\perp) \hat{C}^\dagger_{\bar{\nu} s'}(\Omega,-\vec{K}_\perp) \hat{s}_\text{D}^{n-1} \nonumber \\
&  + \hat{s}_\text{D}^n \hat{D}_{\nu s}(\Omega,\vec{K}_\perp),
\end{align}
\end{subequations}
which holds for any $n \in \mathbb{N}$. By acting on the Rindler vacuum state $|0_\text{L},0_\text{R} \rangle$, Eq.~(\ref{anticommutating_rules_4}) leads to
\begin{subequations}\label{anticommutating_rules_5}
\begin{align}
& \hat{C}_{\nu s}(\Omega,\vec{K}_\perp) \hat{s}_\text{D}^n |0_\text{L},0_\text{R} \rangle = -n s_\nu i e^{-\beta \Omega / 2} \nonumber \\
 & \times \sum_{s'=1}^2 M_{\nu s's} (\Omega, \vec{K}_\perp)  \hat{D}^\dagger_{\bar{\nu} s'}(\Omega,-\vec{K}_\perp) \hat{s}_\text{D}^{n-1}|0_\text{L},0_\text{R} \rangle , \\
& \hat{D}_{\nu s}(\Omega,\vec{K}_\perp) \hat{s}_\text{D}^n|0_\text{L},0_\text{R} \rangle = -n s_\nu i  e^{-\beta \Omega / 2} \nonumber \\
 & \times \sum_{s'=1}^2 M_{\bar{\nu} ss'} (\Omega, -\vec{K}_\perp) \hat{C}^\dagger_{\bar{\nu} s'}(\Omega,-\vec{K}_\perp) \hat{s}_\text{D}^{n-1}|0_\text{L},0_\text{R} \rangle.
\end{align}
\end{subequations}
By multiplying Eq.~(\ref{anticommutating_rules_5}) with $1/n!$ and summing with respect to $n$, one obtains
\begin{subequations}\label{anticommutating_rules_6}
\begin{align}
\hat{C}_{\nu s}(\Omega,\vec{K}_\perp) \exp (\hat{s}_\text{D} ) |0_\text{L},0_\text{R} \rangle =  & -s_\nu i e^{-\beta \Omega / 2}  \sum_{s'=1}^2 M_{\nu s's} (\Omega, \vec{K}_\perp) \nonumber \\
& \times \hat{D}^\dagger_{\bar{\nu} s'}(\Omega,-\vec{K}_\perp) \exp (\hat{s}_\text{D} )|0_\text{L},0_\text{R} \rangle ,\\
\hat{D}_{\nu s}(\Omega,\vec{K}_\perp) \exp (\hat{s}_\text{D} )|0_\text{L},0_\text{R} \rangle  = &  -s_\nu i  e^{-\beta \Omega / 2}\sum_{s'=1}^2 M_{\bar{\nu} ss'} (\Omega, -\vec{K}_\perp) \nonumber \\
& \times  \hat{C}^\dagger_{\bar{\nu} s'}(\Omega,-\vec{K}_\perp)  \exp (\hat{s}_\text{D} )|0_\text{L},0_\text{R} \rangle,
\end{align}
\end{subequations}
which proves that Eq.~(\ref{0_M_0_LR}) is the solution of Eq.~(\ref{Minkowski_vacuum_Rindler}).

We now show how to write Eq.~(\ref{O}) in a more compact form. By computing the sum with respect to $\nu$ and performing the integral variables transformation $\Omega \mapsto -\Omega $ and $\vec{K}_\perp \mapsto -\vec{K}_\perp $ when $\nu = \text{L}$, one obtains
\begin{align}\label{O_2}
\hat{s}_\text{D} = & - i \sum_{s=1}^2 \sum_{s'=1}^2 \left[ - \int_{-\infty}^0 d\Omega \int_{\mathbb{R}^2} d^2 K_\perp  e^{\beta \Omega / 2}  M_{\text{L} s's} (-\Omega, -\vec{K}_\perp) \right.  \nonumber \\ 
&  \times \hat{C}^\dagger_{\text{L} s}(-\Omega,-\vec{K}_\perp)\hat{D}^\dagger_{\text{R} s'}(-\Omega,\vec{K}_\perp)  \nonumber \\ 
&  + \int_0^{+\infty} d\Omega \int_{\mathbb{R}^2} d^2 K_\perp  e^{-\beta \Omega / 2} M_{\text{R}  s's} (\Omega, \vec{K}_\perp)   \nonumber \\ 
& \left. \times  \hat{C}^\dagger_{\text{R} s}(\Omega,\vec{K}_\perp)\hat{D}^\dagger_{\text{L} s'}(\Omega,-\vec{K}_\perp) \right].
\end{align}
By letting $\hat{C}^\dagger_{\text{R} s}(\Omega,\vec{K}_\perp)$ and $\hat{D}^\dagger_{\text{L} s'}(\Omega,-\vec{K}_\perp)$ anticommute [Eq.~(\ref{anticommutating_rules_e})],
\begin{align}\label{O_3}
\hat{s}_\text{D} = & i \sum_{s=1}^2 \sum_{s'=1}^2 \left[ \int_{-\infty}^0 d\Omega \int_{\mathbb{R}^2} d^2 K_\perp  e^{\beta \Omega / 2} \right.  \nonumber \\ 
&  \times  M_{\text{L} s's} (-\Omega, -\vec{K}_\perp) \hat{C}^\dagger_{\text{L} s}(-\Omega,-\vec{K}_\perp)\hat{D}^\dagger_{\text{R} s'}(-\Omega,\vec{K}_\perp)  \nonumber \\ 
&  + \int_0^{+\infty} d\Omega \int_{\mathbb{R}^2} d^2 K_\perp  e^{-\beta \Omega / 2}  M_{\text{R}  s's} (\Omega, \vec{K}_\perp)   \nonumber \\ 
& \left. \times \hat{D}^\dagger_{\text{L} s'}(\Omega,-\vec{K}_\perp) \hat{C}^\dagger_{\text{R} s}(\Omega,\vec{K}_\perp) \right].
\end{align}

Equation (\ref{O_3}) suggests the definition of the operators $\hat{E}(\Theta)$, with $\Theta = (\nu, s, \Omega,\vec{K}_\perp) \in  \{ \text{L}, \text{R} \} \otimes \{ 1, 2\} \otimes \mathbb{R}^3$, such that
\begin{subequations}\label{E_CD}
\begin{align}
& \hat{E}^\dagger(\text{L}, s, \Omega,\vec{K}_\perp) =  \begin{cases}
\sum_{s'=1}^2  M_{\text{R} s's} (\Omega,\vec{K}_\perp) \hat{D}^\dagger_{\text{L} s'}(\Omega,-\vec{K}_\perp) & \text{if } \Omega>0 \\
\sum_{s'=1}^2 M_{\text{L} ss'} (-\Omega,-\vec{K}_\perp) \hat{C}^\dagger_{\text{L} s'}(-\Omega,-\vec{K}_\perp)  & \text{if } \Omega<0
\end{cases},\label{E_CD_a} \\
& \hat{E}^\dagger(\text{R}, s, \Omega,\vec{K}_\perp) = \begin{cases}
\hat{C}^\dagger_{\text{R} s}(\Omega,\vec{K}_\perp)  & \text{if } \Omega>0\\
\hat{D}^\dagger_{\text{R} s}(-\Omega,\vec{K}_\perp) & \text{if } \Omega<0
\end{cases}.\label{E_CD_b}
\end{align}
\end{subequations}
In this way, Eq.~(\ref{O_3}) becomes
\begin{equation}\label{O_4}
\hat{s}_\text{D} = \sum_\theta  f(\theta) \hat{F}^\dagger(\theta),
\end{equation}
with $\theta = (s, \Omega,\vec{K}_\perp) \in  \{ 1, 2\} \otimes \mathbb{R}^3$, 
\begin{subequations}
\begin{align}
& f(s, \Omega,\vec{K}_\perp) = i e^{-\beta |\Omega| / 2} ,\label{f} \\
 & \hat{F}^\dagger(\theta) = \hat{E}^\dagger(\text{L}, \theta) \hat{E}^\dagger(\text{R}, \theta) \label{F}
\end{align}
\end{subequations}
and where $\sum_\theta$ is a generalized sum for the $\theta$ variables consisting in a sum with respect to the discrete variable $s$ and an integral for the continuum variables $\Omega$ and $\vec{K}_\perp$, i.e.,
\begin{align}
& \sum_\theta = \sum_{s=1}^2 \int_{\mathbb{R}} d \Omega \int_{\mathbb{R}^2} d^2 K_\perp, & \theta = (s, \Omega,\vec{K}_\perp).
\end{align}
The operator $\hat{S}_\text{D}$, instead, can be written as
\begin{equation}\label{OO_2}
\hat{S}_\text{D} = \exp \left( 2 \sum_\theta  [ g(\theta) \hat{F}(\theta) ]^\text{A} \right),
\end{equation}
with
\begin{equation}
g(s, \Omega,\vec{K}_\perp) = i \zeta(|\Omega|).
\end{equation}

Since the matrix $M_{\nu ss'}(\Omega, \vec{K}_\perp)$ is unitary, Eq.~(\ref{E_CD}) is invertible. Indeed, by using  Eqs.~(\ref{M_unitary}) and (\ref{E_CD}) one can prove
\begin{subequations}\label{CD_E}
\begin{align}
& \hat{C}^\dagger_{\text{L} s}(\Omega,\vec{K}_\perp) = \sum_{s'=1}^2 M^*_{\text{L} s's} (\Omega,\vec{K}_\perp) \hat{E}^\dagger(\text{L}, s',-\Omega,-\vec{K}_\perp), \\
 & \hat{D}^\dagger_{\text{L} s}(\Omega,\vec{K}_\perp) = \sum_{s'=1}^2 M^*_{\text{R} ss'} (\Omega,-\vec{K}_\perp) \hat{E}^\dagger(\text{L}, s', \Omega,-\vec{K}_\perp),\\
& \hat{C}^\dagger_{\text{R} s}(\Omega,\vec{K}_\perp) = \hat{E}^\dagger(\text{R}, s , \Omega,\vec{K}_\perp),\label{CD_E_c} \\
 & \hat{D}^\dagger_{\text{R} s}(\Omega,\vec{K}_\perp) = \hat{E}^\dagger(\text{R}, s,- \Omega,\vec{K}_\perp),\label{CD_E_d} 
\end{align}
\end{subequations}
for any $\Omega > 0$. For this reason, Eqs.~(\ref{E_CD}) and (\ref{CD_E}) are a one-to-one mapping between $\hat{E}^\dagger(\Theta)$ and the creation operators $\hat{C}^\dagger_{\nu s}(\Omega,\vec{K}_\perp)$ and $\hat{D}^\dagger_{\nu s}(\Omega,\vec{K}_\perp)$.

Notice that, from the definition of the Rindler vacuum $|0_\text{L},0_\text{R} \rangle$ [Eq.~(\ref{Rindler_vacuum})] and the operator $\hat{E}(\Theta)$ [Eq.~(\ref{E_CD})],
\begin{equation} \label{E_annhilator}
\hat{E}(\Theta)|0_\text{L},0_\text{R} \rangle = 0.
\end{equation}
The anticommutation properties for the operator $\hat{E}(\Theta)$ are
\begin{subequations}\label{anticommutating_rules_E}
\begin{align}
& \{ \hat{E}(\Theta), \hat{E}^\dagger(\Theta') \} =  \Delta(\Theta,\Theta'),  \label{anticommutating_rules_E_a}\\
&  \{ \hat{E}(\Theta), \hat{E}(\Theta') \}  = 0 \label{anticommutating_rules_E_b},
\end{align}
\end{subequations}
where $\Delta(\Theta,\Theta')$ is a generalized delta function for the variables $\Theta = (\nu, s, \Omega,\vec{K}_\perp)$ and $\Theta' = (\nu', s', \Omega',\vec{K}'_\perp)$. The function $\Delta(\Theta,\Theta')$ is the product of the Kronecker delta for the discrete variables $\nu$, $\nu'$, $s$ and $s'$ and the Dirac delta function for the continuum variables $\Omega-\Omega'$ and $\vec{K}_\perp - \vec{K}'_\perp$:
\begin{align}
 \Delta((\nu, s, \Omega,\vec{K}_\perp), (\nu', s', \Omega',\vec{K}'_\perp)) = \delta_{\nu\nu'} \delta_{ss'} \delta(\Omega-\Omega') \delta^2(\vec{K}_\perp - \vec{K}'_\perp).
\end{align}
Equation (\ref{anticommutating_rules_E}) can be checked by using Eqs.~(\ref{anticommutating_rules}), (\ref{M_unitary}) and (\ref{E_CD}). As a consequence of Eqs.~(\ref{E_annhilator}) and (\ref{anticommutating_rules_E}), the mapping from $\hat{C}^\dagger_{\nu s}(\Omega,\vec{K}_\perp)$ and $\hat{D}^\dagger_{\nu s}(\Omega,\vec{K}_\perp)$ to $\hat{E}^\dagger(\Theta)$ is canonical.

Equations (\ref{F}) and (\ref{anticommutating_rules_E_b}) lead to
\begin{subequations}\label{commutating_rules_F}
\begin{align}
& [ \hat{F}^\dagger(\theta), \hat{F}^\dagger(\theta') ] = 0,  \label{commutating_rules_F_a}\\
&  \hat{F}^\dagger(\theta) \hat{F}^\dagger(\theta) = 0 \label{commutating_rules_F_b}.
\end{align}
\end{subequations}
The representative of the Minkowski vacuum given by Eqs.~(\ref{0_M_0_LR}) and (\ref{O_4}) and the algebraic properties of the operators $\hat{E}(\Theta)$ and $\hat{F}(\theta)$ [Eqs.~(\ref{E_annhilator}), (\ref{anticommutating_rules_E}) and (\ref{commutating_rules_F})] will be used in the next subsection to derive the statistical operator describing the Minkowski vacuum in the right Rindler frame.

\subsection{Minkowski vacuum in the right Rindler frame}\label{Unruh_effect_for_Dirac_fields}

In the previous subsection we derived the representation of the Minkowski vacuum in both left and right Rindler frames. Here, instead, we will focus only on the right wedge, which describes the accelerated observer with positive acceleration $c^2 a$. By performing a partial trace over the left wedge, we will compute the statistical operator representing the Minkowski vacuum as an element of the right Rindler-Fock space. The result will be a fermionic thermal state, which is at the origin of the Unruh effect for Dirac fields.

In order to perform the partial trace, one needs a basis for the Rindler-Fock space. The single particle space is defined by the creation operators $\hat{C}^\dagger_{\nu s}(\Omega,\vec{K}_\perp)$ and $\hat{D}^\dagger_{\nu s}(\Omega,\vec{K}_\perp)$ acting on the vacuum state $|0_\text{L},0_\text{R} \rangle$. Hence, a basis for single particles and antiparticles in each wedge can be defined through the quantum numbers $s$, $\Omega$ and $\vec{K}_\perp$. Alternatively, one may take advantage of the canonical transformation (\ref{E_CD}) and use the operator $\hat{E}(\Theta)$ and the quantum numbers $\Theta = (\nu, s, \Omega,\vec{K}_\perp)$ to describe single particles and antiparticles of both wedges as
\begin{equation}\label{single_particle_basis}
| \Theta \rangle  = \hat{E}^\dagger(\Theta) |0_\text{L},0_\text{R} \rangle.
\end{equation}
Notice that Eq.~(\ref{single_particle_basis}) is an orthonormal basis for the single particle and antiparticle space. The orthonormality condition can be checked by using Eqs.~(\ref{E_annhilator}) and (\ref{anticommutating_rules_E_a}).

Many-particles states are given by the action of sequences of creation operators $\hat{E}^\dagger(\Theta)$ on the Rindler vacuum. We define the Rindler-Fock state
\begin{equation}\label{many_particle_Rindler_state}
| \mathbf{\Theta} \rangle = \hat{\mathbf{E}}^\dagger (\mathbf{\Theta}) |0_\text{L},0_\text{R} \rangle,
\end{equation}
with
\begin{equation}\label{EE}
\hat{\mathbf{E}}^\dagger (\mathbf{\Theta}) = \prod_{i=1}^{|\mathbf{\Theta}|} \hat{E}^\dagger(\Theta_i)
\end{equation}
and where $\mathbf{\Theta} = \{ \Theta_1, \dots, \Theta_n \}$ is an ordered set of quantum numbers $\Theta_i$ and $|\mathbf{\Theta}|$ the cardinality of the set. By using Eqs.~(\ref{E_annhilator}) and (\ref{anticommutating_rules_E}), one can prove that the scalar product of different states defined by Eq.~(\ref{many_particle_Rindler_state}) is
\begin{equation}\label{many_particle_Rindler_state_product}
 \langle  \mathbf{\Theta} | \mathbf{\Theta}' \rangle =  \sum_{\tau \in S_{|\mathbf{\Theta}|}} \text{sign}(\tau)  \mathbf{\Delta}(\tau(\mathbf{\Theta}), \mathbf{\Theta}'),
\end{equation}
with
\begin{equation}\label{DDelta}
\mathbf{\Delta}(\mathbf{\Theta}, \mathbf{\Theta}') = \delta_{|\mathbf{\Theta}||\mathbf{\Theta}'|}  \prod_{i=1}^{|\mathbf{\Theta}|} \Delta\left( \Theta_i,\Theta'_{i} \right) 
\end{equation}
and where $S_n$ is the space of all permutations of sets with $n$ elements.

Notice that the order of the creation operators $\hat{E}^\dagger(\Theta_i)$ on the right side of Eq.~(\ref{many_particle_Rindler_state}) cannot be ignored because of the anticommuting nature of the Rindler operators $\hat{E}^\dagger(\Theta_i)$ [Eq.~(\ref{anticommutating_rules_E_b})]. Any permutation of quantum numbers $\Theta_i$ leads to the same many-particles state up to a sign. The set of states $| \mathbf{\Theta} \rangle$ cannot be chosen as basis, due to the presence of the sign of permutations appearing in Eq.~(\ref{many_particle_Rindler_state_product}).

To define a basis for the particles space, one has to consider an operator $\mathcal{O}$ that acts on any sequence of quantum numbers $\mathbf{\Theta}$ and rearrange their order by following a fixed ordering rule. The set of states $| \mathcal{O} (\mathbf{\Theta}) \rangle$ form an orthonormal basis for the many-particles space. Indeed the following equation holds
\begin{equation}\label{many_particle_Rindler_basis_product}
 \langle  \mathcal{O} (\mathbf{\Theta}) | \mathcal{O} (\mathbf{\Theta}') \rangle =  \sum_{\tau \in S_{|\mathbf{\Theta}|}}  \mathbf{\Delta}(\tau(\mathbf{\Theta}), \mathbf{\Theta}').
\end{equation}
Notice that in Eq.~(\ref{many_particle_Rindler_basis_product}) the sign of permutations is absent, as opposed to Eq.~(\ref{many_particle_Rindler_state_product}).

The orthonormality condition (\ref{many_particle_Rindler_basis_product}) can be proven in the following way. Firstly notice that the ordering function $\mathcal{O}$ acts on any sequence of quantum numbers $\mathbf{\Theta}$ as a $\mathbf{\Theta}$ dependent permutation. Indeed, for any $\mathbf{\Theta}$, one can define a permutation $\mathcal{P}_\mathbf{\Theta} \in S_{|\mathbf{\Theta}|}$ such that
\begin{equation}\label{P_Theta}
\mathcal{O} (\mathbf{\Theta}) = \mathcal{P}_\mathbf{\Theta} (\mathbf{\Theta}).
\end{equation}
Notice also that the ordering function $\mathcal{O}$ is unaffected by any permutation. Explicitly, this means that
\begin{equation} \label{O_ordering_permutation}
 \mathcal{O} ( \tau(\mathbf{\Theta}) ) = \mathcal{O} (\mathbf{\Theta}),
\end{equation}
for any $\tau \in S_{|\mathbf{\Theta}|}$. By using Eq.~(\ref{P_Theta}) in Eq.~(\ref{O_ordering_permutation}), one can also write
\begin{equation} \label{O_ordering_permutation_2}
\mathcal{P}_\mathbf{\tau(\Theta)} \tau(\mathbf{\Theta}) = \mathcal{P}_\mathbf{\Theta} (\mathbf{\Theta}),
\end{equation}
which means that
\begin{equation}\label{O_ordering_permutation_3}
\text{sign} ( \mathcal{P}_\mathbf{\tau(\Theta)} \tau ) = \text{sign} (\mathcal{P}_\mathbf{\Theta}).
\end{equation}

Equations (\ref{many_particle_Rindler_state_product}) and (\ref{P_Theta}) lead to the scalar product
\begin{equation}\label{many_particle_Rindler_state_product_2}
 \langle  \mathcal{O}(\mathbf{\Theta}) | \mathcal{O}(\mathbf{\Theta}') \rangle  = \sum_{\tau \in S_{|\mathbf{\Theta}|}} \text{sign}(\tau)   \mathbf{\Delta}(\tau \mathcal{P}_\mathbf{\Theta}(\mathbf{\Theta}), \mathcal{P}_{\mathbf{\Theta}'}(\mathbf{\Theta}')).
\end{equation}
Notice that from the definition of $\mathbf{\Delta}(\mathbf{\Theta}, \mathbf{\Theta}')$ [Eq.~(\ref{DDelta})], by rearranging the order of the product index $i \mapsto \tau(i)$ with any permutation $\tau \in S_{|\mathbf{\Theta}|}$ one obtains
\begin{equation}
\mathbf{\Delta}(\tau(\mathbf{\Theta}), \tau(\mathbf{\Theta}')) = \mathbf{\Delta}(\mathbf{\Theta}, \mathbf{\Theta}').
\end{equation}
This can be used in Eq.~(\ref{many_particle_Rindler_state_product_2}) to obtain
\begin{equation}\label{many_particle_Rindler_state_product_3}
\langle  \mathcal{O}(\mathbf{\Theta}) | \mathcal{O}(\mathbf{\Theta}') \rangle  =  \sum_{\tau \in S_{|\mathbf{\Theta}|}} \text{sign}(\tau)   \mathbf{\Delta}(\tau\mathcal{P}_{\mathbf{\Theta}}\mathcal{P}^{-1}_{\mathbf{\Theta}'}(\mathbf{\Theta}), \mathbf{\Theta}').
\end{equation}
By using the fact that the sum $\sum_{\tau \in S_{|\mathbf{\Theta}|}}$ runs over all permutations of $S_{|\mathbf{\Theta}|}$, one can perform the transformation $\tau \mapsto \tau \mathcal{P}_{\mathbf{\Theta}'} \mathcal{P}^{-1}_{\mathbf{\Theta}}$ in Eq.~(\ref{many_particle_Rindler_state_product_3}) and write
\begin{equation}\label{many_particle_Rindler_state_product_4}
\langle  \mathcal{O}(\mathbf{\Theta}) | \mathcal{O}(\mathbf{\Theta}') \rangle = \sum_{\tau \in S_{|\mathbf{\Theta}|}} \text{sign}(\tau \mathcal{P}_{\mathbf{\Theta}'} \mathcal{P}^{-1}_{\mathbf{\Theta}})   \mathbf{\Delta}(\tau(\mathbf{\Theta}), \mathbf{\Theta}').
\end{equation}
Notice that the $\mathbf{\Delta}(\tau(\mathbf{\Theta}), \mathbf{\Theta}')$ function in the right side of Eq.~(\ref{many_particle_Rindler_state_product_4}) is nonvanishing only when $\mathbf{\Theta}' =  \tau(\mathbf{\Theta})$. Hence, Eq.~(\ref{many_particle_Rindler_state_product_4}) reads as
\begin{equation}\label{many_particle_Rindler_state_product_5}
 \langle  \mathcal{O}(\mathbf{\Theta}) | \mathcal{O}(\mathbf{\Theta}') \rangle = \sum_{\tau \in S_{|\mathbf{\Theta}|}} \text{sign}(\tau \mathcal{P}_{\tau(\mathbf{\Theta})} \mathcal{P}^{-1}_{\mathbf{\Theta}})   \mathbf{\Delta}(\tau(\mathbf{\Theta}), \mathbf{\Theta}').
\end{equation}
By using Eq.~(\ref{O_ordering_permutation_3}) in (\ref{many_particle_Rindler_state_product_5}), one obtains Eq.~(\ref{many_particle_Rindler_basis_product}).

Equation (\ref{many_particle_Rindler_basis_product}) is the orthonormality condition for the many-particles states $| \mathcal{O} (\mathbf{\Theta}) \rangle$ defined as
\begin{equation}\label{many_particle_Rindler_basis}
| \mathcal{O} (\mathbf{\Theta}) \rangle =  \hat{\mathbf{E}}^\dagger (\mathcal{O}(\mathbf{\Theta})) |0_\text{L},0_\text{R} \rangle.
\end{equation}
Notice that Eq.~(\ref{many_particle_Rindler_basis}) is symmetric with respect to any permutation of the quantum numbers $\vec{\Theta}_i$ [Eq.~(\ref{O_ordering_permutation})].

Hereafter, we choose any ordering function $\mathcal{O}$ such that for any couple of quantum numbers $\Theta = (\nu, \theta)$ and $\Theta' = (\nu', \theta')$,
\begin{equation}\label{sorting_function}
\mathcal{O} ( \{ (\nu, \theta) , (\nu', \theta') \}  )= \begin{cases}
\mathcal{Q} ( \{ (\nu, \theta) , (\nu', \theta') \}  ) & \text{if } \theta \neq \theta' \\
\mathcal{W} ( \{ (\nu, \theta) , (\nu', \theta') \}  ) & \text{if } \theta = \theta'
 \end{cases},
\end{equation}
where $\mathcal{Q}$ is any ordering function with respect to the non-repeating quantum numbers $\theta = (s, \Omega, \vec{K}_\perp)$. The ordering function $\mathcal{W}$, instead, is with respect to the wedge variable $\nu$. We choose the following definition for $\mathcal{W}$
\begin{subequations} \label{W_sorting}
\begin{align}
& \mathcal{W} ( \{ (\text{L}, \theta) , (\text{R}, \theta) \}  ) =  ( \{ (\text{L}, \theta) , (\text{R}, \theta) \}  ), \\
& \mathcal{W} ( \{ (\text{R}, \theta) , (\text{L}, \theta) \}  ) =  ( \{ (\text{L}, \theta) , (\text{R}, \theta) \}  ).
\end{align}
\end{subequations}
We do not choose any particular definition for $\mathcal{Q}$. However, for completeness, we give a possible example by considering the lexicographical order as
\begin{subequations}
\begin{align}
& \mathcal{Q} ( \{ (\nu, s, \Omega, \vec{K}_\perp) , (\nu', s', \Omega', \vec{K}'_\perp) \}  ) \nonumber \\
 = & \begin{cases} 
 \{ (\nu, s, \Omega, \vec{K}_\perp) , (\nu', s', \Omega', \vec{K}'_\perp) \} & \text{if } s < s' \\
 \{ (\nu', s', \Omega', \vec{K}'_\perp) , (\nu, s, \Omega, \vec{K}_\perp) \} & \text{if } s > s'
 \end{cases}, \\
& \mathcal{Q} ( \{ (\nu, s, \Omega, \vec{K}_\perp) , (\nu', s, \Omega', \vec{K}'_\perp) \}  ) \nonumber \\
 = & \begin{cases} 
 \{ (\nu, s, \Omega, \vec{K}_\perp) , (\nu', s, \Omega', \vec{K}'_\perp) \} & \text{if } \Omega < \Omega' \\
 \{ (\nu', s, \Omega', \vec{K}'_\perp) , (\nu, s, \Omega, \vec{K}_\perp) \} & \text{if } \Omega > \Omega'
 \end{cases}, \\
& \mathcal{Q} ( \{ (\nu, s, \Omega, K_1, K_2) , (\nu', s, \Omega, K'_1, K'_2) \}  ) \nonumber \\
 = & \begin{cases} 
 \{ (\nu, s, \Omega, K_1, K_2) , (\nu', s, \Omega, K'_1, K'_2) \} & \text{if } K_1 < K'_1 \\
 \{ (\nu', s, \Omega, K'_1, K'_2) , (\nu, s, \Omega, K_1, K_2) \} & \text{if } K_1 > K'_1
 \end{cases}, \\
& \mathcal{Q} ( \{ (\nu, s, \Omega, K_1, K_2) , (\nu', s, \Omega, K_1, K'_2) \}  ) \nonumber \\
 = & \begin{cases} 
 \{ (\nu, s, \Omega, K_1, K_2) , (\nu', s, \Omega, K_1, K'_2) \} & \text{if } K_2 < K'_2 \\
 \{ (\nu', s, \Omega, K_1, K'_2) , (\nu, s, \Omega, K_1, K_2) \} & \text{if } K_2 > K'_2
 \end{cases}.
\end{align}
\end{subequations}

We now show how to write the Minkowski vacuum [Eq.~(\ref{0_M_0_LR})] in terms of the many-particles basis (\ref{many_particle_Rindler_basis}). Equation (\ref{O_4}) leads to
\begin{equation}\label{O_n}
\hat{s}_\text{D}^n = \sum_{\theta_1} \dots \sum_{\theta_n}  \prod_{i=1}^n f(\theta_i) \prod_{i=1}^n \hat{F}^\dagger(\theta_i),
\end{equation}
for any $n \in \mathbb{N}$. The operators $\hat{F}^\dagger(\theta_i)$ that appear in Eq.~(\ref{O_n}) are defined by Eq.~(\ref{F}) and can be written in terms of $\hat{\mathbf{E}}^\dagger (\mathbf{\Theta})$ [Eq.~(\ref{EE})] as
\begin{equation}\label{F_EE}
\hat{F}^\dagger(\theta) = \hat{\mathbf{E}}^\dagger ( \{ (\text{L}, \theta), (\text{R}, \theta) \} ).
\end{equation}
Notice that the couple of quantum numbers appearing in Eq.~(\ref{F_EE}) follow the $\mathcal{W}$ order [Eq.~(\ref{W_sorting})]. This means that
\begin{equation}\label{F_EE_2}
\hat{F}^\dagger(\theta) = \hat{\mathbf{E}}^\dagger ( \mathcal{W} ( \{ (\text{L}, \theta), (\text{R}, \theta) \} ) ).
\end{equation}

Consider the chain of operators $\prod_{i=1}^n \hat{F}^\dagger(\theta_i)$ that appears in Eq.~(\ref{O_n}). The operators $\hat{F}^\dagger(\theta_i)$ commute [Eq.~(\ref{commutating_rules_F_a})], and, hence, one may write Eq.~(\ref{O_n}) by following any order for the sequence of $\hat{F}^\dagger(\theta_i)$. Notice also that as a consequence of Eq.~(\ref{commutating_rules_F_b}), no repetition of the quantum numbers $\theta_i$ occurs. Therefore one may choose the $\mathcal{Q}$ order for the sequence of $\hat{F}^\dagger(\theta_i)$. By sorting the $\hat{F}^\dagger(\vec{\Theta}_i)$ operators in Eq.~(\ref{O_n}) with respect to the $\mathcal{Q}$ order and by considering the fact that the $\hat{E}^\dagger(\Theta)$ operators appearing in Eq.~(\ref{F}) already follow the $\mathcal{W}$ order [Eq.~(\ref{F_EE_2})], one derives the identity
\begin{equation}\label{F_EE_3}
\prod_{i=1}^n \hat{F}^\dagger(\theta_i) = \hat{\mathbf{E}}^\dagger \left( \mathcal{O} \left( \bigcup_{i=1}^n \{ (\text{L}, \theta_i), (\text{R}, \theta_i) \} \right) \right).
\end{equation}

By plugging Eq.~(\ref{F_EE_3}) in Eq.~(\ref{O_n}), one obtains
\begin{equation}\label{O_n_2}
\hat{s}_\text{D}^n = \sum_{\theta_1} \dots \sum_{\theta_n}  \prod_{i=1}^n f(\theta_i)  \hat{\mathbf{E}}^\dagger \left( \mathcal{O} \left( \bigcup_{i=1}^n \{ (\text{L}, \theta_i), (\text{R}, \theta_i) \} \right) \right).
\end{equation}
By acting on the Rindler vacuum and by using Eq.~(\ref{many_particle_Rindler_basis}), Eq.~(\ref{O_n_2}) becomes
\begin{equation}\label{O_n_3}
\hat{s}_\text{D}^n |0_\text{L},0_\text{R} \rangle = \sum_{\theta_1} \dots \sum_{\theta_n}  \prod_{i=1}^n f(\theta_i)  \left| \mathcal{O} \left( \bigcup_{i=1}^n \{ (\text{L}, \theta_i), (\text{R}, \theta_i) \} \right)\right\rangle .
\end{equation}
By multiplying Eq.~(\ref{O_n_3}) with $1/n!$ and summing with respect to $n$, one obtains
\begin{align}\label{exp_O_2}
\exp(\hat{s}_\text{D}) |0_\text{L},0_\text{R} \rangle = & |0_\text{L},0_\text{R} \rangle  + \sum_{n=1}^\infty  \frac{1}{n!} \sum_{\theta_1} \dots \sum_{\theta_n}  \prod_{i=1}^n f(\theta_i) \nonumber \\
& \times \left| \mathcal{O} \left( \bigcup_{i=1}^n \{ (\text{L}, \theta_i), (\text{R}, \theta_i) \} \right)\right\rangle,
\end{align}
which provides a representation for the Minkowski vacuum [Eq.~(\ref{0_M_0_LR})] with respect to the basis (\ref{many_particle_Rindler_basis}).

We now compute the partial trace with respect to the left wedge. From Eq.~(\ref{exp_O_2}), one obtains
\begin{align}\label{tr_L_exp_O}
& \text{Tr}_\text{L} \left[ \exp(\hat{s}_\text{D}) | 0_\text{L},0_\text{R} \rangle \langle 0_\text{L},0_\text{R} | \exp(\hat{s}_\text{D})^\dagger \right]  \nonumber \\
  = & |0_\text{R} \rangle \langle 0_\text{R} |+ \sum_{n=1}^\infty \sum_{m=1}^\infty \frac{1}{n!m!} \sum_{\theta_1} \dots \sum_{\theta_n} \sum_{\theta'_1} \dots \sum_{\theta'_m}  \prod_{i=1}^n f(\theta_i)  \prod_{i=1}^m f^*(\theta'_i)   \nonumber \\
 & \times \left\langle  \mathcal{O} \left( \bigcup_{i=1}^n \{ (\text{L}, \theta_i) \} \right) \left|  \mathcal{O} \left( \bigcup_{i=1}^m \{ (\text{L}, \theta'_i) \} \right) \right. \right\rangle  \nonumber \\
& \times  \left| \mathcal{O} \left( \bigcup_{i=1}^n \{ (\text{R}, \theta_i) \} \right) \right\rangle \left\langle \mathcal{O} \left( \bigcup_{i=1}^m \{ (\text{R}, \theta'_i) \} \right) \right|.
\end{align}
The orthonormality condition in the left wedge is
\begin{equation}\label{many_particle_Rindler_basis_product_L}
\left\langle  \mathcal{O} \left( \bigcup_{i=1}^n \{ (\text{L}, \theta_i) \} \right) \left|  \mathcal{O} \left( \bigcup_{i=1}^m \{ (\text{L}, \theta'_i) \} \right) \right. \right\rangle = \delta_{nm}  \sum_{\tau \in S_n} \prod_{i = 1}^n \Delta \left( \left( \text{L}, \theta_{\tau (i)} \right), \left(\text{L}, \theta'_i \right) \right).
\end{equation}
By plugging Eq.~(\ref{many_particle_Rindler_basis_product_L}) in Eq.~(\ref{tr_L_exp_O}) and by computing the sum $\sum_{m=1}^\infty$ and the generalized sums $\sum_{\theta'_1} \dots \sum_{\theta'_m}$, one obtains
\begin{align}\label{tr_L_exp_O_2}
& \text{Tr}_\text{L} \left[ \exp(\hat{s}_\text{D}) | 0_\text{L},0_\text{R} \rangle \langle 0_\text{L},0_\text{R} | \exp(\hat{s}_\text{D})^\dagger \right] \nonumber \\
 = & |0_\text{R} \rangle \langle 0_\text{R} |  +  \sum_{n=1}^\infty \frac{1}{(n!)^2}  \sum_{\tau \in S_n} \sum_{\theta_1} \dots \sum_{\theta_n} \prod_{i=1}^n f(\theta_i)  \prod_{i=1}^n f^*(\theta_{\tau (i)})   \nonumber \\
& \times \left| \mathcal{O} \left( \bigcup_{i=1}^n \{ (\text{R}, \theta_i) \} \right) \right\rangle \left\langle \mathcal{O} \left( \bigcup_{i=1}^n \{ (\text{R}, \theta_{\tau (i)}) \} \right) \right|.
\end{align}
By using Eq.~(\ref{O_ordering_permutation}) and the fact that the cardinality of $S_n$ is $ n!$, Eq.~(\ref{tr_L_exp_O_2}) becomes
\begin{align}\label{tr_L_exp_O_3}
& \text{Tr}_\text{L} \left[ \exp(\hat{s}_\text{D}) | 0_\text{L},0_\text{R} \rangle \langle 0_\text{L},0_\text{R} | \exp(\hat{s}_\text{D})^\dagger \right]  = |0_\text{R} \rangle \langle 0_\text{R} | \nonumber \\
& +  \sum_{n=1}^\infty \frac{1}{n!}  \sum_{\theta_1} \dots \sum_{\theta_n} \prod_{i=1}^n |f(\theta_i)|^2  \left| \mathcal{O} \left( \bigcup_{i=1}^n \{ (\text{R}, \theta_i) \} \right) \right\rangle \left\langle \mathcal{O} \left( \bigcup_{i=1}^n \{ (\text{R}, \theta_i) \} \right) \right|.
\end{align}

The right side of Eq.~(\ref{tr_L_exp_O_3}) is proportional to the thermal state in the right wedge. This can be seen by considering the following eigenstate decomposition of the Hamiltonian operator
\begin{equation}\label{H_R}
\hat{H}_\text{R} = \sum_{n=1}^\infty \frac{1}{n!}  \sum_{\theta_1} \dots \sum_{\theta_n} \left[ \sum_{i=1}^n h_\text{R}(\theta_i) \right]  \left| \mathcal{O} \left( \bigcup_{i=1}^n \{ (\text{R}, \theta_i) \} \right) \right\rangle \left\langle \mathcal{O} \left( \bigcup_{i=1}^n \{ (\text{R}, \theta_i) \} \right) \right|,
\end{equation}
where
\begin{equation}\label{h_R}
h_\text{R}(s, \Omega,\vec{K}_\perp) = \hbar |\Omega|.
\end{equation}
The $1/n!$ factor comes from the repetition of any independent $n$ particles state due to the permutation symmetry (\ref{O_ordering_permutation}). Notice that Eqs.~(\ref{tr_L_exp_O_3}) and (\ref{H_R}) have the same eigenstate decomposition but with different eigenvalues. By comparing Eq.~(\ref{f}) with Eq.~(\ref{h_R}), one can derive the following identity relating the eigenvalues of Eqs.~(\ref{tr_L_exp_O_3}) and (\ref{H_R})
\begin{equation}
 \prod_{i=1}^n |f(\theta_i)|^2 = \exp \left( - \frac{\beta}{\hbar} \sum_{i=1}^n h_\text{R}(\theta_i) \right),
\end{equation}
which means that
\begin{equation}\label{tr_L_exp_O_4}
\text{Tr}_\text{L} \left[ \exp(\hat{s}_\text{D}) | 0_\text{L},0_\text{R} \rangle \langle 0_\text{L},0_\text{R} | \exp(\hat{s}_\text{D})^\dagger \right] =  \exp \left( - \frac{\beta}{\hbar}  \hat{H}_\text{R} \right).
\end{equation}

By using Eqs.~(\ref{0_M_0_LR}) and (\ref{tr_L_exp_O_4}) we prove that 
\begin{equation}\label{tr_L_0_M}
\text{Tr}_\text{L} | 0_\text{M} \rangle \langle 0_\text{M} | \propto  \exp \left( - \frac{\beta}{\hbar} \hat{H}_\text{R} \right),
\end{equation}
which is the fermionic thermal state with temperature $\hbar /(k_\text{B} \beta)$, where $k_\text{B}$ is the Boltzmann constant. Equation (\ref{tr_L_0_M}) represents the Minkowski vacuum seen by the accelerated observer with acceleration $c^2 a$.

\subsection{Spin basis choice}\label{Basis_choice}

The result obtained in Sec.~\ref{Minkowski_vacuum_in_Rindler_spacetime} depends of the basis $\tilde{\mathfrak{W}}_{\nu s}(\Omega, \vec{K}_\perp)$. Indeed, the matrix $ M_{\nu ss'} (\Omega, \vec{K}_\perp)$ appears in the Rindler-Fock representation of the Minkowski vacuum [Eqs.~(\ref{0_M_0_LR}) and (\ref{O})]. From Eq.~(\ref{M}), one can see the relation between $  M_{\nu ss'} (\Omega, \vec{K}_\perp)$ and $\tilde{\mathfrak{W}}_{\nu s}(\Omega, \vec{K}_\perp)$. We find out that different choices for the basis $\tilde{\mathfrak{W}}_{\nu s}(\Omega, \vec{K}_\perp)$ lead to different representations of the Minkowski vacuum in the Rindler spacetime.

In Eq.~(\ref{O}), the matrix $  M_{\nu ss'} (\Omega, \vec{K}_\perp)$ couples modes of one wedge with modes of the opposite wedge. Hence, in the Minkowski vacuum, any solution $\tilde{\mathfrak{W}}_{\text{R} s}(\Omega, \vec{K}_\perp)$ of Eq.~(\ref{W_tilde_constraints}) in the right wedge is coupled with a solution of Eq.~(\ref{W_tilde_constraints}) in the left wedge that is proportional to $\tilde{\mathfrak{W}}_{\text{R} s}(-\Omega, -\vec{K}_\perp)$.

The spin coupling of $| 0_\text{M} \rangle$ is then averaged away by the partial trace over the left wedge in Sec.~\ref{Unruh_effect_for_Dirac_fields}. Indeed, the trace is computed by considering a basis for the left wedge [Eqs.~(\ref{E_CD_a}), (\ref{EE}) and (\ref{many_particle_Rindler_basis})] that absorbs the matrix $  M_{\nu ss'} (\Omega, \vec{K}_\perp)$ in Eq.~(\ref{O_3}) and gives an expression for $| 0_\text{M} \rangle$ without $  M_{\nu ss'} (\Omega, \vec{K}_\perp)$ [Eq.~\ref{O_4}].

Consequently, the result obtained in Sec.~\ref{Unruh_effect_for_Dirac_fields} is independent of the choice for the solutions of Eq.~(\ref{W_tilde_constraints}). Indeed, the thermal state describing the Minkowski vacuum in the right wedge [Eq.~(\ref{tr_L_0_M})] is independent of $\tilde{\mathfrak{W}}_{\nu s}(\Omega, \vec{K}_\perp)$. One can see this by plugging Eqs.~(\ref{E_CD_b}), (\ref{EE}), (\ref{many_particle_Rindler_basis}) and (\ref{H_R}) in Eq.~(\ref{tr_L_0_M}) and noticing that $\tilde{\mathfrak{W}}_{\nu s}(\Omega, \vec{K}_\perp)$ never appears in the explicit form of $\text{Tr}_\text{L} | 0_\text{M} \rangle \langle 0_\text{M} |$.

In this subsection, we go back to the representation of the Minkowski vacuum in both wedges [Eqs.~(\ref{0_M_0_LR}) and (\ref{O})] and we discuss different choices for the basis $\tilde{\mathfrak{W}}_{\nu s}(\Omega, \vec{K}_\perp)$ that lead to different representations of $| 0_\text{M} \rangle$. We study the operator $\hat{s}_\text{D}$ for different choices of $\tilde{\mathfrak{W}}_{\nu s}(\Omega, \vec{K}_\perp)$ and, hence, for different matrices $M_{\nu ss'} (\Omega, \vec{K}_\perp)$. In other words, we consider different outputs of the function $\hat{s}_\text{D} [ M_{\nu ss'} (\Omega, \vec{K}_\perp)]$.

By looking at Eq.~(\ref{O}), one may conclude that the most natural choice for $\tilde{\mathfrak{W}}_{\nu s}(\Omega, \vec{K}_\perp)$ is such that $M_{\nu ss'} (\Omega, \vec{K}_\perp)$ is proportional to the identity. This choice can be made by adopting any spin basis for the $\nu$ wedge and choosing the spin basis in the other wedge $\bar{\nu}$ such that
\begin{equation}
\tilde{\mathfrak{W}}_{\bar{\nu} s}(\Omega, \vec{K}_\perp) \propto \tilde{\mathfrak{W}}_{\nu s}(-\Omega, \vec{K}_\perp).
\end{equation}
In this way, Eq.~(\ref{M}) reads as
\begin{equation}\label{M_propto_identity}
M_{\nu ss'} (\Omega, \vec{K}_\perp) \propto \delta_{ss'}
\end{equation}
and the Minkowski vacuum couples each particle mode of one wedge with the antiparticle mode of same spin number $s$ of the other wedge [Eq.~(\ref{O})].

Possible choices for the unitary matrix $   M_{\nu ss'} (\Omega, \vec{K}_\perp)$ that satisfy Eqs.~(\ref{M_conjugate}) and (\ref{M_propto_identity}) are $\mp \text{sign}(\Omega) i \delta_{ss'}$, and $  \mp s_\nu i \delta_{ss'}$, which, respectively, lead to
\begin{subequations}\label{O_pm_pi2}
\begin{align}
\hat{s}_\text{D} [- \text{sign}(\Omega) i \delta_{ss'}] = &  \sum_{s=1}^2 \int_0^{+\infty} d\Omega \int_{\mathbb{R}^2} d^2 K_\perp e^{-\beta \Omega / 2}   \left[ \hat{C}^\dagger_{s \text{L}}(\Omega,\vec{K}_\perp)\hat{D}^\dagger_{s \text{R}}(\Omega,-\vec{K}_\perp) \right.  \nonumber \\ 
& \left. - \hat{C}^\dagger_{s \text{R}}(\Omega,\vec{K}_\perp) \hat{D}^\dagger_{s \text{L}}(\Omega,-\vec{K}_\perp)  \right] , \\
\hat{s}_\text{D} [ \text{sign}(\Omega) i \delta_{ss'}] = &  \sum_{s=1}^2 \int_0^{+\infty} d\Omega \int_{\mathbb{R}^2} d^2 K_\perp e^{-\beta \Omega / 2}  \left[  - \hat{C}^\dagger_{s \text{L}}(\Omega,\vec{K}_\perp) \hat{D}^\dagger_{s \text{R}}(\Omega,-\vec{K}_\perp)  \right.  \nonumber \\ 
& \left. + \hat{C}^\dagger_{s \text{R}}(\Omega,\vec{K}_\perp)\hat{D}^\dagger_{s \text{L}}(\Omega,-\vec{K}_\perp)\right], \\
\hat{s}_\text{D} [- s_\nu i  \delta_{ss'}] = &  \sum_{s=1}^2 \int_0^{+\infty} d\Omega \int_{\mathbb{R}^2} d^2 K_\perp e^{-\beta \Omega / 2}  \left[ -\hat{C}^\dagger_{s \text{L}}(\Omega,\vec{K}_\perp)\hat{D}^\dagger_{s \text{R}}(\Omega,-\vec{K}_\perp) \right.  \nonumber \\ 
& \left. - \hat{C}^\dagger_{s \text{R}}(\Omega,\vec{K}_\perp) \hat{D}^\dagger_{s \text{L}}(\Omega,-\vec{K}_\perp)  \right] , \\
\hat{s}_\text{D} [ s_\nu i  \delta_{ss'}] = &  \sum_{s=1}^2 \int_0^{+\infty} d\Omega \int_{\mathbb{R}^2} d^2 K_\perp e^{-\beta \Omega / 2}  \left[  \hat{C}^\dagger_{s \text{L}}(\Omega,\vec{K}_\perp) \hat{D}^\dagger_{s \text{R}}(\Omega,-\vec{K}_\perp)  \right.  \nonumber \\ 
& + \left. \hat{C}^\dagger_{s \text{R}}(\Omega,\vec{K}_\perp)\hat{D}^\dagger_{s \text{L}}(\Omega,-\vec{K}_\perp) \right].
\end{align}
\end{subequations}
By letting the creation operators anticommute [Eq.~(\ref{anticommutating_rules_e})], Eq.~(\ref{O_pm_pi2}) becomes equivalent to
\begin{subequations}\label{O_pm_pi2_2}
\begin{align}
\hat{s}_\text{D}[- \text{sign}(\Omega) i \delta_{ss'}] = &  \sum_{s=1}^2 \int_0^{+\infty} d\Omega \int_{\mathbb{R}^2} d^2 K_\perp e^{-\beta \Omega / 2}   \left[ \hat{C}^\dagger_{s \text{L}}(\Omega,\vec{K}_\perp)\hat{D}^\dagger_{s \text{R}}(\Omega,-\vec{K}_\perp) \right.  \nonumber \\ 
& \left. + \hat{D}^\dagger_{s \text{L}}(\Omega,-\vec{K}_\perp)\hat{C}^\dagger_{s \text{R}}(\Omega,\vec{K}_\perp)   \right] , \label{O_pm_pi2_a}\\
\hat{s}_\text{D}[\text{sign}(\Omega) i \delta_{ss'}] = &  \sum_{s=1}^2 \int_0^{+\infty} d\Omega \int_{\mathbb{R}^2} d^2 K_\perp e^{-\beta \Omega / 2}   \left[  \hat{D}^\dagger_{s \text{R}}(\Omega,-\vec{K}_\perp) \hat{C}^\dagger_{s \text{L}}(\Omega,\vec{K}_\perp) \right.  \nonumber \\ 
&  \left.+\hat{C}^\dagger_{s \text{R}}(\Omega,\vec{K}_\perp)\hat{D}^\dagger_{s \text{L}}(\Omega,-\vec{K}_\perp) \right], \label{O_pm_pi2_b} \\
\hat{s}_\text{D} [- s_\nu  i \delta_{ss'}] = &  \sum_{s=1}^2 \int_0^{+\infty} d\Omega \int_{\mathbb{R}^2} d^2 K_\perp e^{-\beta \Omega / 2}   \left[\hat{D}^\dagger_{s \text{R}}(\Omega,-\vec{K}_\perp) \hat{C}^\dagger_{s \text{L}}(\Omega,\vec{K}_\perp)\right.  \nonumber \\ 
& \left. +  \hat{D}^\dagger_{s \text{L}}(\Omega,-\vec{K}_\perp) \hat{C}^\dagger_{s \text{R}}(\Omega,\vec{K}_\perp) \right] ,\label{O_pm_pi2_c} \\
\hat{s}_\text{D} [ s_\nu i \delta_{ss'}] = &  \sum_{s=1}^2 \int_0^{+\infty} d\Omega \int_{\mathbb{R}^2} d^2 K_\perp e^{-\beta \Omega / 2}  \left[ \hat{C}^\dagger_{s \text{L}}(\Omega,\vec{K}_\perp) \hat{D}^\dagger_{s \text{R}}(\Omega,-\vec{K}_\perp)  \right.  \nonumber \\ 
& + \left. \hat{C}^\dagger_{s \text{R}}(\Omega,\vec{K}_\perp)\hat{D}^\dagger_{s \text{L}}(\Omega,-\vec{K}_\perp)\right]. \label{O_pm_pi2_d}
\end{align}
\end{subequations}

Notice that result that we obtained for fermionic fields is very similar to the bosonic case. Indeed, the Minkowski vacuum of scalars in Rindler spacetimes is equal to Eq.~(\ref{0_M_0_LR}), but with $\hat{s}_\text{D}$ replaced by the operator $\hat{o}_\text{S}$ defined by Eq.~(\ref{SS_S_31}), where $\hat{A}_\nu(\Omega,\vec{K}_\perp)$ and $\hat{B}_\nu(\Omega,\vec{K}_\perp)$ are annihilators of scalar particles and antiparticles. Such operators commute. This means that the order between $\hat{A}_\nu(\Omega,\vec{K}_\perp)$ and $\hat{B}_{\bar{\nu}}(\Omega,-\vec{K}_\perp)$ can be switched to give the following equivalent equations
\begin{subequations}\label{Theta_B_2}
\begin{align}
\hat{s}_\text{S} = &  \int_0^{+\infty} d\Omega \int_{\mathbb{R}^2} d^2 K_\perp e^{-\beta \Omega / 2} \nonumber \\ 
& \times \left[\hat{B}^\dagger_\text{R}(\Omega,-\vec{K}_\perp) \hat{A}^\dagger_\text{L}(\Omega,\vec{K}_\perp) + \hat{A}^\dagger_\text{R}(\Omega,\vec{K}_\perp) \hat{B}^\dagger_\text{L}(\Omega,-\vec{K}_\perp) \right], \label{Theta_B_2_a}\\
 \hat{s}_\text{S} = &  \int_0^{+\infty} d\Omega \int_{\mathbb{R}^2} d^2 K_\perp e^{-\beta \Omega / 2}  \nonumber \\ 
& \times \left[ \hat{B}^\dagger_\text{R}(\Omega,-\vec{K}_\perp) \hat{A}^\dagger_\text{L}(\Omega,\vec{K}_\perp)+ \hat{B}^\dagger_\text{L}(\Omega,-\vec{K}_\perp) \hat{A}^\dagger_\text{R}(\Omega,\vec{K}_\perp ) \right] , \label{Theta_B_2_b} \\
 \hat{s}_\text{S} = &  \int_0^{+\infty} d\Omega \int_{\mathbb{R}^2} d^2 K_\perp e^{-\beta \Omega / 2}  \nonumber \\ 
& \times \left[ \hat{A}^\dagger_\text{L}(\Omega,\vec{K}_\perp)\hat{B}^\dagger_\text{R}(\Omega,-\vec{K}_\perp) + \hat{A}^\dagger_\text{R}(\Omega,\vec{K}_\perp ) \hat{B}^\dagger_\text{L}(\Omega,-\vec{K}_\perp) \right] . \label{Theta_B_2_c}
\end{align}
\end{subequations}

For Dirac fields, such an equivalence does not occur because of the anticommuting property of the creation operators [Eq.~(\ref{anticommutating_rules_e})]. Indeed, any swap between creation operators generates a minus sign. However, any of these minus signs can be canceled out by a change of spin basis. One can see this in Eqs.~(\ref{O_pm_pi2_a}), (\ref{O_pm_pi2_b}), (\ref{O_pm_pi2_c}) and (\ref{O_pm_pi2_d}), which are different representations of $| 0_\text{M} \rangle$ that are equivalent up to a change of spin basis. By comparing Eqs.~(\ref{O_pm_pi2_a}), (\ref{O_pm_pi2_b}), (\ref{O_pm_pi2_c}) and (\ref{O_pm_pi2_d}) with Eqs.~(\ref{SS_S_31}), (\ref{Theta_B_2_a}), (\ref{Theta_B_2_b}) and (\ref{Theta_B_2_c}), respectively, one can see a complete analogy between scalar and Dirac fields. 

\subsection{Conclusions} \label{Frame_dependent_content_of_particles_Dirac_Conclusions}

We derived the representation of the Minkowski vacuum $| 0_\text{M} \rangle$ in the Rindler spacetime for Dirac fields [Eqs.~(\ref{0_M_0_LR}) and (\ref{O})]. The result is a two modes squeezed state that pairs particle modes of one wedge with antiparticle modes of the other wedge. At variance with the scalar case, the coupling also occurs with respect to the spin number $s$. The coupling matrix $   M_{\nu ss'} (\Omega, \vec{K}_\perp)$ can be diagonalized by suitable choices for the spin basis of the Rindler-Dirac modes [Eq.~(\ref{O_pm_pi2_2})].

By computing the partial trace of $| 0_\text{M} \rangle \langle 0_\text{M} |$ with respect to the left wedge, we derived the statistical operator representing the Minkowski vacuum in the right wedge. This gives a complete description of the state seen by the accelerated observer with acceleration $c^2 a$. The result is a fermionic thermal state $\exp( - \beta \hat{H}_\text{R}/\hbar)$, with $\beta = 2 \pi/ca$ and $ \hat{H}_\text{R}$ as the Hamiltonian in the right wedge. The consequent thermal distribution of fermionic particles is at the origin of the Unruh effect for Dirac fields.

\chapter{Minkowski particles in accelerated frame}\label{Minkowski_particles_in_accelerated_frame}

\textit{This chapter is based on and contains material from Ref.~\citeRF{PhysRevD.106.045013}.}

\section{Introduction}

In Chap.~\ref{Frame_dependent_content_of_particles}, we derived the representation of the Minkowski vacuum $\Omega_\text{M}$ as seen by the accelerated observer. We showed that for both scalar and Dirac field, the representative of $\Omega_\text{M}$ is the thermal state following bosonic [Secs.~\ref{Unruh_effect_for_scalar_fields_11} and \ref{Unruh_effect_for_scalar_fields_31}] or fermionic thermal [Sec.~\ref{Unruh_effect_for_Dirac_fields}] distribution. This is at the origin of the Unruh effect.

In addition to the Minkowski vacuum, we gave a general prescription to describe any Minkowski particle state from the point of view of the accelerated observer. If the inertial observer Alice prepares a state by means of Minkowski particles, we know how the accelerated observer Rob would describe such a state by means of right Rindler particles. The prescription consists of using the Bogoliubov transformations [Eqs.~(\ref{Bogolyubov_transformation}), (\ref{Rindler_Bogoliubov_transformations}) and (\ref{Bogoliubov_transformations_3_Rindler_4})] and the representation of the Minkowski vacuum in the left and right Rindler frames [Eqs.~(\ref{Rindler_vacuum_to_Minkowski}), (\ref{D}), (\ref{Rindler_vacuum_to_Minkowski_unitary_operator}), (\ref{Rindler_vacuum_to_Minkowski_Dirac}) and Eq.~(\ref{OO})] to represent the Minkowski particle state as an element of the left and right Rindler frame; then one performs the partial trace $\text{Tr}_\text{L}$ to obtain the representation of the state in the right Rindler frame.

An example is given by Eq.~(\ref{single_particle_Minkowski_Rindler_Tr}), which describe a scalar Minkowski single particle as seen by the accelerated observer. The method can also be applied to more complex states such as $n$-particles and states with indefinite number of Minkowski particles. However, the practical implementation of the procedure appears to be tedious or hard to be carried out. In particular, we still lack of a simple way to perform the partial trace $\text{Tr}_\text{L}$ on representatives of Minkowski particles in left and right Rindler frame.

Computing the partial trace $\text{Tr}_\text{L}$ requires a series of non-trivial theoretical properties arising from the transformation of the state from the inertial to the accelerated frame. Here we show that these rules can be formulated in a way such that one can build algorithmically a general expression of states with arbitrary number of particles.

In Sec.~\ref{MinkowskiFock_states_in_accelerated_frames}, we formulate an algorithmic procedure to derive the representation of Minkowski states with arbitrary number of particles as seen by the accelerated observer. The method comprise a general simple way to compute the partial trace $\text{Tr}_\text{L}$. As a result, we obtain the statistical operators representing the states in the right Rindler-Fock space $\mathcal{H}_\text{R}$. For practical purposes, we consider 1+1 massless scalar real fields, but the theory can also be extended to 3+1 massive scalar complex fields.

In Sec.~\ref{Wigner_formulation_of_Minkowski_particle_states_for_accelerated_observers}, we derive an explicit Wigner formulation of Minkowski particle states seen by accelerated observers. The method enables to derive mean values of Rindler observables for Minkowski particles by means of derivatives of the Wigner characteristic function.

\section{Statistical operators}\label{MinkowskiFock_states_in_accelerated_frames}

Here, we consider massless scalar real fields in 1+1 dimensions, presented at the beginning of Sec.~\ref{Massless_scalar_field_in_11_spacetime}. By using the notation of Part~\ref{Relativistic_and_nonrelativistic_quantum_fields}, we identify Minkowski-Fock states by means of wave functions $\phi_n(\textbf{k}_n)$, where, in this case, $\textbf{k}_n = (k_1, \dots, k_n)$ is a collection of $n$ one dimensional momenta. The Minkowski-Fock state is
\begin{equation}\label{free_state_decomposition_11}
| \phi \rangle  = \sum_{n=0}^\infty \int_{\mathbb{R}^n} d^n \textbf{k}_n \tilde{\phi}_n (\textbf{k}_n) | \textbf{k}_n \rangle,
\end{equation}
with
\begin{equation}\label{Minkowski_Fock_state_11}
| \textbf{k}_n \rangle = \begin{cases}
| 0_\text{M} \rangle & \text{if } n=0 \\
\frac{1}{\sqrt{n!}} \prod_{i=1}^n \hat{a}^\dagger (k_i) | 0_\text{M} \rangle  & \text{if } n>0 
\end{cases} .
\end{equation}
Equation (\ref{free_state_decomposition_11}) is the equivalent of Eq.~(\ref{free_state_decomposition}) in 1+1 dimensions.

The representation of the Minkowski-Fock state $| \phi \rangle$ in the left and right Rindler-Fock space $\mathcal{H}_{\text{L}, \text{R}}$ can be obtained by plugging the Bogoliubov transformation (\ref{Bogolyubov_transformation}) in Eq.~(\ref{Minkowski_Fock_state_11}). By replacing the $\hat{a}^\dagger (k_i)$ operator appearing in Eq.~(\ref{Minkowski_Fock_state_11}) with the adjoint of right hand side of Eq.~(\ref{Bogolyubov_transformation}), one obtains
\begin{equation}\label{Minkowski_Fock_state_11_2}
| \phi \rangle  = \left\lbrace \tilde{\phi}_0 + \sum_{n=1}^\infty  \int_{\mathbb{R}^n} d^n \textbf{k}_n  \frac{\tilde{\phi}_n (\textbf{k}_n)}{\sqrt{n!}}  \prod_{i=1}^n \left[ \hat{a}_\text{L}^\dagger(k_i) + \hat{a}_\text{R}^\dagger(k_i)  \right] \right\rbrace | 0_\text{M} \rangle,
\end{equation}
with
\begin{subequations}
\begin{align}
& \hat{a}_\text{L}(k) = \int_{\mathbb{R}} d K \left[ \alpha(k,K)\hat{A}_\text{L}(K) - \beta^*(k,K) \hat{A}^\dagger_\text{L}(K) \right], \\
& \hat{a}_\text{R}(k) = \int_{\mathbb{R}} d K \left[ \alpha^*(k,K) \hat{A}_\text{R}(K) - \beta(k,K) \hat{A}^\dagger_\text{R}(K) \right]\label{A_R}
\end{align}
\end{subequations}
as, respectively, left Rindler and right Rindler operators. By also considering the Rindler-Fock representation of the Minkowski vacuum (\ref{Rindler_vacuum_to_Minkowski}) with $\hat{S}_\text{S}$ given by Eq.~(\ref{D}), Eq.~(\ref{Minkowski_Fock_state_11_2}) can be seen as an element of 
$\mathcal{H}_{\text{L}, \text{R}}$ and leads to a generalization of Eq.~(\ref{single_particle_Minkowski_Rindler}) by including any Minkowski state.

The representation of the Minkowski-Fock state $| \phi \rangle$ in the right-Rindler-Fock space $\mathcal{H}_\text{R}$ can, instead, be obtained by computing the partial trace $\text{Tr}_\text{L}$ of the pure state $| \phi \rangle \langle \phi |$. The resulting statistical operator $\hat{\rho}_\phi = \text{Tr}_\text{L} (| \phi \rangle \langle \phi |)$ describes the Minkowski-Fock state as seen by the accelerated observer.

In this section, we compute the partial trace $\text{Tr}_\text{L} (| \phi \rangle \langle \phi |)$ for any $| \phi \rangle \in \mathcal{H}_\text{M}$. As a result, we derive the explicit expression for the statistical operator $\hat{\rho}_\phi = \text{Tr}_\text{L} (| \phi \rangle \langle \phi |)$ as an element of the right-Rindler-Fock space $\mathcal{H}_\text{R}$. We detail the dependence of $\hat{\rho}_\phi$ with respect to $\phi_n(\textbf{k}_n)$. This provides a comprehensive description of Minkowski-Fock states as seen by accelerated observed.

The method and the result are outlined in Sec.~\ref{Method_and_result}; in that subsection, we show the explicit expression for $\hat{\rho}_\phi$ and a sketch of the proof. A detailed derivation of $\hat{\rho}_\phi$ is provided in Sec.~\ref{Explicit_derivation_of_the_result}, instead.

\subsection{Method and result}\label{Method_and_result}

To derive the explicit expression for $\hat{\rho}_\phi = \text{Tr}_\text{L} (| \phi \rangle \langle \phi |)$ we only make use of a couple of identities that have been derived in Sec.~\ref{Massless_scalar_field_in_11_spacetime}. In particular, we consider Eq.~(\ref{alpha_to_beta}) to relate Bogoliubov coefficients to each other and Eq.~(\ref{vacuum_state_property}) to relate Rindler operators of one wedge acting on the Minkowski vacuum $| 0_\text{M} \rangle$ to Rindler operators of the other wedge acting on $| 0_\text{M} \rangle$. Also, we use the already-known result that $\text{Tr}_\text{L} (| 0_\text{M} \rangle \langle 0_\text{M} |) = \hat{\rho}_0$ is a thermal state in the right-Rindler-Fock space $\mathcal{H}_\text{R}$.

The method is outlined by the following steps.
\begin{enumerate}
\item Consider the right hand side of Eq.~(\ref{Minkowski_Fock_state_11_2}) to put $|\phi\rangle \langle \phi |$ in the form of a combination of chains of left and right Rindler operators acting on the left and right side of $|0_\text{M} \rangle \langle 0_\text{M} |$. \label{prescription_2}
\item Convert all left Rindler operators acting on $|0_\text{M} \rangle \langle 0_\text{M} |$ into right Rindler operators using Eq.~(\ref{vacuum_state_property}), so that $|\phi\rangle \langle \phi |$ is put in the form of a combination of chains of right Rindler operators acting on the left and right side of $| 0_\text{M} \rangle \langle 0_\text{M} |$. \label{prescription_3}
\item Use the Wick theorem \cite{PhysRev.80.268} to normal order the chain of Rindler operators acting on the left and right side of $|0_\text{M} \rangle \langle 0_\text{M} |$.\label{prescription_3_5}
\item Perform the partial trace over the left wedge $\text{Tr}_\text{L}$, so that $\hat{\rho}$ is put in the form of a combination of chains of right Rindler operators acting on the left and right side of $\hat{\rho}_0$. \label{prescription_4}
\end{enumerate}
This procedure will be detailed in Sec.~\ref{Explicit_derivation_of_the_result}.

The result is
\begin{align} \label{varrho_step_4_final}
\hat{\rho} = &  \sum_{n=0}^\infty \sum_{{n'}=0}^\infty \sum_{\mathcal{N} \subseteq [n]} \sum_{\mathcal{N}' \subseteq [n'] }  \sum_{\mathcal{N}_+ \subseteq \mathcal{N}} \sum_{\mathcal{N}'_+ \subseteq \mathcal{N}' } \int_{\mathbb{R}^n} d^n \textbf{k}_n  \int_{\mathbb{R}^{n'}} d^{n'} \textbf{k}'_{n'} \frac{\tilde{\phi}_n (\textbf{k}_n) \tilde{\phi}_{n'}^*(\textbf{k}'_{n'})}{\sqrt{n!n'!}} \nonumber \\
& \times   C_\mathsf{K} \left( \{ k_i \}_{i \in [n] \setminus \mathcal{N}} \right) C_\mathsf{K} \left( \{ k'_{i'} \}_{i' \in [n'] \setminus \mathcal{N}'} \right) \nonumber \\
& \times \prod_{j_- \in\mathcal{N} \setminus \mathcal{N}_+} \hat{A}_-^\dagger(k_{j_-})  \prod_{j_+\in\mathcal{N}_+} \hat{A}_+^\dagger(k_{j_+})   \hat{\rho}_0  \prod_{j'_+\in\mathcal{N}'_+} \hat{A}_+(k'_{j'_+})  \prod_{j'_-\in\mathcal{N}'\setminus\mathcal{N}'_+} \hat{A}_-(k'_{j'_-}),
\end{align}
with $[n] = \{ 1, \dots, n \}$ as the set of the first $n$ natural numbers and
\begin{subequations}\label{Ap_Am}
\begin{align}
&\hat{A}_+(k) =  \int_{\mathbb{R}}dK 2 \sinh \left(\frac{c \beta |K|}{2} \right) \alpha(k,K)\hat{A}^\dagger_\text{R}(K), \\
 & \hat{A}_-(k) =  \int_{\mathbb{R}}dK 2 \sinh \left(\frac{c \beta |K|}{2} \right) \beta^*(k,K)\hat{A}_\text{R}(K).
\end{align}
\end{subequations}
The sums $ \sum_{\mathcal{N} \subseteq [n]} $, $ \sum_{\mathcal{N}' \subseteq [n'] }$, $ \sum_{\mathcal{N}_+ \subseteq \mathcal{N}} $, $ \sum_{\mathcal{N}'_+ \subseteq \mathcal{N}' }$ run over all possible subsets $\mathcal{N} \subseteq [n]$, $\mathcal{N}' \subseteq [n']$, $\mathcal{N}_+ \subseteq \mathcal{N}$ and $\mathcal{N}'_+ \subseteq \mathcal{N}'$. For any ordered sequence $\mathcal{U} = \{ k_i \}_{i=1}^n = \{ k_1, \dots, k_n \}$, the coefficient $C_\mathsf{K} (\mathcal{U})$ is defined to be $0$ when the cardinality of $\mathcal{U}$ is even; conversely, if the cardinality of $\mathcal{U}$ is odd, then $C_\mathsf{K} (\mathcal{U})$ has  the following combinatorial expression
\begin{equation}\label{C_K}
C_\mathsf{K} \left( \{ k_i \}_{i = 1}^n \right)= \frac{1}{2^n} \sum_{\mathsf{P} \in S_n} \prod_{i=1}^{n/2} c_\mathsf{K} \left( k_{\mathsf{P}(2 i - 1)}, k_{\mathsf{P}(2 i)} \right),
\end{equation}
where the sum $\sum_{\mathsf{P} \in S_n}$ runs over all permutations of sets with $n$ elements. The coefficient $c_\mathsf{K} (k_i, k_j)$, instead, is defined by
\begin{equation}
c_\mathsf{K} (k_i, k_j)  = - \int_{\mathbb{R}}dK 2 \sinh \left( \frac{c \beta |K|}{2} \right) \left[ \beta^*(k_i,K) \beta(k_j,K) + \beta(k_i,K) \beta^*(k_j,K) \right].
\end{equation}

A straightforward extension of Eq.~(\ref{varrho_step_4_final}) to massive scalar complex fields in 3+1 dimensions is made possible by noticing that the few ingredients used in steps \ref{prescription_2}-\ref{prescription_4} for 1+1 massless scalar real fields are also present in the 3+1 massive complex case. In particular, the equivalent of Eqs.~(\ref{alpha_to_beta}) and Eq.~(\ref{vacuum_state_property}) are given by Eqs.~(\ref{alpha_to_beta_31}) and Eq.~(\ref{vacuum_state_property_31}), respectively.

\subsection{Explicit derivation of the result}\label{Explicit_derivation_of_the_result}

We now detail the prescription presented by steps \ref{prescription_2}-\ref{prescription_4} in Sec.~\ref{Method_and_result}. Firstly, assume that the chain of $\hat{a}^\dagger (k_i)$ operators appearing in Eq.~(\ref{Minkowski_Fock_state_11}) follows a fixed order. For instance, consider $\prod_{i=1}^n \hat{a}^\dagger (k_i) = \hat{a}^\dagger (k_1) \dots \hat{a}^\dagger (k_n)$. Since these operators commute, any other order works as well. However, we need to make a starting choice about their order to recursively apply Eq.~(\ref{vacuum_state_property}) as detailed by step \ref{prescription_2}. Hence, by convention, assume that the operators $\hat{a}^\dagger (k_i)$ are ordered monotonically with respect to the index $i$.

By choosing the order $\prod_{i=1}^n \hat{a}^\dagger (k_i) = \hat{a}^\dagger (k_1) \dots \hat{a}^\dagger (k_n)$ for the Minkowski operators in Eq.~(\ref{Minkowski_Fock_state_11}), we can only use Eq.~(\ref{vacuum_state_property}) on $\hat{a}^\dagger_\text{L} (k_n)$ as it is the only  $\hat{a}^\dagger_\text{L}$ operator directly acting on $| 0_\text{M} \rangle$ in Eq.~(\ref{Minkowski_Fock_state_11_2}). This leads to
\begin{align}\label{Minkowski_Fock_state_11_3}
| \phi \rangle  = & \left\lbrace \tilde{\phi}_0 +\int_{\mathbb{R}} dk  \tilde{\phi}_1 (k) \left[ \hat{a}_\text{L}^{\prime \dagger}(k) + \hat{a}_\text{R}^\dagger(k)  \right] \right. \nonumber \\
& \left. + \sum_{n=2}^\infty  \int_{\mathbb{R}^n} d^n \textbf{k}_n \frac{\tilde{\phi}_n (\textbf{k}_n)}{\sqrt{n!}}  \prod_{i=1}^{n-1} \left[ \hat{a}_\text{L}^\dagger(k_i) + \hat{a}_\text{R}^\dagger(k_i)  \right]  \hat{a}^{\prime \dagger}(k_n)  \right\rbrace | 0_\text{M} \rangle,
\end{align}
with
\begin{subequations}
\begin{align}
\hat{a}'(k) = & \hat{a}'_\text{L}(k) + \hat{a}_\text{R}(k),\\
\hat{a}'_\text{L}(k) = & \int_{\mathbb{R}}dK \left[\exp \left( \frac{c \beta |K|}{2} \right) \alpha(k,K)\hat{A}^\dagger_\text{R}(K) \right. \nonumber \\
& \left. - \exp \left(-\frac{ c \beta |K|}{2} \right) \beta^*(k,K) \hat{A}_\text{R}(K)\right].\label{A_L}
\end{align}
\end{subequations}

We now use the commutation relation $[\hat{a}_\text{L}^\dagger, \hat{a}^{\prime \dagger}] = 0$, coming from $[\hat{A}_\text{L}, \hat{A}_\text{R}] = [\hat{A}_\text{L}, \hat{A}_\text{R}^\dagger] = 0$, to let $\hat{a}^\dagger_\text{L} (k_{n-1})$ act on $| 0_\text{M} \rangle$, i.e.,
\begin{align}\label{Minkowski_Fock_state_11_4}
& | \phi \rangle  = \left\lbrace \tilde{\phi}_0 +\int_{\mathbb{R}} dk  \tilde{\phi}_1 (k) \left[ \hat{a}_\text{L}^{\prime \dagger}(k) + \hat{a}_\text{R}^\dagger(k)  \right] \right. \nonumber\\
& + \int_{\mathbb{R}^2} d^2 \textbf{k}_2  \frac{\tilde{\phi}_2 (\textbf{k}_2)}{\sqrt{2}} \left[  \hat{a}^{\prime \dagger}(k_2)   \hat{a}_\text{L}^\dagger(k_1) + \hat{a}_\text{R}^\dagger(k_1) \hat{a}^{\prime \dagger}(k_2)  \right] + \sum_{n=3}^\infty \int_{\mathbb{R}^n} d^n \textbf{k}_n  \frac{\tilde{\phi}_n (\textbf{k}_n)}{\sqrt{n!}}   \nonumber \\
& \left. \times \prod_{i=1}^{n-2} \left[ \hat{a}_\text{L}^\dagger(k_i) + \hat{a}_\text{R}^\dagger(k_i)  \right]  \left[ \hat{a}^{\prime \dagger}(k_n) \hat{a}_\text{L}^\dagger(k_{n-1}) + \hat{a}_\text{R}^\dagger(k_{n-1}) \hat{a}^{\prime \dagger}(k_n)  \right]  \right\rbrace | 0_\text{M} \rangle.
\end{align}
This allows us to use Eq.~(\ref{vacuum_state_property}) on $\hat{a}^\dagger_\text{L} (k_{n-1})$ and obtain
\begin{align}\label{Minkowski_Fock_state_11_5}
& | \phi \rangle  = \left\lbrace \tilde{\phi}_0 +\int_{\mathbb{R}} dk  \tilde{\phi}_1 (k) \left[ \hat{a}_\text{L}^{\prime \dagger}(k) + \hat{a}_\text{R}^\dagger(k)  \right] \right. \nonumber\\
& + \int_{\mathbb{R}^2} d^2 \textbf{k}_2  \frac{\tilde{\phi}_2 (\textbf{k}_2)}{\sqrt{2}} \left[  \hat{a}^{\prime \dagger}(k_2)   \hat{a}_\text{L}^{\prime \dagger}(k_1) + \hat{a}_\text{R}^\dagger(k_1) \hat{a}^{\prime \dagger}(k_2)  \right] + \sum_{n=3}^\infty \int_{\mathbb{R}^n} d^n \textbf{k}_n  \frac{\tilde{\phi}_n (\textbf{k}_n)}{\sqrt{n!}}   \nonumber \\
& \left. \times \prod_{i=1}^{n-2} \left[ \hat{a}_\text{L}^\dagger(k_i) + \hat{a}_\text{R}^\dagger(k_i)  \right]  \left[ \hat{a}^{\prime \dagger}(k_n) \hat{a}_\text{L}^{\prime \dagger}(k_{n-1}) + \hat{a}_\text{R}^\dagger(k_{n-1}) \hat{a}^{\prime \dagger}(k_n)  \right]  \right\rbrace | 0_\text{M} \rangle.
\end{align}

The upshot is that left Rindler operators $\hat{a}_\text{L}^\dagger(k_{n-1})$ and $\hat{a}_\text{L}^\dagger(k_n)$ appearing in Eq.~(\ref{Minkowski_Fock_state_11_2}) are replaced with right Rindler operators $\hat{a}_\text{L}^{\prime \dagger}(k_{n-1})$ and $\hat{a}_\text{L}^{\prime \dagger}(k_n)$. The price to be paid is that the order of $\hat{a}_\text{R}^\dagger(k_{n-1})$, $\hat{a}_\text{R}^\dagger(k_n)$, $\hat{a}_\text{L}^{\prime \dagger}(k_{n-1})$ and $\hat{a}_\text{L}^{\prime \dagger}(k_n)$ is not monotonic with respect to the index $i$ anymore.

One can recursively use the same method to replace all left Rindler operators $\hat{a}_\text{L}^\dagger(k_i)$ with right Rindler operators $\hat{a}_\text{L}^{\prime \dagger}(k_i)$. The result is
\begin{align}
| \phi \rangle \langle \phi |  = & \left[ \tilde{\phi}_0 + \sum_{n=1}^\infty \int_{\mathbb{R}^n} d^n \textbf{k}_n \frac{\tilde{\phi}_n (\textbf{k}_n)}{\sqrt{n!}}  \mathsf{K} \left( \prod_{i=1}^n \hat{a}^{\prime \dagger}(k_i) \right) \right] | 0_\text{M} \rangle \langle 0_\text{M} | \nonumber \\
& \times \left[ \tilde{\phi}_0^* + \sum_{n'=1}^\infty \int_{\mathbb{R}^{n'}} d^{n'} \textbf{k}'_{n'} \frac{\tilde{\phi}_{n'}^* (\textbf{k}'_{n'})}{\sqrt{n'!}} \mathsf{K} \left( \prod_{i'=1}^{n'} \hat{a}'(k'_{i'}) \right) \right],
\end{align}
where $\mathsf{K}$ defines an ordering rule for the operators $\hat{a}'_\text{L}(k)$, $\hat{a}_\text{R}(k)$ and their adjoint operators depending on the ordering for the $\hat{a}^\dagger(k_i)$ operators that we used in Eq.~(\ref{Minkowski_Fock_state_11}). Since we chose the monotonic order with respect to the index $i$, the ordering operator $\mathsf{K}$ is defined by
\begin{subequations}
\begin{align}
& \mathsf{K}\left(\hat{a}'_\text{L}(k_i) \hat{a}_\text{R}(k_j)\right) = \mathsf{K}\left(\hat{a}_\text{R}(k_j) \hat{a}'_\text{L}(k_i)\right) = \hat{a}'_\text{L}(k_i) \hat{a}_\text{R}(k_j),\\
& \mathsf{K}\left(\hat{a}^{\prime \dagger}_\text{L}(k_i) \hat{a}^\dagger_\text{R}(k_j)\right) = \mathsf{K}\left(\hat{a}^\dagger_\text{R}(k_j) \hat{a}^{\prime \dagger}_\text{L}(k_i)\right) = \hat{a}^\dagger_\text{R}(k_i) \hat{a}^{\prime \dagger}_\text{L}(k_j),\\
& \mathsf{K}\left(\hat{a}'_\text{L}(k_i) \hat{a}'_\text{L}(k_j)\right) = \begin{cases}
\hat{a}'_\text{L}(k_i) \hat{a}'_\text{L}(k_j) & \text{if } i<j \\
\hat{a}'_\text{L}(k_j) \hat{a}'_\text{L}(k_i) & \text{if } i>j 
\end{cases},\\
& \mathsf{K}\left(\hat{a}_\text{R}(k_i) \hat{a}_\text{R}(k_j)\right) = \begin{cases}
\hat{a}_\text{R}(k_j) \hat{a}_\text{R}(k_i) & \text{if } i<j \\
\hat{a}_\text{R}(k_i) \hat{a}_\text{R}(k_j) & \text{if } i>j 
\end{cases},\\
& \mathsf{K}\left(\hat{a}^{\prime \dagger}_\text{L}(k_i) \hat{a}^{\prime \dagger}_\text{L}(k_j)\right) = \begin{cases}
\hat{a}^{\prime \dagger}_\text{L}(k_j) \hat{a}^{\prime \dagger}_\text{L}(k_i) & \text{if } i<j \\
\hat{a}^{\prime \dagger}_\text{L}(k_i) \hat{a}^{\prime \dagger}_\text{L}(k_j) & \text{if } i>j 
\end{cases},\\
& \mathsf{K} \left(\hat{a}^\dagger_\text{R}(k_i) \hat{a}^\dagger_\text{R}(k_j)\right) = \begin{cases}
\hat{a}^\dagger_\text{R}(k_i) \hat{a}^\dagger_\text{R}(k_j) & \text{if } i<j \\
\hat{a}^\dagger_\text{R}(k_j) \hat{a}^\dagger_\text{R}(k_i) & \text{if } i>j 
\end{cases}.
\end{align}
\end{subequations}
For practical purposes let us write
\begin{align} \label{varrho_step_1}
| \phi \rangle \langle \phi |  = & \sum_{n=0}^\infty \sum_{n'=0}^\infty \int_{\mathbb{R}^n} d^n \textbf{k}_n  \int_{\mathbb{R}^{n'}} d^{n'} \textbf{k}'_{n'} \frac{\tilde{\phi}_n (\textbf{k}_n) \tilde{\phi}_{n'}^* (\textbf{k}'_{n'})}{\sqrt{n!n'!}}  \nonumber \\
& \times \mathsf{K} \left( \prod_{i=1}^n \hat{a}^{\prime \dagger}(k_i) \right) | 0_\text{M} \rangle \langle 0_\text{M} | \mathsf{K} \left( \prod_{i'=1}^{n'} \hat{a}'(k'_{i'}) \right),
\end{align}
and assume that $\prod_{i=1}^0 a_i = 1$ for any sequence $a_i$. 

The operators acting on the left and on the right of $|0_\text{M} \rangle \langle 0_\text{M} |$ in Eq.~(\ref{varrho_step_1}) are combinations of chains of right Rindler operators. Each chain can be rewritten using the Wick theorem by considering $\hat{A}^\dagger_\text{R}(K)$ and $\hat{A}_\text{R}(K)$ as creation and annihilation operators for the normal ordering $\mathsf{N}$, defined by
\begin{equation}
\mathsf{N} \left( \hat{A}^\dagger_\text{R}(K) \hat{A}_\text{R}(K') \right) = \mathsf{N} \left( \hat{A}_\text{R}(K') \hat{A}^\dagger_\text{R}(K)  \right) = \hat{A}^\dagger_\text{R}(K) \hat{A}_\text{R}(K').
\end{equation}
We indicate by $\mathsf{C}_\mathsf{N}$ the real function that act on any chain of right Rindler operators and compute the sum of all the full contractions of such a chain, with the following fundamental contractions
\begin{subequations}
\begin{align}
& \mathsf{C}_\mathsf{N} \left( \hat{A}_\text{R}(K) \hat{A}^\dagger_\text{R}(K') \right) = \delta(K-K'), \\
& \mathsf{C}_\mathsf{N} \left( \hat{A}^\dagger_\text{R}(K) \hat{A}_\text{R}(K') \right) = \mathsf{C}_\mathsf{N} \left( \hat{A}^\dagger_\text{R}(K) \hat{A}^\dagger_\text{R}(K') \right) = \mathsf{C}_\mathsf{N} \left( \hat{A}_\text{R}(K) \hat{A}_\text{R}(K') \right) = 0.
\end{align}
\end{subequations}

The combination between the Wick theorem and the $\mathsf{K}$-ordering gives
\begin{subequations} \label{normal_ordering_partial}
\begin{align}
\mathsf{K} \left( \prod_{i=1}^n \hat{a}'(k_i) \right)  = & \sum_{\mathcal{N} \subseteq [n] } \mathsf{C}_\mathsf{N} \left( \mathsf{K} \left( \prod_{i \in [n] \setminus \mathsf{S}_\mathsf{N}} \hat{a}'(k_i) \right) \right) \mathsf{N} \left( \mathsf{K} \left( \prod_{j \in \mathcal{N}} \hat{a}'(k_j) \right) \right) \nonumber \\
= & \sum_{\mathcal{N} \subseteq [n] } \mathsf{C}_\mathsf{K} \left( \prod_{i \in [n] \setminus \mathcal{N}}  \hat{a}'(k_i) \right) \mathsf{N} \left(  \prod_{j \in \mathcal{N}} \hat{a}'(k_j) \right),\\
\mathsf{K} \left( \prod_{i = 1}^n \hat{a}^{\prime \dagger}(k_i) \right)  = & \sum_{\mathcal{N} \subseteq [n] } \mathsf{C}_\mathsf{N} \left( \mathsf{K} \left( \prod_{i \in [n] \setminus \mathcal{N}} \hat{a}^{\prime \dagger}(k_i) \right) \right) \mathsf{N} \left( \mathsf{K} \left( \prod_{j \in \mathcal{N}} \hat{a}^{\prime \dagger}(k_j) \right) \right) \nonumber \\
= & \sum_{\mathcal{N} \subseteq [n] } \mathsf{C}_\mathsf{K} \left( \prod_{i \in [n] \setminus \mathcal{N}}  \hat{a}^{\prime \dagger}(k_i) \right) \mathsf{N} \left(  \prod_{j \in \mathcal{N}} \hat{a}^{\prime \dagger}(k_j) \right),
\end{align}
\end{subequations}
with $\mathsf{C}_\mathsf{K} = \mathsf{C}_\mathsf{N} \circ \mathsf{K}$ as a new contraction such that
\begin{subequations}\label{fundamental_contractions_A}
\begin{align}
& \mathsf{C}_\mathsf{K} \left( \hat{a}'(k_i) \hat{a}'(k_j) \right) = \mathsf{C}_\mathsf{N} \left( \mathsf{K} \left( \hat{a}'(k_i) \hat{a}'(k_j) \right) \right) = \nonumber \\
& \begin{cases}
 \mathsf{C}_\mathsf{N} \left( \hat{a}'_\text{L}(k_i) \hat{a}'_\text{L}(k_j) + \hat{a}'_\text{L}(k_i) \hat{a}_\text{R}(k_j) + \hat{a}'_\text{L}(k_j) \hat{a}_\text{R}(k_i) + \hat{a}_\text{R}(k_j) \hat{a}_\text{R}(k_i) \right)  & \text{if } i<j \\
\mathsf{C}_\mathsf{N} \left( \hat{a}'_\text{L}(k_j) \hat{a}'_\text{L}(k_i) + \hat{a}'_\text{L}(k_i) \hat{a}_\text{R}(k_j) + \hat{a}'_\text{L}(k_j) \hat{a}_\text{R}(k_i) + \hat{a}_\text{R}(k_i) \hat{a}_\text{R}(k_j) \right)  & \text{if } i>j
\end{cases} \\
& \mathsf{C}_\mathsf{K} \left( \hat{a}^{\prime \dagger}(k_i) \hat{a}^{\prime \dagger}(k_j) \right) = \mathsf{C}_\mathsf{N} \left( \mathsf{K} \left( \hat{a}^{\prime \dagger}(k_i) \hat{a}^{\prime \dagger}(k_j) \right) \right) = \nonumber \\
& \begin{cases}
 \mathsf{C}_\mathsf{N} \left( \hat{a}^{\prime \dagger}_\text{L}(k_j) \hat{a}^{\prime \dagger}_\text{L}(k_i) + \hat{a}^\dagger_\text{R}(k_j) \hat{a}^{\prime \dagger}_\text{L}(k_i) + \hat{a}^\dagger_\text{R}(k_i) \hat{a}^{\prime \dagger}_\text{L}(k_j)  + \hat{a}^\dagger_\text{R}(k_i) \hat{a}^\dagger_\text{R}(k_j) \right)  & \text{if } i<j \\
\mathsf{C}_\mathsf{N} \left( \hat{a}^{\prime \dagger}_\text{L}(k_i)\hat{a}^{\prime \dagger}_\text{L}(k_j)  + \hat{a}^\dagger_\text{R}(k_j) \hat{a}^{\prime \dagger}_\text{L}(k_i) + \hat{a}^\dagger_\text{R}(k_i) \hat{a}^{\prime \dagger}_\text{L}(k_j) + \hat{a}^\dagger_\text{R}(k_j) \hat{a}^\dagger_\text{R}(k_i) \right)  & \text{if } i>j
\end{cases}
\end{align}
\end{subequations}
Owing to Eq.~(\ref{alpha_to_beta}), Eq.~(\ref{fundamental_contractions_A}) has the following more compact form
\begin{subequations}\label{fundamental_contractions_A_2}
\begin{align}
\mathsf{C}_\mathsf{K} \left( \hat{a}'(k_i) \hat{a}'(k_j) \right)  = & - \int_{\mathbb{R}}dK 2 \sinh \left( \frac{c \beta |K|}{2} \right) \nonumber \\
& \times \left[ \beta^*(k_i,K) \beta(k_j,K) + \beta(k_i,K) \beta^*(k_j,K) \right],\\
\mathsf{C}_\mathsf{K} \left( \hat{a}^{\prime \dagger}(k_i) \hat{a}^{\prime \dagger}(k_j) \right) = & \mathsf{C}_\mathsf{K} \left( \hat{a}'(k_i) \hat{a}'(k_j) \right).
\end{align}
\end{subequations}
By using Eq.~(\ref{normal_ordering_partial}) we find a way to put Eq.~(\ref{varrho_step_1}) in normal ordering at the left and right of $|0_\text{M} \rangle \langle 0_\text{M} |$, i.e.,
\begin{align}
|\phi\rangle \langle \phi | = & \sum_{n=0}^\infty \sum_{{n'}=0}^\infty  \sum_{\mathcal{N} \subseteq [n]} \sum_{\mathcal{N}' \subseteq [n'] } \int_{\mathbb{R}^n} d^n \textbf{k}_n  \int_{\mathbb{R}^{n'}} d^{n'} \textbf{k}'_{n'} \tilde{\phi}_n (\textbf{k}_n) \tilde{\phi}_{n'}^* (\textbf{k}'_{n'}) \nonumber \\
& \times  \mathsf{C}_\mathsf{K} \left( \prod_{i \in [n] \setminus \mathcal{N}}   \hat{a}^{\prime \dagger}(k_i) \right)  \mathsf{C}_\mathsf{K} \left(  \prod_{i' \in [n'] \setminus \mathcal{N}'} \hat{a}'(k'_{i'}) \right)  \nonumber \\
& \times \mathsf{N} \left(  \prod_{j \in \mathcal{N}} \hat{a}^{\prime \dagger}(k_j) \right)  |0_\text{M} \rangle \langle 0_\text{M} | \mathsf{N} \left( \prod_{j' \in \mathcal{N}'} \hat{a}'(k'_{j'}) \right).
\end{align}

Step \ref{prescription_4} gives
\begin{align} \label{varrho_step_4_1}
\hat{\rho} = & \text{Tr}_L |\psi\rangle \langle \psi | \nonumber \\
 = & \sum_{n=0}^\infty \sum_{{n'}=0}^\infty \sum_{\mathcal{N} \subseteq [n]} \sum_{\mathcal{N}' \subseteq [n'] } \int_{\mathbb{R}^n} d^n \textbf{k}_n  \int_{\mathbb{R}^{n'}} d^{n'} \textbf{k}'_{n'} \frac{\tilde{\phi}_n (\textbf{k}_n) \tilde{\phi}_{n'}^* (\textbf{k}'_{n'})}{\sqrt{n!n'!}}   \nonumber \\
& \times \mathsf{C}_\mathsf{K} \left( \prod_{i \in [n] \setminus \mathcal{N}}   \hat{a}^{\prime \dagger}(k_i) \right)  \mathsf{C}_\mathsf{K} \left(  \prod_{i' \in [n'] \setminus \mathcal{N}'} \hat{a}'(k'_{i'}) \right)  \nonumber \\
& \times \mathsf{N} \left(  \prod_{j \in \mathcal{N}} \hat{a}^{\prime \dagger}(k_j) \right)  \hat{\rho}_0 \mathsf{N} \left( \prod_{j' \in \mathcal{N}'} \hat{a}'(k'_{j'}) \right).
\end{align}

We can explicitly compute the normal ordering of Eq.~(\ref{varrho_step_4_1}) by giving a new decomposition for $\hat{a}'(k_i)$
\begin{subequations}
\begin{align}
&\hat{a}'(k) = \hat{A}_+(k) + \hat{A}_-(k),\\
&\hat{A}_+(k) =  \int_{\mathbb{R}}dK \left[ \exp \left( \frac{c \beta |K|}{2} \right) \alpha(k,K) - \beta(k,K) \right] \hat{A}^\dagger_\text{R}(K),\label{A_+}\\
&\hat{A}_-(k) =  \int_{\mathbb{R}}dK \left[ - \exp \left(-\frac{c \beta |K|}{2} \right) \beta^*(k,K) + \alpha^*(k,K) \right] \hat{A}_\text{R}(K).\label{A_-}
\end{align}
\end{subequations}
In this way Eq.~(\ref{varrho_step_4_1}) reads as
\begin{align} \label{varrho_step_4_2}
\hat{\rho} = &  \sum_{n=0}^\infty \sum_{{n'}=0}^\infty \sum_{\mathcal{N} \subseteq [n]} \sum_{\mathcal{N}' \subseteq [n'] }  \sum_{\mathcal{N}_+ \subseteq \mathcal{N}} \sum_{\mathcal{N}'_+ \subseteq \mathcal{N}' } \int_{\mathbb{R}^n} d^n \textbf{k}_n  \int_{\mathbb{R}^{n'}} d^{n'} \textbf{k}'_{n'} \nonumber \\
& \times \frac{\tilde{\phi}_n (\textbf{k}_n) \tilde{\phi}_{n'}^* (\textbf{k}'_{n'})}{\sqrt{n!n'!}}   \mathsf{C}_\mathsf{K} \left( \prod_{i \in [n] \setminus \mathcal{N}}  \hat{a}^{\prime \dagger}(k_i) \right) \mathsf{C}_\mathsf{K} \left(  \prod_{i' \in [n'] \setminus \mathcal{N}'} \hat{a}'(k'_{i'}) \right)  \nonumber \\
& \times \prod_{j_- \in\mathcal{N} \setminus \mathcal{N}_+} \hat{A}_-^\dagger(k_{j_-})  \prod_{j_+\in\mathcal{N}_+} \hat{A}_+^\dagger(k_{j_+})   \hat{\rho}_0  \prod_{j'_+\in\mathcal{N}'_+} \hat{A}_+(k'_{j'_+})  \prod_{j'_-\in\mathcal{N}'\setminus\mathcal{N}'_+} \hat{A}_-(k'_{j'_-}).
\end{align}
Owing to Eq.~(\ref{alpha_to_beta}), Eqs.~(\ref{A_+}) and (\ref{A_-}) can be put in the form of Eq.~(\ref{Ap_Am}). Also, by using the definition of contractions and by using Eq.~(\ref{fundamental_contractions_A_2}), one can prove that
\begin{equation}\label{CC_C}
 \mathsf{C}_\mathsf{K} \left( \prod_{i \in [n] \setminus \mathcal{N}}  \hat{a}^{\prime \dagger}(k_i) \right) = C_\mathsf{K} \left( \{ k_i \}_{i \in [n] \setminus \mathcal{N}} \right),
\end{equation}
with $C_\mathsf{K} \left( \{ k_i \}_{i \in [n] \setminus \mathcal{N}} \right)$ defined by Eq.~(\ref{C_K}).

\section{Wigner characteristic functions}\label{Wigner_formulation_of_Minkowski_particle_states_for_accelerated_observers}

Any particle state in the Rindler-Fock space $\mathcal{H}_\text{R}$ is generally defined by a statistical operator $\hat{\rho}$ acting on $\mathcal{H}_\text{R}$. An equivalent representation for $\hat{\rho}$ is provided by its Wigner characteristic function \cite{PhysRev.177.1857, PhysRev.177.1882, Barnett2002}, defined by
\begin{equation} \label{chi_p}
\chi^{(p)}[\xi, \xi^*] = \text{Tr} \left( \hat{\rho} \hat{D}_p[\xi, \xi^*] \right),
\end{equation}
where $\xi = \xi(K)$ is a complex function, $p$ can take values $-1$, $0$ and $+1$ and
\begin{equation}
 \hat{D}_p[\xi, \xi^*] = \exp \left( \int_{\mathbb{R}} dK \left[ \xi(K) \hat{A}^\dagger_R(K)   - \xi^*(K) \hat{A}_R(K) + \frac{p}{2} |\xi(K)|^2\right] \right).
\end{equation}
By using the canonical commutation relation $[\hat{A}_R(K), \hat{A}^\dagger_R(K')] = \delta(K-K')$ and the Baker–Campbell–Hausdorff formula, one can prove that, in particular,
\begin{subequations}
\begin{align}
& \hat{D}_{+1}[\xi, \xi^*] = \exp \left( \int_{\mathbb{R}} dK  \xi(K) \hat{A}^\dagger_R(K)\right) \exp \left(- \int_{\mathbb{R}} dK \xi^*(K) \hat{A}_R(K) \right).\label{D_+1}\\
& \hat{D}_{-1}[\xi, \xi^*] = \exp \left(- \int_{\mathbb{R}} dK \xi^*(K) \hat{A}_R(K) \right) \exp \left( \int_{\mathbb{R}} dK  \xi(K) \hat{A}^\dagger_R(K)\right).\label{D_-1}
\end{align}
\end{subequations}
Notice that is is always possible to derive the characteristic functions $\chi^{(p)}[\xi,\xi^*]$ with $p=\{ -1, 0 , +1 \}$ if at least one of them is known; for instance, one can derive $\chi^{(p)}[\xi,\xi^*]$ by knowing $\chi^{(-1)}[\xi,\xi^*]$ as follows
\begin{equation} \label{chi_p_m1}
\chi^{(p)}[\xi,\xi^*] = \exp \left( \int_{\mathbb{R}} dK \frac{p+1}{2} |\xi(K)|^2 \right) \chi^{(- 1)}[\xi,\xi^*].
\end{equation}

The reason why $\chi^{(p)}[\xi, \xi^*]$ provides a comprehensive description of the state is because any mean value of $\hat{\rho}$ can be computed by means of functional derivatives of $\chi^{(p)}[\xi, \xi^*]$; hence, all the defining information about the state are encoded in the functional dependence of $\chi^{(p)}[\xi, \xi^*]$ with respect to $\xi(K)$. For instance, the mean value of normal ordered operators $ \prod_{i=1}^n \hat{A}_\text{R}^\dagger(K_i) \prod_{j=1}^m \hat{A}_\text{R}(K'_j)$ is given by
\begin{equation}\label{normal_ordered_meanvalue_characteristic}
\text{Tr} \left( \hat{\rho} \prod_{i=1}^n \hat{A}_\text{R}^\dagger(K_i) \prod_{j=1}^m \hat{A}_\text{R}(K'_j) \right) = \left. \prod_{i=1}^n  \frac{\delta}{\delta \xi(K_i)}  \prod_{j=1}^m \left[- \frac{\delta}{\delta \xi^*(K'_j)} \right] \chi^{(+1)} [\xi,\xi^*] \right|_{\xi=0}.
\end{equation}

In a recent paper \cite{Ben-Benjamin_2020}, Ben-Benjamin, Scully, and Unruh reported the characteristic function for Minkowski-Fock states in the right and left Rindler wedges. However, to the best of our knowledge, the case of the right Rindler wedge obtained from the partial trace over the left wedge $\text{Tr}_\text{L}$ is still missing.

In this section, we investigate the general expression of the characteristic function for particle states prepared by an inertial experimenter and seen by an accelerated observer. The aim is to provide a comprehensive description for Minkowski-Fock states in the accelerated frame and a general procedure to derive mean values of Rindler observables.

This section is organized as follows. In Sec.~\ref{Method_and_result_Wigner}, we show the characteristic function $\chi^{(p)}[\xi, \xi^*]$ of a Minkowski-Fock state in the right Rindler frame; we give a list of identities needed to derive $\chi^{(p)}[\xi, \xi^*]$ and a sketch of the proof.  We also give a diagrammatic representation of $\chi^{(p)}[\xi, \xi^*]$ resulting from our combinatorial method. The results are detailed in Sec.~\ref{Examples} with the examples of Minkowski single particle and two particles states. In Sec.~\ref{Comparison_between_inertial_and_accelerated_frame} we compare the Rindler-Wigner function $\chi^{(p)}[\xi, \xi^*]$ with the characteristic function $\chi^{(p)}_\text{M}[\xi, \xi^*]$ describing the state in the Minkowski frame; we study the transformation $\chi^{(p)}_\text{M}[\xi, \xi^*] \mapsto \chi^{(p)}[\xi, \xi^*]$ from the inertial to the accelerated frame. Finally, in Sec.~\ref{Explicit_derivation_of_the_result_Wigner}, we explicitly compute $\chi^{(p)}[\xi, \xi^*]$.

\subsection{Method and result}\label{Method_and_result_Wigner}

Any Minkowski-Fock pure state $| \phi \rangle \in \mathcal{H}_\text{M}$ can be represented in the right Rindler-Fock space $\mathcal{H}_\text{R}$ by means a the statistical operator $\hat{\rho}$. The general procedure to derive $\hat{\rho}$ for any $| \phi \rangle \in \mathcal{H}_\text{M}$ has been provided in Sec.~\ref{MinkowskiFock_states_in_accelerated_frames}. In particular, Eq.~(\ref{varrho_step_4_2}) gives the explicit expression for $\hat{\rho}$ in terms of the wave function $\tilde{\phi}_n (\textbf{k}_n)$. Here, we use this result to derive the characteristic function $\chi^{(p)}[\xi, \xi^*]$ associated to $\hat{\rho}$ and defined by Eq.~(\ref{chi_p}).

We use the properties of the Minkowski vacuum state $|0_\text{M} \rangle$ and its representation in the right Rindler wedge $\hat{\rho}_0$ listed below.
\begin{itemize}
\item It is possible to move right Rindler operators acting on the left of $\hat{\rho}_0$ to its right and the other way round by using the following identity and its adjoint
\begin{equation}\label{ordering_Rindler}
\hat{A}^\dagger_\text{R}(K) \hat{\rho}_0 = e^{c \beta |K|}  \hat{\rho}_0 \hat{A}^\dagger_\text{R}(K).
\end{equation}
Equation (\ref{ordering_Rindler}) can be proven by means of Eq.~(\ref{vacuum_state_property}) as follows
\begin{align}
\hat{A}^\dagger_\text{R}(K) \hat{\rho}_0 = & \text{Tr}_\text{L} \left[ \hat{A}^\dagger_\text{R}(K) |0_M\rangle \langle 0_M| \right] \nonumber \\
= & \exp \left( \frac{c \beta |K|}{2}  \right) \text{Tr}_\text{L} \left[ \hat{A}_\text{L}(K) |0_M\rangle \langle 0_M| \right] \nonumber \\
= & \exp \left( \frac{c \beta |K|}{2} \right) \text{Tr}_\text{L} \left[ |0_M\rangle \langle 0_M| \hat{A}_\text{L}(K) \right]\nonumber \\
= & e^{\beta |K|} \text{Tr}_\text{L} \left[ |0_M\rangle \langle 0_M| \hat{A}_\text{R}^\dagger(K) \right]\nonumber \\
= & e^{\beta |K|}  \hat{\rho}_0 \hat{A}^\dagger_\text{R}(K).
\end{align}

\item The functional derivatives of
\begin{equation}
\chi^{(p)}_0[\xi,\xi^*] = \text{Tr} \left( \hat{\rho}_0 \hat{D}_p[\xi, \xi^*] \right)
\end{equation}
for $p=-1$ with respect to different $\xi(K)$ and $\xi^*(K)$ give the following mean values
\begin{align}\label{multiple_Xi_derivatives_-1}
& \prod_{i=1}^n  \frac{\delta}{\delta \xi(K_i)}  \prod_{j=1}^m \left[- \frac{\delta}{\delta \xi^*(K'_j)} \right] \chi^{(-1)}_0 [\xi,\xi^*] \nonumber \\
= & \text{Tr} \left( \hat{\rho}_0 \prod_{j=1}^m \hat{A}_\text{R}(K'_j) \hat{D}_{-1} [\xi,\xi^*] \prod_{i=1}^n \hat{A}_\text{R}^\dagger(K_i) \right),
\end{align}
as it can be noticed from Eq.~(\ref{D_-1}).

\item $\chi^{(p)}_0[\xi,\xi^*]$ is already known in literature \cite{Barnett2002}, since $\hat{\rho}_0$ has the form of a thermal state. Hence, we know that
\begin{equation} \label{thermal_characteristic_fucntion}
\chi^{(p)}_0[\xi,\xi^*] = \exp \left( \int_{\mathbb{R}} dK |\xi(K)|^2  \left[- n_0(K) + \frac{p-1}{2} \right] \right),
\end{equation}
with $n_0(K) = (e^{\beta|K|}-1)^{-1}$.
\end{itemize}

Starting from this information, we provide a generic procedure to derive $\chi^{(p)}[\xi,\xi^*]$. The prescription is detailed by the following steps:
\begin{enumerate}
\item Reorder the right Rindler operators appearing in Eq.~(\ref{varrho_step_4_2}) by using Eq.~(\ref{ordering_Rindler}) and the canonical commutation rules; choose the rearrangement such that $\hat{\rho}$ is put in the form of a combination of chains of creation operators $ \hat{A}_\text{R}^\dagger(K)$ acting on the left of $\hat{\rho}_0$ and annihilation operators $ \hat{A}_\text{R}(K)$ acting on the right of $\hat{\rho}_0$. \label{prescription_5}
\item Multiply both hand sides of the equation with $\hat{D}_{-1}[\xi,\xi^*]$; use the trace over the right wedge and the cyclic property of the trace in order to see $\chi^{(-1)}[\xi,\xi^*]$ as a combination of terms that have the same form of the right hand side of Eq.~(\ref{multiple_Xi_derivatives_-1}). \label{prescription_6}
\item By using Eq.~(\ref{multiple_Xi_derivatives_-1}), see $\chi^{(- 1)}[\xi,\xi^*]$ as a linear combination of $\xi$-derivatives of $\chi^{(- 1)}_0[\xi,\xi^*]$, which are explicitly obtainable from Eq.~(\ref{thermal_characteristic_fucntion}). \label{prescription_7}
\item extract the final expression of $\chi^{(p)}[\xi,\xi^*]$ from $\chi^{(- 1)}[\xi,\xi^*]$ by means of Eq.~(\ref{chi_p_m1}); this is equivalent to simply replace $\chi^{(- 1)}_0[\xi,\xi^*]$ of step \ref{prescription_7} with $\chi^{(p)}_0[\xi,\xi^*]$. \label{prescription_8}
\end{enumerate}

As a result of this procedure, we obtain
\begin{subequations}\label{chi_final}
\begin{align} 
& \chi^{(p)}[\xi, \xi^*] =  \sum_{n=0}^\infty \sum_{n'=0}^\infty \int_{\mathbb{R}^n} d^n \textbf{k}_n  \int_{\mathbb{R}^{n'}} d^{n'} \textbf{k}'_{n'} \frac{\tilde{\phi}_n (\textbf{k}_n) \tilde{\phi}_{n'}^* (\textbf{k}'_{n'})}{\sqrt{n!n'!}}  \bar{\chi}^{(p)}(\textbf{k}_n , \textbf{k}'_{n'}) [\xi, \xi^*],\label{chi_final_a} \\
 & \bar{\chi}^{(p)}(\textbf{k}_n , \textbf{k}'_{n'})[\xi, \xi^*] = \chi^{(p)}_0[\xi,\xi^*] \sum_{\mathcal{S} \subseteq [n]} \sum_{\mathcal{S}' \subseteq [n'] } C \left( \{ k_i \}_{i \in [n] \setminus \mathcal{S}}, \{ k'_{i'} \}_{i' \in [n'] \setminus \mathcal{S}'} \right)  \nonumber \\
 & \times \prod_{j \in \mathcal{S}}  \{ - L (k_j) [\xi,\xi^*]\}^*  \prod_{j' \in \mathcal{S}'}L (k'_{j'}) [\xi,\xi^*] .\label{chi_bar_final}
\end{align}
\end{subequations}
For any couple of ordered sequence $\mathcal{U} = \{k_i\}_{i=1}^m = \{ k_1, \dots, k_n \}$ and $\mathcal{U}' = \{k'_{j}\}_{j=1}^m = \{k'_1, \dots, k'_m \}$, the coefficients $C(\mathcal{U}, \mathcal{U}') $ have the following combinatorial expression
\begin{equation} \label{c_combinatory}
C \left( \{k_i\}_{i=1}^n,\{k'_{j}\}_{j=1}^m \right) = \delta_{nm} \sum_{\mathsf{P} \in S_n} \prod_{i=1}^n \delta(k_i - k'_{\mathsf{P}(i)}),
\end{equation}
where the sum $\sum_{\mathsf{P} \in S_n} $ runs over all the possible permutations $\mathsf{P}$ for the index $i$; whereas, $L (k) [\xi,\xi^*]  $ is a linear functional of $\xi$ and $\xi^*$ defined as
\begin{equation}\label{L_xi}
L (k) [\xi,\xi^*] = \int_{\mathbb{R}} dK \left[\alpha^*(k,K) \xi(K) - \beta(k,K) \xi^*(K) \right].
\end{equation}
It can be noticed that $L (k) [\xi,\xi^*]$ appears also in Eq.~(\ref{Bogolyubov_transformation}) as a Bogoliubov transformation between Minkowski and Rindler operators:
\begin{equation}\label{Bogoliubov_transformation_L}
\hat{a}(k) = \left\lbrace L (k) \left[\hat{A}^\dagger_\text{L},\hat{A}_\text{L}\right] \right\rbrace^\dagger + L (k) \left[\hat{A}_\text{R},\hat{A}^\dagger_\text{R}\right].
\end{equation}
A detailed derivation of Eq.~(\ref{chi_final}) by means of steps \ref{prescription_5}-\ref{prescription_8} will be provided in Sec.~\ref{Explicit_derivation_of_the_result_Wigner}.

We now show how to put Eq.~(\ref{chi_bar_final}) in a diagrammatic form. A single diagram is defined by the following prescription. Write all components of $\textbf{k}_n $ and $ \textbf{k}'_{n'}$ in two distinct columns and create some pair connections between elements of the left and the right column $k_i \text{ --- } k'_j$. The numerical value associated to this diagram is the product between $\chi^{(p)}_0[\xi,\xi^*]$ and the numerical contributions coming from the elements of the diagram. Each pair connection $k_i \text{ --- } k'_j$ contributes with a delta function between the two momenta $\delta(k_i-k'_j)$. Each unpaired left column element $k_i$ contributes with $\{ - L (k_i) [\xi,\xi^*] \}^*$; whereas, each unconnected right-column element $k'_j$ contributes with $ L (k'_j) [\xi,\xi^*]$.

The full diagrammatic expression for Eq.~(\ref{chi_bar_final}) is
\begin{equation}\label{diagram_Wigner}
\begin{tikzpicture}
\node (0_1) [] {$ \bar{\chi}^{(p)}(\textbf{k}_n , \textbf{k}'_{n'}) = $};

\node (1L_vdots) [right=0cm of 0_1] {$\vdots$};
\node (1R_vdots) [right of=1L_vdots] {$\vdots$};

\node (1L_k_1) [above=0cm of 1L_vdots] {$k_1$};
\node (1L_k_n) [below=0cm of 1L_vdots] {$k_n$};
\node (1R_k_1) [above=0cm of 1R_vdots] {$k'_1$};
\node (1R_k_{n'}) [below=0cm of 1R_vdots] {$k'_{n'}$};

\node (1_2) [right of=1R_vdots] {$+ \sum_{ii'}$};
\node (2L_k_i) [right of=1_2] {$k_i$};
\node (2R_k_i) [right of=2L_k_i] {$k'_{i'}$};

\node (2L_vdots_1) [above=0cm of 2L_k_i] {$\vdots$};
\node (2L_k_1) [above=0cm of 2L_vdots_1] {$k_1$};
\node (2L_vdots_2) [below=0cm of 2L_k_i] {$\vdots$};
\node (2L_k_n) [below=0cm of 2L_vdots_2] {$k_n$};
\node (2R_vdots_1) [above=0cm of 2R_k_i] {$\vdots$};
\node (2R_k_1) [above=0cm of 2R_vdots_1] {$k'_1$};
\node (2R_vdots_2) [below=0cm of 2R_k_i] {$\vdots$};
\node (2R_k_{n'}) [below=0cm of 2R_vdots_2] {$k'_{n'}$};

\node (2_3) [right of=2R_k_i] {$+ \sum_{ii'jj'}$};
\node (3L_vdots) [right of=2_3] {$\vdots$};
\node (3R_vdots) [right of=3L_vdots] {$\vdots$};

\node (3L_k_i) [above=0cm of 3L_vdots] {$k_i$};
\node (3L_vdots_1) [above=0cm of 3L_k_i] {$\vdots$};
\node (3L_k_1) [above=0cm of 3L_vdots_1] {$k_1$};
\node (3L_k_j) [below=0cm of 3L_vdots] {$k_j$};
\node (3L_vdots_2) [below=0cm of 3L_k_j] {$\vdots$};
\node (3L_k_n) [below=0cm of 3L_vdots_2] {$k_n$};
\node (3R_k_i) [above=0cm of 3R_vdots] {$k'_{i'}$};
\node (3R_vdots_1) [above=0cm of 3R_k_i] {$\vdots$};
\node (3R_k_1) [above=0cm of 3R_vdots_1] {$k'_1$};
\node (3R_k_j) [below=0cm of 3R_vdots] {$k'_{j'}$};
\node (3R_vdots_2) [below=0cm of 3R_k_j] {$\vdots$};
\node (3R_k_{n'}) [below=0cm of 3R_vdots_2] {$k'_{n'}$};

\node (3_4) [right of=3R_vdots] {$+ \dots$};

\draw[-] (2L_k_i) to (2R_k_i);
\draw[-] (3L_k_i) to (3R_k_i);
\draw[-] (3L_k_j) to (3R_k_j);
\end{tikzpicture}
\end{equation}

\subsection{Examples: single and two particles states}\label{Examples}

In this subsection, we consider two examples of Minkowski-Fock states to detail the results of Sec.~\ref{Method_and_result_Wigner}. In particular, we consider single and two particles states that are prepared by the inertial experimenter and registered by the accelerated observer. By using Eq.~(\ref{chi_final}) and the diagrammatic expansion (\ref{diagram_Wigner}), we derive the characteristic function $\chi^{(p)}[\xi,\xi^*]$ describing the states in the right Rindler frame.

\subsubsection{Minkowski single particle in the accelerated frame}

A general Minkowski single particle state is defined with respect to its wave function $\tilde{\phi}_1(k) $ as follows
\begin{equation}\label{single_particle_Minkowski_11}
| \phi \rangle = \int_{\mathbb{R}} d k \tilde{\phi}_1(k) \hat{a}^\dagger(k) | 0_\text{M} \rangle,
\end{equation}
with normalization condition
\begin{equation} \label{single_particle_normalization}
\int_{\mathbb{R}} dk \left| \tilde{\phi}_1(k) \right|^2 = 1.
\end{equation}

For such a state, Eq.~(\ref{chi_final_a}) reads as
\begin{equation} \label{single_particle_chi}
\chi^{(p)}[\xi, \xi^*] = \int_{\mathbb{R}} dk \tilde{\phi}_1(k) \int_{\mathbb{R}} dk' \tilde{\phi}_1^*(k') \bar{\chi}^{(p)} (k,k') [\xi, \xi^*]
\end{equation}
and $ \bar{\chi}^{(p)}(k,k')[\xi, \xi^*]$ has the following diagrammatic expression
\begin{equation} \label{single_particle_diagram}
\bar{\chi}^{(p)}(k,k')[\xi, \xi^*] = k \text{ \textcolor{white}{---} } k' + k \text{ --- } k'.
\end{equation}
The explicit version of Eq.~(\ref{single_particle_diagram}) is
\begin{equation} \label{single_particle_chi_bar}
\bar{\chi}^{(p)}(k,k')[\xi, \xi^*] = - \chi^{(p)}_0[\xi,\xi^*] \{L (k) [\xi,\xi^*] \}^* L (k') [\xi,\xi^*] \chi^{(p)}_0[\xi,\xi^*] \delta(k-k').
\end{equation}

Equations (\ref{single_particle_normalization}), (\ref{single_particle_chi}) and (\ref{single_particle_chi_bar}) finally lead to
\begin{equation} \label{single_particle_chi_final}
\chi^{(p)}[\xi, \xi^*] = \left\lbrace -  \left| \int_{\mathbb{R}} dk \tilde{\phi}_1(k)  L^* (k) [\xi,\xi^*]  \right|^2 + 1 \right\rbrace \chi^{(p)}_0[\xi,\xi^*]
\end{equation}
as the explicit expression for the Rindler-Wigner characteristic function of a Minkowski single particle state with wave function $\tilde{\phi}_1(k)$.

\subsubsection{Minkowski two particles in the accelerated frame}

A Minkowski two particles state can be defined by means of a wave function $\tilde{\phi}_2(k_1,k_2)$ which is symmetric with respect to a switch between $k_1$ and $k_2$. The state is
\begin{equation}\label{two_particle}
| \phi \rangle = \int_{\mathbb{R}} dk \int_{\mathbb{R}} dk_2  \frac{\tilde{\phi}_2(k_1,k_2)}{\sqrt{2}} \hat{a}^\dagger (k_1) \hat{a}^\dagger (k_2) | 0_\text{M} \rangle,
\end{equation}
with
\begin{align} \label{two_particle_symmetry_normalization}
& \tilde{\phi}_2(k_1,k_2) =  \tilde{\phi}_2(k_2,k_1), &\int_{\mathbb{R}} dk \int_{\mathbb{R}} dk_2 \left| \tilde{\phi}_2(k_1,k_2) \right|^2 = 1.
\end{align}

For such a state, Eq.~(\ref{chi_final_a}) reads as
\begin{align} \label{two_particle_chi}
\chi^{(p)}[\xi, \xi^*] = & \frac{1}{2}\int_{\mathbb{R}} dk_1 \int_{\mathbb{R}} dk_2 \int_{\mathbb{R}} dk'_1  \int_{\mathbb{R}} dk'_2  \tilde{\phi}_2(k_1,k_2) \tilde{\phi}_2^*(k'_1,k'_2) \nonumber \\
& \times \bar{\chi}^{(p)} ((k_1,k_2),(k'_1,k'_2)) [\xi, \xi^*].
\end{align}
$ \bar{\chi}^{(p)} ((k_1,k_2),(k'_1,k'_2)) [\xi, \xi^*]$ has the following diagrammatic expression
\begin{equation} \label{two_particles_diagram}
\begin{tikzpicture}
\node (0_1) [] {$ \bar{\chi}^{(p)}((k_1,k_2),(k'_1,k'_2))[\xi, \xi^*] = $};

\node (1L_vdots) [right=0cm of 0_1] {};
\node (1R_vdots) [right of=1L_vdots] {};

\node (1L_k_1) [above=0cm of 1L_vdots] {$k_1$};
\node (1L_k_2) [below=0cm of 1L_vdots] {$k_2$};
\node (1R_k_1) [above=0cm of 1R_vdots] {$k'_1$};
\node (1R_k_2) [below=0cm of 1R_vdots] {$k'_2$};

\node (1_2) [right=0cm of 1R_vdots] {$+$};
\node (2L_vdots) [right=0cm of 1_2] {};
\node (2R_vdots) [right of=2L_vdots] {};

\node (2L_k_1) [above=0cm of 2L_vdots] {$k_1$};
\node (2L_k_2) [below=0cm of 2L_vdots] {$k_2$};
\node (2R_k_1) [above=0cm of 2R_vdots] {$k'_1$};
\node (2R_k_2) [below=0cm of 2R_vdots] {$k'_2$};

\node (2_3) [right=0cm of 2R_vdots] {$+$};
\node (3L_vdots) [right=0cm of 2_3] {};
\node (3R_vdots) [right of=3L_vdots] {};

\node (3L_k_1) [above=0cm of 3L_vdots] {$k_1$};
\node (3L_k_2) [below=0cm of 3L_vdots] {$k_2$};
\node (3R_k_1) [above=0cm of 3R_vdots] {$k'_1$};
\node (3R_k_2) [below=0cm of 3R_vdots] {$k'_2$};

\node (3_4) [right=0cm of 3R_vdots] {$+$};
\node (4L_vdots) [right=0cm of 3_4] {};
\node (4R_vdots) [right of=4L_vdots] {};

\node (4L_k_1) [above=0cm of 4L_vdots] {$k_1$};
\node (4L_k_2) [below=0cm of 4L_vdots] {$k_2$};
\node (4R_k_1) [above=0cm of 4R_vdots] {$k'_1$};
\node (4R_k_2) [below=0cm of 4R_vdots] {$k'_2$};

\node (4_5) [below of= 1L_k_2] {$+$};
\node (5L_vdots) [right=0cm of 4_5] {};
\node (5R_vdots) [right of=5L_vdots] {};

\node (5L_k_1) [above=0cm of 5L_vdots] {$k_1$};
\node (5L_k_2) [below=0cm of 5L_vdots] {$k_2$};
\node (5R_k_1) [above=0cm of 5R_vdots] {$k'_1$};
\node (5R_k_2) [below=0cm of 5R_vdots] {$k'_2$};

\node (5_6) [right=0cm of 5R_vdots] {$+$};
\node (6L_vdots) [right=0cm of 5_6] {};
\node (6R_vdots) [right of=6L_vdots] {};

\node (6L_k_1) [above=0cm of 6L_vdots] {$k_1$};
\node (6L_k_2) [below=0cm of 6L_vdots] {$k_2$};
\node (6R_k_1) [above=0cm of 6R_vdots] {$k'_1$};
\node (6R_k_2) [below=0cm of 6R_vdots] {$k'_2$};

\node (6_7) [right=0cm of 6R_vdots] {$+$};
\node (7L_vdots) [right=0cm of 6_7] {};
\node (7R_vdots) [right of=7L_vdots] {};

\node (7L_k_1) [above=0cm of 7L_vdots] {$k_1$};
\node (7L_k_2) [below=0cm of 7L_vdots] {$k_2$};
\node (7R_k_1) [above=0cm of 7R_vdots] {$k'_1$};
\node (7R_k_2) [below=0cm of 7R_vdots] {$k'_2$};

\draw[-] (2L_k_1) to (2R_k_1);
\draw[-] (3L_k_2) to (3R_k_2);
\draw[-] (4L_k_1) to (4R_k_2);
\draw[-] (5L_k_2) to (5R_k_1);
\draw[-] (6L_k_1) to (6R_k_1);
\draw[-] (6L_k_2) to (6R_k_2);
\draw[-] (7L_k_1) to (7R_k_2);
\draw[-] (7L_k_2) to (7R_k_1);

\end{tikzpicture}
\end{equation}

By using the symmetry and the normalization condition (\ref{two_particle_symmetry_normalization}), we can write Eq.~(\ref{two_particle_chi}) as
\begin{align} \label{two_particle_chi_final}
\chi^{(p)}[\xi, \xi^*] = &  \chi^{(p)}_0[\xi, \xi^*] \left\lbrace \frac{1}{2} \left| \int_{\mathbb{R}} dk_1 \int_{\mathbb{R}} dk_2 \tilde{\phi}_2(k_1,k_2)  L^*(k_1) [\xi,\xi^*]  L^*(k_2) [\xi,\xi^*] \right|^2 \right. \nonumber \\
& \left.  - 2 \int_{\mathbb{R}} dk_1 \left| \int_{\mathbb{R}} dk_2 \tilde{\phi}_2(k_1,k_2)  L^*(k_2) [\xi,\xi^*] \right|^2 + 1 \right\rbrace
\end{align}
which results in the characteristic function of the Minkowski two particles state (\ref{two_particle}) in the accelerated frame.

\subsection{Comparison between inertial and accelerated frame}\label{Comparison_between_inertial_and_accelerated_frame}

The explicit form of $\chi^{(p)}[\xi, \xi^*]$ given by Eq.~(\ref{chi_final}) can be compared with the characteristic function of $|\phi\rangle \langle \phi |$ in the Minkowski frame, which, in turn, is defined by
\begin{align} \label{chi_M}
\chi^{(p)}_\text{M}[\xi, \xi^*] = & \text{Tr} \left( |\phi\rangle \langle \phi | \exp \left( \int_{\mathbb{R}} dk \left[ \xi(k) \hat{a}^\dagger(k)  - \xi^*(k) \hat{a}(k) + \frac{p}{2} |\xi(k)|^2\right] \right) \right)
\end{align}
and explicitly reads as
\begin{subequations}\label{chi_M_final}
\begin{align} 
& \chi^{(p)}_\text{M}[\xi, \xi^*] =  \sum_{n=0}^\infty \sum_{n'=0}^\infty \int_{\mathbb{R}^n} d^n \textbf{k}_n  \int_{\mathbb{R}^{n'}} d^{n'} \textbf{k}'_{n'} \frac{\tilde{\phi}_n (\textbf{k}_n) \tilde{\phi}_{n'}^* (\textbf{k}'_{n'})}{\sqrt{n!n'!}}  \bar{\chi}^{(p)}_\text{M}(\textbf{k}_n , \textbf{k}'_{n'})[\xi, \xi^*],\label{chi_M_final_a}\\
 & \bar{\chi}^{(p)}_\text{M}(\textbf{k}_n , \textbf{k}'_{n'})[\xi, \xi^*] = \chi^{(p)}_{0\text{M}}[\xi,\xi^*] \sum_{\mathcal{S} \subseteq [n]} \sum_{\mathcal{S}' \subseteq [n'] } C \left( \{ k_i \}_{i \in [n] \setminus \mathcal{S}}, \{ k'_{i'} \}_{i' \in [n'] \setminus \mathcal{S}'} \right)  \nonumber \\
 & \times \prod_{j \in \mathcal{S}}  [ - \xi^* (k_j) ]  \prod_{j' \in \mathcal{S}'} \xi(k'_{j'}).\label{chi_M_final_b}
\end{align}
\end{subequations}
$\chi^{(p)}_{0\text{M}}[\xi,\xi^*]$ is the characteristic function of the vacuum $|0_\text{M} \rangle \langle 0_\text{M} |$ in the Minkowski frame; it is defined by
\begin{equation} \label{chi_0M}
\chi^{(p)}_{0\text{M}}[\xi, \xi^*] = \text{Tr} \left( |0_\text{M}\rangle \langle 0_\text{M} | \exp \left( \int_{\mathbb{R}} dk \left[ \xi(k) \hat{a}^\dagger(k)  - \xi^*(k) \hat{a}(k) + \frac{p}{2} |\xi(k)|^2\right] \right) \right)
\end{equation}
and has the following explicit form
\begin{equation} \label{chi_0M_final}
\chi^{(p)}_{0\text{M}}[\xi, \xi^*] = \exp \left( \int_{\mathbb{R}} dk \frac{p-1}{2} |\xi(k)|^2 \right).
\end{equation}
The proof for Eq.~(\ref{chi_M_final}) will be provided in Sec.~\ref{Explicit_derivation_of_the_result_Wigner}.

By comparing Eq.~(\ref{chi_M_final}) with Eq.~(\ref{chi_final}) we can deduce how the characteristic function transforms from the Minkowski to the right Rindler frame. As a result, we know how the description of Minkowski particle states chances from the inertial to the accelerated frame. 

Notice that the transformation $\chi^{(p)}_\text{M}[\xi, \xi^*] \mapsto \chi^{(p)}[\xi, \xi^*]$ can be easily computed by performing the substitutions $ \chi^{(p)}_{0\text{M}}[\xi,\xi^*] \mapsto \chi^{(p)}_0[\xi,\xi^*]$ and $\xi(k) \mapsto  L (k) [\xi,\xi^*] $ in Eq.~(\ref{chi_M_final}). This means that the inertial and the accelerated experimenters describe the Minkowski particle state in the same way up to the presence of a thermal background ($ \chi^{(p)}_{0\text{M}}[\xi,\xi^*] \mapsto \chi^{(p)}_0[\xi,\xi^*]$) and a change of variables ($\xi(k) \mapsto  L (k) [\xi,\xi^*] $) that is due to the Bogoliubov transformation (\ref{Bogoliubov_transformation_L}).

\subsection{Explicit derivation of the result}\label{Explicit_derivation_of_the_result_Wigner}

\subsubsection{Proof of Eq.~(\ref{chi_final})}

Here, by following the prescription given by steps \ref{prescription_5}-\ref{prescription_8} of Sec.~\ref{Method_and_result_Wigner} we prove Eq.~(\ref{chi_final}).

By using Eq.~(\ref{ordering_Rindler}) we can manipulate Eq.~(\ref{varrho_step_4_2}) in the following way
\begin{align} \label{varrho_step_5_1}
\hat{\rho} = &  \sum_{n=0}^\infty \sum_{{n'}=0}^\infty \sum_{\mathcal{N} \subseteq [n]} \sum_{\mathcal{N}' \subseteq [n'] }  \sum_{\mathcal{N}_+ \subseteq \mathcal{N}} \sum_{\mathcal{N}'_+ \subseteq \mathcal{N}' } \int_{\mathbb{R}^n} d^n \textbf{k}_n  \int_{\mathbb{R}^{n'}} d^{n'} \textbf{k}'_{n'}  \nonumber \\
& \times \frac{\tilde{\phi}_n (\textbf{k}_n) \tilde{\phi}_{n'}^* (\textbf{k}'_{n'})}{\sqrt{n!n'!}}  \mathsf{C}_\mathsf{K} \left( \prod_{i \in [n] \setminus \mathcal{N}}  \hat{a}^{\prime \dagger}(k_i) \right) \mathsf{C}_\mathsf{K} \left(  \prod_{i' \in [n'] \setminus \mathcal{N}'} \hat{a}'(k'_{i'}) \right)  \nonumber \\
& \times \prod_{j_- \in\mathcal{N} \setminus \mathcal{N}_+} \hat{A}_-^\dagger(k_{j_-})  \prod_{j_+\in\mathcal{N}_+} \hat{A}_+^\dagger(k_{j_+})   \prod_{j'_+\in\mathcal{N}'_+} \hat{A}'_+(k'_{j'_+}) \hat{\rho}_0   \prod_{j'_-\in\mathcal{N}'\setminus\mathcal{N}'_+} \hat{A}_-(k'_{j'_-}).
\end{align}
with
\begin{equation} \label{B_+}
\hat{A}'_+(k) =  \int_{\mathbb{R}}dK 2 \sinh \left(\frac{\beta}{2} |K| \right) e^{-c \beta|K|} \alpha(k,K)\hat{A}^\dagger_R(K).
\end{equation}

By following step \ref{prescription_5} of Sec.~\ref{Method_and_result_Wigner}, we want to put the right hand side of Eq.~(\ref{varrho_step_5_1}) in a normal order for the entire chain of $\hat{A^\dagger}_\pm(k)$, $\hat{A}'_+(k)$ and $\hat{A}_-(k)$ operators. For this reason, we use again the Wick theorem to rearrange the order of the $\hat{A}^\dagger_+ (k)$ and $\hat{A}'_+ (k)$ operators:
\begin{align}
\hat{\rho} = &  \sum_{n=0}^\infty \sum_{{n'}=0}^\infty \sum_{\mathcal{N} \subseteq [n]} \sum_{\mathcal{N}' \subseteq [n'] }  \sum_{\mathcal{N}_+ \subseteq \mathcal{N}} \sum_{\mathcal{N}'_+ \subseteq \mathcal{N}' }  \sum_{\mathcal{M}_+ \subseteq \mathcal{N}_+} \sum_{\mathcal{M}'_+ \subseteq \mathcal{N}'_+ } \int_{\mathbb{R}^n} d^n \textbf{k}_n  \int_{\mathbb{R}^{n'}} d^{n'} \textbf{k}'_{n'}  \nonumber \\
& \times \frac{\tilde{\phi}_n (\textbf{k}_n) \tilde{\phi}_{n'}^* (\textbf{k}'_{n'})}{\sqrt{n!n'!}}  \mathsf{C}_\mathsf{K} \left( \prod_{i \in [n] \setminus \mathcal{N}}  \hat{a}^{\prime \dagger}(k_i) \right) \mathsf{C}_\mathsf{K} \left(  \prod_{i' \in [n'] \setminus \mathcal{N}'} \hat{a}'(k'_{i'}) \right) \nonumber \\
& \times \mathsf{C}_\mathsf{N}\left(  \prod_{l\in\mathcal{N}_+ \setminus \mathcal{M}_+} \hat{A}_+^\dagger(k_{l})   \prod_{l'\in\mathcal{N}'_+ \setminus \mathcal{M}'_+} \hat{A}'_+(k'_{l'}) \right)  \prod_{j_- \in\mathcal{N} \setminus \mathcal{N}_+} \hat{A}_-^\dagger(k_{j_-}) \nonumber \\
& \times \mathsf{N}\left(   \prod_{j_+\in\mathcal{M}_+} \hat{A}_+^\dagger(k_{j_+})   \prod_{j'_+\in\mathcal{M}'_+} \hat{A}'_+(k'_{j'_+}) \right) \hat{\rho}_0   \prod_{j'_-\in\mathcal{N}'\setminus\mathcal{N}'_+} \hat{A}_-(k'_{j'_-})\nonumber \\
 = &  \sum_{n=0}^\infty \sum_{{n'}=0}^\infty \sum_{\mathcal{N} \subseteq [n]} \sum_{\mathcal{N}' \subseteq [n'] }  \sum_{\mathcal{N}_+ \subseteq \mathcal{N}} \sum_{\mathcal{N}'_+ \subseteq \mathcal{N}' }  \sum_{\mathcal{M}_+ \subseteq \mathcal{N}_+} \sum_{\mathcal{M}'_+ \subseteq \mathcal{N}'_+ } \int_{\mathbb{R}^n} d^n \textbf{k}_n  \int_{\mathbb{R}^{n'}} d^{n'} \textbf{k}'_{n'}  \nonumber \\
& \times \frac{\tilde{\phi}_n (\textbf{k}_n) \tilde{\phi}_{n'}^* (\textbf{k}'_{n'})}{\sqrt{n!n'!}}  \mathsf{C}_\mathsf{K} \left( \prod_{i \in [n] \setminus \mathcal{N}}  \hat{a}^{\prime \dagger}(k_i) \right) \mathsf{C}_\mathsf{K} \left(  \prod_{i' \in [n'] \setminus \mathcal{N}'} \hat{a}'(k'_{i'}) \right) \nonumber \\
& \times \mathsf{C}_\mathsf{N}\left(  \prod_{l\in\mathcal{N}_+ \setminus \mathcal{M}_+} \hat{A}_+^\dagger(k_{l})   \prod_{l'\in\mathcal{N}'_+ \setminus \mathcal{M}'_+} \hat{A}'_+(k'_{l'}) \right)  \prod_{j_- \in\mathcal{N} \setminus \mathcal{N}_+} \hat{A}_-^\dagger(k_{j_-}) \nonumber \\
& \times  \prod_{j'_+\in\mathcal{M}'_+} \hat{A}'_+(k'_{j'_+})  \prod_{j_+\in\mathcal{M}_+} \hat{A}_+^\dagger(k_{j_+})    \hat{\rho}_0   \prod_{j'_-\in\mathcal{N}'\setminus\mathcal{N}'_+} \hat{A}_-(k'_{j'_-}).
\end{align}
By defining $\mathcal{M} = \mathcal{N} \setminus(\mathcal{N}_+ \setminus \mathcal{M}_+)$ and $\mathcal{M}' = \mathcal{N}' \setminus(\mathcal{N}'_+ \setminus \mathcal{M}'_+)$, we obtain
\begin{align}
\hat{\rho} = &  \sum_{n=0}^\infty \sum_{{n'}=0}^\infty \sum_{\mathcal{N} \subseteq [n]} \sum_{\mathcal{N}' \subseteq [n'] }  \sum_{\mathcal{M} \subseteq \mathcal{N}} \sum_{\mathcal{M}' \subseteq \mathcal{N}' }  \sum_{\mathcal{M}_+ \subseteq \mathcal{M}} \sum_{\mathcal{M}'_+ \subseteq \mathcal{M}' } \int_{\mathbb{R}^n} d^n \textbf{k}_n  \int_{\mathbb{R}^{n'}} d^{n'} \textbf{k}'_{n'}  \nonumber \\
& \times \frac{\tilde{\phi}_n (\textbf{k}_n) \tilde{\phi}_{n'}^* (\textbf{k}'_{n'})}{\sqrt{n!n'!}}  \mathsf{C}_\mathsf{K} \left( \prod_{i \in [n] \setminus \mathcal{N}}  \hat{a}^{\prime \dagger}(k_i) \right) \mathsf{C}_\mathsf{K} \left(  \prod_{i' \in [n'] \setminus \mathcal{N}'} \hat{a}'(k'_{i'}) \right) \nonumber \\
& \times \mathsf{C}_\mathsf{N}\left(  \prod_{l \in\mathcal{N} \setminus \mathcal{M}} \hat{A}_+^\dagger(k_{l})   \prod_{l'\in\mathcal{N}' \setminus \mathcal{M}'} \hat{A}'_+(k'_{l'}) \right)  \prod_{j_- \in\mathcal{M} \setminus \mathcal{M}_+} \hat{A}_-^\dagger(k_{j_-}) \nonumber \\
& \times  \prod_{j'_+\in\mathcal{M}'_+} \hat{A}'_+(k'_{j'_+})  \prod_{j_+\in\mathcal{M}_+} \hat{A}_+^\dagger(k_{j_+})    \hat{\rho}_0   \prod_{j'_-\in\mathcal{M}'\setminus\mathcal{M}'_+} \hat{A}_-(k'_{j'_-}).
\end{align}
Also, by defining $\mathcal{L} = \mathcal{N} \setminus \mathcal{M}$ and $\mathcal{L}' = \mathcal{N}' \setminus \mathcal{M}'$, we obtain
\begin{align}\label{varrho_step_5_2}
& \hat{\rho} =  \sum_{n=0}^\infty \sum_{{n'}=0}^\infty \sum_{\mathcal{M} \subseteq [n]} \sum_{\mathcal{M}' \subseteq [n'] }  \sum_{\mathcal{L} \subseteq [n] \setminus \mathcal{M}} \sum_{\mathcal{L}' \subseteq [n'] \setminus \mathcal{M}' }  \sum_{\mathcal{M}_+ \subseteq \mathcal{M}} \sum_{\mathcal{M}'_+ \subseteq \mathcal{M}' }  \nonumber \\
& \times \int_{\mathbb{R}^n} d^n \textbf{k}_n  \int_{\mathbb{R}^{n'}} d^{n'} \textbf{k}'_{n'} \frac{\tilde{\phi}_n (\textbf{k}_n) \tilde{\phi}_{n'}^* (\textbf{k}'_{n'})}{\sqrt{n!n'!}}  \mathsf{C}_\mathsf{K} \left( \prod_{i \in [n] \setminus (\mathcal{M} \cup \mathcal{L})}  \hat{a}^{\prime \dagger}(k_i) \right) \nonumber \\
& \times  \mathsf{C}_\mathsf{K} \left(  \prod_{i' \in [n'] \setminus (\mathcal{M}' \cup \mathcal{L}')} \hat{a}'(k'_{i'}) \right) \mathsf{C}_\mathsf{N}\left(  \prod_{l \in\mathcal{L}} \hat{A}_+^\dagger(k_{l})   \prod_{l'\in\mathcal{L}' } \hat{A}'_+(k'_{l'}) \right)  \nonumber \\
& \times  \prod_{j_- \in\mathcal{M} \setminus \mathcal{M}_+} \hat{A}_-^\dagger(k_{j_-}) \prod_{j'_+\in\mathcal{M}'_+} \hat{A}'_+(k'_{j'_+})  \prod_{j_+\in\mathcal{M}_+} \hat{A}_+^\dagger(k_{j_+})    \hat{\rho}_0   \prod_{j'_-\in\mathcal{M}'\setminus\mathcal{M}'_+} \hat{A}_-(k'_{j'_-}).
\end{align}

The full contractions appearing in Eq.~(\ref{varrho_step_5_2}) can be manipulated in a combinatorial way by knowing that
\begin{subequations}
\begin{align}
& \mathsf{C}_\mathsf{N} \left( \hat{A}^\dagger_+(k) \hat{A}^\dagger_+(k') \right) =  \mathsf{C}_\mathsf{N}  \left( \hat{A}'_+(k) \hat{A}'_+(k') \right) = 0, \\
& \mathsf{C}_\mathsf{N}  \left( \hat{A}^\dagger_+(k) \hat{A}'_+(k') \right) =  \int_{\mathbb{R}}dK \left[ 2 \sinh \left(\frac{c \beta |K|}{2} \right) \right]^2 e^{-c \beta|K|}  \alpha^*(k,K) \alpha(k',K) .
\end{align}
\end{subequations}
This allows us to put the right hand side of Eq.~(\ref{varrho_step_5_2}) in a more compact form by means of an unifying contraction $\mathsf{C}_{\mathsf{K},\mathsf{N}}$ between $\hat{a}'(k)$ and $\hat{a}^{\prime \dagger}(k)$ operators defined by
\begin{subequations}
\begin{align}
& \mathsf{C}_{\mathsf{K},\mathsf{N}} \left( \hat{a}'(k) \hat{a}'(k') \right)  = \mathsf{C}_\mathsf{K} \left( \hat{a}'(k) \hat{a}'(k') \right) ,\\
& \mathsf{C}_{\mathsf{K},\mathsf{N}} \left( \hat{a}^{\prime \dagger}(k) \hat{a}^{\prime \dagger}(k') \right) = \mathsf{C}_\mathsf{K} \left( \hat{a}^{\prime \dagger}(k) \hat{a}^{\prime \dagger}(k') \right) ,\\
& \mathsf{C}_{\mathsf{K},\mathsf{N}} \left( \hat{a}^{\prime \dagger}(k) \hat{a}'(k') \right) = \mathsf{C}_\mathsf{N}  \left( \hat{A}^\dagger_+(k) \hat{A}'_+(k') \right).
\end{align}
\end{subequations}
In this way, Eq.~(\ref{varrho_step_5_2}) has the following form
\begin{align}\label{varrho_step_5_3}
& \hat{\rho} =  \sum_{n=0}^\infty \sum_{{n'}=0}^\infty \sum_{\mathcal{M} \subseteq [n]} \sum_{\mathcal{M}' \subseteq [n'] } \sum_{\mathcal{M}_+ \subseteq \mathcal{M}} \sum_{\mathcal{M}'_+ \subseteq \mathcal{M}' } \int_{\mathbb{R}^n} d^n \textbf{k}_n  \int_{\mathbb{R}^{n'}} d^{n'} \textbf{k}'_{n'} \nonumber \\
& \times  \frac{\tilde{\phi}_n (\textbf{k}_n) \tilde{\phi}_{n'}^* (\textbf{k}'_{n'})}{\sqrt{n!n'!}}  \mathsf{C}_{\mathsf{K},\mathsf{N}} \left( \prod_{i \in [n] \setminus \mathcal{M} }  \hat{a}^{\prime \dagger}(k_i)   \prod_{i' \in [n'] \setminus \mathcal{M}' } \hat{a}'(k'_{i'}) \right) \nonumber \\
& \times  \prod_{j_- \in\mathcal{M} \setminus \mathcal{M}_+} \hat{A}_-^\dagger(k_{j_-}) \prod_{j'_+\in\mathcal{M}'_+} \hat{A}'_+(k'_{j'_+})  \prod_{j_+\in\mathcal{M}_+} \hat{A}_+^\dagger(k_{j_+})    \hat{\rho}_0   \prod_{j'_-\in\mathcal{M}'\setminus\mathcal{M}'_+} \hat{A}_-(k'_{j'_-}).
\end{align}
Explicitly, the fundamental contractions of $\mathsf{C}_{\mathsf{K},\mathsf{N}}$ are
\begin{subequations}
\begin{align}
 \mathsf{C}_{\mathsf{K},\mathsf{N}} \left( \hat{a}'(k) \hat{a}'(k') \right)  = & - \int_{\mathbb{R}}dK 2 \sinh \left( \frac{c \beta |K|}{2} \right) \nonumber \\
 & \times \left[ \beta^*(k,K) \beta(k',K) + \beta(k,K) \beta^*(k',K) \right],\\
\mathsf{C}_{\mathsf{K},\mathsf{N}} \left( \hat{a}^{\prime \dagger}(k) \hat{a}^{\prime \dagger}(k') \right) = & \mathsf{C}_{\mathsf{K},\mathsf{N}} \left( \hat{a}'(k) \hat{a}'(k') \right),\\
\mathsf{C}_{\mathsf{K},\mathsf{N}} \left( \hat{a}^{\prime \dagger}(k) \hat{a}'(k') \right) = & \int_{\mathbb{R}}dK  \left[ 2 \sinh \left(\frac{c \beta |K|}{2} \right) \right]^2 e^{-c \beta|K|}  \alpha^*(k,K) \alpha(k',K).
\end{align}
\end{subequations}

Notice that the $\mathsf{C}_{\mathsf{K},\mathsf{N}}$ contraction of chains of $\hat{a}'(k)$ and $\hat{a}^{\prime \dagger}(k)$ operators does only depend of the momenta appearing in the chains. Hence, we write $\mathsf{C}_{\mathsf{K},\mathsf{N}}$ as a function of sequences of momenta as follows
\begin{equation}
\mathsf{C}_{\mathsf{K},\mathsf{N}} \left( \prod_{i = 1 }^n  \hat{a}^{\prime \dagger}(k_i)   \prod_{j = 1 }^m \hat{a}'(k'_j) \right) =  C_{\mathsf{K},\mathsf{N}} ( \{ k_i \}_{i = 1 }^n, \{ k'_j \}_{j = 1 }^m),
\end{equation}
in analogy to Eq.~(\ref{CC_C}). In this way, Eq.~(\ref{varrho_step_5_3}) becomes
\begin{align}
& \hat{\rho} =  \sum_{n=0}^\infty \sum_{{n'}=0}^\infty \sum_{\mathcal{M} \subseteq [n]} \sum_{\mathcal{M}' \subseteq [n'] } \sum_{\mathcal{M}_+ \subseteq \mathcal{M}} \sum_{\mathcal{M}'_+ \subseteq \mathcal{M}' } \int_{\mathbb{R}^n} d^n \textbf{k}_n  \int_{\mathbb{R}^{n'}} d^{n'} \textbf{k}'_{n'} \frac{\tilde{\phi}_n (\textbf{k}_n) \tilde{\phi}_{n'}^* (\textbf{k}'_{n'})}{\sqrt{n!n'!}} \nonumber \\
& \times   C_{\mathsf{K},\mathsf{N}} \left( \{ k_i \}_{i \in [n] \setminus \mathcal{M} } , \{ k'_{i'} \}_{i' \in [n'] \setminus \mathcal{M}' } \right) \nonumber \\
& \times  \prod_{j_- \in\mathcal{M} \setminus \mathcal{M}_+} \hat{A}_-^\dagger(k_{j_-}) \prod_{j'_+\in\mathcal{M}'_+} \hat{A}'_+(k'_{j'_+})  \prod_{j_+\in\mathcal{M}_+} \hat{A}_+^\dagger(k_{j_+})    \hat{\rho}_0   \prod_{j'_-\in\mathcal{M}'\setminus\mathcal{M}'_+} \hat{A}_-(k'_{j'_-}).
\end{align}

By using again Eq.~(\ref{ordering_Rindler}) on $\hat{A}_+^\dagger(k)$ operators we conclude step \ref{prescription_5} and obtain
\begin{align}
& \hat{\rho} =  \sum_{n=0}^\infty \sum_{{n'}=0}^\infty \sum_{\mathcal{M} \subseteq [n]} \sum_{\mathcal{M}' \subseteq [n'] } \sum_{\mathcal{M}_+ \subseteq \mathcal{M}} \sum_{\mathcal{M}'_+ \subseteq \mathcal{M}' } \int_{\mathbb{R}^n} d^n \textbf{k}_n  \int_{\mathbb{R}^{n'}} d^{n'} \textbf{k}'_{n'}  \frac{\tilde{\phi}_n (\textbf{k}_n) \tilde{\phi}_{n'}^* (\textbf{k}'_{n'})}{\sqrt{n!n'!}}  \nonumber \\
& \times C_{\mathsf{K},\mathsf{N}} \left( \{ k_i \}_{i \in [n] \setminus \mathcal{M} } , \{ k'_{i'} \}_{i' \in [n'] \setminus \mathcal{M}' } \right) \nonumber \\
& \times \prod_{j_- \in\mathcal{M} \setminus \mathcal{M}_+} \hat{A}_-^\dagger(k_{j_-})  \prod_{j'_+\in\mathcal{M}'_+} \hat{A}'_+(k'_{j'_+})   \hat{\rho}_0   \prod_{j_+\in\mathcal{M}_+} \hat{A}_+^{\prime \dagger}(k_{j_+})   \prod_{j'_-\in\mathcal{M}'\setminus\mathcal{M}'_+} \hat{A}_-(k'_{j'_-}).
\end{align}

Step \ref{prescription_6} gives
\begin{align}\label{varrho_step_6}
& \chi^{(-1)}[\xi, \xi^*] \nonumber \\
= & \text{Tr} \left( \hat{\rho} \hat{D}_{-1}[\xi, \xi^*] \right) \nonumber \\
 =  & \sum_{n=0}^\infty \sum_{{n'}=0}^\infty \sum_{\mathcal{M} \subseteq [n]} \sum_{\mathcal{M}' \subseteq [n'] } \sum_{\mathcal{M}_+ \subseteq \mathcal{M}} \sum_{\mathcal{M}'_+ \subseteq \mathcal{M}' } \int_{\mathbb{R}^n} d^n \textbf{k}_n  \int_{\mathbb{R}^{n'}} d^{n'} \textbf{k}'_{n'}  \frac{\tilde{\phi}_n (\textbf{k}_n) \tilde{\phi}_{n'}^* (\textbf{k}'_{n'})}{\sqrt{n!n'!}}  \nonumber \\
& \times  C_{\mathsf{K},\mathsf{N}} \left( \{ k_i \}_{i \in [n] \setminus \mathcal{M} } , \{ k'_{i'} \}_{i' \in [n'] \setminus \mathcal{M}' } \right) \text{Tr} \left( \prod_{j_- \in\mathcal{M} \setminus \mathcal{M}_+} \hat{A}_-^\dagger(k_{j_-})  \right. \nonumber \\
& \left. \times  \prod_{j'_+\in\mathcal{M}'_+} \hat{A}'_+(k'_{j'_+})   \hat{\rho}_0  \prod_{j_+\in\mathcal{M}_+} \hat{A}_+^{\prime \dagger}(k_{j_+})   \prod_{j'_-\in\mathcal{M}'\setminus\mathcal{M}'_+} \hat{A}_-(k'_{j'_-}) \hat{D}_{-1}[\xi, \xi^*] \right) \nonumber \\
 =  & \sum_{n=0}^\infty \sum_{{n'}=0}^\infty \sum_{\mathcal{M} \subseteq [n]} \sum_{\mathcal{M}' \subseteq [n'] } \sum_{\mathcal{M}_+ \subseteq \mathcal{M}} \sum_{\mathcal{M}'_+ \subseteq \mathcal{M}' } \int_{\mathbb{R}^n} d^n \textbf{k}_n  \int_{\mathbb{R}^{n'}} d^{n'} \textbf{k}'_{n'} \frac{\tilde{\phi}_n (\textbf{k}_n) \tilde{\phi}_{n'}^* (\textbf{k}'_{n'})}{\sqrt{n!n'!}}  \nonumber \\
& \times  C_{\mathsf{K},\mathsf{N}} \left( \{ k_i \}_{i \in [n] \setminus \mathcal{M} } , \{ k'_{i'} \}_{i' \in [n'] \setminus \mathcal{M}' } \right) \text{Tr} \left(   \hat{\rho}_0   \prod_{j_+\in\mathcal{M}_+} \hat{A}_+^{\prime \dagger}(k_{j_+})  \right. \nonumber \\
& \left. \times  \prod_{j'_-\in\mathcal{M}'\setminus\mathcal{M}'_+} \hat{A}_-(k'_{j'_-}) \hat{D}_{-1}[\xi, \xi^*] \prod_{j_- \in\mathcal{M} \setminus \mathcal{M}_+} \hat{A}_-^\dagger(k_{j_-}) \prod_{j'_+\in\mathcal{M}'_+} \hat{A}'_+(k'_{j'_+})  \right).
\end{align}

As prescribed by step \ref{prescription_7}, we manipulate Eq.~(\ref{varrho_step_6}) by using Eq.~(\ref{multiple_Xi_derivatives_-1}):
\begin{align}\label{varrho_step_7_2}
& \chi^{(-1)}[\xi, \xi^*] \nonumber \\
 =  & \sum_{n=0}^\infty \sum_{{n'}=0}^\infty \sum_{\mathcal{M} \subseteq [n]} \sum_{\mathcal{M}' \subseteq [n'] } \sum_{\mathcal{M}_+ \subseteq \mathcal{M}} \sum_{\mathcal{M}'_+ \subseteq \mathcal{M}' } \int_{\mathbb{R}^n} d^n \textbf{k}_n  \int_{\mathbb{R}^{n'}} d^{n'} \textbf{k}'_{n'} \frac{\tilde{\phi}_n (\textbf{k}_n) \tilde{\phi}_{n'}^* (\textbf{k}'_{n'})}{\sqrt{n!n'!}}  \nonumber \\
& \times  C_{\mathsf{K},\mathsf{N}} \left( \{ k_i \}_{i \in [n] \setminus \mathcal{M} } , \{ k'_{i'} \}_{i' \in [n'] \setminus \mathcal{M}' } \right)  \prod_{j_- \in\mathcal{M} \setminus \mathcal{M}_+} \overrightarrow{\Delta}_-(k_{j_-}) \prod_{j'_+\in\mathcal{M}'_+} \overrightarrow{\Delta}_+(k'_{j'_+})   \nonumber \\
&  \times \prod_{j_+\in\mathcal{M}_+} \left[ -\overrightarrow{\Delta}^*_+(k_{j_+}) \right]  \prod_{j'_-\in\mathcal{M}'\setminus\mathcal{M}'_+} \left[ - \overrightarrow{\Delta}^*_-(k'_{j'_-}) \right] \chi^{(-1)}_0[\xi,\xi^*],
\end{align}
with
\begin{subequations}
\begin{align}
 & \overrightarrow{\Delta}_+(k) = \int_{\mathbb{R}}dK 2 \sinh \left(\frac{c \beta |K|}{2} \right) e^{-c \beta|K|} \alpha(k,K)\frac{\delta}{\delta \xi (K)}, \\
& \overrightarrow{\Delta}_-(k) = \int_{\mathbb{R}}dK 2 \sinh \left(\frac{c \beta |K|}{2} \right) \beta(k,K)\frac{\delta}{\delta \xi (K)}
\end{align}
\end{subequations}
as functional derivatives acting on their right.

By using Eq.~(\ref{thermal_characteristic_fucntion}), we obtain
\begin{align}\label{varrho_step_7_3}
& \chi^{(-1)}[\xi, \xi^*]  \nonumber \\
 =  & \sum_{n=0}^\infty \sum_{{n'}=0}^\infty \sum_{\mathcal{M} \subseteq [n]} \sum_{\mathcal{M}' \subseteq [n'] } \sum_{\mathcal{M}_+ \subseteq \mathcal{M}} \sum_{\mathcal{M}'_+ \subseteq \mathcal{M}' } \int_{\mathbb{R}^n} d^n \textbf{k}_n  \int_{\mathbb{R}^{n'}} d^{n'} \textbf{k}'_{n'} \frac{\tilde{\phi}_n (\textbf{k}_n) \tilde{\phi}_{n'}^* (\textbf{k}'_{n'})}{\sqrt{n!n'!}}  \nonumber \\
& \times  C_{\mathsf{K},\mathsf{N}} \left( \{ k_i \}_{i \in [n] \setminus \mathcal{M} } , \{ k'_{i'} \}_{i' \in [n'] \setminus \mathcal{M}' } \right)  \prod_{j_- \in\mathcal{M} \setminus \mathcal{M}_+} \overrightarrow{\Delta}_-(k_{j_-}) \prod_{j'_+\in\mathcal{M}'_+} \overrightarrow{\Delta}_+(k'_{j'_+})   \nonumber \\
&  \times \prod_{j_+\in\mathcal{M}_+} L_+(k_{j_+}) [\xi] \prod_{j'_-\in\mathcal{M}'\setminus\mathcal{M}'_+} L_-(k'_{j'_-}) [\xi] \chi^{(-1)}_0[\xi,\xi^*],
\end{align}
with
\begin{subequations} \label{L_-B}
\begin{align}
L_+ (k) [\xi] = & \int_{\mathbb{R}}dK 2 \sinh \left(\frac{c \beta |K|}{2} \right) e^{-c \beta|K|} \alpha^*(k,K) (n_0+1) \xi(K) ,
\end{align}
\begin{align}
L_- (k) [\xi]  = & \int_{\mathbb{R}}dK 2 \sinh \left(\frac{c \beta |K|}{2} \right) \beta^*(k,K) (n_0+1) \xi(K) .
\end{align}
\end{subequations}
Equation (\ref{L_-B}) can be computed by means of Eq.~(\ref{alpha_to_beta}) as follows
\begin{subequations}
\begin{align}
L_+ (k) [\xi]  = & \int_{\mathbb{R}}dK 2 \sinh \left(\frac{c \beta |K|}{2} \right) \exp \left(-\frac{c \beta |K|}{2} \right) \beta^*(k,K) (n_0+1) \xi(K) \nonumber \\
 = & \int_{\mathbb{R}}dK \left( 1-e^{-\beta |K|} \right)  \left( \frac{1}{e^{\beta|K|}-1} +1 \right) \beta^*(k,K) \xi(K) \nonumber \\
 = & \int_{\mathbb{R}}dK \frac{e^{\beta |K|}-1}{e^{\beta |K|}}  \frac{e^{\beta|K|}}{e^{\beta|K|}-1} \beta^*(k,K) \xi(K) \nonumber \\
= & \int_{\mathbb{R}}dK \beta^*(k,K) \xi(K),\\
L_- (k) [\xi] = & \int_{\mathbb{R}}dK 2 \sinh \left(\frac{c \beta |K|}{2} \right) \exp \left(-\frac{c \beta |K|}{2} \right) \alpha^*(k,K) (n_0+1) \xi(K) \nonumber \\
 = & \int_{\mathbb{R}}dK \left( 1-e^{-\beta |K|} \right) \left( \frac{1}{e^{\beta|K|}-1} +1 \right) \alpha^*(k,K) \xi(K)\nonumber \\
 = & \int_{\mathbb{R}}dK \frac{e^{\beta |K|}-1}{e^{\beta |K|}}  \frac{e^{\beta|K|}}{e^{\beta|K|}-1} \alpha^*(k,K) \xi(K)\nonumber \\
 = & \int_{\mathbb{R}}dK \alpha^*(k,K) \xi(K).
\end{align}
\end{subequations}

The derivatives $\overrightarrow{\Delta}_\pm(k) $ now have to be evaluated on both  $L_\pm (k) [\xi]$ and  $\chi^{(-1)}_0[\xi,\xi^*]$. In order to simplify the calculation, we define $\overleftarrow{\Delta}_\pm(k) $ as derivatives identical to $\overrightarrow{\Delta}_\pm(k) $ but acting on their left. Moreover, we define $\overleftrightarrow{\Delta}_\pm(k) = \overleftarrow{\Delta}_\pm(k) + \overrightarrow{\Delta}_\pm(k) $. In this way, Eq.~(\ref{varrho_step_7_3}) can be put in a more compact form, i.e.,
\begin{align}\label{varrho_step_7_4}
\chi^{(-1)}[\xi, \xi^*]   =  & \sum_{n=0}^\infty \sum_{{n'}=0}^\infty \sum_{\mathcal{M} \subseteq [n]} \sum_{\mathcal{M}' \subseteq [n'] } \int_{\mathbb{R}^n} d^n \textbf{k}_n  \int_{\mathbb{R}^{n'}} d^{n'} \textbf{k}'_{n'} \frac{\tilde{\phi}_n (\textbf{k}_n) \tilde{\phi}_{n'}^* (\textbf{k}'_{n'})}{\sqrt{n!n'!}}  \nonumber \\
& \times  C_{\mathsf{K},\mathsf{N}} \left( \{ k_i \}_{i \in [n] \setminus \mathcal{M} } , \{ k'_{i'} \}_{i' \in [n'] \setminus \mathcal{M}' } \right)  \prod_{j \in \mathcal{M} } \left\lbrace \overleftrightarrow{\Delta}_-(k_j) + L_+(k_j) [\xi] \right\rbrace  \nonumber \\
& \times  \prod_{j'\in\mathcal{M}'} \left\lbrace \overleftrightarrow{\Delta}_+(k'_{j'})  + L_-(k'_{j'}) [\xi] \right\rbrace  \chi^{(-1)}_0[\xi,\xi^*].
\end{align}
A further simplification can be made by defining $C_\Delta(\mathcal{U} , \mathcal{U}')$ as a contraction with the following fundamental contractions
\begin{subequations} \label{C_Delta}
\begin{align}
C_\Delta(\{k_1,k_2 \} ,\varnothing) =  \overrightarrow{\Delta}_-(k_1)  L_+ (k_2) [\xi] +  \overrightarrow{\Delta}_-(k_2)  L_+ (k_1) [\xi],\\
C_\Delta(\varnothing,\{k_1,k_2 \})  =   \overrightarrow{\Delta}_+(k_1)  L_- (k_2) [\xi] +  \overrightarrow{\Delta}_+(k_2)  L_- (k_1) [\xi],\\
C_\Delta(\{k \},\{k' \})  =  \overrightarrow{\Delta}_-(k)  L_- (k') [\xi] + \overrightarrow{\Delta}_+(k')  L_+ (k) [\xi]
\end{align}
\end{subequations}
and by computing the following identities by means of Eqs.~(\ref{alpha_to_beta}) and (\ref{thermal_characteristic_fucntion})
\begin{subequations}
\begin{align}
& \left\lbrace \overrightarrow{\Delta}_-(k) + L_+ (k) [\xi] \right\rbrace  \chi^{(-1)}_0[\xi,\xi^*] \nonumber \\
 = &\int_{\mathbb{R}}dK  \left[ 2 \sinh \left(\frac{c \beta |K|}{2} \right) \beta(k,K)\frac{\delta}{\delta \xi (K)} + \beta^*(k,K) \xi(K) \right]   \chi^{(-1)}_0[\xi,\xi^*] \nonumber \\
= &\int_{\mathbb{R}}dK  \left\lbrace -2 \sinh \left(\frac{c \beta |K|}{2} \right) \beta(k,K)[n_0(K)+1]\xi ^*(K) + \beta^*(k,K) \xi(K) \right\rbrace   \nonumber \\
& \times \chi^{(-1)}_0[\xi,\xi^*] \nonumber \\
= &\int_{\mathbb{R}}dK  \left[ -  \exp \left(\frac{c \beta |K|}{2} \right) \beta(k,K) \xi ^*(K) + \beta^*(k,K) \xi(K) \right]   \chi^{(-1)}_0[\xi,\xi^*] \nonumber \\
= &\int_{\mathbb{R}}dK  \left[ -  \alpha(k,K) \xi ^*(K) + \beta^*(k,K) \xi(K) \right]   \chi^{(-1)}_0[\xi,\xi^*] \nonumber \\
= & \{ - L (k) [\xi,\xi^*] \}^* \chi^{(-1)}_0[\xi,\xi^*],\\
& \left\lbrace \overrightarrow{\Delta}_+(k) + L_- (k) [\xi] \right\rbrace \chi^{(-1)}_0[\xi,\xi^*] \nonumber \\
= & \int_{\mathbb{R}}dK \left[ 2 \sinh \left(\frac{c \beta |K|}{2} \right) e^{-c \beta|K|} \alpha(k,K)\frac{\delta}{\delta \xi (K)} + \alpha^*(k,K) \xi(K) \right] \chi^{(-1)}_0[\xi,\xi^*] \nonumber \\
= & \int_{\mathbb{R}}dK \left\lbrace -2 \sinh \left(\frac{c \beta |K|}{2} \right) e^{-c \beta|K|} \alpha(k,K)[n_0(K)+1]\xi^* (K) \right. \nonumber \\
& \left. +  \alpha^*(k,K) \xi(K) \right\rbrace  \chi^{(-1)}_0[\xi,\xi^*]\nonumber \\
= &\int_{\mathbb{R}}dK  \left[ - \exp \left(-\frac{c \beta |K|}{2} \right) \alpha(k,K) \xi ^*(K) + \alpha^*(k,K) \xi(K) \right]   \chi^{(-1)}_0[\xi,\xi^*] \nonumber \\
= &\int_{\mathbb{R}}dK  \left[ - \beta(k,K) \xi ^*(K) + \alpha^*(k,K) \xi(K) \right]   \chi^{(-1)}_0[\xi,\xi^*] \nonumber \\
= &  L (k) [\xi,\xi^*] \chi^{(-1)}_0[\xi,\xi^*],
\end{align}
\end{subequations}
where $ L (k) [\xi,\xi^*]$ is defined by Eq.~(\ref{L_xi}). In this way, Eq.~(\ref{varrho_step_7_4}) can be computed as follows
\begin{align}\label{varrho_step_7_5}
& \chi^{(-1)}[\xi, \xi^*]\nonumber\\
= & \sum_{n=0}^\infty \sum_{{n'}=0}^\infty \sum_{\mathcal{M} \subseteq [n]} \sum_{\mathcal{M}' \subseteq [n'] } \sum_{\mathcal{S} \subseteq \mathcal{M}} \sum_{\mathcal{S}' \subseteq \mathcal{M}' } \int_{\mathbb{R}^n} d^n \textbf{k}_n  \int_{\mathbb{R}^{n'}} d^{n'} \textbf{k}'_{n'} \frac{\tilde{\phi}_n (\textbf{k}_n) \tilde{\phi}_{n'}^* (\textbf{k}'_{n'})}{\sqrt{n!n'!}}  \nonumber \\
& \times  C_{\mathsf{K},\mathsf{N}} \left( \{ k_i \}_{i \in [n] \setminus \mathcal{M} } , \{ k'_{i'} \}_{i' \in [n'] \setminus \mathcal{M}' } \right) C_\Delta \left( \{ k_l \}_{l \in \mathcal{M} \setminus \mathcal{S} } , \{ k'_{l'} \}_{l' \in \mathcal{M}' \setminus \mathcal{S}' } \right)  \nonumber \\
& \times  \prod_{j \in \mathcal{S} } \left\lbrace - L(k_j) [\xi,\xi^*] \right\rbrace^*  \prod_{j'\in\mathcal{S}'} L(k'_{j'}) [\xi,\xi^*]  \chi^{(-1)}_0[\xi,\xi^*].
\end{align}

The fundamental contractions of $C_\Delta(\mathbf{U} , \mathbf{U}')$ defined by Eq.~(\ref{C_Delta}) can be computed by means of Eq.~(\ref{alpha_to_beta}) as follows
\begin{subequations}
\begin{align}
&  C_\Delta(\{k_1,k_2 \} ,\varnothing)\nonumber \\
 = &  \int_{\mathbb{R}}dK 2 \sinh \left(\frac{c \beta |K|}{2} \right) [\beta(k_1,K) \beta^*(k_2,K)+\beta^*(k_1,K) \beta(k_2,K)] \nonumber \\
 = &  - C_{\mathsf{K},\mathsf{N}}(\{k_1,k_2 \} ,\varnothing),\\
& C_\Delta(\varnothing,\{k_1,k_2 \})  \nonumber \\
 = &  \int_{\mathbb{R}}dK 2 \sinh \left(\frac{c \beta |K|}{2} \right) e^{-c \beta|K|} [\alpha(k_1,K) \alpha^*(k_2,K)+\alpha^*(k_1,K) \alpha(k_2,K)]\nonumber \\
= & \int_{\mathbb{R}}dK 2 \sinh \left(\frac{c \beta |K|}{2} \right) [\beta(k_1,K) \beta^*(k_2,K)+\beta^*(k_1,K) \beta(k_2,K)] \nonumber \\
 = &  - C_{\mathsf{K},\mathsf{N}}(\varnothing,\{k_1,k_2 \}),\\
& C_\Delta(\{k \},\{k' \}) \nonumber \\
 = &  \int_{\mathbb{R}}dK 2 \sinh \left(\frac{c \beta |K|}{2} \right)[ \beta(k,K) \alpha^*(k',K) +   e^{-c \beta|K|} \beta^*(k,K) \alpha(k',K) ]
\end{align}
\end{subequations}

The contractions $C_{\mathsf{K},\mathsf{N}}(\mathbf{U} , \mathbf{U}')$ and $C_\Delta(\mathbf{U} , \mathbf{U}')$ appearing in Eq.~(\ref{varrho_step_7_5}) can be combined into a single contraction $C(\mathbf{U} , \mathbf{U}') = C_{\mathsf{K},\mathsf{N}}(\mathbf{U} , \mathbf{U}') + C_\Delta(\mathbf{U} , \mathbf{U}')$ which has the following fundamental contractions
\begin{subequations}\label{c}
\begin{align}
& C(\{k_1,k_2 \} ,\varnothing) =  C_{\mathsf{K},\mathsf{N}}(\{k_1,k_2 \} ,\varnothing) + C_\Delta(\{k_1,k_2 \} ,\varnothing)  =  0,\label{c_a}\\
 & C(\varnothing,\{k_1,k_2 \})  =  C_{\mathsf{K},\mathsf{N}}(\varnothing,\{k_1,k_2 \}) + C_\Delta(\varnothing,\{k_1,k_2 \}) =  0,\label{c_b}\\
& C(\{k \},\{k' \})  =  C_{\mathsf{K},\mathsf{N}}(\{k \},\{k' \}) + C_\Delta(\{k \},\{k' \}) \nonumber \\
 = &   \int_{\mathbb{R}}dK 2 \sinh \left(\frac{c \beta |K|}{2} \right)  \left[  2 \sinh \left(\frac{c \beta |K|}{2} \right) e^{-c \beta|K|}  \alpha^*(k,K) \alpha(k',K)\right. \nonumber \\
 & \left. +  \beta(k,K) \alpha^*(k',K) +   e^{-c \beta|K|} \beta^*(k,K) \alpha(k',K)  \right].\label{c_c}
\end{align}
\end{subequations}
The contraction (\ref{c_c}) can be computed by means of Eqs.~(\ref{F_11}), (\ref{Bogolyubov_coefficients}) and (\ref{alpha_to_beta}) as follows
\begin{align}\label{c_c_2}
& C(\{k \},\{k' \}) \nonumber \\
= & \int_{\mathbb{R}}dK 2 \sinh \left(\frac{c \beta |K|}{2} \right)  [\alpha^*(k,K) \beta(k',K) +  \beta(k,K) \alpha^*(k',K) ] \nonumber\\
 = & \int_{\mathbb{R}}dK \theta(kK) \theta(k'K) 2 \sinh \left( \frac{\beta}{2} |K| \right) \frac{|K|}{\sqrt{kk'}} \nonumber\\
 & \times \left[ F(k,K)F(k',-K) + F(k,-K)F(k',K)  \right] \nonumber \\
= &  \frac{\theta(kk')}{\sqrt{kk'}} \left[ \int_{\mathbb{R}}dK \theta(kK) 2 \sinh \left( \frac{\beta}{2} |K| \right) |K|  F(k,K)F(k',-K) \right. \nonumber \\
& \left. + \int_{\mathbb{R}}dK \theta(kK) 2 \sinh \left( \frac{\beta}{2} |K| \right) |K| F(k,-K)F(k',K)  \right] \nonumber \\
= &  \frac{\theta(kk')}{\sqrt{kk'}} \left[ \int_{\mathbb{R}}dK \theta(kK) 2 \sinh \left( \frac{\beta}{2} |K| \right) |K|  F(k,K)F(k',-K) \right. \nonumber \\
& \left. + \int_{\mathbb{R}}dK \theta(-kK) 2 \sinh \left( \frac{\beta}{2} |K| \right) |K| F(k,K)F(k',-K)  \right] \nonumber \\
= &  \frac{\theta(kk')}{\sqrt{kk'}} \int_{\mathbb{R}}dK \left[ \theta(kK)+\theta(-kK) \right] 2 \sinh \left( \frac{\beta}{2} |K| \right) |K|  F(k,K)F(k',-K) \nonumber \\
= &  \frac{\theta(kk')}{\sqrt{kk'}} \int_{\mathbb{R}}dK 2 \sinh \left( \frac{\beta}{2} |K| \right) |K|  F(k,K)F(k',-K) \nonumber \\
= & \frac{\theta(kk')}{\sqrt{kk'}} \int_{\mathbb{R}}dK 2 \sinh \left( \frac{\beta}{2} |K| \right) \frac{|K|}{(2 \pi a)^2} \left| \Gamma \left( \frac{i K}{a} \right) \right|^2 \exp \left( i \frac{K}{a} \ln \left| \frac{k}{k'} \right| \right) \nonumber \\
= & \frac{\theta(kk')}{\sqrt{kk'}} \int_{\mathbb{R}} \frac{dK}{2 \pi a}  \exp \left( i \frac{K}{a} \ln \left| \frac{k}{k'} \right| \right)\nonumber \\
= & \frac{\theta(kk')}{\sqrt{kk'}} \delta \left( \ln \left| \frac{k}{k'} \right| \right) \nonumber \\
= & \frac{\theta(kk')}{|k|} \delta \left( \ln \left| \frac{k}{k'} \right| \right) \nonumber \\
= & \theta(kk') \delta \left( |k| - |k'| \right)\nonumber \\
= & \delta \left( k - k' \right).
\end{align}
Hence, the fundamental contractions of $C(\mathbf{U} , \mathbf{U}')$ are
\begin{align}\label{c_2}
& C(\{k_1,k_2 \} ,\varnothing)  =  0,
 && C(\varnothing,\{k_1,k_2 \})  =  0,
&& C(\{k \},\{k' \})  =  \delta ( k - k' ).
\end{align}
As a consequence of Eq.~(\ref{c_2}), the combinatorial expression for the full contractions $C(\mathbf{U} , \mathbf{U}')$ is precisely given by Eq.~(\ref{c_combinatory}).

Owing to the definition of $C(\mathbf{U} , \mathbf{U}') = C_{\mathsf{K},\mathsf{N}}(\mathbf{U} , \mathbf{U}') + C_\Delta(\mathbf{U} , \mathbf{U}')$, Eq.~(\ref{varrho_step_7_5}) becomes
\begin{align}\label{varrho_step_7_6}
\chi^{(-1)}[\xi, \xi^*] = & \sum_{n=0}^\infty \sum_{{n'}=0}^\infty \sum_{\mathcal{S} \subseteq [n]} \sum_{\mathcal{S}' \subseteq [n'] } \int_{\mathbb{R}^n} d^n \textbf{k}_n  \int_{\mathbb{R}^{n'}} d^{n'} \textbf{k}'_{n'} \frac{\tilde{\phi}_n (\textbf{k}_n) \tilde{\phi}_{n'}^* (\textbf{k}'_{n'})}{\sqrt{n!n'!}}   \nonumber \\
& \times C \left( \{ k_i \}_{i \in [n] \setminus \mathcal{S} } , \{ k'_{i'} \}_{i' \in [n'] \setminus \mathcal{S}' } \right)  \nonumber \\
& \times  \prod_{j \in \mathcal{S} } \left\lbrace - L(k_j) [\xi,\xi^*] \right\rbrace^*  \prod_{j'\in\mathcal{S}'} L(k'_{j'}) [\xi,\xi^*]  \chi^{(-1)}_0[\xi,\xi^*].
\end{align}
In this way we have concluded step \ref{prescription_7}.

Finally, with step \ref{prescription_8}, we obtain Eq.~(\ref{chi_final}).

\subsubsection{Proof of Eq.~(\ref{chi_M_final})}

Now, we prove Eq.~(\ref{chi_M_final}). By plugging Eq.~(\ref{free_state_decomposition_11}) in Eq.~(\ref{chi_M}), we obtain Eq.~(\ref{chi_M_final_a}) with
\begin{align}
& \bar{\chi}^{(p)}_\text{M}(\textbf{k}_n , \textbf{k}'_{n'})[\xi, \xi^*] = \text{Tr} \left( \prod_{i=1}^n \hat{a}^\dagger (k_i) |0_\text{M}\rangle \langle 0_\text{M} | \prod_{i'=1}^{n'} \hat{a} (k'_{i'}) \right. \nonumber \\
& \left. \times \exp \left( \int_{\mathbb{R}} dk \left[ \xi(k) \hat{a}^\dagger(k)  - \xi^*(k) \hat{a}(k) + \frac{p}{2} |\xi(k)|^2\right] \right) \right)
\end{align}

By choosing $p=-1$, we obtain
\begin{align}
\bar{\chi}^{(p)}_\text{M}(\textbf{k}_n , \textbf{k}'_{n'})[\xi, \xi^*] = & \text{Tr} \left( \prod_{i=1}^n \hat{a}^\dagger (k_i) |0_\text{M}\rangle \langle 0_\text{M} | \prod_{i'=1}^{n'} \hat{a} (k'_{i'}) \exp \left( -\int_{\mathbb{R}} dk \xi^*(k) \hat{a}(k) \right) \right. \nonumber \\
& \left. \times  \exp \left( \int_{\mathbb{R}} dk  \xi(k) \hat{a}^\dagger(k)  \right) \right) \nonumber \\
 = & \text{Tr} \left(  |0_\text{M}\rangle \langle 0_\text{M} | \prod_{i'=1}^{n'} \hat{a} (k'_{i'}) \exp \left( -\int_{\mathbb{R}} dk \xi^*(k) \hat{a}(k) \right) \right. \nonumber \\
& \left. \times  \exp \left( \int_{\mathbb{R}} dk  \xi(k) \hat{a}^\dagger(k)  \right) \prod_{i=1}^n \hat{a}^\dagger (k_i) \right) \nonumber \\
 = & \prod_{i=1}^n \frac{\delta}{\delta \xi(k_i)} \prod_{i'=1}^{n'} \left[ - \frac{\delta}{\delta \xi^*(k'_{i'})} \right] \text{Tr} \left(  |0_\text{M}\rangle \langle 0_\text{M} | \right. \nonumber \\
& \left. \times  \exp \left( -\int_{\mathbb{R}} dk \xi^*(k) \hat{a}(k) \right) \exp \left( \int_{\mathbb{R}} dk  \xi(k) \hat{a}^\dagger(k)  \right) \right) \nonumber \\
 = & \prod_{i=1}^n \frac{\delta}{\delta \xi(k_i)} \prod_{i'=1}^{n'} \left[ - \frac{\delta}{\delta \xi^*(k'_{i'})} \right] \chi^{(-1)}_{0\text{M}}[\xi, \xi^*],
\end{align}
which, owing to Eq.~(\ref{chi_0M_final}), result in
\begin{align}
& \bar{\chi}^{(p)}_\text{M}(\textbf{k}_n , \textbf{k}'_{n'})[\xi, \xi^*] \nonumber \\ = & \prod_{i=1}^n \frac{\delta}{\delta \xi(k_i)} \prod_{i'=1}^{n'} \xi(k'_{i'}) \chi^{(-1)}_{0\text{M}}[\xi, \xi^*] \nonumber \\
 = & \sum_{\mathcal{S} \subseteq [n]} \sum_{\mathcal{S}' \subseteq [n'] } C \left( \{ k_i \}_{i \in [n] \setminus \mathcal{S}}, \{ k'_{i'} \}_{i' \in [n'] \setminus \mathcal{S}'} \right) \prod_{i' \in \mathcal{S}'} \xi(k'_{i'}) \prod_{i\in\mathcal{S}} \frac{\delta}{\delta \xi(k_i)} \chi^{(-1)}_{0\text{M}}[\xi, \xi^*] \nonumber \\
 = & \sum_{\mathcal{S} \subseteq [n]} \sum_{\mathcal{S}' \subseteq [n'] } C \left( \{ k_i \}_{i \in [n] \setminus \mathcal{S}}, \{ k'_{i'} \}_{i' \in [n'] \setminus \mathcal{S}'} \right) \prod_{i' \in \mathcal{S}'} \xi(k'_{i'}) \prod_{i\in\mathcal{S}} [ -\xi^*(k_i)] \chi^{(-1)}_{0\text{M}}[\xi, \xi^*].
\end{align}
This proves Eq.~(\ref{chi_M_final_b}) for $p=-1$.

The general case can be proven by using the following identities
\begin{subequations}
\begin{align}
& \chi^{(p)}_\text{M}[\xi,\xi^*] = \exp \left( \int_{\mathbb{R}} dk \frac{p+1}{2} |\xi(k)|^2 \right) \chi^{(- 1)}_\text{M}[\xi,\xi^*], \\
& \chi^{(p)}_{0\text{M}}[\xi,\xi^*] = \exp \left( \int_{\mathbb{R}} dk \frac{p+1}{2} |\xi(k)|^2 \right) \chi^{(- 1)}_{0\text{M}}[\xi,\xi^*].
\end{align}
\end{subequations}

\chapter{Frame dependent nonrelativistic limit}\label{Framedependent_nonrelativistic_limit}

\textit{This chapter is based on and contains material from Ref.~\citeRF{PhysRevD.107.085016}.}

\section{Introduction}\label{Framedependent_nonrelativistic_limit_Introduction}

In Chap.~\ref{Non_relativistic_limit_of_QFT_and_QFTCS}, we studied the nonrelativistic limit of quantum fields in Minkowski [Sec.~\ref{Minkowski_spacetime}] and Rindler [Sec.~\ref{Rindler_frame}] spacetimes. In particular, we considered the nonrelativistic condition as the regime in which energies are very close to the mass energy. As a result, we developed the framework for nonrelativistic phenomena in inertial and accelerated frames when Eqs.~(\ref{non_relativistic_limit}) and (\ref{non_relativistic_limit_curved}), respectively, are satisfied.

Inertial and accelerated observers experience different flows of times due to the inequivalent time coordinates $t$ and $T$. Hence, they are provided with different notions of energy as the generator of time translation. This means that the nature of particle energy is frame dependent. Notice that if two observers experience different energy content of particles, then also the nonrelativistic condition appears to be frame dependent. The aim of the present chapter is to study such a frame dependent effect.

In Chap.~\ref{Frame_dependent_content_of_particles}, we considered the scenario in which both the inertial and the accelerated experimenter observe and interact with the same physical setup. By following the algebraic formulation of QFTCS, we showed that the same state is represented in both frames by different particle representations. Each representation can be mapped to the other by means of a Bogoliubov transformation.

The algebraic QFTCS appears to be the suitable framework to compare the nonrelativistic condition in the two frames. In the present chapter, we will use the results of Chap.~\ref{Frame_dependent_content_of_particles} to show that nonrelativistic particle states of one frame appear as a superposition of relativistic and nonrelativistic particles in the other frame; hence, we will prove that the nonrelativistic limit is frame-dependent. Also, we show that the number of particles is not conserved between the two frames as a consequence of the different particle representations of states.

In Sec.~\ref{Rindler_frame}, we introduced the notion of quasi-inertiality by means of the conditions (\ref{local_limit}). In the nonrelativistic quasi-inertial regime, both the acceleration $\alpha$ and the particle localization are sufficiently constrained to suppress GR corrections to the dynamics coming from, respectively, high values of $\alpha$ and noticeable differences between the Rindler ($g_{\mu\nu}$) and the Minkowski ($\eta_{\mu\nu}$) metric. In the absence of GR perturbations, one may expect that the description of the field is equivalent to the one provided by the flat QFT with some minimal modifications (e.g., Newton potential); hence, the aforementioned frame dependent effect in QFTCS is expected to be suppressed in the quasi-inertial regime. In the present chapter, we explicitly prove this result for both scalar and Dirac fields. In particular, we show that the accelerated observer agrees with the inertial frame about the nonrelativistic nature of particles and the number of particles is conserved.

The chapter is organized as follows. In Sec.~\ref{Inertial_and_non_inertial_frame}, we show that the nonrelativistic limit of one observer is generally incompatible with the nonrelativistic condition in the other frame due to the Bogoliubov transformations mixing relativistic and nonrelativistic modes; also, we discuss the difference between number of particles and antiparticles in the two frames. These frame dependent effects are suppressed in the quasi-inertial regime, as we will prove in Sec.~\ref{Quasi_inertial_regime} for scalar and Dirac fields. Conclusions are drawn in Sec.~\ref{Framedependent_nonrelativistic_limit_Conclusions}.

\section{Inertial and accelerated frames}\label{Inertial_and_non_inertial_frame}

The aim of this section is to show that the nonrelativistic limit in the Minkowski frame is generally non compatible with the nonrelativistic limit in the Rindler frame; we also discuss the particle production between the inertial and the accelerated frame.

Consider an inertial (Alice) and an accelerated (Rob) experimenter performing operations (e.g., preparation and observation) on the same physical state $| \phi \rangle$. Each observer is provided with a representation of $| \phi \rangle$ in terms of Minkowski and Rindler particles, respectively. Now, Alice claims that $| \phi \rangle$ is nonrelativistic if $| \phi \rangle$ is only made of nonrelativistic Minkowski particles---i.e., if $| \phi \rangle$ is defined by Minkowski creators satisfying Eq.~(\ref{non_relativistic_limit}) and acting on the Minkowski vacuum $| 0_\text{M} \rangle$. Conversely, Rob claims that $| \phi \rangle$ is nonrelativistic if it is only populated by nonrelativistic Rindler particles satisfying Eq.~(\ref{non_relativistic_limit_curved}) and created over the Rindler vacuum $| 0_{\text{L},\text{R}} \rangle$. Are these claims compatible to each other? Can both Alice and Rob detect nonrelativistic states at the same time?

The answer to these questions can be deduced from the Bogoliubov transformation relating Minkowski-Fock operators to Rindler-Fock operators. For scalar fields, these transformations are reported in Eq.~(\ref{Rindler_Bogoliubov_transformations}); whereas, for Dirac fields they are shown in Eq.~(\ref{Bogoliubov_transformations_3_Rindler_4}). In both cases, it can be noticed that the integral with respect to Rindler variables include also energies that do not satisfy Eq.~(\ref{non_relativistic_limit_curved}), even when only nonrelativistic Minkowski momenta $\vec{k}$ are considered. In other words, the condition $\hbar |\vec{k}|/mc \lesssim \epsilon^{1/2} $ does not guarantee the absence of relativistic Rindler operators in the right hand side of Eqs.~(\ref{Rindler_Bogoliubov_transformations}) and (\ref{Bogoliubov_transformations_3_Rindler_4}).

This means that the Bogoliubov transformations (\ref{Rindler_Bogoliubov_transformations}) and (\ref{Bogoliubov_transformations_3_Rindler_4}) mix nonrelativistic modes of one frame with relativistic modes of the other. The effect is twofold: (i) The sea of Rindler particles and antiparticles populating the Minkowski vacuum $| 0_\text{M} \rangle $ in the Rindler-Fock space $\mathcal{H}_{\text{L},\text{R}}$ generally appears to include states with relativistic energies; to see this, check Eqs.~(\ref{Rindler_vacuum_to_Minkowski}) and (\ref{Rindler_vacuum_to_Minkowski_unitary_operator}) for the scalar field and Eqs.~(\ref{Rindler_vacuum_to_Minkowski_Dirac}) and (\ref{OO}) for the Dirac field and notice that the integral on the right hand side of Eqs.~(\ref{Rindler_vacuum_to_Minkowski_unitary_operator}) and (\ref{OO}) include Rindler operators that do not satisfy the nonrelativistic condition (\ref{non_relativistic_limit_curved}) as well. (ii) Any nonrelativistic Minkowski particle creator can be responsible for the creation and the destruction of relativistic Rindler particles and antiparticles. These two facts imply that elements of the Minkowski-Fock space $\mathcal{H}_\text{M}$ that are made of nonrelativistic Minkowski particles are generally made of relativistic Rindler particles and antiparticles when seen as elements of the Rindler-Fock space $\mathcal{H}_{\text{L},\text{R}}$. The other way around is also true: not always elements of $\mathcal{H}_{\text{L},\text{R}}$ made of nonrelativistic particles and antiparticles are also made of nonrelativistic particles and antiparticles in $\mathcal{H}_\text{M}$. The conclusion is that the nonrelativistic limit is frame-dependent.

To see an example, consider the Minkowski single particle state of a scalar field defined by Eq.~(\ref{single_particle_Minkowski}). Assume that the state is nonrelativistic from the point of view of the inertial observer; explicitly, this means that the wave function $\tilde{\phi}_1(\vec{k})$ is vanishing for relativistic momenta, i.e.,
\begin{equation}\label{single_particle_nonrelativistic}
\tilde{\phi}_1(\vec{k}) \approx 0, \text{ if } \frac{\hbar |\vec{k}|}{mc} \gg \epsilon^{1/2}.
\end{equation}

The representation of the state $| \phi \rangle$ in the Rindler-Fock space $\mathcal{H}_{\text{L},\text{R}}$ is given by Eq.~(\ref{single_particle_Minkowski_Rindler}), with $\hat{S}_\text{S}$ defined by Eq.~(\ref{Rindler_vacuum_to_Minkowski_unitary_operator}). In the right hand side of  Eq.~(\ref{single_particle_Minkowski_Rindler}), one can see the presence of Rindler operators not satisfying the nonrelativistic condition (\ref{non_relativistic_limit_curved}) acting on the Rindler vacuum $| 0_\text{L}, 0_\text{R} \rangle$.  Independently of presence of the Minkowski single particle, the $\hat{S}_\text{S}$ operator generates a sea of relativistic Rindler particles over the Rindler vacuum  $| 0_\text{L}, 0_\text{R} \rangle$; furthermore, relativistic Rindler operators $\hat{A}_\nu^\dagger(\Omega,\vec{K}_\perp)$ act on the relativistic background $\hat{S}_\text{S} | 0_\text{L}, 0_\text{R} \rangle$ and generate additional relativistic particles.

Provably, these relativistic Rindler operators $\hat{A}_\nu^\dagger(\Omega,\vec{K}_\perp)$ do not always disappear when condition (\ref{single_particle_nonrelativistic}) holds. To see this, consider the approximation
\begin{align} \label{alpha_approximation}
 \alpha_\nu(\vec{k},\Omega, \vec{K}_\perp) \approx & \text{sign}(\Omega)  \frac{1}{2 } \left[  \frac{\pi m c^2 a }{\hbar}  \left| \sinh \left( \frac{\beta \Omega}{2} \right) \right| \right]^{-1/2}  \nonumber \\
& \times \exp \left( \frac{ \Omega}{c a} \left( \frac{\pi}{2} - s_\nu i \frac{\hbar k_3}{m c} \right) \right) \delta^2(\vec{k}_\perp-\vec{K}_\perp).
\end{align}
for the Bogoliubov coefficients (\ref{alpha_final}) in the limit $\hbar |\vec{k}|/mc \lesssim \epsilon^{1/2}$. As a consequence of Eq.~(\ref{alpha_approximation}), the functional
\begin{equation}\label{alpha_phi_tilde}
\alpha_\nu^*[\tilde{\phi}_1](\Omega, \vec{K}_\perp) = \int_{\mathbb{R}^3}d^3k \tilde{\phi}_1(\vec{k})  \alpha_\nu^*(\vec{k},\Omega, \vec{K}_\perp)
\end{equation}
can be approximated by
\begin{align}\label{alpha_phi_tilde_approx}
\alpha_\nu^*[\tilde{\phi}_1](\Omega, \vec{K}_\perp) \approx & \text{sign}(\Omega)  \left[  \frac{2 m c^2 a }{\hbar}  \left| \sinh \left( \frac{\beta \Omega}{2} \right) \right| \right]^{-1/2}  \exp \left( \frac{\beta \Omega}{4} \right)  \delta^2(\vec{k}_\perp-\vec{K}_\perp) \nonumber \\
& \times \mathcal{F}_3^{-1}[\tilde{\phi}_1](\vec{K}_\perp, s_\nu \hbar \Omega/mc^2a),
\end{align}
where
\begin{equation}
\mathcal{F}_3^{-1}[\tilde{\phi}_1](\vec{k}_\perp,x_3)  = \int_{\mathbb{R}}dk_3 \frac{e^{ik_3 x_3}}{\sqrt{2 \pi}} \tilde{\phi}_1(\vec{k})
\end{equation}
is the inverse of the Fourier transform of $\tilde{\phi}_1(\vec{k})$ with respect to the variable $k_3$. If $\mathcal{F}_3^{-1}[\tilde{\phi}_1](\vec{K}_\perp, s_\nu \hbar \Omega/mc^2a)$ is not supported in $|\hbar \Omega/mc^2 - 1| \lesssim \epsilon$, then the Rindler operators $\hat{A}_\nu^\dagger(\Omega,\vec{K}_\perp)$ appearing in Eq.~(\ref{single_particle_Minkowski_Rindler}) are smeared out with functions that have support outside of the nonrelativistic region (\ref{non_relativistic_limit_curved}).

The condition for $\mathcal{F}_3^{-1}[\tilde{\phi}_1](\vec{K}_\perp, s_\nu \hbar \Omega/mc^2a)$ to not have support in the region $|\hbar \Omega/mc^2 - 1| \lesssim \epsilon$ is compatible with condition (\ref{single_particle_nonrelativistic}), in the sense that one can find at least one configuration that satisfy both conditions. To see this, consider the uncertainty principle for Fourier transforms \cite{3167f59422e1481bb877f4a55b23c5be} which states that if $\tilde{\phi}_1(\vec{k})$ and $\mathcal{F}_3^{-1}[\tilde{\phi}_1](\vec{k}_\perp, x_3)$ are supported inside of $|k_3 - \bar{k}_3| \lesssim \Delta k_3$ and $|x_3 - \bar{x}_3| \lesssim \Delta x_3$, respectively, then $\Delta x_3 \Delta k_3 \geq 1$. Assume that Eq.~(\ref{single_particle_nonrelativistic}) holds; a necessary conditions for $\mathcal{F}_3^{-1}[\tilde{\phi}_1](\vec{K}_\perp, s_\nu \hbar \Omega/mc^2a)$ to have support inside of $| \hbar \Omega/mc^2 - 1| \lesssim \epsilon$ is to satisfy the uncertainty principle $\hbar a/mc \lesssim \epsilon^{3/2}$. Consequently, a sufficient condition for $\mathcal{F}_3^{-1}[\tilde{\phi}_1](\vec{K}_\perp, s_\nu \hbar \Omega/mc^2a)$ to not have support inside of $|\hbar \Omega/mc^2 - 1| \lesssim \epsilon$ is given by $\hbar a/mc \gg \epsilon^{3/2}$. We found that when the acceleration is sufficiently high ($\hbar a/mc \gg \epsilon^{3/2}$), the integral appearing in Eq.~(\ref{single_particle_Minkowski_Rindler}) includes relativistic Rindler creators as well, notwithstanding the nonrelativistic condition (\ref{single_particle_nonrelativistic}) for the Minkowski particle.

All the arguments that we used for the single particle case can also be extended to the case of more particles; this leads to the same frame dependent effect for any Minkowski-Fock state. The procedure can also be applied to the Dirac field case, due to the similarities between the Bogoliubov coefficients (\ref{alpha_final}) and (\ref{Bogoliubov_coefficient_2}).

In addition to the frame dependent nature of the nonrelativistic limit, the Bogoliubov transformations (\ref{Rindler_Bogoliubov_transformations}) and (\ref{Bogoliubov_transformations_3_Rindler_4}) give an explanation to the particle production between the two frames. An element of the Minkowski-Fock space $\mathcal{H}_\text{M}$ with $n$ particles and $m$ antiparticles does not appear as an element of the Rindler-Fock space $\mathcal{H}_{\text{L},\text{R}}$ with the same number of particles and antiparticles. This occurs because $| 0_\text{M} \rangle $ is not a vacuum state for $\mathcal{H}_{\text{L},\text{R}}$ and Minkowski particle (antiparticle) creators $\hat{a}^\dagger(\vec{k})$ ($\hat{b}^\dagger(\vec{k})$) annihilate Rindler antiparticles (particles), in addition to creating Rindler particles (antiparticles).

In conclusion, we proved the frame dependent nature of the first quantization scheme. Any Minkowski-Fock state $| \phi \rangle \in \mathcal{H}_\text{M}$ made of nonrelativistic particles can also be seen as an element of Rindler-Fock space $\mathcal{H}_{\text{L},\text{R}}$. The Minkowski vacuum background $| 0_\text{M} \rangle$ is converted into a see of relativistic and nonrelativistic Rindler particles and antiparticles [Eqs.~(\ref{Rindler_vacuum_to_Minkowski}), (\ref{Rindler_vacuum_to_Minkowski_unitary_operator}), (\ref{Rindler_vacuum_to_Minkowski_Dirac}) and (\ref{OO})]; whereas, any Minkowski creator $\hat{a}^\dagger(\vec{k})$ or $\hat{b}^\dagger(\vec{k})$ acting on $| 0_\text{M} \rangle$ is converted into Rindler creation and annihilation operators involving also relativistic modes [Eqs.~(\ref{Rindler_Bogoliubov_transformations}) and (\ref{Bogoliubov_transformations_3_Rindler_4})]. The nonrelativistic limit in the inertial frame is non-equivalent to the nonrelativistic limit in the accelerated frame; also, the number of particles and antiparticles is frame dependent as well.

So far, we have considered a sufficiently large acceleration ($\hbar a/mc \gg \epsilon^{3/2}$). We may expect that the resulting frame dependent effect is suppressed when this condition is not met. Intuitively, in a limit in which the two frames are similar, the nonrelativistic condition and the number of particles become approximately equivalent between the frames. In the following section we prove that such an equivalence precisely occurs in the quasi-inertial regime.

\section{Inertial and quasi-inertial frames}\label{Quasi_inertial_regime}

In Sec.~\ref{Inertial_and_non_inertial_frame}, we showed that the inertial and the accelerated observers, Alice and Rob, do not agree about the particle content of states. In particular, we showed that the nonrelativistic condition and the number of particles differ between the two frames. Here, instead, we investigate the regime in which these frame dependent effects are suppressed. We prove that the suppression occurs when Rob's acceleration is sufficiently low ($\hbar a /mc \lesssim \epsilon^{3/2}$), the nonrelativistic Minkowski-Fock state prepared by Alice is localized in $| a z - 1 | \lesssim \epsilon$ and Rob has only access to the region $| a Z - 1 | \lesssim \epsilon$ in his coordinate frame. In this regime, the nonrelativistic Minkowski-Fock state is nonrelativistic in the accelerated frame as well and the number of particles and antiparticles is conserved.

Additionally, we show that wave functions describing states in the quasi-inertial frame are approximated by the corresponding wave functions in the inertial frame, with the only difference coming from the coordinate transformation relating the two frames. In other words, particle states appear identical by both observers, up to their coordinate representations.

We detail the results by considering scalar Gaussian single particles. The accelerated observer sees a nonrelativistic single particle only when $\alpha$ is sufficiently small and the wave packet in the inertial frame is narrower than the scale length of the curvature, but wider than any relativistic wavelength. We also show that the wave function describing the state in the accelerated frame is approximately Gaussian.

The section is organized in the following way. In Sec.~\ref{Inertial_and_quasiinertial_frame} we define the quasi-inertial condition for the state as a localization condition in $| a z - 1 | \lesssim \epsilon$ for its wave function. In Sec.~\ref{Quasiinertial_observer} we define the notion of quasi-inertial observer as an accelerated experimenter with a sufficiently low acceleration (i.e., $\hbar a /mc \lesssim \epsilon^{3/2}$) and only having access to the region $| a Z - 1 | \lesssim \epsilon$. By combining the results of Secs.~\ref{Inertial_and_quasiinertial_frame} and \ref{Quasiinertial_observer}, we prove that the nonrelativistic limit and the number of particles are approximately the same in both frames. In Sec.~\ref{Wave_function_transformation}, we prove that the wave functions are the same as well, with the only difference coming from the coordinate representations. We detail these results in Sec.~\ref{Gaussian_singleparticle} for scalar Gaussian single particles.

\subsection{Quasi-inertial setup}\label{Inertial_and_quasiinertial_frame}

Here, we investigate the regime in which nonrelativistic Minkowski-Fock states are seen by the accelerated observer as made of nonrelativistic Rindler particles created over the Minkowski vacuum background. We show that a necessary and sufficient condition is given by the localization of the wave function inside the quasiflat region $| a z - 1 | \lesssim \epsilon$. States satisfying this condition are refereed to as quasi-inertial.

\subsubsection{Scalar Minkowski single particle}

Firstly, consider the simple case of a scalar Minkowski single particle, defined by Eq.~(\ref{single_particle_Minkowski}) and represented by Eq.~(\ref{single_particle_Minkowski_Rindler}) in $\mathcal{H}_{\text{L},\text{R}}$. Assume that the particle is nonrelativistic from the point of view of the inertial observer, which means that $\tilde{\phi}_1(\vec{k})$ satisfies Eq.~(\ref{single_particle_nonrelativistic}).

In Sec.~\ref{Inertial_and_non_inertial_frame} we showed that the function $\alpha_\nu^*[\tilde{\phi}_1](\Omega, \vec{K}_\perp)$ defined by Eq.~(\ref{alpha_phi_tilde}) satisfies Eq.~(\ref{alpha_phi_tilde_approx}). This means that $\alpha_\nu^*[\tilde{\phi}_1](\Omega, \vec{K}_\perp)$ has support in $|\hbar \Omega/mc^2 - 1| \lesssim \epsilon$ if and only if the inverse of the Fourier transforms of $\tilde{\phi}_1(\vec{k})$ with respect to $k_3$ is supported in $| s_\nu a x_3 - 1 | \lesssim \epsilon$, in the sense that
\begin{equation}\label{Fourier_phi_tilde_support}
\mathcal{F}_3^{-1}[\tilde{\phi}_1](\vec{k}_\perp, x_3) \approx 0, \text{ if } | s_\nu a x_3 - 1 | \gg \epsilon.
\end{equation}
This gives a necessary and sufficient condition for the $\nu$-Rindler creators acting on $\hat{S}_\text{S} | 0_\text{L}, 0_\text{R} \rangle$ in Eq.~(\ref{single_particle_Minkowski_Rindler}) to be nonrelativistic. Notice that the regions $|\hbar \Omega/mc^2 - 1| \lesssim \epsilon$ and $ \Omega < 0$ are disjoint, which means that if $\alpha_\nu^*[\tilde{\phi}_1](\Omega, \vec{K}_\perp)$ is supported in $|\hbar \Omega/mc^2 - 1| \lesssim \epsilon$, then it is vanishing in $ \Omega < 0$. Consequently, Eq.~(\ref{Fourier_phi_tilde_support}) is a necessary and sufficient condition for relativistic $\nu$-Rindler creators and all Rindler annihilators acting on $\hat{S}_\text{S} | 0_\text{L}, 0_\text{R} \rangle$ in Eq.~(\ref{single_particle_Minkowski_Rindler}) to be smeared out with a vanishing function.

The regions $| s_\nu a x_3 - 1 | \lesssim \epsilon$ and $ s_{\bar{\nu}} x_3 > 0$ are disjoint as well; hence, the function $\mathcal{F}_3^{-1}[\tilde{\phi}_1](\vec{k}_\perp, x_3)$ cannot simultaneously be supported in $| s_\nu a x_3 - 1 | \lesssim \epsilon$ and be non vanishing for some positive values of $ s_{\bar{\nu}} x_3$. The condition for $\mathcal{F}_3^{-1}[\tilde{\phi}_1](\vec{k}_\perp, x_3)$ to be vanishing in $ s_{\bar{\nu}} x_3 > 0$ is equivalent to $\alpha_{\bar{\nu}}[\tilde{\phi}_1](\Omega, \vec{K}_\perp) \approx 0$ for $\Omega > 0$. Consequently, Eq.~(\ref{Fourier_phi_tilde_support}) is a necessary and sufficient condition for relativistic $\nu$-Rindler creators, all $\bar{\nu}$-Rindler creators and all Rindler annihilators acting on $\hat{S}_\text{S} | 0_\text{L}, 0_\text{R} \rangle$ in Eq.~(\ref{single_particle_Minkowski_Rindler}) to have vanishing smearing function.

Equation (\ref{Fourier_phi_tilde_support}) can be proved to be equivalent to a localization condition for $| \phi \rangle$. To see this, consider the single particle wave function in the position space $\phi_1(\vec{x})$ defined by Eq.~(\ref{free_wave_function}) for $n=1$ and $t=0$, i.e.
\begin{equation} \label{free_wave_function_1}
\phi_1 (\vec{x}) = \sqrt{\frac{ mc^2}{(2\pi)^3 \hbar \omega(\vec{k})}} \int_{\mathbb{R}^{3}} d^3 k e^{i\vec{k} \cdot \vec{x}} \tilde{\phi}_1 (\vec{k}),
\end{equation}
and notice that in the nonrelativistic limit (\ref{single_particle_nonrelativistic}), Eq.~(\ref{free_wave_function_1}) can be approximated by
\begin{equation} \label{free_wave_function_1_nonrelativistic}
\phi_1 (\vec{x}) \approx \frac{1}{2\pi} \int_{\mathbb{R}^2} d^2 k_\perp e^{i\vec{k}_\perp \cdot \vec{x}_\perp} \mathcal{F}_3^{-1}[\tilde{\phi}_1](\vec{k}_\perp,z).
\end{equation}
Owing to Eq.~(\ref{free_wave_function_1_nonrelativistic}), the condition (\ref{Fourier_phi_tilde_support}) becomes equivalent to
\begin{equation}\label{Fourier_phi_tilde_support_2}
\phi_1 (\vec{x}) \approx 0, \text{ if } | s_\nu a z - 1 | \gg \epsilon.
\end{equation}

Equation (\ref{Fourier_phi_tilde_support_2}) can be compared with the localization condition expressed by Eq.~(\ref{quasi_inertial_limit_X}). In analogy to Eq.~(\ref{quasi_inertial_limit_X}), Eq.~(\ref{Fourier_phi_tilde_support_2}) describes the situation in which the wave function $\phi_1 (\vec{x})$ is localized in a region in which the metric is quasiflat, i.e., $g_{\mu\nu} \approx \eta_{\mu\nu}$. In Sec.~\ref{Wave_function_transformation}, we will show that when $\nu = \text{R}$, Eq.~(\ref{Fourier_phi_tilde_support_2}) is precisely equivalent to the condition (\ref{local_limit}). In particular, we will show that, in the quasi-inertial regime, the transformation between wave functions is determined by the coordinate transformation (\ref{Rindler_coordinates_transformation_R}). Hence, the localization condition (\ref{Fourier_phi_tilde_support_2}) for the wave function $\phi_1 (\vec{x})$ is equivalent to the localization condition expressed by Eq.~(\ref{quasi_inertial_limit_X}) for the wave function describing $| \phi \rangle$ in the right Rindler wedge.

Equation (\ref{quasi_inertial_limit_a}) is a necessary condition for Eqs.~(\ref{single_particle_nonrelativistic}) and (\ref{Fourier_phi_tilde_support}) due to the uncertainty principle for Fourier transforms, as we have discussed in Sec.~\ref{Inertial_and_non_inertial_frame}. Equation (\ref{local_limit_momentum}), instead, gives a constraint for the Rindler operators acting on $\hat{S}_\text{S} | 0_\text{L}, 0_\text{R} \rangle$ in Eq.~(\ref{single_particle_Minkowski_Rindler}) as a consequence of the nonrelativistic condition (\ref{single_particle_nonrelativistic}) and the $\delta^2(\vec{k}_\perp-\vec{K}_\perp)$ functions appearing in Eq.~(\ref{alpha_final}). Given the analogies between the present configuration and the one described in Sec.~\ref{Rindler_frame} by means of Eq.~(\ref{local_limit}), the setup satisfying Eq.~(\ref{Fourier_phi_tilde_support_2}) is referred to as being quasi-inertial.

\subsubsection{Scalar Minkowski-Fock state}

We can now generalize this result for any Minkowski-Fock state $| \phi \rangle$. In Sec.~\ref{QFT_in_Minkowski_spacetime_scalar}, we gave the definition for wave functions $\phi_n (\textbf{x}_n)$ and $\tilde{\phi}_n (\textbf{k}_n)$ by means of Eqs.~(\ref{free_state_decomposition}) and (\ref{free_wave_function}). For practical purposes, we ignored antiparticles and only focused on Fock states made of particles. However, to consider the Bogoliubov transformations (\ref{Rindler_Bogoliubov_transformations}) and the Rindler-Fock representation of the Minkowski vacuum [Eqs.~(\ref{Rindler_vacuum_to_Minkowski}) and (\ref{Rindler_vacuum_to_Minkowski_unitary_operator})], we have to include antiparticles as well.

We define the wave function of $|\phi \rangle$ in the position space as
\begin{align} \label{free_wave_function_antiparticles}
\phi_{nm} ( \boldsymbol{x}_{n+m}) = & \left( \frac{2 m c^2}{\hbar^2} \right)^{\frac{n+m}{2}} \int_{\mathbb{R}^{3(n+m)}} d^{3(n+m)} \boldsymbol{k}_{n+m} \tilde{\phi}_{nm} (\boldsymbol{k}_{n+m}) \nonumber \\
& \times \prod_{i=1}^n f(\vec{k}_i, 0, \vec{x}_i) \prod_{j=n+1}^{n+m} f(\vec{k}_j, 0, \vec{x}_j),
\end{align}
where $\tilde{\phi}_{nm} (\boldsymbol{k}_{n+m})$ is the wave function in the momentum space, defined from the decomposition of $|\phi \rangle$ with respect to the Minkowski-Fock space $\mathcal{H}_\text{M}$
\begin{equation}\label{free_state_decomposition_antiparticles}
| \phi \rangle  = \hat{c}_\phi | 0_\text{M} \rangle,
\end{equation}
with
\begin{equation}\label{c_phi}
\hat{c}_\phi  = \sum_{n,m=0}^\infty \int_{\mathbb{R}^{3(n+m)}} d^{3(n+m)} \boldsymbol{k}_{n+m} \frac{\tilde{\phi}_{nm} (\boldsymbol{k}_{n+m})}{\sqrt{n!m!}} \prod_{i=1}^n \hat{a}^\dagger(\vec{k}_i)  \prod_{j=n+1}^{n+m} \hat{b}^\dagger(\vec{k}_j) .
\end{equation}
By definition, $\tilde{\phi}_{nm} (\boldsymbol{k}_{n+m})$ is symmetric with respect to the momenta variables $\vec{k}_1, \dots, \vec{k}_n$ and with respect to $\vec{k}_{n+1}, \dots, \vec{k}_{n+m}$. The nonrelativistic condition for $|\phi \rangle$ reads as
\begin{equation}\label{many_particle_nonrelativistic}
\tilde{\phi}_{nm} (\boldsymbol{k}_{n+m}) \approx 0, \text{ if } \frac{\hbar |\vec{k}_i|}{mc} \gg \epsilon^{1/2} \text{ for some } i \in \{ 1, \dots , n+m \},
\end{equation}
which gives the following approximation
\begin{align} \label{free_wave_function_antiparticles_approximation}
\phi_{nm} ( \boldsymbol{x}_{n+m}) \approx & \left( \frac{2 m c^2}{\hbar^2} \right)^{\frac{n+m}{2}} \int_{\mathcal{R}_\text{NR,M}^{3(n+m)}} d^{3(n+m)} \boldsymbol{k}_{n+m} \tilde{\phi}_{nm} (\boldsymbol{k}_{n+m}) \nonumber \\
& \times \prod_{i=1}^n f(\vec{k}_i, 0, \vec{x}_i) \prod_{j=n+1}^{n+m} f(\vec{k}_j, 0, \vec{x}_j),
\end{align}
where $\mathcal{R}_\text{NR,M}$ is the nonrelativistic region defined by the condition (\ref{non_relativistic_limit}), i.e., $\vec{k} \in \mathcal{R}_\text{NR,M}$ if and only if $\hbar | \vec{k}| / mc \lesssim \epsilon^{1/2}$.

The explicit expression for $| \phi \rangle$ as an element of the Rindler-Fock space $\mathcal{H}_{\text{L}, \text{R}}$ can be obtained from Eqs.~(\ref{Rindler_vacuum_to_Minkowski}), (\ref{Rindler_Bogoliubov_transformations_compact}), (\ref{free_state_decomposition_antiparticles}), (\ref{c_phi}) and reads as
\begin{equation}\label{free_state_decomposition_Rindler}
| \phi \rangle  = \hat{C}_\phi \hat{S}_\text{S} | 0_\text{L}, 0_\text{R} \rangle,
\end{equation}
with
\begin{align}\label{C_phi}
\hat{C}_\phi = & \sum_{n,m=0}^\infty \sum_{\boldsymbol{\nu}_{n+m}} \int_{\mathbb{R}^{3(n+m)}} d^{3(n+m)}  \boldsymbol{\theta}_{n+m}  \frac{\tilde{\Phi}_{nm} (\boldsymbol{\theta}_{n+m}, \boldsymbol{\nu}_{n+m})}{\sqrt{n!m!}}  \nonumber \\
& \times\prod_{i=1}^n \hat{\mathcal{A}}^\dagger_{\nu_i}(\vec{\theta}_i) \prod_{j=n+1}^{n+m} \hat{\mathcal{B}}^\dagger_{\nu_j}(\vec{\theta}_j)
\end{align}
and
\begin{align}\label{non_reltaivistic_wavefunction_integration_0}
\tilde{\Phi}_{nm} (\boldsymbol{\theta}_{n+m}, \boldsymbol{\nu}_{n+m}) = & \int_{\mathbb{R}^{3(n+m)}} d^{3(n+m)} \boldsymbol{k}_{n+m} \tilde{\phi}_{nm} (\boldsymbol{k}_{n+m}) \nonumber \\
& \times \prod_{i=1}^n \alpha^*_{\nu_i}(\vec{k}_i,\vec{\theta}_i) \prod_{j=n+1}^{n+m} \alpha^*_{\nu_j}(\vec{k}_j,\vec{\theta}_j),
\end{align}
where the sum $\sum_{\boldsymbol{\nu}_{n+m}}$ in Eq.~(\ref{C_phi}) runs over all the possible $\nu$-variables $\nu \in \{ \text{L}, \text{R} \}$. Notice that the function $\tilde{\Phi}_{nm} (\boldsymbol{\theta}_{n+m}, \boldsymbol{\nu}_{n+m})$ defined in Eq.~(\ref{non_reltaivistic_wavefunction_integration_0}) cannot be regarded as the wave function of $| \phi \rangle$ in the Rindler frame due to the presence of the $\hat{S}_\text{S}$ operator acting on $ | 0_\text{L}, 0_\text{R} \rangle$ in Eq.~(\ref{free_state_decomposition_Rindler}). In other words, $\tilde{\Phi}_{nm} (\boldsymbol{\theta}_{n+m}, \boldsymbol{\nu}_{n+m})$ is not simply the generalization of Eq.~(\ref{wavefunction_F}) for particles and antiparticles. However, in Secs.~\ref{Quasiinertial_observer} and \ref{Wave_function_transformation}, we will show that $\tilde{\Phi}_{nm} (\boldsymbol{\theta}_{n+m}, \boldsymbol{\nu}_{n+m})$ obtains the notion of wave function in the quasi-inertial regime.

The generalization of the quasi-inertial condition (\ref{Fourier_phi_tilde_support_2}) for Minkowski-Fock states with any number of particles and antiparticles is
\begin{equation}\label{Fourier_phi_tilde_support_2_manyparticles}
\phi_{nm} (\boldsymbol{x}_{n+m}) \approx 0, \text{ if } | s_\nu a z_i - 1 | \gg \epsilon \text{ for some } i \in \{ 1, \dots , n+m \}.
\end{equation}
Equation (\ref{Fourier_phi_tilde_support_2_manyparticles}) is a necessary and sufficient condition for $| \phi \rangle$ to not have relativistic $\nu$-Rindler creators, no $\bar{\nu}$-Rindler creators at all and no Rindler annihilators acting on $\hat{S}_\text{S} | 0_\text{L}, 0_\text{R} \rangle$ in Eq.~(\ref{free_state_decomposition_Rindler}). Hence, we find that the operator $\hat{C}_\phi$ is only made of nonrelativistic $\nu$-Rindler creators if and only if the quasi-inertial condition (\ref{Fourier_phi_tilde_support_2_manyparticles}) holds.

Explicitly, we say that Eq.~(\ref{Fourier_phi_tilde_support_2_manyparticles}) is a necessary and sufficient condition for the approximation
\begin{equation}\label{C_nonrelativistic}
\tilde{\Phi}_{nm} (\boldsymbol{\theta}_{n+m}, \boldsymbol{\nu}_{n+m}) \approx 0, \text{ if } \vec{\theta}_i \not\in \mathcal{R}_\text{NR,C} \text{ or } \nu_i = \bar{\nu} \text{ for some } i \in \{ 1, \dots , n+m \},
\end{equation}
where $\mathcal{R}_\text{NR,C}$ is the nonrelativistic region defined by the condition (\ref{non_relativistic_limit_curved}), in the sense that $(\Omega, \vec{K}_\perp) \in \mathcal{R}_\text{NR,C}$ if and only if $\Omega$ satisfies Eq.~(\ref{non_relativistic_limit_curved}). Equation (\ref{C_nonrelativistic}) is equivalent to
\begin{align}\label{C_phi_approximation}
\hat{C}_\phi \approx & \sum_{n,m=0}^\infty \sum_{\boldsymbol{\nu}_{n+m}} \int_{\mathcal{R}_\text{NR,C}^{n+m}} d^{3(n+m)}  \boldsymbol{\theta}_{n+m}  \frac{\tilde{\Phi}_{nm} (\boldsymbol{\theta}_{n+m}, \boldsymbol{\nu}_{n+m})}{\sqrt{n!m!}}   \nonumber \\
& \times \prod_{i=1}^n \hat{A}^\dagger_{\nu_i}(\vec{\theta}_i)\prod_{j=n+1}^{n+m} \hat{B}^\dagger_{\nu_j}(\vec{\theta}_j).
\end{align}
This means that $\hat{C}_\phi$ is approximately only made of creators of nonrelativistic $\nu$-Rindler particles. Therefore, the transformation $\hat{c}_\phi \mapsto \hat{C}_\phi$ conserves the nonrelativistic nature of particles when one switches from the inertial to the accelerated frame.

By comparing Eq.~(\ref{C_phi_approximation}) with Eq.~(\ref{c_phi}) one can notice that $\hat{C}_\phi$ is identical to $\hat{c}_\phi$, up to the function $\tilde{\Phi}_{nm}$ replacing $\tilde{\phi}_{nm}$ and the $\nu$-Rindler creation operators $\hat{A}^\dagger_{\nu_i}(\vec{\theta}_i)$, $\hat{B}^\dagger_{\nu_j}(\vec{\theta}_j)$ replacing the Minkowski operators $\hat{a}^\dagger(\vec{k}_i)$, $\hat{b}^\dagger(\vec{k}_j)$. This implies that the number of particles and antiparticles created by $\hat{C}_\phi$ is the same as $\hat{c}_\phi$. The conclusion is that the transformation $\hat{c}_\phi \mapsto \hat{C}_\phi$ conserves the number of particles and antiparticles, in addition to the nonrelativistic condition.

Finally, notice that if $| \phi \rangle$ satisfies the quasi-inertial condition (\ref{Fourier_phi_tilde_support_2_manyparticles}) for $\nu = \text{L}$, then approximately no right Rindler particles are produced over $\hat{S}_\text{S} | 0_\text{L}, 0_\text{R} \rangle$ as a consequence of Eq.~(\ref{C_phi_approximation}). Conversely, if $\nu = \text{R}$, then all the Minkowski operators $\hat{a}^\dagger(\vec{k}_i)$, $\hat{b}^\dagger(\vec{k}_j)$ are converted into $\hat{A}^\dagger_\text{R}(\vec{\theta}_i)$ and $\hat{B}^\dagger_\text{R}(\vec{\theta}_j)$ operators. Without loss of generality, we assume that the accelerated observer Rob is describe by the right wedge. Hence, the quasi-inertial condition (\ref{Fourier_phi_tilde_support_2_manyparticles}) for $\nu = \text{L}$ leads to the trivial scenario in which Rob can never detect the Minkowski single particle over the Minkowski vacuum. For this reason, hereafter, we only consider Eq.~(\ref{Fourier_phi_tilde_support_2_manyparticles}) for $\nu = \text{R}$, i.e.,
\begin{equation}\label{Fourier_phi_tilde_support_2_manyparticles_R}
\phi_{nm} (\boldsymbol{x}_{n+m}) \approx 0, \text{ if } | a z_i - 1 | \gg \epsilon \text{ for some } i \in \{ 1, \dots , n+m \}.
\end{equation}
Equation (\ref{Fourier_phi_tilde_support_2_manyparticles_R}) is a necessary and sufficient condition for
\begin{equation}\label{C_nonrelativistic_R}
\tilde{\Phi}_{nm} (\boldsymbol{\theta}_{n+m}, \boldsymbol{\nu}_{n+m}) \approx 0, \text{ if } \theta_i \not\in \mathcal{R}_\text{NR,C} \text{ or } \nu_i = \text{L} \text{ for some } i \in \{ 1, \dots , n+m \}
\end{equation}
and, hence,
\begin{equation}\label{C_phi_approximation_R}
\hat{C}_\phi \approx \sum_{n,m=0}^\infty \int_{\boldsymbol{\theta}_{n+m} \in \mathcal{R}_\text{NR,C}^{n+m}} d^{3(n+m)}  \boldsymbol{\theta}_{n+m}  \frac{\tilde{\Phi}_{nm} (\boldsymbol{\theta}_{n+m})}{\sqrt{n!m!}}  \prod_{i=1}^n \hat{A}^\dagger_\text{R}(\vec{\theta}_i)   \prod_{j=n+1}^{n+m} \hat{B}^\dagger_\text{R}(\vec{\theta}_j),
\end{equation}
where $\tilde{\Phi}_{nm} (\boldsymbol{\theta}_{n+m})$ is defined as
\begin{equation}\label{Phi_tilde_R}
\tilde{\Phi}_{nm} (\boldsymbol{\theta}_{n+m})= \tilde{\Phi}_{nm} (\boldsymbol{\theta}_{n+m}, \textbf{R}_{n+m}),
\end{equation}
with
\begin{equation}
\textbf{R}_{n+m} = (\underbrace{\text{R}, \dots, \text{R}}_n, \underbrace{\text{R}, \dots, \text{R}}_m).
\end{equation}

\subsubsection{Numerical check}

These results can be numerically checked as follows. 

Firstly, notice that when $\hbar | \vec{k} | / mc \lesssim \epsilon$, Eq.~(\ref{alpha}) can be approximated as
\begin{equation} \label{alpha_alpha_tilde}
\alpha_\nu(\vec{k},\vec{\theta}) \approx  \int_{\mathbb{R}^3} d^3 x f^*( \vec{k}, 0,  \vec{x} )  \tilde{\alpha}_\nu(\vec{x},\vec{\theta}),
\end{equation}
with
\begin{equation} \label{alpha_tilde}
\tilde{\alpha}_\nu(\vec{x},\vec{\theta}) = \frac{\theta(s_\nu z)}{\hbar}  \left(  \frac{s_\nu \theta_1}{a z} + \frac{m c^2}{\hbar} \right)  \tilde{F}(\vec{\theta},  s_\nu Z_\nu(z)) e^{ i \vec{\theta}_\perp \cdot \vec{x}_\perp}.
\end{equation}

By using the relation between $\phi_{nm}( \boldsymbol{x}_{n+m}) $ and $\tilde{\phi}_{nm} (\boldsymbol{k}_{n+m})$ [Eq.~(\ref{free_wave_function_antiparticles})] and between $\alpha_\nu(\vec{k},\vec{\theta})$ and $\tilde{\alpha}_\nu(\vec{x},\vec{\theta})$ [Eq.~(\ref{alpha_alpha_tilde})], one can approximate Eq.~(\ref{non_reltaivistic_wavefunction_integration_0}) with
\begin{align}\label{non_reltaivistic_wavefunction_integration_0_tilde}
\tilde{\Phi}_{nm} (\boldsymbol{\theta}_{n+m}, \boldsymbol{\nu}_{n+m}) \approx & \left( \frac{2 m c^2}{\hbar^2} \right)^{-\frac{n+m}{2}} \int_{\mathbb{R}^{3(n+m)}} d^{3(n+m)} \boldsymbol{x}_{n+m} \phi_{nm} (\boldsymbol{x}_{n+m})  \nonumber \\
& \times \prod_{i=1}^n \tilde{\alpha}^*_{\nu_i}(\vec{x}_i,\vec{\theta}_i) \prod_{j=n+1}^{n+m} \tilde{\alpha}^*_{\nu_j}(\vec{x}_j,\vec{\theta}_j).
\end{align}
The quasi inertial condition (\ref{Fourier_phi_tilde_support_2_manyparticles_R}) can be used in Eq.~(\ref{non_reltaivistic_wavefunction_integration_0_tilde}) to let the integration with respect to the position variables $\vec{x}_i$ run inside of the region $| a z_i - 1 | \lesssim \epsilon$. The Heaviside theta function appearing in Eq.~(\ref{alpha_tilde}) implies that a necessary condition for the localization condition is that $\tilde{\Phi}_{nm} (\boldsymbol{\theta}_{n+m}, \boldsymbol{\nu}_{n+m})$ is vanishing when at least one $\nu$ variables is equal to $\text{L}$. Therefore, hereafter we only consider the right-wedge wave function $\tilde{\Phi}_{nm} (\boldsymbol{\theta}_{n+m})$ defined by Eq.~(\ref{Phi_tilde_R}).

Due to the quasi inertial condition (\ref{Fourier_phi_tilde_support_2_manyparticles_R}), one may also introduce a cutoff $\delta z$ for any integration variable $z$ in Eq.~(\ref{non_reltaivistic_wavefunction_integration_0_tilde}) and assume that any integration can be approximately performed in $z \in [a^{-1}-\delta z, a^{-1}+\delta z]$, with
\begin{equation}\label{delta_z_constraint}
\delta z \lesssim \epsilon a^{-1},
\end{equation}
instead of the full real axis. By considering such an approximation in Eq.~(\ref{non_reltaivistic_wavefunction_integration_0_tilde}) and by using Eq.~(\ref{free_wave_function_antiparticles}), one obtains
\begin{align}\label{non_reltaivistic_wavefunction_integration_1}
\tilde{\Phi}_{nm} (\boldsymbol{\theta}_{n+m} ) \approx & \int_{\mathbb{R}^{3(n+m)}} d^{3(n+m)} \boldsymbol{k}_{n+m} \tilde{\phi}_{nm} (\boldsymbol{k}_{n+m}) \nonumber \\
& \times \prod_{i=1}^n \alpha^*(\vec{k}_i,\vec{\theta}_i,\delta z)  \prod_{j=n+1}^{n+m} \alpha^*(\vec{k}_j,\vec{\theta}_j,\delta z),
\end{align}
with
\begin{align} \label{alpha_bar}
\alpha(\vec{k},\vec{\theta},\delta z) = & \frac{1}{\hbar} \left(  \theta_1 + \frac{m c^2}{\hbar} \right) \int_{a^{-1}-\delta z}^{a^{-1}+\delta z} d z \int_{\mathbb{R}^2} d^2 x_\perp   \nonumber \\
& \times f^*( \vec{k}, 0,\vec{x} )   \tilde{F}(\vec{\theta},  Z_\text{R}(z)) e^{ i \vec{\theta}_\perp \cdot \vec{x}_\perp}.
\end{align}

By using Eq.~(\ref{free_modes}) and integrating with respect to $\vec{x}_\perp$, Eq.~(\ref{alpha_bar}) becomes
\begin{equation} \label{alpha_bar_2}
\alpha(\vec{k},\vec{\theta},\delta z) =  \delta^2(\vec{k}_\perp-\vec{\theta}_\perp) \chi (\vec{k},\theta_1, \delta z),
\end{equation}
with
\begin{equation} \label{chi_bar_0}
\chi(\vec{k},\Omega,\delta z) = \sqrt{ \frac{\pi}{\hbar \omega(\vec{k})} } \left(  \Omega + \frac{m c^2}{\hbar} \right)   \int_{a^{-1}-\delta z}^{a^{-1}+\delta z} d z   e^{-i k_3 z}  \tilde{F}(\Omega,\vec{k}_\perp, Z_\text{R}(z)),
\end{equation}
which is the equivalent of Eq.~(\ref{chi}) with a cutoff $\delta z$, $\nu=\text{R}$ and $\omega(\vec{k}) \approx mc^2/\hbar$.

We are interested in the behavior of $\chi(\vec{k},\Omega,\delta z)$ with varying $\Omega$. In particular, we want to show that, when constraints (\ref{non_relativistic_limit}), (\ref{quasi_inertial_limit_a}) and (\ref{delta_z_constraint}) hold, $\chi(\vec{k},\Omega,\delta z)$ is not vanishing only for $\Omega$ such that Eq.~(\ref{non_relativistic_limit_curved}) holds. To this end, we perform the coordinate transformation
\begin{equation}\label{coordinate_transformation}
\bar{z} =  \frac{a z -1}{\bar{a}},
\end{equation}
with
\begin{equation}\label{acceleration_adimensional}
\bar{a} = 2^{-1/3} \left(\frac{\hbar a}{m c}\right)^{2/3}
\end{equation}
as an acceleration dependent adimensional variable. We furthermore consider the following adimensional variables
\begin{align}\label{adimensional_variables}
& \vec{\bar{k}} = \frac{\bar{a} \vec{k}}{a} , && \bar{\Omega} = \frac{\hbar \Omega}{m c^2} , && \delta \bar{z} = \frac{a \delta z}{\bar{a}}.
\end{align}
In this way, Eq.~(\ref{chi_bar_0}) reads as
\begin{equation} \label{alpha_bar_chi_bar}
\chi(\vec{k},\Omega,\delta z) =  \frac{\bar{a}}{a}  \sqrt{\frac{ m \delta z}{\hbar}} \exp \left(- i \frac{k_3}{a} \right) \bar{\chi} \left(  \frac{\bar{a} \vec{k}}{a} , \frac{\hbar \Omega}{m c^2} , \frac{a \delta z}{\bar{a}} \right),
\end{equation}
with
\begin{equation} \label{chi_bar}
\bar{\chi}(\vec{\bar{k}}, \bar{\Omega}, \delta \bar{z}) =  \frac{\sqrt{\pi} (\bar{\Omega} + 1)}{\sqrt[4]{1 + 2 \bar{a} \bar{k}^2} \sqrt{\delta \bar{z}}}  \int_{- \delta \bar{z}}^{\delta \bar{z}} d \bar{z}  e^{ -i \bar{k}_3 \bar{z} } \bar{\tilde{F}} (\bar{\Omega}, \vec{\bar{k}}_\perp, \bar{z})
\end{equation}
and
\begin{equation} \label{F_bar}
\bar{\tilde{F}} (\bar{\Omega}, \vec{\bar{k}}_\perp, \bar{z}) = \sqrt{\frac{a}{\bar{a} \hbar}}  c \tilde{F}\left( \frac{m c^2 \bar{\Omega}}{\hbar}, \frac{a \vec{\bar{k}}_\perp }{\bar{a}} ,  Z_\text{R} \left( \frac{\bar{a} \bar{z} + 1}{a}\right) \right)
\end{equation}
as adimensional functions. The variable $ \vec{\bar{k}}_\perp$ appearing in Eq.~(\ref{chi_bar}) is made of the transverse components of $ \vec{\bar{k}}$, i.e.: $ \vec{\bar{k}}_\perp = (\bar{k}_1, \bar{k}_2)$.

\begin{figure}
\center
\includegraphics[]{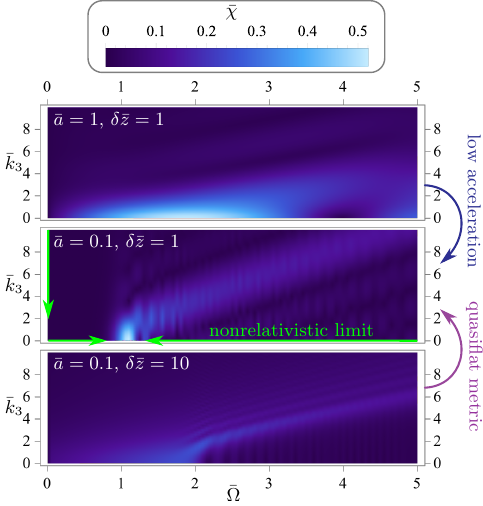}
\caption{Distribution of Rindler energies $\bar{\Omega}$ (horizontal axis) with respect to Minkowski momenta $\bar{k}_3$ (vertical axes). The quantity measured here is $\bar{\chi}(\vec{\bar{k}}, \bar{\Omega}, \delta \bar{z})$, which describes how energy-momentum distributions transform from inertial to accelerated frames [Eqs.~(\ref{non_reltaivistic_wavefunction_integration_1}), (\ref{alpha_bar_2}), (\ref{alpha_bar_chi_bar})]. For simplicity, we ignore the transverse coordinates $x$ and $y$ by choosing $\bar{k}_1=0$ and $\bar{k}_2=0$. The regime of low acceleration ($\bar{a} \ll 1$) and quasiflat metric ($\delta \bar{z} \sim 1$) [Eq.~(\ref{constraints_adimentional})] are indicated with, respectively, blue and purple arrows. In such regime, nonrelativistic Minkowski momenta are paired with nonrelativistic Rindler energies (green arrows). Indeed, when $\bar{k}_3 \lesssim 1$ [Eq.~(\ref{constraints_adimentional})], $\bar{\chi}(\vec{\bar{k}}, \bar{\Omega}, \delta \bar{z})$ is peaked for $\bar{\Omega} \approx 1$ [Eq.~(\ref{non_relativistic_Theta_1_adimensional})]. This means that in the quasi-inertial regime, the accelerated observer agrees with the inertial observer about the nonrelativistic nature of particles created over the vacuum state $| 0_\text{M} \rangle$.}\label{chi_bar_figure}
\end{figure}

Explicitly, Eq.~(\ref{chi_bar}) reads as
\begin{align} \label{chi_bar_explicit}
 \bar{\chi}(\vec{\bar{k}}, \bar{\Omega}, \delta \bar{z}) =  &  \frac{\bar{\Omega} + 1}{2 \pi^{3/2}\sqrt[4]{1 + 2 \bar{a} \bar{k}^2} \sqrt{\bar{a} \delta \bar{z}}} \int_{- \delta \bar{z}}^{\delta \bar{z}} d \bar{z} e^{ -i \bar{k}_3 \bar{z} }   \nonumber \\
 & \times \sqrt{ \left| \sinh \left( \frac{ \pi \bar{\Omega}}{ \sqrt{2 \bar{a}^3} } \right)\right| } K_{i \bar{\Omega} / \sqrt{2 \bar{a}^3} } \left( \sqrt{ \frac{1+ 2 \bar{a} |\vec{\bar{k}}_\perp|^2 }{2 \bar{a}^3}} ( 1 + \bar{a} \bar{z} ) \right).
\end{align}
We numerically compute Eq.~(\ref{chi_bar_explicit}) and plot the results in Fig.~\ref{chi_bar_figure} for different values of $\bar{k}_3$ and $\bar{\Omega}$. We choose $\bar{a} \in \{0.1,1\}$ and $\delta \bar{z} \in \{ 1, 10 \}$ to show the quasi-inertial limit (i.e., $\bar{a} \ll 1$ and $\delta \bar{z} \lesssim 1$).

Conditions (\ref{non_relativistic_limit}), (\ref{quasi_inertial_limit_a}), (\ref{delta_z_constraint}) in the new set of coordinates become
\begin{align}\label{constraints_adimentional}
& \bar{a} \sim \epsilon, && |\vec{\bar{k}}| \lesssim 1, && \delta \bar{z} \lesssim 1,
\end{align}
whereas Eq.~(\ref{non_relativistic_limit_curved}) becomes
\begin{equation}\label{non_relativistic_Theta_1_adimensional}
\frac{|\bar{\Omega} - 1|}{\bar{a}} \lesssim 1.
\end{equation}
In Fig.~\ref{chi_bar_figure}, we show that when the parameter $\bar{a}$ and the coordinates $|\vec{\bar{k}}|$ and $\delta \bar{z}$ are constrained by Eq.~(\ref{constraints_adimentional}), $\bar{\chi}(\vec{\bar{k}}, \bar{\Omega}, \delta \bar{z})$ is not vanishing only for $\bar{\Omega}$ such that Eq.~(\ref{non_relativistic_Theta_1_adimensional}) holds. In other words, the regime of low acceleration ($\bar{a} \ll 1$), quasiflat metric ($\delta \bar{z} \lesssim 1$) and nonrelativistic Minkowski momenta ($|\bar{k}_3| \lesssim 1$) is characterized by a distribution $\bar{\chi}(\vec{\bar{k}}, \bar{\Omega}, \delta \bar{z})$ peaked at nonrelativistic Rindler energies ($\bar{\Omega} \sim 1$). As a result, Eq.~(\ref{non_relativistic_limit_curved}) selects only nonrelativistic frequencies for which $\tilde{\Phi}_{nm} (\boldsymbol{\theta}_{n+m})$ is not vanishing.

\subsubsection{Dirac Minkowski-Fock state}

The discussion we provided for the scalar fields can also be extended to the case of Dirac fields as well. The only differences are given by the Rindler-Dirac operators $\hat{C}_{\nu s}(\Omega,\vec{K}_\perp)$ and $\hat{D}_{\nu s}(\Omega,\vec{K}_\perp)$ replacing the scalar operators $\hat{A}_\nu(\Omega,\vec{K}_\perp)$ and $\hat{B}_\nu(\Omega,\vec{K}_\perp)$, the Bogoliubov transformation (\ref{Bogoliubov_transformations_3_Rindler_4}) replacing Eq.~(\ref{Rindler_Bogoliubov_transformations_compact}) and the $\hat{S}_\text{D}$ operator replacing $\hat{S}_\text{S}$.

For instance, in the case of single particle state $| \psi \rangle$, we can use the definition of wave function $\tilde{\psi}_1(s, \vec{k})$ given by Eq.~(\ref{free_Dirac_state_decomposition}), i.e.,
\begin{equation}\label{free_Dirac_state_decomposition_single}
| \psi \rangle = \sum_{s=1}^2 \int_{\mathbb{R}^3} d^3 k \tilde{\psi}_1 (s,\vec{k}) \hat{c}_s^\dagger(\vec{k}) | 0_\text{L}, 0_\text{R} \rangle.
\end{equation}
$\tilde{\psi}_1(s, \vec{k})$ can be used to define the nonrelativistic condition for the particle in the inertial frame
\begin{equation}\label{single_particle_nonrelativistic_Dirac}
\tilde{\psi}_1 (s,\vec{k}) \approx 0, \text{ if } \frac{\hbar |\vec{k}|}{mc} \gg \epsilon^{1/2},
\end{equation}
which is the equivalent of Eq.~(\ref{single_particle_nonrelativistic}) for Dirac fields. The wave function in the position space, instead, is given by Eq.~(\ref{free_Dirac_wave_function_single}); for $t=0$, we define $\psi_1 (\vec{x}) = \psi_1 (0, \vec{x})$.

The representation of $| \psi \rangle$ in the Rindler frame can be derived by means of the Bogoliubov transformation (\ref{Bogoliubov_transformations_3_Rindler_4}) and the Rindler representation of the Minkowski vacuum (\ref{Rindler_vacuum_to_Minkowski_Dirac}) in analogy to Eq.~(\ref{single_particle_Minkowski_Rindler}). We have that
\begin{equation}\label{single_particle_Minkowski_Rindler_Dirac}
| \psi \rangle =  \hat{C}_\psi \hat{S}_\text{D} | 0_\text{L}, 0_\text{R} \rangle.
\end{equation}
with
\begin{align}\label{C_psi_Dirac}
 & \hat{C}_\psi = \sum_{\nu=\{\text{L},\text{R}\}} \sum_{s=1}^2 \sum_{s'=1}^2  \int_{\mathbb{R}^3} d^3 k  \int_\mathbb{R} d\Omega \int_{\mathbb{R}^2} d^2 K_\perp  \tilde{\psi}_1(s,\vec{k})  \alpha_{\nu}^*(\vec{k},\Omega,\vec{K}_\perp)   \nonumber \\
& \times \tilde{\mathfrak{W}}_{\nu s'}^\dagger(\Omega, \vec{K}_\perp) \tilde{u}_s(\vec{k})\left[ \theta(\Omega) \hat{C}_{\nu s'}^\dagger(\Omega,\vec{K}_\perp)+  \theta(-\Omega)\hat{D}_{\nu s'}(-\Omega,-\vec{K}_\perp) \right]  .
\end{align}

By following the same arguments used for scalar fields, one can prove that Eq.~(\ref{C_psi_Dirac}) can be approximated by
\begin{equation}\label{C_psi_Dirac_approx}
 \hat{C}_\psi \approx \sum_{s=1}^2 \sum_{s'=1}^2  \int_{\mathbb{R}^3} d^3 k  \int_{\mathcal{R}_\text{NR,C}} d^3 \theta  \tilde{\psi}_1(s,\vec{k})  \alpha_{\text{R}}^*(\vec{k},\vec{\theta})  \tilde{\mathfrak{W}}_{\text{R} s'}^\dagger(\vec{\theta}) \tilde{u}_s(\vec{k}) \hat{C}_{\text{R} s'}^\dagger(\vec{\theta})
\end{equation}
if and only if the wave function $\psi_1 (\vec{x})$ satisfies
\begin{equation}\label{Fourier_phi_tilde_support_2_Dirac}
\psi_1 (\vec{x}) \approx 0, \text{ if } | a z - 1 | \gg \epsilon,
\end{equation}
which is the equivalent of Eq.~(\ref{Fourier_phi_tilde_support_2}) for Dirac fields and with $\nu = \text{R}$. The equivalence between the scalar and the Dirac case is due to the similarities between the Bogoliubov coefficients (\ref{alpha_final}) and (\ref{Bogoliubov_coefficient_2}).

Equation (\ref{Fourier_phi_tilde_support_2_Dirac}) is the quasi-inertial condition for the Dirac field $| \psi \rangle$. It gives a necessary and sufficient condition for the $\hat{C}_\psi$ operator to be made exclusivity of nonrelativistic right Rindler creators; additionally, the number of particles created over the Minkowski vacuum $| 0_\text{M} \rangle  = \hat{S}_\text{D} | 0_\text{L}, 0_\text{R} \rangle$ is preserved.

\subsection{Quasi-inertial observer}\label{Quasiinertial_observer}

In Sec.~\ref{Quasi_inertial_regime} we found that Eq.~(\ref{Fourier_phi_tilde_support_2_manyparticles_R}) is necessary and sufficient condition for the operator $\hat{C}_\phi$ to be made of nonrelativistic right Rindler creators and to have the same particle content of $\hat{c}_\phi$ [Eqs.~(\ref{c_phi}) and (\ref{C_phi_approximation_R})]. Notice, however, that Eq.~(\ref{C_phi_approximation_R}) does not imply that $| \phi \rangle$ is nonrelativistic in the Rindler frame and the number of particles are still not conserved if one considers the Minkowski and the Rindler representations of $| \phi \rangle$. Indeed, the operator $\hat{S}_\text{S}$ still generates relativistic particles by acting on the Rindler vacuum $| 0_\text{L}, 0_\text{R} \rangle$. Equation (\ref{Fourier_phi_tilde_support_2_manyparticles_R}) is a condition that guarantees the absence of relativistic operators acting on $\hat{S}_\text{S} | 0_\text{L}, 0_\text{R} \rangle$ in Eq.~(\ref{free_state_decomposition_Rindler}); however $\hat{S}_\text{S} | 0_\text{L}, 0_\text{R} \rangle$ is a relativistic state populated by a non vanishing number of Rindler particles. Moreover, Eq.~(\ref{Fourier_phi_tilde_support_2_manyparticles_R}) is a condition for the state $| \phi \rangle$, whereas the relativistic nature of $\hat{S}_\text{S} | 0_\text{L}, 0_\text{R} \rangle $ and its particle content are independent of the Minkowski particles created over it.

To solve this problem, we assume that the accelerated observer Rob has an acceleration that satisfies Eq.~(\ref{quasi_inertial_limit_a}) and has only access to the region defined by Eq.~(\ref{quasi_inertial_limit_X}), with $Z$ as the right Rindler coordinate describing Rob's frame. We already remarked that for any $Z$ such that Eq.~(\ref{quasi_inertial_limit_X}) holds, the metric $g_{\mu\nu}$ is approximated by $\eta_{\mu\nu}$. For this reason, we say that Rob is a quasi-inertial observer.

The localization condition (\ref{quasi_inertial_limit_X}) defines the set of particles states that can be detected by the quasi-inertial observer. For instance, left-Rindler particles are excluded by such selection, since they are localized beyond the Rindler horizon. The same occurs for right-Rindler particles with frequency $\Omega \lesssim c a$, because they are localized too close to the horizon. This is a consequence of the fact that the $F(\Omega,\vec{K}_\perp,T,\vec{X})$ modes are exponentially vanishing when $\Omega \lesssim c a$, $a Z \gtrsim -1$ and $\hbar a/mc \ll 1$. Indeed, Bessel functions have the following asymptotic behavior
\begin{equation}
K_{i \zeta} (\xi) \sim \frac{e^{-\xi}}{\sqrt{\xi}}
\end{equation}
when $\xi \rightarrow \infty$, and, hence, $F(\Omega,\vec{K}_\perp,T,\vec{X})$ goes to zero as
\begin{equation} \label{F_Rindler_approximation_small_Omega}
F(\Omega,\vec{K}_\perp,T,\vec{X}) \sim \frac{1}{c} \sqrt{\frac{\hbar}{a} \left| \sinh \left( \frac{\pi \Omega}{ca} \right) \right|} \epsilon^{3/4} \exp \left( - \epsilon^{-3/2} \right).
\end{equation}

We define the Fock space $\mathcal{H}_{\text{\sout{QI}}}$ that is generated by left-wedge particles with any frequency $\Omega$ and by right-wedge particles with frequency $\Omega \lesssim c a$. $\mathcal{H}_{\text{\sout{QI}}}$ represents the set of states that cannot be detected by the quasi-inertial observer. We define the partial trace $\text{Tr}_{\text{\sout{QI}}}$ over $\mathcal{H}_{\text{\sout{QI}}}$, which maps any pure state $| \Phi \rangle \in \mathcal{H}_{\text{L}, \text{R}}$ into a statistical operator $\rho = \text{Tr}_{\text{\sout{QI}}} ( | \Phi \rangle \langle \Phi | )$ describing $| \Phi \rangle$ from the point of view of the quasi-inertial observer. We also define the Fock space $\mathcal{H}_\text{QI} = \text{Tr}_{\text{\sout{QI}}} (\mathcal{H}_{\text{L}, \text{R}})$ as the representation space for the quasi-inertial observer. In practice, the quasi-inertial observer is not able to distinguish between any element of $\mathcal{H}_{\text{\sout{QI}}}$ and the vacuum state of $\mathcal{H}_\text{QI}$.

We now consider a state $| \phi \rangle$ that satisfies the quasi-inertial condition (\ref{Fourier_phi_tilde_support_2_manyparticles_R}) and we derive its representation in $\mathcal{H}_\text{QI}$. We start by considering the representation of $| \phi \rangle$ in $\mathcal{H}_{\text{L}, \text{R}}$, which is given by Eq.~(\ref{free_state_decomposition_Rindler}). Notice that the $\hat{S}_\text{S}$ operator defined by Eq.~(\ref{Rindler_vacuum_to_Minkowski_unitary_operator}) can be approximated by
\begin{equation}\label{Rindler_vacuum_to_Minkowski_approximation}
\hat{S}_\text{S} \approx \exp\left(  2 \sum_{\nu=\{\text{L},\text{R}\}} \int_{ \Omega \lesssim ca } d\Omega \int_{\mathbb{R}^2} d^2\vec{K}_\perp \zeta(\Omega)  \left[ \hat{A}^\dagger_\nu(\Omega,\vec{K}_\perp) \hat{B}^\dagger_{\bar{\nu}}(\Omega,-\vec{K}_\perp) \right]^\text{A} \right),
\end{equation}
since $\zeta(\Omega)$ is exponentially small when $ \Omega \gg ca$. Also, notice that when $| \phi \rangle$ satisfies the quasi-inertial condition (\ref{Fourier_phi_tilde_support_2_manyparticles_R}), the operator $\hat{C}_\phi$ is approximated by Eq.~(\ref{C_phi_approximation_R}).

The integration interval in Eq.~(\ref{Rindler_vacuum_to_Minkowski_approximation}) is $\Omega \lesssim ca \ll  m c^2 / \hbar $ [Eq.~(\ref{quasi_inertial_limit_a})], whereas the frequency variables $\theta_1$ in Eq.~(\ref{C_phi_approximation_R}) are constrained by $\theta_1 \approx mc^2/\hbar$. This means that $\hat{C}_\phi$ is left unaffected by the partial trace $\text{Tr}_{\text{\sout{QI}}}$, in the sense that
\begin{equation}\label{partial_trace_C}
\text{Tr}_{\text{\sout{QI}}} ( \hat{C}_\phi \hat{O} ) \approx \hat{C}_\phi \text{Tr}_{\text{\sout{QI}}} ( \hat{O} ),
\end{equation}
whereas $\hat{S}_\text{S}$ satisfies the trace cyclic property
\begin{equation}\label{partial_trace_S}
\text{Tr}_{\text{\sout{QI}}} ( \hat{S}_\text{S} \hat{O} ) \approx \text{Tr}_{\text{\sout{QI}}} ( \hat{O} \hat{S}_\text{S} ).
\end{equation}
Indeed, the particles and antiparticles created by $\hat{C}_\phi$ are not elements of $\mathcal{H}_{\text{\sout{QI}}}$, since $\theta_1 \approx mc^2/\hbar \gg ca$. On the other hand, particles and antiparticles created and annihilated by $\hat{S}_\text{S}$ have frequency $\Omega \lesssim ca$ and, hence, belong to $\mathcal{H}_{\text{\sout{QI}}}$.

Equations (\ref{partial_trace_C}) and (\ref{partial_trace_S}) can be used together with (\ref{free_state_decomposition_Rindler}) to prove that
\begin{equation} \label{partial_trace_phi}
 \text{Tr}_{\text{\sout{QI}}} ( | \phi \rangle \langle \phi | ) \approx \hat{C}_\phi  | 0_\text{QI} \rangle \langle  0_\text{QI} | \hat{C}^\dagger_\phi,
\end{equation}
where
\begin{equation}
| 0_\text{QI} \rangle \langle  0_\text{QI} | = \text{Tr}_{\text{\sout{QI}}} ( | 0_\text{L}, 0_\text{R} \rangle \langle 0_\text{L}, 0_\text{R} | )
\end{equation}
is the vacuum state of $\mathcal{H}_\text{QI}$. Equation (\ref{partial_trace_phi}) states that $| \phi \rangle$ is seen by the quasi-inertial observer as a pure state $| \Phi \rangle$ such that
\begin{equation} \label{Phi_C_phi}
| \Phi \rangle = \hat{C}_\phi  | 0_\text{QI} \rangle.
\end{equation}

In this way, we have proved that $| \phi \rangle$ is seen by the quasi-inertial observer as a nonrelativistic state preserving the number of particles and antiparticles from the Minkowski representation. Indeed, by comparing Eq.~(\ref{Phi_C_phi}) with Eq.~(\ref{free_state_decomposition_antiparticles}) and Eq.~(\ref{C_phi_approximation_R}) with Eq.~(\ref{c_phi}), one notices that the same configuration of nonrelativistic particles and antiparticles are created over the respective vacuum. As said before, the map $\hat{c}_\phi \mapsto \hat{C}_\phi$ preserves the nonrelativistic condition and the number of particles and antiparticles. The conclusion is that the inertial and the quasi-inertial observer agree about the first-quantization description of states.

An analogous result can be obtained for the case of Dirac fields due to the similarities between the operators $\hat{S}_\text{S}$ [Eq.~(\ref{Rindler_vacuum_to_Minkowski_unitary_operator})] and $\hat{S}_\text{D}$ [Eq.~(\ref{OO})] and between the scalar [Eq.~(\ref{F_Rindler_all})] and the Dirac modes [Eq.~(\ref{UV_UV_tilde_conclusion})] in Rindler spacetime.

\subsection{Wave function transformation}\label{Wave_function_transformation}

By means of Eqs.~(\ref{C_phi_approximation_R}) and (\ref{Phi_C_phi}) we proved that the frame dependency of the nonrelativistic limit and the particle production is suppressed in the quasi-inertial regime. Additionally, we can notice that $\tilde{\Phi}_{nm} (\boldsymbol{\theta}_{n+m})$ plays the role of wave function for $| \Phi \rangle$ with respect to the quantum numbers $\boldsymbol{\theta}_{n+m}$, analogously to $\tilde{\phi}_{nm}(\boldsymbol{k}_{n+m})$ in the inertial frame. The transformation between wave functions $\tilde{\phi}_{nm}(\boldsymbol{k}_{n+m}) \mapsto \tilde{\Phi}_{nm}(\boldsymbol{\theta}_{n+m})$ is given by Eq.~(\ref{non_reltaivistic_wavefunction_integration_0}).

The wave function of $| \Phi \rangle$ in the position representation is defined as
\begin{align} \label{free_wave_function_curved}
\Phi_{nm} ( \boldsymbol{X}_{n+m}) = & \left( \frac{2 m c^2}{\hbar^2} \right)^{\frac{n+m}{2}} \int_{\mathbb{R}^{3(n+m)}} d^{3(n+m)} \boldsymbol{\theta}_{n+m} \tilde{\Phi}_{nm} (\boldsymbol{\theta}_{n+m})  \nonumber \\
& \times \prod_{i=1}^n F(\vec{\theta}_i, 0,  \vec{X}_i)  \prod_{j=n+1}^{n+m} F(\vec{\theta}_j, 0, \vec{X}_j),
\end{align}
which is the analogue of Eq.~(\ref{wavefunction_F_right}), but for the quasi inertial frame and with antiparticles included as well. As a consequence of Eq.~(\ref{C_nonrelativistic_R}), Eq.~(\ref{free_wave_function_curved}) can be approximated by
\begin{align} \label{free_wave_function_curved_approximate}
\Phi_{nm} ( \boldsymbol{X}_{n+m}) = & \left( \frac{2 m c^2}{\hbar^2} \right)^{\frac{n+m}{2}} \int_{\mathcal{R}_\text{NR,C}^{3(n+m)}} d^{3(n+m)} \boldsymbol{\theta}_{n+m} \tilde{\Phi}_{nm} (\boldsymbol{\theta}_{n+m})  \nonumber \\
& \times \prod_{i=1}^n F(\vec{\theta}_i, 0,  \vec{X}_i)  \prod_{j=n+1}^{n+m} F(\vec{\theta}_j, 0, \vec{X}_j),
\end{align}

The wave function transformation $\phi_{nm} ( \boldsymbol{x}_{n+m}) \mapsto \Phi_{nm} ( \boldsymbol{X}_{n+m})$ can be derived as follows. Use Eqs.~(\ref{alpha}), (\ref{free_wave_function_antiparticles_approximation}), (\ref{non_reltaivistic_wavefunction_integration_0}) and the definition of $\mathcal{R}_\text{NR,M}$ and $\mathcal{R}_\text{NR,C}$ in Eq.~(\ref{free_wave_function_curved_approximate}) to obtain
\begin{align} \label{free_wave_function_curved_approx}
\Phi_{nm} ( \boldsymbol{X}_{n+m}) \approx & \int_{\mathbb{R}^{3(n+m)}} d^{3(n+m)} \boldsymbol{x}_{n+m} \phi_{nm} (\boldsymbol{x}_{n+m})   \nonumber \\
& \times \prod_{i=1}^n \tilde{\tilde{\alpha}}^*_\text{R}(\vec{x}_i,\vec{X}_i) \prod_{j=n+1}^{n+m} \tilde{\tilde{\alpha}}^*_\text{R}(\vec{x}_j,\vec{X}_j)
\end{align}
with
\begin{equation}\label{alpha_tilde_tilde}
\tilde{\tilde{\alpha}}_\text{R}(\vec{x},\vec{X}) = \frac{\theta(z)}{\hbar } \left(  \frac{1}{a z} + 1 \right) \int_{\theta_1>0} d^3 \theta \theta_1  \tilde{F}(\vec{\theta}, Z_\text{R}(z)) e^{ i \vec{\theta}_\perp \cdot \vec{x}_\perp}   F^*(\vec{\theta}, 0,  \vec{X}).
\end{equation}
Use Eq.~(\ref{F_Rindler}) in Eq.~(\ref{alpha_tilde_tilde}) to have
\begin{equation}\label{alpha_tilde_tilde_2}
\tilde{\tilde{\alpha}}_\text{R}(\vec{x},\vec{X}) = \frac{\theta(z)}{\hbar }  \left(  \frac{1}{a z} + 1 \right) \int_{\theta_1>0} d^3 \theta  \theta_1  \tilde{F}(\vec{\theta}, Z_\text{R}(z)) \tilde{F}(\vec{\theta}, Z) e^{ i \vec{\theta}_\perp \cdot (\vec{x}_\perp - \vec{X}_\perp)}.
\end{equation}

Notice that, as a consequence of Eq.~(\ref{Fourier_phi_tilde_support_2_manyparticles_R}), the functions $\tilde{\tilde{\alpha}}_\text{R}(\vec{x},\vec{X})$ appearing in Eq.~(\ref{free_wave_function_curved_approx}) are smeared out with a function that is vanishing for $\vec{x}$ such that $| a z - 1 | \gg \epsilon $. Hence, we may focus on values of $\tilde{\tilde{\alpha}}_\text{R}(\vec{x},\vec{X})$ for $\vec{x}$ that are inside of $| a z - 1 | \lesssim \epsilon $ and approximate Eq.~(\ref{alpha_tilde_tilde_2}) with
\begin{equation}\label{alpha_tilde_tilde_3}
\tilde{\tilde{\alpha}}_\text{R}(\vec{x},\vec{X}) \approx  \int_{\theta_1>0} d^3 \theta   \frac{2 \theta_1}{\hbar a z}  \tilde{F}(\vec{\theta}, Z_\text{R}(z)) \tilde{F}(\vec{\theta}, Z) e^{ i \vec{\theta}_\perp \cdot (\vec{x}_\perp - \vec{X}_\perp)}.
\end{equation}

It is possible to show that
\begin{equation}\label{alpha_tilde_approx_3}
\int_0^\infty d\theta_1  \frac{2 \theta_1}{\hbar a z} \tilde{F}(\vec{\theta},Z_\text{R}(z))  \tilde{F}(\vec{\theta},Z) = \frac{1}{4 \pi^2} \delta(z - z_\text{R}(Z)).
\end{equation}
A proof for Eq.~(\ref{alpha_tilde_approx_3}) is provided in Appendix \ref{appendix_2}. Equations (\ref{alpha_tilde_tilde_3}) and (\ref{alpha_tilde_approx_3}) lead to
\begin{equation}\label{alpha_tilde_approx_4}
\tilde{\tilde{\alpha}}_\text{R}(\vec{x},\vec{X}) \approx \delta(z - z_\text{R}(X)) \delta^2(\vec{x}_\perp - \vec{X}_\perp),
\end{equation}
which can be used in Eq.~(\ref{free_wave_function_curved_approx}) to obtain
\begin{equation} \label{free_wave_function_curved_approx_2}
\Phi_{nm} ( \boldsymbol{X}_{n+m}) \approx \phi_{nm} (\boldsymbol{x}_\text{R}(\boldsymbol{X}_{n+m})),
\end{equation}
where
\begin{equation} \label{free_wave_function_curved_approx_2_x_nm}
\boldsymbol{x}_\text{R}(\boldsymbol{X}_{n+m}) = (\vec{x}_\text{R}(\vec{X}_1), \dots , \vec{x}_\text{R}(\vec{X}_n), \vec{x}_\text{R}(\vec{X}_{n+1}), \dots , \vec{x}_\text{R}(\vec{X}_{n+m})).
\end{equation}
Equation (\ref{free_wave_function_curved_approx_2}) states that the wave functions in the position representation approximately transform as scalars, in the sense that $\Phi_{nm} ( \boldsymbol{X}_{n+m})$ is identical to $\phi_{nm} ( \boldsymbol{x}_{n+m})$ up to the transformation $\vec{x}_\nu(\vec{X})$ for each spatial coordinate.

\subsection{Example: Gaussian single particle} \label{Gaussian_singleparticle}

We now provide an example of Minkowski single particle state $| \phi \rangle$ to probe the results that we obtained. We assume that $\tilde{\phi}_{nm}$ is vanishing for any $n$ and $m$, except for $n=1$ and $m=0$. We also assume that the wave function $\tilde{\phi}_{10}(\vec{k})$ has a Gaussian form along the $z$ axis and vanishing transverse momentum, i.e.,
\begin{equation}\label{single_particle_Gaussian_Minkowski}
\tilde{\phi}_{10}(\vec{k}) = 2 \pi \tilde{\phi}(k_3) \delta(\vec{k}_\perp),
\end{equation}
with
\begin{equation}\label{single_particle_Gaussian_Minkowski_x}
\tilde{\phi}(k_3) = \frac{\sqrt{\sigma}}{\pi^{1/4} } \exp \left( - \frac{\sigma^2 k_3^2}{2} - i k_3 z_0 \right).
\end{equation}

In the position representation, the wave function $\phi_{10}$ [Eq.~(\ref{free_wave_function_antiparticles})] in
\begin{equation}\label{single_particle_Gaussian_Minkowski_position}
\phi_{10}(\vec{x}) = \phi(z),
\end{equation}
with
\begin{equation}\label{single_particle_Gaussian_Minkowski_position_x}
\phi(z) = \frac{1}{\sqrt{2 \pi}} \int_{\mathbb{R}} dk_3 \sqrt{\frac{mc^2}{\hbar \omega(k_3 \vec{e}_3)}} \tilde{\phi}(k_3) e^{i k_3 z}
\end{equation}
and $\vec{e}_3 = (0,0,1)$. The nonrelativistic limit (\ref{many_particle_nonrelativistic}) leads to
\begin{equation}\label{single_particle_Gaussian_Minkowski_position_x_nonrelativistic}
\phi(z) \approx \frac{1}{\pi^{1/4} \sqrt{\sigma}} \exp \left( - \frac{ (z-z_0)^2}{2 \sigma^2} \right),
\end{equation}
which is a Gaussian wave function with variance $\sigma$.

In the accelerated frame, the wave functions $\tilde{\Phi}_{10}(\Omega,\vec{K}_\perp)$ [Eqs.~(\ref{non_reltaivistic_wavefunction_integration_0}) and (\ref{Phi_tilde_R})] and $\Phi_{10}(\vec{X})$ [Eq.~(\ref{free_wave_function_curved})], respectively, are
\begin{align}
& \tilde{\Phi}_{10}(\Omega,\vec{K}_\perp) = 2 \pi \tilde{\Phi} (\Omega) \delta^2(\vec{K}_\perp),
& \Phi_{10} (\vec{X}) = \Phi (Z),
\end{align}
with
\begin{subequations}\label{single_particle_Gaussian_Rindler_energy_position_x}
\begin{align}
& \tilde{\Phi} (\Omega) = \int_{\mathbb{R}} d k_3 \tilde{\phi}(k_3) \chi_\text{R}^*(k_3 \vec{e}_3,\Omega),\label{single_particle_Gaussian_Rindler_x} \\
& \Phi (Z) = \frac{2 \pi \sqrt{2 m}}{\hbar} \int_0^\infty d\Omega \tilde{\Phi} (\Omega) \tilde{F}(\Omega, 0 ,0 ,  Z)\label{single_particle_Gaussian_Rindler_position_x}
\end{align}
\end{subequations}
and $\chi_\nu(\vec{k},\Omega)$ defined by Eq.~(\ref{alpha_2}).

In order for $| \phi \rangle$ to be nonrelativistic in the inertial frame [Eq.~(\ref{single_particle_nonrelativistic})], we assume that
\begin{equation}\label{Gaussian_non_relativistic_0}
\frac{\hbar}{m c \sigma} \lesssim \epsilon^{1/2},
\end{equation}
which, together with Eq.~(\ref{quasi_inertial_limit_a}), reads as
\begin{equation}\label{Gaussian_non_relativistic}
a \sigma  \gtrsim \epsilon.
\end{equation}
The localized condition (\ref{Fourier_phi_tilde_support_2_manyparticles_R}), instead, requires
\begin{subequations}\label{Gaussian_local}
\begin{align}
& |az_0 - 1| \lesssim \epsilon \label{Gaussian_local_x0} \\
 & a \sigma  \lesssim \epsilon \label{Gaussian_local_sigma}.
\end{align}
\end{subequations}

Hereafter we assume
\begin{align}\label{Gaussian_local_2}
 z_0  = \frac{1}{a},
\end{align}
in order to meet condition (\ref{Gaussian_local_x0}). On the other hand the values of $\sigma$ are constrained by Eqs.~(\ref{Gaussian_non_relativistic}) and (\ref{Gaussian_local_sigma}) and result in
\begin{equation}\label{Gaussian_non_relativistic_local_sigma}
a \sigma  \sim \epsilon.
\end{equation}

We consider the adimensional variables defined by Eqs.~(\ref{coordinate_transformation}), (\ref{acceleration_adimensional}), (\ref{adimensional_variables}), together with
\begin{align}
& \bar{\sigma} = \frac{a \sigma}{\bar{a}}, & \bar{Z} = \frac{a Z}{\bar{a}}
\end{align}
and the following adimensional wave functions
\begin{subequations}
\begin{align}
& \bar{\tilde{\phi}}(\bar{k}_3) =  \sqrt{\frac{a}{\bar{a}}} \exp \left(  i \frac{\bar{k}_3}{\bar{a}} \right) \tilde{\phi} \left( \frac{a \bar{k}_3}{\bar{a}} \right), & \bar{\phi}(\bar{z}) =  \sqrt{\frac{\bar{a}}{a}} \phi \left( \frac{\bar{a} \bar{z} + 1}{a} \right),\\
& \bar{\tilde{\Phi}}(\bar{\Omega}) =  \sqrt{\frac{m c^2}{\hbar}} \tilde{\Phi} \left( \frac{mc^2 \bar{\Omega}}{\hbar} \right),  & \bar{\Phi}(\bar{Z}) =  \sqrt{\frac{\bar{a}}{a}}  \Phi \left( \frac{\bar{a} \bar{Z}}{a} \right).
\end{align}
\end{subequations}

In this way, Eqs.~(\ref{single_particle_Gaussian_Minkowski_x}), (\ref{single_particle_Gaussian_Minkowski_position_x}), (\ref{single_particle_Gaussian_Rindler_energy_position_x}) read as
\begin{subequations}\label{single_particle_x_adimensional}
\begin{align}
\bar{\tilde{\phi}}(\bar{k}_3) = & \frac{\sqrt{\bar{\sigma}}}{\pi^{1/4} } \exp \left( - \frac{\bar{\sigma}^2 \bar{k}_3^2}{2}  \right),\label{single_particle_Gaussian_Minkowski_x_adimensional}\\
\bar{\phi}(\bar{z}) = & \frac{1}{\sqrt{2 \pi}} \int_{\mathbb{R}} d\bar{k}_1 \frac{e^{i \bar{k}_3 \bar{z}} \bar{\tilde{\phi}}(\bar{k}_3)}{\sqrt[4]{ 1 + 2 \bar{a} \bar{k}_3^2 }} ,\label{single_particle_Gaussian_Minkowski_position_x_adimensional}\\
\bar{\tilde{\Phi}} (\bar{\Omega}) = & \frac{1}{\sqrt{\bar{a}}} \int_{\mathbb{R}} d \bar{k}_3   \bar{\tilde{\phi}}(\bar{k}_3) \bar{\chi}_\text{R}^*(\bar{k}_3 \vec{e}_3,\bar{\Omega}), \label{single_particle_Gaussian_Rindler_x_adimensional}\\
\bar{\Phi} (\bar{Z}) = &  \frac{2 \pi}{\sqrt{\bar{a}}} \int_0^\infty d\bar{\Omega} \bar{\tilde{\Phi}} (\bar{\Omega})  \bar{\tilde{F}}( \bar{\Omega}  \vec{e}_3, \bar{z}_\text{R}(\bar{Z}) ), \label{single_particle_Gaussian_Rindler_position_x_adimensional}
\end{align}
\end{subequations}
where
\begin{equation}\label{coordinate_transformation_adimensional}
\bar{z}_\text{R}(\bar{Z}) = \frac{1}{\bar{a}} \left[ a z_\text{R} \left( \frac{\bar{a} \bar{Z}}{a} \right) - 1\right]
\end{equation}
is the coordinate transformation between the adimensional variables $\bar{z}$ and $\bar{Z}$, and where $\bar{\chi}_\nu$ is defined as the adimensional equivalent of $\chi_\nu(\vec{k},\Omega)$ by the following identity
\begin{equation} \label{alpha_bar_chi_bar_nu}
\chi_\nu(\vec{k},\Omega) = \sqrt{\frac{\hbar}{m c^2 a }}  \exp \left(- i \frac{k_3}{a} \right) \bar{\chi}_\nu \left(  \frac{\bar{a} \vec{k}}{a}  , \frac{\hbar \Omega}{m c^2}  \right).
\end{equation}
Moreover, condition (\ref{Gaussian_non_relativistic_local_sigma}) now reads as
\begin{equation}\label{Gaussian_non_relativistic_local_sigma_adimensional}
\bar{\sigma}  \sim 1.
\end{equation}

The explicit form of $\bar{\chi}_\text{R}(\bar{k}_3 \vec{e}_3,\bar{\Omega})$ is given by Eq.~(\ref{chi_derivative_5}). The adimensional equivalent of Eq.~(\ref{chi_derivative_5}) is
\begin{align} \label{chi_nu_bar_explicit}
& \bar{\chi}_\text{R}(\vec{\bar{k}},\bar{\Omega}) = \left[ 4 \pi  \sqrt{1+2 \bar{a} \bar{k}^2} \left| \sinh \left( \frac{\pi \bar{\Omega}}{\sqrt{2 \bar{a}^3}} \right) \right| \right]^{-1/2} \nonumber \\
& \times \exp \left( \frac{ \pi \bar{\Omega}}{(2 \bar{a})^{3/2}}  + i \frac{\bar{k}_3}{\bar{a}} - i \frac{ \bar{\Omega}}{(2 \bar{a})^{3/2}} \ln \left( \frac{\sqrt{1+2 \bar{a} \bar{k}^2}+ \sqrt{2 \bar{a}} \bar{k}_3 }{\sqrt{1+2 \bar{a} \bar{k}^2}-\sqrt{2 \bar{a}} \bar{k}_3} \right) \right).
\end{align}

\begin{figure}
\center
\includegraphics[scale=0.95]{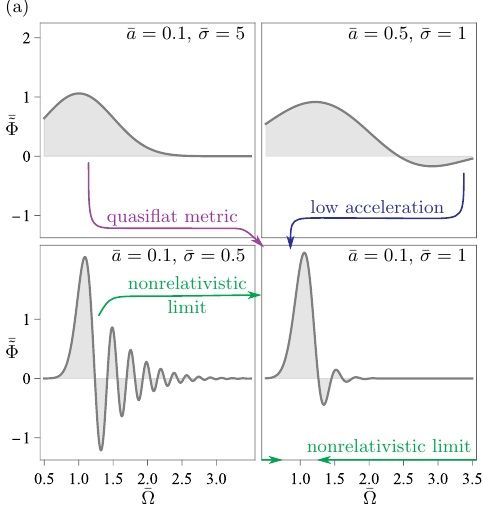}
\includegraphics[scale=0.95]{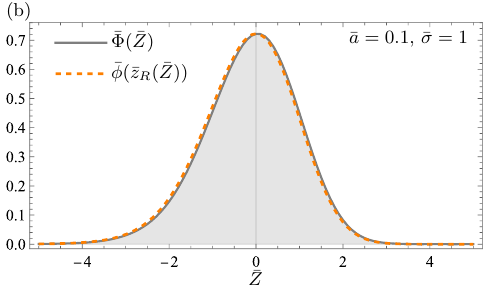}
\caption{Inertial Gaussian single particle wave functions in accelerated frames. In panel (a), we plot the distribution of Rindler frequencies $\bar{\Omega}$ with respect to different acceleration $\bar{a}$ and different variance $\bar{\sigma}$. If $\bar{a} = 0.1$, $\bar{\sigma} = 1$, the wave function $\bar{\tilde{\Phi}}(\bar{\Omega})$ is peaked in $\bar{\Omega} \approx 1$ and, hence, the state is populated by nonrelativistic energies in the accelerated frame [Eq.~(\ref{non_relativistic_Theta_1_adimensional})]. Conversely, relativistic energies appear for other configurations. The reasons are the following: when $\bar{\sigma} = 5$, the particle is not well-localized in the region where the metric is almost flat [Eq.~(\ref{Gaussian_local_sigma})]; when $\bar{\sigma} = 0.5$ the state is populated by relativistic Minkowski momenta [Eq.~(\ref{Gaussian_non_relativistic})]; when $\bar{a}=0.5$ the acceleration is not sufficiently low for the quasi-inertial approximation [Eq.~(\ref{quasi_inertial_limit_a})]. In panel (b), we show the wave function in the position representation $\bar{\Phi}(\bar{Z})$ (gray solid line) for the state seen by the accelerated observer. We choose $\bar{a} = 0.1$ and $\bar{\sigma} = 1$ for the nonrelativistic and quasi-inertial approximation. In such a regime, $\bar{\Phi}(\bar{Z})$ can be approximated by the Minkowski wave function $\bar{\phi}(\bar{z}_\text{R}(\bar{Z}))$ (orange dashed line).}\label{Gaussian_figure}
\end{figure}

By using Eqs.~(\ref{F_tilde_Rindler}), (\ref{F_bar}), (\ref{chi_nu_bar_explicit}) in Eq.~(\ref{single_particle_x_adimensional}), one is able to compute the wave functions $\bar{\tilde{\Phi}} (\bar{\Omega})$ and $\bar{\Phi} (\bar{Z})$ in the accelerated frame. The results are drawn in Fig.~\ref{Gaussian_figure}.

In Fig.~\ref{Gaussian_figure}a, we show that under condition (\ref{Gaussian_non_relativistic_local_sigma_adimensional}) and $\bar{a} \ll 1$, $\bar{\tilde{\Phi}} (\bar{\Omega})$ is not vanishing only for nonrelativistic frequencies ($\bar{\Omega} \approx 1$). This is in agreement with the results of Secs.~\ref{Inertial_and_quasiinertial_frame} and \ref{Quasiinertial_observer}: in the quasi-inertial regime ($\bar{\sigma} \lesssim 1$, $\bar{a} \ll 1$), the accelerated observer detects nonrelativistic particles ($\bar{\Omega} \sim 1$) when the state is nonrelativistic in the inertial frame ($\bar{\sigma} \gtrsim 1$) as well. Conversely, when conditions (\ref{Gaussian_non_relativistic_local_sigma_adimensional}) and $\bar{a} \ll 1$ are not met, relativistic energies are present in the accelerated frame.

In Fig.~\ref{Gaussian_figure}b, we plot the wave function $\bar{\Phi} (\bar{Z})$. We choose a configuration in which condition (\ref{Gaussian_non_relativistic_local_sigma_adimensional}) and $\bar{a} \ll 1$ are met. One can see that $\bar{\Phi} (\bar{Z})$ is approximated by $\bar{\phi}(\bar{z})$, up to the coordinate transformation (\ref{coordinate_transformation_adimensional}). Such a result confirms the prediction of Eq.~(\ref{free_wave_function_curved_approx_2}) for the case of a single Gaussian particle.

\section{Conclusions} \label{Framedependent_nonrelativistic_limit_Conclusions}

We proved the frame-dependence of the nonrelativistic limit. Specifically, we showed that by switching from an inertial to a non-inertial frame, the relativistic nature of quantum states may change: nonrelativistic particles of one frame can be relativistic for the other observer. Also the number of particles may change. 

This can be problematic in the context of non-inertial detectors---e.g., Unruh-DeWitt detectors \cite{PhysRevD.14.870, PhysRevD.29.1047, hawking1980general}. For instance, an atomic detector, that is prepared in the laboratory frame as a nonrelativistic $n$-particles state and then accelerated, cannot be described as a fixed number of nonrelativistic particles in its proper non-inertial frame. The familiar first-quantization description of atomic systems breaks down when one switches from the inertial to the accelerated frame.

We have proposed a solution to such problem by considering a quasi-inertial frame. The observer is defined to have low acceleration in the nonrelativistic limit and can only have access to a region in which the metric is quasiflat. We showed that nonrelativistic states in the inertial frame are also nonrelativistic in the quasi-inertial frame, as opposed to the case of arbitrarily large accelerations. Moreover, the number of particles is preserved when switching from one frame to the other. Finally, we showed how scalar particles wave functions transform from the inertial to the quasi-inertial frame. In particular, we proved that they approximately transform as scalar fields under the coordinate transformation.

\chapter{Experiments: accelerated detectors}\label{Accelerated_non_relativistic_detectors}

\textit{This chapter is based on and contains material from Ref.~\citeRF{PhysRevA.108.022807}.}

\section{Introduction}\label{Introduction_Accelerated_non_relativistic_detectors}

In Chap.~\ref{Frame_dependent_content_of_particles}, we discussed the frame dependent representation of states and the consequent particle production when switching from one frame to the other. In particular, we showed that the representation of states in terms of Minkowski particles generally differ from the representation in the Rindler-Fock space. In principle, this means that the inertial and the accelerated observer would see different particle content for the same field configuration. This theoretical prediction raises the following questions: How can one experimentally probe the different particle content in the two frame? Is it possible to detect a Rindler particle and distinguish it from a Minkowski particle?

In real life experiments, particles are revealed by means of detectors. Due to the difference between particle contents in different frames (e.g., Minkowski and Rindler particles), it appears natural to ask which type of particle each detector is able to reveal: does it detect Minkowski or Rindler particles? The question, however, is not well posed, since Minkowski and Rindler particles actually come from different representations of the same state. In particular, the detection of the Minkowski single particle (\ref{single_particle_Minkowski}) is equivalent to the detection of the corresponding Rindler-Fock state (\ref{single_particle_Minkowski_Rindler}); whereas, the detection of a Rindler particle is equivalent to the detection of a suitable Minkowski-Fock state representing the same state. A detector coupled to Rindler particles in the accelerated frame is equivalent to a detector coupled to Minkowski particles in the inertial frame.

Instead of asking whether the detector interacts with Minkowski or Rindler particles, one should understand how it reacts in the vacuum of each frame. If the detector does not ``click'' in the Minkowski vacuum, then one can say the detector is tuned for the detection of Minkowski particles; whereas, if it does not react in the Rindler vacuum, it can be regarded as a Rindler particle detector. The question now can be formulated as follows: when does the detector is suited for the detection of Minkowski or Rindler particles? By quoting Unruh, ``a particle detector will react to states which have positive frequency with respect to the detectors proper time, not with respect to any universal time'' \cite{PhysRevD.14.870}. Intuitively, this means that detectors with inertial trajectory do not react in the Minkowski vacuum, whereas accelerated detectors are suited for the detection of Rindler particles as they do not react in the Rindler vacuum.

It has been argued that experiments in inertial laboratories hardly provide a test for the particle production and the Unruh effect in accelerated frames, since any noninertial phenomenon can be equivalently described in the inertial frame \cite{Pena:2013zfd}. However, there can be situations in which the description of the phenomenon in one frame appears more natural than in the other. An example is provided by nonrelativistic systems. As pointed out in Sec.~\ref{Framedependent_nonrelativistic_limit_Introduction}, due to the frame dependent notion of time and energy, the nonrelativistic condition is frame dependent as well, in the sense that any physical setup can only be regarded as nonrelativistic in one of the two frames. This gives a preferred choice for the coordinate system and the consequent particle representation.

In QFT, detectors are usually regarded as nonrelativistic systems coupled to the field that the experimenter wants to probe. Intuitively, we may assume that the detector appears as a nonrelativistic system in its own comoving frame. This means that the preferred choice of time coordinate is the proper time. For instance, if the detector is accelerated, the preferred coordinate system is given by the Rindler frame (\ref{Rindler_coordinates_transformation_R}) and both detector and field are naturally described in terms of Rindler-Fock states. Clearly, an equivalent representation can be given in terms of Minkowski particle states; however, the price to be paid is the loss of the nonrelativistic nature of the detector.

Historically, Unruh-DeWitt detectors are the first proposed model of particle detectors in the context of QFTCS \cite{PhysRevD.14.870, hawking1980general}. They are usually described as ideal point-like objects with classical trajectory and quantized internal degrees of freedom coupled with the background field. The first type of model considered by Unruh and DeWitt is nonrelativistic; in particular, Unruh investigated a first quantized particle in a box \cite{PhysRevD.14.870}, whereas DeWitt \cite{hawking1980general} considered a monopole type of detector.

More refined models including the recoil of the accelerated detector and the back reaction of the Unruh effect superseded the original idealized description~\cite{PARENTANI1995227, PhysRevD.57.2403, PhysRevD.103.105023}. Fully relativistic detectors have been considered in Refs.~\cite{PhysRevD.14.870, PARENTANI1995227, PhysRevD.105.125001}. Also, a general description of localized nonrelativistic quantum systems by means of Fermi normal coordinates was presented in Ref.~\cite{PhysRevD.106.025018}.

The Unruh-DeWitt model has been used in different works to describe accelerated atoms as particle detectors \cite{PhysRevLett.91.243004, PhysRevA.74.023807, PhysRevLett.128.163603}. In these cases, the atom is assumed to be nonrelativistic and made of a fixed number of particles. However, to the best of our knowledge, the fact that such a description is frame-dependent has been overlooked. Remarkably, one has to take into account that the laboratory and the atom frames have different representations for the same quantum system.

In this chapter, we give a brief review of the Unruh-DeWitt detectors and the experimental proposals to detect the Unruh effect. Afterwards, we discuss the case of accelerated atomic detector. However, instead of using the idealized Unruh-DeWitt model, we provide a description from first principle based on the results of the previous chapters of this thesis. In particular, we consider the formalism of NRQFTCS and use the nonrelativistic limit of the electron field in Rindler spacetime presented in Sec.~\ref{Nonrelativistic_limit_Dirac_Rindler}.

Due to the accelerating trajectory, the Rindler coordinate system and the Rindler-Fock representation appear as the most natural choices for the description of the detector. We model the interaction between the atom and the electromagnetic background via electron-photon coupling. The atomic system is designed to detect photons with specific spectral energy in the comoving frame.

Unfortunately, we lack of a fully quantum relativistic theory describing the bound states of the atom. In principle, this would require a genuine QFT description of electron and nuclear fields interacting between each other via electromagnetic and nuclear forces and generating a well defined spectrum of atomic bound states. As far as we know, this has not been accomplished yet. Hence, we can only rely on the familiar semiclassical nonrelativistic theory, which needs to be described in the detector comoving frame. We conclude that the characterization of the atomic detector in the Rindler frame in terms of Rindler-Fock particles is not only natural, but also practically necessary. Indeed, the description of the accelerated atom in the Minkowski frame would require a fully relativistic theory of atomic phenomena that we do not have.

The necessity to describe the state in terms of Rindler particles has the following consequence. As shown in Sec.~\ref{Inertial_and_non_inertial_frame}, if the accelerating atom is prepared in the inertial laboratory frame with a fixed number of nonrelativistic electrons, it appears as made of an indefinite number of relativistic and nonrelativistic particles in its proper frame. Hence, the first-quantized description for hydrogen-like atoms cannot be adopted. However, following the results of Sec.~\ref{Quasi_inertial_regime}, we know how to suppress such a frame-dependent effect. For the case of accelerated hydrogen-like atom, we show that this can be accomplished by constraining the atomic ionization and the accelerating electric field.

We identify the physical regimes with nonvanishing atomic excitation probability due to the Unruh electromagnetic background. We recognize the observational limits for the Unruh effect via first-quantized atomic detectors, which appear to be compatible with current technology. Notably, the nonrelativistic energy spectrum of the atom cannot induce coupling with the thermal radiation, even when special relativistic and GR corrections are considered. On the contrary, the coupling with the Unruh radiation arises because of relativistic hyperfine splitting and the Zeeman effect.

This chapter is organized as follows. In Sec.~\ref{UnruhDeWitt_detectors}, we discuss the Unruh-DeWitt detector as a toy model for accelerated detectors; also, we give a brief review of the experimental proposal to test the Unruh effect by means of Unruh-DeWitt-like detectors. In Sec.~\ref{Atomic_detectors}, we study accelerated atoms by applying QFT in the Rindler spacetime and we detail the observational limits to detect Unruh radiation via first-quantized atomic detectors. Conclusions are drawn in Sec.~\ref{Accelerated_non_relativistic_detectors_Conclusions}.

\section{Unruh-DeWitt detectors}\label{UnruhDeWitt_detectors}

Since the theoretical proposal of the Unruh effect \cite{PhysRevD.7.2850, Davies:1974th, PhysRevD.14.870}, there has been a wide interest in noninertial particle detectors to probe this effect. Here, we give a brief introduction about the Unruh-DeWitt detectors, originally introduced in Refs.~\cite{PhysRevD.14.870, hawking1980general}.

In Sec.~\ref{Non_relativistic_models}, we discuss the nonrelativistic models considered by both Unruh and DeWitt; whereas, in Sec.~\ref{Relativistic_model}, we show the relativistic model proposed in Ref.~\cite{PhysRevD.14.870} as an alternative to the nonrelativistic particle detectors. Finally, in Sec.~\ref{Experiments}, we give a brief review about some experimental proposals to detect the Unruh effect.

\subsection{Non relativistic models}\label{Non_relativistic_models}

\subsubsection{Particle in a box}

In his original work \cite{PhysRevD.14.870}, Unruh considered a nonrelativistic particle in a uniformly accelerated box as a noninertial detector. In the comoving accelerated frame $(T,\vec{X})$, the particle detector is prepared in its ground state $| 0_\text{D} \rangle$ and then interacts with the scalar real field $\hat{\Phi}_\text{R}(T,\vec{X})$. The interaction is regulated by the time dependent coupling constant $\varepsilon(T)$ simulating the switching on and off of the detector. Afterwards, if the detector is found in an excited state $| n_\text{D} \rangle$, one says that a quanta of the field has been detected.

The Unruh effect is the prediction that in the accelerated frame, the Minkowski vacuum $| 0_\text{M} \rangle$ of the field $\hat{\Phi}_\text{R}(T,\vec{X})$ is seen as a thermal state [Eq.~(\ref{thermal})]. This means that $| 0_\text{M} \rangle$ is filled with Rindler particles that interact with the particle in the box. For this reason, the excitation of the particle detector is regarded as an experimental verification of the Unruh effect.

The total Hamiltonian describing the time evolution of the field-detector system in the right Rindler frame is $\hat{H}_\text{R} = \hat{H}_{\Phi,\text{R}} + \hat{H}_\text{D,R} + \hat{H}_\text{int,R}$ \cite{PhysRevD.29.1047}. $\hat{H}_{\Phi,\text{R}}$ is the free Hamiltonian for the background field $\hat{\Phi}_\text{R}(T,\vec{X})$ and is equal to the right hand side of Eq.~(\ref{H_R_general}). $\hat{H}_\text{D,R} = E_\text{D} \hat{A}^\dagger_\text{D} \hat{A}_\text{D}$ is the free Hamiltonian of the detector. The interaction Hamiltonian $\hat{H}_\text{int,R}$ couples the detector to the field. Explicitly, in the interaction picture, we have that
\begin{equation}\label{H_int}
\hat{H}_\text{int,R}(T) = \varepsilon(T) \int_{\mathbb{R}^3} d^3 X \sqrt{-g(\vec{X})} \hat{\Phi}_\text{R}(T,\vec{X}) \delta^3(\vec{X} - \hat{\vec{X}}_\text{D}(T))
\end{equation}
Here, $E_\text{D}$ is the energy gap for the detector spectrum and $\hat{A}^\dagger_\text{D}$ and $\hat{A}_\text{D}$ are, respectively, the raising and lowering operators, defined by $\hat{A}^\dagger_\text{D} | n_\text{D} \rangle = \sqrt{n+1} | (n + 1)_\text{D} \rangle$, $\hat{A}_\text{D} | n_\text{D} \rangle = \sqrt{n} | (n - 1)_\text{D} \rangle$ and $\hat{A}_\text{D} | 0_\text{D} \rangle = 0$, for any $n \in \mathbb{N}$. The states $| n_\text{D} \rangle$ are the eigenstates of $\hat{H}_\text{D,R}$ with $| 0_\text{D} \rangle$ as the ground state. The operator $\hat{\vec{X}}_\text{D}$ is the first-quantized position operator for the particle in the box, with eigenstates $| \vec{X}_\text{D} \rangle$. $\hat{\vec{X}}_\text{D}(T)$ is the time evolution of $\hat{\vec{X}}_\text{D}$ with respect to the free Hamiltonian $\hat{H}_\text{D,R}$.

By defining the wave functions $\psi_n(\vec{X}) = \langle \vec{X}_\text{D} | n_\text{D} \rangle$, Eq.~(\ref{H_int}) becomes
\begin{align}\label{H_int_2}
\hat{H}_\text{int,R}(T) = & \varepsilon(T) \sum_{n,m} \int_{\mathbb{R}^3} d^3 X \sqrt{-g(\vec{X})} \hat{\Phi}_\text{R}(T,\vec{X}) \psi_n^*(\vec{X}) \psi_m(\vec{X}) e^{i(n-m) E_\text{D} T / \hbar} \nonumber \\
& \times | n_\text{D} \rangle \langle m_\text{D} |,
\end{align}
where the sum $\sum_{n,m}$ runs over all elements of the detector spectrum.

\subsubsection{Monopole coupling}

If we assume that for all practical purposes the box is well localized in $\vec{X} = 0$, then we can approximate Eq.~(\ref{H_int_2}) as
\begin{equation}\label{H_int_2_approx}
\hat{H}_\text{int,R}(T) \approx \varepsilon(T) \sum_{n,m} \hat{\Phi}_\text{R}(T,\vec{0})  e^{i(n-m) E_\text{D} T / \hbar} | n_\text{D} \rangle \langle m_\text{D} |,
\end{equation}
which results in a monopole coupling between the field and the detector.

The same model was considered by DeWitt in Ref.~\cite{hawking1980general}. In this case, the detector is assumed to follow the classical trajectory $\vec{X}=0$, which, in the inertial frame, is the uniformly accelerating trajectory parameterized by $(t,\vec{x}) = (t_\text{R}(T,\vec{0}),0, 0, z_R(T,\vec{0}))$, with $t_\text{R}(T,\vec{X})$ and $z_R(T,\vec{X})$ defined by Eq.~(\ref{Rindler_coordinate_transformation}). The internal quantum degrees of freedom of the detector couple to the field by means of Eq.~(\ref{H_int_2_approx}) and lead to the detection of Unruh quanta.

The excitation of the Unruh-DeWitt detector can be derived in two equivalent ways: (i) One can use the representation of the Minkowski vacuum as a right-Rindler-Fock state provided by Eq.~(\ref{thermal}) and notice that the time evolution of $\hat{\rho}_0 \otimes | 0_\text{D} \rangle \langle 0_\text{D} |$ with respect to the Hamiltonian $\hat{H}_\text{int,R}(T)$ does not lead to the same initial state $\hat{\rho}_0 \otimes | 0_\text{D} \rangle \langle 0_\text{D} |$; in particular, there is a non vanishing probability for the detector to be found in an excited state. (ii) Instead of using the right-Rindler-Fock representation for the field, one can use the Minkowski-Fock space $\mathcal{H}_\text{M}$; by means of Eq.~(\ref{scalar_transformation_Rindler}), one can represent Eq.~(\ref{H_int_2_approx}) as an operator acting on $\mathcal{H}_\text{M}$; the time evolution of $| 0_\text{M} \rangle \otimes | 0_\text{D} \rangle$ with respect to such an operator has a non vanishing probability to be found in an excited state.

At variance with $| 0_\text{M} \rangle \otimes | 0_\text{D} \rangle$, the time evolution of the state $| 0_\text{R} \rangle \otimes | 0_\text{D} \rangle$, with right Rindler vacuum state $| 0_\text{R} \rangle$, does not lead to excitations of the detector if $\varepsilon(T)$ is sufficiently slowly varying, i.e., the detector is smoothly turned on and off. This makes the Unruh-DeWitt detector defined by Eq.~(\ref{H_int_2_approx}) as an instrument well suited for the detection of Rindler quanta.

Notice that the Lagrangian densities associated to Eqs.~(\ref{H_int}) and (\ref{H_int_2_approx}) are not Lorentz invariant. Physically, this means that the systems are nonrelativistic. Notice also that the Hamiltonian $\hat{H}_\text{R}$ generates the time evolution with respect to the Rindler coordinate $T$. Hence, the unitary operator generated by $\hat{H}_\text{R}$ describes the dynamics from the point of view of an accelerated observer comoving with the detector. To have a proper description from the point of view of an inertial observer comoving with the laboratory, one needs the generator of the Minkowski time $t$. However, this requires a fully relativistic theory that allows for diffeomorphisms between coordinate systems. 

\subsection{Relativistic model}\label{Relativistic_model}

A fully relativistic model was already discussed by Unruh in his original work \cite{PhysRevD.14.870} as an alternative to the nonrelativistic particle in a box. In particular, he considered a particle detector described by two Klein-Gordon complex scalar fields $\hat{\Phi}_{\downarrow \nu}(T,\vec{X})$ and $\hat{\Phi}_{\uparrow \nu}(T,\vec{X})$, with masses $M_\downarrow$ and $M_\uparrow>M_\downarrow$, respectively. $\hat{\Phi}_{\downarrow \nu}(T,\vec{X})$ and $\hat{\Phi}_{\uparrow \nu}(T,\vec{X})$ represent two field states of the detector, with excitation energy $M_\uparrow-M_\downarrow$. The interaction action is given by
\begin{align}\label{H_relativistic}
\hat{S}_\text{int,R} = & \varepsilon \sum_{\nu = \{ \text{L}, \text{R} \}} \int_{\mathbb{R}} d T  \int_{\mathbb{R}^3} d^3 X \sqrt{-g(\vec{X})} \nonumber \\
& \times \left[ \hat{\Phi}_{\downarrow \nu}^\dagger(T,\vec{X})  \hat{\Phi}_{\uparrow \nu}(T,\vec{X}) + \hat{\Phi}_{\downarrow \nu}(T,\vec{X})  \hat{\Phi}_{\uparrow \nu}^\dagger(T,\vec{X}) \right]  \hat{\Phi}_\nu(T,\vec{X}).
\end{align}
Here, we consider both left and right wedge to let $t=0$ and $T=0$ be a common Cauchy surface in the Minkowski and the Rindler frame, respectively, as explained at the beginning of Sec.~\ref{QFT_in_curved_spacetime_Rindler}.

At initial times ($T = -\infty$), the detector is assumed to be a particle of the low mass field $\hat{\Phi}_{\downarrow \nu}(T,\vec{X})$. The ``clicking'' is given by the transition from the $\hat{\Phi}_{\downarrow \nu}(T,\vec{X})$ particle to a $\hat{\Phi}_{\uparrow \nu}(T,\vec{X})$ particle. The detection of the Unruh effect is due to the absorption of an Unruh thermal quanta of $\hat{\Phi}_\text{R}(T,\vec{X})$ with the consequent excitation of the detector via particle transition from $\hat{\Phi}_{\downarrow \nu}(T,\vec{X})$ to $\hat{\Phi}_{\uparrow \nu}(T,\vec{X})$.
 
To obtain the representative of Eq.~(\ref{H_relativistic}) in the Minkowski frame, one needs to consider the transformation of the fields from the Rindler to the Minkowski frame. The equivalent of Eq.~(\ref{scalar_transformation_Rindler}) for the detector fields is
\begin{align}\label{scalar_transformation_Rindler_detector}
& \hat{\Phi}_{\downarrow \nu} (T,\vec{X}) = \hat{\phi}_\downarrow (t_\nu(T,\vec{X}), \vec{x}_\nu(T,\vec{X})), & \hat{\Phi}_{\uparrow \nu} (T,\vec{X}) = \hat{\phi}_\uparrow (t_\nu(T,\vec{X}), \vec{x}_\nu(T,\vec{X})),
\end{align}
where $\hat{\phi}_\downarrow (t, \vec{x})$ and $\hat{\phi}_\uparrow (t, \vec{x})$ describe the detector in the inertial frame.

Notice that the Rindler coordinate system $(T, \vec{X})$ only covers the left and right wedge in the Minkowski frame, respectively defined by $z < -c |t|$ and $z > c |t|$. The future ($ct>|z|$) and the past ($ct<-|z|$) wedges, instead, are excluded by the Rindler frame. Hence, by using Eq.~(\ref{scalar_transformation_Rindler_detector}) in Eq.~(\ref{H_relativistic}) and by performing the coordinate transformation $(T, \vec{X}) \mapsto (t, \vec{x}) = (t_\nu(T,\vec{X}), \vec{x}_\nu(T,\vec{X}))$ for each wedge, one obtains
\begin{equation}\label{H_relativistic_approx_3_exact}
\hat{S}_\text{int,R} = \varepsilon \int_{\mathbb{R}} d t \int_{|z|>c|t|} d^3 x  \left[ \hat{\phi}_\downarrow^\dagger (t,\vec{x})  \hat{\phi}_\uparrow(t,\vec{x}) + \hat{\phi}_\downarrow(t,\vec{x})  \hat{\phi}_\uparrow^\dagger(t,\vec{x}) \right]  \hat{\phi}(t,\vec{x}).
\end{equation}

The assumption that the particle detector is well localized in $|\vec{X}| \lesssim \lambda $ in the right wedge can be implemented by restricting the action of the field operators  $\hat{\Phi}_{\downarrow \nu}(T,\vec{X})$ and $\hat{\Phi}_{\uparrow \nu}(T,\vec{X})$ to states that allow for the following approximation
\begin{equation}\label{Phi_detector_approx}
\hat{\Phi}_{\downarrow \nu}(T,\vec{X}) \approx 0 \text{ and } \hat{\Phi}_{\uparrow \nu}(T,\vec{X}) \approx 0 \text{ if } \nu = \text{L} \text{ or } |\vec{X}| \gg \lambda.
\end{equation}
Equation (\ref{Phi_detector_approx}) can be seen as an approximation for the matrix elements of the field operators $\hat{\Phi}_{\downarrow \nu}(T,\vec{X})$ and $\hat{\Phi}_{\uparrow \nu}(T,\vec{X})$ with respect to the localized states. The masses of the fields $M_\downarrow$ and $M_\uparrow$ must be sufficiently large to let the semiclassical trajectory $\vec{X} \approx 0$ hold; in particular, a Compton length much smaller than $\lambda$ is required. Also, an accelerating external force in the Minkowski frame must be introduced to let the detector have a stationary motion in the Rindler frame.

By assuming that $\lambda \lesssim a^{-1}$, the localization condition for the Minkowski fields is
\begin{equation}\label{Phi_detector_approx_Minkowski}
\hat{\phi}_\downarrow(t, \vec{x}) \approx 0 \text{ and } \hat{\phi}_\uparrow(t, \vec{x}) \approx 0 \text{ if } |\vec{x} - \vec{x}(t) | \gg \lambda,
\end{equation}
where $\vec{x}(t) = (0, 0, z(t))$ is the accelerating trajectory in the inertial frame, with $z(t) = a^{-1} \cosh ( \sinh^{-1}( cat )  ) = a^{-1} \sqrt{1 + (cat)^2}$. Notice that, as a consequence of Eq.~(\ref{scalar_transformation_Rindler_detector}), Eq.~(\ref{Phi_detector_approx}) is a necessary condition for Eq.~(\ref{Phi_detector_approx_Minkowski}).

When the detector localization condition (\ref{Phi_detector_approx_Minkowski}) holds, the spatial integral in the right hand side of Eq.~(\ref{H_relativistic_approx_3_exact}) can be extended to the entire space $\mathbb{R}^3$, including the future and the past wedges, i.e.,
\begin{equation}\label{H_relativistic_approx_3}
\hat{S}_\text{int,R}  \approx  \varepsilon \int_{\mathbb{R}} d t \int_{\mathbb{R}^3} d^3 x  \left[ \hat{\phi}_\downarrow^\dagger (t,\vec{x})  \hat{\phi}_\uparrow(t,\vec{x}) + \hat{\phi}_\downarrow(t,\vec{x})  \hat{\phi}_\uparrow^\dagger(t,\vec{x}) \right]  \hat{\phi}(t,\vec{x}).
\end{equation}

This result can be understood as follows. The operator $\hat{S}_\text{int,R}$ appearing on the left hand side of Eq.~(\ref{H_relativistic_approx_3}) and defined by Eq.~(\ref{H_relativistic}) gives the time evolution of the detector-field system with respect to the coordinate time $T$; physically, it describes how the accelerated observer sees the system evolving from the asymptotic time $T=-\infty$ to $T=+\infty$. However, in principle, $\hat{S}_\text{int,R}$ is not associated to the notion of time evolution in the Minkowski frame; hence, it does not tell how an inertial observer in the laboratory sees the fields evolving from $t=-\infty$ to $t=+\infty$. The genuine action describing such an evolution is 
\begin{equation}\label{H_relativistic_M}
\hat{S}_\text{int,M}  =  \varepsilon \int_{\mathbb{R}} d t \int_{\mathbb{R}^3} d^3 x  \left[ \hat{\phi}_\downarrow^\dagger (t,\vec{x})  \hat{\phi}_\uparrow(t,\vec{x}) + \hat{\phi}_\downarrow(t,\vec{x})  \hat{\phi}_\uparrow^\dagger(t,\vec{x}) \right]  \hat{\phi}(t,\vec{x}),
\end{equation}
which is equal to the right hand side of Eq.~(\ref{H_relativistic_approx_3}). Consequently, Eq.~(\ref{H_relativistic_approx_3}) gives an equivalence between the two actions, i.e.,
\begin{equation}\label{H_relativistic_RM}
\hat{S}_\text{int,R} \approx \hat{S}_\text{int,M}.
\end{equation}
This is a result of the Lorentz invariance of the relativistic action and the localization condition (\ref{Phi_detector_approx_Minkowski}).

Equation (\ref{H_relativistic_RM}) states that both the inertial and the accelerated observer describe the asymptotic evolution of the localized accelerated detector by means of the same interaction Hamiltonian. Hence, they agree about the results of the detector measurements. In particular, if the detector is prepared as a particle of the low mass field $\hat{\Phi}_{\downarrow \nu}(T,\vec{X})$, then the probability to be found excited (i.e., as a particle of $\hat{\Phi}_{\uparrow \nu}(T,\vec{X})$) after a large time of interaction is the same in the two frame. For this reason, all the predictions that are made in one frame are also valid in the other frame. 

This result can be applied to the nonrelativistic case as well. By taking the nonrelativistic limit of the detector fields $\hat{\Phi}_{\downarrow \nu}(T,\vec{X})$ and $\hat{\Phi}_{\uparrow \nu}(T,\vec{X})$ in the Rindler frame, one obtains a model that is similar to the one shown in Sec.~\ref{Non_relativistic_models}. By considering the fully relativistic theory we know that the predictions made in Sec.~\ref{Non_relativistic_models} also apply to the inertial frame. However, we cannot use the nonrelativistic theory to reach such a conclusion. Instead, we need to start from the fully relativistic theory to obtain Eq.~(\ref{H_relativistic_RM}). Then, one can focus on the accelerated frame and take the nonrelativistic limit to derive physical predictions is such a frame; the validity of these predictions for both observers is a priori guaranteed by the equivalence (\ref{H_relativistic_RM}).

\subsection{Experiments}\label{Experiments}

There have been a series of experimental proposal to test the Unruh effect.

For instance, the depolarization of electrons in storage rings \cite{Sokolov:1963zn} has been explained in terms of Unruh effect in rotational frames \cite{BELL1983131}. In the experimental setup, the electron plays the role of Unruh-DeWitt detector, with its spinorial degrees of freedom coupled to the background electromagnetic field. It has been argued that, in addition to the spin, also the vertical fluctuation needs to be considered \cite{BELL1987488}. Such a degree of freedom has been considered in an experimental proposal about an electron in a Penning trap \cite{Rogers:1988zz} as well.

There are proposals that suggest the use of ultraintense lasers \cite{PhysRevLett.83.256, Schutzhold}. The growing interest is mainly motivated by the ever-improving experimental set-ups, that allow to reach very high accelerations \cite{Brodin_2008}.

Besides electrons, uniformly accelerated protons have been considered as well. They can be seen as Unruh-DeWitt detectors due to acceleration-induced weak-interaction decay \cite{PhysRevD.56.953, PhysRevLett.87.151301, PhysRevD.67.065002} and photon emission \cite{PhysRevD.46.3450, Higuchi:1992we}.

Accelerated atomic detectors have also been considered in more recent works \cite{PhysRevLett.91.243004, PhysRevA.74.023807}. In these articles, the authors consider a two-level atom accelerated through a microwave cavity. The coupling between the atom and the electromagnetic field is non-adiabatic due to the sharp boundaries of the cavity. Such a non-adiabatic condition enhances the acceleration-induced radiation.

In the next section, we will consider accelerated atoms as Unruh-DeWitt detectors. However, differently from the existing literature, we will derive the description of such detectors from first principles by considering the nonrelativistic limit of a Dirac field in Rindler spacetime.

\section{Atomic detectors}\label{Atomic_detectors}

In Sec.~\ref{UnruhDeWitt_detectors}, we discussed the Unruh-DeWitt detectors. In particular, we considered the nonrelativistic [Sec.~\ref{Non_relativistic_models}] and relativistic [Sec.~\ref{Relativistic_model}] models proposed by Unruh \cite{PhysRevD.14.870} and DeWitt \cite{hawking1980general} in their original works. These particle detectors give a simplified representation of physical objects used in real life experiments.

In this section, we consider accelerated atoms as a particle detector to probe the Unruh effect. In literature, both the nonrelativistic \cite{PhysRevLett.91.243004, PhysRevA.74.023807, PhysRevLett.128.163603} and the relativistic \cite{PARENTANI1995227} Unruh-DeWitt model have been used to study these type of detectors. At variance with these works, however, we provide a model for accelerated atomic detectors based on the nonrelativistic limit of electron fields in curved spacetimes. In this way, contrary to Sec.~\ref{Non_relativistic_models}, we outline relativistic corrections and fundamental subtleties, including the results of Chap.~\ref{Framedependent_nonrelativistic_limit}, that have been overlooked in the nonrelativistic theory. Also, at variance with the relativistic two-scalar-field model of Sec.~\ref{Relativistic_model}, here, we provide a description for the particle detector by means of Dirac theory in Rindler spacetime for hydrogen-like atoms.

As already remarked in Sec.~\ref{Introduction_Accelerated_non_relativistic_detectors}, the nonrelativistic limit represents nowadays the only regime for a fully satisfactory description of hydrogen-like atoms. Indeed, we still lack of an explicit energy spectrum for the fundamental quantum fields forming the atomic system. More generally, the fully-relativistic theory of quantum fields is not very suited for the description of bound states. For this reason, we consider the nonrelativistic limit of relativistic hydrogen-like atoms in the accelerating comoving frame.

The framework for the nonrelativistic limit of Dirac fields can be found in Sec.~\ref{Nonrelativistic_limit_Dirac_Rindler}. This method allows for a comprehensive study of the nonrelativistic regime for hydrogen-like atoms while including fundamental subtleties that come from the relativistic theory, such as the frame-dependence of the nonrelativistic condition and the particles number that has been outlined in Chap.~\ref{Framedependent_nonrelativistic_limit}.

Despite the atom is initially prepared in the inertial laboratory frame as a nonrelativistic bound state with a fixed number of electrons and nuclear particles, in the comoving accelerated frame, the energy and the number of the quantum particles are different. The single electron appears as a superposition of states with varying energy and particles number and the electronic and nuclear structure is radically modified. The frame-dependent nature of particles---at the origin of the Unruh effect---not only alters the background electromagnetic vacuum but also the electron and nuclear fields.

Such a frame-dependence poses limits to adopting the familiar first-quantized description of the hydrogen-like atom in its proper frame and brings up difficulties in understanding light-matter interaction with noninertial observers. Lowering the acceleration suppresses the effect on the electrons and the other nuclear particles [Sec.~\ref{Quasi_inertial_regime}]; however, this may also suppress the Unruh background electromagnetic vacuum, with the consequent decrease in the temperature [Eq.~(\ref{tempterature_Unruh})]. For a non-vanishing measurement of the Unruh effect, one needs an energy gap $\Delta E$ such that $\Delta E \lesssim k_{B} T_\text{U}$. Hence, the atomic spectrum must have a sufficiently fine structure to absorb the low-energy Unruh thermal photons. Is it possible to suppress the frame-dependent effect on the electron while still detecting the electromagnetic thermal background?

We give a positive answer to the previous question by a rigorous analysis based on QFTCS. Notwithstanding the suppressed frame-dependent effect for electrons, the hyperfine splitting provides the energy gap to reveal the Unruh radiation. We identify a specific parameter region in terms of the nuclear charge number $\mathcal{Z}$ and the electric field $E$ for the detection via first-quantized atomic detectors.

The section is organized as follows. In Sec.~\ref{Electron_field_in_inertial_and_accelerated_frame}, we use QFTCS to describe free electrons in inertial and accelerated frames; by using the results of Chap.~\ref{Framedependent_nonrelativistic_limit}, we show the frame-dependent particle content of the field and how to suppress the effect. In Sec~\ref{Avoiding_complete_ionization}, we discuss the atomic stability by studying the interaction with the accelerating field and the nuclear electric field. In Sec.~\ref{Detecting_Unruh_radiation_via_Hyperfine_splitting}, we investigate the physical regimes to detect Unruh radiation and we show that the relativistic hyperfine splitting is responsible for the coupling between the atom and the Unruh radiation.

\subsection{Electron field in inertial and accelerated frame}\label{Electron_field_in_inertial_and_accelerated_frame}

We assume that the atom is ionized with $1$ electron and $\mathcal{Z}>1$ protons and that the electron is prepared in the laboratory frame as a nonrelativistic particle. We consider a uniform electric field $\vec{E} = E \vec{e}_3$, with $E>0$, that produces an acceleration $\alpha$ along $\vec{e}_3 = (0,0,1)$ such that $\alpha = (\mathcal{Z}-1) e E / M$, with $e$ the elementary charge and $M$ the atomic mass.  Both the nucleus and electric field $\vec{E}$ are treated classically, whereas the electron is treated via quantum field theory of Dirac fields.

In Secs.~\ref{QFT_in_Minkowski_spacetime_Dirac} and \ref{Rindler_Dirac_modes}, we studied the Dirac field in Minkowski and Rindler spacetimes. Here, we adopt the same formalism to describe the electron field in the inertial and the accelerated frame, respectively. By using the interaction picture, we separate the free field theory from the interaction Lagrangian. In the inertial frame, the decomposition of the free electron field $\hat{\psi}(t,\vec{x})$ in free modes is given by Eqs.~(\ref{free_Dirac_field}), (\ref{free_Dirac_field_modes}) and (\ref{Dirac_uv_tilde}); whereas, in the accelerated frame the decomposition of the field $\Psi_\nu(T,\vec{X})$ is given by Eqs.~(\ref{free_Dirac_field_Rindler}), (\ref{UV_UV_tilde_conclusion}) and (\ref{W_tilde_W_tilde}).

In Sec.~\ref{Frame_dependent_content_of_particles_Dirac}, we derived the frame-dependent particle content of the Dirac field. In particular, we computed the Bogoliubov transformation (\ref{Bogoliubov_transformations_3_Rindler_4}) relating Minkowski to Rindler operators. The representation of the single Minkowski-Dirac particle $| \psi \rangle$ in the Rindler-Fock space $\mathcal{H}_{\text{L},\text{R}}$ was reported in Eq.~(\ref{single_particle_Minkowski_Rindler_Dirac}), with $\hat{C}_\psi$ and $ \hat{S}_\text{D}$ respectively define by Eqs.~(\ref{C_psi_Dirac}) and (\ref{OO}). We can use this result to describe the single electron as seen by an accelerated observer comoving with the atom.

In Sec.~\ref{Inertial_and_non_inertial_frame}, we showed the frame dependent nature of the particles. By using Eq.~(\ref{single_particle_Minkowski_Rindler_Dirac}), one can see that any nonrelativistic single particle prepared in the inertial frame appears as a superposition of states with varying energy and particle number in the accelerated frame. This can be applied to the case of noninertial hydrogen-like atom prepared in the laboratory frame by an inertial experimenter. The frame-dependent particle content of the electron field is responsible for the appearance of electron states with varying energy and particle number in the accelerated frame.

In Sec.~\ref{Quasi_inertial_regime} we showed that the frame-dependent effect is suppressed when the acceleration $\alpha$ is sufficiently low and the particle state is localized in the approximately Minkowskian region of the Rindler spacetime (i.e., where $g_{\mu\nu} \approx \eta_{\mu\nu}$). Any nonrelativistic Minkowski single particle appears as a nonrelativistic Rindler particle in the accelerated frame if $\alpha$ satisfies Eq.~(\ref{quasi_inertial_limit_a}) and the localization in $\vec{x}$ in such that
\begin{equation}\label{local}
|a z - 1| \lesssim \epsilon,
\end{equation}
where $\epsilon=\hbar \Omega/mc^2$ is the nonrelativistic parameter defined as the ratio between the nonrelativistic energy $\hbar \Omega$ and the mass energy $mc^2$. The resulting Rindler single particle is created over the Unruh background $|0_\text{M} \rangle$, which is in a superposition of Rindler particles. These background particles are mostly localized far from the region (\ref{local}) and close to the Rindler horizon. Hence, they can be ignored for the local detection of the Unruh effect.

\subsection{Avoiding complete ionization}\label{Avoiding_complete_ionization}

In this subsection, we discuss the interaction between the electron and the classic electromagnetic field. We obtain the conditions under which the atom is not completely ionized by the electric field $\vec{E}$. We show that such conditions not only guarantee the atomic stability, but also suppress the frame-dependent effect described in the previous subsections.

The classic electromagnetic field acting on the electron comes from the potential energy $V_\text{ext}$ due to the external electric field $\vec{E}$ and the potential energy $V_\text{nuc}$ due to the nuclear Coulomb interaction. In the comoving frame, $V_\text{ext}$ is
\begin{equation}\label{V_ext}
V_\text{ext}(Z) = \frac{1-e^{-2 a Z}}{2} \frac{e E}{a}.
\end{equation}
The nuclear potential energy $V_\text{nuc}$ is
\begin{equation}
V_\text{nuc}(R) = - \epsilon_\text{QED}^{1/2} \frac{\hbar c}{R},
\end{equation}
where $R=|\vec{X}|$ is the radial coordinate, $\epsilon_\text{QED} = (\mathcal{Z} \alpha_0)^2$ is the quantum electrodynamics (QED) coupling and $\alpha_0$ the fine-structure constant.

Equation (\ref{V_ext}) can be proved as follows. The electromagnetic tensor associated to the electric field $\vec{E}$ in the inertial frame is
\begin{equation}\label{em_tensor}
F^{\mu \nu}=\frac{E}{c} \begin{pmatrix}
& 0 & 0 & 0 & 1 \\
& 0 & 0 & 0 & 0\\
& 0 & 0 & 0 & 0\\
& -1 & 0 & 0 & 0
\end{pmatrix}.
\end{equation}
The Jacobian matrix associated to the coordinate transformation (\ref{Rindler_coordinates_transformation_R}) is
\begin{equation}\label{Jacobian}
\frac{\partial X^\mu}{\partial x^\nu} = \begin{pmatrix}
& e^{- a Z} \cosh(caT) & 0 & 0 & c^{-1} e^{- a Z} \sinh(caT)\\
& 0 & 1 & 0 & 0\\
& 0 & 0 & 1 & 0\\
& c e^{- a Z} \sinh(caT) & 0 & 0 & e^{- a Z} \cosh(caT)
\end{pmatrix}.
\end{equation}
Both Eqs.~(\ref{em_tensor}) and (\ref{Jacobian}) can be used to compute the components of the electromagnetic tensor in the noninertial frame $(T,\vec{X})$ as
\begin{equation}\label{em_tensor_Rindler}
\frac{\partial X^\mu}{\partial x^\alpha} \frac{\partial X^\nu}{\partial x^\beta} F^{\alpha \beta} = e^{- 2 a Z} \frac{E}{c} \begin{pmatrix}
& 0 & 0 & 0 & 1 \\
& 0 & 0 & 0 & 0\\
& 0 & 0 & 0 & 0\\
& -1 & 0 & 0 & 0
\end{pmatrix}.
\end{equation}
Equation (\ref{em_tensor_Rindler}) states that in the accelerated frame, $\vec{E}$ appears as an electric field along $Z$ with magnitude $e^{- 2 a Z} E$. The consequent potential energy $V_\text{ext}$ that is vanishing for $Z = 0$ is Eq.~(\ref{V_ext}).

The electron is pulled away from its orbit by $V_\text{ext}$ while it is dragged by the accelerating nucleus via $V_\text{nuc}$. If $E$ is sufficiently large, the electron escapes from the nuclear Coulomb barrier via quantum tunneling, compromising the atomic stability. To avoid complete ionization, we require a small $E$ such that
\begin{equation}\label{E_sufficiently_small}
|V_\text{ext}(R_0)| \ll |E^{(0)}_0|,
\end{equation}
with $E^{(0)}_0 = -\epsilon_\text{QED} \mu c^2/2 $ as the ground state of $V_\text{nuc}$, $\mu=(m + M_\text{N})/m M_\text{N} \approx m$ as the reduced mass, $M_\text{N} \approx M$ as the nuclear mass, $R_0=a_0/\mathcal{Z}$ as the atomic radius and $a_0 = \hbar/m c \alpha_0$ as the Bohr radius. Hence, we assume that the external force $V_\text{ext}$ simply perturbs the spectrum of $V_\text{nuc}$ via Stark effect.

Equation (\ref{E_sufficiently_small}) reads as
\begin{equation}\label{a_sufficiently_small_local}
a R_0 \ll \epsilon_\text{QED} \frac{(\mathcal{Z}-1)m}{2 M},
\end{equation}
or, equivalently,
\begin{equation} \label{Z_E_constraint_1_3}
E \ll \frac{(\mathcal{Z} \alpha_0)^3}{2} \frac{m^2 c^3}{\hbar e}.
\end{equation}
Notice that the electron is localized inside the region $R\lesssim R_0$ since $V_\text{nuc}$ dominates over $V_\text{ext}$. Notice also that $\epsilon_\text{QED} \ll 1$ and $(\mathcal{Z}-1) m \ll M$. Hence, from Eqs.~(\ref{V_ext}) and (\ref{a_sufficiently_small_local}), one concludes that the electron is localized where the electric field is approximately uniform.

Notice that $\epsilon_\text{QED} mc^2$ is the order of the nonrelativistic atomic energies. This is a consequence of the fact that the spectrum of $V_\text{nuc}$ is
\begin{equation}\label{atom_spectrum}
E^{(0)}_n = - \frac{\epsilon_\text{QED}}{2 (n+1)^2} \mu c^2.
\end{equation}
By comparing Eq.~(\ref{a_sufficiently_small_local}) with Eq.~(\ref{local}), one finds out that the localization condition (\ref{local}) is already met by configurations that satisfy Eq.~(\ref{a_sufficiently_small_local}). Furthermore, Eq.~(\ref{a_sufficiently_small_local}) leads to
\begin{equation}\label{a_sufficiently_small}
\frac{\hbar a}{mc} \ll \epsilon^{3/2}_\text{QED} \frac{(\mathcal{Z}-1)m}{2M},
\end{equation}
which is a sufficient condition for Eq.~(\ref{quasi_inertial_limit_a}). By constraining $E$ and $\mathcal{Z}$ accordingly to Eq.~(\ref{Z_E_constraint_1_3}), one guarantees the atom stability and the first-quantization electron description in the accelerated frame. The atom does not ionize and the electron appears as a nonrelativistic single particle in both frames.

Equation (\ref{E_sufficiently_small}) guarantees a lifetime $\tau$ for the atom that is exponentially increasing for decreasing electric field. Indeed, by using the WKB approximation, one can find the following ionization rate \cite{landau1991quantum}
\begin{equation}\label{tau}
\frac{1}{\tau} \approx  \frac{16}{\hbar R_0 e E}  \left( E^{(0)}_0 \right)^2 \exp \left( \frac{4 E^{(0)}_0}{3 R_0 e E} \right).
\end{equation}

\subsection{Detecting Unruh radiation via Hyperfine splitting}\label{Detecting_Unruh_radiation_via_Hyperfine_splitting}

In this subsection, we consider the interaction between the accelerated atom and the electromagnetic Unruh background. We show the conditions under which the coupling between electron and Unruh radiation produce measurable effects. We study the spectrum of the relativistic hydrogen-like atom in Rindler spacetime with uniform external electric field and we show that the coupling is induced by the hyperfine splitting. Finally, we plot the regime of parameters for the observability of the Unruh effect.

In the accelerated frame, as a consequence of the Unruh effect, the electromagnetic background appears as a thermal bath [Eq.~(\ref{thermal})], with temperature given by Eq.~(\ref{tempterature_Unruh}). The electron can be excited by absorbing a photon with the energy $\Delta E_n = E_n - E_0$ of the $n$-th electronic transition. The event is detectable if
\begin{equation}\label{Delta_E_n_order_a}
\Delta E_n \lesssim k_B T_\text{U},
\end{equation}
and if the atom has a sufficiently large lifetime such that
\begin{equation}\label{upper_limit_tau}
\tau \gtrsim \frac{\hbar}{\Delta E_n}.
\end{equation}

Equation (\ref{Delta_E_n_order_a}) guarantees a nonvanishing probability for the electron to interact with photons described by the following Boltzmann distribution
\begin{equation}\label{Boltzmann_distribution}
P_\text{B} = \frac{1}{e^{\Delta E_n/k_\text{B} T_\text{U}} - 1}.
\end{equation}
A more refined constraint than Eq.~(\ref{Delta_E_n_order_a}) can be imposed by assuming a lower bound for $P_\text{B}$, i.e.,
\begin{equation}\label{lower_limit}
P_\text{B} < P_\text{min},
\end{equation}
with $P_\text{min}<1$. Equation (\ref{upper_limit_tau}), instead, ensures that the absorption spectrum of the atom is narrow around $\Delta E_n$. Given the exponential growth of the atom lifetime for smaller $E$ [Eq.~(\ref{tau})], it is safe to assume that Eq.~(\ref{upper_limit_tau}) gives an almost exact lower limit for $\tau$, i.e.,
\begin{equation}\label{upper_limit_tau_2}
\tau > \frac{\hbar}{\Delta E_n}.
\end{equation}

The states and energies of the spectrum $E_n$ are the solutions of the Dirac equation in Rindler spacetime for hydrogen-like atoms with the interaction potential $V_\text{ext}$. They can be computed perturbatively by considering the nonrelativistic hydrogen-like spectrum $E^{(0)}_n$ [Eq.~(\ref{atom_spectrum})] perturbed by $V_\text{ext}$ and by relativistic corrections coming from the Rindler-Dirac equation.

The energies gaps of the unperturbed Hamiltonian $\Delta E^{(0)}_n = E^{(0)}_n - E^{(0)}_0$ in Eq.~(\ref{atom_spectrum}) are of the order
\begin{equation}\label{Delta_E_0_n_order}
\Delta E^{(0)}_n \sim \epsilon_\text{QED} mc^2.
\end{equation}
By plugging Eq.~(\ref{Delta_E_0_n_order}) in Eq.~(\ref{Delta_E_n_order_a}), one finds that the lower bound for the electric field is
\begin{equation}
E \gtrsim \frac{2\pi (\mathcal{Z} \alpha_0)^2}{\mathcal{Z}-1} \frac{m M c^3}{ \hbar e},
\end{equation}
which is way larger than the upper bound (\ref{Z_E_constraint_1_3}). Hence, $\Delta E^{(0)}_n$ does not induce coupling with the electromagnetic background for any stable configuration.

Perturbations of $E^{(0)}_n$ do not significantly change the energies gaps, unless they break the spin degeneracy of the atomic ground state. In that case, the first level $E^{(0)}_0$ splits into the actual ground state $E_0$ and the first excited state $E_1$, with $\Delta E = E_1 - E_0 \ll \epsilon_\text{QED} mc^2$.

Provably, the Rindler-Dirac equation for the hydrogen-like atom with potentials $V_\text{nuc}$ and $V_\text{ext}$ have a degenerate minimum energy level. Hence, the external electron field $V_\text{ext}$ and the special and general relativity corrections do not break the spin degeneracy of $E^{(0)}_0$.

To see this, consider the full Rindler-Dirac equation with potentials $V_\text{nuc}$ and $V_\text{ext}$ describing the electron in the accelerated frame, i.e.,
\begin{equation} \label{Dirac_perturbed}
i \hbar \partial_0  \Psi  = H   \Psi,
\end{equation}
with the following Hamiltonian
\begin{align}
H = & - i \hbar c^2 \gamma^0 \gamma^3 \partial_3 - \frac{i}{2} \hbar \alpha \gamma^0 \gamma^3  + e^{aZ}\gamma^0(- i \hbar c^2  \gamma^1 \partial_1 - i \hbar c^2 \gamma^2 \partial_2 + m c^3 ) \nonumber \\
& + V_\text{nuc} + V_\text{ext}.
\end{align}
Notice that $H$ is symmetric with respect to the following unitary operators
\begin{align}
& U_1 = i c \gamma^0 P_1 \gamma^2 \gamma^3, & U_2 = i c \gamma^0 \gamma^1 P_2 \gamma^3,
\end{align}
where $P_i$ is the parity operator for the $i$-th coordinate, i.e., $P_1: X \mapsto -X$ and $P_2: Y \mapsto -Y$. Indeed, one can prove that $H$ commutes with $U_1$ and $U_2$, i.e.,
\begin{align}\label{U_H_commutation}
& [U_1, H] = 0, & [U_2, H] = 0,
\end{align}
by using the following anticommutative properties
\begin{subequations}\label{P_gamma_identities}
\begin{align}
& \{ P_1, \partial_1 \} = 0, & \{ P_2, \partial_2 \} = 0,\label{P_identities} \\ 
& \{ \gamma^\mu, \gamma^\nu \} = - 2 \eta^{\mu\nu}.\label{gamma_identities}
\end{align}
\end{subequations}
Equation (\ref{gamma_identities}), the hermiticity of $\gamma^0$, $P_1$ and $P_2$ and the antihermiticity of $\gamma^i$ can be used to prove the unitarity of $U_1$ and $U_2$. Furthermore, Eq.~(\ref{gamma_identities}) leads to the following anticommutative relation
\begin{equation}\label{U_anticommutation}
\{ U_1, U_2 \} = 0.
\end{equation}
Equation (\ref{U_H_commutation}) implies that $H$ and $U_1$ are simultaneously diagonalizable. The same occurs for $H$ and $U_2$. However, $U_1$ and $U_2$ are not compatible, since, as a consequence of Eq.~(\ref{U_anticommutation}), they do not commute.

Consider a state $\Psi$ which is simultaneously eigenstate of $H$ and $U_2$. The non-compatibility between $U_1$ and $U_2$ implies that $\Psi$ is not eigenstate of $U_1$. Hence, $U_1 \Psi$ is a different state from $\Psi$. However, $U_1 \Psi$ is still eigenstate of $H$ with the same energy of $\Psi$. We find that the Hamiltonian $H$ is at least two-degenerate for each energy level. This proves that the spin degeneracy of the first energy level is not lifted by the energy potential $V_\text{ext}$ nor by the special and general relativity corrections to the accelerated hydrogen-like atom.

One has to look at the hyperfine structure to see a split of $E^{(0)}_0$ due to quantum electrodynamics corrections. The electron-nucleus interaction via spin-spin coupling generates the following energy gap \cite{bethe2013quantum}
\begin{equation}\label{Delta_E_hf}
\Delta E_\text{hf} = \begin{cases}
\frac{1}{3 \pi} (2I + 1) Z^3 \alpha_0^4 g \frac{m^2c^2}{M_\text{P}} & \text{if } I \neq 0 \\
0 & \text{if } I = 0
\end{cases},
\end{equation}
with $M_\text{P}$ as the proton mass. $I$ is the quantum number such that $|\vec{I}|^2 = I(I+1)$, where $\vec{I}$ is the nucleus spin. $g$ is the effective $g$-factor defined as follows: $\vec{\mu} = (g \hbar e/2M_\text{P}) \vec{I}$, where $\vec{\mu}$ is the magnetic moment of the nucleus resulting from its spin.

Notice that the selection rule that forbids transitions between levels with vanishing azimuthal quantum number $\ell = 0$ breaks down due to the Stark effect. Hence, the absorption of photons coupled to the hyperfine structure is allowed.

By plugging Eq.~(\ref{Delta_E_hf}) in Eq.~(\ref{lower_limit}), one finds that the atomic hyperfine structure produces a measurable Boltzmann distribution when $I \neq 0$ and when
\begin{equation}\label{hf_measurable_Unruh}
\left[ \exp \left( \frac{2 (2I + 1) \mathcal{Z}^3 \alpha_0^4 g}{3(\mathcal{Z}-1)} \frac{M m^2 c^3}{M_\text{P} \hbar e E} \right) - 1 \right]^{-1} < P_\text{min} .
\end{equation}
Furthermore, the atom has a sufficiently long lifetime when [Eqs.~(\ref{upper_limit_tau_2}) and (\ref{Delta_E_hf})]
\begin{equation}\label{upper_limit_hf}
\frac{ \hbar e E}{m^2 c^3} \exp \left( \frac{2 (\mathcal{Z} \alpha_0)^3 }{3} \frac{ m^2 c^3}{\hbar e E} \right) > \frac{12 \pi \mathcal{Z}^2 \alpha_0}{(2I + 1) g} \frac{M_\text{P}}{m} .
\end{equation}

\begin{figure}
\center
\includegraphics{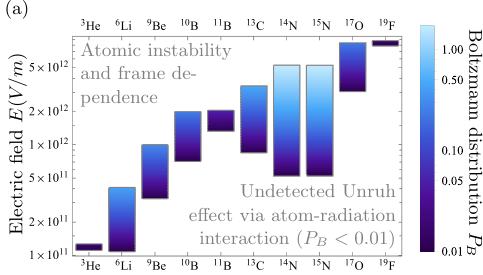}
\includegraphics{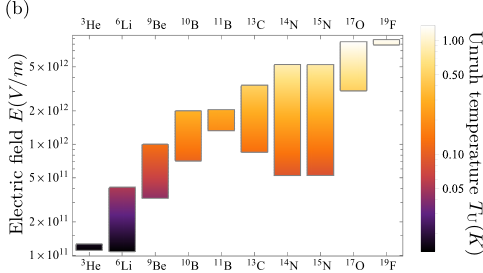}
\caption{Observation window for the Unruh effect via first-quantized atomic detectors. The constrained variable is the accelerating electric field $E$ for each nuclear configuration $\mathcal{Z}$, $M$, $I$, $g$ \cite{STONE200575}. The upper limit for $E$ [Eq.~(\ref{upper_limit_hf})] guarantees the stability of the atomic bound state and the first-quantized description of the electron in the accelerated frame. Above this limit, the electron escapes from the Coulomb potential via tunneling and it appears as a superposition of states with different energies and number of particles in the accelerated frame. The lower limit [Eq.~(\ref{hf_measurable_Unruh})], instead, ensures the detection of the Unruh effect from the electromagnetic thermal background via light-matter interaction. Below such a limit, the hyperfine structure of the atom produces an energies gap that is too large for the Boltzmann distribution to be detected. In (a) we show the Boltzmann distribution for the atom-radiation interaction [Eq.~(\ref{Boltzmann_distribution_hp})]. In (b), we show the Unruh temperature $T_\text{U}$ of the electromagnetic background in the accelerated frame for different configurations [Eq.~(\ref{T_E_Z})].} \label{Fig1}
\end{figure}

Equations (\ref{hf_measurable_Unruh}) and (\ref{upper_limit_hf}) define the regime of parameters $E$, $\mathcal{Z}$, $M$, $I$, $g$ for the detection of the Unruh effect via first-quantized atomic detectors. The results are shown in Fig.~\ref{Fig1}, where for some nuclear configurations we plot the range of validity for the electric field $E$. In Fig.~\ref{Fig1}, we also plot the Boltzmann distribution
\begin{equation}\label{Boltzmann_distribution_hp}
P_\text{B} = \left[ \exp \left( \frac{(2I + 1) \mathcal{Z}^3 \alpha_0^4 g}{3(\mathcal{Z}-1)} \frac{M m^2 c^3}{M_\text{P} \hbar e E} \right) - 1 \right]^{-1}
\end{equation}
and the Unruh temperature
\begin{equation}\label{T_E_Z}
 T_\text{U} = \frac{\mathcal{Z} - 1 }{2 \pi } \frac{ \hbar e E }{k_\text{B} M c}.
\end{equation}
To overcome background noises, a sufficiently large $T_\text{U}$ is needed and requires high ionization $\mathcal{Z}$.

\section{Conclusions}\label{Accelerated_non_relativistic_detectors_Conclusions}

Various experimental proposals have been reported to test the Unruh effect. The proposals include the depolarization of electrons in storage rings \cite{BELL1983131, BELL1987488}, Penning traps \cite{Rogers:1988zz}, ultraintense lasers \cite{PhysRevLett.83.256, Schutzhold}. The growing interest is motivated by the ever-improving experimental equipment that allows to reach high accelerations \cite{Brodin_2008}. Besides electrons, also uniformly accelerated protons have been considered as Unruh-DeWitt detectors via acceleration-induced weak-interaction decay \cite{PhysRevD.56.953, PhysRevLett.87.151301, PhysRevD.67.065002} and photon emission \cite{PhysRevD.46.3450, Higuchi:1992we}.

In this chapter, we analyzed the electron in the accelerated atom by the nonrelativistic limit of a Dirac field in Rindler spacetime. While considering hyperfine splitting, we addressed three problems: (i) the instability of the atom due to a strong accelerating field; (ii) the frame-dependent nature of the electron; (iii) the detectability of the Unruh effect due to the electromagnetic radiation.
We have shown that (i) and (ii) impose an upper boundary condition for the electric field accelerating the ionized hydrogen-like atom. Taking into account (iii), we determined an observation window in the $E$ and $\mathcal{Z}$ plane [Fig.~\ref{Fig1}]. Surprisingly, quantitative estimates unveil that the effect can be detected with electric fields within the reach of the modern technologies of high-power lasers and nuclear magnetic resonance.

As a concluding remark, we discuss the possibility to extend our results by including the Zeeman effect. The presence of a uniform magnetic field $\vec{B}$ parallel to the accelerating electric field $\vec{E}$ has the following effects: (i) it does not affect the accelerating trajectory of the atom; (ii) it induces an energy splitting at the ground states similarly to the hyperfine splitting. Hence, for atoms that have vanishing hyperfine splitting (e.g., nuclei with zero spin), only the Zeeman effect produces a measurable energy gap that couples to the Unruh radiation. We remark that, at variance with the hyperfine splitting, the strength of the magnetic field $\vec{B}$ can be controlled and induces an arbitrarily small energy gap. Consequently no lower bound for the electric field $\vec{E}$ occurs.

\part{Localization}\label{Localization}

\chapter{Localization in inertial frame}\label{Localization_in_Quantum_Field_Theory}

\textit{This chapter is based on and contains material from Ref.~\citeRF{localization_QFT}.}

\section{Introduction}

Much work has been done on the problem of localization in QFT. The difficulties are explainable by means of some no-go theorems. Notably, the Hegerfeldt theorem \cite{PhysRevD.10.3320} forbids any notion of localization that assumes causal propagation of wave functions and orthogonality condition between states in disjoint regions of space. The Reeh-Schlieder theorem \cite{Reeh:1961ujh}, instead, shows that the causal and local structure of fields does not guarantee the independence between the preparation of states and the measurement of observables in disjoint regions.

In NRQM, the notion of localization is notoriously given in terms of wave functions and position operator and follows Born's interpretation of quantum mechanics. States are localized in the support of their wave functions, whereas second-quantized observables are localized in $\vec{x}$ if they are generated by creators and annihilators of particles in $\vec{x}$. Also, states are orthogonal if the supports of their wave functions are disjoint and, hence, if they are localized in different regions. In this case, the Hegerfeldt theorem does not lead to contradicting results, as the relativistic postulate of causality is not needed in NRQM.

This notion of localization was then extended to QFT by Newton, Wigner and Fulling \cite{RevModPhys.21.400, fulling_1989}. The so-called Newton-Wigner localization is based on the orthogonality condition between states in disjoint regions and other natural requirements that make it conceptually equivalent to the Born scheme. At variance with Born, however, Newton and Wigner worked in the context of relativistic theories. Hence, the superluminal propagation of wave functions predicted by Hegerfeldt gives unsatisfactory results and leads to the idea that the Newton-Wigner scheme is not suited for a genuine description of local phenomena in QFT. 

A fundamental notion of localization is, instead, provided by the AQFT formalism. As already detailed in Sec.~\ref{Algebraic_approach_in_QFT_and_QFTCS}, the algebraic approach to QFT is based on the definition of local algebras by means of quantum fields in spacetime points. This gives a natural definition of local observables and local preparation of states. 

At variance with the Newton-Wigner scheme, the AQFT framework provides a genuine notion of localization and it faithfully describes local experiments in isolated laboratories. The main argument is that the instantaneous propagation of Newton-Wigner wave functions is in conflict with relativity. Conversely, in AQFT, the microcausality condition appears as an axiom of the theory and is expressed by the commutativity of quantum fields in spacelike separated points.

Unfortunately, the microcausality condition does not guarantee the independence between the preparation of states and the measurement of observables in spacelike separated regions. The Reeh-Schlieder theorem demonstrates that states that are localized with respect to the AQFT scheme are not necessarily strictly localized \cite{10.1063/1.1703731, 10.1063/1.1703925}. Explicitly, this means that the outcome of measurements in a region $\mathcal{O}_\text{B}$ may depend on the local preparation of states in $\mathcal{O}_\text{A}$ even if $\mathcal{O}_\text{A}$ and $\mathcal{O}_\text{B}$ are causally disconnected.  However, it has been argued that such a nonlocal effect does not violate causality since it only comes from selective nonunitary preparations of states \cite{Redhead1995-REDMAA-2, CLIFTON20011, VALENTE2014147, RevModPhys.90.045003}.

At variance with the AQFT, the Newton-Wigner scheme is not affected by the Reeh-Schlieder nonlocality. In particular, any state localized with respect to the Newton-Wigner scheme is also strictly localized and does not affect measurements in any other disjoint space regions. In this case, we say that the strict localization property is always satisfied.

Also, as a consequence of a corollary of the Reeh-Schlieder theorem \cite{Redhead1995-REDMAA-2, RevModPhys.90.045003}, no local creator and annihilator operator can be defined in the AQFT formalism. This means that, contrary to the Newton-Wigner scheme, the AQFT scheme is not characterized by local Fock spaces and local vacua.

All the incompatibilities between the two localization schemes disappear in the nonrelativistic limit. In particular, it has been proven that any operator that is localized in $\vec{x}$ with respect to the Newton-Wigner scheme approximates to an operator localized in $\vec{x}$ with respect to the AQFT scheme \cite{PhysRevD.90.065032, Papageorgiou_2019}. This result is in agreement with the fact that the Born-Newton-Wigner and the AQFT schemes are suited for the description of phenomena in, respectively, the nonrelativistic and the relativistic regime.

As a consequence of the convergence to the AQFT, the Newton-Wigner scheme acquires a fundamental notion of localization in the nonrelativistic limit. Furthermore, the Reeh-Schlieder nonlocal effect is suppressed and any state locally prepared in a space region $\mathcal{V}_1$ is also strictly localized in $\mathcal{V}_1$, even if the preparation is a selective nonunitary operation.

In addition to the Newton-Wigner and the AQFT scheme, we study the modal localization scheme, which was already introduced in Sec.~\ref{QFT_in_Minkowski_spacetime_Dirac} by means of the modal representation of single particle states. In Parts \ref{Relativistic_and_nonrelativistic_quantum_fields} and \ref{Inequivalent_particle_representations_and_Unruh_effect} of the thesis, we used the definition of Minkowski particles as positive frequency solutions of the field equation and we implicitly assumed that these states are localized in the support of the corresponding modes. This is generally inaccurate, since a genuine notion of localization is only given by the AQFT scheme which is not compatible with the modal scheme. However, here, we show that in the nonrelativistic limit the two localization schemes converge. This implies that the modal scheme is actually suited for a genuine description of localized states but only in the nonrelativistic regime.

This chapter is intended to be an introductory review on the problem of localization in QFT. In particular, we consider a Minkowski spacetime and we study the Newton-Wigner, the AQFT and the modal schemes. We compare their features and show their convergence in the nonrelativistic limit. We only focus on real scalar fields, as the only important elements of the theory are captured by quantum fields without internal degrees of freedom.

The chapter is organized as follows. In Secs.~\ref{NewtonWigner_localization_scheme}, \ref{AQFT_localization_scheme} and \ref{Modal_localization_scheme} we define the Newton-Wigner, the AQFT and the modal schemes, respectively. Their features are then shown and compared to each other in Sec.~\ref{Comparison_between_localization_schemes}. In Sec.~\ref{Localization_in_NRQM}, we study the localization in NRQM; in particular, we define the Born scheme and we show the convergence of all localization schemes in the nonrelativistic limit. Conclusions are drawn in Sec.~\ref{Localization_in_Quantum_Field_Theory_Conclusions}.

\section{Newton-Wigner scheme}\label{NewtonWigner_localization_scheme}

In their paper \cite{RevModPhys.21.400}, Newton and Wigner addressed the problem of localization of particles in Relativistic Quantum Mechanics (RQM) by deriving the position observable and its eigenstates from first principles. They showed that the definition of localization is uniquely determined by some natural requirements. They assumed the following general theoretic postulates on the basis of which a particle can be considered localized at time $t=0$ in $\vec{x}$: (i) the superposition of localized states is localized as well; (ii) the set of localized states in $\vec{x}$ is invariant under rotations and time and space reflections with $\vec{x}$ as a fixed point; (iii) states localized in different spatial positions $\vec{x} \neq \vec{x}'$ are orthogonal; (iv) some regularity conditions of mathematical good behavior. From these assumptions, Newton and Wigner derived the definition of a unique position operator $\hat{\vec{x}}_\text{NW}$ and localized states $|  \vec{x}_\text{NW} \rangle$. The operator was then second quantized by Fulling \cite{fulling_1989}, who reformulated the theory in the context of QFT.

Newton and Wigner started from the representation of the spinless elementary particles (i.e., Klein-Gordon single particles) via irreducible representation of the Poincare group (i.e., energy, momentum and angular momentum). Then, they studied the case of particles with spin and finite mass. The uniqueness of the position operator satisfying the natural transformation conditions in RQM with arbitrary spin was later discussed in \cite{Weidlich_Mitra} and led to the same conclusions as Newton and Wigner.

The Newton-Wigner scheme in QFT predicts a phenomenon of superluminal spreading \cite{PhysRev.139.B963} that is in contrast with the relativistic notion of causality. This is a consequence of the Hegerfeldt theorem \cite{PhysRevD.10.3320, PhysRevD.22.377}, whose only hypotheses are the positivity of the energy of relativistic particles and the orthogonality condition of states localized in disjoint regions. Due to the violation of causality, the Newton-Wigner scheme is not regarded as fundamental in nature. Conversely, in Sec.~\ref{AQFT_localization_scheme}, we will see that the notion of locality provided by AQFT does not lead to superluminal signaling and can be regarded as a genuine localization scheme in QFT.

The present section is organized as follows. In Sec.~\ref{NewtonWigner_scheme_in_RQM}, we briefly show the results of Newton and Wigner's work in the context of RQM. Then, in Sec.~\ref{NewtonWigner_scheme_in_QFT} we consider the second-quantized version of the position operator to define the Newton-Wigner scheme in QFT. Lastly, in Sec.~\ref{Hegerfeldt_theorem}, we review the literature about the Hegerfeldt theorem and its consequences on the problem of localization.

\subsection{Newton-Wigner scheme in RQM}\label{NewtonWigner_scheme_in_RQM}

Here, we give the definition of Newton-Wigner position operator and wave functions in the context of RQM.

States with defined momentum $| \vec{k} \rangle$ are defined as eigenstates of the momentum operator $\hat{\vec{k}}$, satisfying the following orthonormalization property
\begin{equation}\label{relativistic_orthonormalization}
\langle \vec{k} | \vec{k}' \rangle = \delta^3(\vec{k}-\vec{k}').
\end{equation}
Starting from the definition of $| \vec{k} \rangle$, Newton and Wigner derived the unique state satisfying conditions (i)-(iv) as 
\begin{equation}\label{x_state}
|  \vec{x}_\text{NW} \rangle = \int_{\mathbb{R}^3} d^3 k \frac{e^{-i \vec{k} \cdot \vec{x}}}{\sqrt{(2 \pi)^3}} | \vec{k} \rangle.
\end{equation}

Equation (\ref{x_state}) describes the state localized in $\vec{x}$ at time $t=0$ according to the Newton-Wigner scheme in RQM. This provides the definition of the position operator
\begin{equation}\label{x_NW}
\hat{\vec{x}}_\text{NW} = \int_{\mathbb{R}^3} d^3 x \vec{x} | \vec{x}_\text{NW} \rangle \langle \vec{x}_\text{NW} |,
\end{equation}
whose eigenstates are $|  \vec{x}_\text{NW} \rangle$ with eigenvalues $\vec{x}$. Also, for any state $| \phi \rangle$, Newton and Wigner defined the wave function in position space as
\begin{equation}\label{psi_NW}
\phi_\text{NW}(t,\vec{x}) = \langle \vec{x}_\text{NW} | \phi(t) \rangle = \int_{\mathbb{R}^3} d^3 k \frac{ e^{-i \omega(\vec{k}) t+i \vec{k} \cdot \vec{x}}}{\sqrt{(2 \pi)^3}} \tilde{\phi}(\vec{k}),
\end{equation}
with $\omega(\vec{k})$ as the frequency of $\vec{k}$ [Eq.~(\ref{omega_k})] and with $\tilde{\phi}(\vec{k})$ as the wave function in momentum space, defined as
\begin{equation}
\tilde{\phi}(\vec{k}) = \langle \vec{k} | \phi \rangle.
\end{equation}
The inner product between states can be written in terms of their wave function in position space as
\begin{equation}\label{psi_NW_product}
\langle \phi | \phi' \rangle = \int_{\mathbb{R}^3} d^3 x \phi_\text{NW}^*(t,\vec{x}) \phi'_\text{NW}(t,\vec{x}),
\end{equation}
which is the familiar $L^2(\mathbb{R}^3)$ scalar product.

Notice that in NRQM, wave functions in momentum space are related to wave functions in position space by means of the Fourier transform. The same occurs in RQM between the wave functions $\tilde{\phi}(\vec{k})$ and $\phi_\text{NW}(0,\vec{x})$ [Eq.~(\ref{psi_NW})]. This analogy leads to the equivalence between the Newton-Wigner and the Born localization schemes, which will be detailed in Sec.~\ref{Born_localization_is_NewtonWigner_localization}.

Newton and Wigner already pointed out in their original work \cite{RevModPhys.21.400} that the position operator $\hat{\vec{x}}_\text{NW}$ is not relativistically covariant. For any Lorentz boost $\Lambda_{\vec{v}}: (t,\vec{x}) \mapsto (t',\vec{x}')$, the state that is localized in (say) $\vec{x}=0$ at $t=0$ is not localized in $\vec{x}' = 0$ at $t'=0$. Hence, two inertial observers do not share the same notion of localization. This is an important argument against the Newton-Wigner localization program, since Lorentz transformed frames are physically equivalent in relativistic theories.

Furthermore, the Newton-Wigner localization is found not to be preserved in time. Specifically, a particle localized in a bounded region at $t=0$ will develop infinite tails at immediately later times $t \neq 0$, exceeding the light cone of the initial region \cite{PhysRev.139.B963}. The phenomenon of superluminal spreading of the wave functions was then proved to occur for a more general class of localization schemes. The only condition is a nonconstant Hamiltonian that is a semibounded function of the particle momentum (e.g., Klein-Gordon particles) \cite{PhysRevD.22.377}. This model-independent result goes under the name of Hegerfeldt theorem and will be discussed in Sec.~\ref{Hegerfeldt_theorem}.

The non-covariant behavior of the position operator $\hat{\vec{x}}_\text{NW}$ and the acausal spreading of the wave functions make the Newton-Wigner localization unsatisfactory for a fully relativistic theory. The solution to this problem will be found by noticing that the operator $\hat{\vec{x}}_\text{NW}$ does not entail any fundamental notion of locality; conversely, it is a mathematical artifice that comes from the nonrelativistic theory. Only in the nonrelativistic limit of RQM, the Newton-Wigner scheme obtains a genuine notion of locality. This result will be shown in Sec.~\ref{Comparison_with_the_relativistic_theory}.

\subsection{Newton-Wigner scheme in QFT}\label{NewtonWigner_scheme_in_QFT}

In Sec.~\ref{NewtonWigner_scheme_in_RQM}, we worked in the context of RQM and used the definition of first-quantized Newton-Wigner position operator $\hat{\vec{x}}_\text{NW}$ to define localized states and wave functions in position space. Here, we apply these results to the framework of QFT. In particular, we define localized states and observables by means of a second-quantized version of $\hat{\vec{x}}_\text{NW}$. The method is based on the natural embedding of RQM in QFT as the theory of single particle states of the corresponding quantum fields.

For any scalar field $\hat{\phi}(t,\vec{x})$, the corresponding single particle state with defined momentum $\vec{k}$ is defined as
\begin{equation}\label{single_particle_field_embedding}
| \vec{k} \rangle = \hat{a}^\dagger(\vec{k}) | 0_\text{M} \rangle ,
\end{equation}
where $\hat{a}^\dagger(\vec{k})$ is a creator operator satisfying the canonical commutation identities (\ref{Minkowski_canonical_commutation}). Single particle states with defined position are defined by Eq.~(\ref{x_state}). Owing to Eq.~(\ref{single_particle_field_embedding}), Eq.~(\ref{x_state}) is equivalent to
\begin{equation}
| \vec{x}_\text{NW} \rangle = \hat{a}_\text{NW}^\dagger(\vec{x}) | 0_\text{M} \rangle,
\end{equation}
where
\begin{equation}\label{a_NW}
\hat{a}_\text{NW}(\vec{x}) = \int_{\mathbb{R}^3} d^3 k \frac{e^{i \vec{k} \cdot \vec{x}}}{\sqrt{(2 \pi)^3 }} \hat{a}(\vec{k})
\end{equation}
is the inverse of the Fourier transform of the annihilation operator $\hat{a}(\vec{k})$.

Notice that there is a one-to-one mapping between the operators $\hat{a}_\text{NW}(\vec{x})$ and $\hat{a}(\vec{k})$ for varying $\vec{x}$ and $\vec{k}$. Hence, the entire Minkowski-Fock algebra $\mathfrak{A}_\text{M}$ is generated by $\hat{a}_\text{NW}(\vec{x})$ with varying $\vec{x}$, in the sense that any operator acting on the Minkowski-Fock space $\mathcal{H}_\text{M}$ can be written as a linear combination of products of $\hat{a}_\text{NW}(\vec{x})$ and $\hat{a}_\text{NW}^\dagger(\vec{x})$ operators.

Hereafter, the algebra generated by $\hat{a}_\text{NW}(\vec{x})$ with fixed $\vec{x}$ is denoted as $\mathfrak{A}_\text{M}^\text{NW}(\vec{x})$. We say that $\mathfrak{A}_\text{M}^\text{NW}(\vec{x})$ is a local algebra with respect to the Newton-Wigner scheme. Any element of $\mathfrak{A}_\text{M}^\text{NW}(\vec{x})$ is an operator that is localized in $\vec{x}$. Conversely, any local state $| \phi \rangle$ is the result of local operations on the vacuum background $| 0_\text{M} \rangle$. Hence, $| \phi \rangle$ is said to be localized in $\vec{x}$ if there is a local operator $\hat{O} \in \mathfrak{A}_\text{M}^\text{NW}(\vec{x})$ such that $| \phi \rangle = \hat{O} | 0_\text{M} \rangle$.

The definition of localized states and observables can also be generalized to extended regions. For any region $\mathcal{V} \subset \mathbb{R}^3$, we define $\mathfrak{A}_\text{M}^\text{NW}(\mathcal{V})$ as the local algebra in $\mathcal{V}$ generated by the operators $\hat{a}_\text{NW}(\vec{x})$ with $\vec{x} \in \mathcal{V}$. We say that the operator $\hat{O}$ and the state $| \phi \rangle = \hat{O} | 0_\text{M} \rangle$ are localized in $\mathcal{V}$ if $\hat{O}$ is an element of $\mathfrak{A}_\text{M}^\text{NW}(\mathcal{V})$.

The definitions of local operators and states provided here come from a second-quantized generalization of $\hat{\vec{x}}_\text{NW}$. Notice that, by embedding the relativistic theory of single particles, the Newton-Wigner scheme in QFT inherits all the issues concerning the localization of states described by Sec.~\ref{NewtonWigner_scheme_in_RQM}. This includes the instantaneous propagation of localized states and the consequent violation of causality, which will be discussed in the next subsection.

\subsection{Hegerfeldt theorem}\label{Hegerfeldt_theorem}

In his original work \cite{PhysRevD.10.3320}, Hegerfeldt showed that the phenomenon of instantaneous spreading for a relativistic particle does not occur only in the Newton-Wigner scenario. An alternative proof was later provided by Perez and Wilde \cite{PhysRevD.16.315}. Hegerfeldt and Ruijsenaars recognized that relativity is not needed to prove the results, while positivity of the energy and translation invariance suffice to give the instantaneous spreading \cite{PhysRevD.22.377}. Then, Hegerfeldt recognized that translation invariance is also not needed and, hence, the role of positivity of energy appears to be crucial in the instantaneous spreading of the wave function \cite{10.1007/BFb0106784}. However, when translation invariance is not considered, the localized particle either develops infinite tails immediately after or stays in its support indefinitely.

In \cite{PhysRevD.22.377}, the authors show that any particle confined in a bounded region can be found in spacelike separated regions at later times if the Hamiltonian is a nonconstant semibounded function of the momentum and translation invariant. Under stronger assumptions, the spreading of the wave function is over all of space. The conditions considered by Hegerfeldt and Ruijsenaars are met in RQM and QFT, where the energy of particles $\omega(\vec{k})$ is a function of the momentum $\vec{k}$ and is always positive.

The generality of the results is given by the fact that no specific definition of localization has been considered. To prove that no state can be localized in a finite region for a finite time interval, Hegerfeldt and Ruijsenaars only assumed that states localized in disjoint regions are orthogonal to each other. Also, to show that the spreading is over all of space, the authors assumes the existence of a positive operator $\hat{N}(\mathcal{V})$ for any space region $\mathcal{V} \subset \mathbb{R}^3$, such that $\langle \phi | \hat{N}(\mathcal{V}) | \phi \rangle \in [0,1]$ is the probability of finding the particle in $\mathcal{V}$. For instance, in the case of Newton-Wigner localization, $\hat{N}(\mathcal{V})$ is defined as
\begin{equation}\label{N_NewtonWigner}
\hat{N}(\mathcal{V}) = \frac{1}{|\mathcal{V}|} \int_\mathcal{V} d^3 x \hat{a}_\text{NW}^\dagger(\vec{x}) \hat{a}_\text{NW}(\vec{x}),
\end{equation}
where $|\mathcal{V}|$ is the volume of $\mathcal{V}$.

The apparent contradiction with the causal nature of the Klein-Gordon equation (or any other hyperbolic equation satisfying finite propagation speed, e.g., Maxwell equation, Dirac equation) can be argued in the following way \cite{Afanasev:1996nm}. Given any positive frequency solution of the Klein-Gordon equation $\phi(t,\vec{x})$ such that $\phi(0,\vec{x}) = 0$ for any $\vec{x}$ outside $\mathcal{V}$, then one finds that $\phi(t,\vec{x}) = 0$ in any region spacelike distant from $\mathcal{V}$ only if $\partial_t \phi(0,\vec{x}) = 0$, which does not occur for positive frequency solutions. This means that the localization in a finite region for a finite time is only possible for superpositions of positive and negative frequency solutions of the Klein-Gordon equation, which are excluded by the hypotheses of the Hegerfeldt theorem.

\section{AQFT scheme}\label{AQFT_localization_scheme}

In Sec.~\ref{NewtonWigner_localization_scheme} we reviewed the Newton-Wigner approach to the problem of localization in QFT. We remarked that the assumption made by Newton and Wigner are included in the hypotheses of the Hegerfeldt theorem. The results of the theorem are incompatible with the causality principle, as they imply a superluminal propagation of the localization condition. The paradox can be resolved by noticing that, in QFT, the spacetime coordinates $x^\mu$ appear as variables of the fields $\hat{\phi}(x^\mu)$ and the causality condition is defined via commutativity of spacelike separated fields.

In the framework of AQFT, any spacetime event $\mathcal{E}$ is provided with a local algebra $\mathfrak{A}(\mathcal{E})$ generated by the field $\hat{\phi}(x^\mu)$ with $x^\mu$ as the Minkowski coordinate representing $\mathcal{E}$ [Sec.~\ref{Algebraic_approach_in_QFT_and_QFTCS}]. More generally, for any spacetime region $\mathcal{O}$, the local algebra $\mathfrak{A}(\mathcal{O})$ is generated by the field $\hat{\phi}(x^\mu)$ smeared out with test functions that are supported in the Minkowski coordinate region $\mathcal{O}_\text{M} \subset \mathbb{R}^4$ representing $\mathcal{O}$. The operator $\hat{O}$ is said to be localized in $\mathcal{O}$ with respect to the AQFT scheme if $\hat{O}$ is an element of $\mathfrak{A}(\mathcal{O})$. We also define localized states by means of the notion of preparation over the vacuum $| 0_\text{M} \rangle$. The state $| \phi \rangle$ is said to be localized in $\mathcal{O}$ if it is the result of local operations on $| 0_\text{M} \rangle$. Explicitly, this means that $| \phi \rangle = \hat{O} | 0_\text{M} \rangle$, with $\hat{O} \in \mathfrak{A}(\mathcal{O})$.

The causality condition states that if $\mathcal{O}_\text{A}$ and $\mathcal{O}_\text{B}$ are spacelike separated regions, the corresponding local algebras $\mathfrak{A}(\mathcal{O}_\text{A})$ and $\mathfrak{A}(\mathcal{O}_\text{B})$ mutually commute. For the particular case of the real scalar field $\hat{\phi}(x^\mu)$, this is a consequence of the canonical commutation relation (\ref{phi_commutation}). From Eqs.~(\ref{phi_commutation}) and (\ref{PauliJordan_function}), we find that $[ \hat{\phi}(t,\vec{x}), \hat{\phi}(t',\vec{x}') ] = 0$ if $(t,\vec{x})$ and $(t',\vec{x}')$ are spacelike separated. The commutativity of quantum fields imposes statistical independence of measurements in spacelike separated regions $\mathcal{O}_\text{A}$ and $\mathcal{O}_\text{B}$, in the sense that measurements in $\mathcal{O}_\text{A}$ and $\mathcal{O}_\text{B}$ do not influence each other. This leads to the notion of microcausality of fields, which is included in the axioms of AQFT \cite{haag1992local}.

The microcausality axiom only ensures independence of measurements in spacelike separated regions $\mathcal{O}_\text{A}$ and $\mathcal{O}_\text{B}$. However, experiments also include other types of operations, such as the preparation of states. Hence, one can be interested in studying how and when operations in $\mathcal{O}_\text{A}$ can be considered independent of experiments made in any other spacelike separated region $\mathcal{O}_\text{B}$. For instance, one may ask if the local preparation of a state in $\mathcal{O}_\text{A}$ influences measurement in $\mathcal{O}_\text{B}$.

The localization program in AQFT is crucially affected by the Reeh-Schlieder theorem \cite{Reeh:1961ujh}, which predicts the presence of nonlocal quantum correlations in the vacuum $| 0_\text{M} \rangle$ \cite{haag1992local, Redhead1995-REDMAA-2, PhysRevA.58.135}. One of the consequences of the Reeh-Schlieder theorem is that measurements made in $\mathcal{O}_\text{B}$ are able to distinguish the vacuum $| 0_\text{M} \rangle$ from some states $| \phi \rangle$ localized in $\mathcal{O}_\text{A}$, even if $\mathcal{O}_\text{B}$ is spacelike separated from $\mathcal{O}_\text{A}$. Notwithstanding this apparent incompatibility with causality, it can be shown that the Reeh-Schlieder nonlocality cannot be used for superluminal signaling.

Another consequence of the theorem is the fact that one cannot use local fields to construct the operator $\hat{N}(\mathcal{V})$ that gives the probability to find a particle in space region $\mathcal{V}$ \cite{Redhead1995-REDMAA-2, RevModPhys.90.045003}. This implies that Hegerfeldt formulation of locality is not compatible with the algebraic notion of locality in AQFT.

The explicit hypotheses and statement of the Reeh-Schlieder theorem will be given in Sec.~\ref{ReehSchlieder_theorem}. Conversely, the solution to the apparent violation of causality will be discussed in Sec.~\ref{solving_the_paradox}.

\subsection{Reeh-Schlieder theorem}\label{ReehSchlieder_theorem}

In this subsection, we show the hypotheses and statement of the Reeh-Schlieder theorem. We discuss the dependency of spacelike separated operations and the nonlocality of number operators as consequences of the theorem.

Besides the microcausality of fields, the other axioms for AQFT in flat spacetime are
\begin{enumerate}
\item Isotony: any observable in $\mathcal{O}$ can also be measured in a larger region $\mathcal{O}'$, hence, $\mathfrak{A}(\mathcal{O}) \subset \mathfrak{A}(\mathcal{O}')$ if $\mathcal{O} \subset \mathcal{O}'$;
\item Relativistic covariance: each Poincar\'e transformation $\rho$ is provided with a unitary representation $\hat{U}(\rho)$ such that $\hat{U}(\rho) \mathfrak{A}(\mathcal{O}) \hat{U}^\dagger(\rho) = \mathfrak{A}(\rho(\mathcal{O}))$ with the vacuum $|0_\text{M} \rangle$ as the uniquely invariant state;
\item Spectrum condition: the spectrum of the generators $\hat{P}^\mu$ of the translation are such that $P^0 \geq 0$ (i.e., the energy is nonnegative) and $(P^0)^2 \geq |\vec{P}|^2$ (i.e., the spectrum  of the energy-momentum is confined to the forward light cone, capturing the notion of luminal and subluminal propagation of physical effect);
\item Weak Additivity: for any region $\mathcal{O} \subseteq \mathcal{M}$, with $\mathcal{M}$ as the Minkowski spacetime, $\mathfrak{A}(\mathcal{M})$ is the smallest algebra containing $\bigcup_{\alpha^\mu \in \mathbb{R}^4} \mathfrak{A}(\mathcal{O}_\alpha)$, where $\mathcal{O}_\alpha$ is the region $\mathcal{O}$ translated by $\alpha^\mu$.
\end{enumerate}

Assumptions 2-4 are used to prove the Reeh-Schlieder theorem \cite{Reeh:1961ujh, haag1992local}. The theorem states that the vacuum $|0_\text{M} \rangle$ is cyclic for any local algebra $\mathfrak{A}(\mathcal{O})$, in the sense that for any region $\mathcal{O}$, for any state $| \phi \rangle$ and for any $\epsilon > 0$, there exist an operator $\hat{O} \in \mathfrak{A}(\mathcal{O})$ such that $\parallel \hat{O} |0_\text{M} \rangle - | \phi \rangle \parallel < \epsilon $, where $\parallel \cdot \parallel$ is the norm in the Hilbert space. This means that one can approximate any state of the global Hilbert space with arbitrary precision by applying an element of any local algebra $\mathfrak{A}(\mathcal{O})$ to the vacuum $|0_\text{M} \rangle$. Such a nonlocal effect is the result of entangled correlations in the vacuum \cite{haag1992local, Redhead1995-REDMAA-2, PhysRevA.58.135}.

By operating in any bounded spacetime region $\mathcal{O}_\text{A}$, one is able to produce any global state $| \phi \rangle$ that may, in principle, differ from $|0_\text{M} \rangle$ in another spacelike separated region $\mathcal{O}_\text{B}$. Even if $\mathcal{O}_\text{A}$ and $\mathcal{O}_\text{B}$ are not causally connected, the restriction of $| \phi \rangle$ in $\mathfrak{A}(\mathcal{O}_\text{B})$ may be different from the restriction of $|0_\text{M} \rangle$ in $\mathfrak{A}(\mathcal{O}_\text{B})$. This result seems to be incompatible with the notion of causality. However, the contradiction is resolved by noticing that the nonlocal effect cannot be used for superluminal signaling. A more detailed discussion will be provided in Sec.~\ref{solving_the_paradox}.

A corollary to the Reeh Schlieder theorem is that the vacuum is a separating state in any local algebra $\mathfrak{A}(\mathcal{O})$, in the sense that for any $\hat{O} \in \mathfrak{A}(\mathcal{O})$, if $\hat{O}$ annihilates the vacuum (i.e., $\hat{O} | 0_\text{M} \rangle = 0$), then $\hat{O}$ is trivial (i.e., $\hat{O}=0$) \cite{Redhead1995-REDMAA-2, RevModPhys.90.045003}. The consequence is that local number operators do not exist. As already anticipated by the previous section, the number operator $\hat{N}(\mathcal{V})$ is inevitably nonlocal with respect to the AQFT scheme. This also applies to the Newton-Wigner number operator [Eq.~(\ref{N_NewtonWigner})].

\subsection{Apparent violation of causality}\label{solving_the_paradox}

In Sec.~\ref{ReehSchlieder_theorem}, we introduced the apparent violation of causality due to the Reeh Schlieder theorem. To see the problem in a physical scenario, consider two observers, Alice and Bob, which are localized in two spacelike separated regions, $\mathcal{O}_\text{A}$ and $\mathcal{O}_\text{B}$. Alice prepares a state $| \phi \rangle = \hat{O}_\text{A} | 0_\text{M} \rangle$ by means of a local operator $\hat{O}_\text{A} \in \mathfrak{A}(\mathcal{O}_\text{A})$ acting on the vacuum $| 0_\text{M} \rangle$; whereas Bob performs local measurements by means of the observable $\hat{O}_\text{B} \in \mathfrak{A}(\mathcal{O}_\text{B})$. As a consequence of the Reeh Schlieder theorem, we find that there are some cases in which
\begin{equation}\label{local_operator_measurament}
\langle \phi | \hat{O}_\text{B} | \phi \rangle \neq \langle 0_\text{M} | \hat{O}_\text{B} | 0_\text{M} \rangle.
\end{equation}
Equation (\ref{local_operator_measurament}) implies that the preparation of the local state $| \phi \rangle =  \hat{O}_\text{A} | 0_\text{M} \rangle$ in $\mathcal{O}_\text{A}$ can be detected by Bob as a result of measurements of the local observable $\hat{O}_\text{B}$. This seems to be incompatible with the notion of causality since Alice and Bob are spacelike separated.

The problem has been addressed by different authors \cite{Redhead1995-REDMAA-2, CLIFTON20011, VALENTE2014147, RevModPhys.90.045003} and led to the conclusion that the violation of causality is only apparent. The solution is given by noticing that a global change of the state is only due to selective operations that cannot be used for superluminal signaling. This argument will be detailed in the present subsection.

Firstly, notice that Eq.~(\ref{local_operator_measurament}) does not hold if $\hat{O}_\text{A}$ is unitary. Indeed, by using the unitarity condition $\hat{O}_\text{A}^\dagger \hat{O}_\text{A} = 1$ and the microcausal commutation relation $[\hat{O}_\text{A}, \hat{O}_\text{B}] = 0$, we obtain
\begin{equation}\label{local_unitary_operator_measurament_A}
\langle 0_\text{M} | \hat{O}_\text{A}^\dagger \hat{O}_\text{B} \hat{O}_\text{A} | 0_\text{M} \rangle = \langle 0_\text{M} | \hat{O}_\text{A}^\dagger \hat{O}_\text{A} \hat{O}_\text{B} | 0_\text{M} \rangle = \langle 0_\text{M} | \hat{O}_\text{B} | 0_\text{M} \rangle.
\end{equation}
Explicitly, this means that
\begin{equation}\label{KnightLicht_property}
\langle \phi | \hat{O}_\text{B} | \phi \rangle = \langle 0_\text{M} | \hat{O}_\text{B}  | 0_\text{M} \rangle.
\end{equation}

By following Knight and Licht \cite{10.1063/1.1703731, 10.1063/1.1703925}, we say that the state $| \phi \rangle$ satisfies the strictly localization property if it gives the same expectation values as the vacuum for all measurements in the causal complement of $\mathcal{O}_\text{A}$. Equivalently, we say that $| \phi \rangle$ is strictly localized in $\mathcal{O}_\text{A}$ if Eq.~(\ref{KnightLicht_property}) holds for any $\hat{O}_\text{B} \in \mathfrak{A}(\mathcal{O}_\text{B})$ and for any region $\mathcal{O}_\text{B}$ spacelike separated from $\mathcal{O}_\text{A}$. As a result of Eq.~(\ref{local_unitary_operator_measurament_A}), we know that any local unitary operator $\hat{U}_\text{A} \in \mathfrak{A}(\mathcal{O}_\text{A})$ produces a strictly localized state $| \phi \rangle = \hat{U}_\text{A} | 0_\text{M} \rangle$ by acting on the vacuum $| 0_\text{M} \rangle$.

In general, the modification of quantum states due to the interaction with experimental instruments (e.g., emitters) is represented by a unitary evolution $| 0_\text{M} \rangle \mapsto \hat{U}_\text{int} | 0_\text{M} \rangle$. However, one can argue that this is not the only way to prepare local states. For instance, one can use the following procedure: (i) let the device interact with the vacuum $| 0_\text{M} \rangle $ to unitarily prepare the state $ \hat{U}_\text{int} | 0_\text{M} \rangle$; (ii) perform the projective measurement $\hat{P}_i$ over a set of subspaces $\mathcal{H}_i$ of the global Hilbert space; (iii) reject all the states that are not elements of (say) $\mathcal{H}_0$. In this way, the experimenter is sure that the resulting state is an element of $\mathcal{H}_0$. The overall operation is said to be selective due to the experimenter's choice of selecting a subensemble after the measurement.

In this scenario, the preparation of the state in $\mathcal{O}_\text{A}$ affects observations in the spacelike separated region $\mathcal{O}_\text{B}$. To see this, consider a local observable $\hat{O}_\text{B} \in \mathfrak{A}(\mathcal{O}_\text{B})$ and assume that $\hat{U}_\text{int} \in \mathfrak{A}(\mathcal{O}_\text{A})$ and $\hat{P}_0 \in \mathfrak{A}(\mathcal{O}_\text{A})$. The normalized state after the preparation is $| \phi \rangle = \hat{O}_\text{A} | 0_\text{M} \rangle$, with
\begin{equation}\label{psi_local_unitary_nonselective}
\hat{O}_\text{A} = \frac{\hat{P}_0 \hat{U}_\text{int}}{\sqrt{\langle 0_\text{M} | \hat{U}_\text{int}^\dagger \hat{P}_0 \hat{U}_\text{int} | 0_\text{M} \rangle}}
\end{equation}
as a local operator in $\mathcal{O}_\text{A}$. The mean value of $\hat{O}_\text{B}$ is
\begin{equation}\label{local_unitary_nonselective_operator_measurament_A}
\langle \phi | \hat{O}_\text{B} | \phi \rangle = \frac{\langle 0_\text{M} | \hat{U}_\text{int}^\dagger \hat{P}_0 \hat{U}_\text{int} \hat{O}_\text{B}  | 0_\text{M} \rangle}{\langle 0_\text{M} | \hat{U}_\text{int}^\dagger \hat{P}_0 \hat{U}_\text{int} | 0_\text{M} \rangle},
\end{equation}
which is different from $\langle 0_\text{M} | \hat{O}_\text{B} | 0_\text{M} \rangle$. Hence, in this scenario, Eq.~(\ref{local_operator_measurament}) holds and the state $| \phi \rangle$ is not strictly localized in $\mathcal{O}_\text{A}$. Notice that the inequality $ \hat{P}_0 \neq 1$ is crucial for the proof of Eq.~(\ref{local_operator_measurament}).\footnote{The results of Eqs.~(\ref{local_unitary_operator_measurament_A}) and (\ref{local_unitary_nonselective_operator_measurament_A}) can be extended to the case of general quantum operations with local Kraus operators $\hat{K}_i \in \mathfrak{A}(\mathcal{O}_\text{A})$ \cite{kraus1983states, CLIFTON20011, VALENTE2014147}. The statistical operator describing the state after the operation is
\begin{equation}\label{rho_Kraus}
\hat{\rho} = \frac{\sum_i \hat{K}_i |0_\text{M} \rangle \langle 0_\text{M} | \hat{K}_i^\dagger}{\sum_i \langle 0_\text{M} | \hat{K}_i^\dagger \hat{K}_i |0_\text{M} \rangle}.
\end{equation}
The operation is said to be nonselective only when
\begin{equation}\label{Kraus_nonselective}
\sum_i \hat{K}_i^\dagger \hat{K}_i = 1.
\end{equation}
By using the ciclicity of the trace and the commutation relation between $\hat{K}_i \in \mathfrak{A}(\mathcal{O}_\text{A})$ and $\hat{O} \in \mathfrak{A}(\mathcal{O}_\text{B})$, one can prove that
\begin{equation}\label{Tr_rho_A}
\text{Tr} (\hat{\rho} \hat{O}) = \frac{\sum_i \text{Tr} (\hat{K}_i^\dagger \hat{K}_i|0_\text{M} \rangle \langle 0_\text{M} | \hat{O})}{\sum_i \langle 0_\text{M} | \hat{K}_i^\dagger \hat{K}_i |0_\text{M} \rangle}.
\end{equation}
The right hand side of Eq.~(\ref{Tr_rho_A}) is equal to $\text{Tr} (|0_\text{M} \rangle \langle 0_\text{M} | \hat{O})$ if and only if the Kraus operators satisfy Eq.~(\ref{Kraus_nonselective}).}

We found that only nonselective local operations in $\mathcal{O}_\text{A}$ do not change the vacuum in the causal complement of $\mathcal{O}_\text{A}$. Conversely, the Reeh-Schlieder nonlocal effect and the consequent apparent violation of causality occur when the state is prepared by means of selective operations on $| 0_\text{M} \rangle$. The defining feature of this type of operations is the experimenter's decision to only consider the subspace $\mathcal{H}_0$ and reject all states that give negative results in measuring the effect $\hat{P}_0$. Crucially, the outcomes of the projective measurements are random and only the observer knows when the state has been successfully prepared. This information can only be shared by means of a classical communication. Hence, causality is not violated.

To see that selective operations cannot be used for superluminal signaling, consider again the two experimenters, Alice and Bob, localized in two spacelike separated regions $\mathcal{O}_\text{A}$ and $\mathcal{O}_\text{B}$. Alice performs a selective operation in $\mathcal{O}_\text{A}$ to prepare a state, whereas Bob measures the observable $\hat{O} \in \mathfrak{A}(\mathcal{O}_\text{B})$ in $\mathcal{O}_\text{B}$. In order to prepare the desired state, Alice repeats the operation multiple times and excludes the cases in which the outcome of her selective measurements are unsuccessful, i.e., when the desired state has not been successfully prepared. At this point, Alice is biased, as she knows which operation was successful and which was not. Bob, in principle, is ignorant about the outcome of Alice's operations and, hence, does not know when to perform the measurement with the correct state. He can only acquire this information in two possible ways: (i) by performing Alice's projective measurement to verify if the state is the correct one; however this is only possible if Bob has access to Alice's algebra and, hence, if they are not spacelike separated; (ii) by letting Alice share her information via classical communication, which follows relativistic causality and forbids superluminal signaling.

To connect with the literature, we agree with Clifton, Halvorson and Valente \cite{CLIFTON20011, VALENTE2014147} who recognized that the problematic operations are selective. However, we give a different argument on why no violation of causality occurs even in the case of selective operations. For Clifton and Halvorson \cite{CLIFTON20011}, these operations do not retain full physical meaning, but are partly affected by the purely conceptual operation of selecting subensembles. In other words, the selective component of the operation is regarded as mathematical and nonphysical. This leads to the interpretation of quantum states as partly epistemic entities, where each update of state after a measurement only represents a change of knowledge of the experimenter based on the outcome of the measurement. Conversely, Valente \cite{VALENTE2014147} avoided any interpretation of states, while giving arguments to support the thesis that superluminal signaling of selective operations cannot be controlled. We also showed how these operations cannot be used to instantly send information to another experimenter; however, we used a different argument.

Our approach is inspired by the quantum teleportation technique \cite{PhysRevLett.70.1895}, where a maximally entangled state is used to teleport a quantum state. In that case, no violation of causality occurs because a classical channel has to be employed to transmit information about the outcomes of Alice's measurement. This is in complete analogy with the scenario of the Reeh-Schlieder apparent paradox described here. Hence, we used the argument that the nonlocal correlations due to entanglement is compatible with the prohibitions of superluminal causation. QFT admits correlations between spacelike separated regions \cite{SUMMERS1985257, Summers:1987fn} while causality is not violated.

The result is also directly connected to the EPR experiment \cite{PhysRev.47.777, CLIFTON20011}, where nonlocal quantum correlations are used to globally change a state via local measurements. The wave function collapse in the EPR scenario cannot be exploited to instantly send information to another experimenter.  Equivalently, in the Reeh-Schlieder scenario, one uses the vacuum correlations to produce a nonlocal effect, which, however, does not lead to superluminal signaling.

\section{Modal scheme}\label{Modal_localization_scheme}

In Sec.~\ref{QFT_in_Minkowski_spacetime_scalar}, we derived the representation of single particle states as positive frequency modes of the Klein-Gordon equation (\ref{Klein_Gordon}). States with defined momentum $| \vec{k} \rangle$ are represented by the free modes (\ref{free_modes}), which satisfy the orthonormality condition (\ref{KG_scalar_product_orthonormality_f}) with $( \phi, \phi' )_\text{KG}$ as the Klein-Gordon product (\ref{KG_scalar_product}). More generally, any Minkowski-Fock state $| \phi \rangle \in \mathcal{H}_\text{M}$ is represented by the wave functions in momentum space $\tilde{\phi}_n (\textbf{k}_n)$ and in position space $\phi_n (t, \textbf{x}_n)$, respectively defined by Eqs.~(\ref{free_state_decomposition}) and (\ref{free_wave_function}). To not get confused with the notation of Sec.~\ref{NewtonWigner_localization_scheme}, we say that $\phi_\text{NW}(t,\vec{x})$ is a Newton-Wigner wave function and $\phi_n (t, \textbf{x}_n)$ is a modal wave function.

In Sec.~\ref{QFT_in_Minkowski_spacetime_scalar}, we also remarked that $\phi_n (t, \textbf{x}_n)$ does not entail any genuine notion of localization in QFT, at variance with the AQFT scheme. To see this, one can use the same argument provided for the Newton-Wigner scheme. In particular, one can refer to the superluminal spreading of the modal wave functions $\phi_n (t, \textbf{x}_n)$ to claim that the modal scheme is not suitable for the description of localized relativistic states. Such an instantaneous spreading can be proven by noticing that $\phi_n (t, \textbf{x}_n)$ is a linear combination of products of positive frequency modes. Hence, if the support of $\phi_n (t, \textbf{x}_n)$ is compact at a fixed time $t$, then its time derivative $\partial_0 \phi_n (t, \textbf{x}_n)$ is not compactly supported at the same time $t$ \cite{Afanasev:1996nm}. Consequently, the modal wave function instantly develops infinite tails.

In QFT, the function $\phi_n (t, \textbf{x}_n)$ cannot be associated to the probability to find the $n$ particles in $\textbf{x}_n = (\vec{x}_1, \dots, \vec{x}_n)$. However, this is not true in NRQM. In Secs.~\ref{Comparison_with_the_relativistic_theory} and \ref{Convergence_of_the_modal_scheme_to_the_Born_scheme}, we will show that both the AQFT and the modal scheme converge to the same localization scheme when states and observables are restricted to the nonrelativistic regime.  This means that the modal wave functions acquire a fundamental notion of localization only in the nonrelativistic limit.

In the remaining part of this section, we formulate the modal localization scheme in terms of localized states and observables. By definition, the state $| \phi \rangle$ is said to be localized in a volume $\mathcal{V}$ at time $t$ with respect to the modal scheme if the support of $\phi_n (t, \textbf{x}_n)$ is in $\mathcal{V}^n$, in the sense that $\phi_n (t, \textbf{x}_n) = 0$ when there is an $l \in \{ 1, \dots, n \}$ such that $\vec{x}_l \notin \mathcal{V}$.

We now show that there is a natural definition of localized operators based on the localization of states with respect to the modal scheme. We start by considering the Minkowski-Fock representation of any state $| \phi \rangle \in \mathcal{H}_\text{M}$ given by Eqs.~(\ref{free_state_decomposition_antiparticles}) and (\ref{c_phi}). As specified at the beginning of the chapter, we consider a real field $\hat{\phi}(t,\vec{x})$ which does not include antiparticles; hence, we assume that $m=0$ in Eq.~(\ref{c_phi}).

The identity relating $\tilde{\phi}_n (\textbf{k}_n)$ to $\phi_n (t, \textbf{x}_n)$ is Eq.~(\ref{free_wave_function}), which can be inverted by means of a Fourier transform as
\begin{equation} \label{free_wave_function_inverse}
\tilde{\phi}_n (\textbf{k}_n) = \left[ \frac{\hbar}{(2 \pi)^3 m c^2} \right]^{n/2} \int_{\mathbb{R}^{3n}} d^{3n} \textbf{x}_n  \phi_n (0, \textbf{x}_n) \prod_{l=1}^n \left[ \sqrt{\omega(\vec{k}_l)}  e^{- i \vec{k}_l \cdot \vec{x}_l} \right].
\end{equation}
Equation (\ref{free_wave_function_inverse}) can be plugged in Eq.~(\ref{c_phi}) to obtain
\begin{equation}\label{c_phi_space}
\hat{c}_\phi  =  \sum_{n=0}^\infty \frac{1}{\sqrt{n!}}   \int_{\mathbb{R}^{3n}} d^{3n} \textbf{x}_n  \phi_n (0, \textbf{x}_n)  \prod_{l=1}^n \hat{a}_\text{mod}^\dagger (\vec{x}_l),
\end{equation}
with
\begin{equation}\label{a_mod_a}
\hat{a}_\text{mod} (\vec{x})  = \int_{\mathbb{R}^3} d^3 k  \sqrt{\frac{\hbar \omega(\vec{k})}{(2 \pi)^3 mc^2}}  e^{i \vec{k} \cdot \vec{x}} \hat{a}(\vec{k}) .
\end{equation}

For each $\vec{x}$ we indicate the algebra generated by the operator $\hat{a}_\text{mod} (\vec{x})$ and its adjoint as $\mathfrak{A}_\text{M}^\text{mod}(\vec{x})$. For extended regions of space $\mathcal{V} \subseteq \mathbb{R}^3$, we define $\mathfrak{A}_\text{M}^\text{mod}(\mathcal{V})$ as the algebra generated by the operators $\hat{a}_\text{mod} (\vec{x})$ with $\vec{x} \in \mathcal{V}$. By using Eq.~(\ref{c_phi_space}) and the definition of localized states with respect to the modal scheme, we find that $\hat{c}_\phi $ is an element of $ \mathfrak{A}_\text{M}^\text{mod}(\mathcal{V})$, with $\mathcal{V}$ as the region in which the state $| \phi \rangle = \hat{c}_\phi | 0_\text{M} \rangle$ is localized at $t=0$. This naturally leads to the definition of local operators as elements of $ \mathfrak{A}_\text{M}^\text{mod}(\mathcal{V})$ and the identification of $ \mathfrak{A}_\text{M}^\text{mod}(\mathcal{V})$ as a local algebra with respect to the modal scheme.

Notice that, due to the invertibility of Eq.~(\ref{a_mod_a}), any operator $\hat{O}$ acting on the Minkowski-Fock space $\mathcal{H}_\text{M}$ admits a region $\mathcal{V} \subseteq \mathbb{R}^3$ such that $\hat{O} \in \mathfrak{A}_\text{M}^\text{mod}(\mathcal{V})$. If, in particular, $\mathcal{V} = \mathbb{R}^3$, then the operator is said to be global (i.e., nonlocal) with respect to the modal scheme.

\section{Comparison between localization schemes}\label{Comparison_between_localization_schemes}

\begin{table}
\begin{center}
\centering
\begin{tabular}{| >{\raggedright\arraybackslash}m{12.5em} || >{\centering\arraybackslash}m{5.5em} | >{\centering\arraybackslash}m{6em} |  >{\centering\arraybackslash}m{3.5em} |}
\hline
 & Newton-Wi-gner scheme & AQFT scheme & modal scheme\\
\hline
\hline
Relativistic covariance and causality hold & No & Yes & No \\
\hline
The variable $\vec{x}$ is a genuine position coordinate & No & Yes & Yes \\
\hline
Operators in disjoint spatial regions commute & Yes & Yes & No \\
\hline
States in disjoint spatial regions are orthogonal & Yes & No & No \\
\hline
The global Hilbert space factorizes into local Hilbert spaces: $\mathcal{H}_\text{M} = \bigotimes_i \mathcal{H}_\text{M}(\mathcal{V}_i) $ & Yes & Yes & No \\
\hline
The global vacuum $| 0_\text{M} \rangle$ is entangled across the local Hilbert spaces $ \mathcal{H}_\text{M}(\mathcal{V}_i) $ & No & Yes & - \\
\hline
The local Hilbert spaces $ \mathcal{H}_\text{M}(\mathcal{V}_i) $ are Fock spaces with local vacua $| 0_\text{M}(\mathcal{V}_i) \rangle$ & Yes & No & - \\
\hline
The global vacuum factorizes into the local vacua: $| 0_\text{M} \rangle = \bigotimes_i | 0_\text{M}(\mathcal{V}_i) \rangle $ & Yes & - & - \\
\hline
Localized states live in local Hilbert spaces [Eq.~(\ref{NW_localization_decomposition})] & Yes & - & - \\
\hline
The strict localization property [Eq.~(\ref{KnightLicht_property})] at $t=0$ is guaranteed & Yes & Only for local nonselective preparations & No \\
\hline
\end{tabular}
\end{center}
\caption{Summary table of the differences between the Newton-Wigner, the AQFT and the modal localization schemes.}\label{Comparison_between_localization_schemes_Table} 
\end{table}

In the previous sections of this chapter, we introduced three different localization schemes in QFT. Here, we compare them and we detail the relevant differences. A summary of the discussion can be found in Table \ref{Comparison_between_localization_schemes_Table}.

\subsection{Newton-Wigner and AQFT scheme}\label{Comparison_between_the_NewtonWigner_and_the_AQFT_schemes}

\subsubsection{Fundamental differences}

There are significant conceptual differences between the Newton-Wigner localization and the AQFT localization schemes. The former is based on the orthogonality between states localized in different regions and leads to frame-dependent (i.e., noncovariant) features and superluminal phenomena. Conversely, the AQFT localization scheme is causal and covariant and, hence, it is regarded as fundamental in nature. In particular, the commutativity of fields in spacelike separated regions guarantees the independence of measurements. Furthermore, if a state is localized in one region $\mathcal{V}$ of space at $t=0$, there is no mean by which one can instantaneously send information outside the light cone of $\mathcal{V}$.

In the Newton-Wigner scheme, the variable $\vec{x}$ appears as a result of second-quantizing the position operator $\hat{\vec{x}}_\text{NW} $. Conversely, in the AQFT scheme, the variable $\vec{x}$ is associated the coordinate system representing the underling spacetime and, hence, entails a genuine notion of localization.

Other differences between the Newton-Wigner and the AQFT schemes can be obtained by considering the respective local algebras $\mathfrak{A}_\text{M}^\text{NW}(\mathcal{V})$ and $\mathfrak{A}(\mathcal{O})$. However, notice that $\mathcal{V}$ is a subset of $\mathbb{R}^3$, whereas $\mathcal{O}$ is a region of spacetime. Hence, a direct comparison between the two schemes can only be made if we restrict $\mathcal{O}$ to a space region at $t=0$. This is possible due to the dynamical structure of the field $\hat{\phi}(x^\mu)$. Indeed, as a consequence of the Klein-Gordon equation (\ref{Klein_Gordon}) and the hyperbolic nature of the spacetime, any region $\mathcal{O}$ admits a minimally extended Cauchy region $\mathcal{C}_\mathcal{O}$ inside the hypersurface $t=0$ such that the field operators $\hat{\phi}(x^\mu)$ inside $\mathcal{O}$ can be written in terms of field operators $\hat{\phi}(x^\mu)$ and $\hat{\pi}(x^\mu)$ [Eq.~(\ref{Pi_conjugate})] inside $\mathcal{C}_\mathcal{O}$. Explicitly, this means that $\mathfrak{A}(\mathcal{O}) \subseteq \mathfrak{A}_\text{M}^\text{AQFT}(\mathcal{C}_\mathcal{O})$, where $\mathfrak{A}_\text{M}^\text{AQFT}(\mathcal{V})$ is the algebra generated by the field operators $\hat{\phi}(0,\vec{x})$ and $\hat{\pi}(0,\vec{x})$ with varying $\vec{x} \in \mathcal{V}$. We also define the local algebra in a space point $\mathfrak{A}_\text{M}^\text{AQFT}(\vec{x})$ as the one generated by the fields $\hat{\phi}(0,\vec{x})$ and $\hat{\pi}(0,\vec{x})$ with fixed $\vec{x}$. The Newton-Wigner and the AQFT localization schemes can now be directly compared by means of the algebras $\mathfrak{A}_\text{M}^\text{NW}(\mathcal{V})$ and $\mathfrak{A}_\text{M}^\text{AQFT}(\mathcal{V})$, or, equivalently, by means of $\mathfrak{A}_\text{M}^\text{NW}(\vec{x})$ and $\mathfrak{A}_\text{M}^\text{AQFT}(\vec{x})$.

By comparing the algebras $\mathfrak{A}_\text{M}^\text{NW}(\vec{x})$ and $\mathfrak{A}_\text{M}^\text{AQFT}(\vec{x})$, one can notice that the two notions of locality are not compatible, in the sense that if an operator is Newton-Wigner localized in $\vec{x}$, it cannot be localized with respect to the AQFT scheme as well. Explicitly, we are saying that $\mathfrak{A}_\text{M}^\text{NW}(\vec{x}) \neq \mathfrak{A}_\text{M}^\text{AQFT}(\vec{x})$.

This can be proved by reminding that operators in $\mathfrak{A}_\text{M}^\text{NW}(\vec{x})$ are generated by the Newton-Wigner annihilation operator $\hat{a}_\text{NW}(\vec{x})$ and its adjoint, whereas the algebra $\mathfrak{A}_\text{M}^\text{AQFT}(\vec{x})$ is generated by the field operators $\hat{\phi}(0,\vec{x})$ and $\hat{\pi}(0,\vec{x})$. By using Eqs.~(\ref{Pi_conjugate}), (\ref{free_field}), (\ref{free_modes}) and (\ref{a_NW}), we obtain
\begin{equation}\label{a_tilde_phi_Pi}
\hat{a}_\text{NW}(\vec{x}) = \int_{\mathbb{R}^3} d^3 x' \left[ f_{\hat{\phi}\mapsto \text{NW}}(\vec{x} - \vec{x}') \hat{\phi}(0, \vec{x}') + f_{\hat{\pi}\mapsto \text{NW}}(\vec{x} - \vec{x}') \hat{\pi}(0, \vec{x}')  \right],
\end{equation}
with
\begin{align}\label{f_phi_Pi}
& f_{\hat{\phi}\mapsto \text{NW}}(\vec{x}) = \int_{\mathbb{R}^3} d^3 k \frac{\sqrt{\omega (\vec{k})} e^{i \vec{k} \cdot \vec{x}}}{\sqrt{2 \hbar} (2 \pi)^3}, & f_{\hat{\pi}\mapsto \text{NW}}(\vec{x}) = \int_{\mathbb{R}^3} d^3 k \frac{-i e^{i \vec{k} \cdot \vec{x}}}{\sqrt{2 \hbar \omega (\vec{k})} (2 \pi)^3}.
\end{align}
From Eq.~(\ref{f_phi_Pi}), it is possible to notice that $f_{\hat{\phi}\mapsto \text{NW}}(\vec{x}) $ and $f_{\hat{\pi}\mapsto \text{NW}}(\vec{x}) $ are supported in the whole space $\mathbb{R}^3$. Consequently, the right hand side of Eq.~(\ref{a_tilde_phi_Pi}) is nonlocal with respect to the AQFT scheme. This means that $\hat{a}_\text{NW}(\vec{x}) \notin \mathfrak{A}_\text{M}^\text{AQFT}(\vec{x})$ and, hence, $\mathfrak{A}_\text{M}^\text{NW}(\vec{x}) \neq \mathfrak{A}_\text{M}^\text{AQFT}(\vec{x})$.

\subsubsection{Local particle content}

An important difference between the two schemes is given by the notion of the vacuum as locally and globally devoid of quanta \cite{Fleming2000-FLERMN}. The Newton-Wigner operators $\hat{a}_\text{NW}(\vec{x}) $ [Eq.~(\ref{a_NW})] annihilate the vacuum, i.e., $\hat{a}_\text{NW}(\vec{x}) | 0_\text{M} \rangle = 0 $, and can be used to define a local number density operator $\hat{N}(\mathcal{V})$ [Eq.~(\ref{N_NewtonWigner})]. Conversely, the corollary of the Reeh-Schlieder theorem forbids the definition of such an operator in the AQFT localization scheme. In that case, the vacuum is not locally devoid of quanta, but only globally.

Notice that the Newton-Wigner operators $\hat{a}_\text{NW}(\vec{x}) $ satisfy the canonical commutation identity
\begin{align}\label{a_NW_commutation}
& [\hat{a}_\text{NW}(\vec{x}), \hat{a}_\text{NW}^\dagger(\vec{x}') ] = \delta^3(\vec{x}-\vec{x}'), & [\hat{a}_\text{NW}(\vec{x}), \hat{a}_\text{NW}(\vec{x}') ] = 0.
\end{align}
Hence, $\hat{a}_\text{NW}(\vec{x}) $ can be interpreted as a local annihilation operator in the Newton-Wigner scheme. Conversely, due to the corollary of the Reeh-Schlieder theorem, local creation and annihilation operators do not exist in the AQFT scheme.

The existence of local creation and annihilation operators in the Newton-Wigner scheme ensures that the global Fock space factorized into local Fock spaces $\mathcal{H}_\text{M} = \bigotimes_i \mathcal{H}_\text{M}^\text{NW}(\mathcal{V}_i)$, where $\{ \mathcal{V}_i \}$ is any partition of $\mathbb{R}^3$. The vacuum of each Fock space $\mathcal{H}_\text{M}^\text{NW}(\mathcal{V}_i)$ will be denoted as $| 0_{\text{M}}^\text{NW}(\mathcal{V}_i) \rangle$ and is defined by $\hat{a}_\text{NW}(\vec{x}) | 0_{\text{M}}^\text{NW}(\mathcal{V}_i) \rangle = 0$, for any $\vec{x} \in \mathcal{V}_i$. From the definition of $| 0_{\text{M}}^\text{NW}(\mathcal{V}_i) \rangle$ and the fact that the Minkowski vacuum $| 0_\text{M} \rangle$ is always annihilated by $\hat{a}_\text{NW}(\vec{x})$, we find that $| 0_\text{M} \rangle$ is equal to the product state of the local vacua, i.e., $| 0_{\text{M}} \rangle = \bigotimes_i | 0_{\text{M}}(\mathcal{V}_i) \rangle$, and, hence, it is not entangled across the local Hilbert spaces $\mathcal{H}_\text{M}^\text{NW}(\mathcal{V}_i)$. 

In the AQFT scheme, the global Hilbert space $\mathcal{H}_\text{M}$ can be factorized by means of the local field operators $\hat{\phi}(0,\vec{x})$ and $\hat{\pi}(0,\vec{x})$ and their commutation relations (\ref{phi_commutation_Lagrangian}), which lead to $\mathcal{H}_\text{M} = \bigotimes_i \mathcal{H}_\text{M}^\text{AQFT}(\mathcal{V}_i)$.\footnote{We remark that the factorization $\mathcal{H}_\text{M} = \bigotimes_i \mathcal{H}_\text{M}^\text{AQFT}(\mathcal{V}_i)$ is not mathematically precise and can be considered valid only in some sort of limit. In particular, in the rigorous context of AQFT, the microcausality condition does not guarantee the factorization of the global algebra into the local algebras $\mathfrak{A} = \bigotimes_i \mathfrak{A}_\text{M}^\text{AQFT}(\mathcal{V}_i)$.

However, a weaker version of $\mathfrak{A} = \bigotimes_i \mathfrak{A}_\text{M}^\text{AQFT}(\mathcal{V}_i)$ can be found in those theories that satisfy the so-called split property \cite{Doplicher1984}. The assumption is that for any couple of spacetime regions $\mathcal{O}$ and $\mathcal{O}' \supset \mathcal{O} $ there is a type I von Neumann algebra $\mathfrak{R}$ such that $\mathfrak{A}(\mathcal{O}) \subset \mathfrak{R} \subset \mathfrak{A}(\mathcal{O}')$. The split property has been proven in a variety of models, including free massive scalar field \cite{Buchholz1974}, Dirac, Maxwell, free massless scalar fields \cite{Horujy1979} and free massive fermion fields \cite{Summers1982}.

A weak notion of independence via tensor product of local Hilbert spaces and algebras is present in quantum field theories with split property \cite{10.1007/978-3-030-38941-3_1}. In particular, for any regions $\mathcal{O}_\text{A}$, $\mathcal{O}'_\text{A}$ and $\mathcal{O}_\text{B}$ such that $\mathcal{O}_\text{A} \subset \mathcal{O}'_\text{A}$ and $\mathcal{O}_\text{B}$ is spacelike separated from $\mathcal{O}'_\text{A}$, the following isomorphism holds
\begin{equation}
\mathfrak{A}(\mathcal{O}_\text{A}) \vee \mathfrak{A}(\mathcal{O}_\text{B}) \cong \mathfrak{A}(\mathcal{O}_\text{A}) \otimes \mathfrak{A}(\mathcal{O}_\text{B}), 
\end{equation}
where the left-hand side is the algebra generated by sums and products of elements in $\mathfrak{A}(\mathcal{O}_\text{A}) $ and $ \mathfrak{A}(\mathcal{O}_\text{B}) $ and the right-hand side is the spatial tensor product of the algebras.

The notion of independence via tensor product is weak because one can consider any $\mathcal{O}_\text{A}'$ arbitrarily close to $\mathcal{O}_\text{A}$, but never equal. In other words, the region $\mathcal{O}_\text{A}'$ ensures that $\mathcal{O}_\text{A}$ and $\mathcal{O}_\text{B}$ do not touch at their border; however, one can consider the limiting case in which the two regions $\mathcal{O}_\text{A}$ and $\mathcal{O}_\text{B}$ are arbitrary close. Hence, the factorization $\mathfrak{A} = \bigotimes_i \mathfrak{A}_\text{M}^\text{AQFT}(\mathcal{V}_i)$ can only by formalized in such a limit.} However, a factorization of $\mathcal{H}_\text{M}$ into local Fock space is not possible, due to the nonexistence of local creators and annihilators. Consequently, the local Hilbert spaces $\mathcal{H}_\text{M}^\text{AQFT}(\mathcal{V}_i)$ cannot be Fock spaces and the global vacuum $| 0_\text{M} \rangle$ cannot factorize into local vacua. More precisely, $| 0_\text{M} \rangle$ does not factorize into any set of local states, since it is entangled across the local Hilbert spaces $\mathfrak{A}_\text{M}^\text{AQFT}(\mathcal{V}_i)$ \cite{haag1992local, Redhead1995-REDMAA-2, PhysRevA.58.135}.

\subsubsection{Independence via tensor product of local Hilbert spaces and algebras}

In quantum physics, the independence of physical phenomena is represented by the factorization of states and observables. In the usual prescription, two distinct laboratories, A and B, are supplied with their own Hilbert space $\mathcal{H}_\text{A}$ and $\mathcal{H}_\text{B}$, the respective experimenters prepare the states $| \psi_\text{A} \rangle \in \mathcal{H}_\text{A}$ and $| \psi_\text{B} \rangle \in \mathcal{H}_\text{B}$ and perform the measurement of the observable $\hat{O}_\text{A}$ and $\hat{O}_\text{B}$. The global Hilbert space, state and observable are the respective tensor product $\mathcal{H}_\text{A} \otimes \mathcal{H}_\text{B}$, $|\psi_\text{A} \rangle \otimes | \psi_\text{B} \rangle$ and $\hat{O}_\text{A} \otimes \hat{O}_\text{B}$.

A similar factorization also occurs in the Newton-Wigner and the AQFT schemes. In particular, the global Hilbert space factorizes into local Hilbert spaces, i.e., $\mathcal{H}_\text{M} = \bigotimes_i \mathcal{H}_\text{M}(\mathcal{V}_i) $. Hence, the two laboratories A and B can be represented by local fields in the respective regions $\mathcal{V}_\text{A}$ and $\mathcal{V}_\text{B}$. Here, we use a unifying notation for both schemes by indicating local Hilbert spaces as $\mathcal{H}_\text{M}(\mathcal{V}_i) $. Depending on the circumstances, if we are referring to the Newton-Wigner scheme, then $\mathcal{H}_\text{M}(\mathcal{V}_i) =  \mathcal{H}_\text{M}^\text{NW}(\mathcal{V})$; conversely, for the AQFT scheme, $\mathcal{H}_\text{M}(\mathcal{V}_i) =  \mathcal{H}_\text{M}^\text{AQFT}(\mathcal{V})$.

The factorization of $\mathcal{H}_\text{M}$ into $\mathcal{H}_\text{M}(\mathcal{V}_\text{A}) \otimes \mathcal{H}_\text{M}(\mathcal{V}_\text{B}) \otimes \dots $ allows experimenters in $\mathcal{V}_\text{A}$ and $\mathcal{V}_\text{B}$ to independently prepare and measure states in their own bounded regions. The fact that the experimenter in $\mathcal{V}_\text{A}$ is able to perform measurements independently from $\mathcal{V}_\text{B}$ is made possible by local operators in $\mathcal{V}_\text{A}$ which act as an identity on $\mathcal{H}(\mathcal{V}_\text{B})$.

We remark that the only preparations in $\mathcal{V}_\text{A}$ that are guaranteed to not affect measurements in $\mathcal{V}_\text{B}$ are nonselective. Notwithstanding the factorization $\mathcal{H}_\text{M} = \mathcal{H}_\text{M}(\mathcal{V}_\text{A}) \otimes \mathcal{H}_\text{M}(\mathcal{V}_\text{B}) \otimes \dots $, selective operations may still lead to nonlocal effects as a consequence of the nonunitary state update. The problem has been discussed in Sec.~\ref{solving_the_paradox} for the case of the AQFT scheme. In particular, we showed that the strict localization property of states is not always satisfied as a result of the Reeh-Schlieder theorem.

In this subsection, we will demonstrate that the Newton-Wigner scheme is not affected by these nonlocal effects. In particular, we will show that in the Newton-Wigner scheme the strict localization property is always satisfied and, hence, local measurements in $\mathcal{V}_\text{B}$ are independent of selective preparations of states in $\mathcal{V}_\text{A}$.

\subsubsection{Intrinsic notion of localization}

Due to the factorization of the global Fock state $\mathcal{H}_\text{M} = \bigotimes_i \mathcal{H}_\text{M}^\text{NW}(\mathcal{V}_i)$ and the global vacuum $| 0_{\text{M}} \rangle = \bigotimes_i | 0_{\text{M}}(\mathcal{V}_i) \rangle$ in the Newton-Wigner scheme, we find that any state that is localized in  $\mathcal{V}$ can be written as
\begin{equation}\label{NW_localization_decomposition}
| \phi \rangle = \hat{O} | 0_{\text{M}}(\mathcal{V}) \rangle \otimes \left[ \bigotimes_i | 0_{\text{M}}(\mathcal{V}_i) \rangle \right],
\end{equation}
where, in this case, $\{ \mathcal{V}_i \}$ is a partition of $\mathbb{R}^3 \setminus \mathcal{V}$ and $\hat{O}$ is an operator acting on $ \mathcal{H}_\text{M}^\text{NW}(\mathcal{V})$. The same factorization does not occur for localized states in the AQFT scheme, because the global vacuum does not factorize in $\mathcal{H}_\text{M} = \bigotimes_i \mathcal{H}_\text{M}^\text{AQFT}(\mathcal{V}_i)$.

Equation (\ref{NW_localization_decomposition}) gives a definition of localized states in terms of a local Hilbert space $ \mathcal{H}_\text{M}^\text{NW}(\mathcal{V})$ as the domain of the state. Intuitively, we say that the state $| \phi \rangle$ lives in $ \mathcal{H}_\text{M}^\text{NW}(\mathcal{V})$, while it appears indistinguishable from the vacuum outside the region $\mathcal{V}$. Such a notion of localization can be compared to the one provided in Sec.~\ref{NewtonWigner_scheme_in_QFT} by means of local operators $\hat{O} \in \mathfrak{A}_\text{M}^\text{NW}(\mathcal{V})$ acting on the vacuum $| 0_\text{M} \rangle$. The physical interpretation was that the local state is the result of local operations occurring in $\mathcal{V}$ over the vacuum background $| 0_\text{M} \rangle$. Conversely, Eq.~(\ref{NW_localization_decomposition}) gives a notion of localization that is independent of the preparation of the state.

The intrinsic notion of localization provided by Eq.~(\ref{NW_localization_decomposition}) is missing in the AQFT scheme, which can therefore only rely on the interpretation of localized states in terms of local preparations over the vacuum $| 0_\text{M} \rangle$. In that case, the local state $| \phi \rangle = \hat{O} | 0_\text{M} \rangle$ with $\hat{O} \in \mathfrak{A}_\text{M}^\text{AQFT}(\mathcal{V})$ cannot be said to live inside the local Hilbert space $ \mathcal{H}_\text{M}^\text{AQFT}(\mathcal{V})$.

\subsubsection{Strict localization property and Alice-Bob scenario}

As a consequence of Eq.~(\ref{NW_localization_decomposition}) the strict localization property is always satisfied in the Newton-Wigner scheme, in the sense that any state $| \phi \rangle $ localized in $\mathcal{V}_\text{A}$ with respect to the Newton-Wigner scheme always appears indistinguishable from the vacuum $| 0_\text{M} \rangle $ in any other separated region $\mathcal{V}_\text{B}$. Explicitly this means that for any $| \phi \rangle = \hat{O}_\text{A} | 0_\text{M} \rangle $, with $\hat{O}_\text{A} \in \mathfrak{A}_\text{M}^\text{NW}(\mathcal{V}_\text{A})$, and for any observable $\hat{O}_\text{B} \in \mathfrak{A}_\text{M}^\text{NW}(\mathcal{V}_\text{B})$, Eq.~(\ref{KnightLicht_property}) holds. The proof comes from the factorization of $| \phi \rangle $ and $\hat{O}_\text{B}$ in $\mathcal{H}_\text{M} =  \mathcal{H}_\text{M}^\text{NW}(\mathcal{V}_\text{A}) \otimes \mathcal{H}_\text{M}^\text{NW}(\mathcal{V}_\text{B}) \otimes \dots$ and from the normalization condition $1 = \langle \phi | \phi \rangle = \langle 0_\text{M}(\mathcal{V}_\text{A}) | \hat{O}_\text{A}^\dagger \hat{O}_\text{A} | 0_\text{M} (\mathcal{V}_\text{A}) \rangle$.

The result may be understood in terms of the Alice-Bob scenario presented in Sec.~\ref{solving_the_paradox} for the AQFT scheme. An experimenter (Alice) prepares a state over the vacuum $| 0_\text{M} \rangle$ by means of local Newton-Wigner operators $\hat{a}_\text{NW}(\vec{x}) \in \mathfrak{A}_\text{M}^\text{NW}(\mathcal{V}_\text{A})$. Another experimenter (Bob) performs measurements in a separated region by means of local Newton-Wigner operators $\hat{a}_\text{NW}(\vec{x}) \in \mathfrak{A}_\text{M}^\text{NW}(\mathcal{V}_\text{B})$. From Eq.~(\ref{KnightLicht_property}), we deduce that the outcomes of Bob's measurements are independent of the preparation of the state by Alice.

Hereafter, we will refer to this scenario as the Newton-Wigner Alice-Bob experiment to not get confused with AQFT Alice-Bob experiment presented in Sec.~\ref{solving_the_paradox}. The discussion of Sec.~\ref{solving_the_paradox} led to the conclusion that not all the states that are localized with respect to the AQFT scheme are also strictly localized, at variance with the Newton-Wigner scheme.

The two Alice-Bob scenarios lead to different results. One may ask which one would be applicable in real experiments. We have already remarked that the AQFT localization scheme is fundamental and entails causal processes. Hence, we may be prone to consider the AQFT Alice-Bob experiment as the most relevant one, while the Newton-Wigner Alice-Bob scenario should not be understood as having a genuine notion of localization. The processes considered in the Newton-Wigner case are physically realizable, in the sense that the state prepared by Alice and the observable used by Bob exist; however, they can hardly be interpreted as genuinely local. If, for instance, Alice uses an emitter to produce the state over the vacuum, the correct way to describe the QFT interaction between the device and the field is by means of local unitary evolution, with the AQFT notion of localization. This would motivate the idea of considering the AQFT Alice-Bob scenario as the one genuinely describing two macroscopic experimenter living in disjoint regions of space.

\subsubsection{Orthogonality condition}

By means of Eq.~(\ref{a_tilde_phi_Pi}) we found that the Newton-Wigner and the AQFT schemes are incompatible. This seems to contradict the idea of generality advocated by Newton and Wigner in Ref.~\cite{RevModPhys.21.400}. In particular, the two authors only considered a minimal set of physically motivated postulates to define the notion of localization in RQM. At least one of the postulates for the Newton-Wigner localization must have been ignored in the AQFT scheme.

The missing assumption is the orthogonality of states in different spatial positions. To see this, consider the states $| \phi_\text{A} \rangle = \hat{\phi}(0,\vec{x}_\text{A}) | 0_\text{M} \rangle \in \mathfrak{A}_\text{M}^\text{AQFT}(\vec{x}_\text{A}) $ and $| \phi_\text{B} \rangle = \hat{\phi}(0,\vec{x}_\text{B}) | 0_\text{M} \rangle  \in \mathfrak{A}_\text{M}^\text{AQFT}(\vec{x}_\text{B})$, which are respectively localized in $\vec{x}_\text{A}$ and $\vec{x}_\text{B}$ according to the AQFT scheme. Assume that the two points are different, $\vec{x}_\text{A} \neq \vec{x}_\text{B}$, and, hence, the states are localized in disjoint regions. By following Newton and Wigner's assumptions, one would expect that $\langle \phi_\text{A} | \phi_\text{B} \rangle = 0$; however, this is not true. The inequality $\langle \phi_\text{A} | \phi_\text{B} \rangle \neq 0$ can be checked by computing the $2$-point correlation function
\begin{equation}
\langle 0_\text{M} | \hat{\phi}(0,\vec{x}_\text{A}) \hat{\phi}(0,\vec{x}_\text{B}) | 0_\text{M} \rangle  = \frac{\hbar}{(2\pi)^3 }  \int_{\mathbb{R}^3} d^3 k \frac{e^{ i\vec{k} \cdot (\vec{x}_\text{A}-\vec{x}_\text{B})}}{2 \omega(\vec{k})},
\end{equation}
which is different form zero.

We recognize that the orthogonality condition is not met by the AQFT localization. Consequently, the probability transition associated to the two spatially separated states $|\langle \phi_\text{A} | \phi_\text{B} \rangle|^2$ is different form zero. The result is apparently paradoxical, as it seems that there is a nonvanishing probability for a local state to be found in another disjoint region \cite{Fleming2000-FLERMN}. The paradox is resolved if we assume that in AQFT the definition of localized states can only be given in terms of local preparations over the vacuum $| 0_\text{M} \rangle$.

At the beginning of Sec.~\ref{AQFT_localization_scheme}, we said that $| \phi \rangle$ is a localized state with respect to the AQFT scheme if it is the result of local operations on $| 0_\text{M} \rangle$. The same definition was also provided for the Newton-Wigner scheme in Sec.~\ref{NewtonWigner_scheme_in_QFT}. Then, we found that an intrinsic notion of localization naturally occurs due to Eq.~(\ref{NW_localization_decomposition}), which provides a definition of localized states as elements of the local algebras $ \mathcal{H}_\text{M}^\text{NW}(\mathcal{V})$. This notion of localization only appears in the Newton-Wigner scheme. Conversely, in the AQFT scheme, Eq.~(\ref{NW_localization_decomposition}) does not hold because local vacuum states do not exist; hence, the definition of localized states can only be provided in terms of local preparations over the vacuum $| 0_\text{M} \rangle$.

The quantity $|\langle \psi_\text{A} | \psi_\text{B} \rangle|^2$ should be interpreted as the probability for a state locally prepared in $\vec{x}_\text{A}$ to turn into a state that can be locally prepared in $\vec{x}_\text{B}$. The fact that $\langle \psi_\text{A} | \psi_\text{B} \rangle$ is different from zero implies that $| \psi_\text{B} \rangle$ can be obtained as an outcome of the projective measurement $| \psi_\text{B} \rangle \langle \psi_\text{B} | $ on $| \psi_\text{A} \rangle$, i.e., $| \psi_\text{B} \rangle \propto | \psi_\text{B} \rangle \langle \psi_\text{B} |\psi_\text{A} \rangle$. This means that $| \psi_\text{B} \rangle$ may be prepared in both of the following ways: either (i) via local operation $\hat{\phi}(0,\vec{x}_\text{B})$ on $| 0_\text{M} \rangle$ in $\vec{x}_\text{B}$ or (ii) via local operation $\hat{\phi}(0,\vec{x}_\text{A})$ on $| 0_\text{M} \rangle$ in $\vec{x}_\text{A} $ followed by the projective measurement $| \psi_\text{B} \rangle \langle \psi_\text{B} | $. The apparent paradox comes from the unexpected compatibility between (i) and (ii) notwithstanding the fact that $\vec{x}_\text{A} $ and $\vec{x}_\text{B} $ are different points. However, notice that the operator $| \psi_\text{B} \rangle \langle \psi_\text{B} | $ is nonlocal, i.e., $| \psi_\text{B} \rangle \langle \psi_\text{B} | \notin \mathfrak{A}_\text{M}^\text{AQFT}(\vec{x}_\text{B})$. Due to the nonlocality of the projective operation, one should not be surprised by the compatibility between (i) and (ii).

\subsection{Newton-Wigner and modal scheme}\label{Comparison_between_NewtonWigner_and_modal_schemes}

In this subsection, we detail the differences between the Newton-Wigner and the modal scheme.

As remarked in Sec.~\ref{Comparison_between_the_NewtonWigner_and_the_AQFT_schemes}, the variable $\vec{x}$ in the Newton-Wigner scheme is not a space coordinate and, hence, it does not entail any genuine notion of position. Conversely, in the modal scheme, $\vec{x}$ appears as a space coordinate for the positive frequency modes $f(\vec{k},t,\vec{x})$ that are solutions of the Klein-Gordon equation (\ref{Klein_Gordon}). The representatives $f(\vec{k},t,\vec{x})$ inherit the fundamental notion of spacetime event $(t,\vec{x})$ from the QFT framework. Hence, in analogy to the AQFT scheme, we say that the variable $\vec{x}$ entails a genuine notion of position.

A feature that both localization schemes share is the acausal spreading of the wave functions, which was discussed in Secs.~\ref{NewtonWigner_localization_scheme} and \ref{Modal_localization_scheme}. In particular, the superluminal effect in the Newton-Wigner is a result of the Hegerfeldt theorem [Sec.~\ref{Hegerfeldt_theorem}], which is a no-go theorem for localization schemes that simultaneously satisfy (i) causality, (ii) positivity of energy and (iii) orthogonality condition for states in disjoint spatial regions. The acausal spreading of the modal wave functions $\phi_n (t, \textbf{x}_n)$, instead, was proved in Sec.~\ref{Modal_localization_scheme} by means of the non-localizability of positive frequency modes for finite intervals of time. The Hegerfeldt theorem cannot be applied in this case because the assumption (iii) is missing.

The lack of assumption (iii) can be proved by considering two single particle states $| \phi_\text{A} \rangle = \hat{a}_\text{mod}^\dagger (\vec{x}_\text{A}) | 0_\text{M} \rangle$ and $| \phi_\text{B} \rangle = \hat{a}_\text{mod}^\dagger (\vec{x}_\text{B}) | 0_\text{M} \rangle$ respectively localized in $\vec{x}_\text{A}$ and $\vec{x}_\text{B}$. By using Eq.~(\ref{a_mod_a}) and the commutation relations (\ref{Minkowski_canonical_commutation}), we obtain
\begin{equation}
\langle \phi_\text{A} | \phi_\text{B} \rangle  = \int_{\mathbb{R}^3} d^3 k  \frac{\hbar \omega(\vec{k})}{mc^2}  \frac{e^{i \vec{k} \cdot (\vec{x}_\text{A} - \vec{x_\text{B}})}}{(2 \pi)^3} , 
\end{equation}
which means that $\langle \phi_\text{A} | \phi_\text{B} \rangle$ different from zero even if $\vec{x}_\text{A} \neq \vec{x}_\text{B}$. At variance with the Newton-Wigner scheme, the modal scheme admits non-orthogonal states that are localized in disjoint spatial regions.

By using again Eq.~(\ref{a_mod_a}) and the commutation relations (\ref{Minkowski_canonical_commutation}) one can also prove that
\begin{subequations}\label{a_mod_commutation}
\begin{align}
& [\hat{a}_\text{mod}(\vec{x}), \hat{a}_\text{mod}^\dagger(\vec{x}') ] = \int_{\mathbb{R}^3} d^3 k  \frac{\hbar \omega(\vec{k})}{ mc^2}  \frac{e^{i \vec{k} \cdot (\vec{x} - \vec{x}')}}{(2 \pi)^3} , \label{a_mod_commutation_a}\\
& [\hat{a}_\text{mod}(\vec{x}), \hat{a}_\text{mod}(\vec{x}') ] = 0.
\end{align}
\end{subequations}
Equation (\ref{a_mod_commutation_a}) implies that the modal operators $\hat{a}_\text{mod}(\vec{x})$ and $\hat{a}_\text{mod}^\dagger(\vec{x})$ cannot be interpreted as local annihilation and creation operators, at variance with the Newton-Wigner operators $\hat{a}_\text{NW}(\vec{x})$ and $\hat{a}_\text{NW}^\dagger(\vec{x})$. It also implies that operators localized in disjoint spatial regions generally do not commute. Explicitly, this means that there are operators $\hat{O}_\text{A} \in \mathfrak{A}_\text{M}^\text{mod}(\vec{x}_\text{A})$ and $\hat{O}_\text{B} \in \mathfrak{A}_\text{M}^\text{mod}(\vec{x}_\text{B})$ such that
\begin{equation}\label{O_A_O_B_neq_0}
[\hat{O}_\text{A}, \hat{O}_\text{B}] \neq 0, 
\end{equation}
even if $\vec{x}_\text{A} \neq \vec{x}_\text{B}$.

As a consequence of Eq.~(\ref{O_A_O_B_neq_0}), the global Hilbert space does not factorize into local Hilbert spaces. This means that the modal localization scheme lacks of the notion of independence via tensor product of local Hilbert spaces. Also, local Fock spaces do not exist and the global vacuum cannot factorize into local vacua, since local Hilbert spaces are nonexistent in the first place. Consequently, the strict localization property is not guaranteed in the modal scheme, at variance with the Newton-Wigner scheme.

All of these differences show that the two localization schemes are incompatible. More generally, it is possible to demonstrate that any operator or state that is localized with respect to one scheme it is not localized with respect to the other. The proof is similar to the one provided in the previous subsection for the Newton-Wigner and the AQFT scheme. Consider the operators $\hat{a}_\text{mod} (\vec{x}) $ and $\hat{a}_\text{NW}(\vec{x})$, which generate the respective local algebras $\mathfrak{A}_\text{M}^\text{mod}(\vec{x})$ and $\mathfrak{A}_\text{M}^\text{NW}(\vec{x})$. Use their definitions [Eqs.~(\ref{a_NW}) and (\ref{a_mod_a})] to compute
\begin{equation}\label{a_mod_a_NW}
\hat{a}_\text{mod} (\vec{x})  =  \int_{\mathbb{R}^3} d^3 x' f_{\text{NW} \mapsto \text{mod}}(\vec{x}-\vec{x}')  \hat{a}_\text{NW}(\vec{x}'),
\end{equation}
with
\begin{equation}\label{f_NW_mod}
f_{\text{NW} \mapsto \text{mod}}(\vec{x}) = \int_{\mathbb{R}^3} d^3 k  \sqrt{\frac{\hbar \omega(\vec{k})}{mc^2}} \frac{e^{i \vec{k} \cdot \vec{x}}}{(2 \pi)^3}.
\end{equation}
Notice that the support of $f_{\text{NW} \mapsto \text{mod}}(\vec{x})$ is $\mathbb{R}^3$, which means that $\hat{a}_\text{mod} (\vec{x}) $ is nonlocal with respect to the Newton-Wigner scheme. This proves that $\mathfrak{A}_\text{M}^\text{mod}(\vec{x}) \neq \mathfrak{A}_\text{M}^\text{NW}(\vec{x})$ with the consequent incompatibility between the two schemes.

\subsection{AQFT and modal scheme}\label{Comparison_between_modal_and_AQFT_schemes}

In Secs.~\ref{Comparison_between_the_NewtonWigner_and_the_AQFT_schemes} and \ref{Comparison_between_NewtonWigner_and_modal_schemes}, we detailed the relevant features of the AQFT and the modal scheme, respectively, and we made a comparison with the Newton-Wigner scheme. In this subsection, instead, we use the results of Secs.~\ref{Comparison_between_the_NewtonWigner_and_the_AQFT_schemes} and \ref{Comparison_between_NewtonWigner_and_modal_schemes} to show the differences between the AQFT and the modal scheme.

The incompatibility between the two schemes can be proved by comparing the respective algebras $\mathfrak{A}_\text{M}^\text{AQFT}(\vec{x})$ and $\mathfrak{A}_\text{M}^\text{mod}(\vec{x})$. By plugging Eq.~(\ref{a_tilde_phi_Pi}) in Eq.~(\ref{a_mod_a_NW}) we obtain
\begin{equation}
\hat{a}_\text{mod} (\vec{x})  = \int_{\mathbb{R}^3} d^3 x' \left[ f_{\hat{\phi}\mapsto \text{mod}}(\vec{x} - \vec{x}') \hat{\phi}(0, \vec{x}') + f_{\hat{\pi}\mapsto \text{mod}}(\vec{x} - \vec{x}') \hat{\pi}(0, \vec{x}')  \right],
\end{equation}
with
\begin{align}
& f_{\hat{\phi}\mapsto \text{mod}}(\vec{x}) = \int_{\mathbb{R}^3} d^3 k  \frac{\omega (\vec{k}) e^{i \vec{k} \cdot \vec{x}}}{ (2 \pi)^3 \sqrt{2 m c^2}}, & f_{\hat{\pi}\mapsto \text{mod}}(\vec{x}) = \frac{-i}{\sqrt{2 m c^2}} \delta^3(\vec{x}).
\end{align}
The fact that $f_{\hat{\phi}\mapsto \text{mod}}(\vec{x})$ has support in the entire space $\mathbb{R}^3$ implies that the modal operators $\hat{a}_\text{mod} (\vec{x})$ are nonlocal with respect to the AQFT scheme, i.e., $\hat{a}_\text{mod} (\vec{x}) \notin \mathfrak{A}_\text{M}^\text{AQFT}(\vec{x})$, which means that $\mathfrak{A}_\text{M}^\text{mod}(\vec{x}) \neq \mathfrak{A}_\text{M}^\text{AQFT}(\vec{x})$.

As remarked in Secs.~\ref{Comparison_between_the_NewtonWigner_and_the_AQFT_schemes} and \ref{Comparison_between_NewtonWigner_and_modal_schemes}, both the AQFT and the modal scheme are characterized by a genuine position coordinate $\vec{x}$ representing the underling Minkowski spacetime. However, at variance with the AQFT scheme, the modal scheme cannot be considered fundamental in nature. This is due to the acausal effects produced by the superluminal spreading of the wave functions. Also, the microcausality axiom does not hold, as it can be noticed from Eq.~(\ref{O_A_O_B_neq_0}). The non commutativity of operators in disjoint spatial regions does not guarantee the statistical independence of measurements in those regions. For these reasons, the modal scheme does not give a genuine notion of localization.

Due to Eq.~(\ref{O_A_O_B_neq_0}), the strict localization property is not always satisfied, which means that Eq.~(\ref{KnightLicht_property}) does not hold for any $\hat{O}_\text{B} \in \mathfrak{A}_\text{M}^\text{mod}(\mathcal{V}_\text{B})$ and any $| \phi \rangle = \hat{O}_\text{A} | 0_\text{A} \rangle$, with $\hat{O}_\text{A} \in \mathfrak{A}_\text{M}^\text{mod}(\mathcal{V}_\text{A})$ and $\mathcal{V}_\text{A} \cap \mathcal{V}_\text{B} = \varnothing$. This also occurs in the AQFT scheme with $\hat{O}_\text{A} \in \mathfrak{A}_\text{M}^\text{AQFT}(\mathcal{V}_\text{A})$ and $\hat{O}_\text{B} \in \mathfrak{A}_\text{M}^\text{AQFT}(\mathcal{V}_\text{B})$, as a consequence of the Reeh-Schlieder theorem [Sec.~\ref{ReehSchlieder_theorem}]. However, in Sec.~\ref{solving_the_paradox}, we showed that the unitarity of the local operator $\hat{O}_\text{B}$ guarantees the validity of the strict localization property (\ref{KnightLicht_property}) in the AQFT scheme. Crucially, the commutation relation $[\hat{O}_\text{A}, \hat{O}_\text{B}] = 0$ and the definition of unitary operators were used to derive Eq.~(\ref{KnightLicht_property}). In the case of the modal localization scheme, the operators $\hat{O}_\text{A}$ and $\hat{O}_\text{B}$ do not commute, which means that Eq.~(\ref{KnightLicht_property}) is not guaranteed anymore.

\section{Localization in the nonrelativistic regime}\label{Localization_in_NRQM}

In the previous section we detailed three localization schemes for the fully relativistic QFT. Among them, only the AQFT scheme gives a genuine notion of localization. In particular, any local experiment can only be faithfully described in the framework of AQFT. The Newton-Wigner and the modal scheme, instead, appear more as mathematical artifices not suited for a genuine description of local phenomena.

In this section, we consider the nonrelativistic limit of QFT and we show that the three localization scheme converge to each other. Hence, in such a regime, the Newton-Wigner and the modal scheme acquire the fundamental notion of localization entailed by the AQFT framework. We say that the two schemes are ``fundamentalized'' by the nonrelativistic limit.

\begin{figure}
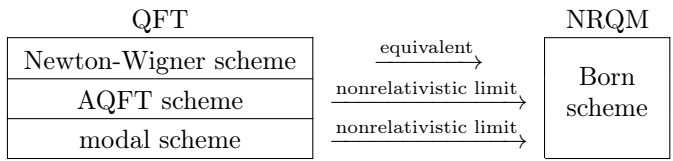

\begin{center}
\begin{tabular}{| c | c | c |}
\multicolumn{1}{c}{QFT} &  \multicolumn{1}{l}{} & \multicolumn{1}{c}{NRQM} \\
\cline{1-1} \cline{3-3}
Newton-Wigner scheme & $\xrightarrow[]{\text{equivalent}}$  & \multirow{3}{3.5em}{\centering Born scheme} \\
\cline{1-1}
AQFT scheme & $\xrightarrow[]{\text{nonrelativistic limit}}$ &  \\
\cline{1-1}
modal scheme & $\xrightarrow[]{\text{nonrelativistic limit}}$ &  \\
\cline{1-1} \cline{3-3}
\end{tabular}
\end{center}
\caption{Localization schemes in the relativistic (QFT) and the nonrelativistic (NRQM) theory. The Newton-Wigner and the Born scheme are equivalent, whereas the AQFT and the modal scheme converge to the Born scheme in the nonrelativistic limit.} \label{Localization_in_NRQM_Table}
\end{figure}

To obtain this result, we study NRQM and the notion of localization prescribed by the nonrelativistic theory. We remark that in NRQM the fundamental objects are the first-quantized position and momentum operator, $\hat{x}^i$ and $\hat{k}^i$. The notion of localization in NRQM is based on the definition of $\hat{x}^i$ and on the Born interpretation of quantum mechanics, according to which the modulo square of wave functions gives the probability density to find particles. We demonstrate that such a localization program is equivalent to the Newton-Wigner scheme as they both rely on local creators and annihilators. Then, by following Refs.~\cite{PhysRevD.90.065032, Papageorgiou_2019}, we show that both the AQFT and the modal scheme converge to the Born scheme in the nonrelativistic limit. These results are summarized by Fig.~\ref{Localization_in_NRQM_Table}.

Due to the converge between the Newton-Wigner and the AQFT, we prove that the nonlocal effect described in Sec.~\ref{solving_the_paradox} is suppressed by the nonrelativistic limit. In particular, we show that any state localized in a space region $\mathcal{V}_\text{A}$ is also strictly localized in $\mathcal{V}_\text{A}$, in the sense that it does not affect any measurement conducted in some disjoint region $\mathcal{V}_\text{B}$.

Similarly to Sec.~\ref{Comparison_between_the_NewtonWigner_and_the_AQFT_schemes}, we detail this result by considering an Alice-Bob scenario, in which Alice prepares the state $| \psi \rangle$ in $\mathcal{V}_\text{A}$ and Bob measures $\hat{O}_\text{B}$ in $\mathcal{V}_\text{B}$. At variance with Sec.~\ref{Comparison_between_the_NewtonWigner_and_the_AQFT_schemes}, here, $| \psi \rangle$ and $\hat{O}_\text{B}$ are nonrelativistic and, hence, can be equivalently localized with respect to any scheme. The nonrelativistic Alice-Bob scenario inherits from the AQFT scheme the fundamental notion of localization, in the sense that, regardless of the scheme used to describe the experiment, one always obtains an approximately genuine description of the local phenomena in the two regions $\mathcal{V}_\text{A}$ and $\mathcal{V}_\text{B}$. Also, the strict localization property of the Newton-Wigner Alice-Bob scenario emerges as an independence between the preparation of $| \psi \rangle$ and the measurements of $\hat{O}_\text{B}$.

The section is organized as follows. In Sec.~\ref{Born_localization}, we present the Born scheme, which gives the familiar description of localized states in NRQM. In Sec.~\ref{Born_localization_is_NewtonWigner_localization}, we show the equivalence between the Newton-Wigner and the Born scheme; whereas, in Secs.~\ref{Comparison_with_the_relativistic_theory} and \ref{Convergence_of_the_modal_scheme_to_the_Born_scheme} we demonstrate the convergence of the AQFT and the modal scheme, respectively, to the Born-Newton-Wigner scheme. In Sec.~\ref{Strict_localization_in_the_nonrelativistic_limit} we use this convergence to show that the Reeh-Schlieder nonlocal effect is suppressed in the nonrelativistic limit and the strict localization property always holds; we detail this result by considering the Alice-Bob scenario in the nonrelativistic regime.

\subsection{Born localization scheme}\label{Born_localization}

In NRQM, states are localized according to the Born localization principle which assumes that the probability density of finding the system in any space point is the square of the amplitude of its wave function. Hence, particles are localized in the support of their wave functions and they are orthogonal to each other if the localization occurs in disjoint spatial regions. 

In the first-quantized theory, the algebra is generated by the observables position $\hat{x}^i$ and and momentum $\hat{k}^i$, satisfying the canonical commutation relation
\begin{equation}\label{x_p_CCR}
[\hat{x}_i, \hat{k}_j] = i \delta_{i j},
\end{equation}
and by eventual internal degrees of freedom (e.g., spin), which we will ignore for the sake of simplicity. The wave function $\psi(\vec{x}) $ of any state $| \psi \rangle$ can be derived from the eigenstates of $\hat{\vec{x}}$, such that $\psi(\vec{x}) = \langle \vec{x}_\text{B} | \psi \rangle $, where $\hat{x}_i | \vec{x}_\text{B} \rangle = x_i | \vec{x}_\text{B} \rangle$.

Wave functions in the momentum space can be obtained by means of states with defined momentum $| \vec{k} \rangle$, which are defined by $ \hat{k}_i | \vec{k} \rangle = k_i | \vec{k} \rangle$. The identity relating $| \vec{k} \rangle$ to the states with defined position $| \vec{x}_\text{B} \rangle$ is
\begin{equation}\label{x_state_B}
|  \vec{x}_\text{B} \rangle = \int_{\mathbb{R}^3} d^3 k \frac{e^{-i \vec{k} \cdot \vec{x}}}{\sqrt{(2 \pi)^3}} | \vec{k} \rangle,
\end{equation}
which is the Fourier transform of $| \vec{k} \rangle$. One can use Eq.~(\ref{x_state_B}) to switch from the representation of states in the position space to their representation in the momentum space.

From the normalization of $| \psi \rangle$ (i.e., $\langle \psi | \psi \rangle = 1$) and the orthogonality condition $\langle \vec{x}_\text{B} | \vec{x}_\text{B}' \rangle = \delta^3 (\vec{x}-\vec{x}')$, one obtains the familiar result for wave functions
\begin{align}\label{psi_normalization_product}
& \int_{\mathbb{R}^3} d^3 x | \psi(\vec{x}) |^2 = 1, & \langle \psi | \psi' \rangle =  \int_{\mathbb{R}^3} d^3 x \psi^*(\vec{x}) \psi'(\vec{x}) .
\end{align}
Equation (\ref{psi_normalization_product}) captures the idea that $\psi(\vec{x})$ is the probability amplitude of finding the particle in $\vec{x}$, with the consequent interpretation of the support of $\psi(\vec{x})$ as the region of localization for the particle. For any couple of states $| \psi \rangle$ and $| \psi' \rangle$, if $\psi(\vec{x})$ and $\psi'(\vec{x})$ have disjoint support, they are orthogonal to each other.

In the second-quantized theory, the state $| \vec{x}_\text{B} \rangle $ appears as a single particle with defined position. It is defined as
\begin{equation}
| \vec{x}_\text{B} \rangle = \hat{a}_\text{B}^\dagger(\vec{x}) | 0 \rangle,
\end{equation}
with $\hat{a}_\text{B}^\dagger(\vec{x})$ as the creator of the particle in $\vec{x}$ and $| 0 \rangle$ as the vacuum state. All the relevant features of the Born localization scheme in second quantization are inherited from the first-quantized theory. This includes the definition of localized states in terms of compactly supported wave functions and the orthogonality condition for states that are localized in disjoint regions.

By definition, the operators $\hat{a}_\text{B}(\vec{x})$ and $\hat{a}_\text{B}^\dagger(\vec{x})$ satisfy the canonical commutation relation
\begin{align}\label{a_commutation}
& [\hat{a}_\text{B}(\vec{x}), \hat{a}_\text{B}^\dagger(\vec{x}') ] = \delta^3(\vec{x}-\vec{x}'), & [\hat{a}_\text{B}(\vec{x}), \hat{a}_\text{B}(\vec{x}') ] = 0.
\end{align}
As a result of Eq.~(\ref{a_commutation}), the global Fock space factorizes into local Fock spaces $\mathcal{H} = \bigotimes_i \mathcal{H}(\mathcal{V}_i)$ and the global vacuum factorizes into the local vacua $| 0 \rangle = \bigotimes_i | 0 (\mathcal{V}_i) \rangle $. Any state $| \psi \rangle$ localized in $\mathcal{V}$ is equivalently represented by an element of $\mathcal{H}(\mathcal{V})$ such that
\begin{equation}
| \psi \rangle = | \psi (\mathcal{V})  \rangle \otimes \left[ \bigotimes_i | 0 (\mathcal{V}_i) \rangle \right],
\end{equation}
where $| \psi (\mathcal{V}) \rangle$ is the element of $\mathcal{H}(\mathcal{V})$ and $\{ \mathcal{V}_i \}$ is a partition of $\mathbb{R}^3 \setminus \mathcal{V}$. In this sense, we say that the localized state $| \psi \rangle$ lives in the local Fock space $\mathcal{H}(\mathcal{V})$.

\subsection{Equivalence between the Newton-Wigner and the Born scheme}\label{Born_localization_is_NewtonWigner_localization}

In Sec.~\ref{NewtonWigner_localization_scheme}, we presented the Newton-Wigner scheme as an attempt to formalize the notion of localization in RQM and in QFT. Nonrelativistic theories, instead, are described by the Born scheme, which was introduced in Sec.~\ref{Born_localization}.

By comparing the two localization schemes, it is straightforward to see that they are equivalent. In particular, they are both based on the existence of local creation and annihilation operators. All the features found for the Newton-Wigner scheme in Secs.~\ref{NewtonWigner_localization_scheme} and \ref{Comparison_between_the_NewtonWigner_and_the_AQFT_schemes} also apply to the Born localization scheme.

Conceptually, the only difference is given by the regime in which they are defined. The Born scheme was originally introduced in nonrelativistic theories (i.e., NRQM), whereas the Newton-Wigner scheme was conceived in relativistic physics (i.e., RQM and QFT). The original attempt by Newton and Wigner was precisely to recover the Born interpretation of localized states in the relativistic regime. Consequently, the assumptions postulated by the authors are also met by the Born scheme in NRQM. The results of their work \cite{RevModPhys.21.400} are not only applicable in the relativistic theory but can also be understood in the context of NRQM.

By seeing NRQM as the nonrelativistic limit of the corresponding field theory, one can embed the Born scheme into QFT. This leads to a complete equivalence between the Newton-Wigner and the Born scheme in the nonrelativistic regime of quantum fields.

The equivalence is made possible by the fact that both theories are provided with an unifying notion of single particles with defined momentum $| \vec{k} \rangle$. In NRQM, these states are defined as the eigenstates of the first-quantized operator $\hat{\vec{k}}$, whereas in QFT they are associated to a basis of positive frequency solutions of the Klein-Gordon equation (\ref{Klein_Gordon}). The two definitions of $| \vec{k} \rangle$ are unified by the idea that they both represent the same physical object.

By following the Newton-Wigner approach, we define the single particle localized in $\vec{x}$ by means of Eq.~(\ref{x_state}). The same definition applies to the state $| \vec{x}_\text{B} \rangle$ with respect to the Born localization scheme. By comparing Eq.~(\ref{x_state}) with Eq.~(\ref{x_state_B}) we find that $| \vec{x}_\text{B} \rangle = | \vec{x}_\text{NW} \rangle$. This proves the equivalence between the two localization schemes. 

\subsection{Convergence of the AQFT to the Born scheme}\label{Comparison_with_the_relativistic_theory}

In Sec.~\ref{Born_localization}, we remarked that NRQM is characterized by the Born notion of localization. NRQM, however, is not regarded as a fundamental theory of physics and it only comes from nonrelativistic approximations of QFT. Hence, one expects that the Born scheme actually emerges as the nonrelativistic limit of a more fundamental notion of localization properly defined in QFT.

In Sec.~\ref{Born_localization_is_NewtonWigner_localization}, we showed that the Born localization scheme is equivalent to the Newton-Wigner scheme. In Sec.~\ref{Comparison_between_localization_schemes}, we remarked that QFT has more than one localization scheme and that the Newton-Wigner scheme is in conflict with the notion of localization in AQFT. Between the two schemes, the latter is more fundamental than the former for the following reasons: (i) The genuine notion of position in QFT is given by the Minkowski spacetime upon which the algebra of fields is constructed, while the Newton-Wigner position operator is by no means associated to spacetime events on a manifold; (ii) The AQFT localization scheme is based on the microcausality of fields which forbids violation of causality and superluminal signaling, at variance with the Newton-Wigner which is affected by the instantaneous spreading of wave functions.

In summary, the AQFT scheme gives the fundamental notion of localization in QFT, while the Born scheme defines the localization in NRQM. The convergence between the two schemes is expected in the nonrelativistic limit as a consequence of the equivalence between NRQM and QFT in such a limit. In this section, we show that the Newton-Wigner and the AQFT scheme converge in the nonrelativistic regime. Due to the equivalence between the Newton-Wigner and the Born scheme, this also proves the convergence between the Born and the AQFT scheme.

\subsubsection{Classical versus quantum position}

Before showing the convergence between the two localization schemes, we want to discuss conceptual differences that seem to make them incompatible at any limit. We already remarked that, in AQFT scheme, the variable $\vec{x}$ labeling the fields $\hat{\phi}(0,\vec{x})$ and $\hat{\pi}(0,\vec{x})$ are coordinates representing classical events in the Minkowski spacetime; in this sense we say that the notion of localization in AQFT is classical. Conversely, in NRQM, the variable $\vec{x}$ is used as an index for second-quantized operators generated by the first-quantized position observable $\hat{\vec{x}}$; hence, the notion of localization is quantum. This leads to the apparent incompatibility between the two notions of localization. Why is the position quantum in NRQM and classical in QFT?

An answer to this question can be found by comparing the Galilean and the Poincar\'e group which are at the foundation of the nonrelativistic and the relativistic physics. In NRQM, the operators $\hat{k}^i$ and $m \hat{x}^i$ are, respectively, the generators of the translations and the Galilean boosts. Conversely, in relativistic theories, the Poincar\'e group is defined by the generators of translations $\hat{P}^\mu$, rotations $\hat{J}^i$ and Lorentzian boosts $\hat{K}^i$. It has been proven that in the nonrelativistic limit, the Poincar\'e group converges to the centrally-extended Galilean group and that the generator of Lorentzian boosts $\hat{K}^i$ is approximated by the generator of Galilean boosts $m \hat{x}^i$ \cite{4860f44e-649d-341b-9a70-b912b6531bea, 10.1063/1.524962}. Intuitively, this can be seen by noticing that for small momenta $|\vec{k}| \ll m c / \hbar$, Lorentzian boosts effectively act as Galilean boosts by transforming $\vec{k}$ linearly.

The upshot is that the operator $\hat{x}^i$ should not be interpreted as a quantized version of the Minkowski coordinate $x^i$, but as the limit of the Lorentzian boost operator $\hat{K}^i$ divided by the mass $m$. In the centrally-extended Galilean algebra, the operators $\hat{x}^i$ and $\hat{k}^i$ satisfy the canonical commutation relation (\ref{x_p_CCR}), which leads to the correct transformation rule for the position operator under space translation, i.e.,
\begin{equation}\label{x_translation}
\exp(i \vec{a} \cdot \hat{\vec{k}}) \hat{x}^i \exp(- i \vec{a} \cdot \hat{\vec{k}}) = \hat{x}^i + a^i.
\end{equation}
Consequently, the operator $\hat{x}^i$ plays the dual role of position observable and Galilean boost generator.

The interpretation of $\hat{x}^i$ as a position observable is only valid in the nonrelativistic Galilean theory. The lack of fully relativistic nature in $\hat{x}^i$ is noticeable from the noncovariant and acausal features described in Sec.~\ref{NewtonWigner_localization_scheme}. Notwithstanding the correct behavior under spatial translation [Eq.~(\ref{x_translation})], the operator $\hat{x}^i$ does not properly transform under Lorentz boost and, hence, cannot be seen as a representative of the Poincar\'e group.

We now know why the Born-Newton-Wigner operator $\hat{x}^i$ emerges as a position operator in NRQM. However, in the relativistic theory we already had classical Minkowski coordinates $x^\mu$ assuming the role of position variable. Are they still somehow present in NRQM or do they disappear in the nonrelativistic limit? The question is conceptually relevant, because, contrary to the operator $\hat{x}^i$, the coordinates $x^\mu$ have a fundamental notion of localization.

Clearly, we cannot directly compare the classical variable $x^\mu$ with the quantum operator $\hat{x}^i$. Instead, we need to consider the second-quantized operators of NRQM labeled by $\vec{x}$ and compare them with field operators of QFT in the hypersurface $t=0$. We remark that second-quantized operators in $\vec{x}$ are generated by the Newton-Wigner operators $\hat{a}_\text{NW}(\vec{x})$, whereas field operators in $(0,\vec{x})$ are generated by $\hat{\phi}(0,\vec{x})$ and $\hat{\pi}(0,\vec{x})$. Alternately, one can consider
\begin{equation}\label{alpha_tilde_phi_Pi}
\hat{a}_\text{AQFT}(\vec{x}) = \sqrt{\frac{m c^2}{2 \hbar^2}} \hat{\phi}(0,\vec{x}) - \frac{i}{\sqrt{2 m c^2}} \hat{\pi}(0,\vec{x}),
\end{equation}
and its adjoin as generators of the local algebra $\mathfrak{A}_\text{M}^\text{AQFT}(\vec{x})$. The inverse of Eq.~(\ref{alpha_tilde_phi_Pi}) is
\begin{subequations}\label{alpha_tilde_phi_Pi_inverse}
\begin{align}
& \hat{\phi}(0,\vec{x})  = \frac{\hbar}{\sqrt{2 m c^2}} \left[ \hat{a}_\text{AQFT}(\vec{x}) + \hat{a}_\text{AQFT}^\dagger(\vec{x}) \right], \\
&  \hat{\pi}(0,\vec{x})  =  i \sqrt{\frac{m c^2}{2}} \left[ \hat{a}_\text{AQFT}(\vec{x}) - \hat{a}_\text{AQFT}^\dagger(\vec{x}) \right].
\end{align}
\end{subequations}

A priori, the variable $\vec{x}$ appearing in $\hat{a}_\text{NW}(\vec{x})$ and in $\hat{a}_\text{AQFT}(\vec{x})$ have different meaning. In the case of $\hat{a}_\text{NW}(\vec{x})$, $\vec{x}$ appears as an index resulting from the second quantization prescription; whereas, in $\hat{a}_\text{AQFT}(\vec{x})$, $\vec{x}$ is a genuine coordinate representing a spacetime event. However, it has been proven that in the nonrelativistic limit, the two fields $\hat{a}_\text{AQFT}(\vec{x})$ and $\hat{a}(\vec{x})$ converge \cite{PhysRevD.90.065032, Papageorgiou_2019}. Consequently, we see that the Minkowski coordinate $\vec{x}$ does not disappear in the nonrelativistic limit, but remains as an index for the annihilator field $\hat{a}(\vec{x})$.

\subsubsection{Convergence between Newton-Wigner and AQFT operators}

The convergence between $\hat{a}_\text{AQFT}(\vec{x})$ and $\hat{a}(\vec{x})$ has two consequences. On one hand, we see that the genuine Minkowski coordinate $\vec{x}$ emerges in NRQM as an index for the annihilator field $\hat{a}(\vec{x})$. On the other hand, we have the proof that the Newton-Wigner and the AQFT localization schemes converge in the nonrelativistic limit. Indeed, any operator that is localized in $\vec{x}$ with respect to the Newton-Wigner scheme can be approximated by operators which are localized in $\vec{x}$ with respect to the AQFT scheme. This means that the fundamental notion of localization owned by the Minkowski coordinate $\vec{x}$ is approximately shared with the Newton-Wigner position operator $\hat{\vec{x}}$. In other words, the nonrelativistic limit is able to ``fundamentalize'' the Newton-Wigner localization.

To show the convergence between $\hat{a}_\text{AQFT}(\vec{x})$ and $\hat{a}(\vec{x})$, different approaches have been considered, including the use of coarse-grained operators \cite{PhysRevD.90.065032} and the restriction of the Hilbert space to a bandlimited subspace \cite{Papageorgiou_2019}. These methods are based on the definition of a minimum resolution in space and a maximum energy scale by means of the Compton wavelength $\lambda_\text{C} = \hbar/mc$. 

In Ref.~\cite{PhysRevD.90.065032}, the minimum experimental resolution of nonrelativistic phenomena is described via coarse-graining modeling. In such a regime, coarse-grained operators are assumed to appear indistinguishable from their fine-grained counterparts. The convergence between the Newton-Wigner and the AQFT localization schemes can be realized by noticing that the kernels $f_{\hat{\phi}\mapsto \text{NW}}(\vec{x}) $ and $f_{\hat{\pi}\mapsto \text{NW}}(\vec{x}) $ appearing in Eq.~(\ref{a_tilde_phi_Pi}) decay exponentially as $\exp(- |\vec{x} |/\lambda_\text{C})$ when $\vec{x}$ is outside the minimum spatial resolution, i.e., $|\vec{x} | \gg \lambda_\text{C}$.

Explicitly, the coarse-grained versions of $\hat{a}_\text{AQFT}(\vec{x})$ and $\hat{a}_\text{NW}(\vec{x})$ are defined as
\begin{subequations}
\begin{align}
& \hat{a}_{\text{AQFT}, \vec{j},\Lambda} = \int_{\mathbb{R}^3} d^3 x G_\Lambda (D \vec{j} - \vec{x}) \hat{a}_\text{AQFT}(\vec{x}), \\
& \hat{a}_{\text{NW},\vec{j},\Lambda} = \int_{\mathbb{R}^3} d^3 x G_\Lambda (D \vec{j} - \vec{x}) \hat{a}_\text{NW}(\vec{x}),
\end{align}
\end{subequations}
with $\vec{j} \in \mathbb{Z}^3$ as grid coordinates, $D$ as the spatial separation of the grid points and 
\begin{equation}
G_\Lambda (\vec{x}) = \frac{1}{(2 \pi \Lambda^2 )^{1/4}} \exp \left( - \frac{|\vec{x}|^2}{4 \Lambda^2} \right)
\end{equation}
as the Gaussian smearing function with spatial resolution $\Lambda \ll D$.  The approximation $\hat{a}_{\text{AQFT}, \vec{j},\Lambda}  \approx \hat{a}_{\text{NW},\vec{j},\Lambda} $ for $\Lambda \gg \lambda_\text{C}$ is proven in Ref.~\cite{PhysRevD.90.065032} and leads to the convergence between the Newton-Winger and the AQFT schemes in the nonrelativistic limit.

At variance with Ref.~\cite{PhysRevD.90.065032}, the method adopted in Ref.~\cite{Papageorgiou_2019} is based on the definition of the bandlimited subspace $\mathcal{H}_\text{M}^\Lambda$ as the Fock space of particles with momenta lower than the cutoff $1/\Lambda$. By restricting $\hat{a}_\text{AQFT}(\vec{x})$ and $\hat{a}_\text{NW}(\vec{x})$ to $\mathcal{H}_\text{M}^\Lambda$ with $\Lambda \gg \lambda_\text{C}$, the authors derive the approximation $\left. \hat{a}_\text{NW}(\vec{x}) \right|_{\mathcal{H}_\text{M}^\Lambda} \approx \left. \hat{a}_\text{AQFT}(\vec{x}) \right|_{\mathcal{H}_\text{M}^\Lambda}$ at first order in $\lambda_\text{C}/\Lambda \ll 1$.

The proof is based on computing the Bogoliubov transformation between the operators $\hat{a}_\text{AQFT}(\vec{x})$ and $\hat{a}(\vec{k})$, i.e.,
\begin{equation}\label{alpha_tilde_a}
\hat{a}_\text{AQFT}(\vec{x}) = \int_{\mathbb{R}^3} d^3 k \left[ f_{\hat{a} \mapsto \text{AQFT}}(\vec{x},\vec{k}) \hat{a}(\vec{k}) + f_{\hat{a}^\dagger \mapsto \text{AQFT}}(\vec{x},\vec{k}) \hat{a}^\dagger(\vec{k})  \right],
\end{equation}
with
\begin{subequations}\label{f_a_alpha_tilde}
\begin{align}
& f_{\hat{a} \mapsto \text{AQFT}}(\vec{x},\vec{k}) = \frac{e^{i \vec{k} \cdot \vec{x}}}{2 \sqrt{(2 \pi)^3}}  \left[ \sqrt{\frac{m c^2}{ \hbar \omega(\vec{k})}} +  \sqrt{\frac{ \hbar \omega(\vec{k})}{m c^2}}  \right], \\
& f_{\hat{a}^\dagger \mapsto \text{AQFT}}(\vec{x},\vec{k}) = \frac{e^{-i \vec{k} \cdot \vec{x}}}{2 \sqrt{(2 \pi)^3}}  \left[ \sqrt{\frac{m c^2}{ \hbar \omega(\vec{k})}} - \sqrt{\frac{ \hbar \omega(\vec{k})}{m c^2}}  \right].
\end{align}
\end{subequations}
The restriction of Eqs.~(\ref{a_NW}) and (\ref{f_a_alpha_tilde}) to the bandlimited subspace $\mathcal{H}_\text{M}^\Lambda$ is
\begin{subequations}\label{alpha_tilde_a_a_tilde_a_bandlimited}
\begin{align}
& \left. \hat{a}_\text{NW}(\vec{x})  \right|_{\mathcal{H}_\text{M}^\Lambda} = \int_{|\vec{k}|<1/\Lambda} d^3 k   f_{\hat{a} \mapsto \text{NW}}(\vec{x},\vec{k}) \hat{a}(\vec{k}),\label{a_tilde_a_bandlimited}\\
& \left. \hat{a}_\text{AQFT}(\vec{x})  \right|_{\mathcal{H}_\text{M}^\Lambda} = \int_{|\vec{k}|<1/\Lambda} d^3 k   \left[ f_{\hat{a} \mapsto \text{AQFT}}(\vec{x},\vec{k}) \hat{a}(\vec{k}) + f_{\hat{a}^\dagger \mapsto \text{AQFT}}(\vec{x},\vec{k}) \hat{a}^\dagger(\vec{k})  \right],\label{alpha_tilde_a_bandlimited}
\end{align}
\end{subequations}
with
\begin{equation}\label{f_a_NW}
f_{\hat{a} \mapsto \text{NW}}(\vec{x},\vec{k}) = \frac{e^{i \vec{k} \cdot \vec{x}}}{\sqrt{(2\pi)^3}}.
\end{equation}
Notice that
\begin{subequations}\label{Kernels_approximation}
\begin{align}
& f_{\hat{a} \mapsto \text{AQFT}}(\vec{x},\vec{k}) \approx f_{\hat{a} \mapsto \text{NW}}(\vec{x},\vec{k}) \text{ if } |\vec{k}| \ll \lambda_\text{C}^{-1},\\
&  f_{\hat{a}^\dagger \mapsto \text{AQFT}}(\vec{x},\vec{k}) \approx 0 \text{ if } |\vec{k}| \ll \lambda_\text{C}^{-1}.
\end{align}
\end{subequations}
Hence, by expanding Eqs.~(\ref{alpha_tilde_a_a_tilde_a_bandlimited}) to the first order in $\lambda_\text{C}/\Lambda \ll 1$, we obtain $\left. \hat{a}_\text{NW}(\vec{x}) \right|_{\mathcal{H}_\text{M}^\Lambda} \approx \left. \hat{a}_\text{AQFT}(\vec{x}) \right|_{\mathcal{H}_\text{M}^\Lambda}$, which leads to the convergence between the two localization schemes in the nonrelativistic regime.

In Ref.~\cite{Papageorgiou_2019}, the Bogoliubov transformation (\ref{alpha_tilde_a_a_tilde_a_bandlimited}) is also expanded up to the second order in $\lambda_\text{C}/\Lambda \ll 1$. This gives corrective terms that spoil the nonlocality of $\hat{a}_\text{NW}(\vec{x})$ with respect to the AQFT scheme at the first nontrivial order.

\subsection{Convergence of the modal to the Born scheme}\label{Convergence_of_the_modal_scheme_to_the_Born_scheme}

In this subsection, we show the convergence between the modal and the Newton-Wigner scheme in the nonrelativistic limit. Due to the equivalence between the Newton-Wigner and Born scheme [Sec.~\ref{Born_localization_is_NewtonWigner_localization}], we implicitly show the convergence between the modal and the Born scheme.

We follow the strategy of Ref.~\cite{Papageorgiou_2019} that we already used in Sec.~\ref{Comparison_with_the_relativistic_theory} for the case of the AQFT schemes. Firstly, we restrict Eq.~(\ref{a_mod_a}) to the bandlimited subspace $\mathcal{H}_\text{M}^\Lambda$ to obtain
\begin{equation}\label{a_mod_a_Lambda}
\left. \hat{a}_\text{mod} (\vec{x})  \right|_{\mathcal{H}_\text{M}^\Lambda}  = \int_{|\vec{k}|<1/\Lambda} d^3 k  \left[ f_{\hat{a} \mapsto \text{AQFT}}(\vec{x},\vec{k}) - f_{\hat{a}^\dagger \mapsto \text{AQFT}}^*(\vec{x},\vec{k}) \right] \hat{a}(\vec{k}),
\end{equation}
where $f_{\hat{a} \mapsto \text{AQFT}}(\vec{x},\vec{k})$ and $f_{\hat{a}^\dagger \mapsto \text{AQFT}}(\vec{x},\vec{k})$ are defined in Eq.~(\ref{f_a_alpha_tilde}). Then, we use Eqs.~(\ref{a_tilde_a_bandlimited}), (\ref{Kernels_approximation}) and (\ref{a_mod_a_Lambda}) to derive the approximation $\left. \hat{a}_\text{NW}(\vec{x}) \right|_{\mathcal{H}_\text{M}^\Lambda} \approx \left. \hat{a}_\text{mod}(\vec{x}) \right|_{\mathcal{H}_\text{M}^\Lambda}$ when $\lambda_\text{C}/\Lambda \ll 1$. This implies that any element of $\mathfrak{A}_\text{M}^\text{mod}(\vec{x})$ can be approximated to an element of $\mathfrak{A}_\text{M}^\text{NW}(\vec{x})$ and that the two localization schemes converge. Due to the equivalence between the Newton-Wigner and Born scheme, we also proved the convergence between the modal and the Born scheme.

We remark that the Born scheme converges to the AQFT scheme as well [Sec.~\ref{Comparison_with_the_relativistic_theory}]. Hence, in this subsection, we have also indirectly proven the convergence between the modal and the AQFT scheme. To have a direct proof, compare Eq.~(\ref{alpha_tilde_a_bandlimited}) with Eq.~(\ref{a_mod_a_Lambda}) and use Eq.~(\ref{Kernels_approximation}) to show that $\left. \hat{a}_\text{AQFT}(\vec{x}) \right|_{\mathcal{H}_\text{M}^\Lambda} \approx \left. \hat{a}_\text{mod}(\vec{x}) \right|_{\mathcal{H}_\text{M}^\Lambda}$. As a consequence of this convergence, we find that the modal scheme acquires a genuine notion of localization in the nonrelativistic regime.

\subsection{The strict localization property in the nonrelativistic limit}\label{Strict_localization_in_the_nonrelativistic_limit}

In Sec.~\ref{solving_the_paradox} we showed that, as a consequence of the Reeh-Schlieder theorem, the AQFT scheme does not always satisfy the strict localization property. This means that the outcome of experiments in any space region $\mathcal{V}_\text{B}$ may depend on the preparation of states in on other disjoint region $\mathcal{V}_\text{A}$. 

At variance with the AQFT scheme, the Newton-Wigner scheme always satisfies the strict localization property [Sec.~\ref{Comparison_between_the_NewtonWigner_and_the_AQFT_schemes}]. However, real life experiments can only be faithfully represented by the AQFT scheme, which is the only one providing a genuine notion of localization. Hence, the strict localization property satisfied by the Newton-Wigner scheme does not generally occur in genuinely local experiments.

The incompatibility between the two schemes disappears in the nonrelativistic limit [Sec.~\ref{Comparison_with_the_relativistic_theory}]. In such a regime, the Newton-Wigner scheme acquires a genuine notion of localization from the AQFT and local experiments are expected to satisfy the strict localization property.

In this subsection, we will show that the nonlocal effects predicted by the AQFT scheme do not occur in the nonrelativistic limit of QFT. Intuitively, the result can be deduced from noticing that Reeh-Schlieder quantum correlations of the vacuum are exponentially suppressed in nonrelativistic scales \cite{SUMMERS1985257}. However, a more detailed proof can be given by using the results of Sec.~\ref{Comparison_with_the_relativistic_theory}. 

The local algebra $\mathfrak{A}_\text{M}^\text{AQFT}(\vec{x})$ is generated by the local fields $\hat{a}_\text{AQFT}(\vec{x})$, which are indistinguishable from the Newton-Wigner fields $\hat{a}_\text{NW}(\vec{x})$ in the nonrelativistic limit. Explicitly, this means that any operator $\hat{O} \in \mathfrak{A}_\text{M}^\text{AQFT}(\vec{x})$ generated by momentum operators $\hat{a}(\vec{k})$ satisfying $\lambda_\text{C} |\vec{k}| \ll 1$ can be approximated to the operator $\hat{O}_\text{NW} \in \mathfrak{A}_\text{M}^\text{NW}(\vec{x})$ obtained by replacing $\hat{a}_\text{AQFT}(\vec{x})$ with $\hat{a}_\text{NW}(\vec{x})$.

Hence, any nonrelativistic state $| \phi \rangle = \hat{O}_\text{A} | 0_\text{M} \rangle$ with $\hat{O}_\text{A} \in \mathfrak{A}_\text{M}^\text{AQFT}(\vec{x}_\text{A})$ and any nonrelativistic observable $\hat{O}_\text{B} \in \mathfrak{A}_\text{M}^\text{AQFT}(\vec{x}_\text{B})$ are approximated by some $| \phi_\text{NW} \rangle = \hat{O}_{\text{NW},\text{A}} | 0_\text{M} \rangle$ and $\hat{O}_{\text{NW},\text{B}} \in \mathfrak{A}_\text{M}^\text{NW}(\vec{x}_\text{B})$,  with $\hat{O}_{\text{NW},\text{A}} \in \mathfrak{A}_\text{M}^\text{NW}(\vec{x}_\text{A})$. The state $| \phi_\text{NW} \rangle$ and the operator $\hat{O}_{\text{NW},\text{B}}$ satisfy the strict localization property
\begin{equation}\label{KnightLicht_property_NW}
\langle \phi_\text{NW} | \hat{O}_{\text{NW},\text{B}} | \phi_\text{NW} \rangle = \langle 0_\text{M} | \hat{O}_{\text{NW},\text{B}}  | 0_\text{M} \rangle.
\end{equation}
when $\vec{x}_\text{A} \neq \vec{x}_\text{B}$. This means that $| \phi_\text{NW} \rangle$ gives the same outcome as the vacuum $| 0_\text{M} \rangle$ when measuring $\hat{O}_{\text{NW},\text{B}}$.

In summary, any state $| \phi \rangle$ that is localized in $\vec{x}_\text{A}$ with respect to the AQFT scheme is approximately localized in $\vec{x}_\text{A}$ with respect to the Newton-Wigner scheme and hence it appears indistinguishable from the vacuum in $\vec{x}_\text{B} \neq \vec{x}_\text{A}$. This means that $| \phi \rangle$ is approximately strictly localized.

To give a practical example, consider the two Alice-Bob scenarios described in Sec.~\ref{Comparison_between_the_NewtonWigner_and_the_AQFT_schemes}. Alice is an experimenter that locally prepares a state in the region $\vec{x}_\text{A}$, while Bob performs local measurements in $\vec{x}_\text{B}$. Depending on the localization scheme, the outcome of Bob's measurements may or may not be influenced by the preparation of the state by Alice.

In the nonrelativistic limit, the two localization scheme converge. This leads to an equivalence between the two Alice-Bob experiments. In this unifying scenario, the preparation and the measurement in disjoint region appear independent, in agreement with the Newton-Wigner Alice-Bob experiment. Also, the fundamental notion of localization inherited from the AQFT Alice-Bob scenario guarantees the applicability of the results for genuinely local experiments.

\section{Conclusions}\label{Localization_in_Quantum_Field_Theory_Conclusions}

Different localization schemes have been considered for QFT in Minkowski spacetime. Among them, only the AQFT framework is able to provide a relativistically consistent notion of localization for states and observables. The Newton-Wigner scheme, instead, is inspired by the nonrelativistic theory and it is based on local creators and annihilators resulting from the definition of a second-quantized position operator. Finally, the modal scheme comes from the modal representation of particles as positive frequency solutions of the Klein-Gordon equation.

Even if the Newton-Wigner and the modal schemes are not suited for the description of relativistic local phenomena, they become indistinguishable from the AQFT scheme in the nonrelativistic limit. Due to the fundamental nature of the notion of localization entailed by the AQFT scheme, we say that the Newton-Wigner and the modal schemes are fundamentalized in the nonrelativistic limit.

The fundamentalization of the Newton-Wigner scheme and its equivalence to the Born theory imply that the familiar description of local states in NRQM by means of wave functions and position operator leads to correct laboratory predictions, but only in the nonrelativistic regime. Also, due to the fundamentalization of the modal scheme, we find that nonrelativistic states are approximately localized in the support of their representative modes. This provides a justification for the notion of localization that has been considered throughout Parts \ref{Relativistic_and_nonrelativistic_quantum_fields} and \ref{Inequivalent_particle_representations_and_Unruh_effect} of the thesis.

\chapter{Localization in accelerated frame}\label{Localization_in_accelerated_frame}

\textit{This chapter is based on and contains material from Ref.~\citeRF{localization_QFTCS}.}

\section{Introduction}\label{Localization_in_accelerated_frame_Introduction}

In Chap.~\ref{Localization_in_Quantum_Field_Theory}, we reviewed the problem of localization in QFT. In particular we considered the Minkowski spacetime and discussed different notions of localization from the point of view of inertial experimenters. In this chapter, instead, we consider accelerated observes as well. In particular we will assume that at least one of the two experimenters---between experimenter A preparing the states and experimenter B performing measurements---is accelerated. Hence, the aim of this chapter is to extend the results of Chap.~\ref{Localization_in_Quantum_Field_Theory} to the QFTCS framework.

We immediately find difficulties related to the not unique notion of vacuum states. In QFT, there is only one vacuum $| 0_\text{M} \rangle$, which is defined as the only Poincar\'e invariant state. On the other hand, in QFTCS, there can be more than one unitarily nonequivalent vacuum states, each of them associated to a notion of global time via timelike Killing vector fields. For any localization scheme in QFT, the Minkowski vacuum $| 0_\text{M} \rangle$ represents the background over which states are prepared. Conversely, in QFTCS, one can choose between more background states.

The Rindler vacuum $| 0_\text{L}, 0_\text{R} \rangle$ is defined as the state annihilated by the operators $\hat{A}_\nu(\vec{\theta})$ [Eq.~(\ref{Rindler_vacuum_scalar})] and it is not equivalent to the Minkowski vacuum $| 0_\text{M} \rangle$ [Eq.~(\ref{Rindler_vacuum_to_Minkowski})]. Hence, inertial and noninertial experimenters may choose between two different background states. The natural choice for each observer would be given by the vacuum of the respective frame, i.e., $| 0_\text{M} \rangle$ for inertial experimenters and $| 0_\text{L}, 0_\text{R} \rangle$ for accelerated experimenters. However, one can also consider the case in which laboratory phenomena are described as occurring over the Minkowski vacuum background $| 0_\text{M} \rangle$, regardless of the motion of the observers. Hereafter, we note by $| \Omega \rangle$ any background state; depending on the situation, $| \Omega \rangle$ is equal to $| 0_\text{M} \rangle$ or $| 0_\text{L}, 0_\text{R} \rangle$.

\begin{table}
\begin{center}
\begin{tabular}{|l|| >{\centering\arraybackslash}m{9em} | >{\centering\arraybackslash}m{8em} | >{\centering\arraybackslash}m{7em} |}
\hline
& Experimenter A (preparation) & Experimenter B (observation) & Background vacuum state \\
\hline
\hline
ABM & inertial (Alice) & inertial (Bob) & Minkowski $|0_\text{M} \rangle$ \\
\hline
RaRbR & accelerated (Rachel) & accelerated (Rob) & Rindler $|0_\text{L},0_\text{R} \rangle$ \\
\hline
RaRbM & accelerated (Rachel) & accelerated (Rob) & Minkowski $|0_\text{M} \rangle$ \\
\hline
ARbM & inertial (Alice) & accelerated (Rob) & Minkowski $|0_\text{M} \rangle$ \\
\hline
\end{tabular}
\end{center}
\caption{Different measurement setups. The experimenter A prepares a local state over the background state and the experimenter B performs local measurement. Each experimenter may be inertial or accelerated and the background state can be either the Minkowski ($|0_\text{M} \rangle$) or the Rindler ($|0_\text{L},0_\text{R} \rangle$) vacuum.}\label{scenarios_Table}
\end{table}

Overall, we consider four different scenarios, which are summarized by Table \ref{scenarios_Table}. Each of them is labeled by an acronym. The ABM scenario was already detailed in Chap.~\ref{Localization_in_Quantum_Field_Theory} since it is characterized by inertial experimenters (Alice and Bob) and by the Minkowski vacuum $|0_\text{M} \rangle$ as the background state. Conversely, the RaRbR and the RaRbM scenarios include only accelerated observers (Rachel and Rob); the difference between RaRbR and RaRbM is given by the background state $| \Omega \rangle$ which is equal to the Rindler ($|0_\text{L},0_\text{R} \rangle$) or the Minkowski ($|0_\text{M} \rangle$) vacuum, respectively. The RaRbR scenario, then, appears conceptually equivalent to the ABM, as the experimenters and the background state are defined by the same frame in each scenario; the only difference between the ABM and the RaRbR is given by the respective field equations [Eqs.~(\ref{Klein_Gordon}) and (\ref{Rindler_Klein_Gordon})]. Finally, in the ARbM scenario, the experimenter A is inertial, whereas the experimenter B is accelerated; the local preparation of states is performed over the Minkowski vacuum $|0_\text{M} \rangle$.
 
In Chap.~\ref{Localization_in_Quantum_Field_Theory}, we explained how the inertial experimenter Alice prepares local states over the Minkowski vacuum $|0_\text{M} \rangle$ and in which sense the inertial experimenter Bob performs local measurements. To understand the RaRbR, RaRbM and ARbM scenarios, we also need a notion of localization from the point of view of accelerated observers and with respect to different background states. In this chapter, we show how to describe local measurements by Rob and local preparation of states over any background state $| \Omega \rangle$ by Rachel.

We are particularly interested in deriving the dependence between outcomes of measurements by the experimenter B and the local preparation of states by the experimenter A. In Chap.~\ref{Localization_in_Quantum_Field_Theory}, we found that, in the ABM scenario, the independence between the two operations is not always guaranteed. In particular, we defined the strict localization property [Eq.~(\ref{KnightLicht_property})] and we showed that it is not always satisfied by some localization schemes.

Here, we need to extend the definition of the strict Knight-Licht localization to include the vacuum Rindler $|0_\text{L},0_\text{R} \rangle$ as a possible background states $| \Omega \rangle$ in addition to the Minkowski vacuum $|0_\text{M} \rangle$. We say that a state $| \phi \rangle $ is strictly localized in $\mathcal{O}_\text{A}$ over $| \Omega \rangle$ if it gives the same expectation values as $| \Omega \rangle$ for all measurements in any region $\mathcal{O}_\text{B}$ spacelike separated from $\mathcal{O}_\text{A}$, i.e.,
\begin{equation}\label{KnightLicht_property_Omega}
\langle \phi | \hat{O}_\text{B} | \phi \rangle = \langle \Omega | \hat{O}_\text{B}  | \Omega \rangle,
\end{equation}
with $\hat{O}_\text{B}$ as any observable localized in $\mathcal{O}_\text{B}$. We also say that the strict localization property is always satisfied if for any $| \phi \rangle $ localized in $\mathcal{O}_\text{A}$ and for any $\hat{O}_\text{B}$ localized in $\mathcal{O}_\text{B}$, Eq.~(\ref{KnightLicht_property_Omega}) holds.

As in Chap.~\ref{Localization_in_Quantum_Field_Theory}, we consider the Newton-Wigner, the AQFT and the modal schemes. We extend their definition for accelerated observers and for any background state.

We start from the Newton-Wigner scheme for massless scalar real fields in 1+1 Rindler spacetime. The reason why we cannot consider the corresponding theory in 3+1 dimensions is because the definition of Newton-Wigner position states (\ref{x_state}) and position operators (\ref{x_NW}) requires the existence of momentum particle states $| \vec{k} \rangle$. Rindler scalar fields in 3+1 dimensions lack translation symmetry along the direction of the acceleration; hence, positive frequency modes with defined momentum along $Z$ do not exist. In 1+1 dimensions, instead, the Minkowski and the Rindler massless scalar real fields are described by the same field equations [Eqs.~(\ref{Klein_Gordon_11}), (\ref{Rindler_Klein_Gordon_R_11}) and (\ref{Rindler_Klein_Gordon_L_11})] and, hence, they share the existence of momentum particle states, i.e., $| k \rangle$ and $| K \rangle$.

The 1+1 Rindler-Newton-Wigner scheme inherits all the properties from the Newton-Wigner scheme in Minkowski spacetime, including the factorization of the global Hilbert space $\mathcal{H}_{\text{L},\text{R}}$ into local Fock spaces $\mathcal{H}_\nu^\text{NW}(\mathcal{V}_i)$ and the factorization of the global vacuum $|0_\text{L},0_\text{R} \rangle$ into the local vacua $|0_\nu (\mathcal{V}_i) \rangle$. However, these local Hilbert spaces differ from the corresponding local Minkowski-Fock spaces $\mathcal{H}_\text{M}^\text{NW}(\mathcal{V}_i)$. This is due to the fact that the Minkowski and the Rindler Newton-Wigner schemes are incompatible, in the sense that any operator that is localized with respect to one of the two schemes is not localized with respect to the other.

In Sec.~\ref{NewtonWigner_localization_scheme}, we remarked that the Newton-Wigner position operator is not relativistically covariant and that the notion of localization is not preserved by special relativistic transformations. Here, we find that it is not preserved by GR diffeomorphisms either, due to the incompatibilities between different frames. This gives a further motivation to disregard the Newton-Wigner scheme as a faithful description of local phenomena in the QFTCS regime.

In Sec.~\ref{Comparison_between_the_NewtonWigner_and_the_AQFT_schemes}, we showed that the strict localization property is always satisfied in the Newton-Wigner ABM scenario. Here, we show that this is also true for the Newton-Wigner RaRbR experiment. Conversely, in the Newton-Wigner RaRbM scenario, only local nonselective preparations lead to strictly localized states. This is due to the fact that the Minkowski vacuum $|0_\text{M} \rangle$ does not factorize into local vacua with respect to the factorization $\mathcal{H}_{\text{L},\text{R}} = \bigotimes_i \mathcal{H}_\nu^\text{NW}(\mathcal{V}_i)$.

In the Newton-Wigner ARbM scenario, instead, the strict localization property is not even guaranteed for unitarily prepared local states, due to the incompatibility between Alice and Rob's notion of Newton-Wigner localization. As a result, we find that the Newton-Wigner scheme is incompatible with causality since Alice can send information to Rob by means of such a nonlocal effect. This violation of causality occurs exactly at the instant of time $t=0$ and it is based on the Bogoliubov transformation (\ref{Bogolyubov_transformation}) relating the two frames; hence, this acausal effect is different from the one described in Sec.~\ref{NewtonWigner_localization_scheme} by means of the Hegerfeldt theorem. 

After discussing the 1+1 Rindler-Newton-Wigner scheme, we go back to 3+1 dimensions and we study the AQFT scheme. By following the algebraic approach to QFT and QFTCS [Sec.~\ref{Algebraic_approach_in_QFT_and_QFTCS}], we show that the definition given in Sec.~\ref{AQFT_localization_scheme} for the inertial frame also applies to the accelerated frame. The argument is that the local algebras $\mathfrak{A}(\mathcal{O})$ and $\mathfrak{A}(\mathcal{E})$ are defined with respect to spacetime regions $\mathcal{O}$ and events $\mathcal{E}$, which are frame independent objects without an intrinsic notion of coordinates.

Similarly, we prove that the Minkowski and the Rindler AQFT schemes are compatible, in the sense that states and observables that are localized in any event $\mathcal{E}$ with respect to the Minkowski AQFT scheme are also localized in $\mathcal{E}$ with respect to the Rindler AQFT scheme. In this way, we show the covariant behavior of the AQFT localization under coordinate diffeomorphism in agreement with the GR theory. As a result, we obtain an additional confirmation that the AQFT scheme gives a genuine notion of localization.

Due to the equivalence between the Minkowski and the Rindler AQFT schemes, we find that all the scenarios considered here (i.e., ABM, RaRbR, RaRbM and ARbM) share the same features. This also includes the strict localization property which only applies to states that are prepared via nonselective operations [Sec.~\ref{solving_the_paradox}].

In addition to the AQFT scheme, we study the modal scheme in the Rindler frame. In particular, we use the representation of Rindler-Fock states as positive frequency solutions of the Rindler-Klein-Gordon equation [Sec.~\ref{QFT_in_curved_spacetime_Rindler_scalar}]. We follow the same approach of Sec.~\ref{Modal_localization_scheme} to define localized states and observables in the Rindler frame. We also extend this definition to include any background state $| \Omega \rangle$.

We find that the Minkowski and the Rindler modal schemes are incompatible to each other, in the sense that there is no state or observable that is simultaneously localized with respect to both schemes. This leads to the conclusion that the modal scheme does not satisfy the GR notion of physical equivalence between frames and, hence, cannot be adopted for a genuine description of local phenomena in the QFTCS regime.

An additional incompatibility with relativity is given by the violation of causality. This is due to the instantaneous spreading of wave functions. Also, the strict localization property is not guaranteed for nonselective preparations of states. Such an issue occurs in both the Minkowski and the Rindler frame. Hence, in all the scenarios considered here (i.e., ABM, RaRbR, RaRbM and ARbM), the preparation of the state by experimenter A may influence the outcome of measurements by B and violate relativistic causality.

The noncovariant and acausal features make the modal scheme unsuitable for the description of local phenomena in the relativistic regime. In the nonrelativistic limit, instead, the modal scheme acquires a genuine notion of locality due to the convergence to the AQFT scheme. Such a convergence give rise to a unified localization scheme in NRQFTCS, which will be detailed at the end of this chapter.

We firstly find that the NRQFTCS scheme over the Rindler vacuum $|0_\text{L},0_\text{R} \rangle$ is equivalent to the Born scheme. In particular, we show that the global Hilbert space factorizes into local Fock spaces and $|0_\text{L},0_\text{R} \rangle$ factorizes into the local vacua. As a consequence, we find that the strict localization property is always satisfied in the nonrelativistic RaRbR scenario.

Conversely, in the nonrelativistic RaRbM and ARbM scenarios, the strict localization property is only guaranteed for nonselective preparations of the state. In the RaRbM scenario, this is due to the fact that the background Minkowski vacuum $|0_\text{M} \rangle$ is entangled between the local Rindler-Fock spaces. In ARbM case, instead, the cause is the incompatibility between the Minkowski and the Rindler nonrelativistic schemes, which, in turn, is due to the incompatibility between the nonrelativistic limits in the respective frames [Sec.~\ref{Inertial_and_non_inertial_frame}].

The chapter is organized as follows. In Sec.~\ref{NewtonWigner_localization_in_11_conformally_flat_spacetimes}, we study the Newton-Wigner scheme for Rindler massless scalar real fields in 1+1 dimensions. In Secs.~\ref{AQFT_localization_scheme_in_curved_spacetime} and \ref{Modal_localization_scheme_Rindler}, we consider Rindler scalar fields in 3+1 dimensions and we detail the AQFT and the modal scheme, respectively. A comparison between the two localization schemes is given in Sec.~\ref{Comparison_between_localization_schemes_Rindler}. Their nonrelativistic limit, instead, is detailed in Sec.~\ref{Localization_in_NRQFTCS}. Conclusions are drawn in Sec.~\ref{Localization_in_accelerated_frame_Conclusions}.

\section{Newton-Wigner scheme in 1+1 spacetime} \label{NewtonWigner_localization_in_11_conformally_flat_spacetimes}

In Sec.~\ref{NewtonWigner_localization_scheme}, we introduced the Newton-Wigner scheme in the Minkowski frame.
Here, instead, we want to discuss such a localization scheme in the Rindler frame.

In the attempt to generalize definition of the Minkowski-Newton-Wigner scheme in curved spacetime, we immediately find a problem. By following Newton and Wigner \cite{RevModPhys.21.400}, we need single particle states with defined momentum $| \vec{k} \rangle$ to derive states with defined position [Eq.~(\ref{x_state})] and the corresponding Newton-Wigner position operator [Eq.~(\ref{x_NW})]. However, the existence of states like $| \vec{k} \rangle$ are not guaranteed in curved spacetimes, due to a possible lack of translational symmetry in the Lagrangian theory. 

Remarkably, massless scalar real fields in 1+1 spacetime are described by the same field equation in both the Minkowski [Eq.~(\ref{Klein_Gordon_11})] and the Rindler [Eqs.~(\ref{Rindler_Klein_Gordon_R_11}) and (\ref{Rindler_Klein_Gordon_L_11})] frames. Hence, the Rindler field $\hat{\Phi}_\nu(T,X)$ is symmetric with respect to spatial translations and admits the free modes $f(K,T,X)$ [Eq.~(\ref{f_11})] as solutions of the Rindler-Klein-Gordon equation. The momentum states $| K, \nu \rangle = \hat{A}_\nu^\dagger(K) |0_\text{L},0_\text{R} \rangle $ can be defined as the $\nu$-Rindler single particles associated to the positive frequency modes $f(K,T,X)$. From $| K , \nu \rangle$, one can formulate the Rindler-Newton-Wigner localization scheme. 

By following Sec.~\ref{NewtonWigner_localization_scheme}, we define the Newton-Wigner annihilation operators as the anti-Fourier transform of the annihilators of particles with defined momentum, i.e.,
\begin{equation}\label{A_NW_X_A_K}
\hat{A}_{\text{NW},\nu}(X) = \int_{\mathbb{R}} dK \frac{e^{i K X}}{\sqrt{2\pi}} \hat{A}_\nu (K).
\end{equation}
These operators generate the local algebras $\mathfrak{A}_\nu^{\text{NW}}(X)$ in wedges $\nu = \{ \text{L}, \text{R} \}$ and space points $X \in \mathbb{R}$. Similarly, one can define the local algebras $\mathfrak{A}_\nu^{\text{NW}}(\mathcal{V})$ in space regions $\mathcal{V} \subset \mathbb{R}$. The operator $ \hat{O}$ is said to be Rindler-Newton-Wigner localized in the wedge $\nu$ and the region $\mathcal{V}$ if $\hat{O} \in \mathfrak{A}_\nu^{\text{NW}}(\mathcal{V})$. The state $| \phi \rangle = \hat{O} |\Omega \rangle $ is Rindler-Newton-Wigner localized in $\nu$ and $\mathcal{V}$ over the background $|\Omega \rangle $ if $\hat{O} \in \mathfrak{A}_\nu^{\text{NW}}(\mathcal{V})$.

In the following subsections we study the RaRbR, the RaRbM and the ARbM scenarios introduced in Sec.~\ref{Localization_in_accelerated_frame_Introduction}. The ABM scenario, instead, was already discussed in Sec.~\ref{NewtonWigner_localization_scheme}.

\subsection{Newton-Wigner RaRbR scenario}

By definition, the operator $\hat{A}_{\text{NW},\nu}(X)$ annihilates the vacuum $|0_\text{L},0_\text{R} \rangle$ and has the same algebraic properties as $\hat{a}_\text{NW}(\vec{x})$ from the Minkowski spacetime. Hence, all the features of the Newton-Wigner scheme detailed in Secs.~\ref{NewtonWigner_localization_scheme} and \ref{Comparison_between_localization_schemes} also apply here when $|0_\text{L},0_\text{R} \rangle$ is chosen as the background, i.e., $|\Omega \rangle = |0_\text{L},0_\text{R} \rangle$.

In particular, we find that the Rindler-Newton-Wigner operators satisfy the canonical commutation relation
\begin{align}\label{A_NW_commutation}
& [\hat{A}_{\text{NW},\nu}(X), \hat{A}_{\text{NW},\nu'}^\dagger(X') ] = \delta_{\nu\nu'} \delta(X-X'), & [\hat{A}_{\text{NW},\nu}(X), \hat{A}_{\text{NW},\nu'}(X') ] = 0.
\end{align}
As a consequence of Eq.~(\ref{A_NW_commutation}), operators localized in disjoint regions with respect to the Rindler-Newton-Wigner scheme commute. Explicitly, this means the for any $\hat{O}_\text{A} \in \mathfrak{A}_{\nu_\text{A}}^{\text{NW}}(\mathcal{V}_\text{A})$ and $\hat{O}_\text{B} \in \mathfrak{A}_{\nu_\text{B}}^{\text{NW}}(\mathcal{V}_\text{B})$, we have that $[\hat{O}_\text{A}, \hat{O}_\text{B}] = 0$ if $\nu_\text{A} \neq \nu_\text{B}$ or if $\mathcal{V}_\text{A}$ and $\mathcal{V}_\text{B}$ are disjoint. Also, due to the canonical commutation rule (\ref{A_NW_commutation}), the global Rindler-Fock space $\mathcal{H}_{\text{L},\text{R}}$ factorizes into local Rindler-Fock spaces $\mathcal{H}_\nu^\text{NW}(\mathcal{V}_i)$ upon which elements of $ \mathfrak{A}_\nu^{\text{NW}}(\mathcal{V}_i)$ act. We note by $| 0_\nu (\mathcal{V}) \rangle$ the vacuum of $\mathcal{H}_\nu^\text{NW}(\mathcal{V})$. The factorization of the global Rindler vacuum $|0_\text{L},0_\text{R} \rangle$ in $\mathcal{H}_{\text{L},\text{R}} = \bigotimes_\nu \bigotimes_i \mathcal{H}_\nu^\text{NW}(\mathcal{V}_i)$ is $|0_\text{L},0_\text{R} \rangle = \bigotimes_\nu \bigotimes_i | 0_\nu (\mathcal{V}_i) \rangle$.

As a consequence of the factorization of the Rindler-Fock space $\mathcal{H}_{\text{L},\text{R}} = \bigotimes_\nu \bigotimes_i \mathcal{H}_\nu^\text{NW}(\mathcal{V}_i)$ and the Rindler vacuum $|0_\text{L},0_\text{R} \rangle = \bigotimes_\nu \bigotimes_i | 0_\nu (\mathcal{V}_i) \rangle$, we find that for any couple of operators $\hat{O}_\text{A} \in \mathfrak{A}_{\nu_\text{A}}^{\text{NW}}(\mathcal{V}_\text{A})$ and $\hat{O}_\text{B} \in \mathfrak{A}_{\nu_\text{B}}^{\text{NW}}(\mathcal{V}_\text{B})$, the local state $| \phi \rangle = \hat{O}_\text{A} |0_\text{L},0_\text{R} \rangle $ satisfies Eq.~(\ref{KnightLicht_property_Omega}) with $| \Omega \rangle = |0_\text{L},0_\text{R} \rangle$ if $\nu_\text{A} \neq \nu_\text{B}$ or if $\mathcal{V}_\text{A}$ and $\mathcal{V}_\text{B}$ are disjoint. Hence, any state localized with respect to the Rindler-Newton-Wigner scheme is also strictly localized over the Rindler vacuum.

This result can be understood in terms of the Newton-Wigner RaRbR scenario introduced in Sec.~\ref{Localization_in_accelerated_frame_Introduction}. An accelerated experimenter (Rachel) prepares the state $| \phi \rangle$ over the Rindler vacuum $|0_\text{L},0_\text{R} \rangle $ by means of Newton-Wigner local operators in $\mathfrak{A}_{\nu_\text{A}}^{\text{NW}}(\mathcal{V}_\text{A})$. The accelerated observer Rob performs measurements by using elements of $\mathfrak{A}_{\nu_\text{B}}^{\text{NW}}(\mathcal{V}_\text{B})$. We find that the local preparation of $| \phi \rangle$ in $\nu_\text{A}$ and $\mathcal{V}_\text{A}$ do not influence measurements in $\nu_\text{B}$ and $\mathcal{V}_\text{B}$. This result is equivalent to the one obtained for the Newton-Wigner ABM scenario in Sec.~\ref{Comparison_between_the_NewtonWigner_and_the_AQFT_schemes}.

\subsection{Newton-Wigner RaRbM scenario}\label{RaRbM_scenario}

The factorization $|0_\text{L},0_\text{R} \rangle = \bigotimes_\nu \bigotimes_i | 0_\nu (\mathcal{V}_i) \rangle$ only holds for the Rindler vacuum. For instance, in the case of the Minkowski vacuum, we have that $|0_\text{M} \rangle \neq \bigotimes_\nu \bigotimes_i | 0_\nu (\mathcal{V}_i) \rangle$, as a consequence of Eq.~(\ref{Rindler_vacuum_to_Minkowski}), with $\hat{S}_\text{S}$ defined by Eq.~(\ref{D}). The Minkowski vacuum factorizes with respect to the Minkowski local Fock spaces $\mathcal{H}_\text{M}^\text{NW}(\mathcal{V}_i)$ [Sec.~\ref{Comparison_between_the_NewtonWigner_and_the_AQFT_schemes}]; however, it does not factorize with respect to $\mathcal{H}_{\text{L},\text{R}} = \bigotimes_\nu \bigotimes_i \mathcal{H}_\nu^\text{NW}(\mathcal{V}_i)$. Consequently, Eq.~(\ref{KnightLicht_property_Omega}) does not necessarily hold when $| \Omega \rangle = |0_\text{M} \rangle$, even if $| \phi \rangle = \hat{O}_\text{A} | 0_\text{M} \rangle$ and if the operators $\hat{O}_\text{A} $ and $\hat{O}_\text{B}$ are Rindler-Newton-Wigner localized in two disjoint regions. 

However, one can use Eq.~(\ref{local_unitary_operator_measurament_A}) to show that the unitarity preparation of $| \phi \rangle $ is a sufficient condition for Eq.~(\ref{KnightLicht_property_Omega}). In particular, if $\hat{O}_\text{A}$ is a unitary element of $\mathfrak{A}_{\nu_\text{A}}^{\text{NW}}(\mathcal{V}_\text{A})$ and if $\hat{O}_\text{B}$ is an element of $\mathfrak{A}_{\nu_\text{B}}^{\text{NW}}(\mathcal{V}_\text{B})$, with $\nu_\text{B} \neq \nu_\text{A}$ or with $\mathcal{V}_\text{B}$ disjoint from $\mathcal{V}_\text{A}$, then Eq.~(\ref{local_unitary_operator_measurament_A}) holds, since $[\hat{O}_\text{A}, \hat{O}_\text{B}] = 0$ and $\hat{O}_\text{A}^\dagger \hat{O}_\text{A} = 1$. As a result, we find that when the Minkowski vacuum $|0_\text{M} \rangle$ is chosen as a background state, the strict localization property in the Rindler-Newton-Wigner scheme is only guaranteed for unitary preparations of states.

We can interpret this results in the context of the RaRbM scenario, which is equivalent to the RaRbR scenario, except for the Minkowski vacuum $| 0_\text{M} \rangle$ replacing the Rindler vacuum $|0_\text{L},0_\text{R} \rangle$ as the background state. We find that, in the Newton-Wigner RaRbM experiment, selective preparations by Rachel may be detected by Rob; whereas nonselective operations on the vacuum $|0_\text{M} \rangle$ suffice to not influence measurements in the other disjoint region.

The effect looks very similar to the Reeh-Schlieder nonlocality discussed in Sec.~\ref{solving_the_paradox} for two reasons: (i) The local preparation of states affects local measurements in a separated region, except when only nonselective preparations are considered; (ii) The origin of the effect is ascribed to a background state that does not factorize into local vacua. However, here, we did not consider the AQFT scheme, but the Newton-Wigner localization.

\subsection{Newton-Wigner ARbM scenario}\label{NewtonWigner_ARbM_scenario}

The last scenario that we want to detail is the ARbM. In this case, states are prepared by an inertial observer (Alice) via Minkowski-Newton-Wigner operators $\hat{a}_\text{NW}(x)$, which are defined as in Eq.~(\ref{a_NW}) but in one spatial dimension, i.e.,
\begin{equation}\label{a_NW_11}
\hat{a}_\text{NW}(x) = \int_{\mathbb{R}} d k \frac{e^{i k x}}{\sqrt{2 \pi}} \hat{a}(k).
\end{equation}
The identity relating $\hat{a}_\text{NW}(x)$ to $\hat{A}_{\text{NW},\nu}(X)$ can be computed by means of Eqs.~(\ref{Bogolyubov_transformation}), (\ref{A_NW_X_A_K}) and (\ref{a_NW_11}), which give
\begin{align}\label{NW_operators_MR}
\hat{a}_\text{NW}(x) = & \int_{\mathbb{R}} dX  \left[  \tilde{\alpha}_\text{L}(x,X)  \hat{A}_\text{NW,L}(X) - \tilde{\beta}_\text{L}^*(x,X) \hat{A}^\dagger_\text{NW,L}(X) \right. \nonumber \\
& \left. +  \tilde{\alpha}_\text{R}^*(x,X) \hat{A}_\text{NW,R}(X) - \tilde{\beta}_\text{R}(x,X) \hat{A}^\dagger_\text{NW,R}(X) \right],
\end{align} 
with
\begin{subequations}\label{alpha_tilde_beta_tilde}
\begin{align}
& \tilde{\alpha}_\nu(x,X) = \int_{\mathbb{R}} dk \int_{\mathbb{R}} dK \frac{e^{-s_\nu i k x + s_\nu i K X}}{2\pi} \alpha(k,K), \\
& \tilde{\beta}_\nu(x,X) = \int_{\mathbb{R}} dk \int_{\mathbb{R}} dK \frac{e^{s_\nu i k x + s_\nu i K X}}{2\pi} \beta(k,K).
\end{align}
\end{subequations}

By using Eqs.~(\ref{F_k_K}) and (\ref{Bogolyubov_coefficients}) in Eq.~(\ref{alpha_tilde_beta_tilde}), we find that the support of the kernel functions $\tilde{\alpha}_\nu(x,X)$ and $\tilde{\beta}_\nu(x,X)$ with respect to the variable $X$ is the entire real axis $\mathbb{R}$. Hence, the Minkowski-Newton-Wigner operator $\hat{a}_\text{NW}(x)$ is not localized with respect to the Rindler-Newton-Wigner scheme. Due to the nontrivial Bogoliubov transformation relating the two sets of creators and annihilators, we find that the Minkowski and the Rindler Newton-Wigner schemes are incompatibles. Hence, the notion of Newton-Wigner localization is noncovariant with respect to GR diffeomorphisms. 

In Sec.~\ref{Comparison_between_the_NewtonWigner_and_the_AQFT_schemes}, we discussed the Newton-Wigner ABM scenario, where both preparation of states and measurements are preformed by means of operators localized with respect to the Minkowski-Newton-Wigner scheme and over the Minkowski vacuum $| 0_\text{M} \rangle$. The independence between preparation and measurement is a consequence of the factorization of the global Minkowski-Fock space $\mathcal{H}_\text{M}$ into local Fock spaces $\mathcal{H}_\text{M}^\text{NW}(\mathcal{V}_i)$ with local vacua $| 0_\text{M} (\mathcal{V}_i)\rangle$ and the global vacuum $| 0_\text{M} \rangle$ into the local vacua $| 0_\text{M} (\mathcal{V}_i)\rangle$.

In the ARbM scenario, the inertial observer Bob is replaced by the accelerated observer Rob, who has access to the algebra $\mathfrak{A}_{\nu_\text{B}}^\text{NW} (\mathcal{V}_\text{B})$. Due to the incompatibility between the Minkowski and the Rindler Newton-Wigner schemes, any operator $\hat{O}_\text{B} \in \mathfrak{A}_{\nu_\text{B}}^\text{NW} (\mathcal{V}_\text{B})$ measured by Rob is actually global with respect to the Minkowski-Newton-Wigner scheme, i.e., $\hat{O}_\text{B} \in \mathfrak{A}_\text{M}^\text{NW} (\mathbb{R})$. For this reason, the preparation of the state $| \phi \rangle = \hat{O}_\text{A} |0_\text{M} \rangle $ by Alice, with $\hat{O}_\text{A}\in \mathfrak{A}_\text{M}^\text{NW} (\mathcal{V}_\text{A})$, may influence the measurement of $\hat{O}_\text{B}$ by Rob.

In Sec.~\ref{Example_single_particle_state_NW}, we will show an explicit example in which such a nonlocal effect occurs. In particular, we will consider a Minkowski single particle localized in the left wedge with respect to the Newton-Wigner scheme and we will demonstrate that its preparation affects measurements of Rindler observables in the right wedge.

At variance with the AQFT ABM scenario [Sec.~\ref{solving_the_paradox}] and the Newton-Wigner RaRbM scenario [Sec.~\ref{RaRbM_scenario}], here, the independence between local preparations of states and local measurements of observables by the two experimenters is not guaranteed by the unitarity of $\hat{O}_\text{A}$. Hence, we find that nonselective local preparations of states can be detected in other disjoint regions.

In Sec.~\ref{solving_the_paradox}, we argued that selective operations cannot be used to instantly send information to other disjoint regions, whereas nonselective operations are guaranteed to satisfy the strict localization property in the AQFT ABM scenario. Here, we find that the strict localization property is not guaranteed by nonselective operations in the Newton-Wigner ARbM experiment. Hence, the nonlocal effect predicted in this case is in contradiction with causality. In principle, Alice can unitarily prepare a local state to instantly send information to Rob if they are localized with respect to the Newton-Wigner scheme of their respective frame.

We highlight the novelty of this subsection by recalling the features of the Newton-Wigner localization in QFT. In Sec.~\ref{NewtonWigner_localization_scheme}, we pointed out that noncovariance under Lorentz transformation and superluminal spreading of the wave function are unavoidable features of the Newton-Wigner scheme. However, the theory is not problematic at all if one fixes the Lorentz frame and only considers the spacelike hypersurface $t = 0$. We remark that the acausal effect of the Hegerfeldt theorem [Sec.~\ref{Hegerfeldt_theorem}] can only be detected at times different from $t = 0$, though arbitrary close from it.

Here, instead, we only consider the initial hypersurface $t = 0$ and no Lorentz boost, since at that time the accelerated observer has zero velocity in the inertial frame. Instead, we consider noninertial coordinate transformations. We show the noncovariance under diffeomorphism and we derive the violation of causality occurring at the instant $t = 0$. Hence, we conclude that the Newton-Wigner scheme lack GR covariance in addition to special relativistic covariance and that acausal effects already occur at $t = 0$ for an inertial and a noninertial observer.

\section{AQFT scheme}\label{AQFT_localization_scheme_in_curved_spacetime}

In Sec.~\ref{Algebraic_approach_in_QFT_and_QFTCS}, we discussed the algebraic approach to QFT. A natural generalization to QFTCS---with particular focus on accelerated frames---was obtained by associating local algebras to spacetime events, without using any particular choice of coordinate systems. Local field operators in each frame---i.e., $\hat{\phi}(x^\mu)$ and $\hat{\Phi}_\nu(X^\mu)$, respectively---emerge as a coordinatization to the local algebraic structure and, hence, they provide different representations of the same theory. Such a unifying property of AQFT is explicitly expressed by Eq.~(\ref{scalar_transformation_Rindler}), which tells how field operators transform between different frames via global coordinate transformation.

The same line of argument can be used to naturally extend the AQFT localization scheme [Sec.~\ref{AQFT_localization_scheme}] to curved spacetimes and, in particular, to the accelerated frames. By following Sec.~\ref{AQFT_localization_scheme}, we say that the observable $\hat{O}$ and the state $| \Phi \rangle = \hat{O} | \Omega \rangle$ are localized in $\mathcal{O}$ over the background $| \Omega \rangle$ if $\hat{O}$ belongs to the local algebra $\mathfrak{A}(\mathcal{O})$. Notice that the only difference with the QFT case is given by the freedom of choosing any background state $| \Omega \rangle$, which, in principle, may differ from the Minkowski vacuum $|0_\text{M} \rangle$.

To see an example, consider the Rindler spacetime coordinate $(T,\vec{X})$ and the real scalar field $\hat{O} = \hat{\Phi}_\nu(T,\vec{X})$ evaluated at the point $(T,\vec{X})$ and at the wedge $\nu$. The observable $\hat{\Phi}_\nu(T,\vec{X})$ is an element of $\mathfrak{A}(\mathcal{E})$, where $\mathcal{E}$ is the spacetime event represented by $(T,\vec{X})$ and $\nu$ in the Rindler coordinate system. Hence, we say that the observable $\hat{O}$ is localized in $\mathcal{E}$ with respect to the AQFT scheme. Also, the state $| \Phi \rangle = \hat{O} | \Omega \rangle$ is localized in $\mathcal{E}$ over the background $| \Omega \rangle$.

Notice that, as a consequence of Eq.~(\ref{scalar_transformation_Rindler}), the operator $\hat{O} = \hat{\Phi}_\nu(T,\vec{X})$ can be written in terms of the Minkowski scalar field as $\hat{\phi} (t_\nu(T,\vec{X}), \vec{x}_\nu(T,\vec{X}))$. The Minkowski coordinate point $(t,\vec{x}) = (t_\nu(T,\vec{X}), \vec{x}_\nu(T,\vec{X}))$ represents the same event $\mathcal{E}$ identified by the Rindler coordinate $(T,\vec{X})$ and the wedge $\nu$. We find that if an observable is localized in an event $\mathcal{E}$ with respect to an accelerated observer, it is also localized in $\mathcal{E}$ with respect to an inertial observer. Hence, the notion of localization in AQFT appears to be frame independent, in the sense that inertial and accelerated observers agree on the region in which states and observables are localized. As we remarked above, this is due to the fact that local algebras are defined with respect to frame independent spacetime events.

The frame independent notion of the AQFT scheme can be described in terms of local preparation of states over the background $| \Omega \rangle$ and local measurements of observables. We consider an inertial and an accelerated experimenter, Alice and Rachel, that prepare states in each of their frames. If they are both localized in the same spacetime region, they share the same algebra of operators and may, in principle, prepare the same state. The only difference is given by the respective coordinate systems by means of which they describe the spacetime. Analogously, an inertial and an accelerated experimenter, Bob and Rob, can perform the same measurement if they are localized in the same region of spacetime.

We now study the dependence between outcomes of measurements by Rob and the local preparation of states by Alice or Rachel. In analogy to what we saw in Sec.~\ref{solving_the_paradox}, the strict localization property is only guaranteed for nonselective operations. In particular, Eq.~(\ref{KnightLicht_property_Omega}) with $| \phi \rangle = \hat{O}_\text{A} | \Omega \rangle$ is satisfied when $\hat{O}_\text{A}$ is unitary and the operators $\hat{O}_\text{A} \in \mathfrak{A}(\mathcal{O}_\text{A})$ and $\hat{O}_\text{B} \in \mathfrak{A}(\mathcal{O}_\text{B})$ are localized in spacelike separated regions. Selective preparation of states over $| \Omega \rangle$, instead, lead to nonlocal effects such that
\begin{equation}\label{not_KnightLicht_property_Omega}
\langle \phi | \hat{O}_\text{B} | \phi \rangle \neq \langle \Omega | \hat{O}_\text{B}  | \Omega \rangle,
\end{equation}
even if $\hat{O}_\text{A} \in \mathfrak{A}(\mathcal{O}_\text{A})$ and $\hat{O}_\text{B} \in \mathfrak{A}(\mathcal{O}_\text{B})$ are localized in spacelike separated regions. As argued in Sec.~\ref{solving_the_paradox}, this nonlocality cannot be used for superluminal signaling.

Such a result is independent of the nature of the experimenters in $\mathcal{O}_\text{A}$ and $\mathcal{O}_\text{B}$. They can be both inertial, both accelerated or one inertial and the other one accelerated; in all of these cases, nonlocal effects occur if states are prepared via selective operations. This is a consequence of the compatibility between the AQFT schemes in the two frames, i.e., the fact that inertial and accelerated observers agree on the localization of states and observables. By using the notation of Table \ref{scenarios_Table}, we say that, in the AQFT scheme, all the proposed scenarios (i.e., ABM, RaRbR, RaRbM and ARbM) are guaranteed to satisfy the strict localization property only for nonselective preparations of states.

An example of selectively prepared state that does not satisfy the strict localization property is given by the Minkowski single particle state. In Sec.~\ref{Example_single_particle_state_AQFT}, we will explicitly prove that the preparation of such a state in the left wedge over the Minkowski vacuum $| 0_\text{M} \rangle$ affects measurements of Rindler observables in the right wedge.

In Sec.~\ref{Strict_localization_in_the_nonrelativistic_limit}, we found that such a nonlocal effect is suppressed in the nonrelativistic limit of the ABM scenario. We may wonder if this also occurs in the other scenarios. In Secs.~\ref{Localization_in_NRQFTCS} we will show that a similar suppression only occurs when the nonrelativistic limit of the two experimenters is with respect to the frame associated to the vacuum background; hence, only in the nonrelativistic ABM and RaRbR scenarios the strict localization property is always satisfied.

\section{Modal scheme}\label{Modal_localization_scheme_Rindler}

In Sec.~\ref{Modal_localization_scheme}, we defined the modal localization scheme in QFT. The approach was based on the representation of Minkowski single particle states as positive frequency modes of the field equation [Sec.~\ref{QFT_in_Minkowski_spacetime_scalar}]. A similar representation was provided for Rindler-Fock states as well [Sec.~\ref{QFT_in_curved_spacetime_Rindler_scalar}]. Here, we use such a modal representation of Rindler particles and the prescription of Sec.~\ref{Modal_localization_scheme} to define the modal scheme in the Rindler frame.

The modes $F_\nu(\Omega,\vec{K}_\perp,T,\vec{X}) $ [Eqs.~(\ref{F_Rindler_all}) and (\ref{F_F_nu})] are representatives of the $\nu$-Rindler single particles with frequency $\Omega$ and transverse momentum $\vec{K}_\perp$. They are solutions of the Rindler-Klein-Gordon equation and are orthonormal with respect to the Klein-Gordon product (\ref{KG_scalar_product}), according to Eqs.~(\ref{Rindler_Klein_Gordon_product}) and (\ref{KG_scalar_curved_product_orthonormality_F}). Any general Rindler-Fock state $| \Phi \rangle \in \mathcal{H}_{\text{L},\text{R}}$ is represented by the wave functions in momentum space $\tilde{\Phi}_n (\bm{\nu}_n,\bm{\theta}_n)$ [Eq.~(\ref{general_Fock_expansion_Rindler})] and in position space $\Phi_n (T, \bm{\nu}_n, \textbf{X}_n)$ [Eq.~(\ref{wavefunction_F})]. 

The representation of Rindler particle states in terms of positive frequency modes provides the definition of the modal localization scheme over the Rindler vacuum $| 0_\text{L}, 0_\text{R} \rangle$ as the background state. In particular, we say that the state $| \Phi \rangle \in \mathcal{H}_{\text{L},\text{R}}$ is localized in the wedge $\nu$ and in the volume $\mathcal{V}$ at $T=0$ over the Rindler vacuum $| 0_\text{L}, 0_\text{R} \rangle$ and with respect to the Rindler modal scheme if $\Phi_n (0, \bm{\nu}_n, \textbf{X}_n)$ is supported in $\nu^n \otimes \mathcal{V}^n$, in the sense that $\Phi_n (0, \bm{\nu}_n, \textbf{X}_n) = 0$ when there is an $l \in \{ 1, \dots, n \}$ such that $\nu_l \neq \nu$ or $\vec{X}_l \notin \mathcal{V}$.

The localization of Rindler-Fock states over the Rindler vacuum $| 0_\text{L}, 0_\text{R} \rangle$ provides a natural definition of local operators. By following the algebraic approach [Sec.~\ref{Algebraic_approach}], we know that any state $| \Phi \rangle $ of the Rindler-Fock space $ \mathcal{H}_{\text{L},\text{R}}$ can be---at least approximately---written as $| \Phi \rangle = \hat{O} | 0_\text{L}, 0_\text{R} \rangle$. Naturally, we say that $\hat{O}$ is localized in the wedge $\nu$ and in the volume $\mathcal{V}$ with respect to the Rindler modal scheme if $| \Phi \rangle $ is localized in $\nu$ and $\mathcal{V}$. This provides the definition of local algebras $\mathfrak{A}_\nu^\text{mod}(\mathcal{V})$ generated by operators localized in $\nu$ and $\mathcal{V}$.

The algebra of operators localized in $\nu$ and in the space point $\vec{X}$ can be obtained by inverting Eq.~(\ref{wavefunction_F}) as
\begin{equation}\label{wavefunction_F_inverse}
\tilde{\Phi}_n (\bm{\nu}_n,\bm{\theta}_n)  =  \left( \frac{2 m c^2}{\hbar^2} \right)^{n/2}  \int_{\mathbb{R}^{3n}} d^{3n} \textbf{X}_n  \Phi_n (0, \bm{\nu}_n, \textbf{X}_n) \prod_{l=1}^n  \mathcal{F}_{\nu_l}^*(\vec{\theta}_l,0,\vec{X}_l),
\end{equation}
with
\begin{equation}\label{FF_nu_F_nu}
\mathcal{F}_\nu (\Omega, \vec{K}_\perp, T ,\vec{X}) = \frac{\hbar \Omega}{m c^2} F_\nu (\Omega, \vec{K}_\perp, T ,\vec{X}).
\end{equation}
Equation (\ref{wavefunction_F_inverse}) is a result of
\begin{equation}\label{FF_nu_F_nu_delta}
\int_{\mathbb{R}^3} d^3X  \frac{2 m c^2}{ \hbar^2 } F_\nu(\vec{\theta},0,\vec{X})  \mathcal{F}_\nu^*(\vec{\theta}',0,\vec{X}) = \delta^3(\vec{\theta}-\vec{\theta}'),
\end{equation}
which can be derived from Eqs.~(\ref{F_Rindler}), (\ref{F_F_nu}), (\ref{FF_nu_F_nu}) and
\begin{equation}\label{alpha_tilde_approx_3_inverse}
\int_0^\infty dz  \frac{2 \Omega}{\hbar a z}  \tilde{F}(\Omega, \vec{K}_\perp,Z_\text{R}(z)) \tilde{F}(\Omega', \vec{K}_\perp,Z_\text{R}(z)) = \frac{1}{4 \pi^2} \delta(\Omega-\Omega').
\end{equation}
The proof for Eq.~(\ref{alpha_tilde_approx_3_inverse}) is shown in Appendix \ref{Proof_of_alpha_tilde_approx_3_inverse}.

By using Eq.~(\ref{wavefunction_F_inverse}) in Eq.~(\ref{general_Fock_expansion_Rindler}) we obtain
\begin{equation}
| \Psi \rangle  =  \sum_{n=0}^\infty \frac{1}{\sqrt{n!}} \sum_{\bm{\nu}_n}   \int_{\mathbb{R}^{3n}} d^{3n} \textbf{X}_n  \Phi_n (0, \bm{\nu}_n, \textbf{X}_n)  \prod_{l=1}^n \hat{A}_{\text{mod},\nu_l}^\dagger (\vec{X}_l) | 0_\text{L}, 0_\text{R} \rangle,
\end{equation}
with
\begin{equation}\label{a_mod_a_nu}
\hat{A}_{\text{mod},\nu} (\vec{X})  = \int_{\theta_1>0} d^3 \theta \frac{\sqrt{2 m c^2}}{\hbar} \mathcal{F}_\nu(\vec{\theta},0,\vec{X}) \hat{A}_\nu(\vec{\theta}).
\end{equation}
By definition, $| \Psi \rangle $ is localized in the wedge $\nu$ and in the region $\mathcal{V}$ over $| 0_\text{L}, 0_\text{R} \rangle$ with respect to the Rindler modal scheme if $\Phi_n (0, \bm{\nu}_n, \textbf{X}_n) $ is supported in  $\nu^n \otimes \mathcal{V}^n$. This gives a natural definition for local operators. We say that the operator $\hat{O}$ is localized in the wedge $\nu$ at the space coordinate $\vec{X}$ if it belongs to the algebra $\mathfrak{A}_\nu^\text{mod}(\vec{X})$ generated by $\hat{A}_{\text{mod},\nu} (\vec{X})$. We also define the local algebra $\mathfrak{A}_\nu^\text{mod}(\mathcal{V})$ generated by the operators $\hat{A}_{\text{mod},\nu} (\vec{X})$ with $\vec{X} \in \mathcal{V}$.

By means of the local algebras $\mathfrak{A}_\nu^\text{mod}(\vec{X})$ and $\mathfrak{A}_\nu^\text{mod}(\mathcal{V})$, it is possible to extend the definition of localized states to include any background $| \Omega \rangle$ that differs from the Rindler vacuum $| 0_\text{L}, 0_\text{R} \rangle$. In particular, we say that the state $| \Phi \rangle = \hat{O} | \Omega \rangle$ is localized in $\nu$ and $\vec{X}$ over the background $| \Omega \rangle$ with respect to the Rindler modal scheme if $\hat{O}$ is an element of $\mathfrak{A}_\nu^\text{mod}(\vec{X})$.

It can be noticed that the Rindler modal scheme is not compatible with the Minkowski modal scheme, in the sense that no state or operator is localized in the same spacetime event with respect to both schemes. Mathematically, this means that the algebras $\mathfrak{A}_\text{M}^\text{mod}(\vec{x})$ and $\mathfrak{A}_\nu^\text{mod}(\vec{X})$ do not coincide, even if $\vec{x}$ and $\vec{X}$ represent the same point in the $\nu$ wedge, i.e., $\vec{x} = \vec{x}_\nu(0,\vec{X})$. This is a consequence of the nontrivial Bogoliubov transformation relating the Rindler modal operators $\hat{A}_{\text{mod},\nu} (\vec{X}) $ [Eq.~(\ref{a_mod_a_nu})] to their Minkowski counterpart $\hat{a}_\text{mod} (\vec{x})$ [Eq.~(\ref{a_mod_a})], i.e.,
\begin{align}\label{a_mod_M_A_mod_nu}
\hat{a}_\text{mod} (\vec{x})  = & \sum_{\nu=\{\text{L},\text{R}\}} \int_{\mathbb{R}^3} d^3X  \left[ f_{\text{mod},\nu \mapsto \text{mod}}^+ (\vec{x},\vec{X})  \hat{A}_{\text{mod},\nu} (\vec{X}) \right. \nonumber \\
& \left. + f_{\text{mod},\nu \mapsto \text{mod}}^- (\vec{x},\vec{X}) \hat{A}_{\text{mod},\nu}^\dagger (\vec{X}) \right],
\end{align}
with
\begin{equation}
f_{\text{mod},\nu \mapsto \text{mod}}^\pm (\vec{x},\vec{X}) = \int_{\mathbb{R}^3} d^3 k  \int_{\theta_1>0} d^3 \theta \sqrt{\frac{ \omega(\vec{k})}{4 \pi^3 \hbar}}  e^{i \vec{k} \cdot \vec{x}}  \alpha_\nu(\vec{k},\pm \vec{\theta})   F_\nu^*(\pm \vec{\theta},0,\vec{X}).
\end{equation}
Equation (\ref{a_mod_M_A_mod_nu}) can be derived from Eq.~(\ref{Rindler_Bogoliubov_transformations}) and from the inverse of (\ref{a_mod_a_nu}), i.e.,
\begin{equation}\label{a_mod_a_nu_inverse}
\hat{A}_\nu(\vec{\theta}) = \int_{\mathbb{R}^3} d^3X  \frac{\sqrt{2 m c^2}}{ \hbar } F_\nu^*(\vec{\theta},0,\vec{X}) \hat{A}_{\text{mod},\nu} (\vec{X}),
\end{equation}
which, in turn, can be proven by using Eq.~(\ref{FF_nu_F_nu_delta}). As a result, we find that the modal localization scheme does not satisfy GR covariance under diffeomorphism and cannot be used to describe local physical phenomena in the QFTCS regime.

In addition to the violation of the GR covariance under diffeomorphism, one can prove the incompatibility between the Rindler modal scheme and the relativistic causality by noticing that operators localized in different regions of the space do not generally commute. To see this, use Eq.~(\ref{a_mod_a_nu}) to obtain
\begin{subequations}\label{A_mod_commutation}
\begin{align}
& [ \hat{A}_{\text{mod},\nu} (\vec{X}), \hat{A}_{\text{mod},\nu'}^\dagger (\vec{X}')] = \delta_{\nu\nu'} \int_{\theta_1>0} d^3 \theta \frac{2 m c^2}{\hbar^2} \mathcal{F}_\nu(\vec{\theta},0,\vec{X}) \mathcal{F}_\nu^*(\vec{\theta},0,\vec{X}'), \\
 & [ \hat{A}_{\text{mod},\nu} (\vec{X}), \hat{A}_{\text{mod},\nu'} (\vec{X}')]=0,
\end{align}
\end{subequations}
which implies that the operators $\hat{A}_{\text{mod},\nu} (\vec{X}) $ and $\hat{A}_{\text{mod},\nu}^\dagger (\vec{X}') $ do not commute with each other, even if $\vec{X}$ is different from $\vec{X}'$. Consequently, operators localized in different regions with respect to the Rindler modal scheme are not guaranteed to commute. Also, by means of Eqs.~(\ref{a_mod_M_A_mod_nu}) and (\ref{A_mod_commutation}) it can be proven that $\hat{a}_\text{mod} (\vec{x}) $ does not commute with $\hat{A}_{\text{mod},\nu} (\vec{X}) $ nor $\hat{A}_{\text{mod},\nu}^\dagger (\vec{X}) $, even if the Minkowski coordinate $\vec{x}$ represents the same event of the $\nu$-Rindler coordinate $\vec{X}$. Hence, we find that $[\hat{O}_\text{A}, \hat{O}_\text{B}] \neq 0$ for some couples of operators $\hat{O}_\text{A} \in \mathfrak{A}_\text{M}^\text{mod}(\mathcal{V}_\text{A})$ and $\hat{O}_\text{B} \in \mathfrak{A}_\nu^\text{mod}(\mathcal{V}_\text{B})$, with $\mathcal{V}_\text{A}$ and $\mathcal{V}_\text{B}$ representing disjoint regions of the spacetime.

We already saw this feature in the case of the Minkowski modal scheme [Sec.~\ref{Comparison_between_NewtonWigner_and_modal_schemes}]. As consequence, we found that the strict localization property is not guaranteed. The result can be interpreted in therms of the modal ABM scenario, where an inertial experimenter (Alice) prepares a state localized in $\mathcal{O}_\text{A}$ with respect to the Minkowski modal scheme over the Minkowski vacuum $| 0_\text{M} \rangle$ and another inertial observer (Bob) performs local measurements in $\mathcal{O}_\text{B}$. When the state is not strictly localized in $\mathcal{O}_\text{A}$, its preparation may affect the results of Bob's measurements.

The same result can be extended to the case in which experimenter A (or B) uses the Rindler modal scheme, or when both experimenters use the Rindler modal scheme over any background state. Consequently, in all the scenarios defined by Table \ref{scenarios_Table} (i.e., ABM, RaRbR, RaRbM and ARbM) the strict localization property is not guaranteed and preparations of states by the experimenter A can influence measurements performed by the experimenter B.

Crucially, the strict localization property is not guaranteed for nonselective preparations either. To see this, assume that the state $| \phi \rangle = \hat{O}_\text{A} | \Omega \rangle$ is localized in $\mathcal{O}_\text{A}$ over $| \Omega \rangle$, with $\hat{O}_\text{A}$ satisfying the unitary condition $\hat{O}_\text{A}^\dagger \hat{O}_\text{A} = 1$, and consider any operator $\hat{O}_\text{B}$ measured by the experimenter B in $\mathcal{O}_\text{B}$. By using the noncommutative identity for operators localized in different regions [Eq.~(\ref{A_mod_commutation})], we find that
\begin{equation}
\langle \phi | \hat{O}_\text{B} | \phi \rangle = \langle \Omega | \hat{O}_\text{A}^\dagger \hat{O}_\text{B} \hat{O}_\text{A} | \Omega \rangle \neq \langle \Omega | \hat{O}_\text{A}^\dagger \hat{O}_\text{A}  \hat{O}_\text{B}| \Omega \rangle  = \langle \Omega |  \hat{O}_\text{B}| \Omega \rangle,
\end{equation}
which means that the unitary preparation of $| \phi \rangle$ by experimenter A is able to influence measurements by experimenter B in $\hat{O}_\text{B}$. This leads to a violation of causality [Sec.~\ref{solving_the_paradox}].

An additional acausal effect is given by the instantaneous spreading of wave functions, which was already noticed for the modal scheme in QFT [Sec.~\ref{Modal_localization_scheme}]. This issue is also present in the Rindler frame and is due to the fact that the modes $F_\nu (\Omega, \vec{K}_\perp, T ,\vec{X})$ have positive frequency [Eqs.~(\ref{F_Rindler}) and (\ref{F_F_nu})], which means that $ \partial_0 \Phi_n (T, \bm{\nu}_n, \textbf{X}_n) |_{T=0}$ and $\Phi_n (0, \bm{\nu}_n, \textbf{X}_n)$ cannot be simultaneously compactly supported \cite{Afanasev:1996nm}. For this reason, the modal scheme is not consistent with the notion of relativistic causality.

The modal scheme does not entail any genuine notion of localization in QFTCS since it violates relativistic causality and covariance under diffeomorphism. Only the AQFT framework is able to provide a valid description for special relativistic and GR local phenomena. In Sec.~\ref{The_modal_scheme_converges_to_the_AQFT_scheme}, we will show that the modal scheme converges to the AQFT scheme in the nonrelativistic limit. Hence, we find that the nonrelativistic limit is able to fundamentalize the modal scheme.

\section{Comparison between localization schemes}\label{Comparison_between_localization_schemes_Rindler}

\begin{table}
\begin{center}
\centering
\begin{tabular}{| >{\raggedright\arraybackslash}m{17em} || >{\centering\arraybackslash}m{10em} |  >{\centering\arraybackslash}m{3.5em} |}
\hline
 & AQFT scheme & modal scheme\\
\hline
\hline
GR covariance and causality hold  & Yes & No \\
\hline
Operators in disjoint spatial regions commute  & Yes & No \\
\hline
The strict localization property [Eq.~(\ref{KnightLicht_property_Omega})] at $t=T=0$ is guaranteed & Only for local nonse- lective preparations & No \\
\hline
\end{tabular}
\end{center}
\caption{Summary table of the differences between the AQFT and the modal localization schemes in the Minkowski and the Rindler frame.}\label{Comparison_between_localization_schemes_QFTCS_Table} 
\end{table}

In this section, we compare the AQFT and the modal localization schemes presented in Secs.~\ref{AQFT_localization_scheme_in_curved_spacetime} and \ref{Modal_localization_scheme_Rindler}, respectively. The results are summarized in Table \ref{Comparison_between_localization_schemes_QFTCS_Table}.

In Sec.~\ref{AQFT_localization_scheme_in_curved_spacetime}, by showing the compatibility between the Minkowski and the Rindler AQFT schemes, we proved that the GR covariance holds. The modal scheme, instead, lacks this property as the local algebras $\mathfrak{A}_\text{M}^\text{mod}(\vec{x})$ and $\mathfrak{A}_\nu^\text{mod}(\vec{X})$ are not equivalent, even if $\vec{x}$ and $\vec{X}$ represent the same event in the $\nu$ wedge, i.e., $\vec{x} = \vec{x}_\nu(0,\vec{X})$. Also, the relativistic causality in the AQFT schemes is ensured by the microcausality axiom and by the fact that nonselectively prepared local states are always strictly localized. In the modal scheme, instead, wave functions instantly propagate and the strict localization property is not guaranteed even for unitarily prepared states, due to the noncommutativity relation between operators localized in disjoint regions. For these reasons we conclude that, between the two schemes, only the AQFT scheme gives a genuine description of local phenomena.

A direct comparison between the AQFT and the modal scheme in the Rindler frame can be obtained by defining the local algebras in spatial regions with respect to the AQFT scheme, in analogy to the algebras $\mathfrak{A}_\text{M}^\text{AQFT}(\vec{x})$ introduced in Sec.~\ref{Comparison_between_the_NewtonWigner_and_the_AQFT_schemes} for the Minkowski frame. In particular, we define $\mathfrak{A}_\nu^\text{AQFT}(\vec{X})$ as the algebra generated by the field $\hat{\Phi}_\nu(0,\vec{X})$ and its conjugate $\hat{\Pi}_\nu(0,\vec{X}) =  - \partial_0 \hat{\Phi}_\nu(T,\vec{X})|_{T=0}$. The algebra $\mathfrak{A}_\nu^\text{AQFT}(\mathcal{V})$, instead, is for the space region $\mathcal{V}$.

Notice that the local algebra in the Rindler frame $\mathfrak{A}_\nu^\text{AQFT}(\vec{X})$ is equivalent to the algebra $\mathfrak{A}_\text{M}^\text{AQFT}(\vec{x})$ in the Minkowski frame if $\vec{x} = \vec{x}_\nu(0, \vec{X})$. This can be proven by deriving the one-to-one map between the couple of fields $\hat{\phi}(0,\vec{X}), \hat{\pi}(0,\vec{x})$ and the couple $\hat{\Phi}_\nu(0,\vec{X}), \hat{\Pi}_\nu(0,\vec{x}) $ which can be obtained by means of Eq.~(\ref{scalar_transformation_Rindler}) and the chain rule (\ref{chain_rule_Rindler}) as follows
\begin{align}
& \hat{\Phi}_\nu (0,\vec{X}) = \hat{\phi} (0, \vec{x}_\nu(0,\vec{X})), & \hat{\Pi}_\nu(0,\vec{X})  = s_\nu a z_\nu(Z) \hat{\pi} (0, \vec{x}_\nu(0,\vec{X})).
\end{align}

The incompatibility between the modal and the AQFT scheme can be proved by showing that operators that are localized with respect to the one of the two schemes are not localized with respect to the other. To see this, consider a Rindler single particle creator, defined as
\begin{equation}\label{A_mod_Phi_1}
\hat{A}_{\text{mod},\nu}^\dagger [\Phi] = \int_{\theta_1>0} \tilde{\Phi} (\vec{\theta})  \hat{A}_\nu^\dagger(\vec{\theta}).
\end{equation}
By means of Eq.~(\ref{a_mod_a_nu_inverse}), one can also consider
\begin{equation}\label{A_mod_Phi}
\hat{A}_{\text{mod},\nu}^\dagger [\Phi] = \int_{\mathbb{R}^3} d^3X \Phi (\vec{X})  \hat{A}_{\text{mod},\nu}^\dagger(\vec{X}),
\end{equation}
with $\Phi (\vec{X}) = \Phi (0, \vec{X})$ and
\begin{equation}
\Phi (T, \vec{X}) =  \frac{\sqrt{2 m c^2}}{\hbar}  \tilde{\Phi} (\vec{\theta})  F_\nu(\vec{\theta},T,\vec{X})
\end{equation}
as the wave function in position space. The operator $\hat{A}_{\text{mod},\nu}^\dagger [\Phi]$ is localized in the wedge $\nu$ and in the region $\mathcal{V}$ with respect to the Rindler modal scheme (i.e., $\hat{A}_{\text{mod},\nu}^\dagger [\Phi] \in \mathfrak{A}_\nu^\text{mod}(\mathcal{V})$) if $\Phi (\vec{X})$  is supported in $\mathcal{V}$.

Consider an operator $\hat{A}_{\text{mod},\nu}^\dagger [\Phi]$ such that $\tilde{\Phi} (\vec{\theta})$ is real. By using the orthonormality condition (\ref{KG_scalar_curved_product_orthonormality_F}) in Eq.~(\ref{Rindler_scalar_decomposition}), we can write $\hat{A}_{\text{mod},\nu}^\dagger [\Phi]$ in terms of the Rindler fields $\hat{\Phi}_\nu(0,\vec{X}), \hat{\Pi}_\nu(0,\vec{x}) $ as
\begin{align}\label{A_Phi_1_2}
\hat{A}_{\text{mod},\nu}^\dagger [\Phi] = & -\int_{\theta_1>0} \tilde{\Phi} (\vec{\theta}) (\hat{\Phi}_\nu, F_\nu^*(\vec{\theta}))_\text{KG} \nonumber\\
= & - \frac{\hbar}{\sqrt{2 m c^2}} (  \hat{\Phi}_\nu, \Phi^* )_\text{KG}\nonumber\\
= & - \frac{i}{\sqrt{2 m c^2}} \int_{\mathbb{R}^3} d^3X \left[ \hat{\Phi}_\nu(0,\vec{X}) \left. \partial_0 \Phi^*(T,\vec{X})\right|_{T=0} \right. \nonumber \\
& \left. + c^2 \Phi^*(\vec{X})  \hat{\Pi}_\nu(0,\vec{X}) \right].
\end{align}
If $\Phi (\vec{X})$ is supported in $\mathcal{V}$, the operator $\hat{A}_\text{mod}^\dagger [\Phi]$ is localized with respect to the modal scheme. However, it is not localized with respect to the AQFT scheme due to $\partial_0 \Phi(T,\vec{X})|_{T=0}$ acting as a smearing function for $\hat{\Phi}_\nu(0,\vec{X})$ in Eq.~(\ref{A_Phi_1_2}). The function $\Phi(T,\vec{X})$ is a positive frequency solution of the field equation; therefore, if it is supported in $\mathcal{V}$, its time derivative $\partial_0 \Phi(T,\vec{X})$ is not.

In Sec.~\ref{The_modal_scheme_converges_to_the_AQFT_scheme}, we will show how the two localization schemes converge in the nonrelativistic limit. This will imply that the modal operators $\hat{A}_{\text{mod},\nu}^\dagger [\Phi] \in \mathfrak{A}_\nu^\text{mod} (\mathcal{V})$ can be approximated by some local field operators $\hat{A}_{\text{AQFT},\nu}^\dagger [\Phi] \in \mathfrak{A}_\nu^\text{AQFT} (\mathcal{V})$. The intuition is that if $\Phi(\vec{X})$ is supported in $\mathcal{V}$, then its time derivative $\partial_0 \Phi(T,\vec{X}) |_{T=0}$ is approximately vanishing outside $\mathcal{V}$. Consequently, the right hand side of Eq.~(\ref{A_Phi_1_2}) is made of field operators approximately smeared over $\mathcal{V}$ and, hence, the single particle modal operator $\hat{A}_\text{mod}^\dagger [\Phi]$ is approximately local in $\mathcal{V}$ with respect to the AQFT scheme.

\section{Localization in the nonrelativistic regime}\label{Localization_in_NRQFTCS}

In the previous sections, we considered different localization schemes in the Rindler frames over a general background state $| \Omega \rangle$. In Sec.~\ref{Comparison_between_localization_schemes_Rindler}, we showed that the AQFT and the modal schemes are incompatible.

\begin{figure}
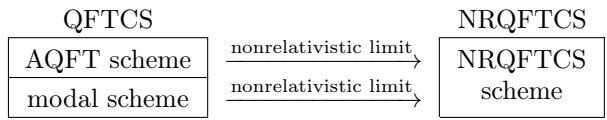

\begin{center}
\begin{tabular}{| c | c | c |}
\multicolumn{1}{c}{QFTCS} &  \multicolumn{1}{l}{} & \multicolumn{1}{c}{NRQFTCS} \\
\cline{1-1} \cline{3-3}
\centering AQFT scheme & $\xrightarrow[]{\text{nonrelativistic limit}}$  & \multirow{2}{5em}{\centering NRQFTCS scheme} \\
\cline{1-1}
\centering modal scheme & $\xrightarrow[]{\text{nonrelativistic limit}}$ &  \\
\cline{1-1} \cline{3-3}
\end{tabular}
\end{center}
\caption{Localization schemes in the relativistic (QFTCS) and the nonrelativistic (NRQFTCS) theory in curved spacetime. The AQFT and the modal scheme converge to a unified localization scheme in NRQFTCS.} \label{Localization_in_NRQFTCS_Table}
\end{figure}

In this section, we consider the nonrelativistic limit of these theories and we show that the two schemes converge [Fig.~\ref{Localization_in_NRQFTCS_Table}]. Then we detail the features of this unified localization scheme in NRQFTCS.

In Sec.~\ref{Modal_localization_scheme_Rindler}, we introduced the modal scheme in Rindler spacetime based on the modal representation of states [Sec.~\ref{QFT_in_curved_spacetime_Rindler_scalar}]. In Sec.~\ref{Rindler_frame_Scalar_field}, we showed that the space of modal wave functions is approximated by the $L^2(\mathbb{R}^3)$ Hilbert space when the nonrelativistic condition is satisfied. However, we did not discussed the probabilistic content of the modal wave functions. Here, we show that if $| \Omega \rangle = |0_\text{L},0_\text{R} \rangle$, the modal wave functions acquire the Born probabilistic notion of finding particles in each point. This is a consequence of the convergence to the AQFT scheme---which entails a genuine notion of locality---and the factorization of the Rindler nonrelativistic global Hilbert space $\mathcal{H}_{\text{L},\text{R}}^\epsilon$ into Rindler nonrelativistic local Fock spaces $\mathcal{H}_\nu^\epsilon (\mathcal{V}_i)$.

Due to the emergence of the Born localization scheme in the Rindler frame, we find that the nonrelativistic RaRbR scenario is formally equivalent to the nonrelativistic ABM scenario discussed in Sec.~\ref{Localization_in_NRQM}. In particular, in analogy to Sec.~\ref{Strict_localization_in_the_nonrelativistic_limit}, here, the preparation of nonrelativistic states by an accelerated experimenter over the Rindler vacuum $|0_\text{L},0_\text{R} \rangle$ does not affect nonrelativistic measurements performed by another accelerated observer in a separated region.

Conversely, if the background $| \Omega \rangle$ is not the vacuum $|0_\text{L},0_\text{R} \rangle$, the independence between preparation and measurement in disjoint regions does not generally hold. In particular, in the RaRbM scenario, the background Minkowski vacuum $|0_\text{M} \rangle$ is entangled between the Rindler nonrelativistic local Fock spaces and, hence, it produces nonlocal effects that are similar to the Reeh-Schlieder nonlocality discussed in Sec.~\ref{solving_the_paradox}.

Finally, in the ARbM scenario, the two experimenters (Alice and Rob) do not share the same notion of nonrelativistic limit and, hence, their nonrelativistic localization schemes are incompatible. This leads to a persistence of the Reeh-Schlieder nonlocal effect in nonrelativistic ARbM experiments.

The section is organized as follows. In Sec~\ref{The_modal_scheme_converges_to_the_AQFT_scheme}, we prove the convergence between the modal and the AQFT scheme in the nonrelativistic limit. In Sec.~\ref{Born_scheme_in_NRQFTCS} we show that the Rindler nonrelativistic limit has the same features of the Born scheme already discussed in the context of NRQM. In Secs.~\ref{The_modal_scheme_over_vacuum_background_converges_to_the_Born_scheme}, \ref{The_modal_scheme_over_general_background_does_not_converge_to_the_Born_scheme} and \ref{AliceRob_nonlocality_is_not_suppressed_by_the_nonrelativistic_limit}, we respectively study the RaRbR, RaRbM and ARbM scenarios, with particular focus on the strict localization property.

\subsection{Convergence between the modal scheme and the AQFT scheme}\label{The_modal_scheme_converges_to_the_AQFT_scheme}

In this subsection, we show the convergence between the Rindler modal and the AQFT scheme in the nonrelativistic limit. Specifically, we prove that operators that are localized with respect to the Rindler modal scheme are also localized with respect to the AQFT scheme.

We firstly adopt a method similar to the one used in Secs.~\ref{Comparison_with_the_relativistic_theory} and \ref{Convergence_of_the_modal_scheme_to_the_Born_scheme}. In particular we define the space of Rindler nonrelativistic particles $\mathcal{H}_{\text{L},\text{R}}^\epsilon$ as the subspace of the Rindler-Fock space $\mathcal{H}_{\text{L},\text{R}}$ that contains particles with frequencies satisfying the nonrelativistic condition (\ref{non_relativistic_limit_curved}). Also, we define the operator
\begin{equation}\label{alpha_tilde_Phi_Pi}
\hat{A}_{\text{AQFT},\nu}(\vec{X}) = \sqrt{\frac{m c^2}{2 \hbar^2}} \hat{\Phi}_\nu(0,\vec{X}) - \frac{i}{\sqrt{2 m c^2}} \hat{\Pi}_\nu(0,\vec{X}),
\end{equation}
which generates the local field algebra $\mathfrak{A}_\nu^\text{AQFT}(\vec{X})$.

The right hand side of Eq.~(\ref{alpha_tilde_Phi_Pi}) can be written in terms of Rindler-Fock operators as
\begin{align}\label{alpha_tilde_Phi_Pi_2}
\hat{A}_{\text{AQFT},\nu}(\vec{X}) = & \int_0^{\infty} d\Omega \int_{\mathbb{R}^2} d^2 K_\perp \left[ f_{\hat{A} \mapsto \text{AQFT},\nu}(\vec{X},\Omega,\vec{K}_\perp) \hat{A}_\nu(\Omega,\vec{K}_\perp)  \right. \nonumber \\
& \left. + f_{\hat{A}^\dagger \mapsto \text{AQFT},\nu}(\vec{X},\Omega,\vec{K}_\perp) \hat{A}_\nu^\dagger(\Omega,\vec{K}_\perp) \right],
\end{align}
with
\begin{subequations}\label{f_A_alpha_tilde}
\begin{align}
& f_{\hat{A} \mapsto \text{AQFT},\nu}(\vec{X},\Omega,\vec{K}_\perp) = \left( \sqrt{\frac{m c^2}{2 \hbar^2}} + \frac{\Omega}{\sqrt{2 m c^2}} \right) F_\nu(\Omega,\vec{K}_\perp,0, \vec{X}) , \\
& f_{\hat{A}^\dagger \mapsto \text{AQFT},\nu}(\vec{X},\Omega,\vec{K}_\perp) =  \left( \sqrt{\frac{m c^2}{2 \hbar^2}} - \frac{\Omega}{\sqrt{2 m c^2}} \right) F_\nu^*(\Omega,\vec{K}_\perp,0, \vec{X}).
\end{align}
\end{subequations}
Notice that when $\Omega$ satisfies the nonrelativistic condition (\ref{non_relativistic_limit_curved}), Eq.~(\ref{f_A_alpha_tilde}) can be approximated by $f_{\hat{A} \mapsto \text{AQFT},\nu}(\vec{X},\Omega,\vec{K}_\perp) \approx (\sqrt{2m c^2}/ \hbar) \mathcal{F}_\nu(\Omega,\vec{K}_\perp,0,\vec{X}) $ and $f_{\hat{A}^\dagger \mapsto \text{AQFT},\nu}(\vec{X},\Omega,\vec{K}_\perp) \approx 0$. Hence, by restricting the operators $\hat{A}_{\text{AQFT},\nu}(\vec{X})$ and $\hat{A}_{\text{mod},\nu} (\vec{X})$ to $\mathcal{H}_{\text{L},\text{R}}^\epsilon$, we find that
\begin{equation}\label{A_AQFT_A_mod_nonrelativistic}
\left. \hat{A}_{\text{AQFT},\nu}(\vec{X}) \right|_{\mathcal{H}_{\text{L},\text{R}}^\epsilon} \approx \left. \hat{A}_{\text{mod},\nu} (\vec{X}) \right|_{\mathcal{H}_{\text{L},\text{R}}^\epsilon},
\end{equation}
which means that any element of the modal algebra $\mathfrak{A}_\nu^\text{mod}(\vec{X})$ can be approximated to an element of $\mathfrak{A}_\nu^\text{AQFT}(\vec{X})$ when restricted to $\mathcal{H}_{\text{L},\text{R}}^\epsilon$.

The Hilbert space $\mathcal{H}_{\text{L},\text{R}}^\epsilon $ describes nonrelativistic Rindler particles that are prepared over the Rindler vacuum $| 0_\text{L}, 0_\text{R} \rangle$. Hence, the results of Eq.~(\ref{A_AQFT_A_mod_nonrelativistic}) can be used in the nonrelativistic RaRbR scenario, where all the experimenters are accelerated and the background state $| \Omega \rangle$ is precisely the Rindler vacuum $| 0_\text{L}, 0_\text{R} \rangle$. However, we also want to consider scenarios in which $| \Omega \rangle$ is different from $| 0_\text{L}, 0_\text{R} \rangle$ (e.g., RaRbM scenario) or where one of the two observers uses nonrelativistic Minkowski operators to prepare states (e.g., ARbM scenario). To include all of these cases we now adopt an algebraic approach to the nonrelativistic limit. 

Starting from the AQFT scheme, we introduce the local algebra $\mathfrak{A}_\nu^{\text{AQFT},\epsilon} (\mathcal{V}) $ as the subset of $\mathfrak{A}_\nu^\text{AQFT} (\mathcal{V}) $ that contains combinations of products of Rindler operators $\hat{A}_\nu(\Omega,\vec{K}_\perp)$ satisfying the nonrelativistic condition (\ref{non_relativistic_limit_curved}). By definition, the algebra $\mathfrak{A}_\nu^{\text{AQFT},\epsilon} (\mathcal{V}) $ is generated by operators of the form of
\begin{equation}\label{A_AQFT_Phi_1}
\hat{A}_{\text{AQFT},\nu}^\dagger [\Phi_\epsilon] = \int_{\mathbb{R}^3} d^3X \Phi_\epsilon (\vec{X})  \hat{A}_{\text{AQFT},\nu}^\dagger(\vec{X}),
\end{equation}
where $\Phi_\epsilon (\vec{X}) $ is a function supported in $\mathcal{V}$ that satisfies the following nonrelativistic condition
\begin{subequations}\label{nonrelativistic_condition_AQFT}
\begin{align}
& \int_{\mathbb{R}^3} d^3X f_{\hat{A} \mapsto \text{AQFT},\nu}^*(\vec{X},\Omega,\vec{K}_\perp) \Phi_\epsilon (\vec{X}) \approx 0 \text{ if } \left|  \frac{\hbar \Omega}{m c^2} - 1 \right| \gg \epsilon, \\
& \int_{\mathbb{R}^3} d^3X f_{\hat{A}^\dagger \mapsto \text{AQFT},\nu}^*(\vec{X},\Omega,\vec{K}_\perp) \Phi_\epsilon (\vec{X})  \approx 0 \text{ if } \left|  \frac{\hbar \Omega}{m c^2} - 1 \right| \gg \epsilon.
\end{align}
\end{subequations}
Equation (\ref{nonrelativistic_condition_AQFT}) can be obtained by plugging Eq.~(\ref{alpha_tilde_Phi_Pi_2}) in Eq.~(\ref{A_AQFT_Phi_1}) and by assuming that $\hat{A}_{\text{AQFT},\nu}^\dagger [\Phi_\epsilon]$ is a linear combination of Rindler operators $\hat{A}_\nu(\Omega,\vec{K}_\perp)$ satisfying the nonrelativistic condition (\ref{non_relativistic_limit_curved}).

Analogously, we define the local algebra $\mathfrak{A}_\nu^{\text{mod},\epsilon} (\mathcal{V}) $ as the subset of $\mathfrak{A}_\nu^\text{mod} (\mathcal{V}) $ that contains combinations of products of Rindler operators $\hat{A}_\nu(\Omega,\vec{K}_\perp)$ satisfying the nonrelativistic condition (\ref{non_relativistic_limit_curved}). Explicitly, this means that $\mathfrak{A}_\nu^{\text{mod},\epsilon} (\mathcal{V}) $ is generated by the operators $\hat{A}_{\text{mod},\nu}^\dagger [\Phi_\epsilon]$ [Eq.~(\ref{A_mod_Phi})] with $\Phi_\epsilon (\vec{X})$ supported in $\mathcal{V}$ and satisfying
\begin{equation}\label{nonrelativistic_condition_mod}
\int_{\mathbb{R}^3} d^3X \mathcal{F}_\nu^*(\Omega,\vec{K}_\perp,0,\vec{X}) \Phi_\epsilon (\vec{X})  \approx 0 \text{ if } \left|  \frac{\hbar \Omega}{m c^2} - 1 \right| \gg \epsilon.
\end{equation}
Equation (\ref{nonrelativistic_condition_mod}) is a consequence of Eq.~(\ref{A_mod_Phi_1}) and the nonrelativistic condition for $\hat{A}_{\text{mod},\nu}^\dagger [\Phi_\epsilon]$.

The convergence between the two algebras $\mathfrak{A}_\nu^{\text{AQFT},\epsilon} (\mathcal{V}) $ and $\mathfrak{A}_\nu^{\text{mod},\epsilon} (\mathcal{V}) $ can be proved by noticing that the nonrelativistic conditions (\ref{nonrelativistic_condition_AQFT}) and (\ref{nonrelativistic_condition_mod}) are equivalent, in the sense that any function $\Phi_\epsilon (\vec{X})$ satisfying Eq.~(\ref{nonrelativistic_condition_AQFT}) satisfies Eq.~(\ref{nonrelativistic_condition_mod}) as well and the other way round. When both equations hold we have that
\begin{align}\label{A_AQFT_Phi_1_A_mod_Phi}
\hat{A}_{\text{AQFT},\nu}^\dagger [\Phi_\epsilon] = & \int_{\mathbb{R}^3} d^3X \Phi_\epsilon (\vec{X})  \hat{A}_{\text{AQFT},\nu}^\dagger(\vec{X}) \nonumber \\
\approx & \int_{\mathbb{R}^3} d^3X \Phi_\epsilon (\vec{X}) \left. \hat{A}_{\text{AQFT},\nu}^\dagger(\vec{X}) \right|_{\mathcal{H}_{\text{L},\text{R}}^\epsilon} \nonumber \\
\approx & \int_{\mathbb{R}^3} d^3X \Phi_\epsilon (\vec{X}) \left. \hat{A}_{\text{mod},\nu}^\dagger(\vec{X}) \right|_{\mathcal{H}_{\text{L},\text{R}}^\epsilon} \nonumber \\
\approx & \int_{\mathbb{R}^3} d^3X \Phi_\epsilon (\vec{X}) \hat{A}_{\text{mod},\nu}^\dagger(\vec{X}) \nonumber \\
= & \hat{A}_{\text{mod},\nu}^\dagger [\Phi_\epsilon].
\end{align}
Hence, any operator in  $\mathfrak{A}_\nu^{\text{AQFT},\epsilon} (\mathcal{V}) $ approximates to an element of $\mathfrak{A}_\nu^{\text{mod},\epsilon} (\mathcal{V}) $.

Due to the convergence between $\mathfrak{A}_\nu^{\text{AQFT},\epsilon} (\mathcal{V}) $ and $\mathfrak{A}_\nu^{\text{mod},\epsilon} (\mathcal{V}) $, we define a unified algebra $\mathfrak{A}_\nu^\epsilon (\mathcal{V}) $. Any element of $\mathfrak{A}_\nu^\epsilon (\mathcal{V}) $ is localized in $\nu$ and $\mathcal{V}$ with respect to both schemes and is nonrelativistic from the point of view of accelerated observers. We say that the localized state $| \Phi \rangle = \hat{O} | \Omega \rangle$ is nonrelativistic with respect to the Rindler frame and over the background state $| \Omega \rangle$ if $\hat{O} \in \mathfrak{A}_\nu^\epsilon (\mathcal{V}) $, which means that the state has been prepared over $| \Omega \rangle$ by the accelerated observer Rachel via the nonrelativistic local operator $\hat{O} \in \mathfrak{A}_\nu^\epsilon (\mathcal{V}) $.

\subsection{Born scheme in NRQFTCS}\label{Born_scheme_in_NRQFTCS}

In this subsection, we show that the notion of Born localization naturally emerges in the nonrelativistic limit of Rindler spacetime. In particular we demonstrate that, in such a regime, the Hilbert space $\mathcal{H}_{\text{L},\text{R}}^\epsilon$ factorizes into local Fock spaces $\mathcal{H}_\nu^\epsilon (\mathcal{V}_i)$ and the Rindler vacuum $| 0_\text{L}, 0_\text{R} \rangle $ factorizes into the local vacua $| 0_\nu (\mathcal{V}_i) \rangle $. Also, the orthogonality condition for states that are localized in disjoint regions holds and the modal wave function acquires the Born notion of probability to find particles in space points.

The factorization of $\mathcal{H}_{\text{L},\text{R}}^\epsilon$ into local Fock spaces $\mathcal{H}_\nu^\epsilon (\mathcal{V}_i)$ can be obtained by using the definition of $\mathfrak{A}_\nu^\epsilon (\mathcal{V}) $ as the algebra generated by the operators $\hat{A}_{\text{mod},\nu} [\Phi_\epsilon]$, with $\Phi_\epsilon(\vec{X})$ supported in $\mathcal{V}$ and satisfying the nonrelativistic condition
\begin{equation}\label{A_mod_Phi_1_nonrelativistic}
\tilde{\Phi}_\epsilon (\vec{\theta}) \approx 0 \text{ if } \left| \frac{\hbar \theta_1}{mc^2} - 1 \right| \gg \epsilon,
\end{equation}
with
\begin{equation}\label{wavefunction_F_inverse_epsilon}
\tilde{\Phi}_\epsilon (\vec{\theta})  =  \frac{\sqrt{2 m c^2}}{\hbar}  \int_{\mathbb{R}^3} d^3 X  \Phi_\epsilon (\vec{X})  \mathcal{F}_\nu^*(\vec{\theta},0,\vec{X}).
\end{equation}

The commutation relation between two nonrelativistic local operators is
\begin{subequations}
\begin{align}
& \left[ \hat{A}_{\text{mod},\nu} [\Phi_\epsilon], \hat{A}_{\text{mod},\nu'}^\dagger [\Phi'_\epsilon] \right] = \delta_{\nu\nu'} \int_{\theta_1>0} d^3 \theta \tilde{\Phi}_\epsilon^* (\vec{\theta})  \tilde{\Phi}'_\epsilon (\vec{\theta}) \label{A_mod_Phi_1_commutation_a} \\
& \left[ \hat{A}_{\text{mod},\nu} [\Phi_\epsilon], \hat{A}_{\text{mod},\nu'} [\Phi'_\epsilon] \right] = 0.
\end{align}
\end{subequations}
By means of Eqs.~(\ref{FF_nu_F_nu}), (\ref{A_mod_Phi_1_nonrelativistic}) and (\ref{wavefunction_F_inverse_epsilon}), one can approximate Eq.~(\ref{A_mod_Phi_1_commutation_a}) with
\begin{align}\label{A_mod_Phi_1_commutation_a_approx}
 \left[ \hat{A}_{\text{mod},\nu} [\Phi_\epsilon], \hat{A}_{\text{mod},\nu'}^\dagger [\Phi'_\epsilon] \right] \approx &  \delta_{\nu\nu'} \int_{| \hbar \theta_1/mc^2 - 1 | <\epsilon} d^3 \theta \tilde{\Phi}_\epsilon^* (\vec{\theta})  \tilde{\Phi}'_\epsilon (\vec{\theta}) \nonumber \\
 = & \delta_{\nu\nu'} \int_{\mathbb{R}^3} d^3 X \int_{\mathbb{R}^3} d^3 X' \int_{| \hbar \theta_1/mc^2 - 1 | <\epsilon} d^3 \theta \frac{2 m c^2}{\hbar^2} \nonumber \\
& \times  \Phi_\epsilon^* (\vec{X}) \Phi'_\epsilon (\vec{X}') \mathcal{F}_\nu(\vec{\theta},0,\vec{X}) \mathcal{F}_\nu^*(\vec{\theta},0,\vec{X}') \nonumber \\
 \approx & \delta_{\nu\nu'} \int_{\mathbb{R}^3} d^3 X \int_{\mathbb{R}^3} d^3 X' \int_{| \hbar \theta_1/mc^2 - 1 | <\epsilon} d^3 \theta \frac{2 m c^2}{\hbar^2} \nonumber \\
& \times  \Phi_\epsilon^* (\vec{X}) \Phi'_\epsilon (\vec{X}') F_\nu(\vec{\theta},0,\vec{X}) \mathcal{F}_\nu^*(\vec{\theta},0,\vec{X}') \nonumber \\
 \approx & \delta_{\nu\nu'} \int_{\mathbb{R}^3} d^3 X \int_{\mathbb{R}^3} d^3 X' \int_{\theta_1>0} d^3 \theta \frac{2 m c^2}{\hbar^2} \nonumber \\
& \times  \Phi_\epsilon^* (\vec{X}) \Phi'_\epsilon (\vec{X}') F_\nu(\vec{\theta},0,\vec{X}) \mathcal{F}_\nu^*(\vec{\theta},0,\vec{X}'),
\end{align}
which leads to
\begin{equation}\label{A_mod_Phi_1_commutation_a_approx_3}
 \left[ \hat{A}_{\text{mod},\nu} [\Phi_\epsilon], \hat{A}_{\text{mod},\nu'}^\dagger [\Phi'_\epsilon] \right] \approx \delta_{\nu\nu'} \int_{\mathbb{R}^3} d^3 X \Phi_\epsilon^* (\vec{X}) \Phi'_\epsilon ( \vec{X}),
\end{equation}
owing to
\begin{equation}\label{FF_nu_F_nu_delta_inverse}
\int_{\theta_1>0} d^3\theta  \frac{2 m c^2}{ \hbar^2 } F_\nu(\vec{\theta},0,\vec{X})  \mathcal{F}_\nu^*(\vec{\theta},0,\vec{X}') = \delta^3(\vec{X}-\vec{X}').
\end{equation}
Equation (\ref{FF_nu_F_nu_delta_inverse}) can be proven from Eqs.~(\ref{F_Rindler}), (\ref{F_F_nu}), (\ref{alpha_tilde_approx_3}) and (\ref{FF_nu_F_nu}).

Due to the localization of the wave functions $\Phi_\epsilon (\vec{X})$ and $\Phi_\epsilon' (\vec{X})$ in, respectively, $\mathcal{V}$ and $\mathcal{V}'$, we find that the right hand side of Eq.~(\ref{A_mod_Phi_1_commutation_a_approx_3}) is vanishing if $\mathcal{V}$ and $\mathcal{V}'$ are disjoint. Hence, nonrelativistic operators localized in disjoint regions commute. Consequently, the Hilbert space $\mathcal{H}_{\text{L},\text{R}}^\epsilon$ factorizes into local Hilbert spaces $\mathcal{H}_\nu^\epsilon (\mathcal{V}_i)$ upon which elements of $\mathfrak{A}_\nu^\epsilon (\mathcal{V}_i) $ act. Also, if $\Phi_\epsilon ( \vec{X})$ and $\Phi_\epsilon' ( \vec{X})$ are orthonormal with respect to the $L^2(\mathbb{R}^3)$ scalar product, then $\hat{A}_{\text{mod},\nu} [\Phi_\epsilon]$ and $\hat{A}_{\text{mod},\nu'}^\dagger [\Phi'_\epsilon]$ approximately satisfy the canonical commutation relations. This means that the local Hilbert spaces $\mathcal{H}_\nu^\epsilon (\mathcal{V})$ and $\mathcal{H}_\nu^\epsilon (\mathcal{V}')$ are also Fock spaces.

We note by $| 0_\nu (\mathcal{V}) \rangle $ the local vacuum of $\mathcal{H}_\nu^\epsilon (\mathcal{V})$. Notice that the global vacuum $| 0_\text{L}, 0_\text{R} \rangle $ is annihilated by any $ \hat{A}_{\text{mod},\nu} [\Phi_\epsilon] \in \mathfrak{A}_\nu^\epsilon (\mathcal{V})$. Hence, $| 0_\text{L}, 0_\text{R} \rangle $ factorizes into $\bigotimes_\nu \bigotimes_i | 0_\nu (\mathcal{V}_i) \rangle $ in $\mathcal{H}_{\text{L},\text{R}}^\epsilon = \bigotimes_\nu \bigotimes_i \mathcal{H}_\nu^\epsilon (\mathcal{V}) $. This means that the Rindler vacuum $| 0_\text{L}, 0_\text{R} \rangle $ is equal to $ [ \bigotimes_\nu \bigotimes_i | 0_\nu (\mathcal{V}_i) \rangle ] \otimes  \dots $ as an element of $\mathcal{H}_{\text{L},\text{R}} = \mathcal{H}_{\text{L},\text{R}}^\epsilon \otimes \dots$. Here, the dots are associated to the relativistic subspace of $\mathcal{H}_{\text{L},\text{R}}$ containing states that do not satisfy the nonrelativistic condition (\ref{non_relativistic_limit_curved}).

As a consequence of the factorizations $\mathcal{H}_{\text{L},\text{R}}^\epsilon = \bigotimes_\nu \bigotimes_i \mathcal{H}_\nu^\epsilon (\mathcal{V}_i) $ and $| 0_\text{L}, 0_\text{R} \rangle  = [ \bigotimes_\nu \bigotimes_i | 0_\nu (\mathcal{V}_i) \rangle ] \otimes  \dots  $, we find that any couple of nonrelativistic states localized over the Rindler vacuum $| 0_\text{L}, 0_\text{R} \rangle $ in different regions are orthogonal. Explicitly, this means that for any couple of operators $\hat{O}_\text{A} \in  \mathfrak{A}_{\nu_\text{A}}^\epsilon (\mathcal{V}_\text{A})$ and $\hat{O}_\text{B} \in  \mathfrak{A}_{\nu_\text{B}}^\epsilon (\mathcal{V}_\text{B})$, the states $| \Phi_\text{A} \rangle = \hat{O}_\text{A} | 0_\text{L}, 0_\text{R} \rangle $ and $| \Phi_\text{B} \rangle = \hat{O}_\text{B} | 0_\text{L}, 0_\text{R} \rangle $ are orthogonal if $\nu_\text{A} \neq \nu_\text{B}$ or if $\mathcal{V}_\text{A} \cap \mathcal{V}_\text{B} = \varnothing$.

In Sec.~\ref{QFT_in_curved_spacetime_Rindler_scalar} we provided the description of any Rindler single particle state $| \Phi \rangle \in \mathcal{H}_{\text{L},\text{R}}$ in terms of the modal wave function $\Phi_1(T,\nu,\vec{X})$ such that $| \Phi \rangle = \sum_{\nu \in \{\text{L},\text{R}\}} \hat{A}_{\text{mod},\nu} [\Phi_1(0,\nu)] | 0_\text{L}, 0_\text{R} \rangle$. As a consequence of the convergence between the AQFT and the modal scheme, the modal wave function $\Phi_1(T,\nu,\vec{X})$ acquires a genuine notion of localization. Also, due to the emergence to the Born scheme, we find that $|\Phi_1(T,\nu,\vec{X})|^2$ gives the probability for the particle to be found in $\vec{X}$ at time $T$. The same result can be obtained for general Fock states.

\subsection{Nonrelativistic RaRbR scenario}\label{The_modal_scheme_over_vacuum_background_converges_to_the_Born_scheme}

In this subsection, we consider the RaRbR scenario in which a state $| \Phi \rangle = \hat{O}_\text{A} | 0_\text{L}, 0_\text{R} \rangle$ is locally prepared by an accelerated experimenter (Rachel) over the Rindler vacuum background $| 0_\text{L}, 0_\text{R} \rangle$ and a second accelerated observer (Rob) performs local measurements of the observable $\hat{O}_\text{B}$. The operators $ \hat{O}_\text{A}$ and $ \hat{O}_\text{B}$ are, respectively, localized in the wedges $\nu_\text{A}$ and $\nu_\text{B}$ and in the space regions $\mathcal{V}_\text{A}$ and $\mathcal{V}_\text{B}$. We assume that the state prepared by Rachel and the observable measured by Rob are nonrelativistic with respect to the Rindler frame, which means that $\hat{O}_\text{A} \in \mathfrak{A}_{\nu_\text{A}}^\epsilon (\mathcal{V}_\text{A}) $ and $\hat{O}_\text{B} \in \mathfrak{A}_{\nu_\text{B}}^\epsilon (\mathcal{V}_\text{B}) $.

As a consequence of the factorizations $\mathcal{H}_{\text{L},\text{R}}^\epsilon = \bigotimes_\nu \bigotimes_i \mathcal{H}_\nu^\epsilon (\mathcal{V}_i) $ and $| 0_\text{L}, 0_\text{R} \rangle  = [ \bigotimes_\nu \bigotimes_i | 0_\nu (\mathcal{V}_i) \rangle ] \otimes  \dots  $, the strict localization property is always satisfied in the nonrelativistic RaRbR scenario. In particular, we find that
\begin{equation}
\langle \Phi | \hat{O}_\text{B} | \Phi \rangle = \langle 0_\text{L}, 0_\text{R} | \hat{O}_\text{B}  | 0_\text{L}, 0_\text{R} \rangle
\end{equation}
when $\nu_\text{A}$ is different from $\nu_\text{B}$ or when $\mathcal{V}_\text{A}$ and $\mathcal{V}_\text{B}$ are disjoint. This means that Rob's measurements are never influenced by Rachel's preparations of states.

Such a result is in analogy to what we found for the ABM scenario in Sec.~\ref{Strict_localization_in_the_nonrelativistic_limit}. We conclude that nonlocal effects are always suppressed if the nonrelativistic limit and the background vacuum state are associated to the same frame.

\subsection{Nonrelativistic RaRbM scenario} \label{The_modal_scheme_over_general_background_does_not_converge_to_the_Born_scheme}

In this subsection, we study the RaRbM scenario in the nonrelativistic limit. At variance with Sec.~\ref{The_modal_scheme_over_vacuum_background_converges_to_the_Born_scheme}, here, the background state is the Minkowski vacuum $| 0_\text{M} \rangle$. Hence, the state prepared by Rachel is $| \Phi \rangle = \hat{O}_\text{A} | 0_\text{M} \rangle$ with $\hat{O}_\text{A} \in \mathfrak{A}_{\nu_\text{A}}^\epsilon (\mathcal{V}_\text{A}) $. The observable measured by Rob, instead, is still $\hat{O}_\text{B} \in \mathfrak{A}_{\nu_\text{B}}^\epsilon (\mathcal{V}_\text{B}) $.

In Sec.~\ref{Comparison_between_the_NewtonWigner_and_the_AQFT_schemes}, we showed that the Minkowski vacuum $| 0_\text{M} \rangle$ factorizes into $\bigotimes_i | 0_\text{M} (\mathcal{V}_i) \rangle$ with respect to the Minkowski-Newton-Wigner local Fock spaces $\mathcal{H}_\text{M}^\text{NW}(\mathcal{V}_i)$. However, here, we are interested in the factorization with respect to $\mathcal{H}_{\text{L},\text{R}}^\epsilon = \bigotimes_\nu \bigotimes_i \mathcal{H}_\nu^\epsilon (\mathcal{V}_i) $. Due to Eqs.~(\ref{Rindler_vacuum_to_Minkowski}) and (\ref{Rindler_vacuum_to_Minkowski_unitary_operator}) and due to the factorization of the Rindler vacuum $| 0_\text{L}, 0_\text{R} \rangle  = [\bigotimes_\nu \bigotimes_i | 0_\nu (\mathcal{V}_i) \rangle] \otimes \dots $, we find that the Minkowski vacuum $| 0_\text{M} \rangle$ is entangled between the nonrelativistic local Fock spaces $\mathcal{H}_\nu^\epsilon (\mathcal{V}_i) $.

As a consequence of this entanglement, we find that, in general,
\begin{equation}
\langle \Phi | \hat{O}_\text{B} | \Phi \rangle \neq \langle 0_\text{M} | \hat{O}_\text{B}  | 0_\text{M} \rangle
\end{equation}
even if $\mathcal{V}_\text{A}$ and $\mathcal{V}_\text{B}$ are disjoint. However, the identity
\begin{equation}
\langle \Phi | \hat{O}_\text{B} | \Phi \rangle = \langle 0_\text{M} | \hat{O}_\text{B}  | 0_\text{M} \rangle
\end{equation}
is guaranteed when $\hat{O}_\text{A}$ is unitary and either the two wedges $\nu_\text{A} $ and $ \nu_\text{B}$ are different or the two regions $\mathcal{V}_\text{A}$ and $\mathcal{V}_\text{B}$ are disjoint. This is due the fact that $\hat{O}_\text{A}$ and $\hat{O}_\text{B}$ commute and, hence, Eq.~(\ref{local_unitary_operator_measurament_A}) holds.

As a result, we find that the strict localization property in the nonrelativistic RaRbM scenario is guaranteed for unitary preparations of the state. By only performing nonselective operations on the vacuum $|0_\text{M} \rangle$, Rachel does not influence Rob's measurements in the other disjoint region. Conversely, selective preparations of the state lead to nonlocal effects.

The nonlocality shown here is similar to the Reeh-Schlieder effect discussed in Sec.~\ref{solving_the_paradox}. Indeed, the origin of this effect is ascribed to the entanglement between local spaces induced by the background vacuum state. These correlations are suppressed in the nonrelativistic limit with respect to the Minkowski frame [Sec.~\ref{Strict_localization_in_the_nonrelativistic_limit}]. However, here, we are considering the nonrelativistic regime of accelerated observers and we find that no suppression occurs in the RaRbM scenario.

\subsection{Nonrelativistic ARbM scenario} \label{AliceRob_nonlocality_is_not_suppressed_by_the_nonrelativistic_limit}

In Sec.~\ref{Localization_in_NRQM}, we detailed the nonrelativistic ABM scenario, in which all experimenters are inertial and the background state is the Minkowski vacuum $| 0_\text{M} \rangle$. In that case, physical phenomena are described by means of the Hilbert space $\mathcal{H}_\text{M}^\Lambda$.

In the Minkowski frame, the nonrelativistic local algebras are defined with respect to the nonrelativistic condition (\ref{non_relativistic_limit}). In particular, $\mathfrak{A}_\text{M}^\epsilon(\mathcal{V})$ is defined as the subalgebra of $\mathfrak{A}_\text{M}^\text{NW}(\mathcal{V})$ generated by operators of the form of
\begin{equation}
\hat{a}_\text{NW}^\dagger [\phi_\epsilon] = \int_{\mathbb{R}^3} d^3x \phi_\epsilon (\vec{x})  \hat{a}_\text{NW}^\dagger(\vec{x}),
\end{equation}
where $\phi_\epsilon (\vec{x})$ is supported in $\mathcal{V}$ and its Fourier transform $\tilde{\phi}_\epsilon (\vec{k})$ is supported in the nonrelativistic region (\ref{non_relativistic_limit}). Due to the convergence between the different localization schemes in the nonrelativistic limit, one can use $\mathfrak{A}_\text{M}^\text{AQFT}(\mathcal{V})$ or $\mathfrak{A}_\text{M}^\text{mod}(\mathcal{V})$ and $\hat{a}_\text{AQFT}^\dagger(\vec{x})$ or $\hat{a}_\text{mod}^\dagger(\vec{x})$ instead of $\mathfrak{A}_\text{M}^\text{NW}(\mathcal{V})$ and $ \hat{a}_\text{NW}^\dagger(\vec{x})$ to define $\mathfrak{A}_\text{M}^\epsilon(\mathcal{V})$.

Any state $| \phi \rangle$ of the nonrelativistic Hilbert space $\mathcal{H}_\text{M}^\Lambda$, with $\Lambda \sim \epsilon^{-1/2} \lambda_\text{C}$, localized in $\mathcal{V}_\text{A}$ can be identified by an operator $\hat{O}_\text{A} \in \mathfrak{A}_\text{M}^\epsilon(\mathcal{V}_\text{A})$ such that $| \phi \rangle = \hat{O}_\text{A} | 0_\text{M} \rangle$. In Sec.~\ref{Strict_localization_in_the_nonrelativistic_limit}, we found that $| \phi \rangle$ is also strictly localized in $\mathcal{V}_\text{A}$, in the sense that, for any $\hat{O}_\text{B} \in \mathfrak{A}_\text{M}^\epsilon(\mathcal{V}_\text{B})$, with $\mathcal{V}_\text{B}$ disjoint from $\mathcal{V}_\text{A}$, Eq.~(\ref{KnightLicht_property}) holds.

At variance with Sec.~\ref{Localization_in_NRQM}, here, we consider the nonrelativistic ARbM scenario, in which the observer performing the measurement (i.e., experimenter B) is accelerated (Rob). Hence, the operator $\hat{O}_\text{B}$ is an element of $\mathfrak{A}_{\nu_\text{B}}^\epsilon(\mathcal{V}_\text{B})$ instead of $\mathfrak{A}_\text{M}^\epsilon(\mathcal{V}_\text{B})$, which means that $\hat{O}_\text{B}$ is a linear combination of products of $\hat{A}_{\text{mod},\nu_\text{B}} [\Phi_\epsilon]$ operators, with $\Phi_\epsilon(\vec{X})$ supported in $\mathcal{V}_\text{B}$ and satisfying the nonrelativistic condition in the Rindler frame (\ref{A_mod_Phi_1_nonrelativistic}).

In Sec~\ref{Inertial_and_non_inertial_frame}, we showed that the nonrelativistic condition in the two frames are not compatible. This means that if $\Phi(\vec{X})$ satisfies the nonrelativistic condition in the Rindler frame (\ref{A_mod_Phi_1_nonrelativistic}), then, for any $\mathcal{V}'_\text{B} \subseteq \mathbb{R}^3$, the operator $\hat{A}_{\text{mod},\nu} [\Phi_\epsilon]$ is not an element of the Minkowski nonrelativistic local algebra $\mathfrak{A}_\text{M}^\epsilon(\mathcal{V}'_\text{B})$. Hence, the Minkowski and the Rindler nonrelativistic schemes are incompatible.

Due to the incompatibility between the two schemes, Eq.~(\ref{KnightLicht_property}) does not generally hold. This means that the preparation of the state $| \phi \rangle = \hat{O}_\text{A} |0_\text{M} \rangle $ by Alice, with $\hat{O}_\text{A} \in \mathfrak{A}_\text{M}^\epsilon(\mathcal{V}_\text{A})$, may influence the measurement of $\hat{O}_\text{B}$ by Rob, even if $\mathcal{V}_\text{A}$ and $\mathcal{V}_\text{B}$ represent different regions of the spacetime.

However, unitary preparations of $| \phi \rangle$ guarantee the independence between the local preparation of the state and the local measurements of the observable $\hat{O}_\text{B}$. This is a consequence of the microcausality axiom that ensures that operators localized in different region of spacetimes with respect to the AQFT scheme commute. Since the algebras $\mathfrak{A}_\text{M}^\epsilon(\mathcal{V}_\text{A})$ and $\mathfrak{A}_{\nu_\text{B}}^\epsilon(\mathcal{V}_\text{B})$ are subalgebras of, respectively, $\mathfrak{A}_\text{M}^\text{AQFT}(\mathcal{V}_\text{A})$ and $\mathfrak{A}_{\nu_\text{B}}^\text{AQFT}(\mathcal{V}_\text{B})$, we find that $\hat{O}_\text{A}$ and $\hat{O}_\text{B}$ commute if $\mathcal{V}_\text{A}$ and $\mathcal{V}_\text{B}$ represent different regions of the spacetime. Then, one can use Eq.~(\ref{local_unitary_operator_measurament_A}) to prove that, if $\hat{O}_\text{A}$ is unitary, Eq.~(\ref{KnightLicht_property}) holds.

In conclusion, we find that, at variance with the ABM and the RaRbR scenarios, in the ARbM scenario the Reeh-Schlieder nonlocal effects are not suppressed. This is a consequence of the incompatibility between the nonrelativistic limit in the two frames and the consequent incompatibility between the respective nonrelativistic localization schemes.

\section{Conclusions}\label{Localization_in_accelerated_frame_Conclusions}

The algebraic approach to QFTCS provides an exact description of local physical phenomena in a regime in which relativistic energies and noninertial effects cannot be ignored. At variance with the modal representation of particle states, the AQFT localization scheme is frame independent and does not violate relativistic causality. Only in the nonrelativistic regime the two schemes appear indistinguishable from each other.

Notwithstanding the relativistic causal nature of the AQFT scheme, nonlocal Bell-like instantaneous effects occur when states are selectively prepared over any background $| \Omega \rangle$ in confined regions $\mathcal{V}_\text{A}$. In particular, local observations in regions $\mathcal{V}_\text{B}$ disjoint from $\mathcal{V}_\text{A}$ are influenced by the nonunitary preparations of states in $\mathcal{V}_\text{A}$. In Sec.~\ref{Strict_localization_in_the_nonrelativistic_limit} we already showed that such a nonlocal effect is suppressed in NRQM. In the present chapter, we reach the conclusion that the suppression generally occurs in the nonrelativistic regime of the frame associated to the vacuum background $| \Omega \rangle$.

As already pointed out in Sec.~\ref{Framedependent_nonrelativistic_limit_Introduction}, different frames are characterized by different time coordinates, which lead to different notions of vacuum state and nonrelativistic limit. Hence, it is not surprising that for any (inertial or accelerated) observer there is a preferred background state ($| 0_\text{M} \rangle$ or $| 0_\text{L}, 0_\text{R} \rangle$) and a preferred nonrelativistic condition (Eq.~(\ref{non_relativistic_limit}) or Eq.~(\ref{non_relativistic_limit_curved})). Nonlocal effects are completely suppressed when all of these frame dependent elements match up, i.e., when the vacuum background state and the nonrelativistic condition are associated to the same frame. The nonrelativistic ABM and the nonrelativistic RaRbR scenarios fall into this category and are associated to, respectively, the Minkowski and the Rindler frame.

In the RaRbM and the ARbM scenarios, instead, the nonlocal effects are not suppressed, due to the aforementioned elements not matching up. Specifically, in the RaRbM scenario, the background state is the Minkowski vacuum $| 0_\text{M} \rangle$, whereas the nonrelativistic condition is associated to the Rindler frame. The origin of the nonlocal effect needs to be ascribed to the entanglement of $| 0_\text{M} \rangle$ between the Rindler nonrelativistic local Fock spaces. Conversely, in the ARbM scenario, states are prepared by the inertial observer Alice via nonrelativistic Minkowski operators over the Minkowski vacuum $| 0_\text{M} \rangle$; whereas, local measurements are performed by the accelerated observer Rob via nonrelativistic Rindler operators. From Alice's point of view, Rob uses relativistic observables, since her notion of nonrelativistic limit is different than Rob's. Hence, the Reeh-Schlieder nonlocality is not suppressed.

These theoretical results may find practical applications in the context of, e.g., nonrelativistic emitters and detectors. If both instruments are inertial and operate over the Minkowski vacuum $| 0_\text{M} \rangle$, no nonlocal effect is expected to be detected. The same occurs if they are both accelerated over the Rindler vacuum background $| 0_\text{L}, 0_\text{R} \rangle$. Conversely, if at least one of them is accelerated and the background state is the Minkowski vacuum $| 0_\text{M} \rangle$, then nonlocal effects may appear and can be eventually measured.

By means of this nonlocality, it is possible to probe the nature of the background state in the presence of gravitational fields (e.g., Earth gravity). Due to the Einstein's equivalence principle, any standing experimenter affected by a gravitational field can be locally represented by the Rindler frame, whereas a free falling observer is described by the Minkowski frame. In this scenario, we still do not have experimental evidence about the nature of the background state: is it a Minkowski or a Rindler vacuum? To answer this question, one may use nonrelativistic emitters and detectors that follow different trajectories and measure eventual nonlocal effects.

\chapter{Single particle beyond Rindler horizon}\label{Single_particle_beyond_Rindler_horizon}

\textit{This chapter is based on and contains material from Ref.~\citeRF{PhysRevA.107.L030203}.}

In classical physics, accelerated observers cannot detect signals emitted beyond the Rindler horizon. One can argue if this is also true in quantum physics. Here, we ask if inertial Minkowski particle states localized beyond the horizon can be revealed by monitoring variations in the Rindler particle distribution from the thermal background (\ref{thermal}).

In Chap.~\ref{Localization_in_accelerated_frame}, we found that nonunitary preparations of states over the Minkowski vacuum $| 0_\text{M} \rangle$ by an inertial experimenter (Alice) may influence the outcome of measurements carried out by an accelerated observer (Rob), even if the two experimenters are localized in different wedges. We obtained such a result for both the Newton-Wigner [Sec.~\ref{NewtonWigner_ARbM_scenario}] and the AQFT [Sec.~\ref{AQFT_localization_scheme_in_curved_spacetime}] localization schemes. In the former case, the nonlocal effect is due to the noncovariant behavior of the scheme under coordinate transformations. In the letter case, instead, Rob's measurements are only affected by the selective nature of the preparation of the state by Alice; no violation of causality occurs because a classical communication between Alice and Rob is required [Sec.~\ref{solving_the_paradox}].

In this chapter, we consider a Minkowski single particle state $| \phi \rangle$ prepared by Alice in the left wedge. By definition, the preparation of $| \phi \rangle$ over the Minkowski vacuum $| 0_\text{M} \rangle$ is selective. To see this, consider Eq.~(\ref{single_particle_Minkowski}) and notice that the operator acting on $| 0_\text{M} \rangle$ is not unitary. Hence, the accelerated observer Rob, localized in the right wedge, is expected to detect the presence of the particle beyond the horizon. 

In Secs.~\ref{Example_single_particle_state_NW} and \ref{Example_single_particle_state_AQFT}, we use, respectively, the Newton-Wigner and the AQFT schemes to see these results. This chapter can, then, be intended as a practical application of the ARbM scenarios originally described in Secs.~\ref{NewtonWigner_ARbM_scenario} and \ref{AQFT_localization_scheme_in_curved_spacetime}.

\section{Newton-Wigner scheme in 1+1 spacetime}\label{Example_single_particle_state_NW}

In this section, we consider the Newton-Wigner ARbM scenario, which was presented in Sec.~\ref{NewtonWigner_ARbM_scenario}. We give some quantitative results by studying the case in which a single particle state $| \phi \rangle$ is prepared by Alice in the inertial frame and local measurements of particle distributions are carried out in the right Rindler frame by Rob.

We explicitly demonstrate that the strict localization property is not satisfied. In particular, we show that the preparation of the Minkowski single particle $| \phi \rangle$ by Alice influences the outcome of Rob's measurements. Such a nonlocal effect occurs even if $| \phi \rangle$ is localized beyond the Rindler horizon. In other words, we show that Minkowski single particle states localized beyond the horizon with respect to the Newton-Wigner scheme modify the Unruh thermal distribution in the accelerated frame.

The state $| \phi \rangle$ is defined by Eq.~(\ref{single_particle_Minkowski_11}) with $\tilde{\phi}_1(k)$ as its wave function in the momentum space. Alternatively, $| \phi \rangle$ can be defined in terms of its Newton-Wigner wave function $\phi_\text{NW}(x)$ by
\begin{equation} \label{one_state}
|\phi\rangle = \int_{\mathbb{R}} dx \phi_\text{NW}(x) \hat{a}_\text{NW}^\dagger (x) |0_\text{M} \rangle,
\end{equation}
with
\begin{equation}\label{phi_NW_phi_tilde}
\phi_\text{NW}(x) = \int_{\mathbb{R}} dk \frac{e^{i k x}}{\sqrt{2 \pi}} \tilde{\phi}_1 (k).
\end{equation}

By taking the partial trace of $|\phi\rangle \langle \phi|$ over the left wedge, we obtain the statistical operator $\hat{\rho}$ representing the state in the right Rindler frame [Sec.~\ref{MinkowskiFock_states_in_accelerated_frames}]. An equivalent representation can be obtained by considering its right-Rindler-Wigner characteristic function $\chi^{(p)}[\xi, \xi^*]$, explicitly shown in Eq.~(\ref{single_particle_chi_final}).

The function $\chi^{(p)}[\xi, \xi^*]$ can be used to compute the right Rindler particle distribution, defined as the mean value of $\hat{n}_\text{R}(X) = \hat{A}_\text{NW,R}^\dagger(X) \hat{A}_\text{NW,R} (X)$. Owing to Eq.~(\ref{normal_ordered_meanvalue_characteristic}), we obtain
\begin{equation} \label{n_postion}
\left\langle \hat{n}_\text{R}(X) \right\rangle_{\hat{\rho}} =  - \int_{\mathbb{R}} dK \int_{\mathbb{R}} dK' \frac{e^{i (K'-K) X}}{2\pi} \left. \frac{\delta}{\delta \xi(K)}\frac{\delta}{\delta \xi^*(K')}\chi^{(+1)}[\xi,\xi^*] \right|_{\xi=0}
\end{equation}
as the mean value of $\hat{n}_\text{R}(X)$ with respect to $\hat{\rho}$. Similarly, one can derive the mean value of $\hat{n}_\text{R}(X)$ with respect to the Minkowski vacuum $\hat{\rho}_0$ as
\begin{equation} \label{n_0_postion}
\left\langle \hat{n}_\text{R}(X) \right\rangle_{\hat{\rho}_0} =  - \int_{\mathbb{R}} dK \int_{\mathbb{R}} dK' \frac{e^{i (K'-K) X}}{2\pi} \left. \frac{\delta}{\delta \xi(K)}\frac{\delta}{\delta \xi^*(K')}\chi_0^{(+1)}[\xi,\xi^*] \right|_{\xi=0}.
\end{equation}

We define the difference in the right Rindler density function between the Minkowski single particle $\hat{\rho}$ and the Minkowski vacuum state $\hat{\rho}_0$ as
\begin{equation} \label{Delta_n}
\Delta n_\text{R}(X) = \left\langle \hat{n}_\text{R}(X) \right\rangle_{\hat{\rho}} - \left\langle \hat{n}_\text{R}(X) \right\rangle_{\hat{\rho}_0}.
\end{equation}
If Rob measures a value of $\Delta n_\text{R}(X)$ different from zero, then he can infer that the preparation of the Minkowski particle has influenced the outcome of the measurement.

The explicit expression for the right hand side of Eq.~(\ref{Delta_n}) can be derived by plugging Eq.~(\ref{single_particle_chi_final}) in Eq.~(\ref{n_postion}) and by using Eqs.~(\ref{thermal_characteristic_fucntion}), (\ref{L_xi}) and (\ref{n_0_postion}), which lead to
\begin{equation}\label{Delta_n_explicit}
\left\langle \hat{n}_\text{R}(X) \right\rangle_{\hat{\rho}} =  n_+(X) + n_-(X),
\end{equation}
with
\begin{subequations}
\begin{align} 
& n_+(X) = \left| \int_{\mathbb{R}} dk\int_{\mathbb{R}} dK  \tilde{\phi}_1(k) \frac{e^{i K X}}{\sqrt{2\pi}}  \alpha(k,K) \right|^2, \\
 & n_-(X) = \left| \int_{\mathbb{R}} dk\int_{\mathbb{R}} dK  \tilde{\phi}_1(k) \frac{e^{-i K X}}{\sqrt{2\pi}} \beta^*(k,K) \right|^2.
\end{align}
\end{subequations}
By using Eqs.~(\ref{F_11}) and (\ref{Bogolyubov_coefficients}), we obtain
\begin{align}
n_\pm(X) = & \left| \int_{\mathbb{R}} dk\int_{\mathbb{R}} dK  \tilde{\phi}_1(k) \frac{e^{\pm i K X}}{(2\pi)^{3/2} a}  \theta(kK) \sqrt{\frac{K}{k}} \Gamma \left( \mp \frac{i K}{a} \right) \right. \nonumber \\
& \times \left. \exp \left(\pm i \frac{K}{a} \ln \left( \frac{|k|}{a} \right) \pm \text{sign} \left( k \right) \frac{c \beta K}{4} \right) \right|^2.
\end{align}
By inverting Eq.~(\ref{phi_NW_phi_tilde}), we can relate the distributions $n_\pm(X)$ to the Newton-Wigner wave function $\phi_\text{NW}(x)$ as
\begin{align}\label{n_pm_X_2}
n_\pm(X) = & \left| \int_{\mathbb{R}} dx \int_{\mathbb{R}} dk \int_{\mathbb{R}} dK  \phi_\text{NW}(x) \frac{e^{-i k x \pm i K X}}{(2\pi)^2 a}  \theta(kK) \sqrt{\frac{K}{k}} \Gamma \left( \mp \frac{i K}{a} \right) \right. \nonumber \\
& \times \left. \exp \left(\pm i \frac{K}{a} \ln \left( \frac{|k|}{a} \right) \pm \text{sign} \left( k \right) \frac{c \beta K}{4} \right) \right|^2.
\end{align}

The right hand side of Eq.~(\ref{n_pm_X_2}) can be computed by considering the variable transformation $k \mapsto X' =  \ln ( \text{sign}(K) k /a ) / a$, which leads to
\begin{align}\label{n_pm_X_3}
n_\pm(X) = & \left| \int_{\mathbb{R}} dx \int_{\mathbb{R}} dX' \int_{\mathbb{R}} dK \phi_\text{NW}(x) \frac{e^{\pm i K X \pm i K X' }}{(2\pi)^2} \sqrt{a|K|} \Gamma \left( \mp \frac{i K}{a} \right) \right. \nonumber \\
& \times \left. \exp \left( \frac{a X'}{2} - \text{sign}(K) i a x e^{a X'} \pm \frac{c \beta |K|}{4} \right) \right|^2.
\end{align}
Due to Eq.~(\ref{F_k_K}), Eq.~(\ref{n_pm_X_3}) is equivalent to
\begin{align}
n_\pm(X) = & \left| \int_{\mathbb{R}} dx \int_{\mathbb{R}} dK \phi_\text{NW}(x) \frac{e^{\pm i K X  }}{2\pi} \sqrt{a|K|} \Gamma \left( \mp \frac{i K}{a} \right) \right. \nonumber \\
& \times \left. \exp \left( \pm \frac{c \beta |K|}{4} \right) F \left(  - \text{sign}(K) a^2 x ,  \mp  K + i \frac{a}{2} \right) \right|^2.
\end{align}
By using Eq.~(\ref{F_11}), we obtain
\begin{align}\label{n_pm_X_4}
n_\pm(X) = & \left| \int_{\mathbb{R}} dx \int_{\mathbb{R}} dK \phi_\text{NW}(x) \frac{e^{\pm i K X  }}{(2\pi)^2 a} \sqrt{\frac{|K|}{|x|}} \Gamma \left( \mp \frac{i K}{a} \right)  \Gamma \left( \frac{1}{2} \pm \frac{i K}{a} \right) \right. \nonumber \\
& \times \left.  \exp \left( \mp i \frac{K}{a} \ln (a |x|) \pm \theta ( x ) \frac{c \beta |K|}{2}  - \text{sign} ( K x ) i \frac{\pi}{4} \right) \right|^2.
\end{align}

Finally, by considering the variable transformations $x \mapsto \xi = -a X_\text{L}(x)$ if $x<0$, $x \mapsto \xi = a X_\text{R}(x)$ if $x>0$ and $K \mapsto \kappa = \mp K/a$ in Eq.~(\ref{n_pm_X_4}), we obtain
\begin{equation}\label{n_pm_2}
n_\pm (X) =  | \theta(\pm 1) \phi_\text{R}(X) + \phi_{\text{R} \pm}(X) + \phi_{\text{L} \pm}(X) |^2 ,
\end{equation}
with
\begin{subequations}
\begin{align}
\phi_\nu(X) = & \sqrt{a |x_\nu(X)|} \phi_\text{NW}(x_\nu(X)), \label{psi_LR}\\
\phi_{\nu \pm}(X) = & \int_{\mathbb{R}} d\xi \phi_\nu\left(s_\nu\frac{\xi}{a} \right) f_{\nu \pm}( \xi - aX ), \label{psi_LR_pm}\\
f_{\nu \pm}( \xi ) = & \int_{\mathbb{R}} d\kappa \frac{e^{i \kappa \xi}}{2 \pi} \left[ - \theta(s_\nu) \theta(\pm 1) + \frac{1}{2 \pi}\sqrt{|\kappa|} \Gamma (i\kappa) \Gamma \left(\frac{1}{2} - i\kappa \right)\right. \nonumber \\
& \times \left. \exp \left( \pm \theta(s_\nu) \pi |\kappa| \pm  s_\nu \text{sign}(\kappa) i \frac{\pi}{4} \right) \right].\label{f_LR}
\end{align}
\end{subequations}

The function $\phi_\nu(X)$ can be interpreted as the transformed version of the wave function $\phi_\text{NW}(x)$ from the Minkowski to the $\nu$-Rindler frame, such that the infinitesimal distribution $|\phi_\text{NW}(x)|^2 dx$ is conserved under the coordinate transformation $x \mapsto X=X_\nu(x)$, i.e., $\left| \phi_\text{NW}(x) \right|^2 dx = \left| \phi_\nu(X) \right|^2 dX$.

Notice that the functions $\phi_\text{L}(X)$ and $\phi_{\text{L} \pm}(X)$ are constructed from $\phi_\text{NW}(x)$ with negative values of $x$; conversely, $\phi_\text{R}(X)$ and $\phi_{\text{R} \pm}(X)$ depend of $\phi_\text{NW}(x)$ with $x>0$. Hence, within the decomposition $\left\langle \hat{n}_\text{R}(X) \right\rangle_{\hat{\rho}} = |\phi_\text{R}(X) + \phi_{\text{L} +}(X) + \phi_{\text{R} +}(X)|^2 + |\phi_{\text{L} -}(X) + \phi_{\text{R} -}(X)|^2$, the functions $\phi_{\text{L}\pm}(X)$ derive from the left-wedge part of $\phi_\text{NW}(x)$, whereas $\phi_\text{R}(X)$ and $\phi_{\text{R} \pm}(X)$ come from the right-wedge part of $\phi_\text{NW}(x)$.

\begin{figure}[]
\center
\includegraphics[]{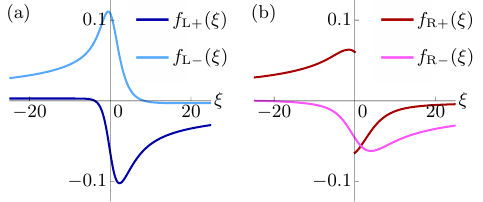}
\caption{Profile of $f_{\nu \pm}( \xi )$ defined by Eq.~(\ref{f_LR}) which have been numerically derived.}\label{f_figure}
\end{figure}

The functions $f_{\nu \pm}( \xi )$ are shown in Fig.~\ref{f_figure}. Notice that they are localized around $\xi = 0$. This means that $\Delta n_\text{R}(X)$ receives most contributions from $\phi_\text{L} (X')$ and $\phi_\text{R} (X')$ such that, respectively, $|X'+ X| \lesssim a^{-1}$ and $|X' - X| \lesssim a^{-1}$ or, equivalently, from $\phi_\text{NW} (x)$ such that $|x+x_\text{R}(X)| \lesssim a^{-1}$ and $|x-x_\text{R}(X)| \lesssim a^{-1}$.

The state $| \phi \rangle$ is said to be localized in the left wedge with respect to the Newton-Wigner scheme if $\phi_\text{NW}(x)$ is such that
\begin{equation}\label{localized_in_left_wedge_NW}
\phi_\text{NW}(x) \approx 0 \text{ if } x>0.
\end{equation}
When Eq.~(\ref{localized_in_left_wedge_NW}) holds, the functions $\phi_\text{R}(X)$ and $\phi_{\text{R} \pm}(X)$ are vanishing. Hence, $\Delta n_\text{R}(X)$ receives contributions only from $\phi_{\text{L} \pm}(X)$, in the sense that
\begin{equation}\label{Delta_n_localized_in_left_wedge_NW}
\Delta n_\text{R}(X) \approx \left| \phi_{\text{L}+}(X) \right| ^2 + \left| \phi_{\text{L}-}(X) \right|^2.
\end{equation}
Also, most of the contributions come from $\phi_\text{NW} (x)$, with $x= - x_\text{R}(X)$ as the specular point of the right Rindler coordinate $X$ with respect to the horizon.

The right hand side of Eq.~(\ref{Delta_n_localized_in_left_wedge_NW}) is generally nonvanishing. Hence, notwithstanding the localization of the particle in the left wedge with respect to the Newton-Wigner scheme, the mean value of $\hat{n}_\text{R}(X)$ can be different from the Rindler particle distribution of the Minkowski vacuum. In other words, the presence of the Minkowski particle influences the outcomes of the measurements carried out by Rob. 

At variance with Eq.~(\ref{localized_in_left_wedge_NW}), we say that the particle is localized in the right wedge with respect to the Newton-Wigner scheme when
\begin{equation}
\phi_\text{NW}(x) \approx 0 \text{ if } x<0.
\end{equation}
Notice from Fig.~\ref{f_figure}b that $ |f_{\text{R} \pm}(\xi)| \ll 1$. Therefore, if $\phi_\text{NW}(x)$ is localized in a region $\mathcal{R} = [x_0 - \Delta x, x_0 + \Delta x ]$ that belongs to the right wedge and is way narrower than $a^{-1}$ (i.e., $x_0 \gtrsim a^{-1}$ and $\Delta x \ll a^{-1}$), then the functions $\phi_{\text{R} \pm}(X)$ are expected to be negligible with respect to $\phi_\text{R}(X)$. The same occurs if the distance between $\mathcal{R}$ and the Minkowski origin is way larger than $a^{-1}$ (i.e., $x_0 \gg a^{-1}$), since the transformed region $X_\text{R} (\mathcal{R})$ becomes way narrower than $a^{-1}$. In summary, $|\phi_{\text{R} \pm}(X)| \ll |\phi_\text{R}(X)|$ when $x_0 \gtrsim a^{-1}$ and $\Delta x \ll a^{-1}$ or when $x_0 \gg a^{-1}$. Also, when these conditions hold, the functions $\phi_{\text{L}\pm}(X)$ are negligible and, hence,
\begin{equation}\label{Delta_n_localized_in_right_wedge_NW}
\Delta n_\text{R}(X) \approx \left| \phi_\text{R}(X) \right|^2.
\end{equation}
In other words, $\phi_\text{R}(X)$ appears as the dominant term in Eq.~(\ref{n_pm_2}), with $\phi_{\nu\pm}(X)$ as small corrective terms.

\begin{figure}[]
\center
\includegraphics[]{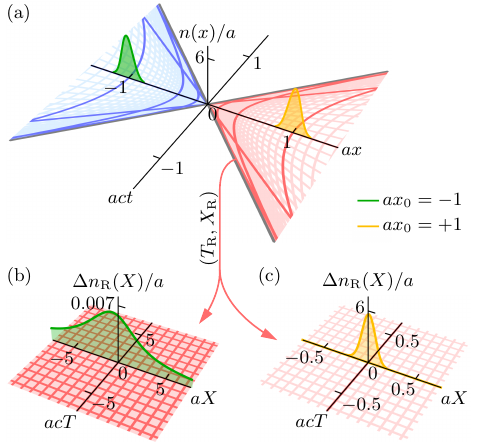}
\caption{Representation of the coordinate transformation $(t,x) \mapsto (T,X) = (T_\nu(t,x),X_\nu(t,x))$ [Eq.~(\ref{Rindler_coordinate_transformation_11})] and the Newton-Wigner density function transformation $n(x) \mapsto \Delta n_\text{R}(X)$ from the inertial to the accelerated frame. The yellow and the green lines are associated to single particle states with Gaussian wave function $\phi_\text{NW}(x)$ [Eq.~(\ref{Gaussian_psi_x})] and with parameters $a \sigma = 1$ and $ax_0 = \pm 1$. The left and right Rindler wedges are shown in panel (a) by means of constant $T$ and $X$ lines. The two wedges are delimited by the Rindler horizons (gray). The profile of the Minkowski-Newton-Wigner density $n(x)$ [Eq.~(\ref{n})] for the two wave packets is drawn. In panels (b) and (c), we show the constant $T$ and $X$ lines in the accelerated frame and the profile of $\Delta n_\text{R}(X)$, defined by Eq.~(\ref{Delta_n}). The value of $\Delta n_\text{R}(X)$ gives the variation of Rindler particle density distribution from the thermal Minkowski vacuum background. In panel (b), $\Delta n_\text{R}(X)$ is not negligible, even if $\phi_\text{NW}(x)$ is localized in the left wedge. On the other hand, in panel (c), $\Delta n_\text{R}(X)$ is larger and narrower, as we expect from wave functions localized within the right wedge.}\label{rindler_figure}
\end{figure}

We now detail these arguments by considering the specific example of a normalized Gaussian wave function, i.e.,
\begin{equation} \label{Gaussian_psi_x}
\phi_\text{NW}(x) = \frac{1}{\sqrt[4]{\pi}\sqrt{\sigma}} \exp \left( - \frac{(x-x_0)^2}{2 \sigma^2} \right).
\end{equation}
In Fig.~\ref{rindler_figure}, we show the distribution $\Delta n_\text{R}(X)$ in comparison with the Minkowski-Newton-Wigner density function, defined as
\begin{equation} \label{n}
n(x) =   \langle \psi | \hat{a}_\text{NW}^\dagger (x) \hat{a}_\text{NW}(x)| \psi \rangle .
\end{equation}

\begin{figure}[]
\center
\includegraphics[]{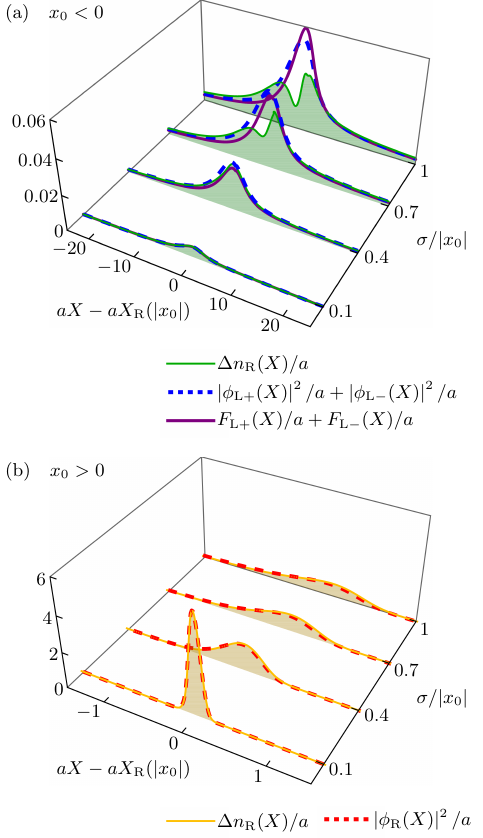}
\caption{Profile of $\Delta n_\text{R} (X)$ [Eq.~(\ref{Delta_n})] for different Gaussian single particle states [Eq.~(\ref{Gaussian_psi_x})]. In panel (a), we consider $x_0<0$ and different values of $\sigma/|x_0|$. $\left| \phi_{\text{L}+}(X) \right| ^2 + \left| \phi_{\text{L}-}(X) \right|^2$ is the dominant contribution to $\Delta n_\text{R} (X)$ when $\sigma/|x_0| \rightarrow 0$ and $x_0<0$. We also plot the function $F_{\text{L}+}(X)+F_{\text{L}-}(X)$ defined by Eq.~(\ref{F_LRpm}). In panel (b), we show the profile of $\Delta n_\text{R} (X)$ and $\left| \phi_\text{R}(X) \right|^2$ for configurations with $x_0>0$. In this case, the dominant contribution to $\Delta n_\text{R} (X)$ is $\left| \phi_\text{R}(X) \right|^2$, which converges to a delta function [Eq.~(\ref{components_limits_psi_R})].}\label{Delta_n_R_figure}
\end{figure}

It is possible to prove that the functions $\phi_\text{R}(X)$, $\phi_{\nu \pm}(X)$ and $\Delta n_\text{R}(X)$ are invariant under the transformation $x_0 \mapsto x_0/\gamma$, $\sigma \mapsto \sigma / \gamma$, $a X \mapsto a X - \ln \gamma$ for any $\gamma>0$. Hence, they can be put in a form that only depends of $\text{sign}(x_0)$, $\sigma/|x_0|$ and $X-X_\text{R}(|x_0|)$ instead of $\sigma$, $x_0$ and $X$. This feature is used in Fig.~\ref{Delta_n_R_figure}, where we show the profile of $\Delta n_\text{R}(X)$ for different $\sigma/x_0$.

The limit of well-localized wave functions is identified by $\sigma/|x_0| \rightarrow 0$. In Fig.~\ref{Delta_n_R_figure}, we show how Eqs.~(\ref{Delta_n_localized_in_left_wedge_NW}) and (\ref{Delta_n_localized_in_right_wedge_NW}) hold, respectively, for $x_0<0$ and $x_0>0$. More detailed limits can be obtained by computing
\begin{align} \label{limit_small_sigma_x_0_R}
& \lim_{\sigma/|x_0| \rightarrow 0} | \phi_\text{R}(X) |^2\nonumber \\
 = &  a e^{ a X - \ln (a |x_0|)   } \lim_{\sigma/|x_0| \rightarrow 0} \frac{|x_0|}{\sqrt{\pi} \sigma} \exp \left(  - \left(\frac{|x_0|}{\sigma} \right)^2 \left[   e^{a X - \ln (a |x_0|)}- \text{sign}(x_0) \right]^2 \right) \nonumber \\
 = & \frac{e^{a X}}{|x_0|} \delta \left( e^{a X - \ln (a |x_0|)} - \text{sign}(x_0) \right) \nonumber \\
 = & \theta(x_0) \delta(X-X_\text{R}(x_0)),
\end{align}
and
\begin{align} \label{limit_small_sigma_x_0_R_2}
 \lim_{\sigma/|x_0| \rightarrow 0} \sqrt{\frac{|x_0|}{\sigma}}\phi_\nu \left( s_\nu \frac{\xi}{a} \right) 
 = & \lim_{\sigma/|x_0| \rightarrow 0} \frac{\sqrt{|x_0|}}{\sqrt[4]{\pi}\sigma} \exp \left( \frac{\xi}{2} - \frac{x_0^2}{2 \sigma^2} \left(\frac{ s_\nu e^{\xi}}{a x_0} - 1\right)^2  \right)  \nonumber \\
 = & \sqrt{\frac{2\sqrt{\pi}}{|x_0|}} \exp \left( \frac{\xi}{2} \right)   \delta \left(\frac{ s_\nu e^{\xi}}{a x_0} - 1 \right)  \nonumber \\
= & \theta(s_\nu x_0) \sqrt{ 2 \sqrt{\pi} a} \delta (\xi -  a X_\text{R} (s_\nu x_0)),
\end{align}
which lead to
\begin{subequations}\label{components_limits}
\begin{align}
| \phi_\text{R}(X) |^2  \approx \theta(x_0) \delta(X-X_\text{R}(x_0)) ,\label{components_limits_psi_R} \\
|\phi_{\nu \pm}(X)|^2  \approx \theta(s_\nu x_0) F_{\nu \pm}(X),\label{components_limits_psi_LRpm}
\end{align}
\end{subequations}
with
\begin{equation}\label{F_LRpm}
F_{\nu \pm} (X) = \frac{\sigma}{|x_0|}  2 \sqrt{\pi} a f^2_{\nu \pm}(a X_\text{R}(|x_0|)- a X).
\end{equation}

From Eq.~(\ref{components_limits}) we obtain the explicit limit $\sigma/|x_0| \rightarrow 0$ of $\Delta n_\text{R}(X)$ for Gaussian wave functions. If $x_0<0$, $\Delta n_\text{R}(X) \rightarrow 0$ with leading term $F_{\text{L}+}(X) + F_{\text{L}-}(X)$ which is proportional to $ \sigma / |x_0| $. When the degree of localization of the particle increases, the probability of detection in the right wedge decreases. If $x_0>0$, instead, the distributional limit of $\Delta n_\text{R}(X)$ is $\delta(X-X_\text{R}(x_0))$. The single particle appears perfectly localized in both inertial and accelerated frame at the same position, up to the coordinate transformation.

In conclusion, we find that Minkowski single particle states localized beyond the horizon with respect to the Newton-Wigner scheme modify the Unruh thermal distribution in the accelerated frame. As a result, it seems that, contrary to classical predictions, accelerated observers are able to detect particles that are emitted beyond the horizon. However, we remark that the Newton-Wigner scheme is not fundamental and, hence, it does not give a genuine description of relativistic local phenomena. In Sec.~\ref{Example_single_particle_state_AQFT}, we will consider again the example of a single particle state detected beyond the horizon but in the framework of AQFT. This will give physical predictions about the non-strict localization of single particles.

\section{AQFT scheme} \label{Example_single_particle_state_AQFT}

In Sec.~\ref{Example_single_particle_state_NW}, we found that if a Minkowski single particle is localized in the left wedge with respect to the Newton-Wigner scheme, its presence can be revealed by an accelerated observer that measures the distribution of Rindler particles in the right wedge. In Sec.~\ref{NewtonWigner_ARbM_scenario}, we argued that the origin of this effect is to be found in the noncovariant behavior of the Newton-Wigner localization.

At variance with the Newton-Wigner scheme, the AQFT localization is compatible with the GR postulate of physical equivalence between different frames [Sec.~\ref{AQFT_localization_scheme_in_curved_spacetime}]. Hence, the framework in which local experiments in the QFTCS regime should be described is precisely AQFT. In this section, we show that the nonlocal effect that we saw in Sec.~\ref{Example_single_particle_state_NW} for the Newton-Wigner scheme appears in the AQFT scheme as well.

We consider localized single particle states defined as $| \phi \rangle = \hat{O}_\text{A} | 0_\text{M} \rangle$ with
\begin{equation}\label{single_particle_AQFT}
\hat{O}_\text{A}  =  \int_{\mathcal{V}} d^3 x \phi_\text{AQFT}(\vec{x}) \hat{\phi}(0,\vec{x}),
\end{equation}
where $\phi_\text{AQFT}(\vec{x})$ is supported in $\mathcal{V}$. The connection between Eq.~(\ref{single_particle_Minkowski}) and $| \phi \rangle = \hat{O}_\text{A} | 0_\text{M} \rangle$ can be obtained from
\begin{equation}\label{single_particle_AQFT_NW}
\int_{\mathcal{V}} d^3 x \phi_\text{AQFT}(\vec{x}) \hat{\phi}(0,\vec{x}) | 0_\text{M} \rangle = \int_{\mathbb{R}^3} d^3 k \tilde{\phi}_1(\vec{k}) \hat{a}^\dagger(\vec{k}) | 0_\text{M} \rangle,
\end{equation}
with
\begin{equation}
\tilde{\phi}_1(\vec{k}) = \int_{\mathcal{V}} d^3 x \phi_\text{AQFT}(\vec{x}) f^*(\vec{k},0,\vec{x})
\end{equation}
as the single particle wave function in the momentum space.

We assume that the particle is prepared by an inertial experiment (Alice) in the left wedge and an accelerated observer (Rob) carries out measurements of an observable $\hat{O}_\text{B} \in \mathfrak{A}^\text{AQFT}_\text{R}$ in the right wedge. As an example, we may consider $\hat{O}_\text{B} = \hat{O}_\text{B}(\vec{X}, \vec{X}') = \hat{\Phi}_\text{R} (0, \vec{X}) \hat{\Phi}_\text{R} (0, \vec{X}')$ as the $2$-point correlation operator for the right Rindler field.

The mean value of $\hat{O}_\text{B}$ with respect to $| \phi \rangle$ is
\begin{align}
\langle \phi | \hat{O}_\text{B} | \phi \rangle = & \int_{\mathcal{V}} d^3 x \int_{\mathcal{V}} d^3 x' \phi_\text{AQFT}^*(\vec{x})  \phi_\text{AQFT}(\vec{x}') \nonumber \\
& \times \langle 0_\text{M} | \hat{\phi}(0,\vec{x}) \hat{\Phi}_\text{R} (0, \vec{X}) \hat{\Phi}_\text{R} (0, \vec{X}') \hat{\phi}(0,\vec{x}') | 0_\text{M} \rangle.
\end{align}
By using Eq.~(\ref{scalar_transformation_Rindler}), we obtain
\begin{equation}\label{correlations_AQFT_Rindler_2}
\langle \phi | \hat{O}_\text{B} | \phi \rangle = \int_{\mathcal{V}} d^3 x \int_{\mathcal{V}} d^3 x' \phi_\text{AQFT}^*(\vec{x})  \phi_\text{AQFT}(\vec{x}') G_4(\vec{x}, \vec{x}_\text{R}(0,\vec{X}) , \vec{x}_\text{R}(0,\vec{X}'),\vec{x}'),
\end{equation}
with
\begin{equation}\label{G_4_point}
G_4(\vec{x}_1, \vec{x}_2, \vec{x}_3, \vec{x}_4) = \langle 0_\text{M} | \hat{\phi}(0,\vec{x}_1) \hat{\phi} (0, \vec{x}_2) \hat{\phi} (0, \vec{x}_3) \hat{\phi}(0,\vec{x}_4) | 0_\text{M} \rangle
\end{equation}
as the $4$-point correlation function in the Minkowski frame. Equation (\ref{G_4_point}) can be computed by using the Wick theorem \cite{PhysRev.80.268}, which leads to
\begin{align}\label{G_4_point_2}
G_4(\vec{x}_1, \vec{x}_2, \vec{x}_3, \vec{x}_4) = & G_2(\vec{x}_1, \vec{x}_2)G_2(\vec{x}_3, \vec{x}_4)  +  G_2(\vec{x}_1, \vec{x}_3)G_2(\vec{x}_2, \vec{x}_4) \nonumber \\
& +  G_2(\vec{x}_1, \vec{x}_4)G_2(\vec{x}_2, \vec{x}_3) ,
\end{align}
where 
\begin{equation}\label{G_2_point}
G_2(\vec{x}_1, \vec{x}_2) = \langle 0_\text{M} | \hat{\phi}(0,\vec{x}_1) \hat{\phi} (0, \vec{x}_2)  | 0_\text{M} \rangle
\end{equation}
is the $2$-point correlation function.

By plugging Eq.~(\ref{G_4_point_2}) in Eq.~(\ref{correlations_AQFT_Rindler_2}) and by using the normalization condition of $| \phi \rangle$, we obtain
\begin{equation}\label{correlations_AQFT_Rindler_3}
\langle \phi | \hat{O}_\text{B}(\vec{X}, \vec{X}') | \phi \rangle - \langle 0_\text{M} | \hat{O}_\text{B}(\vec{X}, \vec{X}') | 0_\text{M} \rangle = \Delta O_\text{B}(\vec{X}, \vec{X}'),
\end{equation}
with
\begin{align}\label{correlations_AQFT_Rindler_4}
& \Delta O_\text{B}(\vec{X}, \vec{X}') = \int_{\mathcal{V}} d^3 x \int_{\mathcal{V}} d^3 x' \phi_\text{AQFT}^*(\vec{x})  \phi_\text{AQFT}(\vec{x}') \nonumber \\
& \times \left[  G_2(\vec{x}, \vec{x}_\text{R}(0,\vec{X}) )G_2(\vec{x}_\text{R}(0,\vec{X}') , \vec{x}') +  G_2(\vec{x}, \vec{x}_\text{R}(0,\vec{X}') )G_2(\vec{x}_\text{R}(0,\vec{X}) , \vec{x}')  \right].
\end{align}
By showing that $\Delta O_\text{B}(\vec{X}, \vec{X}')$ is different from zero, we demonstrate that the preparation of $| \phi \rangle$ in the left wedge influences the measurements in the right wedge. 

Notice that the $2$-point function $G_2(\vec{x}_1, \vec{x}_2)$ is given by the Wightman propagator (\ref{twopoint_function_Minkowski}) as
\begin{equation}\label{twopoint_function_Minkowski_2}
G_2(\vec{x}_1, \vec{x}_2) = \frac{\hbar}{(2 \pi)^3} \int_{\mathbb{R}^3} \frac{d^3 k}{2 \omega(\vec{k})} e^{ i\vec{k} \cdot (\vec{x}_1-\vec{x}_2)}.
\end{equation}
By plugging Eq.~(\ref{twopoint_function_Minkowski_2}) in Eq.~(\ref{correlations_AQFT_Rindler_4}), we obtain
\begin{align}\label{correlations_AQFT_Rindler_5}
& \Delta O_\text{B}(\vec{X}, \vec{X}') = \frac{\hbar^2}{(2 \pi)^6}  \int_{\mathcal{V}} d^3 x \int_{\mathcal{V}} d^3 x' \int_{\mathbb{R}^3} d^3 k  \int_{\mathbb{R}^3} d^3 k'\frac{  \phi_\text{AQFT}^*(\vec{x})  \phi_\text{AQFT}(\vec{x}')}{4 \omega(\vec{k})\omega(\vec{k}')}   \nonumber \\
& \times \left[ e^{i \vec{k} \cdot (\vec{x} - \vec{x}_\text{R}(0,\vec{X}) ) + i \vec{k}' \cdot (\vec{x}_\text{R}(0,\vec{X}') - \vec{x}') }  + e^{i \vec{k} \cdot (\vec{x} - \vec{x}_\text{R}(0,\vec{X}') ) + i \vec{k}' \cdot (\vec{x}_\text{R}(0,\vec{X}) - \vec{x}') } \right],
\end{align}
which is generally different from zero, even when $\mathcal{V}$ belongs to the region of space with negative values of $z$. 

To see that $\Delta O_\text{B}(\vec{X}, \vec{X}') $ can be nonvanishing when $\mathcal{V}$ is in the left wedge, consider the example of a rectangle wave function $\phi_\text{AQFT} (\vec{x}) = N \theta( \sigma - |z + z_0| )$, where $N$ is the normalization factor, $z_0$ is a positive real number and $\sigma $ is positive and smaller than $z_0$. The support of $\phi_\text{AQFT} (\vec{x})$ is $\mathcal{V} = \mathbb{R}^2\otimes [-z_0-\sigma, -z_0+\sigma] $. In this case, Eq.~(\ref{correlations_AQFT_Rindler_5}) reads as
\begin{align}
\Delta O_\text{B}(\vec{X}, \vec{X}') = & \frac{ \hbar^2}{(2 \pi)^2}  \int_{|z + z_0|<\sigma} d z \int_{{|z' + z_0|<\sigma}} d z' \int_{\mathbb{R}} d k  \int_{\mathbb{R}} d k'\frac{N^2}{2 \omega(k \vec{e}_3)\omega(k' \vec{e}_3)}   \nonumber \\
& \times  e^{i k (z - z_\text{R}(Z) )+ i k' (z_\text{R}(Z') - z') },
\end{align}
with $\vec{e}_3 = (0,0,1)$ as the versor along $z$.

The integrals with respect to $k$ and $k'$ can be computed by using the modified Bessel function of the second kind $K_\zeta(\xi)$, which leads to
\begin{align}\label{correlations_AQFT_Rindler_6}
\Delta O_\text{B}(\vec{X}, \vec{X}') = & \frac{ \hbar^2 N^2}{2 \pi^2 c^2}  \int_{|z + z_0|<\sigma} d z \int_{{|z' + z_0|<\sigma}} d z' \nonumber \\ 
& \times K_0 \left( \frac{|z - z_\text{R}(Z)|}{\lambda_\text{C}} \right)  K_0 \left( \frac{|z_\text{R}(Z') - z'|}{\lambda_\text{C}} \right),
\end{align}
with $\lambda_\text{C} = \hbar/mc$ as the Compton wavelength. The integrands appearing in Eq.~(\ref{correlations_AQFT_Rindler_6}) are always positive; hence we find that $\Delta O_\text{B}(\vec{X}, \vec{X}') \neq 0$ for any couple $\vec{X}$ and $\vec{X}'$.

As a result, we showed that the preparation of the particle beyond the Rindler horizon can be revealed by the accelerated observer. However, as we remarked in Secs.~\ref{solving_the_paradox} and \ref{AQFT_localization_scheme_in_curved_spacetime}, such a nonlocal effect does not violate relativistic causality. The reason is that the preparation of the state $| \phi \rangle$ is selective, which means that it is partly affected by the experimenter choice to disregard other unsuccessful state preparations. The selective nature of $| \phi \rangle$ can be noticed from the operator $\hat{O}_\text{A}$ [Eq.~(\ref{single_particle_AQFT})], which is clearly not unitary. As already explained in Sec.~\ref{solving_the_paradox}, the experiment requires a classical communication between the inertial experimenter preparing the state and the accelerated observer carrying out the measurement. Such a classical communication restores causality.

\chapter*{Conclusions}
\addcontentsline{toc}{part}{Conclusions}
\section*{Results}
\addcontentsline{toc}{section}{Results}
\markboth{CONCLUSIONS}{RESULTS}

Here, we summarize the main results of the thesis.

\begin{itemize}
\item We defined NRQFTCS as the regime suited to describe nonrelativistic quantum systems in gravitational fields or in noninertial motion. We started from the formulation of QFTCS in terms of fields and particles affected by a non-Minkowskian metric. By restricting to states with energy very close to their mass, we found that all the fundamental features of the nonrelativistic quantum physics emerge in such a regime. Specifically, we obtained the following results.
\begin{itemize}
\item In Chap.~\ref{Non_relativistic_limit_of_QFT_and_QFTCS}, we demonstrated that states can be described by means of wave functions as elements of the $\mathbb{C}^n \otimes L^2(\mathbb{R}^3)$ Hilbert space---up to a metric dependent measure---and their time evolution is given by some Schrödinger equations.
\item In Chap.~\ref{Localization_in_accelerated_frame}, we found that the wave functions acquire the Born probabilistic interpretation to find the particles in space points.
\end{itemize}

\item NRQFTCS inherits some of the features from QFTCS that makes it conceptually and physically different from the nonrelativistic limit of QFT in Minkowski spacetime.
\begin{itemize}
\item In Chap.~\ref{Non_relativistic_limit_of_QFT_and_QFTCS}, we derived the GR corrections to the quantum theory of particles affected by Newtonian gravity. We showed that an accelerated observer (equivalently, an observer in the presence of a gravitational field) can distinguish a scalar field from a Dirac field by particle-gravity interaction.
\item In Chap.~\ref{Framedependent_nonrelativistic_limit}, we studied how the frame dependent content of particles in QFTCS affects the NRQFTCS regime. We found that an inertial and an accelerated observer agree about the first-quantized description of nonrelativistic quantum systems only when the relative acceleration is sufficiently constrained.
\item In Chap.~\ref{Localization_in_accelerated_frame}, we studied the effects of the frame dependent notions of vacuum state and nonrelativistic limit in the context of localized experiments in NRQFTCS. We showed possible violations of the Born postulate of independence between preparations of states and measurements of observables in disjoint regions of space. Such a nonlocal effect occurs in ``hybrid'' scenarios, in which either the background state is different from the vacuum associated to the experimenter's frame, or one of the two observers is inertial, while the other is accelerated.
\end{itemize}

\item The frame dependence of the nonrelativistic limit and numbers of particles was considered in Chap.~\ref{Accelerated_non_relativistic_detectors} for the particular example of accelerated atomic detectors. We gave a comprehensive description of accelerated atoms by means of the nonrelativistic limit of quantum Dirac fields in Rindler spacetime. We found the conditions under which the frame dependent effect is suppressed, allowing for a first-quantized characterization of the system in the accelerated frame. Also, by constraining the ionization of the atom and the accelerating electric field, we identified the configurations that can be experimentally implemented to detect the Unruh effect via hyperfine coupling.

\item The dependence between preparations of states and measurements of observables in disjoint regions of space was explicitly detailed in Chap.~\ref{Single_particle_beyond_Rindler_horizon}. There, we considered the example of a single particle prepared by an inertial observer beyond the Rindler horizon and detected by an accelerated observer in the Rindler frame.

\item In Chap.~\ref{Minkowski_particles_in_accelerated_frame}, we provided an algorithmic procedure to represent Minkowski states as elements of the right Rindler frame. As a result, we found how accelerated observers perceive particles that are prepared by inertial experimenters. Also, we derived their representation in terms of Wigner characteristic functions.

\end{itemize}

\section*{Future directions}
\addcontentsline{toc}{section}{Future directions}
\markboth{CONCLUSIONS}{FUTURE DIRECTIONS}

The results of the thesis may have both experimental and theoretical future applications.

\begin{itemize}
\item Throughout the thesis, we focused on inertial and uniformly accelerated frames. However, the theory may be extended to more general stationary spacetimes, that are provided with a well-defined notion of particle energies \cite{Wald:1995yp}. This includes, e.g., rotating frames, which can be implemented in circular accelerators of quantum systems. Also, due to the Einstein equivalence principle that locally relates gravitational fields to accelerated frames, the results of the thesis may be extended to GR and astrophysical scenarios (e.g., in the presence of Earth's gravity or close to a Black Hole).

\item As already pointed out in Chap.~\ref{Non_relativistic_limit_of_QFT_and_QFTCS}, the regime of quantum systems affected by weak Newtonian gravity has already been probed by a series of experiments on ultracold neutrons \cite{article1, PhysRevD.67.102002, article2, article3, article4, PhysRevLett.112.071101, Kamiya:2014qia}. GR corrections are expected to be revealed by the growing instrumental precision and require the implementation of NRQFTCS. Hence, the theoretical framework developed here may find practical applications in the measurements of (arbitrary strong) gravitational effects in nonrelativistic quantum systems. For sufficiently high precision in the experiments, this can potentially become an innovative resource for gravitation wave detection.

\item In Chap.~\ref{Minkowski_particles_in_accelerated_frame}, we reported a general procedure to characterize Minkowski particle states as seen by accelerated observers. We believe that this method can find applications in different scenarios, where many particles are prepared by inertial experimenters and are detected by accelerated devices.
 
\item In Chap.~\ref{Accelerated_non_relativistic_detectors}, we used the nonrelativistic limit of quantum Dirac field in Rindler spacetime to describe accelerated atoms from first principles. We found specific configurations, in terms of the ionization of the atom and the magnitude of the accelerating electric field, suited for the detection of the Unruh effect by hyperfine splitting. The experiment can be extended  by including the Zeeman effect, which, in absence of hyperfine structure, induces an arbitrarily small energy splitting at the ground states. Also, the same theoretical framework can be applied to other experimentally relevant scenarios. As an example, rotating and oscillating charged particles in electromagnetic fields may be described in the context of NRQFTCS. Their relevance to the experimental physics is due to the high power nowadays achievable by ultraintense lasers.

\item Notably, QFTCS is characterized by the inequivalence between vacuum states in different frames. Inertial and accelerated observers do not agree about the nature of the respective vacuum. The same applies to free-falling and standing experimenters affected by a gravitational field. Then, a natural question arises: in everyday laboratory experiments on the Earth, which kind of vacuum do we experience? Unfortunately, we still do not have a definite answer to this question. To solve the problem, one may consider to experimentally implement the RaRbR and the RaRbM scenarios discussed in Chap.~\ref{Localization_in_accelerated_frame}. This requires the use of two nonrelativistic devices, one preparing some states and the other one measuring observables; they are both standing in the laboratory frame while affected by the Earth's gravity. Depending on the nature of the background state, different outcomes of the measurements are expected. 

\item The ARbM scenario discussed in Chaps.~\ref{Localization_in_accelerated_frame} and \ref{Single_particle_beyond_Rindler_horizon} gives predictions about noninertial nonrelativistic detectors that can be used to reveal Reeh-Schlieder nonlocal effects. Equivalently, the theory may be applied to nonrelativistic detectors affected by gravitational fields.

\end{itemize}

\appendix
\part*{Appendices}
\addcontentsline{toc}{part}{Appendices}

\chapter{Bessel functions}\label{Bessel_functions}

This chapter is dedicated to the modified Bessel function of the second kind $K_\zeta (\xi)$ considered throughout the thesis. We study the functions $\tilde{F}(\Omega,\vec{K}_\perp,Z)$ and $\mathfrak{K}(\Omega,\vec{K}_\perp, Z)$ [Eqs.~(\ref{F_tilde_Rindler}) and (\ref{K_pm})] as well.

\section{Integral representation}\label{Bessel_functions_Integral_representation}

Here, we use the following integral representation for $K_\zeta (\xi)$ with positive argument \cite{bateman_erdelyi_1955}
\begin{align}\label{Bessel_integral_representation}
& K_\zeta (\xi) = \int_0^\infty d\tau e^{- \xi \cosh(\tau) } \cosh(\zeta \tau), & \xi > 0.
\end{align}
From Eq.~(\ref{Bessel_integral_representation}) it is straightforward to prove Eq.~(\ref{Bessel_conjugate}). In this section, we will also prove Eqs.~(\ref{Bessel_orthonormal}) and (\ref{Bessel_integral_representation_final}).

Alternately to Eq.~(\ref{Bessel_integral_representation}) one may use the following integrals \cite{Oriti}
\begin{equation}
f_\zeta^\pm(\xi) = \frac{1}{2} \int_{\mathbb{R}} d\tau \exp\left( i \xi \sinh(\tau) \pm \zeta \left( - i \frac{\pi}{2} + \tau\right) \right).
\end{equation}
When $\xi > 0$, both functions $f_\zeta^\pm(\xi)$ are a representation of $K_\zeta (\xi)$,
\begin{align}\label{Bessel_integral_representation_2}
& K_\zeta (\xi) = f_\zeta^+(\xi) = f_\zeta^-(\xi), & \xi > 0.
\end{align}
Equation (\ref{Bessel_integral_representation_2}) can be proven by using Eq.~(\ref{Bessel_integral_representation}) and by performing a contour integral of $\exp(- \xi \cosh(\tau) + \zeta \tau) $ with respect to $\tau$ along the rectangle with vertexes $-\infty $, $+\infty$, $+\infty \mp i \pi/2$ and $-\infty \mp i \pi/2$, respectively for $f_\zeta^\pm(\xi)$.

Notice that, for any $\xi > 0$,
\begin{align}\label{Bessel_integral_representation_3_a}
K_\zeta (\xi)  = & \frac{ e^{i \pi \zeta} f_\zeta^+(\xi) - e^{-i \pi \zeta} f_\zeta^-(\xi)}{2 i \sin(\pi \zeta)} \nonumber \\
 = & \frac{1}{2 \sin(\pi \zeta)} \int_{\mathbb{R}} d\tau e^{i \xi \sinh(\tau) } \sin \left( \zeta \left( \frac{\pi}{2} - i \tau  \right) \right)
\end{align}
and that, for any $\xi > 0$,
\begin{align}\label{Bessel_integral_representation_3_b}
0 = &  \frac{  f_\zeta^-(\xi) -  f_\zeta^+(\xi)}{2 i \sin(\pi \zeta)} \nonumber \\
 = & \frac{1}{2 \sin(\pi \zeta)} \int_{\mathbb{R}} d\tau e^{i \xi \sinh(\tau) } \sin \left( \zeta \left( \frac{\pi}{2} + i \tau  \right) \right) \nonumber \\
 = & \frac{1}{2 \sin(\pi \zeta)} \int_{\mathbb{R}} d\tau e^{-i \xi \sinh(\tau) } \sin \left( \zeta \left( \frac{\pi}{2} - i \tau \right) \right).
\end{align}
By considering both Eqs.~(\ref{Bessel_integral_representation_3_a}) and (\ref{Bessel_integral_representation_3_b}) one obtains the following identity that holds for any $\xi \in \mathbb{R}$
\begin{equation}\label{Bessel_integral_representation_4}
\theta(\xi) K_\zeta (\xi)  = \frac{1}{2 \sin(\pi \zeta)} \int_{\mathbb{R}} d\tau e^{i \xi \sinh(\tau) } \sin \left( \zeta \left( \frac{\pi}{2} - i \tau \right) \right).
\end{equation}
Equation (\ref{Bessel_integral_representation_4}) can be used to prove both Eq.~(\ref{Bessel_orthonormal}) and Eq.~(\ref{Bessel_integral_representation_final}) as we will show in the following subsections.

\subsection{Proof of Eq.~(\ref{Bessel_orthonormal})}\label{Proof_of_Bessel_orthonormal}

Regarding Eq.~(\ref{Bessel_orthonormal}), the proof is given by
\begin{align}
& \int_0^\infty d\xi \left[  K_{-i \zeta-1/2}(\xi) K_{i \zeta'-1/2}(\xi) + K_{i \zeta-1/2}(\xi) K_{-i \zeta'-1/2}(\xi) \right] \nonumber \\
= & \frac{1}{4} \int_{\mathbb{R}} d\tau  \int_{\mathbb{R}} d\tau' \left\lbrace \left[ \sin\left(  -i \pi \zeta - \frac{\pi}{2} \right)   \sin\left(  i \pi \zeta' - \frac{\pi}{2} \right) \right]^{-1}  \right.  \nonumber \\
& \times \sin \left(   -i \frac{\pi \zeta}{2}  - \zeta \tau  -  \frac{\pi}{4} + i \frac{\tau}{2}  \right) \sin \left(  i \frac{\pi \zeta'}{2} + \zeta' \tau' - \frac{\pi}{4} + i \frac{\tau'}{2}  \right) \nonumber \\
& + \left[ \sin\left(  i \pi \zeta - \frac{\pi}{2} \right)   \sin\left( -i \pi \zeta' - \frac{\pi}{2} \right) \right]^{-1} \sin \left(   i \frac{\pi \zeta}{2}  + \zeta \tau  -  \frac{\pi}{4} + i \frac{\tau}{2}  \right) \nonumber \\
& \left. \times  \sin \left( - i \frac{\pi \zeta'}{2} - \zeta' \tau' - \frac{\pi}{4} + i \frac{\tau'}{2}  \right) \right\rbrace \int_{\mathbb{R}} d\xi e^{i \xi [\sinh(\tau) + \sinh(\tau')] } \nonumber \\
= &-\frac{\pi}{8 \cosh ( \pi \zeta )   \cosh (  \pi \zeta' )} \int_{\mathbb{R}} d\tau  \int_{\mathbb{R}} d\tau' \left[ \exp \left( \pi \frac{\zeta - \zeta'}{2}   - i (\zeta \tau - \zeta' \tau')  - i \frac{\pi}{2}  \right.   \right. \nonumber \\
&  \left. - \frac{\tau+\tau'}{2}  \right) + \exp \left(- \pi \frac{\zeta - \zeta'}{2}  + i (\zeta \tau - \zeta' \tau')  + i \frac{\pi}{2} + \frac{\tau+\tau'}{2}  \right)- \exp \left( \pi \frac{\zeta + \zeta'}{2} \right. \nonumber \\
&  \left.  - i (\zeta \tau + \zeta' \tau')  - \frac{\tau-\tau'}{2}  \right)  - \exp \left(-\pi \frac{\zeta + \zeta'}{2}   + i (\zeta \tau + \zeta' \tau')  + \frac{\tau-\tau'}{2}  \right) \nonumber \\
&+  \exp \left( - \pi \frac{\zeta - \zeta'}{2}  + i (\zeta \tau - \zeta' \tau')  - i \frac{\pi}{2}   - \frac{\tau+\tau'}{2}  \right) + \exp \left(\pi \frac{\zeta - \zeta'}{2} \right.\nonumber \\
& \left.  - i (\zeta \tau - \zeta' \tau')  + i \frac{\pi}{2}   + \frac{\tau+\tau'}{2}  \right) - \exp \left( - \pi \frac{\zeta + \zeta'}{2}    + i (\zeta \tau + \zeta' \tau') \right.  \nonumber \\
& \left. \left.  - \frac{\tau-\tau'}{2}  \right)  - \exp \left(\pi \frac{\zeta + \zeta'}{2}  - i (\zeta \tau + \zeta' \tau')    + \frac{\tau-\tau'}{2}  \right) \right]  \delta (\sinh(\tau) + \sinh(\tau')) \nonumber \\
= & \frac{\pi}{8 \cosh ( \pi \zeta )   \cosh (  \pi \zeta' )} \int_{\mathbb{R}} d\tau \frac{1}{\cosh (\tau)} \nonumber \\
& \times \left[  \exp \left( \pi \frac{\zeta + \zeta'}{2}   - i (\zeta - \zeta')\tau   - \tau \right)   + \exp \left(-\pi \frac{\zeta + \zeta'}{2}   + i (\zeta - \zeta' ) \tau   + \tau  \right) \right.\nonumber \\
&  \left.  + \exp \left( - \pi \frac{\zeta + \zeta'}{2}    + i (\zeta - \zeta') \tau  - \tau  \right)   + \exp \left(\pi \frac{\zeta + \zeta'}{2}  - i (\zeta - \zeta' ) \tau    + \tau  \right) \right]\nonumber \\
= & \frac{\pi}{4 \cosh ( \pi \zeta )   \cosh (  \pi \zeta' )}\nonumber \\
& \times \int_{\mathbb{R}} d\tau  \left[  \exp \left( \pi \frac{\zeta + \zeta'}{2}   - i (\zeta - \zeta')\tau  \right) + \exp \left(-\pi \frac{\zeta + \zeta'}{2}   + i (\zeta - \zeta' ) \tau    \right) \right] \nonumber \\
 = &  \pi^2 \frac{e^{\pi \zeta}+e^{-\pi \zeta}}{ 2 \cosh^2(\pi \zeta)}\delta(\zeta-\zeta') \nonumber \\
 = &  \frac{\pi^2 \delta(\zeta-\zeta')}{\cosh(\pi \zeta)}.
\end{align}

\subsection{Proof of Eq.~(\ref{Bessel_integral_representation_final})}\label{Proof_of_Bessel_integral_representation_final}

Equation (\ref{Bessel_integral_representation_final}), instead, can be proved by using Eq.~(\ref{Bessel_integral_representation_4}) to compute the following Fourier transform
\begin{align}\label{Bessel_integral_representation_5}
 \int_{\mathbb{R}} d\xi \theta(\xi) e^{- i \xi \sinh(\tau)} K_\zeta (\xi) 
  = &  \int_{\mathbb{R}} d\tau' \int_{\mathbb{R}} d\xi \frac{e^{i \xi [\sinh(\tau') - \sinh(\tau)] }}{2 \sin(\pi \zeta)} \sin \left( \zeta \left( \frac{\pi}{2} - i \tau' \right) \right) \nonumber \\
  = &   \int_{\mathbb{R}} d\tau' \frac{\pi \delta (\sinh(\tau') - \sinh(\tau))}{ \sin(\pi \zeta)} \sin \left( \zeta \left( \frac{\pi}{2} - i \tau' \right) \right) \nonumber \\
  = &  \frac{\pi}{ \sin(\pi \zeta) \cosh(\tau)} \sin \left( \zeta \left( \frac{\pi}{2} - i \tau \right) \right).
\end{align}

\section{Approximations for $\tilde{F}(\Omega,\vec{K}_\perp,Z)$}

We consider the function $ \bar{\tilde{F}} (\bar{\Omega},\vec{\bar{k}}_\perp, \bar{z})$ defined by Eq.~(\ref{F_bar}). In the limit $|\bar{\Omega}| \gg  \bar{a}^{3/2}$, $\bar{\tilde{F}} (\bar{\Omega}, \vec{\bar{k}}_\perp, \bar{z})$ can be approximated as \cite{Dunster1990BesselFO, olver2014asymptotics}
\begin{equation}\label{F_tilde_Rindler_Hankel_2_bar}
\bar{\tilde{F}} (\bar{\Omega},\vec{\bar{k}}_\perp, \bar{z})  \approx \frac{1}{2 \pi} \sqrt[6]{\frac{2}{\bar{\Omega}^2}} \sqrt[4]{\frac{\zeta(z(\bar{\Omega},\vec{\bar{k}}_\perp,\bar{z}))}{1-z^2(\bar{\Omega},\vec{\bar{k}}_\perp,\bar{z})} } \text{Ai} \left( - \frac{1}{ \bar{a}} \sqrt[3]{\frac{\bar{\Omega}^2}{2}} \zeta(z(\bar{\Omega},\vec{\bar{k}}_\perp,\bar{z})) \right),
\end{equation}
with
\begin{subequations}\label{z_zeta_z}
\begin{equation}
z(\bar{\Omega},\vec{\bar{k}}_\perp,\bar{z}) = \frac{1}{|\bar{\Omega}|} \sqrt{ 1+ 2 \bar{a} |\vec{\bar{k}}_\perp|^2 }  (1 + \bar{a} \bar{z}) ,
\end{equation}
\begin{equation}\label{zeta_z}
\begin{cases}
\frac{2}{3} \zeta^{3/2}(z) = \ln \left( \frac{1+\sqrt{1-z^2}}{z} \right) - \sqrt{1-z^2}, & \text{if } 0 \leq z \leq 1,  \\
\frac{2}{3} [-\zeta(z)]^{3/2} = \sqrt{z^2-1} - \arccos \left( \frac{1}{z} \right), & \text{if }  z \geq 1 
\end{cases}
\end{equation}
\end{subequations}
and where $\text{Ai}(\xi)$ is the Airy function. 

When $|\bar{\Omega} - 1| \ll 1$, $\bar{a}|\vec{\bar{k}}| \ll 1$ and $\bar{a} |\bar{z}| \ll 1$ the variables $z(\bar{\Omega},\vec{\bar{k}}_\perp,\bar{z})$ and $\zeta(z)$ can be approximated by \cite{olver2014asymptotics}
\begin{subequations}\label{z_zeta_z_approximation_Puiseux}
\begin{align}
& z(\bar{\Omega},\vec{\bar{k}}_\perp,\bar{z}) \approx  1  +  \bar{a} (|\vec{\bar{k}}_\perp|^2 + \bar{z}) - (|\bar{\Omega}|  -1) ,\\
&   \zeta(z(\bar{\Omega},\vec{\bar{k}}_\perp,\bar{z})) \approx -\sqrt[3]{2} [ \bar{a} (|\vec{\bar{k}}_\perp|^2 + \bar{z}) - (|\bar{\Omega}|  -1) ] \label{zeta_z_approximation_Puiseux}
\end{align}
\end{subequations}
and, hence, Eq.~(\ref{F_tilde_Rindler_Hankel_2_bar}) can be approximated by
\begin{equation}\label{F_tilde_Rindler_Hankel_2_nonrelativistic}
\bar{\tilde{F}} (\bar{\Omega},\vec{\bar{k}}_\perp, \bar{z})  \approx \frac{1}{2 \pi} \text{Ai} \left( |\vec{\bar{k}}_\perp|^2 + \bar{z} - \frac{|\bar{\Omega}|  -1}{\bar{a}} \right).
\end{equation}

When the argument of the Airy function diverges at $\xi \rightarrow \infty$, one finds the following asymptotic behavior
\begin{equation}\label{Ai_infty}
\text{Ai}(\xi) \sim \frac{1}{\sqrt[4]{\xi}} \exp \left( -\frac{2}{3} \xi^{3/2} \right).
\end{equation}
When the argument of the Airy function diverges at $\xi \rightarrow -\infty$, instead, modulus and phase of $\text{Ai}(\xi)$ have the following asymptotic leading terms
\begin{equation}\label{Airy_asymptotic_minus_infinity}
\text{Ai}(\xi) \approx \frac{1}{\sqrt{\pi}\sqrt[4]{-\xi}} \sin \left( \frac{\pi}{4} + \frac{2}{3}(-\xi)^{3/2} \right).
\end{equation}

\subsection{Proof of compatibility between nonrelativistic and quasi-inertial conditions}\label{Proof_of_compatibility_between_nonrelativistic_and_quasiinertial_conditions}

In this subsection, we show the existence of states that simultaneously satisfy Eqs.~(\ref{non_relativistic_limit_curved_wavefunction}), (\ref{quasi_inertial_limit_X_wavefunction}) and (\ref{local_limit_momentum_wavefunction}) for $n=1$, while the acceleration $\alpha$ satisfies Eq.~(\ref{quasi_inertial_limit_a}). The generalization to the case $n > 1$ is straightforward. In this way, we demonstrate the compatibility between the nonrelativistic and the quasi-inertial conditions.

In particular, here, we prove that if $a$ satisfies Eq.~(\ref{quasi_inertial_limit_a}), then there are some wave functions $\tilde{\Phi}_1(\Omega,\vec{K}_\perp)$ such that: (i) the support of
\begin{equation}
\Phi_1(T,\vec{X}) = \int_0^{\infty} d\Omega \int_{\mathbb{R}^2} d^2 K_\perp \tilde{\Phi}_1(\Omega,\vec{K}_\perp) F (\Omega,\vec{K}_\perp, T, \vec{X})
\end{equation}
lies approximately in (\ref{quasi_inertial_limit_X}); (ii) the support of $\tilde{\Phi}_1(\Omega,\vec{K}_\perp)$ lies approximately in (\ref{non_relativistic_limit_curved}) and (\ref{local_limit_momentum}). We start by considering a class of wave functions $\tilde{\Phi}_1(\Omega,\vec{K}_\perp)$ satisfying condition (ii) and we demonstrate that some of them satisfy condition (i) as well, i.e., $\Phi_1(T,\vec{X}) \approx 0$ when $\vec{X}$ lies outside of Eq.~(\ref{quasi_inertial_limit_X}). 

Firstly, notice that the spatial points $\vec{X} = (X,Y,Z)$ satisfying $|\bar{z}| \gtrsim \bar{a}^{-1}$, with $\bar{z}$ defined by $Z = Z_\text{R}((\bar{a} \bar{z}+ 1)/a)$, lead to trivial results. Indeed, in the limits $\bar{z} \gtrsim \bar{a}^{-1}$, $|\vec{\bar{k}}| \lesssim 1$, $|\bar{\Omega} - 1| \lesssim \bar{a} $ and $\bar{a} \ll 1$ the function $\bar{\tilde{F}} (\bar{\Omega}, \vec{\bar{k}}_\perp, \bar{z})$ is exponentially vanishing [Eq.~(\ref{F_tilde_Rindler_Hankel_2_bar})]. Conversely, for negative values of $\bar{z}$, the limit $|\bar{z}| \gtrsim \bar{a}^{-1}$ is not associated to any right wedge coordinate $\vec{X}$.

We now consider a point $\vec{X}$ such that $1 \ll |\bar{z}| \ll \bar{a}^{-1}$. This is the condition for $\vec{X}$ to lie outside of Eq.~(\ref{quasi_inertial_limit_X}), but close from it. In this case, one can use the approximations (\ref{Ai_infty}) and (\ref{Airy_asymptotic_minus_infinity}) depending on the sign of $\bar{z}$. If $\bar{z} > 0$, Eq.~(\ref{Ai_infty}) holds and implies that $\Phi_1(T,\vec{X}) \approx 0$; conversely, if $\bar{z} < 0$, one can use Eq.~(\ref{Airy_asymptotic_minus_infinity}) to show the rapidly oscillating behavior of the Airy function that may lead to $\Phi_1(T,\vec{X}) \approx 0$. To see this, plug Eq.~(\ref{Airy_asymptotic_minus_infinity}) in Eq.~(\ref{F_tilde_Rindler_Hankel_2_nonrelativistic}) and let $\bar{z}  \rightarrow - \infty$ to obtain
\begin{align}\label{F_tilde_Rindler_Hankel_2_nonrelativistic_x_bar_m_infty}
\bar{\tilde{F}} (\bar{\Omega},\vec{\bar{k}}_\perp, \bar{z}) \approx & \frac{1}{2 \pi^{3/2}\sqrt[4]{|\bar{z}}|}  \sin \left( \frac{2}{3}|\bar{z}|^{3/2}   -  \sqrt{|\bar{z}|}  \left( |\vec{\bar{k}}_\perp|^2 - \frac{|\bar{\Omega}|  -1}{\bar{a}} \right) + \frac{\pi}{4} \right. \nonumber \\
& \left. + \mathcal{O}\left( \frac{1}{\sqrt{|\bar{z}|}} \right) \right).
\end{align}
One can notice that the integration of $\bar{\tilde{F}} (\bar{\Omega},\vec{\bar{k}}_\perp, \bar{z})$ with a function of $(\bar{\Omega},\vec{\bar{k}}_\perp)$ that is slowly varying in $|\vec{\bar{k}}_\perp| \lesssim 1$ and $|\bar{\Omega}| - 1 \lesssim \bar{a}$ leads to vanishing results. Condition (ii) implies that $\tilde{\Phi}_1(\Omega,\vec{K}_\perp)$ is allowed to be slowly varying in $|\vec{\bar{k}}_\perp| \lesssim 1$ and $|\bar{\Omega}| - 1 \lesssim \bar{a}$. Consequently, there are some $\tilde{\Phi}_1(\Omega,\vec{K}_\perp)$ such that $\Phi_1(T,\vec{X}) \approx 0$ when $\vec{X}$ lies outside of Eq.~(\ref{quasi_inertial_limit_X}).

\subsection{Proof of compatibility between nonrelativistic and high acceleration conditions}\label{Proof_of_compatibility_between_nonrelativistic_and_high_acceleration_conditions}

Here we use the same arguments of Sec.~\ref{Proof_of_compatibility_between_nonrelativistic_and_quasiinertial_conditions} to prove that the high acceleration condition (\ref{local_limit_alternative}) is compatible with the nonrelativistic limit (\ref{non_relativistic_limit_curved}).

Conditions (\ref{non_relativistic_limit_curved}) and (\ref{local_limit_alternative}) can be written in terms of the variables $\bar{\Omega}$, $\bar{a}$, $\vec{\bar{k}}$ and $\bar{z}$ as follows
\begin{align}\label{constraints_adimentional_high_acceleration}
& \frac{|\bar{\Omega} - 1|}{\bar{a}} \lesssim \epsilon_\text{HA}^{-1/3}\epsilon^{1/3}, && \bar{a} \lesssim \epsilon_\text{HA}^{1/3}\epsilon^{2/3}, && |\vec{\bar{k}}| \lesssim\epsilon_\text{HA}^{-1/6}\epsilon^{1/6}, &&  \bar{z} \lesssim \epsilon_\text{HA}^{2/3}\epsilon^{-2/3}.
\end{align}
We consider a point $\vec{X}$ such that $\epsilon_\text{HA}^{2/3} \epsilon^{-2/3} \ll |\bar{z}| \ll \bar{a}^{-1}$. One can use the same arguments of Sec.~\ref{Proof_of_compatibility_between_nonrelativistic_and_quasiinertial_conditions} to prove that $\Phi_1(T,\vec{X}) \approx 0$ when $\bar{a} \lesssim \epsilon_\text{HA}^{1/3}\epsilon^{2/3}$ and when $\tilde{\Phi}_1(\Omega,\vec{K}_\perp)$ is supported and slowly varying in $|\bar{\Omega} - 1|\lesssim \epsilon^{1/3}\epsilon_\text{HA}^{-1/3} \bar{a} $ and $|\vec{\bar{k}}| \lesssim \epsilon^{1/6} \epsilon_\text{HA}^{-1/6}$.

\subsection{Proof of Eq.~(\ref{F_tilde_Rindler_Hankel_2})}\label{Proof_of_F_tilde_Rindler_Hankel_2}

The nonrelativistic and quasi-inertial conditions (\ref{non_relativistic_limit_curved}) and (\ref{local_limit}) can be expressed in terms of variables $\bar{\Omega}$, $\bar{a}$, $\vec{\bar{k}}$ and $\bar{z}$ by means of Eqs.~(\ref{constraints_adimentional}) and (\ref{non_relativistic_Theta_1_adimensional}). When Eqs.~(\ref{constraints_adimentional}) and (\ref{non_relativistic_Theta_1_adimensional}) are satisfied, one can use the approximation (\ref{F_tilde_Rindler_Hankel_2_nonrelativistic}), which proves Eq.~(\ref{F_tilde_Rindler_Hankel_2}).

Equation (\ref{F_tilde_Rindler_Hankel_2}) can also be proven in the high acceleration regime (\ref{local_limit_alternative}). The use of the approximation (\ref{F_tilde_Rindler_Hankel_2_nonrelativistic}) is justified by Eqs.~(\ref{constraints_adimentional_high_acceleration}), which is the nonrelativistic and the high acceleration conditions (\ref{non_relativistic_limit_curved}) and (\ref{local_limit_alternative}) written in terms of the variables $\bar{\Omega}$, $\bar{a}$, $\vec{\bar{k}}$ and $\bar{z}$.

\section{Approximations for $\mathfrak{K}(\Omega,\vec{K}_\perp, Z)$}

Here we prove Eq.~(\ref{Dirac_representation_reduction_nonrelativistic_limit_general_Rindler}) in the quasi-inertial limit (\ref{local_limit}) and Eq.~(\ref{Dirac_representation_reduction_nonrelativistic_limit_general_Rindler_high_a}) in the high acceleration limit (\ref{local_limit_alternative}).

By using Eq.~(\ref{local_limit_momentum}), $\mathfrak{G}_\text{R} (\vec{K}_\perp)$ defined by Eq.~(\ref{Gamma}) can be approximated by
\begin{equation}\label{Gamma_quasiinertial}
\mathfrak{G}_\text{R} (\vec{K}_\perp) = - i c \gamma^0 + \mathcal{O}(\epsilon) \gamma^0 + \mathcal{O}(\epsilon^{1/2}) \gamma^0 \gamma^1 + \mathcal{O}(\epsilon^{1/2}) \gamma^0 \gamma^2
\end{equation}
and, hence, Eq.~(\ref{W_tilde_W_tilde}) by
\begin{align}\label{W_tilde_W_tilde_quasiinertial}
\tilde{W}_{\text{R} s}(\Omega,\vec{K}_\perp,Z_\text{R}(z)) = & \left[ \mathfrak{K} (\Omega,\vec{K}_\perp,Z) - i c \mathfrak{K} (-\Omega,\vec{K}_\perp,Z)  \gamma^0 + \mathcal{O}(\epsilon) c \gamma^0 \right. \nonumber \\
& \left. + \mathcal{O}(\epsilon^{1/2}) c \gamma^0 \gamma^1 + \mathcal{O}(\epsilon^{1/2}) c \gamma^0 \gamma^2 \right] \mathfrak{W}_{\text{R} s}(\Omega,\vec{K}_\perp).
\end{align}
Equation (\ref{W_tilde_W_tilde_quasiinertial}) can be decomposed with respect to the spinorial basis $ \mathfrak{u}_s$ and $ \mathfrak{v}_s$ as follows
\begin{subequations}\label{W_tilde_W_tilde_quasiinertial_decomposition}
\begin{align}
\mathfrak{u}_s^\dagger \tilde{W}_{\text{R} s'}(\Omega,\vec{K}_\perp,Z_\text{R}(z)) = & \left[ \mathfrak{K} (\Omega,\vec{K}_\perp,Z) - i \mathfrak{K} (-\Omega,\vec{K}_\perp,Z) + \mathcal{O}(\epsilon) \right] \nonumber \\
&   \times \mathfrak{u}_s^\dagger  \mathfrak{W}_{\text{R} s'}(\Omega,\vec{K}_\perp) +  \mathcal{O}(\epsilon^{1/2}) \mathfrak{v}_{\bar{s}}^\dagger \mathfrak{W}_{\text{R} s'}(\Omega,\vec{K}_\perp), \\
\mathfrak{v}_s^\dagger \tilde{W}_{\text{R} s'}(\Omega,\vec{K}_\perp,Z_\text{R}(z)) = & \left[ \mathfrak{K} (\Omega,\vec{K}_\perp,Z) + i \mathfrak{K} (-\Omega,\vec{K}_\perp,Z)  + \mathcal{O}(\epsilon) \right]  \nonumber \\
& \times \mathfrak{v}_s^\dagger  \mathfrak{W}_{\text{R} s'}(\Omega,\vec{K}_\perp) +  \mathcal{O}(\epsilon^{1/2}) \mathfrak{u}_{\bar{s}}^\dagger \mathfrak{W}_{\text{R} s'}(\Omega,\vec{K}_\perp), \label{W_tilde_W_tilde_quasiinertial_decomposition_b}
\end{align}
\end{subequations}
where $\bar{s}$ is defined as the opposite of $s$, in the sense that $\bar{s} = 2$ if $s = 1$ and $\bar{s} = 1$ if $s = 2$.

\subsection{Proof of Eq.~(\ref{Dirac_representation_reduction_nonrelativistic_limit_general_Rindler})}\label{Proof_of_Dirac_representation_reduction_nonrelativistic_limit_general_Rindler}

Now, notice that in the nonrelativistic limit (\ref{non_relativistic_limit_curved}) and quasi-inertial regime (\ref{local_limit}), Eq.~(\ref{K_recurrence}) can be approximated by
\begin{equation}\label{K_recurrence_nonrelativistic_quasiinertial}
 \left(\frac{\hbar}{m c}\partial_3 -  i \right) \mathfrak{K}(\Omega,\vec{K}_\perp, Z) = - \mathfrak{K} (- \Omega,\vec{K}_\perp, Z) + \mathcal{O}(\epsilon).
\end{equation}
By using Eqs.~(\ref{F_tilde_Rindler}), (\ref{F_tilde_Rindler_Hankel_2}) and (\ref{K_pm_nonrelativistic_quasiinertial}), one can also prove that
\begin{equation}\label{derivative_K_nonrelativistic}
\partial_3 \mathfrak{K}(\Omega,\vec{K}_\perp,Z) \sim \left(\frac{m c}{\hbar a} \right)^{2/3} a \mathfrak{K}(\Omega,\vec{K}_\perp,Z),
\end{equation}
which means that
\begin{equation}\label{derivative_K_nonrelativistic_quasiinertial}
\frac{\hbar}{m c} \partial_3 \mathfrak{K}(\Omega,\vec{K}_\perp,Z) = \mathcal{O}(\epsilon^{1/2}).
\end{equation}
By combining Eqs.~(\ref{K_recurrence_nonrelativistic_quasiinertial}) and (\ref{derivative_K_nonrelativistic_quasiinertial}), one obtains
\begin{equation}\label{K_recurrence_nonrelativistic_quasiinertial_2}
\mathfrak{K}(\Omega,\vec{K}_\perp, Z) + i \mathfrak{K} (- \Omega,\vec{K}_\perp, Z) = \mathcal{O}(\epsilon^{1/2}).
\end{equation}

By plugging Eq.~(\ref{K_recurrence_nonrelativistic_quasiinertial_2}) in Eq.~(\ref{W_tilde_W_tilde_quasiinertial_decomposition}), one obtains
\begin{subequations}\label{W_tilde_W_tilde_quasiinertial_decomposition_2}
\begin{align}
\mathfrak{u}_s^\dagger \tilde{W}_{\text{R} s'}(\Omega,\vec{K}_\perp,Z_\text{R}(z)) = & \left[ 2 \mathfrak{K} (\Omega,\vec{K}_\perp,Z) + \mathcal{O}(\epsilon^{1/2}) \right] \mathfrak{u}_s^\dagger  \mathfrak{W}_{\text{R} s'}(\Omega,\vec{K}_\perp) \nonumber \\
& +  \mathcal{O}(\epsilon^{1/2}) \mathfrak{v}_{\bar{s}}^\dagger \mathfrak{W}_{\text{R} s'}(\Omega,\vec{K}_\perp), \\
\mathfrak{v}_s^\dagger \tilde{W}_{\text{R} s'}(\Omega,\vec{K}_\perp,Z_\text{R}(z)) = & \mathcal{O}(\epsilon^{1/2}) \mathfrak{v}_s^\dagger  \mathfrak{W}_{\text{R} s'}(\Omega,\vec{K}_\perp) +  \mathcal{O}(\epsilon^{1/2}) \mathfrak{u}_{\bar{s}}^\dagger \mathfrak{W}_{\text{R} s'}(\Omega,\vec{K}_\perp).
\end{align}
\end{subequations}
Notice that
\begin{equation}\label{u_s_u_v_sim}
\mathfrak{u}_s^\dagger  \mathfrak{W}_{\text{R} s'}(\Omega,\vec{K}_\perp) \sim \mathfrak{v}_r^\dagger  \mathfrak{W}_{\text{R} r'}(\Omega,\vec{K}_\perp).
\end{equation}
Equations (\ref{W_tilde_W_tilde_quasiinertial_decomposition_2}) and (\ref{u_s_u_v_sim}) prove Eq.~(\ref{Dirac_representation_reduction_nonrelativistic_limit_general_Rindler}).

\subsection{Proof of Eq.~(\ref{Dirac_representation_reduction_nonrelativistic_limit_general_Rindler_high_a})}\label{Proof_of_Dirac_representation_reduction_nonrelativistic_limit_general_Rindler_high_a}

In the high acceleration regime (\ref{local_limit_alternative}), Eq.~(\ref{K_recurrence}) can be approximated by
\begin{equation}\label{K_recurrence_nonrelativistic_highacceleration}
 \left(\frac{\hbar}{m c}\partial_3 -  i \right) \mathfrak{K}(\Omega,\vec{K}_\perp, Z) = - \mathfrak{K} (- \Omega,\vec{K}_\perp, Z) + \mathcal{O}(\epsilon_\text{HA})
\end{equation}
and $\partial_3 \mathfrak{K}(\Omega,\vec{K}_\perp,Z)$ can be approximated by Eq.~(\ref{derivative_K_nonrelativistic}). This means that
\begin{equation}\label{derivative_K_nonrelativistic_highacceleration}
\frac{\hbar}{m c} \partial_3 \mathfrak{K}(\Omega,\vec{K}_\perp,Z) = \mathcal{O}(\epsilon_\text{HA}^{1/6}\epsilon^{1/3})
\end{equation}
and, hence,
\begin{equation}\label{K_recurrence_nonrelativistic_highacceleration_2}
\mathfrak{K}(\Omega,\vec{K}_\perp, Z) + i \mathfrak{K} (- \Omega,\vec{K}_\perp, Z) = \mathcal{O}(\epsilon_\text{HA}^{1/6}\epsilon^{1/3}).
\end{equation}

By plugging Eq.~(\ref{K_recurrence_nonrelativistic_highacceleration_2}) in Eq.~(\ref{W_tilde_W_tilde_quasiinertial_decomposition}), one obtains
\begin{subequations}\label{W_tilde_W_tilde_highacceleration_decomposition_2}
\begin{align}
\mathfrak{u}_s^\dagger \tilde{W}_{\text{R} s'}(\Omega,\vec{K}_\perp,Z_\text{R}(z)) = & \left[ 2 \mathfrak{K} (\Omega,\vec{K}_\perp,Z) + \mathcal{O}(\epsilon_\text{HA}^{1/6}\epsilon^{1/3}) \right] \mathfrak{u}_s^\dagger  \mathfrak{W}_{\text{R} s'}(\Omega,\vec{K}_\perp) \nonumber \\
& +  \mathcal{O}(\epsilon^{1/2}) \mathfrak{v}_{\bar{s}}^\dagger \mathfrak{W}_{\text{R} s'}(\Omega,\vec{K}_\perp), \\
\mathfrak{v}_s^\dagger \tilde{W}_{\text{R} s'}(\Omega,\vec{K}_\perp,Z_\text{R}(z)) = & \mathcal{O}(\epsilon_\text{HA}^{1/6}\epsilon^{1/3}) \mathfrak{v}_s^\dagger  \mathfrak{W}_{\text{R} s'}(\Omega,\vec{K}_\perp)  \nonumber \\
& +  \mathcal{O}(\epsilon^{1/2}) \mathfrak{u}_{\bar{s}}^\dagger \mathfrak{W}_{\text{R} s'}(\Omega,\vec{K}_\perp).
\end{align}
\end{subequations}
which, together with Eq.~(\ref{u_s_u_v_sim}) proves Eq.~(\ref{Dirac_representation_reduction_nonrelativistic_limit_general_Rindler_high_a}).

\section{Kontorovich–Lebedev transform}

The Bessel functions $K_{i \zeta}(\xi)$ can be used to define the Kontorovich–Lebedev transform as
\begin{equation}\label{KontorovichLebedev}
\mathcal{K} [\varphi] (\zeta) = \frac{2 \zeta}{\pi^2} \sinh(\pi \zeta) \int_0^\infty \frac{d\xi}{\xi}  K_{i \zeta}(\xi)  \varphi(\xi)
\end{equation}
for any function $\varphi(\xi)$ in $\xi>0$. The inverse of Eq.~(\ref{KontorovichLebedev}) is
\begin{equation}
\mathcal{K}^{-1} [\varphi] (\xi) = \int_0^\infty d\zeta  K_{i \zeta}(\xi)  \varphi(\zeta).
\end{equation}
The Kontorovich–Lebedev transform can be used to prove Eqs.~(\ref{alpha_tilde_approx_3}) and (\ref{alpha_tilde_approx_3_inverse}).

\subsection{Proof of Eq.~(\ref{alpha_tilde_approx_3})} \label{appendix_2}

The proof of Eq.~(\ref{alpha_tilde_approx_3}) follows from considering any function $\varphi(\xi)$ in $\xi>0$ and the following integral
\begin{align} \label{Kontorovich_Lebedev_0}
&\int_0^\infty dx \int_0^\infty d\theta_1  \frac{2 \theta_1}{\hbar a z}  \tilde{F}(\vec{\theta},Z_\text{R}(z)) \tilde{F}(\vec{\theta},Z) \varphi\left( \kappa(\vec{\theta}_\perp) z \right)\nonumber \\
 = & \frac{1}{2\pi^4 (c a)^2} \int_0^\infty dx \int_0^\infty d\theta_1  \frac{\theta_1}{z} \sinh \left( \frac{\beta \theta_1}{2} \right)   K_{i \theta_1 /ca} \left( \kappa(\vec{\theta}_\perp) z \right)  \nonumber \\
&  \times K_{i \theta_1 /ca} \left( \kappa(\vec{\theta}_\perp) \frac{e^{aX}}{a} \right)   \varphi\left( \kappa(\vec{\theta}_\perp) z \right).
\end{align}

By using the coordinate transformation
\begin{align}
& \zeta = \frac{\theta_1}{ca}, & \xi = \kappa(\vec{\theta}_\perp) z,
\end{align}
Eq.~(\ref{Kontorovich_Lebedev_0}) reads as
\begin{align} \label{Kontorovich_Lebedev_1}
&\int_0^\infty dx \int_0^\infty d\theta_1  \frac{2 \theta_1}{\hbar a z}  \tilde{F}(\vec{\theta},Z_\text{R}(z)) \tilde{F}(\vec{\theta},Z)  \varphi\left( \kappa(\vec{\theta}_\perp) z \right) \nonumber \\
=  &  \frac{1}{2 \pi^4} \int_0^\infty d\xi \int_0^\infty d\zeta \frac{\zeta}{ \xi} \sinh ( \pi \zeta )  K_{i \zeta} (\xi)  K_{i \zeta} \left( \kappa(\vec{\theta}_\perp) \frac{e^{aX}}{a} \right) \varphi (\xi).
\end{align}
Equation (\ref{Kontorovich_Lebedev_1}) can also be written in the following way
\begin{align}\label{Kontorovich_Lebedev_2}
&\int_0^\infty dx \int_0^\infty d\theta_1  \frac{2 \theta_1}{\hbar a z}  \tilde{F}(\vec{\theta},Z_\text{R}(z))  \tilde{F}(\vec{\theta},Z) \varphi\left( \kappa(\vec{\theta}_\perp) z \right) \nonumber \\
= & \frac{1}{4 \pi^2}  \mathcal{K}^{-1} [\mathcal{K} [\varphi]] \left( \kappa(\vec{\theta}_\perp) \frac{e^{aX}}{a} \right).
\end{align}
Since $\mathcal{K}^{-1}$ is the inverse of $\mathcal{K}$, Eq.~(\ref{Kontorovich_Lebedev_2}) reads as
\begin{equation}\label{Kontorovich_Lebedev_3}
\int_0^\infty dx \int_0^\infty d\theta_1  \frac{2 \theta_1}{\hbar a z}  \tilde{F}(\vec{\theta},Z_\text{R}(z)) \tilde{F}(\vec{\theta},Z)  \varphi\left( \kappa(\vec{\theta}_\perp) z \right) = \frac{1}{4 \pi^2}  \varphi \left( \kappa(\vec{\theta}_\perp) \frac{e^{aX}}{a} \right).
\end{equation}
Since Eq.~(\ref{Kontorovich_Lebedev_3}) holds for any $\varphi$, we have proven Eq.~(\ref{alpha_tilde_approx_3}).

\subsection{Proof of Eq.~(\ref{alpha_tilde_approx_3_inverse})} \label{Proof_of_alpha_tilde_approx_3_inverse}

The method of Sec.~\ref{appendix_2} can also be used to prove Eq.~(\ref{alpha_tilde_approx_3_inverse}). We consider any function $\varphi(\zeta)$ and we compute  
\begin{align} \label{Kontorovich_Lebedev_2_0}
&\int_0^\infty d\Omega \int_0^\infty \frac{dx  2 \Omega'}{\hbar a z} \sqrt{\frac{\sinh \left( \frac{\beta \Omega'}{2} \right)}{\sinh \left( \frac{\beta \Omega}{2} \right)}}  \tilde{F}(\Omega, \vec{K}_\perp,Z_\text{R}(z)) \tilde{F}(\Omega', \vec{K}_\perp,Z_\text{R}(z)) \varphi\left( \frac{\Omega}{ca} \right)\nonumber \\
 = & \frac{1}{2\pi^4 (c a)^2} \int_0^\infty d\Omega \int_0^\infty dx  \frac{\Omega'}{z} \sinh \left( \frac{\beta \Omega'}{2} \right)   K_{i \Omega /ca} \left( \kappa(\vec{K}_\perp) z \right)  \nonumber \\
&  \times K_{i \Omega' /ca} \left( \kappa(\vec{K}_\perp) z \right)   \varphi\left(  \frac{\Omega}{ca}\right).
\end{align}

The coordinate transformation
\begin{align}
& \zeta = \frac{\Omega}{ca}, & \xi = \kappa(\vec{K}_\perp) z
\end{align}
leads to
\begin{align} \label{Kontorovich_Lebedev_2_1}
& \int_0^\infty d\Omega \int_0^\infty \frac{dx  2 \Omega'}{\hbar a z} \sqrt{\frac{\sinh \left( \frac{\beta \Omega'}{2} \right)}{\sinh \left( \frac{\beta \Omega}{2} \right)}}  \tilde{F}(\Omega, \vec{K}_\perp,Z_\text{R}(z)) \tilde{F}(\Omega', \vec{K}_\perp,Z_\text{R}(z)) \varphi\left( \frac{\Omega}{ca} \right)\nonumber  \\
=  &  \frac{1}{2 \pi^4 ca}  \int_0^\infty d\xi \int_0^\infty d\zeta \frac{\Omega'}{ \xi} \sinh \left( \frac{\beta \Omega'}{2}  \right)  K_{i \zeta} (\xi)  K_{i \Omega'/ca} (\xi) \varphi (\zeta)\nonumber \\
= & \frac{1}{4 \pi^2}  \mathcal{K} [\mathcal{K}^{-1} [\varphi]] \left( \frac{\Omega'}{ca} \right)\nonumber \\
= & \frac{1}{4 \pi^2}  \varphi \left( \frac{\Omega'}{ca} \right),
\end{align}
which proves Eq.~(\ref{alpha_tilde_approx_3_inverse}).

\chapter{Two mode squeezed states}

\section{Massless scalar real field in 1+1 spacetime}\label{Two_mode_squeezed_state_Massless_scalar_field_in_11_spacetime}

We consider the regularized theory with discrete quantum numbers $K$ as elements of the grid with width $\Delta K$, i.e., $K = n \Delta K$ for some $n \in \mathbb{Z}$. The annihilator of a $\nu$-wedge particle with discrete momentum $K$ is $\hat{A}_{\nu K}$ and satisfies the discrete commutation relation
\begin{align}\label{commutation_descrete}
& [ \hat{A}_{\nu K}, \hat{A}_{\nu' K'}^\dagger ] = \delta_{\nu \nu'} \delta_{KK'}, & [ \hat{A}_{\nu K}, \hat{A}_{\nu' K'} ] = 0.
\end{align}

This is the familiar scenario of a particle in a box. The continuum theory is recovered by considering the limit $\Delta K \rightarrow 0$.

To discretize the continuum theory of Sec.~\ref{Massless_scalar_field_in_11_spacetime}, consider the substitution
\begin{align}\label{discrete_substitutions}
& \int_{\mathbb{R}} d K \mapsto \Delta K \sum_{K \in \mathbb{Z} \Delta K}, & \hat{A}_\nu(K) \mapsto \frac{\hat{A}_{\nu K}}{\sqrt{\Delta K}}.
\end{align}
By performing the substitution (\ref{discrete_substitutions}) in Eqs.~(\ref{SS_S}) and (\ref{D}), we obtain the discrete versions of the operators $\hat{s}_\text{S}$ and $\hat{S}_\text{S}$, i.e.,
\begin{subequations}
\begin{align} 
& \hat{s}_\text{S} = \sum_{K \in \mathbb{Z} \Delta K} \exp \left( -\frac{c \beta |K|}{2} \right) \hat{A}^\dagger_{\text{L}K} \hat{A}^\dagger_{\text{R}K}.\label{SS_S_discrete}\\
& \hat{S}_\text{S} = \exp \left( 2 \sum_{K \in \mathbb{Z} \Delta K} \zeta(c |K|) \left[ \hat{A}^\dagger_{\text{L}K} \hat{A}^\dagger_{\text{R}K}\right]^\text{A} \right),\label{D_discrete}
\end{align}
\end{subequations}
which lead to
\begin{subequations}
\begin{align} 
& \exp(\hat{s}_\text{S}) = \prod_{K \in \mathbb{Z} \Delta K} \exp \left( \exp \left( -\frac{c \beta |K|}{2} \right) \hat{A}^\dagger_{\text{L}K} \hat{A}^\dagger_{\text{R}K} \right).\label{SS_S_discrete_2}\\
& \hat{S}_\text{S} = \prod_{K \in \mathbb{Z} \Delta K}  \exp \left(  2 \zeta(c |K|) \left[ \hat{A}^\dagger_{\text{L}K} \hat{A}^\dagger_{\text{R}K}\right]^\text{A} \right).\label{D_discrete_2}
\end{align}
\end{subequations}

Now, we want to use the operator ordering theorem \cite{Barnett2002}, which states that for any set of operators $\hat{K}_3$ and $\hat{K}_\pm$ satisfying the commutation relations
\begin{align}\label{K_commutation}
& [\hat{K}_3, \hat{K}_\pm] = \pm \hat{K}_\pm, & [ \hat{K}_+, \hat{K}_- ] = - 2 \hat{K}_3
\end{align}
and for any $\zeta \in \mathbb{R}$ and $\varphi \in [0, 2 \pi)$, the following identity holds
\begin{align}\label{operator_ordering_theorem}
\exp ( \zeta (e^{i \varphi} \hat{K}_+ - e^{-i \varphi} \hat{K}_- ) ) = & \exp ( e^{i \varphi} \tanh(\zeta) \hat{K}_+) \exp ( - 2 \ln(\cosh(\zeta)) \hat{K}_3) \nonumber \\
& \exp ( - e^{-i \varphi} \tanh(\zeta) \hat{K}_-).
\end{align}

In our case, by choosing $\hat{K}_3 = (\hat{A}^\dagger_{\text{L}K} \hat{A}_{\text{L}K} + \hat{A}_{\text{R}K} \hat{A}^\dagger_{\text{R}K}) / 2$, $\hat{K}_+ = \hat{A}^\dagger_{\text{L}K} \hat{A}^\dagger_{\text{R}K}$, $\hat{K}_- = \hat{A}_{\text{L}K} \hat{A}_{\text{R}K}$, it is possible to check Eq.~(\ref{K_commutation}) by means of the commutation relations (\ref{commutation_descrete}). Hence, by choosing $\zeta = \zeta(c |K|)$ and $\varphi = 0$, Eq.~(\ref{operator_ordering_theorem}) gives
\begin{align}\label{operator_ordering_theorem_scalar_11}
&  \exp \left( 2 \zeta(c |K|) \left[ \hat{A}^\dagger_{\text{L}K} \hat{A}^\dagger_{\text{R}K}\right]^\text{A} \right) = \exp \left( \tanh(\zeta(c |K|)) \hat{A}^\dagger_{\text{L}K} \hat{A}^\dagger_{\text{R}K}\right) \nonumber \\
 & \times \exp \left( - \ln(\cosh(\zeta(c |K|))) \left( \hat{A}^\dagger_{\text{L}K} \hat{A}_{\text{L}K} + \hat{A}_{\text{R}K} \hat{A}^\dagger_{\text{R}K} \right) \right)\nonumber \\
 & \times  \exp \left( -\tanh(\zeta(c |K|)) \hat{A}_{\text{L}K} \hat{A}_{\text{R}K} \right).
\end{align}
By using Eq.~(\ref{zeta_omega}) in Eq.~(\ref{operator_ordering_theorem_scalar_11}), we obtain
\begin{align}\label{operator_ordering_theorem_scalar_11_2}
&  \exp \left( 2 \zeta(c |K|) \left[ \hat{A}^\dagger_{\text{L}K} \hat{A}^\dagger_{\text{R}K}\right]^\text{A} \right) = \exp \left( \exp \left( -\frac{c \beta |K|}{2} \right) \hat{A}^\dagger_{\text{L}K} \hat{A}^\dagger_{\text{R}K}\right) \nonumber \\
 & \times \exp \left( \frac{1}{2} \ln \left( 1 - e^{-c \beta |K|}  \right) \left( \hat{A}^\dagger_{\text{L}K} \hat{A}_{\text{L}K} + \hat{A}_{\text{R}K} \hat{A}^\dagger_{\text{R}K} \right) \right)\nonumber \\
 & \times  \exp \left( -\exp \left( -\frac{c \beta |K|}{2} \right)  \hat{A}_{\text{L}K} \hat{A}_{\text{R}K} \right).
\end{align}

Notice that the Rindler vacuum $| 0_\text{L}, 0_\text{R} \rangle$ is eigenstate of $ \hat{A}^\dagger_{\text{L}K} \hat{A}_{\text{L}K} + \hat{A}_{\text{R}K} \hat{A}^\dagger_{\text{R}K} $ with eigenvalue $1$, as a consequence of the commutation relation (\ref{commutation_descrete}) and that $ \hat{A}^\dagger_{\nu K}$ annihilates $| 0_\text{L}, 0_\text{R} \rangle$. Hence, by acting with Eq.~(\ref{operator_ordering_theorem_scalar_11_2}) on $| 0_\text{L}, 0_\text{R} \rangle$, we prove that
\begin{align}\label{operator_ordering_theorem_scalar_11_3}
&  \exp \left( 2 \zeta(c |K|) \left[ \hat{A}^\dagger_{\text{L}K} \hat{A}^\dagger_{\text{R}K}\right]^\text{A} \right) | 0_\text{L}, 0_\text{R} \rangle  \nonumber \\
=  & \exp \left( \frac{1}{2} \ln \left( 1 - e^{-c \beta |K|}  \right)  + \exp \left( -\frac{c \beta |K|}{2} \right) \hat{A}^\dagger_{\text{L}K} \hat{A}^\dagger_{\text{R}K}\right) | 0_\text{L}, 0_\text{R} \rangle.
\end{align}
By multiplying with respect to $K \in \mathbb{Z} \Delta K$, we obtain 
\begin{align}\label{operator_ordering_theorem_scalar_11_4}
& \prod_{K \in \mathbb{Z} \Delta K}  \exp \left( 2 \zeta(c |K|) \left[ \hat{A}^\dagger_{\text{L}K} \hat{A}^\dagger_{\text{R}K}\right]^\text{A} \right) | 0_\text{L}, 0_\text{R} \rangle  \nonumber \\
=  & \prod_{K \in \mathbb{Z} \Delta K} \exp \left( \frac{1}{2} \ln \left( 1 - e^{-c \beta |K|}  \right)  + \exp \left( -\frac{c \beta |K|}{2} \right) \hat{A}^\dagger_{\text{L}K} \hat{A}^\dagger_{\text{R}K}\right) | 0_\text{L}, 0_\text{R} \rangle
\end{align}
and, hence,
\begin{align}\label{operator_ordering_theorem_scalar_11_5}
& \exp \left( 2 \sum_{K \in \mathbb{Z} \Delta K}  \zeta(c |K|) \left[ \hat{A}^\dagger_{\text{L}K} \hat{A}^\dagger_{\text{R}K}\right]^\text{A} \right) | 0_\text{L}, 0_\text{R} \rangle  \nonumber \\
=  &  \exp \left( \frac{1}{2} \sum_{K \in \mathbb{Z} \Delta K}  \ln \left( 1 - e^{-c \beta |K|}  \right)  + \sum_{K \in \mathbb{Z} \Delta K}   \exp \left( -\frac{c \beta |K|}{2} \right) \hat{A}^\dagger_{\text{L}K} \hat{A}^\dagger_{\text{R}K}\right) | 0_\text{L}, 0_\text{R} \rangle.
\end{align}

Now, notice that the left hand side of Eq.~(\ref{operator_ordering_theorem_scalar_11_5}) is exactly $\hat{S}_\text{S} | 0_\text{L}, 0_\text{R} \rangle $ in the discrete theory, whereas the right hand side is proportional to $\exp(\hat{s}_\text{S}) | 0_\text{L}, 0_\text{R} \rangle$. In other words, we proved that $\exp(\hat{s}_\text{S}) | 0_\text{L}, 0_\text{R} \rangle$ is proportional to $\hat{S}_\text{S} | 0_\text{L}, 0_\text{R} \rangle $ in the discrete theory, i.e.,
\begin{equation}\label{S_S_N_SS_S}
\hat{S}_\text{S} | 0_\text{L}, 0_\text{R} \rangle = \mathcal{N} \exp(\hat{s}_\text{S}) | 0_\text{L}, 0_\text{R} \rangle,
\end{equation}
with
\begin{equation}\label{N_vacuum}
\mathcal{N} = \exp \left( \frac{1}{2} \sum_{K \in \mathbb{Z} \Delta K}  \ln \left( 1 - e^{-c \beta |K|}  \right) \right) 
\end{equation}
as normalization factor for $\exp(\hat{s}_\text{S}) | 0_\text{L}, 0_\text{R} \rangle$.

Notice that in the continuum limit, $\mathcal{N} \rightarrow 0$. Indeed, by inverting the substitution (\ref{discrete_substitutions}) and by considering the IR cutoff $\Delta K$ in Eq.~(\ref{N_vacuum}), we obtain
\begin{equation}\label{N_vacuum_continuum}
\mathcal{N} \approx \exp \left( \frac{1}{2 \Delta K} \int_{|K|>\Delta K} d K  \ln \left( 1 - e^{-c \beta |K|}  \right) \right).
\end{equation}
The negative divergence of exponential argument in Eq.~(\ref{N_vacuum_continuum}) is due to the limit of the factor $1/\Delta K \rightarrow \infty$ and the negative IR divergence of the integral, which is of the order of $-1/\Delta K$. Hence, $\mathcal{N}$ goes to zero as fast as $\mathcal{N} \sim \exp(- 1 / \Delta K^2)$. Consequently, the vector $\exp(\hat{s}_\text{S}) | 0_\text{L}, 0_\text{R} \rangle$ is not normalizable in the continuum limit, unless one considers Eq.~(\ref{S_S_N_SS_S}) \textit{before} taking the limit $\Delta K \rightarrow 0$ and use the fact that $\hat{S}_\text{S} | 0_\text{L}, 0_\text{R} \rangle$ is already normalized in the continuum theory.

\section{Scalar field}\label{Two_mode_squeezed_state_Frame_dependent_content_of_particles_scalar}

Similar results can be provided for massive scalar complex fields in 3+1 dimensions.

At variance with massless scalar real fields in 1+1 dimensions, here we consider the quantum numbers $\vec{\theta} = (\Omega, \vec{K}_\perp)$ labeling Rindler modes in 3+1 dimensions. Hence, we define the grid of discrete $\vec{\theta}$ variables separated by $\Delta \vec{\theta} = (\Delta \Omega, \Delta K_1, \Delta K_2)$. Elements of the grid $\vec{\theta}$ are such that $\vec{\theta} = \vec{n} \circ \Delta \vec{\theta}  = (n_1 \Delta \Omega, n_1 \Delta K_1, n_1 \Delta K_2)$ for some $\vec{n} \in \mathbb{Z}^3$ and where $\circ$ is the Hadamard product. The continuum limit consists of $\Delta \Omega \rightarrow 0, \Delta K_1 \rightarrow 0, \Delta K_2 \rightarrow 0$.

We also consider the discrete annihilation operators $\hat{A}_{\nu \vec{\Theta}}$ and $\hat{B}_{\nu \vec{\theta}}$ satisfying the commutation rule
\begin{align}\label{commutation_descrete_31}
& [ \hat{A}_{\nu \vec{\theta}}, \hat{A}_{\nu' \vec{\theta}'}^\dagger ] = [ \hat{B}_{\nu \vec{\theta}}, \hat{B}_{\nu' \vec{\theta}'}^\dagger ] = \delta_{\nu \nu'} \delta_{\vec{\theta}\vec{\theta}'}, \\
 & [ \hat{A}_{\nu \vec{\theta}}, \hat{A}_{\nu' \vec{\theta}'} ] = [ \hat{B}_{\nu \vec{\theta}}, \hat{B}_{\nu' \vec{\theta}'} ] = [ \hat{A}_{\nu \vec{\theta}}, \hat{B}_{\nu' \vec{\theta}'} ] = [ \hat{A}_{\nu \vec{\theta}}, \hat{B}_{\nu' \vec{\theta}'}^\dagger ] = 0.
\end{align}

The discretization of the scalar theory is implemented by the substitution
\begin{align}\label{discrete_substitutions_31}
& \int_{\theta_1 > 0} d^3\theta \mapsto \Delta^3 \theta \sum_{\vec{\theta} \in \mathbb{Z}^3 \circ \Delta \vec{\theta}}, && \hat{A}_\nu(\vec{\theta}) \mapsto \frac{\hat{A}_{\nu \vec{\theta}}}{\sqrt{\Delta^3 \theta}}, && \hat{B}_\nu(\vec{\theta}) \mapsto \frac{\hat{B}_{\nu \vec{\theta}}}{\sqrt{\Delta^3 \theta}},
\end{align} 
with $\Delta^3 \theta = \Delta \Omega \Delta K_1 \Delta K_2$, in analogy to Eq.~(\ref{discrete_substitutions}).

In this framework, it is possible to prove that the state $\hat{S}_\text{S} | 0_\text{L}, 0_\text{R} \rangle$ with $\hat{S}_\text{S}$ given by Eq.~\ref{D} is the normalized version of $\exp(\hat{s}_\text{S}) | 0_\text{L}, 0_\text{R} \rangle$ with $\hat{s}_\text{S}$ given by Eq.~(\ref{SS_S}). The discretization of the space of the variables $\vec{\theta} = (\Omega, \vec{K}_\perp)$ serves as a regularization procedure to define $\exp(\hat{s}_\text{S}) | 0_\text{L}, 0_\text{R} \rangle$ as a normalizable state.

The proof for the equivalent of Eq.~(\ref{S_S_N_SS_S}) in the 3+1 massive complex case can be obtained in the same fashion as in Sec.~\ref{Two_mode_squeezed_state_Massless_scalar_field_in_11_spacetime}. The only difference is given by the presence of more degrees of freedoms including transverse momentum $\vec{K}_\perp$ and charge number encoded in the separation between particle and antiparticle operators, whereas the massless energy $c|K|$ is replaced by $\Omega$. In summary, one uses $\hat{K}_3 = (\hat{A}^\dagger_{\text{L}\vec{\theta}} \hat{A}_{\text{L}\vec{\theta}} + \hat{A}_{\text{R}\vec{\theta}} \hat{A}^\dagger_{\text{R}\vec{\theta}} + \hat{B}^\dagger_{\text{L}\vec{\theta}} \hat{B}_{\text{L}\vec{\theta}} + \hat{B}_{\text{R}\vec{\theta}} \hat{B}^\dagger_{\text{R}\vec{\theta}}) / 2$, $\hat{K}_+ = \hat{A}^\dagger_{\text{L}\vec{\theta}} \hat{A}^\dagger_{\text{R}\vec{\theta}} + \hat{B}^\dagger_{\text{L}\vec{\theta}} \hat{B}^\dagger_{\text{R}\vec{\theta}}$, $\hat{K}_- = \hat{A}_{\text{L}\vec{\theta}} \hat{A}_{\text{R}\vec{\theta}} + \hat{B}_{\text{L}\vec{\theta}} \hat{B}_{\text{R}\vec{\theta}}$, $\zeta = \zeta(\Omega)$ and $\varphi = 0$; all sums and products are with respect to the quantum numbers $\vec{\theta} = (\Omega, \vec{K}_\perp)$.

\section{Dirac field}\label{Two_mode_squeezed_state_Frame_dependent_content_of_particles_Dirac}

Here, we consider Dirac fields and we prove that by normalizing the vector $\exp (\hat{s}_\text{D} ) | 0_\text{L}, 0_\text{R} \rangle$ with $\hat{s}_\text{D} $ given by Eq.~(\ref{O}), one obtains $\hat{S}_\text{D} | 0_\text{L}, 0_\text{R} \rangle$ with $\hat{S}_\text{D} $ given by Eq.~(\ref{OO}). The proof requires a discretization of the theory as a regulator for the normalization of $\exp (\hat{s}_\text{D} ) | 0_\text{L}, 0_\text{R} \rangle$. 

At variance with the scalar theory, here we need spinorial degrees of freedom as well. Luckily, the spin variable $s$ is already discrete, hence the discretization of the variables $\Omega$ and $\vec{K}_\perp$ is carried out in the same way as in Sec.~\ref{Two_mode_squeezed_state_Frame_dependent_content_of_particles_scalar}.

We consider the annihilator operators defined by Eq.~(\ref{E_CD}). Their discrete counterparts are indicated by $\hat{E}_{\nu\theta}$, where $\theta = (s, \Omega, \vec{K}_\perp)$ is a collection of discrete variables. The anticommutation relations are
\begin{align}\label{anticommutation_descrete}
& \{ \hat{E}_{\nu \theta}, \hat{E}_{\nu' \theta'}^\dagger \} = \delta_{\nu \nu'} \delta_{\theta \theta'}, & \{ \hat{E}_{\nu \theta}, \hat{E}_{\nu' \theta'} \} = 0.
\end{align}
This is a clear difference with the scalar case and must be taken into account in the proof.

The decomposition of the exponential operators $\exp(\hat{s}_\text{D})$ and $\hat{S}_\text{D}$ into a product of exponents is guaranteed by the fact that couples of anticommuting fermionic operators commute. Hence, in the discrete case, Eqs.~(\ref{O_4}) and (\ref{OO_2}) give
\begin{subequations}
\begin{align}
& \exp( \hat{s}_\text{D} ) = \prod_\theta \exp \left( f(\theta) \hat{F}^\dagger_\theta \right), \\
& \hat{S}_\text{D} = \prod_\theta \exp \left( 2  [ g(\theta) \hat{F}_\theta ]^\text{A} \right),
\end{align}
\end{subequations}
with $\hat{F}_\theta = \hat{E}_{\text{L} \theta}^\dagger \hat{E}_{\text{R} \theta}^\dagger$ and where the product is evaluated along the entire grid for $\theta$. This is a consequence of the commutation relation $[ \hat{F}_\theta, \hat{F}_{\theta'}] = 0$ which is the discrete version of Eq.~(\ref{commutating_rules_F_a}) and can be proved from the anticommutation relations (\ref{anticommutation_descrete}).

We define $\hat{K}_3 = (\hat{E}^\dagger_{\text{L}\theta} \hat{E}_{\text{L}\theta} - \hat{E}_{\text{R}\theta} \hat{E}^\dagger_{\text{R}\theta}) / 2$, $\hat{K}_+ = \hat{E}^\dagger_{\text{L}\theta} \hat{E}^\dagger_{\text{R}\theta}$, $\hat{K}_- =  \hat{E}_{\text{L}\theta}\hat{E}_{\text{R}\theta}$. Also, we choose $\zeta = \zeta(\Omega)$ and $\varphi = \pi$. One can prove that the commutation relations (\ref{K_commutation}) are satisfied by means of the anticommutation relations (\ref{anticommutation_descrete}) and by knowing that $\hat{E}_{\nu \theta}\hat{E}_{\nu \theta} = [\hat{E}_{\nu \theta}, \hat{E}_{\nu \theta}]/2 + \{ \hat{E}_{\nu \theta}, \hat{E}_{\nu \theta} \}/2 = 0$. The proof is detailed by
\begin{subequations}
\begin{align}
2 \hat{K}_3 \hat{K}_+ = & \hat{E}^\dagger_{\text{L}\theta} \hat{E}_{\text{L}\theta} \hat{E}^\dagger_{\text{L}\theta} \hat{E}^\dagger_{\text{R}\theta} - \hat{E}_{\text{R}\theta} \hat{E}^\dagger_{\text{R}\theta} \hat{E}^\dagger_{\text{L}\theta} \hat{E}^\dagger_{\text{R}\theta} \nonumber \\
= & \hat{E}^\dagger_{\text{L}\theta} \hat{E}^\dagger_{\text{R}\theta} - \hat{E}^\dagger_{\text{L}\theta}  \hat{E}^\dagger_{\text{L}\theta}\hat{E}_{\text{L}\theta} \hat{E}^\dagger_{\text{R}\theta} + \hat{E}_{\text{R}\theta} \hat{E}^\dagger_{\text{L}\theta} \hat{E}^\dagger_{\text{R}\theta} \hat{E}^\dagger_{\text{R}\theta} \nonumber \\
= & \hat{E}^\dagger_{\text{L}\theta} \hat{E}^\dagger_{\text{R}\theta} \nonumber \\
= & \hat{K}_+  , \\
2 \hat{K}_+ \hat{K}_3 = & \hat{E}^\dagger_{\text{L}\theta} \hat{E}^\dagger_{\text{R}\theta} \hat{E}^\dagger_{\text{L}\theta} \hat{E}_{\text{L}\theta} - \hat{E}^\dagger_{\text{L}\theta} \hat{E}^\dagger_{\text{R}\theta} \hat{E}_{\text{R}\theta} \hat{E}^\dagger_{\text{R}\theta} \nonumber \\
= & - \hat{E}^\dagger_{\text{L}\theta}  \hat{E}^\dagger_{\text{L}\theta} \hat{E}^\dagger_{\text{R}\theta} \hat{E}_{\text{L}\theta} +   \hat{E}^\dagger_{\text{R}\theta}\hat{E}^\dagger_{\text{L}\theta} + \hat{E}^\dagger_{\text{L}\theta} \hat{E}_{\text{R}\theta} \hat{E}^\dagger_{\text{R}\theta} \hat{E}^\dagger_{\text{R}\theta} \nonumber \\
= & -\hat{E}^\dagger_{\text{L}\theta} \hat{E}^\dagger_{\text{R}\theta} \nonumber \\
= & -\hat{K}_+ ,  \\
2 \hat{K}_3 \hat{K}_- = & \hat{E}^\dagger_{\text{L}\theta} \hat{E}_{\text{L}\theta} \hat{E}_{\text{L}\theta}\hat{E}_{\text{R}\theta} - \hat{E}_{\text{R}\theta} \hat{E}^\dagger_{\text{R}\theta} \hat{E}_{\text{L}\theta}\hat{E}_{\text{R}\theta}\nonumber \\
= & \hat{E}_{\text{R}\theta} \hat{E}^\dagger_{\text{R}\theta} \hat{E}_{\text{R}\theta}\hat{E}_{\text{L}\theta}\nonumber \\
= & \hat{E}_{\text{R}\theta} \hat{E}_{\text{L}\theta} -  \hat{E}_{\text{R}\theta} \hat{E}_{\text{R}\theta}\hat{E}^\dagger_{\text{R}\theta} \hat{E}_{\text{L}\theta}\nonumber \\
= &- \hat{E}_{\text{L}\theta} \hat{E}_{\text{R}\theta}\nonumber \\
= & -\hat{K}_-,\\
2 \hat{K}_- \hat{K}_3 = &\hat{E}_{\text{L}\theta}\hat{E}_{\text{R}\theta}  \hat{E}^\dagger_{\text{L}\theta} \hat{E}_{\text{L}\theta} -  \hat{E}_{\text{L}\theta}\hat{E}_{\text{R}\theta}\hat{E}_{\text{R}\theta} \hat{E}^\dagger_{\text{R}\theta}\nonumber \\
= & \hat{E}_{\text{L}\theta}\hat{E}_{\text{R}\theta} - \hat{E}_{\text{L}\theta}\hat{E}_{\text{R}\theta} \hat{E}_{\text{L}\theta} \hat{E}^\dagger_{\text{L}\theta}\nonumber \\
= & \hat{E}_{\text{L}\theta}\hat{E}_{\text{R}\theta} + \hat{E}_{\text{L}\theta}\hat{E}_{\text{L}\theta}\hat{E}_{\text{R}\theta}  \hat{E}^\dagger_{\text{L}\theta}\nonumber \\
= & \hat{E}_{\text{L}\theta}\hat{E}_{\text{R}\theta} \nonumber \\
= & \hat{K}_- , \\
\hat{K}_+ \hat{K}_- = & \hat{E}^\dagger_{\text{L}\theta} \hat{E}^\dagger_{\text{R}\theta} \hat{E}_{\text{L}\theta}\hat{E}_{\text{R}\theta} \nonumber \\
 = & -\hat{E}^\dagger_{\text{L}\theta} \hat{E}_{\text{L}\theta} \hat{E}^\dagger_{\text{R}\theta}\hat{E}_{\text{R}\theta}  \nonumber \\
 = & - \hat{E}^\dagger_{\text{R}\theta}\hat{E}_{\text{R}\theta} +  \hat{E}_{\text{L}\theta} \hat{E}^\dagger_{\text{L}\theta} \hat{E}^\dagger_{\text{R}\theta}\hat{E}_{\text{R}\theta}  \nonumber \\
 = & - \hat{E}^\dagger_{\text{R}\theta}\hat{E}_{\text{R}\theta} +  \hat{E}_{\text{L}\theta} \hat{E}^\dagger_{\text{L}\theta} -  \hat{E}_{\text{L}\theta} \hat{E}^\dagger_{\text{L}\theta}\hat{E}_{\text{R}\theta} \hat{E}^\dagger_{\text{R}\theta}  \nonumber \\
 = & \hat{E}_{\text{R}\theta} \hat{E}^\dagger_{\text{R}\theta} -  \hat{E}^\dagger_{\text{L}\theta} \hat{E}_{\text{L}\theta} +  \hat{E}_{\text{L}\theta} \hat{E}_{\text{R}\theta} \hat{E}^\dagger_{\text{L}\theta} \hat{E}^\dagger_{\text{R}\theta}  \nonumber \\
 = & -2 \hat{K}_3 +  \hat{K}_- \hat{K}_+.
\end{align}
\end{subequations}

Now, by following the same arguments of the scalar case, one can prove that $\hat{S}_\text{D} | 0_\text{L}, 0_\text{R} \rangle$ with $\hat{S}_\text{D}$ given by Eq.~(\ref{OO_2}) is the normalized version of $\exp(\hat{s}_\text{D}) | 0_\text{L}, 0_\text{R} \rangle$ with $\hat{s}_\text{D}$ given by Eq.~(\ref{O_4}).

\bibliographystyleRF{ieeetr}
\bibliographyRF{bibliography}

\bibliographystyle{ieeetr}
\bibliography{bibliography}

\end{document}